Dominik Hangleiter

# Sampling and the complexity of nature

## Assessing, testing, and challenging the computational power of quantum devices

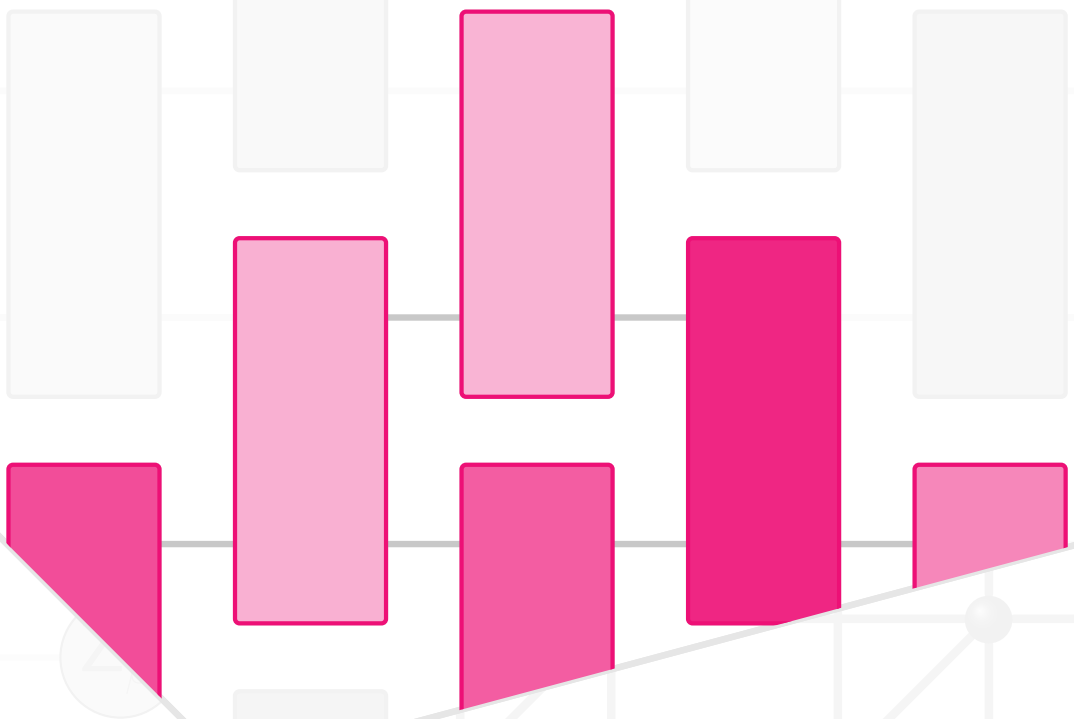

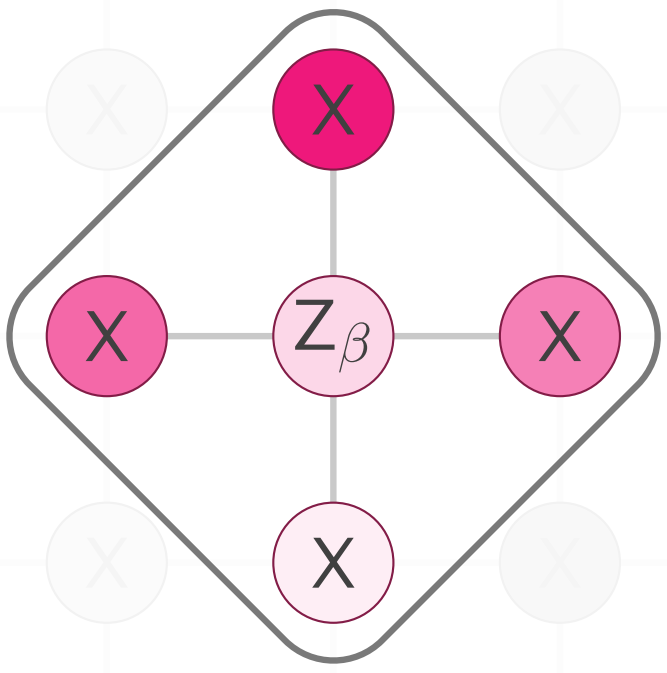

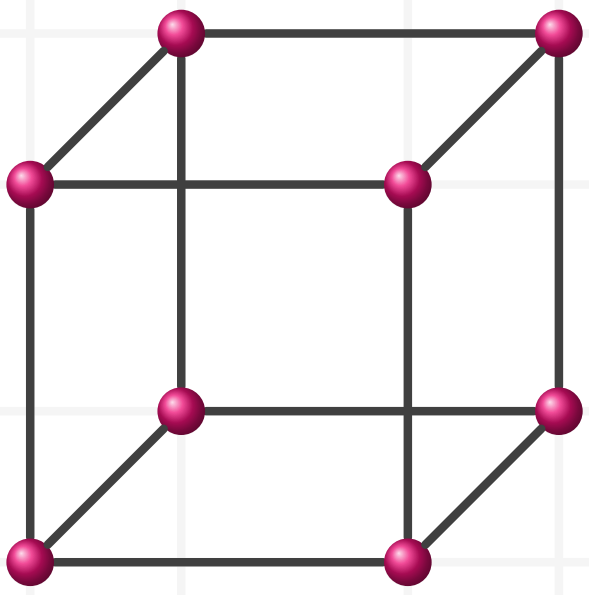

Im Fachbereich Physik der Freien Universität Berlin eingereichte Dissertation zur Erlangung des Grades eines Doktors der Naturwissenschaften.

Berlin, im Mai 2020

**Gutachter:innen**
Prof. Dr. Jens Eisert
Prof. Dr. Christiane Koch

**Datum der Disputation**
24. November 2020

# Preface

In this thesis, I present and put into context some of the results of my research in the past four years.

Fortunately, research is not anymore the lone art that it used to be in the old days and we are sorely reminded of this in the days of isolation. I have had the pleasure to work together, to discuss, to spend time with colleagues and friends that made this time a joy and enrichment that I would not want to miss. As such, this thesis should not be regarded as the work of a single author. Rather, it is a scriptum of joint publications as well as thoughts that did not make it into a publication, inspired by and building on numerous scientific discussions, conference talks, and coffee-break chats. I try to pay reverence to those facets of research in the margins, highlighting passages that have already been published, and acknowledging discussions that inspired, shaped, or clarified a thought.

Only in hindsight did I realize that sampling and the quantum sign problem is a unifying theme of much of my work in the past years. Not only does the sign problem explain the origin of complexity in quantum random sampling experiments the verification and physicalization of which I studied in the first part of my thesis, its basis dependence is also the main theme of my work on Quantum Monte Carlo algorithms. I like to think of the quantum sign problem as delineating a computational boundary between quantum and classical sampling algorithms and in this thesis I hope to convey some of the many aspects of this theme to you.

To achieve this, in this thesis, I will not only present the results but also provide an accessible introduction into the general mindset and the methods used in the study of classical and quantum sampling. With this, I hope to not only transport the scientific content but also to take you on a journey led by the fascination that the puzzles of quantum computing exert.

*Dominik Hangleiter*
*Berlin, May 2020*

# Author publications

## Peer-reviewed publications

## Preprints

# Contents

# Introduction | 1

Quantum mechanics is an intriguing theory. It predicts phenomena lying beyond what we can comprehend or could have imagined before its inception. Particles can tunnel through barriers. Particles behave like waves and waves like particles. While most quantum phenomena surface on the microscopic level, quantum phenomena can also reach macroscopic scale. For example, at extremely low temperatures, many particles can condense into a single macroscopic quantum state – a Bose-Einstein condensate [And+95; Dav+95; Bra+95]. Likewise, there are currents and liquids that flow infinitely without resistance [BCS57] as well as the suppression of such currents by disorder only [BAA06; Imb16; Sch+15; Cho+16; Aba+19].

The properties and features of quantum mechanics are counterintuitive and challlenge reseachers still today – a century after its inception. Among them is *entanglement* between particles, a feature that Einstein famously ridiculed as "spooky action at a distance", as well as the famous *measurement problem*, which arises due to the fact that the predictions of quantum theory depend intrinsically on how a quantum measurement is made. Experiments only have outcomes if measured and, what is more, those outcomes are not deterministic as they used to be in the days of classical mechanics. Those features can be exploited in protocols that achieve tasks fit for science fiction movies such as the teleportation of a quantum state across large distances [NC00; Pan+01; Urs+04].

But quantum mechanics is not only a theory, a set of mathematical equations. It also ranges among the best confirmed theories that are available to us in order to describe nature. The validity of quantum theory has been confirmed in a plethora of instances [Aas+13; HOM87; And+95], across wide regimes of applicability [Rie+18], and with unprecedented precision [Lud+15]. Undamped currents are not a fantasy in a theorist's ideal world. Indeed, superconductors can be controlled to an enormous degree of precision in laboratories around the world [Rou+17; Aru+19]. More than 70 years after its discovery in theory [Bos24; Ein25], Bose-Einstein condensation was confirmed [And+95; Dav+95; Bra+95], and its replication has become a standard experimental procedure today [Gre+02]. By exploiting the phenomena and features of quantum mechanics one can build highly precise measurement apparata such as detectors of spatial oscillations via squeezed light used in gravitational-wave detection [Aas+13] and high-precision clocks [UHH02; Lud+15], making the confirmation of quantum mechanics happen in a frequency measurement every trillionth of a second, with an accuracy of $\sim 10^{-14}$ Hz.

Among the most recent breakthroughs, confirming the validity of quantum mechanics as a description of nature once again, are the recent Bell tests [Hen+15; Sha+15] which close the last loopholes that remained after the initial Bell violation by Freedman and Clauser [FC72] and the experiments of Aspect *et al.* [ADR82; AGR82] that closed the locality loophole. Those loopholes gave room for a sceptic to doubt that there is no local and realistic description of nature, as famously discovered by Einstein, Podolsky, and Rosen [EPR35] and converted to a testable criterion by Bell [Bel64].

Today, saying that some phenomenon is predicted by quantum mechanics is therefore nearly tantamount to saying that nature exhibits this phenomenon.

But in which situations can we actually make such predictions? Vice versa:

**What is the complexity of nature?**

In this thesis, I take on a pragmatic perspective and investigate the predictability of nature via quantum theory computationally. I will study the complexity of simulating quantum systems using classical computing resources.

## 1.1 A computing perspective on nature

One of the phenomena predicted by quantum theory was first witnessed in the famous Hong-Ou-Mandel experiment [HOM87]: *photon bunching*. This phenomenon will allow me to phrase the questions I ask in this thesis in concrete terms and provides physical context to computational ideas. It will serve as an example, a guide and a source of intuition for the concepts involved.

So let me start out with a description of the Hong-Ou-Mandel experiment.

### The Hong-Ou-Mandel experiment

In their famous experiment, Hong, Ou, and Mandel [HOM87] produced coherent pairs of individual photons via so-called *parametric down conversion* by shining a laser beam onto a nonlinear crystal. The photon pairs then interfered on a $50 : 50$ beam splitter as shown in Fig. 1.1, and the output ports of the beam splitter were measured using photomultipliers. In particular, Hong, Ou, and Mandel [HOM87] observed the fraction of so-called coincidence counts, that is, the fraction of experiments in which on each of the two output ports a photon was detected.

Let us now work through what happens in the experiment. Each incoming photon can either be reflected or transmitted through the beam splitter so that all together there are four possible outcome states (see Fig. 1.1)

i. The first photon is reflected, the second is transmitted,
ii. both photons are transmitted,
iii. both photons are reflected,
iv. the first photon is transmitted, the second is reflected.

If the photons are distinguishable, all of these cases happen with equal probability given by $1/4$ as the photons do not interact. For example, this is the case if the photons have different frequencies or if there is a time delay between their arrival at the beam splitter.

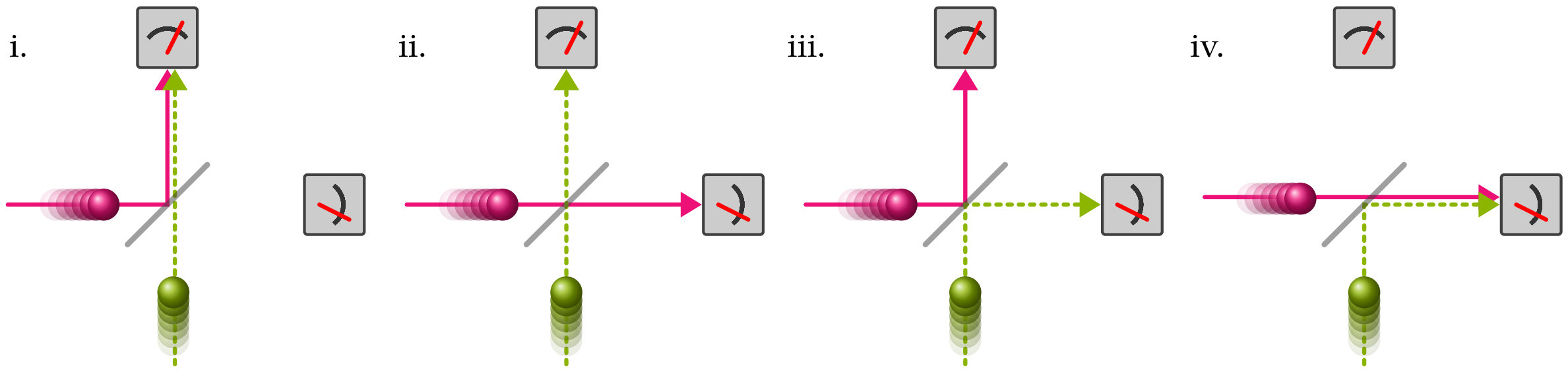

**Figure 1.1:** When two photons incedent meet on a beam splitter they can take four possible paths (i.–iv.). Each of these possibilities is equally probable whenever the photons can be distinguished, for instance, by their frequency or the time of their incidence on the beam splitter, represented here by the different color and position of the incoming particles.

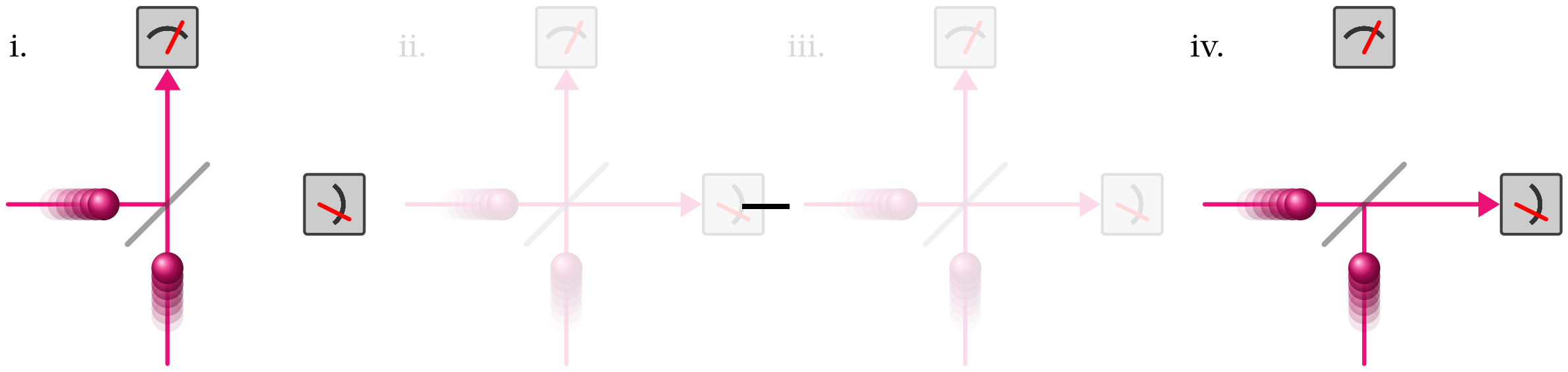

**Figure 1.2:** If the photons are indistinguishable, the two paths ending up in the state $|1,1\rangle$ interfere destructively so that the probability of a coincidence measurement vanishes.

If the photons are indistinguishable they will interfere. For indistinguishable photons, the experiment can be described in terms of two optical modes with annihilation (creation) operators $a^{(\dagger)}$ and $b^{(\dagger)}$ corresponding to the two input arms of the interferometer. Those input modes are transformed at the beam splitter into output modes (corresponding to the output arms of the interferometer) according to the unitary transformation

$$\begin{pmatrix} a^\dagger \\ b^\dagger \end{pmatrix} \mapsto \frac{1}{\sqrt{2}} \begin{pmatrix} 1 & 1 \\ 1 & -1 \end{pmatrix} \begin{pmatrix} a^\dagger \\ b^\dagger \end{pmatrix} = \frac{1}{\sqrt{2}} \begin{pmatrix} a^\dagger + b^\dagger \\ a^\dagger - b^\dagger \end{pmatrix}, \tag{1.1}$$

which mixes the input modes with probability 1/2. An initial state $|1,1\rangle = a^\dagger b^\dagger |0,0\rangle$, where $|0,0\rangle$ denotes the vaccuum state of the two modes, thus transforms as

$$a^\dagger b^\dagger |0,0\rangle \mapsto \frac{1}{2}(a^\dagger + b^\dagger)(a^\dagger - b^\dagger)|0,0\rangle \tag{1.2}$$

$$= \frac{1}{2}(a^\dagger a^\dagger + b^\dagger a^\dagger - a^\dagger b^\dagger - b^\dagger b^\dagger)|0,0\rangle \tag{1.3}$$

$$\overset{[a^\dagger, b^\dagger]=0}{=} \frac{1}{2}(a^\dagger a^\dagger - b^\dagger b^\dagger)|0,0\rangle = \frac{1}{\sqrt{2}}\left(|2,0\rangle - |0,2\rangle\right). \tag{1.4}$$

This calculation incorporates indistinguishability of the photons in the guise of the canonical commutation relations of $a^\dagger$ and $b^\dagger$.

Notice that one of the paths with output state $|1,1\rangle$, namely the one in which both photons are reflected, acquires a $(-1)$-phase factor. This phase factor may be interpreted mathematically as ensuring unitarity of the beam-splitter transformation or physically in terms of the fact that reflection at a surface with higher refractive index causes a $\pi$ phase shift. This results in *destructive interference* of the possibilities ii. (both photons are transmitted) and iii. (both photons are reflected). As a result only two out of the four possibility are actually realized when the two photons are indistinguishable, namely those in which the photons resurface in the same mode (see Fig. 1.2). The Hong-Ou-Mandel phenomenon is therefore also called *photon bunching*.

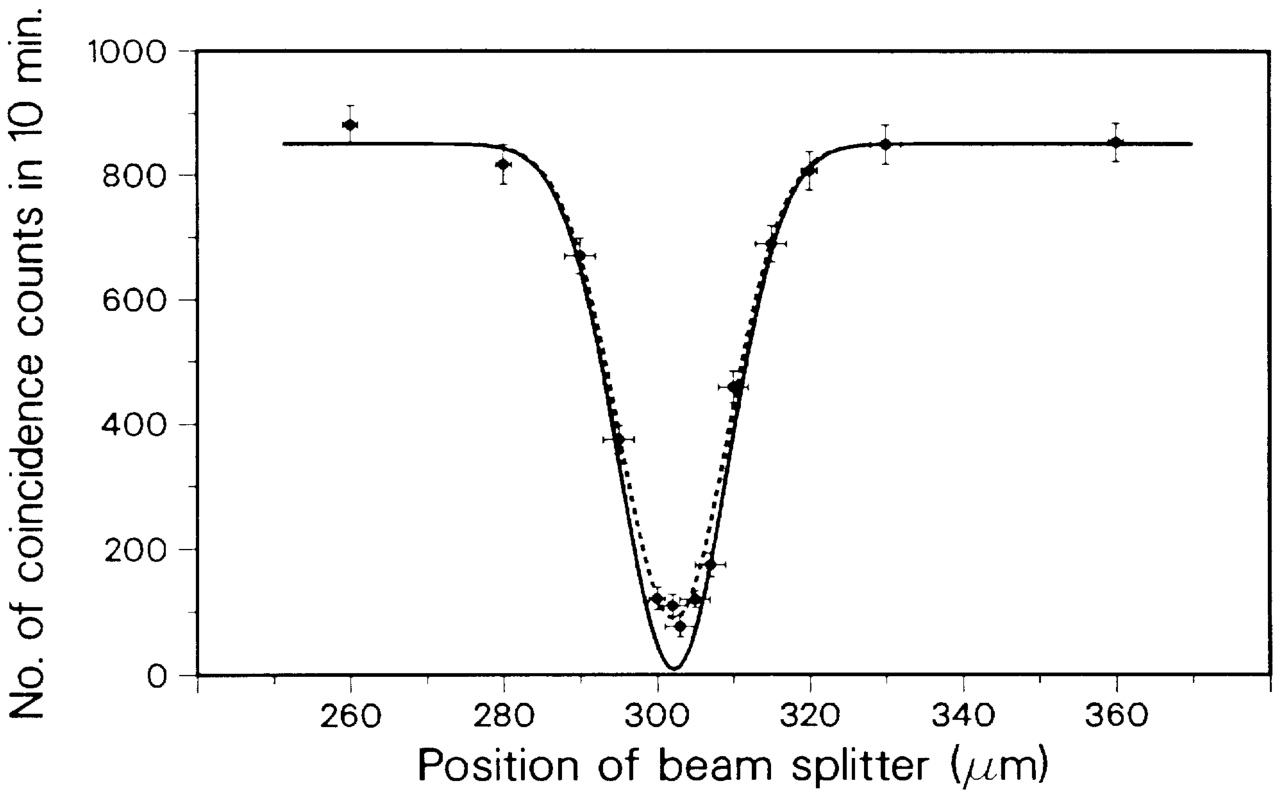

**Figure 1.3:** In their experiment Hong, Ou, and Mandel [HOM87] measured the coincidence counts (measurement outcome $(1,1)$) of the two photodetectors as a function of the position of the beam splitter, which changes the relative delay between the incidence of the photons on the beam splitter. At relative delay 0 the photons are indistinguishable; as the beam splitter position is moved away in either direction they become gradually more distinguishable, marking the transition to the regime in which the photons can interfere. Figure reproduced from [HOM87, Fig. 2].

Hong, Ou, and Mandel [HOM87] observed photon bunching in their experiment as they measured the probability of obtaining the $(1,1)$ coincidence measurement outcome on the photodetectors. For perfectly indistinguishably photons, this probability should be zero as we saw above. In their experiment Hong, Ou, and Mandel manipulated the degree of distinguishability between the photons. They did so by changing the position of the beam splitter, thereby introducing a relative delay between the incidence of the photons on the beam splitter. As the beam splitter position was varied, they observed the dip in the fraction coincidence counts that is displayed in Fig. 1.3, confirming quantum theory in yet another instance.

## The complexity of physics: simulating quantum experiments

The Hong-Ou-Mandel experiment highlights a key feature of quantum experiments: in contrast to classical physics, quantum experiments are intrinsically *random*. In quantum theory any measurement amounts to sampling from a probability distribution with probabilities determined by the Born rule. In the case of measuring photon counts after letting photons interfere in an interferometer, the possible measurement outcomes are all states with a fixed photon number. In the Hong-Ou-Mandel experiment those are the possible outcomes of the two photon counting measurements as a pair of photons is injected into the interferometer – $(2,0), (1,1)$ and $(0,2)$. The Born rule assigns a probability to measuring each outcome. Simulating the experiment then amounts to sampling from this probability distribution.

It is such sampling experiments that lie at the heart of this thesis. As we will see, understanding quantum sampling experiments from a computing perspective will allow us to shed some light on the intrinsic complexity of nature.

**Sampling experiments**

The focus of this thesis will be the question, under which conditions *sampling experiments* such as the Hong-Ou-Mandel experiment can be simulated on a classical computer.

As the photons interfere destructively, that is, as the canonical commutation relations between the photon creation and annihilation operators become relevant, the outcome probability distribution acquires an intricate structure. When letting two photons interfere as in the Hong-Ou-Mandel experiment, this structure is manifest in the photon bunching phenomenon whose output probabilities can be easily computed as we just saw. But as more and more photons are evolved in a larger number of modes that are coupled by a network of beamsplitters (see Fig. 1.4), the probability distribution will become extremely hard to compute. Every time a photon is reflected, it acquires a phase; if it is transmitted it does not. Working out the combinatorial number of possible paths by which a particular output pattern of photon numbers is produced becomes increasingly challenging.

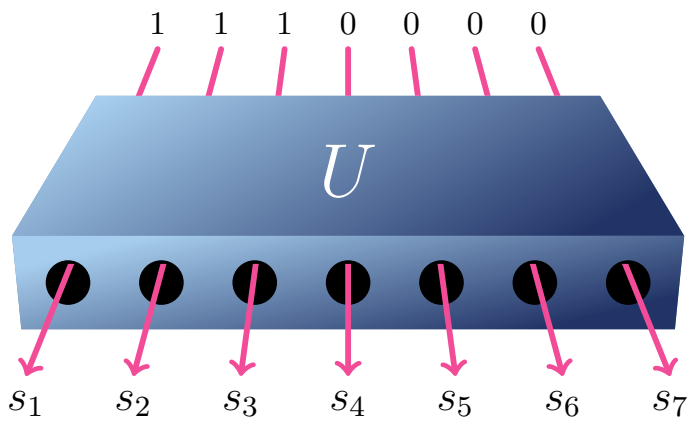

**Figure 1.4:** In a generalized Hong-Ou-Mandel or *boson sampling* experiment $m$ modes, $n$ of which host a single photon, are transformed by a network of beam splitters and phase shifters that can be described by a unitary $U$. In the final stage of the experiment, the output photon pattern $(s_1, \ldots, s_m)$ is measured using photon detectors. Simulating such a boson sampling experiment is computationally intractable [Aar+16]. The figure is an adapted version of [Gog+13, Fig. 1].

This intuition can be made rigorous: it can be proven that the output probabilities of the experiment just sketched are extremely hard to compute [Sch08]. This is due to the fact that the output probabilities of many photons sent through a large network of interferometers can be expressed as the *permanent* of a matrix [Sch08] with entries that are multiples of $-1$ and $+1$[1]. Like the more commonly known determinant the permanent is a polynomial of the matrix entries. It directly quantifies the constructive and destructive interference of all paths through the interferometer that reach a particular output state. As the number of paths increases combinatorially with the number of photons and modes, computing the permanent becomes an extremely hard problem. In fact, even approximating the permanent is as hard as any problem in the complexity class #P, for which no efficient classical algorithm exists [Val79].

1: More generally, when allowing for beam splitters with arbitrary ratio as well as *phase shifters*, the output probabilities are permanents of complex matrices; see Sec. 2.4 for details.

Predicting features of such an experiment – such as the probability of obtaining a specific outcome – is therefore computationally intractable. But even more is true. The less demanding task of simulating the experiment, that is, the task of reproducing the statistics of such a large-scale Hong-Ou-Mandel experiment via sampling in so-called *boson sampling* is computationally intractable, too [AA13].

## The physics of complexity: interference and entanglement

The complexity of both predicting outcome probabilities of a boson sampling experiment and simulating the experiment is only one paradigmatic example of a situation in which quantum mechanical systems are computationally intractable.

In fact, complexity is abundant in quantum theory. Its equations – the static and dynamic Schrödinger equation – while simple at first sight, are notoriously difficult to tackle even for a moderate number of interacting particles. This has long been an obstacle to better understanding the physics of interacting many-body systems. A paradigmatic example of quantum systems as described by a Hamiltonian that are not solvable include the Hubbard model [Hub63] which has been proposed as a simple model of electrons in a solid state and is hoped to help explain conductivity and magnetism on a microscopic level. The Hubbard model only involves nearest-neighbour hopping and density-density (on-site) interactions. And yet, in dynamically evolving Hubbard system, not even simple observables such as the average occupation of individual sites in a lattice can be computed.

Computational intractability is a generic feature of interacting quantum many-body systems both in terms of their static properties such as the expected value of a two-point correlation function in a ground state, and in terms of their dynamical properties such as the dynamical evolution of that expected value after initializing the system in some reference state. In part, this is due to the sheer size of the state spaces involved – state spaces that grow exponentially with the system size.

But classical random processes as simple as many coin flips also have an exponential-size state space. To better understand the computational intractability intrinsic to quantum mechanical systems it is instructive to analyse where computational methods devised for large-scale quantum systems fail.

One particularly successful class of methods are so-called *Quantum Monte Carlo methods* wherein classical pseudo-random sampling from an auxiliary distribution is employed in order to approximate equilibrium properties of quantum systems. Quantum Monte Carlo is particularly useful for large-scale quantum systems as it can handle up to hundreds of particles in some situations [Tro+10]. In other situations, however, Monte Carlo techniques are severely restricted in their applicability due to the so-called *Monte Carlo sign problem*[2] [Hir+82; Sor+89; Loh+90]. The sign problem can be seen as the very manifestion of destructive interference in simulations of physical systems. One particular example thereof is the #P hardness of computing outcome probabilities of Hong-Ou-Mandel-type experiments. The probability of obtaining a particular pattern of photons on the output of the interferometer can be estimated via Monte Carlo sampling from a distribution over the possible photon paths leading to that pattern. But to compute the output probability, not only the number of photon paths leading to a particular outcome need to be approximated as in the classical setting, but their precise interference pattern. While the former is feasible using Monte Carlo methods, the latter quickly becomes intractable. Roughly speaking, what makes Monte Carlo methods fail is destructive interference between different paths leading to the same outcome.

2: For many-electron systems, variants of Quantum Monte Carlo can deal with half-filled lattices comprising up to 50 sites [DLA19].

An example of a numerical method used to study both equilibrium and nonequilibrium physics are *tensor-network methods*. Tensor networks are an ansatz class for quantum states with little entanglement in lattice systems. Tensor-network methods are used both to find ground states via variational minimization of the energy functional and to simulate the dynamical evolution of a locally interacting quantum system by iteratively applying the discretized time evolution operator. Tensor network methods have been extremely successful to overcome the sign problem of Quantum Monte Carlo in some situations to simulate exotic states of matter [Cor+10;

CHM15; Cor16; Lia+17] as well as ground state properties of realistic materials [Boo+19; Ksh+20a] for reasonably large system sizes. However, tensor networks are only efficient descriptions of quantum states if their entanglement is local. This criterion is satisfied for ground states of local Hamiltonians but fails, for example, when simulating time evolution as the entanglement builds up dynamically [CDC19; HC19; Ksh+20b]. While tensor networks can efficiently describe states with little entanglement, computing properties of such states – expectation values of local observables or outcome probabilities – is possible efficiently only in one dimension. This task becomes #P-hard in higher dimensions [Sch+07; Haf+20][3]. Roughly speaking, what makes tensor-network algorithms fail is therefore an intricate entanglement structure in the simulated quantum state.

3: In Ref. [Haf+20] we show that computing expectation values in tensor network states is *generically hard*, improving upon the result by Schuch et al. [Sch+07] which proved hardness in the worst case.

**The origin of quantum complexity**

A central goal of this thesis is to shed light on the role played by interference and nontrivial (two-dimensional) entanglement structure in making quantum systems intractable for classical computations.

Going yet a step further we may ask whether fact that we cannot predict many features of nature at a microscopic level is an intrinsic feature of the world around us or merely an artifact of our quantum description of nature?

## 1.2 Harnessing quantum complexity computationally

> *"We can give up on our rule about what the computer was, we can say: Let the computer itself be built of quantum mechanical elements which obey quantum mechanical laws."*
>
> —*Richard Feynman [Fey82, p. 474]*

So far, we have looked at physics from a computing perspective and wondered about the intrinsic complexity of nature and our quantum description thereof. Let us now flip the coin and look at computing from a physics perspective.

**The computational power of nature**

Can we harness the intrinsic complexity of quantum nature to perform useful computations?

### The computational power of quantum systems

Among the first to ask this question was Feynman [Fey82, p. 475]: "What, in other words, is the universal quantum simulator?" Already Feynman conjectured that such a 'universal quantum simulator' should exist; quantum systems should be efficiently intersimulatable in the sense that there should exist quantum machines able to mimick any quantum system that occurs in nature. More than ten years later[4], this claim was proven to be true by Lloyd [Llo96] in a rigorous sense.

4: ... and in spite of the fact that Feynman claimed: "I believe it's rather simple to answer that question (...) but I just haven't done it." (ibid.)

Using a quantum machine that interacts in a spatially and temporally local way such that, locally, one can realize arbitrary unitaries up to any desired accuracy, it is possible to simulate the time evolution of any locally interacting quantum system with a well-controlled error. This is achieved by discretizing the time evolution of such a system into spatially and temporally local pieces or operations via the so-called *Trotter decomposition*. To achieve a given target error $\epsilon$ on the full time evolution requires only a number of local operations that scales inverse polynomially with $\epsilon$. The operations can then be realized on the quantum machine to the effect that it 'mimicks' or simulates the time evolution of the target system. Such a machine constitute a *quantum computer* as envisioned by Benioff [Ben80] and made concrete by Deutsch [Deu85].

In a quantum computer, sequences of unitary transformations – *quantum gates* – are applied locally to *qubits*, that is, quantum two-level systems, arranged spatially in a lattice. This makes a quantum computer a reasonable model of computing for which algorithms can be devised. Benioff [Ben80; Ben82] realized that, in particular, a quantum computer could be used to efficiently simulate Turing machines, the paradigmatic model of classical computation. As the example of universal quantum simulation suggests, quantum computers are more powerful than classical computers, however.

This intuition has found its rigorous manifestation in several algorithms which outperform classical computers on certain tasks. The Deutsch-Jozsa algorithm [DJ92] determines whether a function on $n$ bits with a single output bit is constant or balanced[5], promised one is the case, in a time that is independent of $n$. Any classical algorithm would require $2^n/2+1$ many evaluations of the function to achieve the same task. Grover's algorithm [Gro96] performs database search in a time that scales as the square root of the database size, while a classical algorithm takes linear time in the worst case. Finally, Shor's algorithm [Sho94; Sho97] is able to find the unique prime factors of a large number in polynomial time, while this task is expected to require exponential time on a classical computer.

5: A Boolean function $f:\{0,1\}^n \to \{0,1\}$ is constant if $f(x)=f(y)\,\forall x,y \in \{0,1\}$; it is balanced if it returns 1 on exactly one half of the inputs (and 0 on the other half).

A possible concern one might have about quantum computers is whether they are really digital devices in which finite precision is sufficient for a computation to be successful. In contrast, in analog models of computation [VSD86], infinite precision is required to obtain the answer as errors accumulate in an uncontrolled way[6]. In a landmark paper Bernstein and Vazirani [BV97] have shown that quantum computers in fact tolerate finite precision in the application of the individual quantum gates and are therefore a genuine and realistic model of computing. Crucial to making the idea of a quantum computer conceivably work in practice is the further possibility to correct errors which inevitably occur in a physical application of individual quantum gates. *Quantum error correction* [Sho96; AB08; CTV17] ensures that local errors can be mitigated so that the output of the computation is ensured to be correct notwithstanding.

6: Such models of computation are therefore not physical. Analog computing would even allow to solve problems harder than any problems in the complexity class NP.

Making the original ideas of Benioff [Ben80], Feynman [Fey82], Deutsch [Deu85] and Lloyd [Llo96] practical is the goal of the field of quantum simulation and computing. Today, there already is strong evidence that the dream of a universal quantum computer can come true in the not-too-far future. Since the landmark results by Lloyd [Llo96] and Shor [Sho94], special purpose analogue quantum simulators have been developed in a plethora of experimental platforms, ranging from ultracold atoms trapped in an optical-lattice potential [Gre+02; BDZ08], Rydberg atoms in optical tweezers [Ber+17b] to superconducting qubits [HTK12; Bar+15; Aru+19] and trapped ions [BR12; Fri+18; Zha+17]. Already for more than ten years such special purpose analogue quantum simulators have been able to qualitatively simulate variants of the Hubbard model[7] [Jak+98], variants of the Heisenberg model [Fri+18], and other intractable Hamiltonians with unprecedented precision and flexibility at scales of up to tens of thousands of atoms [Tro+10]. While much smaller still, universal quantum devices are advancing with a rapid pace. Moving beyond the proof-of-principle demonstrations of quantum algorithms on small scales [Van+00; Van+01; Aru+20], a prerequisite for scalable universal quantum computers is quantum error correction, first steps towards which are being made at the moment [Gon+19; Ngu+20]. The largest fully programmable quantum computer today is the 53-qubit Sycamore superconducting qubit chip developed by the Google AI Quantum team [Aru+19]. This chip can realize computations that are just beyond the edge of what can be simulated using classical computing resources. The search for applications of noisy intermediate-scale quantum (NISQ) devices [Pre18] has now reached industry. Quantum computing has thus expanded from an area of academic interest into the subject of news headlines around the world.

7: Remember that the Hubbard model is numerically intractable on classical computers. A specific example of a quantum simulation using the Hubbard model studied the existence of the Higgs mechanism in two dimensions [End+12], a question which could not be resolved using numerical simulations [Sac99; AA02; PAA11; PS12; PP12; Liu+15].

## The Church-Turing thesis and high-complexity regimes of nature

As quantum computing is on the verge of becoming practically relevent, the question asked by investors and basic research scientists alike – albeit with very different motivation – is: Is it possible to realize a scalable and universal quantum computer?

This question becomes the more pressing as it would contradict an empirical observation about physically realizable computational models, namely, that all such “reasonable” models that have been conceived of in the past century are complexity-theoretically equivalent to a Turing machine model. That is, those models of computation are mutually inter-simulatable with a polynomial resource overhead. As mentioned above, models of computation are deemed physically realizable or “reasonable” [BV97] if the operations can tolerate small errors that arise due to finite precision, or more generally speaking, noise occurring in the physical application of the gate operations. Formally this is captured in what has been named the *complexity-theoretic Church-Turing thesis*[8] [VSD86; BV97].

8: The complexity-theoretic Church-Turing thesis is a modern extension of the original sentiment expressed by Church and Turing about the *computability* of functions. This (weaker) thesis asserts that all functions computable by a “reasonable” model of computation are computable by a Turing machine. Quantum computers *do not* violate the original Church-Turing thesis.

**The complexity-theoretic Church-Turing thesis [BV97]**

Any “reasonable” model of computation can be *efficiently* simulated on a probabilistic Turing machine.

Given the evidence for their computational power presented above, quantum computers are expected to violate the complexity-theoretic Church-Turing thesis. If it turned out that quantum computers were indeed not only conceptually but also *physically possible* in that it would be possible to actually build such a computer, this would falsify the complexity-theoretic Church-Turing thesis. Quantum computers would then be the first computational model that could not be efficiently simulated on a Turing machine, that is, a classical computer, in the century-long success story of digital computers.

Now, in contrast to most statements in complexity theory, the complexity-theoretic Church-Turing thesis is an *empirical statement* about the physical possibility of certain machines. A violation of the thesis therefore requires empirical evidence gathered in an experiment – an experiment in which the physical possibility of quantum computers is demonstrated. A violation of the complexity-theoretic Church-Turing thesis by a real quantum computer would be a watershed moment in the history of computation. Viewed from a computer-science perspective, it would *falsify* a commonly held belief about the nature of physical computing machines (given distinct beliefs about the nature of the problems for which quantum computers provide a speedup).

Viewed from a physics perspective, it would *confirm* that quantum mechanics is also valid in a regime in which it had not been tested as of yet – in the regime of high complexity. Above, I have argued that quantum theory is among the best confirmed theories and it has been tested in numerous instances in many regimes of applicability, ranging from single-photon effects [HOM87] to the verification of entanglement between mechanical oscillators [Rie+18] and macroscopic quantum states such as Bose-Einstein condensates [And+95; Dav+95; Bra+95]. Just as the size of the involved particles can act as a definition of 'regime', so might the physical complexity of the involved systems play that role. But due to its intrinsic complexity, it is unclear whether or not it is at all possible to test quantum mechanics in the high-complexity limit [AV13]: by the very definition of the high-complexity regime, predictions to which experimental results can be compared in order to confirm quantum theory cannot be obtained using classical computing power.

Such *physical complexity* is harnessed *computationally* in a large-scale quantum computer. A verified demonstration of the superior computational power of a quantum computer would therefore not only violate the complexity-theoretic Church-Turing thesis, but, by the same token, also confirm quantum mechanics in the limit of high complexity.

## 1.3 Demonstrating quantum computational supremacy

Acknowledging the recent debate, I use the term "quantum (computational) supremacy" strictly in its technical meaning as established by Preskill [Pre13].

Violating the complexity-theoretic Church-Turing thesis, that is, demonstrating a so-called "computational supremacy" of quantum computing devices over classical computers in an experimental demonstration has become a key milestone in the field of quantum computing [Pre13]. But what kind of experiment would achieve the milestone of quantum supremacy in a way that convinces computer scientists, physicists and quantum computing investors alike?

For a computer scientist, such an experiment must demonstrate the physical possibility of a computing device which violates the complexity-theoretic Church-Turing thesis. To achieve such a violation, it must be the case that the *scaling of resources* required for a classical computer to simulate the task performed in such a demonstration is larger than any polynomial. For a physicist, confirming quantum theory in the high-complexity limit requires not only that any task that is intractable on a classical computer be performed but also that the outcome is *compared to the predictions* of quantum theory. For an investor, an experiment must not only show that quantum computers are physically possible or that quantum mechanics is true in the high-complexity limit. A quantum supremacy experiment must also demonstrate that such a device is worth investing in. Given the large anticipated investements as compared to the relatively small costs of classical computing, a quantum computer should be able to *solve tasks* which no classical computer can possibly solve.

The cartoon I just painted illustrates the three commonly held requirements on a convincing demonstration of quantum supremacy:

i. the demonstrated speedup must be superpolynomial,
ii. it must be verified against the predictions of quantum mechanics, and
iii. it must outperform the best available classical simulation.

All of these requirements are extremely challenging on different levels. The central complexity-theoretic challenge is to prove a superpolynomial speedup of quantum computers over classical computers, a challenge that has remained elusive for several decades now. Next, given the intrinsic complexity of the task by the first requirement, a direct verification using only classical computing resources seems impossible at first sight. The final challenge is to actually build a large-scale quantum computer that is able to outperform the classical supercomputers available today.

The neatest way to achieve a fair compromise between those requirements, and in particular so, the second requirement, would be via factoring using Shor's algorithm. This is because factoring is believed to be a problem for which no efficient classical algorithm exists. In fact, a large part of the presently applied public-key cryptography is based on the hardness of factoring. Factoring is particularly suited to public-key cryptography because it is believed to define a so-called one-way function, that is, a function which can be computed easily (the product of two large prime numbers) but which is extremely difficult to invert (finding those numbers given their product). Vice versa, this means that verifying a successful implementation of Shor's algorithm is simple: simply multiply the output and compare to the input.

While proof-of-principle demonstrations of Shor's algorithm have been achieved [Van+01], factoring a large 2048 bit number as is used for public encryption via RSA [IEEE 1363] is estimated to require a large-scale, error-corrected universal quantum computer using roughly 20 million qubits [HRS17; OC17; GM19; GE19]. This is out of reach in the near term. Available today are noisy universal quantum devices with up to roughly 50 qubits [Aru+19; Zha+17; Ber+17b], reaching up to potentially hundreds of qubits in the next years, as well as special-purpose quantum simulators which allow for somewhat larger system sizes but lose universal programmability.

What is more, the evidence for the hardness of factoring stands on shaky grounds from the point of view of complexity theory. While there is strong empirical evidence that factoring is practically impossible for large instances, the hardness of factoring does not have any theoretical underpinning. In particular, an efficient factoring algorithm would not bear any complexity-theoretic consequences such as the falsity of the famous $\mathsf{P} \neq \mathsf{NP}$ conjecture whatsoever. While empirical evidence may be good enough for all practical purposes, a sceptic might attribute the apparent hardness of factoring to the lack of ingenuity in finding good algorithms. We should therefore strive to find a task for the hardness of which there also is strong theoretical evidence.

In order to demonstrate quantum computational supremacy we must therefore find a task that satisfies the requirements i.-iii. as convincingly as possible *and* is feasible on the near-term quantum computing architectures. This is a primary goal of mine for this thesis.

### Demonstrating quantum supremacy via quantum random sampling

The kind of task that any quantum experiment naturally solves is a sampling task: it produces samples from a probability distribution governed by the Born rule. A breakthrough observation in the quest to demonstrate quantum computational supremacy is that certain *quantum random sampling* tasks, while inherently simple to perform on a quantum device, are immensely difficult for classical computers [BJS10; AA13]. In quantum random sampling, the task is to simulate – to sample from – the output distribution of a *randomly chosen* quantum state.

We have already seen a glimpse of the classical complexity of quantum sampling in our case study of the Hong-Ou-Mandel experiment. We discussed that randomness is an intrinsic feature of quantum experiments. What is more, the output probability distributions of generic quantum processes have an intrinsically complex structure. This structure is manifest in the fact that computing the output probabilities of quantum devices is an extremely hard task generically. We saw this by the example of computing the output probabilities of a generalized Hong-Ou-Mandel experiment, a boson sampler.

Using an involved technical machinery it has been shown that the generic hardness of *approximating outcome probabilities* of a boson sampler implies the hardness of *sampling from the outcome distribution* of such a device. Simulating a boson sampler classicallyis infeasible for roughly 60 photons in 3600 modes [Nev+17; CC18]. On the other hand, letting photons interfere is a comparably simple task for few photons [Spr+13; Til+13; Bro+13; Cre+13] and has been achieved for up to 20 photons [Wan+19]. Those results are the more exciting as it is not possible to realize an arbitrary quantum computation on a boson sampler – it is not universal for quantum computing. They demonstrate the possibility to outperform classical computers using *subuniversal*, that is, special-purpose quantum devices.

Quantum random sampling schemes have the additional feature that they not only highlight a path towards demonstrating quantum supremacy. If supplemented with adequate measures of quality, they also allow to *benchmark* and *calibrate* quantum devices in the development phase. This is because the task involves sampling from a randomly chosen instance of a computation which the device is designed to perform. Exploiting concentration-of-measure phenomena, by testing the quantum device on random problem instances, one can benchmark and gain confidence in its correct functioning for *most* instances. Thus, quantum random sampling is not only useful as a means to answer foundational empirical questions in physics and computer science, but also as a useful method in the *engineering* of those devices.

In particular, it is possible to extend subuniversal quantum random sampling schemes such as boson sampling to universal devices [Boi+18]. This allows to even further push down the number of qubits or particles necessary to outperform classical computers: Simulating randomly chosen quantum computations is infeasible for present-day supercomputers on as little as roughly 50 qubits [Boi+18]. The improvement of the complexity-theory behind such sampling tasks together with the rapid development of near-term quantum devices in the past years has recently culminated in an impressive experimental demonstration of universal quantum random sampling on 53 qubits [Aru+19]. This demonstration is (arguably [Ped+19]) just beyond the edge of what can be classically simulated and can therefore be considered a first demonstration of quantum supremacy.

## 1.4 Sampling and the complexity of nature

In the run-up to this first demonstration over the past years several loopholes of quantum random sampling needed to be closed. On the one hand, the argument for the *classical intractability* of quantum random sampling was (and still is) based on complexity-theoretic conjectures about the hardness of certain computational tasks. On the other hand, a pressing issue was, and still is, whether and if so how it might be possible to *verify* quantum sampling tasks in the absence of a classical simulation. The question of verification is particularly challenging for quantum sampling schemes as, unlike in typical computational problems, not only a single outcome has to be checked in its correctness, but rather *the distribution* of many samples from the device. I sketch the setting for demonstrating quantum supremacy via quantum random sampling in Figure 1.5.

And even today, key questions about the underlying physics as well as the complexity-theoretic foundations of quantum sampling remain open. We have a reasonable understanding of the complexity-theoretic reasons for why quantum sampling is classically intractable. But what are the physical effects and mechanisms that govern this complexity? What can we learn about the apparent complexity and computational power of special-purpose quantum simulators that have already outperformed the best classical algorithms for ten years [Tro+12] using the complexity-theoretic tools developed for quantum sampling?

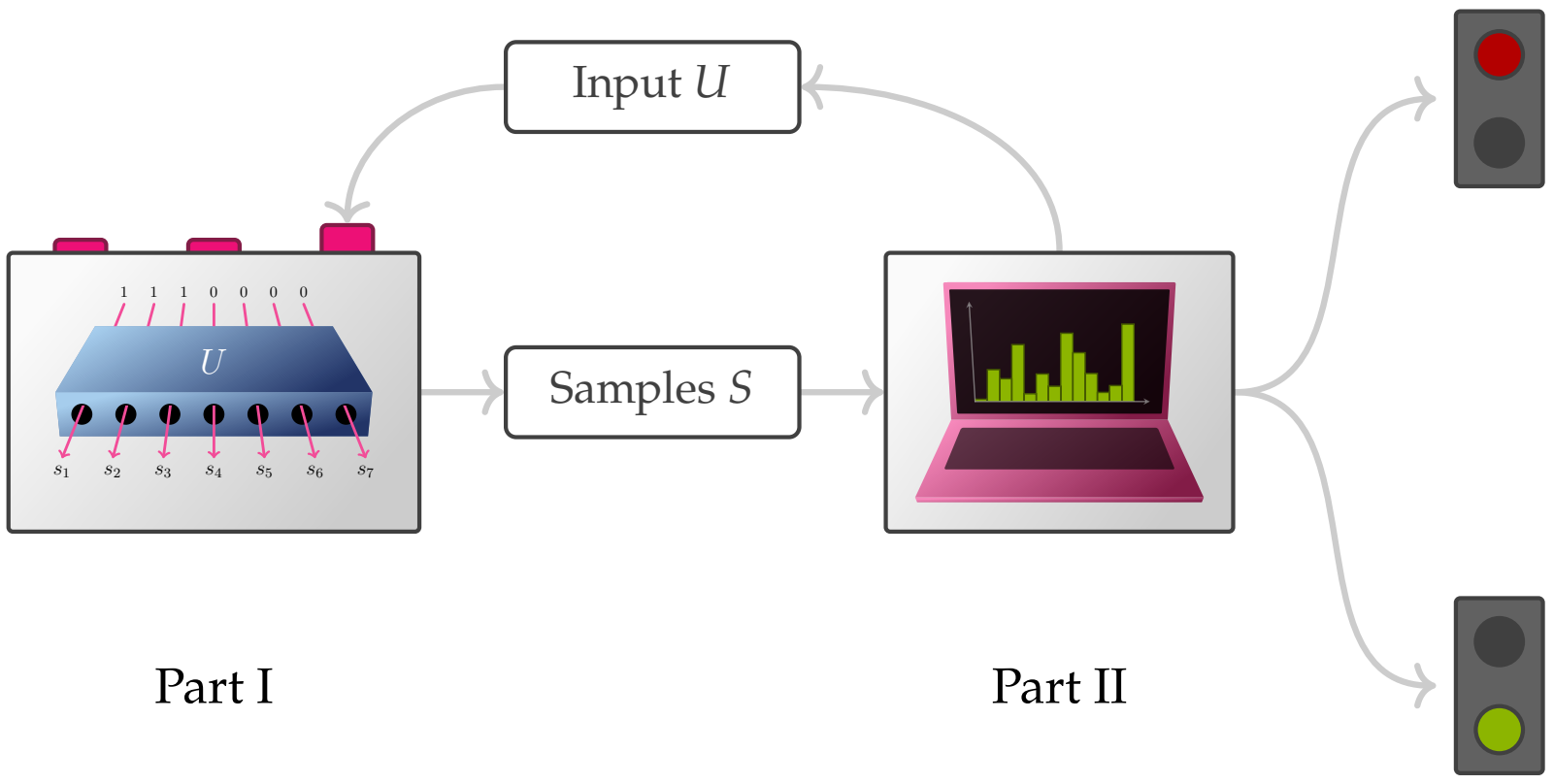

**Figure 1.5:** In a quantum random sampling experiment, the task is to sample from the output distribution of a randomly chosen computation $U$. An unambiguous supremacy can only be achieved if, given the samples $S$, we can *verify* that the targeted task has indeed been achived. In Part I of this thesis, we will take a closer look at the complexity-theoretic arguments and practically feasible schemes for quantum random circuit sampling. In Part II, we will discuss the possbility of and practical schemes for certifying quantum sampling devices.

In this thesis, I will answer some of those questions. Using the theory of computational complexity, I will *assess* the computational power of natural quantum simulators and close complexity-theoretic loopholes in the hardness argument (Part I). Using methods from property testing and quantum system identification, I will shed light on the question, how and under which conditions quantum sampling devices can be *tested* or verified in regimes that are not simulable on classical computers (Part II). Finally, I will try to understand the physical mechanisms governing the computational boundary between classical and quantum computing devices by *challenging* their computational power using tools from computational physics (Part III).

## The computational complexity of quantum sampling

In the first part of this thesis I will elaborate on the complexity-theoretic underpinnings of the computational complexity of quantum random sampling. We will see that the complexity of sampling can be traced back to the possibility of destructive interference. As we witnessed when looking at the Hong-Ou-Mandel experiment, destructive interference leads to an intricate structure of the output probability distribution of a quantum experiment, which is reflected in the computational hardness of computing output probabilities. We will refer to this root of quantum complexity as the *quantum sign problem*.

The argument that relates the complexity of sampling and the complexity of computing output probabilities is based on powerful results from the theory of computational complexity. Nevertheless, some loopholes remain. The argument requires a particular structure of the output distribution of a quantum random sampler, so-called *anticoncentration*, as well as fine-grained *average-case hardness* of computing those probabilities. For many schemes those properties cannot be verified but need to be conjectured. To be safe against a sceptic of quantum supremacy those loopholes must be closed. In Chapter 3 we will see how those loopholes can be (partially) closed for a range of sampling schemes. In particular, we will review methods for showing average-case hardness. We will show an anticoncentration result for a large class of quantum circuit families, namely, such families that form a 2-design. Building on this result, we then identify a simple 'recipe' following which theoretically sound quantum random sampling schemes can be constructed.

Finally, we will try and relate the paradigmatic hardness-of-sampling results back to physics. Recall that a generalized Hong-Ou-Mandel experiment – boson sampling – is computationally intractable in the rigorous sense described here. But at the same time such an experiment is computationally restricted and useful only for special purposes. In Chapter 4, we develop random sampling schemes that come with the same complexity-theoretic underpinning but are tailored to special-purpose natural quantum simulators such as cold atoms in optical lattices and trapped ions . This effort is intrinsically motivated by the question whether the practical advantage that such special-purpose quantum simulators can be proven in a complexity-theoretic sense. But it is also practically driven by the hope to harness the practical advantage of such large-scale quantum simulators in order to demonstrate quantum supremacy on thousands of qubits that lies well beyond what can be simulated classically today.

## Verifying quantum sampling schemes

In the second part, we will turn to a question that is particularly pressing in any quest to demonstrate quantum computational supremacy: We must verify that the targeted task has indeed been achieved. As in the complexity-theoretic argument for classical intractability, we must measure with particularly stringent standards when it comes to verifying the output of a realistic device so as to be safeguarded against a sceptic. This task is particularly challenging given not only the computational hardness of producing samples from the anticipated device, but also because sampling tasks are intrinsically difficult to verify. In fact, it is not even clear whether such verification is possible at all. In Chapter 5, I will elaborate those questions and set the stage by discussing what we could possibly mean by 'verification'. The requirements on verification might be very different depending on the setting at hand. For example, if a quantum device is accessed remotely via the cloud, we should conceive of it as a black box and certify its functioning based on this model. After all, the provider of the device might want to cheat their user, say, to save money. This is different in a laboratory setting in which all components of the device are well-controlled and characterized. In this spirit, I will introduce different settings in which the verification task can be meaningfully phrased.

In Chapter 6, I will show that doubt is well-justified: Verifying that a quantum sampling device functions correctly in a black-box setting requires infeasibly many runs of that device. This sets a challenge that any verification of a quantum sampler must cope with

In Chapter 7, I will show how this no-go result can be circumvented by making some assumptions or by significantly increasing the complexity of the verification protocol. Coming from a physics perspective, we will focus on a laboratory setting in which the experimenter has a good understanding of the functioning of individual components of a large-scale quantum device. For this setting, I will develop particularly simple and experimentally feasible efficient means to verify quantum sampling devices, and in particular, the quantum simulation scheme proposed in Chapter 4.

I conclude Parts I and II with an outlook on the field of quantum supremacy and pose open questions on the verification and simulation of quantum random sampling schemes as well as possible applications thereof in Chapter 8.

## Simulating quantum systems using classical sampling

Finally, we will return to the question where the hardness of quantum sampling originates. To shed some light on this question, we will try and push the limits of classical simulation algorithms for properties of quantum systems. Among the best such algorithms are so-called Quantum Monte Carlo algorithms – classical algorithms that are based on sampling. In Chapter 9, I will introduce such classical sampling algorithms and sketch their application in Quantum Monte Carlo to estimate properties of quantum systems. It turns out that the quantum sign problem – the complexity-theoretic root of the hardness of quantum sampling – finds its practical manifestation in Quantum Monte Carlo methods as *the Monte Carlo sign problem*.

Quantum Monte Carlo constitutes a concrete framework within which the sign problem can be meaningfully studied. The starting point for this endeavour is the basic observation that the sign problem heavily depends on our *representation* of the target quantum system to be simulated in terms of the basis in which a Hamiltonian is expressed. Studying the sign problem from this perspective will not only illuminate fundamental questions regarding the roots of the complexity of quantum systems. It will also find practical applications in improved classical simulation algorithms for quantum systems.

In Chapter 10, I will introduce computationally meaningful measures for assessing the sign problem in Quantum Monte Carlo simulations. Indeed, generically, estimating the severity of the sign problem in a given simulation is as hard as performing that simulation. Within this framework, we can optimize the sign problem over meaningful basis sets. We can then ask the fundamental question: What is the intrinsic degree to which a given simulation suffers from the sign problem? This question finds its practical bearing in the question: How far can we *ease the sign problem* of a given system?

In the subsequent chapters, we will further examine those questions. In a first step, we numerically optimize the sign problem with respect to local basis choices for physically relevant, simple models (Chapter 11). In a second step, we will identify fundamental obstacles to this endeavour by studying the computational complexity of finding the sign-optimal basis (Chapter 12). Ironically, we will find that determining the degree to which a given system has an *intrinsic sign problem* is a computationally intractable task in itself.

In this last result, the hardness of simulating quantum devices, which we studied in the first chapters of this thesis, comes back with a vengeance: it prohibits finding the optimal way of simulating quantum systems.

# Part I

## The computational complexity of quantum sampling

# The computational complexity of quantum sampling

# 2

The central theme of this thesis is the notion of sampling. In the introduction, we have already seen a glimpse of how sampling can become a computationally difficult task for quantum systems: in the Hong-Ou-Mandel experiment, quantum interference resulted in cancellations of computational paths, which leads to an intricate structure of the output probability distribution. I already mentioned in passing that such intricate structure of the probability distribution is reflected in the computational hardness of computing the probabilities of individual outcomes. But how is the task of computing individual probabilities related to the task of sampling from an entire distribution? Intuitively, if computing probabilities is hard, then sampling must be, too.

In this chapter, we will make those intuitions rigorous. We will see which ingredients come together in a strategy to provide strong complexity-theoretic evidence for the hardness of sampling from or *weakly simulating* certain circuit families. This will include both results about the computational complexity of approximately computing the output probabilities of or *strongly simulating* quantum circuits and hypothetical algorithms for this task.

Those results will constitute the complexity-theoretic underpinning of experimental prescriptions designed to demonstrate quantum computational supremacy, that is, to experimentally violate the extended Church-Turing thesis. To do this, we must hold ourselves to a particularly stringent standard of evidence. While many large-scale quantum simulators have demonstrated practical advantages over existing classical algorithms, in order to have high confidence in the claim that quantum computing devices cannot be efficiently simulated in a Turing-machine model, we must show that *no classical algorithm* which we could possibly run on a classical computer will ever be able to solve the targeted task efficiently.

### Basics of computational complexity theory

We have come into the terrain of theoretical computer science. There, classes of problems, so called complexity classes, are studied with respect to their *computational complexity*, that is, the runtime that a (typically classical) algorithm designed to solve problem instances from such a class would have in the worst case. Most of those complexity classes are in fact not *efficiently* solvable on a classical computer and thus involve *hypothetical algorithms* as well as algorithms that can be performed practically. We can still discern distinct problem classes defined by certain resource restrictions that differ significantly in their computational complexity. Understanding the relations between different complexity classes, that is, separations and inclusions between them is the main subject of study in the theory of computational complexity. For convenience, most often *decision problems* are considered, where the task is to decide whether a given string[1] $x \in \{0,1\}^*$ is in a so-called *language* $L \subset \{0,1\}^*$, which is a set of bitstrings. A machine that computes a Boolean function $f_L : \{0,1\}^* \longrightarrow \{0,1\}$ such that $f_L(x) = 1 \Leftrightarrow x \in L$ decides $L$.

1: We write the set of all finite-length bit strings as $\{0,1\}^* \equiv \bigcup_{n \in \mathbb{N}} \{0,1\}^n$.

The central concept of computational complexity theory is that of an *algorithm*. In particular, an algorithm can compute a Boolean function $f : \{0,1\}^* \longrightarrow \{0,1\}$ for arbitrary-length inputs. Abstractly speaking, an algorithm is a set of rules according to which a machine acts on any given input. In the case of classical algorithms, formalized as a Turing machine, those rules may involve *reading* bits of the input or a scratch pad and *writing* bits to that scratch pad, *choosing* a new rule according to which to continue, or *stopping* and outputting either 0 or 1 [AB09]. We say that an algorithm is efficient if its runtime scales polynomially in the input size[2] $|x|$.

2: For $x \in \{0,1\}^*$ we define $|x|$ to be the length of $x$ in binary representation.

On an actual silicon-chip computer, those rules can be realized by certain elementary logic operations that are applied sequentially (or in parallel) to some of the input registers (bits) at a time. The elementary logical operations might act on a single register or bit such as the `NOT` operation[3], on two such as `OR` and `AND`[4] or even more registers. A set of such operations is said to be universal if an arbitrary Boolean function $f : \{0,1\}^n \longrightarrow \{0,1\}$ can be expressed as a classical circuit using $\mathsf{poly}(n)$ many input registers. Examples of such sets are $\{\mathtt{AND}, \mathtt{NOT}\}$ and the singleton[5] $\{\mathtt{NAND}\}$. Using a sequence of universal logical operations one can therefore express any other elementary logical operation. For any given input size $n$, we call a network of elementary logic gates acting on the input registers a *classical circuit*, and hence the corresponding model of computation the circuit model. A circuit $C_n$, potentially using additional auxiliary registers, in particular effectively computes a Boolean function on its input registers. On input $x \in \{0,1\}^n$, its outcome $C_n(x) \in \{0,1\}$ is given by its value on a single – say, the first – output register. The size of a circuit $|C_n|$ is given by the number of gates in it.

3: For $a \in \{0,1\}$ $\mathtt{NOT}(a) = a \oplus 1$, where we denote addition modulo 2 by $\oplus$.

4: For $a, b \in \{0,1\}$ $\mathtt{OR}(a,b) = 0$ if and only if $a = b = 0$. $\mathtt{AND}(a,b) = 1$ if and only if $a = b = 1$.

5: $\mathtt{NAND}(a,b) = \mathtt{NOT}(\mathtt{AND}(a,b))$.

Notice that any given circuit takes inputs of a fixed size $n$, while of an algorithm we demand that it works for any input size. We can turn a family of circuits $\{C_n\}_{n \in \mathbb{N}}$ into a meaningful algorithm[6] by supplementing it with an efficient instance-generating procedure that given the input size $n$ efficiently produces a description of $C_n$, which is then run on the input $x \in \{0,1\}^n$. We call circuit families for which such a procedure is possible *uniform circuit families*. Uniform circuit families thus realise a classical algorithm in the circuit model.

6: Indeed, if we ask merely for the existence of a circuit family as opposed to an efficient algorithm then this allows us to solve undecidable problems using polynomial-size circuits.

The fundamental class of problems in computational complexity theory is the class $\mathsf{P}$, the class of problems which can be solved efficiently on a deterministic classical computer[7].

7: I summarize the most important technical concepts used in this thesis, and in particular, the complexity classes that appear, on the last pages of this thesis. See also `https://complexityzoo.uwaterloo.ca`.

**Definition 2.1** ($\mathsf{P}$) *A language $L \subset \{0,1\}^*$ is in the class $\mathsf{P}$ if there exists a classical algorithm that given $x \in \{0,1\}^*$ as an input decides whether $x \in L$ in polynomial runtime in $|x|$.*

Relations between complexity classes are typically studied with respect to polynomial reductions – so called *Turing reductions* where access to a machine in $\mathsf{P}$ is granted. A key problem in the theory of computational complexity is that the relation between different complexity classes defined with very different resource restrictions in mind is inherently hard to pin down. Very basic relations between complexity classes are therefore conjectured and cannot be proven. The most basic and at the same time most fundamental separation in complexity theory is the belief that $\mathsf{P} \neq \mathsf{NP}$. While P is the class of problems which can be efficiently *computed* on a classical computer, NP is the class of problems which can be efficiently *verified*.

**Definition 2.2** (NP [AB09]) *A language $L \subset \{0,1\}^*$ is in the class* NP *if there exists a polynomial $p : \mathbb{N} \to \mathbb{N}$ and a polynomial-time classical algorithm $V$ (called the verifier for $L$) such that for every $x \in \{0,1\}^*$,*

$$x \in L \quad \Leftrightarrow \quad \exists y \in \{0,1\}^{p(|x|)} : V(x,y) = 1. \tag{2.1}$$

*We call $y$ the* proof *of $x$.*

When gathering evidence for a separation between quantum and classical computation, quantum and classical sampling in particular, we therefore want to try and keep as close to problems that have been well-studied such as the conjecture $\mathsf{P} \neq \mathsf{NP}$. The main challenge is that, at the same time, the computational task must be such that it can realistically be realized on near-term quantum devices in as easy and error resilient a way as possible.

## 2.1 Where look for a quantum-classical separation?

Our starting point from the perspective of computational complexity theory shall be ever so slight differences between randomized quantum and classical computation[8]. We formalize randomized classical and quantum computations in terms of decision problems as complexity classes BPP and BQP.

8: In this section, I will follow a train of thought which to the best of my knowledge is due to Scott Aaronson @ `https://www.scottaaronson.com/blog/?p=3427`.

**Definition 2.3** (Classical and quantum computation) *Define the class* BPP *(*BQP*) as the class of all languages $L \subset \{0,1\}^*$ for which there exists a uniform polynomial-time randomized classical (quantum) circuit family $\{C_n\}_{n\in\mathbb{N}}$ such that for all $n \in \mathbb{N}$ and all inputs $x \in \{0,1\}^n$*

$$x \in L \quad \Rightarrow \quad \Pr[C_n(x) = 1] \geq 2/3, \tag{2.2}$$

$$x \notin L \quad \Rightarrow \quad \Pr[C_n(x) = 1] \leq 1/3, \tag{2.3}$$

*where the probability is taken over the internal randomness of the circuit.*

What are *randomized* classical and quantum algorithms?

**Randomized and reversible classical circuits** To define randomized classical algorithms, we assume access to a source of randomness. A randomized classical algorithm is one in which some of the input bits are allowed to be uniformly random bits. On input $x$ its output is given by a function Boolean $f_x$ that takes a uniformly random bit string $r \in \{0,1\}^{\mathsf{poly}(|x|)}$ as an input. Clearly, randomized algorithms are at least as powerful as deterministic ones as such a function can simply disregard the random inputs, giving rise to a deterministic algorithm. In many practical situations, randomized algorithms turn out to be much more efficient than deterministic algorithms, however. A central question in the theory of computational complexity is whether one can *derandomize* such algorithms efficiently, that is, whether $\mathsf{P} = \mathsf{BPP}$, an equality that is generally believed true.

While classical logical gates are not generally *reversible* in that the mapping from input to output is not injective, it turns out that one can realize any classical circuit using reversible operations [Tof80; FT82]. In other words, there are sets of reversible operations such as the three-bit Toffoli or controlled-controlled-NOT gate TOF [Tof80] such that an arbitrary Boolean function can be expressed using the outcome of a single register in a computation involving TOF. The Toffoli gate can also be expressed as a permutation matrix mapping the three-bit space $\{0,1\}^3$ to itself[9]

$$\mathtt{TOF} = \begin{pmatrix} 1 & 0 & 0 & 0 & 0 & 0 & 0 & 0 \\ 0 & 1 & 0 & 0 & 0 & 0 & 0 & 0 \\ 0 & 0 & 1 & 0 & 0 & 0 & 0 & 0 \\ 0 & 0 & 0 & 1 & 0 & 0 & 0 & 0 \\ 0 & 0 & 0 & 0 & 1 & 0 & 0 & 0 \\ 0 & 0 & 0 & 0 & 0 & 1 & 0 & 0 \\ 0 & 0 & 0 & 0 & 0 & 0 & 0 & 1 \\ 0 & 0 & 0 & 0 & 0 & 0 & 1 & 0 \end{pmatrix}. \tag{2.5}$$

9: Equivalently

$$\mathtt{TOF}(a,b,c) = \begin{cases} (a,b,\bar{c}) & \text{if } a = b = 1 \\ (a,b,c) & \text{else.} \end{cases} \tag{2.4}$$

The Toffoli gate is universal by itself which can easily be seen from the observation that the last output bit of $\mathtt{TOF}(a,b,1)$ gives the NAND of the two input bits $a$ and $b$.

By taking the leap to reversible classical computation we have already made half-way on the way to quantum computation. Indeed, the question about the possibility of reversible classical computation was originally motivated from a physics perspective [FT82] noticing that the laws of physics are reversible. Hence, so the thought, a physical model of computation should be, too.

**Quantum circuits** Quantum circuits are a generalization of reversible classical circuits. There are several key differences between reversible classical circuits and quantum circuits, however. A quantum circuit acts on *qubits* the state space of which is given by $\mathbb{C}^2$. The elementary operations or quantum gates are unitary matrices acting on a $k$-qubit input space $(\mathbb{C}^2)^{\otimes k}$. A classical circuit acting on $m = \mathsf{poly}(n)$ qubit registers produces not a single bit string as an output but a quantum state in $(\mathbb{C}^2)^{\otimes m}$, which only upon a quantum measurement in some basis – typically the standard $Z$ basis – produces a bit string as an output. Throughout this thesis, we say that a quantum computation accepts if the outcome of such a measurement is the all-zero string. Notice that classical computation is a special case of quantum computation: If we restrict ourselves to state preparations and measurements in the computational basis and permutation matrices in that basis(which are in particular unitary), then we recover classical computation.

A gate set is universal for quantum computation if for $n \geq n_0$ the subgroup generated by $\mathcal{C}$ is dense in the $n$-qubit special unitary group $SU(2^n)$. A paradigmatic quantum universal gate set is given by the Hadamard gate $H$, the $T$ gate and the classical CNOT operation, which

are defined as

$$H = \frac{1}{\sqrt{2}}\begin{pmatrix} 1 & 1 \\ 1 & -1 \end{pmatrix}, \quad T = \begin{pmatrix} 1 & 0 \\ 0 & e^{i\pi/4} \end{pmatrix}, \quad \texttt{CNOT} = \begin{pmatrix} 1 & 0 & 0 & 0 \\ 0 & 1 & 0 & 0 \\ 0 & 0 & 0 & 1 \\ 0 & 0 & 1 & 0 \end{pmatrix}. \tag{2.6}$$

A gate set is said to be *computationally universal* if an arbitrary quantum circuit acting on $n$-qubits and using $t$ gates can be simulated up to error $\epsilon$ with overhead $\mathsf{polylog}(n, t, 1/\epsilon)$ in terms of both the number of registers and gates [Aha03]. With polynomial overhead in $n$ and $t$ computational universality therefore tolerates errors on the order $2^{-\mathsf{poly}(n,t)}$. In contrast to universality of the group generated by a gate set, computational universality thus focuses on the aspect of *efficiency* in terms of the number of gates required to approximately *synthesize*, that is, express a unitary.

A computationally universal gate set that will serve us well in due course is $\{H, \texttt{TOF}\}$. This gate set is universal for $n$-qubit computations when acting on $n + 1$ many qubits[10] [Aha03].

10: The reason for this is, of course, that what is missing for this gate set to be fully universal is imaginary numbers. Those can be simulated using a single additional qubit as $\mathbb{C} \approx \mathbb{R}^2$.

Notice that while classical universality is an *exact notion* since the number of Boolean functions is finite, quantum universality incurs approximations since the group $U(2^n)$ is infinite. The famous Solovay-Kitaev theorem solves this problem for quantum gates acting on qubits with efficiently computable entries.

**Theorem 2.1** (Solovay-Kitaev [DN06]) *Let $\mathcal{G} \subset SU(2)$ be a universal finite gate set containing its own inverses. For every $U \in SU(2)$ with efficiently computable entries there is finite sequence of gates $S$ from $\mathcal{G}$ that approximates $U$ up to error $\epsilon$ for which both its length and the time to find it scale as* $\mathsf{polylog}(1/\epsilon)$.

Quantum computations are intrinsically randomized – the probability that an $n$-qubit quantum computation $C_n$ applied to an input state $|x\rangle \in \mathbb{C}^n$ accepts, that is, that after a measurement one obtains the all-zero outcome, is given by the Born rule

$$\Pr[C_n(x) = 1] = |\langle 0|C_n|x\rangle|^2. \tag{2.7}$$

In contrast, classical circuits are naturally deterministic: given an input the output is inevitably and fully determined by the rules of classical logic. Only by artificially introducing randomness into the circuit can we construct a randomized classical algorithm using elementary logic gates. A randomized circuit for a Boolean function $f : \{0,1\}^n \times \{0,1\}^m \to \{0,1\}$ acts on both the problem input $x \in \{0,1\}^n$ and a length-$m$ uniformly random bit string $r \in \{0,1\}^m$ with $m = \mathsf{poly}(n)$.

## Computing acceptance probabilities of randomized algorithms

A key but very subtle difference between randomized classical and quantum computations presents itself in the guise of the probability that such computations accept. This difference will be a lever allowing us to separate the two types of algorithms in terms of their computational power.

**Classical acceptance probabilities** The acceptance probability of a classical randomized circuit $C_n(x)$ computing a Boolean function $f_x$ is given by the fraction of random inputs $r \in \{0,1\}^m$ for $m \in \mathsf{poly}(n)$ on which it accepts:

$$\Pr[C_n(x) = 1] = \frac{1}{2^m} \sum_{r \in \{0,1\}^m} f_x(r). \tag{2.8}$$

Computing the acceptance probability of classical circuits is therefore clearly a #P-complete problem.

**Definition 2.4** (#P [AB09]) *Define the* function class #P *as the class of all functions* $f : \{0,1\}^* \to \mathbb{N}$ *for which there exists a polynomial-time classical algorithm C such that there exists a polynomial* $p : \mathbb{N} \to \mathbb{N}$

$$f(x) = \left|\left\{y \in \{0,1\}^{p(|x|)} : C(x,y) = 1\right\}\right|. \tag{2.9}$$

In other words, #P functions count the number of accepting inputs to a polynomial-time computation $C$. In turn we can view the decision class NP (Def. 2.2) as asking to decide whether there exists *any input* such that a computation $C$ accepts. #P functions, that is, functions which can be computed in polynomial time, can thus be viewed as counting the number of solutions an NP-verifier accepts.

A paradigmatic NP-complete[11] problem is 3SAT, which asks whether a Boolean formula in conjunctive normal form containing three literals per clause, that is, a formula $F : \{0,1\}^n \to \{0,1\}$ of the type

11: We say that a problem is *complete* for a complexity class X if it is both as hard as any problem in X and in X.

$$F(x) = (x_1 \vee x_3 \vee \overline{x}_5) \wedge (x_7 \vee \overline{x}_2 \vee \overline{x}_3) \wedge \cdots, \tag{2.10}$$

where $\overline{x} = x \oplus 1$ denotes the negation of $x$ is *satisfiable*. This is the case if there exitss a string $x_0 \in \{0,1\}^n$ such that $F(x_0) = 1$. The paradigmatic #P-complete problem is #SAT – the problem to count the number of accepting inputs to a 3SAT formula.

**Quantum acceptance probabilities** Similarly, we can express the ouput probabilities of a quantum circuit $C_n$ on input $x$ that uses $m = \mathsf{poly}(n)$ registers via a function $g_x : \{0,1\}^m \to \{+1,-1\}$ as[12]

12: I recommend the paper by Montanaro [Mon17] as an introduction to Boolean functions and their relation to quantum success probabilities that is both accessible and insightful.

$$\Pr[C_n(x) = 1] = \frac{1}{2^m} \sum_{y \in \{0,1\}^m} g_x(y). \tag{2.11}$$

This is easily seen using the fact that the gate set comprising the Hadamard and the Toffoli gate is universal for quantum computing[13]. In this gate set, we can express the all-zero amplitude of an $n$-qubit computation $C_n = C_t \cdots C_1$ using $t$ quantum gates $C_1, \ldots, C_t$ [Daw+05]

13: As discussed above, since Hadamard and Toffoli are computationally universal [Shi02; Aha03] the success probability of an arbitrary polynomial size computation can be expressed as the success probability of such a circuit up to an error $\epsilon$ with an overhead of $\mathsf{polylog}(1/\epsilon)$. This means that we can obtain a $O(2^{-\mathsf{poly}(n)})$ approximation of this success probability. We will shortly come to a more detailed discussion of such approximations and the question how hard it is to compute them.

$$\langle 0|C_n|x\rangle = \sum_{\lambda_1,\ldots,\lambda_t} \langle 0|C_t|\lambda_1\rangle \cdots \langle \lambda_t|C_1|x\rangle \tag{2.12}$$

$$= \frac{1}{\sqrt{2^h}} \sum_{y \in \{0,1\}} s_x(y), \tag{2.13}$$

in terms of the number $h$ of Hadamard gates and a signed function $s_x$. This is because the matrix elements of the Toffoli gate are binary and those of the Hadamard gate are $\pm 1/\sqrt{2}$ so that each entry of the matrix product $C_t \cdots C_1$ is a sum of numbers $(\pm 1) \cdot 2^{-h/2}$. We thus obtain that

$$|\langle 0|C_n|x\rangle|^2 = \frac{1}{2^h}\left|\sum_{y\in\{0,1\}} s_x(y)\right|^2 = \frac{1}{2^h}\sum_{y,z}\underbrace{s_x(y)s_x(z)}_{g_x(y,z)}. \tag{2.14}$$

Conversely, given a signed function $s : \{0,1\}^n \to \{+1,-1\}$ described by a polynomial-size classical circuit $C_s$, we can find a quantum circuit $Q_s$ with acceptance probability (cf. (2.11)) proportional to $\sum_{x\in\{0,1\}^n} s(x)$. By universality of TOF for classical computation, we can find a reversible classical implementation $S$ of the circuit $C_s$ that acts as[14] [BMS16]

$$S|0^a\rangle|x\rangle|0\rangle = |0^a\rangle|x\rangle|f_s(x)\rangle, \tag{2.15}$$

where $f_s(x) = (s(x)+1)/2 \in \{0,1\}$, potentially using $a = \mathsf{poly}(n)$ auxiliary qubits. Now we apply $S$ in superposition as

$$Q_s = (1_a \otimes H^{\otimes n} \otimes XH)S(1_a \otimes H^{\otimes n} \otimes HX) \tag{2.16}$$

resulting in the acceptance amplitude[15]

$$\langle 0|Q_s|0\rangle = \frac{1}{2^n}\sum_y s(y). \tag{2.17}$$

14: We write $0^a = \underbrace{00\cdots 0}_{a}$.

15: For ease of notation, throughout this thesis I use $|0\rangle$, $|0^n\rangle$ and $|0\rangle^{\otimes n}$ interchangeably to denote a product reference state in the computational basis for adequate $n$. In Eq. (2.17) $|0\rangle = |0^{a+n+1}\rangle$.

Notice the subtle difference in the range of the function $g_x$ versus the range of the function $f_x$ arising in classical computation: while $f_x$ is Boolean, $g_x$ takes values in $\{+1,-1\}$. We can view this difference between Boolean and signed functions as a signature of *quantum interference* as it allows for the possibility of cancelling paths famously demonstrated in the Hong-Ou-Mandel experiment which we discussed in detail in the introduction (Chapter 1).

But we can easily translate back and forth between signed and Boolean functions via the map $g'_x(y) := \frac{1}{2}(g_x(y)+1)$ and reexpress

$$\sum_y g_x(y) = \left|\{y : g'_x(y) = 1\}\right| - \left|\{y : g'_x(y) = 0\}\right|. \tag{2.18}$$

Notice that $g'_x$ is again a Boolean #P function. The sum (2.11) can be viewed as the difference between the accepting paths of the function $g'_x$ and its rejecting paths or, in other words, the *gap* of the function, which for a Boolean function $f : \{0,1\}^n \to \{0,1\}$ is defined as

$$\mathrm{gap}(f) = \left|\{y : f(y) = 1\}\right| - \left|\{y : f(y) = 0\}\right|, \tag{2.19}$$

which we normalize to

$$\mathrm{ngap}(f) = \frac{1}{2^n}\mathrm{gap}(f). \tag{2.20}$$

This is why computing those functions is complete for a class called GapP.

**Definition 2.5** (GapP [FFK94]) *Define the* function class GapP *as the class of all functions $f : \{0,1\}^* \to \mathbb{Z}$ for which there exist $g, h \in$ #P such that*

$f = g - h$.

Altogether, we have found that success probabilities of classical circuits are given by the fraction of accepting paths of #P functions, while the success probabilities of quantum circuits can be expressed as the absolute value of normalized gaps of GapP functions as

$$|\langle 0|C_n|0\rangle|^2 = |\operatorname{ngap}(g_0)|^2. \tag{2.21}$$

Again, classical computations are a strict subset of quantum computations as by definition every #P function is also in GapP function. As we will see shortly, the difference in computational power between quantum and classical computations can be traced back to the complexity of approximating functions in the respective class.

How are GapP and #P related in terms of their computational complexity? We have already seen a simple mapping between the two, which implies that computing GapP and #P functions is equivalent under Turing reductions[16] which we write as

$$\mathsf{P}^{\mathsf{GapP}} = \mathsf{P}^{\#\mathsf{P}}. \tag{2.22}$$

16: We write a complexity class X in the exponent of another class Y to mean that a machine in Y can call an *oracle* with access to a machine solving arbitrary problems in the class X at unit time cost.

So in this sense the two classes are very similar. But they actually turn out to be very distinct once we turn to the hardness of *approximating* the respective sums (2.8) and (2.11) up to a multiplicative error $c$.

## 2.2 Approximating GapP

Here, and in the following we distiniguish the following notions of approximation: We say that for $c \in (0, 1]$ an estimator $s$ is a *c-multiplicative approximation* of the value $S$ if

$$cS \leq s \leq S/c. \tag{2.23}$$

We say that for $r > 0$ it is a *r-relative approximation* if

$$(1 - r)S \leq s \leq (1 + r)S, \tag{2.24}$$

and an *$\epsilon$-additive approximation* for $\epsilon > 0$ if

$$|S - s| \leq \epsilon. \tag{2.25}$$

To see why there might be a difference in approximability, notice that a #P sum over $n$-bit strings takes on values between 0 and $2^n$. Typically, the values will therefore be on the order of $2^n$ so that a constant relative error is also of that order. Conversely, GapP sums take on values between $-2^n$ and $+2^n$, but as the corresponding #P function takes on an exponentially large value, the value of the GapP function is the difference between two such exponentially large numbers. This difference will in general be much smaller than each individual value so that the allowed error for relative-error approximation is, too.

Importantly, a relative-error approximation of a quantity is *guaranteed* to get the sign of that quantity right. In contrast to #P functions which always have a positive sign, a relative-error approximation of a GapP function already teaches us correct sign information. That information can in particular be used to decide whether at least half of the paths of a #P function accept, a problem complete for the class PP. And in fact, we can use this information to learn the *exact value* of any GapP function if allowing for *any* multiplicative error.

**Lemma 2.2** (Approximating GapP) *Let $f$ be a* #P *function. Then approximating* $\mathrm{gap}(f)$ *up to* any *constant multiplicative error is* GapP*-hard.*

*Proof.* For the proof, let $f : \{0,1\}^n \mapsto \{0,1\}$ be a function that can be computed by a polynomial-size circuit. Assume that we can approximate $\mathrm{gap}(f)$ for such a Boolean function $f$ up to a constant multiplicative error $c \in (0,1)$ in polynomial time. Let $s$ be the sign of the estimate of $\mathrm{gap}(f)$ returned by the approximation algorithm. Using the approximation algorithm, we can then also estimate the gap of $f_s : \{0,1\}^{n+1} \longrightarrow \{0,1\}$ with

$$f_s(x,y) := \begin{cases} f(x) & \text{if } y = 0 \\ \frac{1}{2}(1-s) & \text{if } y = 1 \end{cases} \tag{2.26}$$

up to the same multiplicative error as, clearly, $f_s$ can still computed by a polynomial-size circuit: conditioned on the additional register, we either apply $f$ or output $(1+s)/2$. The gap of $f_s$ is given by

$$\mathrm{gap}(f_s) = \mathrm{gap}(f) - s2^n, \tag{2.27}$$

as the extra condition ensures that half of all length-$(n+1)$ bit strings contribute to the positive or negative contribution to the gap.

We now write $\mathrm{gap}(f) = b_0 b_1 \dots b_{n-1}$ in terms of bits $b_i \in \{0,1\}$ in its binary representation. By comparing the sign of the approximation to $\mathrm{gap}(f)$ with the sign of the approximation to we can decide whether the first (that is, the most significant) bit $b_0$ of $\mathrm{gap}(f)$ is 1 or 0 since multiplicative-error approximation will always return the correct sign. If the first bit is 0 then $2^n$ will be larger than $\mathrm{gap}(f)$ and hence the sign of $\mathrm{gap}(f_s)$ is flipped. Likewise, if the first bit of $\mathrm{gap}(f)$ is 1 then $2^n$ will be as large as the gap and hence the sign of $\mathrm{gap}(f_s)$ does not change. In this case, $\mathrm{gap}(f_s) = 0$ and therefore the approximation algorithm must return the correct result since $c \cdot 0 = 0$ for any $c$. Hence, if $\mathrm{sign}(\mathrm{gap}(f_s)) = \mathrm{sign}(\mathrm{gap}(f))$ then the first bit is 1, otherwise it is 0. If this bit is 1, we halt as $2^n$ is the maximal possible value of $\mathrm{gap}(f)$. Otherwise we proceed to learn the subsequent bits.

The idea of the learning algorithm is to iteratively construct a function $f_s^{(i)}$ such that by comparing the sign of the gap of $f_s^{(i)}$ with the sign of the gap of $f$, we can learn the $i^{\text{th}}$ bit of that gap. To learn the second bit of $\mathrm{gap}(f)$, we want to construct a function $f_s^{(2)}$ such that $\mathrm{gap}(f_s^{(2)}) = \mathrm{gap}(f) - s2^{n-1}$. To construct this function, we let the outcome be $f(x)$ on half of the inputs, that is, for $y = 0$. On the other half, for which $y = 1$, we then set the outcome of a $3/4$ fraction of input bit strings $x_1 x_2 \dots x_n$ to be $(1-s)/2$. This can be achieved by requiring that $f_s^{(2)}(x,y) = (1-s)/2$ if $y = 1$ and the first two bits of the input string $x$ satisfy either $x_1 = 0$, capturing a $1/2$ fraction of bit strings, or $x_1 x_2 = 01$, capturing another $1/4$ fraction of bit strings. This yields a gap of $\mathrm{gap}(f) + s(1/2 + 1/4 - 1/4)2^n = s2^{n-1}$. From the sign of this we can then learn the second bit of $\mathrm{gap}(f)$.

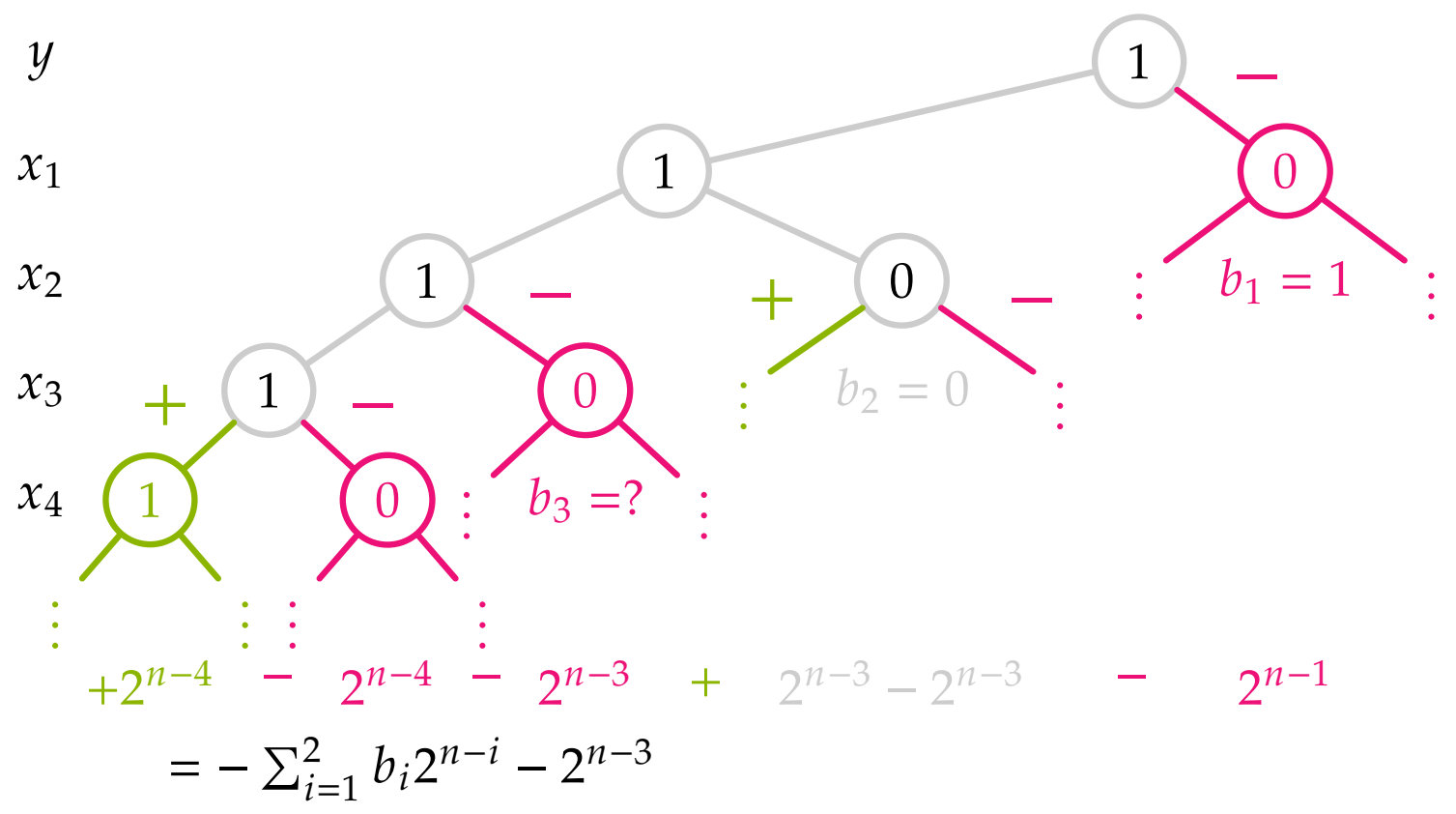

**Figure 2.1:** We construct the function $f_s^{(i)}$ such that in the $i^{\text{th}}$ step of the algorithm, its gap is given by $\text{gap}(f) - s\sum_{k=1}^{i-1} b_k 2^{n-k} - s2^{n-i}$. Here, we show this in terms of the tree of possible inputs in the third step of the algorithm in which the first and second (nonsignificant) bit of $\text{gap}(f)$ are given by $b_1 = 1$, and $b_2 = 0$. All leaves belonging to a branch labeled with a $\pm$ contribute with this sign to the gap of $f_s^{(i)}$ resulting in the shown total contribution (cf. Eq. (2.29)).

We proceed by setting the outcome of the correct fraction of inputs to be $(1-s)/2$. In the $i^{\text{th}}$ step of the learning procedure we apply the $\mathsf{GapP}$ approximation algorithm to the function $f_s^{(i)} : \{0,1\}^{n+1} \to \{0,1\}$ defined by

$$f_s^{(i)}(x,y) = \begin{cases} f(x) & \text{if } y = 0 \\ \frac{1}{2}(1-s) & \text{if } y = 1 \text{ and if} \\ & \text{for } k \in \{1,\dots,i-1\}: \\ & \quad \{(x_1x_2\dots x_k = 1\dots 10 \wedge b_k = 1) \\ & \vee(x_1x_2\dots x_{k+1} = 1\dots 110 \wedge b_k = 0)\} \\ & \vee(x_1\dots x_i = 11\dots 10) \\ & \vee(x_1 \dots x_{i+1} = 11\dots 10) \\ 1 \oplus \frac{1}{2}(1-s) & \text{else,} \end{cases} \tag{2.28}$$

where we denote the $k^{\text{th}}$ bit of $\text{gap}(f)$ by $b_{k-1}$ and binary addition modulo 2 by $\oplus : (a,b) \mapsto a \oplus b := (a+b) \mod 2$. Those functions are still efficiently computable as there are at most $O(n)$ many conditions. The gap of $f_s^{(i)}$ is then given by

$$\text{gap}(f_s^{(i)}) = \text{gap}(f) - \sum_{k=1}^{i-1} b_k 2^k - s2^i, \tag{2.29}$$

so that in this step we learn the $i^{\text{th}}$ bit of $\text{gap}(f)$. The algorithm ends after evaluating $f_s^{(n-1)}$. □

If for a function class $\mathsf{X}$ we define $\mathsf{Apx}_{\cdot c}\mathsf{X}$ as the class of problems to approximate $\sum_x f(x)$ up to a multiplicative error $c$ for $f \in \mathsf{X}$, then we have now found that for any $c \in (0,1)$

$$\mathsf{P}^{\mathsf{Apx}_{\cdot c}\mathsf{GapP}} = \mathsf{P}^{\mathsf{GapP}}. \tag{2.30}$$

The attentive reader will have noticed that in our discussion of the hardness of approximating $\mathsf{GapP}$ using the sign information we have glossed over the fact that, of course, success probabilities of quantum circuits are *nonnegative*. And indeed, it seems unlikely that those success probabilities are hard to approximate up to *any constant multiplicative error*.

Nevertheless, using a similar proof strategy one can prove GapP-hardness of approximations for the square of the output amplitudes of quantum circuits [TD04; AA13; GG14; FM17]. This strategy notices that not only do multiplicative-error approximations get the sign correct, but certainly also the instances in which the true value is exactly zero.

**Lemma 2.3** (Approximating quantum output probabilities [BMS16]) *Let $f$ be a #P function. Then approximating* $\mathrm{ngap}(f)^2$ *up to any relative error* $\epsilon < 1/2$ *is* GapP*-complete.*

*Proof sketch.* [17] In the first step, we show that there is a polynomial-size circuit $C$ computing a function $f_c$ such that $\mathrm{ngap}(f_c) = (\mathrm{ngap}(f) - c)/2$ . To this end we make use of the following: for any $m \in O(n)$ and $c \in [-1, 1]$ such that $c = 2k/2^m$ there is a polynomial-size circuit $D_c$ computing a function $g$ such that $\mathrm{ngap}(g) = -c$. Now consider the polynomial-size circuit $Q_c$ acting on $m + n + 1$ registers which executes either $C$ or $D_c$ depending on the control register. This circuit computes a function $f_c$ as desired.

17: We proceed similarly to the proof of Lemma 2.2 above, following Bremner, Montanaro, and Shepherd [BMS16, Proposition 8]. The idea of the proof is to estimate $\mathrm{ngap}(f)$ by using the fact that given a guess $c$ an algorithm that ouputs relative-error approximation to $|\mathrm{ngap}(f) - c|$ can certify the correctness of $c$.

Assume we have an efficient algorithm $\mathcal{A}$ that given a polynomial-size circuit $C$ approximates $|\langle 0|H^{\otimes n} C H^{\otimes n}|0\rangle|^2 = \mathrm{ngap}(f_c)$ up to relative error $\epsilon < 1$. On input $Q_c$ this machine can certify whether $\mathrm{ngap}(f) = c$. We now use $\mathcal{A}$ to estimate $\mathrm{ngap}(f)$ until we have found its exact value. At each step, we have a guess $c_i$ for $c$, starting with $c_0 = 0$. We use $\mathcal{A}$ to output an estimate $d_i$ to $|\mathrm{ngap}(f) - c_i|$ and then apply it again to $|\mathrm{ngap}(f) - (c_i \pm d_i)|$ obtaining $d_i^{\pm}$. Define $c_{i+1} = c_i + d_i$ if $d_i^+ \leq d_i^-$ and as $c_i - d_i$ otherwise.

The algorithms acts contractively: assuming $c < \mathrm{ngap}(f)$ we find that an estimate $d = (1 + \gamma)|c - \mathrm{ngap}(f)|$ for some $|\gamma| < \epsilon$ satisfies

$$|c + d - \mathrm{ngap}(f)| = |\gamma(\mathrm{ngap}(f) - c)| \leq \epsilon |c - \mathrm{ngap}(f)|, \tag{2.31}$$

and a similar inequality holds for $c - d$ if $c > \mathrm{ngap}(f)$. Consequently, since $\mathrm{ngap}(f)$ is an integer multiple of $2/2^n$, if the correct choice of $c \pm d$ is made in each step, the algorithm halts after $O(n)$ many steps.

It remains to be shown that this is indeed the case. This can be seen from the equivalence

$$(1 + \epsilon)|c + d - \mathrm{ngap}(f)| < (1 - \epsilon)|c - d - \mathrm{ngap}(f)| \tag{2.32}$$

$$\Leftrightarrow (1 + \epsilon)|\gamma| < (1 - \epsilon)|2 + \gamma|, \tag{2.33}$$

which holds for $|\gamma| \leq \epsilon < 1/2$.

The same argument immediately holds for $|\mathrm{ngap}(f)|$ as now we have not used the sign of $\mathrm{ngap}(f)$. □

## 2.3 Approximating #P: Stockmeyer's algorithm

For many #P-complete problems such as computing the value of the permanent of a matrix taking values in $\{0, 1\}$, there are efficient randomized approximation schemes, so called FPRAS [JSV04]. Many such algorithms for approximate counting are based on Markov-chain Monte Carlo methods [JVV86; JS90], which we will discuss in more detail in Chapter 9. The property that those algorithms exploit is the fact that each element of the sum (2.8) is nonnegative. Thus, the sum can be estimated by importance sampling. Insofar, the intricate sign structure of GapP functions is what makes their relative-error approximation via such sampling algorithms hard. We will return to investigate Monte Carlo algorithms and the question of how their applicability can be extended in more detail in Part III.

Going beyond specific problems, in this section, we will get to know a powerful general result on the approximability of such functions. More specifically, we will get to know a computationally restricted hypothetical algorithm due to Stockmeyer [Sto83]. Stockmeyer's algorithm lies in the so-called third level of the polynomial hierarchy. This class is much more powerful than NP, for example, but much less powerful than #P. Stockmeyer's algorithm is able to approximately count the number of accepting paths of #P functions (up to small multiplicative errors) even though it is not able to exactly compute this number. It thus provides a rigorous foundation for the distinction between the approximability of GapP and #P. In the next section (Sec. 2.4), we will leverage the power of this algorithm to derive rigorous separations between classical and quantum *sampling algorithms*.

Before we are able to make those statements precise, however, we need to dive a little further into the depths of computational complexity theory and define what is called the *polynomial hierarchy*. Stockmeyer's algorithm lies within the polynomial hierarchy but at the same time the hierarchy is contained in the class #P (informally speaking).

## The polynomial hierarchy

In the introduction, we have already seen the most important classes in the theory of computational complexity, namely, P and NP. It is no exaggeration to say that the conjecture that $\mathsf{P} \subsetneq \mathsf{NP}$ is indeed one of the if not *the* most tested and studied unproven statements that scientists across a range of disciplines are confident in. It is a generalization of this statement that will form the complexity-theoretic grounding of claims to quantum supremacy. This generalization – the *polynomial hierachy* – comprises an infinite hierarchy of complexity classes that, so the conjecture, are strict subsets of one another. Considering hypothetical algorithms within and outside of this hierarchy will also allow us to understand the computational complexity of approximating #P functions.

To appreciate this generalization let us add to our definition of NP a definition of its sibling called coNP.

**Definition 2.6** (coNP [AB09]) *A language $L \subset \{0,1\}^*$ is in* coNP *if there exists a polynomial $p : \mathbb{N} \to \mathbb{N}$ and a uniform polynomial-time circuit family $\{C_n\}_{n\geq 1}$ (called the* verifier *of $L$) such that for every $x \in \{0,1\}^*$*

$$x \in L \quad \Leftrightarrow \quad \forall u \in \{0,1\}^{p(|x|)} \; s.t. \; C_{|x|}(x,u) = 1. \tag{2.34}$$

Notice that while in NP we asked for *the existence of* a certificate to prove the truth of a statement, in coNP, in contrast, a counterexample is sufficient to certify its falsity. Hence, we can equivalently define coNP as $\mathsf{coNP} = \{L : \overline{L} \in \mathsf{NP}\}$, where by $\overline{L} = \{0,1\}^* \setminus L$ we define the complement of $L$. Nevertheless, coNP is *not* the complement of NP [AB09] and in fact the two have a non-empty intersection. It is easy to see that if P = NP then coNP = NP = P[18].

18: In the contrapositive, proving that coNP ≠ NP implies that P ≠ NP.

Building on those classes one can in fact construct an entire hierarchy of complexity classes that is called the *polynomial hierarchy*. The basic idea behind this hierarchy lies in the subtle difference between the ∃ and ∀ quantifiers in the definitions of NP and coNP, respectively: if we add alternating ∃ and ∀ quantifiers to our definitions of NP and coNP, respectively, surely we will strictly increase the computational power of the resulting complexity classes. And indeed, for the lowest level this intuition is well tested, it amounts to the conjecture P ≠ NP.

**Definition 2.7** (The polynomial hierarchy [AB09]) *For $i \in \mathbb{N}$ a language $L \subset \{0,1\}^*$ is in $\Sigma_i^p$ if there exists a polynomial $q$ and a uniform polynomial-time circuit family $\{C_n\}_{n\geq 1}$ such that*

$$x \in L \Leftrightarrow \exists u_1 \in \{0,1\}^{q(|x|)} \forall u_2 \in \{0,1\}^{q(|x|)} \cdots Q_i u_i \in \{0,1\}^{q(|x|)} : \\ C_{|x|}(x, u_1, \ldots, u_i) = 1, \tag{2.35}$$

*where $Q_i$ denotes a ∀ or ∃ quantifier depending on whether $i$ is even or odd, respectively. The* polynomial hierarchy *is the set* $\mathsf{PH} = \cup_i \Sigma_i^p$.

Clearly $\Sigma_i^p \subset \Sigma_{i+1}^p$. The intuition appealed to above is made firmer in the conjecture that "the polynomial hierarchy is infinite", that is, every level strictly contains the previous levels. Stated in other words, the conjecture is that "the polynomial hierarchy does not collapse". An alternative formulation of the polynomial hierarchy is as a hierarchy of complexity classes where in each level an additional NP-oracle is added, see Figure 2.2. In particular, we can express $\Sigma_1^p = \mathsf{NP}$, $\Sigma_2^p = \mathsf{NP}^{\mathsf{NP}}, \ldots$ and the hierarchy of classes (see Fig. 2.2)

$$\Delta_i^p = \mathsf{P}^{\Sigma_{i-1}^p}. \tag{2.36}$$

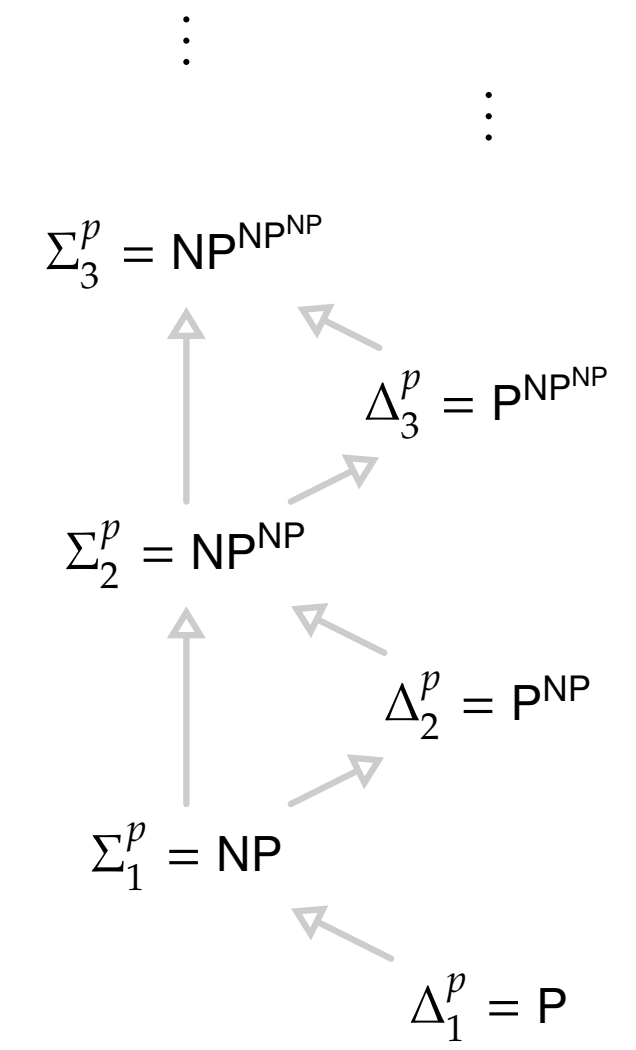

**Figure 2.2:** The polynomial hierarchy is a hierarchy of complexity classes that can be seen as arising from one other by subsequent oracularization. We only show the left part of the entire ladder. Inverting the order of ∃ and ∀ quantifiers in Def. 2.7, its right part arises as coNP, $\mathsf{coNP}^{\mathsf{coNP}}$, etc.

## Stockmeyer's approximate counting algorithm

Indeed, it is no surprise that, given access to NP oracles one can solve an enormously rich class of computational problems. Nevertheless, it is quite surprising that one can efficiently approximate exponentially large sums up to any inverse polynomial *multiplicative error*. Stockmyer's approximate counting algorithm [Sto83] achieves this task in a low level of the polynomial hierarchy – the third level. We are now ready to state this result.

**Theorem 2.4** ([Sto83; AA13] ) *Given a Boolean function $f : \{0,1\}^n \to \{0,1\}$, let*

$$p = \Pr_{x \in \{0,1\}^n}[f(x) = 1] = \frac{1}{2^n} \sum_{x \in \{0,1\}^n} f(x). \tag{2.37}$$

*Then for all $c \geq 1 + 1/\mathrm{poly}(n)$, there exists a $\Delta_3^p$ machine that approximates $p$ to within a multiplicative factor of $c$.*

For the sake of completeness, let me sketch the proof of Stockmeyer's [Sto83] theorem.

*Proof.* Let $a(x) = |\mathrm{Acc}_C(x)|$ be the function that given an input $x \in \{0,1\}^*$ counts the number of accepting computations collected in the set $\mathrm{Acc}_C(x)$ of some polynomial-time computation $C$. The basic idea of the proof is to use random *hash functions*[19] $h : \{0,1\}^t \longrightarrow \{0,1\}^m$ with $m$ much smaller than $t$ and to check for collisions, that is, points $z \neq z' \in \mathrm{Acc}_C(x) \subset \{0,1\}^m$ for which

19: The idea behind a hash function is to map data sets of large sizes into a few number of categories or *hashes* so as to perform data storage and retrieval more efficiently. Here, they prove to be a useful proof technique.

$$h(z) = h(z'). \tag{2.38}$$

The idea beyond the proof is the following: if the set $\mathrm{Acc}_C(x)$ is small, the probability that a collision occurs is small, too. If it is large, that probability will be large. To determine whether there is a collision we will make use of an NP oracle.

To be more precise, define the following Hash predicate.

**Definition 2.8** $\mathtt{Hash}(x, m)$:
*There exist functions $h_1, \dots, h_m : \{0,1\}^t \longrightarrow \{0,1\}^m$ such that*

1. *Each $h_i$ is computed by a Boolean circuit of size $\le mt$.*
2. $\forall z \in \mathrm{Acc}_C(x)\, \exists i : \forall z' \in \mathrm{Acc}_C(x) \setminus \{z\}$

$$h_i(z) \neq h_i(z')\ . \tag{2.39}$$

The Hash predicate is true if there exists a collection of hash functions such that for each pair of points in $\mathrm{Acc}_C(x)$ at least one of the hash functions does not have a collision at that point. In this sense, those functions then *separate* $\mathrm{Acc}_C(x)$ within itself.

The remaining part of the proof deals with carefully bounding the probability of Hash being true. The key ingredient for this is the *coding lemma* due to Sipser [Sip83]. By log we denote the logarithm to basis 2.

**Lemma 2.5** (Coding lemma [Sip83]) *Let $A \subset \{0,1\}^n, k = |A|, m = 1 + \lceil \log k \rceil$. If $H$ is a collection of $m$ randomly chosen linear transformations $h : \{0,1\}^n \longrightarrow \{0,1\}^m$ then*

$$\begin{aligned} &\Pr_H[\forall x \in A\, \exists h \in H\, \forall x \neq y \in A : \ h(x) \neq h(y)] \\ &\quad = \Pr_H[H \textit{ separates } A \textit{ within } A] \geq 1/2 \end{aligned} \tag{2.40}$$

In particular, as it gives a lower bound to the probability over randomly chosen families $H$, the coding lemma states that there exists such a collection $H$ separating $A$ within $A$.

Let us now apply this fact to the set $\mathrm{Acc}_C(x)$.

**Lemma 2.6** *There exists a constant $c$ such that*

$$|\mathrm{Acc}_C(x)| \leq 2^{m-c} \quad \Rightarrow \quad \mathtt{Hash}(x, m) \tag{2.41}$$

$$|\mathrm{Acc}_C(x)| \leq m 2^{m} \quad \Leftarrow \quad \mathtt{Hash}(x, m) \tag{2.42}$$

*Moreover* $\mathtt{Hash}(x, m) \in \Sigma_2^p$.

*Proof.* (2.41): By the coding lemma, if $\mathrm{Acc}_C(x) \leq 2^{m-c}$ then there exists an $H$ that separates $\mathrm{Acc}_C(x)$ within itself of size $\ell = 1 + m - c$, i.e., $\mathtt{Hash}(x, \ell)$. In particular, for $c = 1$, $\mathtt{Hash}(x, m)$, and since $\mathtt{Hash}(x, \ell) \Rightarrow \mathtt{Hash}(x, \ell + k)$ for all $k \geq 0$ the statement holds.

(2.42): If $\mathsf{Hash}(x, m)$ is true, then by its definition for each $z \in \mathrm{Acc}_C(x)$ there exists a unique pair $(h, h(z))$. Since there are at most $|H| \cdot 2^m = m2^m$ unique such pairs the statement follows.

To see that $\mathrm{Hash}(x, m) \in \Sigma_2^p$ observe that the quantifier $\exists i$ can be removed and replaced by a deterministic search as it only has polynomial range $m = O(\mathsf{poly}(|x|))$. What remains is that we ask whether there exists a collection of functions $h_1, \ldots, h_m$ such that for all $z, z' \in \mathrm{Acc}_C(x)$ $h_i(z) \neq h_i(z')$. This task can be solved by a machine in $\Sigma_2^p = \mathsf{NP}^{\mathsf{NP}}$ as it has the characteristic $\exists\forall$ form of alternating quantifiers. In other words, an NP machine can determine the existence of such functions by calling an additional NP oracle asking the question: "Are there two distinct $z$ and $z'$ such that $h(z) = h(z')$ and both are in $\mathrm{Acc}_C(x)$?" Clearly, this task is again in NP, since a pair $z, z'$ for which the statement is true provides an efficient proof for it. □

Hence a polynomial-time machine in P making calls to an oracle for Hash can determine the minimal $m$ such that $\mathrm{Hash}(x, m)$ is true via a binary search of the type considered in the proof of Lemma 2.2. We then find

$$2^{m-c-1} \leq |\mathrm{Acc}_C(x)| \leq m2^m\,, \tag{2.43}$$

so that we have obtained a $m2^{c+1}$ multiplicative approximation of $|\mathrm{Acc}_C(x)|$.

This approximation can be further amplified[20] to $(m2^{c+1})^{1/\alpha} = 1 + O(1/\alpha)$ for some $\alpha = \mathsf{poly}(n)$ by executing this search $\alpha$ times in series for the function $f(x)^\alpha$. In total, we have obtained a $1 + 1/\mathsf{poly}(n)$-multiplicative approximation within the third level $\Delta_3^p = \mathsf{P}^{\mathsf{NP}^{\mathsf{NP}}}$ of the polynomial hierarchy. □

20: R.J. Lipton: *Stockmeyer's Approximate Counting Method*, Blog post @ `https://rjlipton.wordpress.com/2009/08/27/stockmeyers-approximate-counting-method`. (August 27, 2009.)

## The complexity of Stockmeyer's algorithm

Stockmeyer's Theorem 2.4 characterizes the complexity of approximately counting up to an inverse polynomially small multiplicative error: this task lies within the third level $\Delta_3^p \subset \mathsf{PH}$ of the polynomial hierarchy. But where does this complexity class lie in relation to exactly computing a #P sum? For the answer, we refer to a final fact in complexity theory, namely that exactly computing #P functions lets one solve *any task* in PH.

**Theorem 2.7** (Toda's theorem [Tod91])

$$\mathsf{PH} \subset \mathsf{P}^{\#\mathsf{P}}. \tag{2.44}$$

Multiplicative approximations therefore significantly ease the task of counting #P sums. Conversely, we have already seen above in Eq. (2.22) and Lemma 2.2 that GapP does not change its complexity under multiplicative approximations so that $\mathsf{P}^{\mathsf{GapP}} = \mathsf{P}^{\#\mathsf{P}} \supset \mathsf{PH}$ remains above the polynomial hierarchy. Altogether, we find that for any constant $c > 0$ the following inclusions hold

$$P^{\mathsf{Apx}_c\#\mathsf{P}} \subset \Delta_3^p \subsetneq \mathsf{PH} \subset \mathsf{P}^{\mathsf{GapP}} = \mathsf{P}^{\mathsf{Apx}_c\mathsf{GapP}}, \tag{2.45}$$

where the separation $\Delta_3^p \subsetneq \mathsf{PH}$ marks the conjectured non-collapse of the polynomial hierarchy to a low level. The same inclusions hold true for $c < 1/2$ when restricted to GapP-functions with nonnegative gap.

**Approximating GapP vs. #P: the quantum sign problem**

We conclude that Stockmeyer's algorithm (Theorem 2.4) together with Lemma 2.2 provides strong evidence for an exponential separation between the hardness of approximating GapP and #P functions.

This substantial difference in the classical hardness of approximating quantum vs. classical acceptance probabilities is often termed the *sign problem* of quantum mechanics.

We will return to a more thorough investigation of the sign problem in Chapters 9–12. There, we will study a practical manifestation of the sign problem and the question, whether it is intrinsic to a given problem or artificially introduced by our lack of imagination when describing quantum systems.

---

We have now carved out a substantial difference in complexity between quantum and classical randomized algorithms in terms of the computational complexity of approximating the respective acceptance probability. But neither type of algorithm is able to multiplicatively approximate its own acceptance probability. Nevertheless, this difference in complexity serves as an important tool using which we can attenuate harder-to-grasp differences in the runtime of actual classical and quantum algorithms. Following this route, we will arrive at a (conditional) exponential separation for sampling tasks.

## Sampling versus approximating outcome probabilities

We are now in the position to leverage the difference in complexity for approximating the output probabilities of randomized quantum and classical algorithms to a more tangible task that can actually be carried out on quantum devices. More specifically, our goal in this section will be to prove that not only is there an exponential quantum/classical divide in *approximating output probabilities* of computations, but also that this divide reappears when it actually comes to *performing such computations*. Randomized algorithms, indeed, seem naturally suited to attenuate a hypothetical quantum/classical divide. While of randomized classical computations we merely conceive as deterministic computation supplemented with random inputs, quantum computations are naturally probabilistic: Whenever we perform a quantum measurement described by measurement operators $\{M_x\}_{x\in\mathcal{X}}$ for some index set $\mathcal{X}$ with the property that all $M_x \geq 0$ and $\sum_{x\in\mathcal{X}} M_x = 1$ to obtain a classical outcome from a quantum state $\rho$, we sample from a probability distribution determined by the Born rule

$$p_x(\rho) = \mathrm{Tr}[M_x \rho]\,. \tag{2.46}$$

Quite naturally, quantum computations are therefore nothing but sampling algorithms for a distribution which is determined by the Born rule. Of course, in a useful computation it is desirable for the distributions to be rather peaked on a deterministic outcome of the computation. Nevertheless, for the sake of attenuating the differences between classical and quantum computation, we might as well allow for arbitrary distributions. Rather, we will consider the task of sampling for its own sake, not caring about a specific outcome of the computation. To be able to apply the machinery of complexity theory and Stockmeyer's algorithm in particular, it will prove useful to consider the task of sampling from *randomly chosen quantum computations*. This will allow us to talk about problem classes – from which a random instance is drawn – and their hardness rather than specific instances about which one can rarely make a complexity-theoretic statement.

This brings us to the heart of the proof of quantum supremacy: the relation between the tasks of approximating outcome probabilities of classical and quantum sampling algorithms and the task of actually performing those algorithms. In the proof, we will leverage the separation between classical and quantum computations with regards to approximating success probabilities to results about simulating such computations classically. The key ingredient that we will use when relating the very different tasks of sampling from a distribution and computing its output probabilities will be Stockmeyer's algorithm.

Indeed, we observe that Stockmeyer's counting theorem 2.4 can be directly applied to estimate the acceptance probability, and in fact all output probabilities, of so-called *derandomizable sampling algorithms*, which are deterministic algorithms with random inputs as discussed above [AA13, Def. 24].

**Definition 2.9** (Derandomizable sampling) *A* derandomizable sampling algorithm *is an algorithm $\mathcal{A}$ that takes as an input a particular instance $y \in \{0,1\}^n$ of a problem, an error bound $\epsilon > 0$ in terms of a binary* $\mathsf{poly}(|y|)$*-bit approximation $0^{1/\epsilon}$ thereof, as well as a uniformly random string $r \in \{0,1\}^{\mathsf{poly}(|y|)}$ and outputs a random bit string $x = \mathcal{A}_{y,\epsilon}(r)$ distributed according to*

$$p_{y,\epsilon}(x) = \Pr_r[\mathcal{A}_{y,\epsilon}(r) = x]\,. \tag{2.47}$$

If $\mathcal{A}_{y,\epsilon}$ is such a derandomizable algorithm we can use Stockmeyer's algorithm to estimate its output probabilities (2.47). To do so, we define its input function as

$$\begin{aligned} f_{y,\epsilon} : \{0,1\}^{\mathsf{poly}(|y|)} &\longrightarrow \{0,1\} \\ r &\mapsto \begin{cases} 1 & \text{if } \mathcal{A}_{y,\epsilon}(r) = x \\ 0 & \text{else} \end{cases} . \end{aligned} \tag{2.48}$$

The output of Stockmeyer's approximation algorithm will then be a $1 - 1/\mathsf{poly}(|y|)$-multiplicative approximation to the probability

$$\frac{1}{2^{\mathsf{poly}(|y|)}} \sum_{r \in \{0,1\}^{\mathsf{poly}(|y|)}} f_{y,\epsilon}(r) = p_{y,\epsilon}(x). \tag{2.49}$$

This provides the sought for connection between sampling and approximation of probabilities that forms the basis of the proofs of sampling hardness below.

## Some history

The basic idea of relating the complexity of sampling to the complexity of computing probabilities goes back to Terhal and DiVincenzo [TD04]. Terhal and DiVincenzo [TD04] showed not only that an efficient approximation algorithm for the outcome probabilities of quantum circuits would imply a collapse of the polynomial hierarchy. They also proved that an efficient algorithm to compute all marginal probabilities of a probability distribution implies an efficient sampling algorithm for that distribution, an idea that was later formalized by Nest [Nes11]. Conversely, they were the first to explore the complexity-theoretic consequences of the existence of efficient algorithms for sampling from the output distribution of certain very restricted families of quantum circuits, showing that the existence of such algorithms would imply that $\mathsf{BQP} \subset \mathsf{AM}$, an unlikely consequence[21].

21: The complexity class AM is the class of decision problems for which there exists a two-round interactive protocol in which a verifier generates a challenge and a prover returns a response that can be evaluated efficiently in BPP.

In introducing nonadaptive quantum circuit families Terhal and DiVincenzo [TD04] implicitly introduced the idea of *postselection*: even if certain outcomes are exponentially unlikely the presence of a success flag can often have surprising implications. Those implications were further studied by Aaronson [Aar05], who proved that the power of postselected quantum computing is equal to that of exact probabilistic classical computation PP.

**Definition 2.10** (PP [AB09] ) *A language $L$ is in the complexity class* PP *if for $x \in \{0,1\}^*$ there exists a polynomial $p : \mathbb{N} \to \mathbb{N}$ and a* #P *function $f_x : \{0,1\}^{p(|x|)} \to \{0,1\}$ such that*

$$x \in L \Leftrightarrow \mathbb{E}_y[f_x(y)] \geq 1/2. \tag{2.50}$$

In other words, PP asks to decide whether the fraction of inputs accepted by a polynomial-size computation is larger than 1/2. It is easy to see that $\mathsf{P}^{\mathsf{PP}} = \mathsf{P}^{\#\mathsf{P}}$ since computing a #P sum clearly lets you decide a PP task, and vice versa, one can compute the sum bit by bit using few calls to a PP oracle.

Shepherd and Bremner [SB09] and Bremner, Jozsa, and Shepherd [BJS10] then introduced commuting quantum computations – instantaneous quantum polynomial time (IQP) – as a particularly simple paradigm for quantum computing. They first related the hypothetical existence of efficient sampling algorithms for the output distributions of such computations to a collapse of the polynomial hierarchy (PH). Their proof made use of the older result by Aaronson [Aar05] and Toda's theorem [Tod91] on the relation between PP and PH. At about the same time Aaronson and Arkhipov [AA13] provided a similar proof for the classical hardness of sampling from the output distribution of yet another restricted circuit family, namely, boson sampling. While Bremner, Jozsa, and Shepherd [BJS10] could only make statements about near-exact sampling algorithms based on worst-case hardness results for the strong simulation of IQP circuits, Aaronson and Arkhipov [AA13] leveraged average-case hardness of matrix permanents [Val79; Lip91] to prove the hardness of *approximate sampling* in total-variation distance for the case of boson sampling, albeit with additional conjectures. A modern version of their proof has been nicely formulated for the case of IQP by Bremner, Montanaro, and Shepherd [BMS16] and it is this version of the proof that will form the basis of my presentation in this chapter.

## 2.4 Quantum random sampling

In what follows, we will provide strong complexity-theoretic evidence for the classical hardness of performing random computations from certain families of quantum circuits. Such families may be defined by random choices of individual quantum gates within a fixed architecture. Given a problem size $n$, the task is then to sample from the output distribution $p(C)$ of a circuit $C$ chosen at random from a circuit family $\mathcal{C}_n$ and applied to a reference state $|0\rangle$. The probability of obtaining outcomes $S \in \Omega$ from the corresponding sample space $\Omega$ is then given by[22]

$$p_S(C) = |\langle S|C|0\rangle|^2. \qquad (2.51)$$

22: Depending on whether the emphasis lies on the probability distribution over the circuits $C$ or the outcomes $S$ of a fixed unitary, I will at times use $p_C(S)$ and at other times $p_S(C)$ for the outcome probabilities.

**Universal circuit sampling** For example, we can construct a family of circuits starting from a set $\mathcal{G}$ comprising two-qubit gates. A depth-$N$ circuit $C \in \mathcal{C}_{\mathcal{G},N}$ acting on $n$ qubits might then be constructed by choosing a random gate in $G \in \mathcal{G}$ and the pair of qubits it is applied to at random [BHH16]. Alternatively, we could imagine a parallel architecture in which each layer of the circuit comprises random gates from $\mathcal{G}$ applied in parallel (see Fig. 2.3(a)), or even a much more architecture-specific prescription such as the one of Boixo et al. [Boi+18].

A particularly important family of random circuits are random universal circuits, as defined by a universal gate set $\mathcal{G}$ comprising one- and two-qubit gates.

**IQP circuit sampling** Another prominent family of circuits are IQP circuits [SB09]. An IQP circuit is a commuting quantum circuit diagonal in the Hadamard basis, that is, a circuit $C = H^{\otimes n} D H^{\otimes n}$, where $D$ is diagonal in the standard basis; see Fig. 2.3(b). Examples of IQP circuit families are defined by diagonal circuits comprised of diagonal 2-qubit gates with arbitrary phases on the diagonal [NKM14], circuits of $Z$, $CZ$, and $CCZ$ gates [BMS16] and the time evolution under Ising Hamiltonians with arbitrary edge weights [BMS16], and generalizations thereof to arbitrary multi-qubit interactions – so-called $X$-programs [SB09].

The families most important to this thesis are the ones introduced by Bremner, Montanaro, and Shepherd [BMS16]. An instance $C_f$ of the first family is defined by a degree-3 Boolean polynomial $f : \{0,1\}^n \to \{0,1\}$ over the field $\mathbb{F}_2 = (\{0,1\}, \oplus, \cdot)$.

$$f(x) = \sum_{i,j,k} \alpha_{i,j,k} x_i x_j x_k + \sum_{i,j} \beta_{i,j} x_i x_j + \sum_i \gamma_i x_i, \qquad (2.52)$$

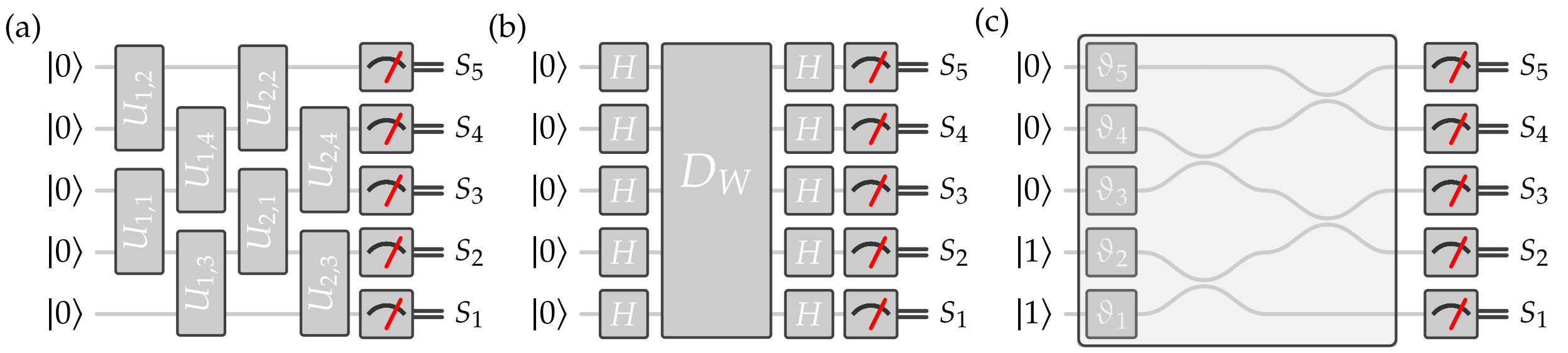

**Figure 2.3:** Circuit diagrams for *(a)* parallel random universal circuits with random gates $U_{i,j} \in \mathscr{G}$ acting on qubits $j, j+1$ in the $i$th layer of the circuit, *(b)* IQP circuits which act diagonally as $D_W$ definied by weights $w_{i,j}$ in the $X$-basis, and *(c)* boson sampling, where passive linear optics comprising beam splitters and phase shifters are applied to a Fock input state $|1^n\rangle|0^{m-n}\rangle$ and then measured in the Fock basis with outcomes $S_i$.

with Boolean coefficients $\alpha_{ijk}, \beta_{ij}, \gamma_i \in \{0,1\}$ denoting whether or not a $CCZ$, $CZ$ and $Z$ gate is applied to qubits $(i,j,k)$, $(i,j)$ and $i$, respectively.

*The following paragraphs have been published in Ref. [Han+19].*

The second family is defined by a set of angles $A$, e.g., $A = \{0, \pi/8, \dots, 7\pi/8\}$. An instance $C_W$ with $W := (w_{i,j})_{i,j=1,\dots,n}$ and $w_{i,j} \in A$ drawn uniformly at random, is then given by the following prescription

$$C_W = H^{\otimes n} D_W H^{\otimes n} = \exp\left[\mathrm{i}\left(\sum_{i<j} w_{i,j} X_i X_j + \sum_i w_{i,i} X_i\right)\right], \tag{2.53}$$

where $X_i$ is the Pauli-$X$ matrix acting on site $i$[23]. In other words, on every edge $(i,j)$ of the complete graph on $n$ qubits a gate $\exp(\mathrm{i} w_{i,j} X_i X_j)$ with edge weight $w_{i,j}$ and on every vertex $i$ a gate $\exp(\mathrm{i} w_{i,i} X_i)$ with vertex weight $w_{i,i}$ is performed.

23: On an $n$-fold tensor-product Hilbert space $\mathscr{H}^{\otimes n}$, for $X : \mathscr{H} \to \mathscr{H}$ we define a local operator

$$X_i := \underbrace{1 \otimes \cdots \otimes 1}_{i-1} \otimes X \otimes \underbrace{1 \otimes \cdots \otimes 1}_{n-i},$$

acting on the $i^{\text{th}}$ copy of $\mathscr{H}$.

**Boson sampling** The boson sampling scheme is a generalization of the Hong-Ou-Mandel experiment, which we discussed in the introduction. Recall that in this experiment, single photons interfered on a beam splitter. In the boson sampling problem, we increase the number of photons: $n \geq 1$ photons are injected into the first $n$ of $m \in \mathsf{poly}(n)$ modes which are transformed in a linear-optical network comprising beam splitters and phase shifters via a mode transformation given by a Haar-random unitary $U \in U(m)$ and then measured in the Fock basis (see Fig. 2.3(c)). As unitary mode transformations conserve the photon number, the sample space of boson sampling is given by

$$\Phi_{m,n} := \left\{(s_1, \dots, s_m) : \sum_{j=1}^{m} s_j = n\right\}, \tag{2.54}$$

i.e., the set of all sequences of non-negative integers of length $m$ which sum to $n$. Its output distribution is

$$p_S(U) \equiv P_{\mathrm{bs},U}(S) := |\langle S|\varphi(U)|1_n\rangle|^2. \tag{2.55}$$

Here, the state vector $|S\rangle$ is the Fock space vector corresponding to a measurement outcome $S \in \Phi_{m,n}$, $|1_n\rangle$ is the initial state vector with $1_n := (1, \dots, 1, 0, \dots, 0)$, and $\varphi(U)$ the Fock space (metaplectic) representation of the mode transformation $U$.

**Quantum random sampling schemes**

In a random circuit sampling scheme we consider a family of quantum circuits $\mathscr{C} = \{\mathscr{C}_n\}_{n\geq 1}$ from which a random $n$-qubit instance $C \in \mathscr{C}_n$ is drawn uniformly at random. The task is to sample from the output distribution $p_C$ of $C$ as defined by $p_S(C) = |\langle S|C|0\rangle|^2$ for outcomes $S \in \Omega$.

Equivalently, we can conceive of this as defining a measure $\mu$ over the entire $n$-qubit unitary group $U(2^n)$.

Since the first quantum random sampling schemes – IQP sampling [BJS10] and boson sampling [AA13] – were conceived, many more proposals for random circuit sampling schemes have been put forward. A theoretically particularly clear proposal is Fourier sampling [FU15], which is a qubit analogue of boson sampling. The one clean qubit (DQC1) model is a model in which all but one qubit are initialized in the maximally mixed state [Fuj+14; MFF14; Mor17]. This model is motivated by mixed-state quantum computations, which is suitable to describe, for instance, nuclear magnetic resonance quantum processors [Neg+05]. Other proposals are motivated by the study of certain practically relevant circuit architectures such as Clifford circuits with magic-state inputs [YJS19], Clifford circuits which are conjugated by arbitrary product unitaries [BFK18], permutations of distinguishable particles in specific conditions [Aar+16], and architectures motivated by large-scale quantum simulators [GWD17; Ber+18]. Finally, certain models were also proposed with the goal to close loopholes such as the necessity to certify the correct implementation of a quantum supremacy experiment [Han+17; MSM17], a question we will return to in Part II, or to make such an experiment more error-tolerant [Fuj16; KD19].

## Strongly simulating random quantum computations

For the examples presented above, we know that approximating output probabilities is a $\mathsf{GapP}$-hard task and thus just as hard as for arbitrary quantum computations.

Generally, and this is in particular true for universal random circuits, the output probabilities of a circuit family $\mathscr{C}$ are $\mathsf{GapP}$-hard to approximate if the circuit family generates the whole of $\mathsf{BQP}$ after so-called postselection [FM17]. In a postselection argument we compare two probabilistic complexity classes by allowing ourselves the ability to restrict attention to a certain subset of desired outcomes even if that subset has exponentially small probability. A postselected class $\mathsf{postA}$ is defined as a class of decision problems which we can solve by using a computation within $\mathsf{A}$ and postselecting on certain outcomes with a bounded error [FM17].

**Definition 2.11** (Postselected class [FM17]) *A language $L$ is in the class $\mathsf{postA}$ if there exists a uniform family of circuits $\{C_x\}$ associated with $\mathsf{A}$ for which there are a single output register $O_x$ and a $\mathsf{poly}(|x|)$-size postselection register $P_x$*

*i. if* $x \in L$ *then* $\Pr(O_x = 1|P_x = 00\dots0) \geq 2/3$, *and*
*ii. if* $x \notin L$ *then* $\Pr(O_x = 1|P_x = 00\dots0) \leq 1/3$ .

As mentioned above, Aaronson [Aar05] showed that postBQP = PP and hence $\mathsf{P}^{\mathsf{postBQP}} = \mathsf{P}^{\mathsf{PP}} \supset \mathsf{PH}$. Building on this result, Fujii and Morimae [FM17] showed that if postA = postBQP then a machine that approximates the output probabilities of circuits associated with A up to a multiplicative error $1/\sqrt{2} < c < 1$ can be used to decide any problem in PP and hence any problem in GapP. This is because the postA = postBQP condition ensures that A is rich enough to encode the output probabilities of arbitrary quantum computations and hence gaps of #P functions.

Coming from another route, one can show that estimating the success probability of a universal quantum circuit up to relative error[24] 1/4 can encode hard instances of the so-called Jones polynomial [Kup15b; GG14; MB17] as well as so-called Tutte polynomials [Kup15b; GG14] and certain Ising model partition functions [Boi+18]. Expressing the output probabilities in terms of such quantities, which have been studied in some detail in the classical literature will also prove to be extremely useful once we get to approximate sampling hardness.

24: Notice that achieving a relative error 1/4 is slightly more demanding than a multiplicative erorr $1/\sqrt{2}$.

But how do computationally restricted circuit families such as IQP circuits and boson sampling fare? As it turns out the property that computing the output probabilities is GapP-hard is more rule than exception in random circuit families [FU15; Fuj+14; MFF14; Mor17; Fuj+18; BFK18; MSM17; GWD17; Ber+18].

**IQP circuits** As a particularly neat example of this, for IQP circuits, one finds [BMS16] not only that postIQP = postBQP[25] but also that the output amplitudes

$$\langle 0|C_W|0\rangle = \frac{1}{2^n} Z_W, \tag{2.56}$$

can be expressed as imaginary-temperature partition functions of random Ising models [BMS16; FM17]:

$$Z_W \equiv \sum_{z\in\{-1,+1\}^n} \exp\left[\mathrm{i}\left(\sum_{i<j} w_{i,j} z_i z_j + \sum_i w_{i,i} z_i\right)\right]. \tag{2.57}$$

25: This can be shown using a gadget to implement the Hadamard gate via teleportation, the idea being that what IQP circuits are lacking for universality is the possibility to switch between $X$ and $Z$ bases. By measuring a single output line one can teleport a Hadamard gate to an arbitrary position in the circuit using gate teleportation [BMS16; Mon17].

The modulus square $|Z_W|^2$ of such partition functions has been shown to be GapP-hard to approximate up to a relative error $1/4 + o(1)$ [GG14; FM17].

For an IQP circuit $C_f$ defined by a Boolean degree-3 polynomial $f$ with coefficient vectors $\alpha, \beta, \gamma$ (cf. Eq. (2.52)) one finds that the all-zero amplitude is given by the gap of $f$[26]

$$\langle 0|H^{\otimes n} C_f H^{\otimes}|0\rangle = \frac{1}{2^n}\sum_{x,y}\langle y|C_f|x\rangle \tag{2.61}$$

$$= \frac{1}{2^n}\sum_x (-1)^{f(x)} = \mathrm{ngap}(f). \tag{2.62}$$

26: To see this notice that

$$Z_i|x\rangle = (-1)^{x_i}, \tag{2.58}$$
$$CZ_{i,j}|x\rangle = (-1)^{x_i x_j}, \tag{2.59}$$
$$CCZ_{i,j,k}|x\rangle = (-1)^{x_i x_j x_k}. \tag{2.60}$$

We already saw above that approximating the gaps of arbitrary #P functions $f$ up to multiplicative errors $1/\sqrt{2}$ is GapP-complete. This remains true when restricting the function $f$ to a degree-3 Boolean polynomial over the field $\mathbb{F}_2$. Recall the construction leading to Eq. (2.17) and Lemma 2.3. By postselected universality of the IQP circuits $C_f$ via the Hadamard gadget[27], there is a circuit $C_f$ using polynomially many auxiliary qubits with success probability proportional to that of $Q_s$ with easily computable normalization factor $|\langle 0|C_f|0\rangle|^2 = |\langle 0|Q_s|0\rangle|^2/2^h$, where $h$ is the number of teleported Hadamard gates [BMS16].

27: The only component that *IQP* circuits $C_f$ lack to exactly simulate the circuit $Q_s$ defined in Eq. (2.16) that uses the Hadamard gates as well as the classical `TOF` gate is the possibility to switch between $X$ and $Z$ basis via the Hadamard gate (see note 25 above).

*The following paragraph has previously been published in Ref. [Han+19].*

**Boson sampling** The output distribution $P_{\mathrm{bs},U}$ of a boson sampling experiment can be expressed as [Sch08]

$$P_{\mathrm{bs},U}(S) = \frac{|\operatorname{Perm}(U_S)|^2}{\prod_{j=1}^{m}(s_j!)}, \tag{2.63}$$

in terms of the permanent of the matrix $U_S \in \mathbb{C}^{n\times n}$ constructed from $U$ by discarding all but the first $n$ columns of $U$ and then, for all $j \in [m] \equiv \{1,2,\ldots,m\}$[28], taking $s_j$ copies of the $j^{\text{th}}$ row of that matrix (deviating from Aaronson and Arkhipov's notation [AA13]). The permanent for a matrix $X = (x_{j,k}) \in \mathbb{C}^{n\times n}$ is defined similarly to the determinant but without the negative signs as

28: For any $j \in \mathbb{Z}^+$ we employ the short hand notation $[j] := \{1,\ldots,j\}$ for the range.

$$\operatorname{Perm}(X) := \sum_{\tau\in \mathrm{Sym}([n])} \prod_{j=1}^{n} x_{j,\tau(j)}, \tag{2.64}$$

where $\mathrm{Sym}([n])$ is the symmetric group acting on $[n]$. It is a known fact that exactly computing the permanent of even a binary matrix is a problem that is #P-hard [Val79], while its close cousin, the determinant, is computable in polynomial time. Aaronson and Arkhipov [AA13, Thm. 4.3] extend this famous result of Valiant [Val79] to the case of multiplicative-error approximation for the *modulus of the permanent*. More precisely, they show that for any $c \in [1/\mathsf{poly}(n), 1]$, approximating $\operatorname{Perm}(X)^2$ up to multiplicative error $c$ for $X \in \mathbb{R}^{n\times n}$ remains #P-hard by a reduction similar to the one in Lemma 2.2 on multiplicative-error GapP-hardness of computing the modulus of the gap of a #P function.

## 2.5 The complexity of quantum random sampling

We are now in a position to bring together the abstract ideas about the relation between the computational complexity of sampling from a distribution and computing its probabilities. We will prove that under certain conditions on the random circuit family $\mathscr{C}$, sampling from the output distribution of a random instance $C \in \mathscr{C}$ acting on a fixed number of qubits cannot be done in classical polynomial time in the description size of $C$ (which is polynomial in the number of qubits it acts on).

On an abstract level, the basic idea of the proof is to exploit the fact that approximating output probabilities of unitaries in $\mathscr{C}$ is GapP-hard in this input size. We will then argue that *if there was an efficient (derandomizable) sampling algorithm* for a random $C \in \mathscr{C}$ then we could approximate its success probability using Stockmeyer's algorithm. But if we could do that, because Stockmeyer's algorithm lies within $\Delta_3^p \subset \mathsf{PH}$ and approximating success probabilities of $\mathscr{C}$ is $\mathsf{GapP}$-hard, then this would imply that $\Delta_3^p \supset \mathsf{P}^{\mathsf{GapP}} \supset \mathsf{PH}$ so that the polynomial hierarchy collapses to its third level – a highly unlikely consequence.

The most basic ingredient to this proof is that, in fact, approximating the success probability of circuits in $\mathscr{C}$, i.e., the probability of obtaining the all-zero outcome, is $\mathsf{GapP}$-hard *in the worst case* or, in more technical terms, that approximating the class of problems

$$\mathsf{Gap}\mathscr{C} := \left\{ f : \exists C \in \mathscr{C} : \mathrm{gap}(f) = |\langle 0|C|0\rangle|^2 \right\} \subset \mathsf{GapP}, \tag{2.65}$$

is as hard as the full power of $\mathsf{GapP}$, i.e., that for some constant[29] $c > 0$

29: We will get to the details of this below.

$$\mathsf{P}^{\mathsf{GapP}} = \mathsf{P}^{\mathsf{Apx.}_c\mathsf{Gap}\mathscr{C}}. \tag{2.66}$$

## Proof of exact sampling hardness: worst-case hardness

Based on such worst-case hardness we can then prove the following theorem.

**Theorem 2.8** (Exact sampling hardness) *Let $\mathscr{C}$ be a family quantum circuits such that there exists a constant $c \in (0,1]$ for which approximating the output probabilities up to multiplicative error $c$ is $\mathsf{GapP}$-hard. If there was an* exact derandomizable sampling algorithm *for circuits in $\mathscr{C}$ then the polynomial hierarchy would collapse to $\Delta_3^p$.*

*Proof.* Suppose there was a classical derandomizable sampling algorithm $\mathscr{A}$ that, given as an input an efficient description of a circuit $C \in \mathscr{C}$ could efficiently sample from its output distribution $p(C)$ as defined in Eq. (2.51). Then we can apply Stockmeyer's algorithm (Theorem 2.4) to the function $f_{C,0}$ defined in Eq. (2.48). In time $\mathsf{poly}(1/c)$ within the third level $\Delta_3^p$ of the polynomial hierarchy the output of this procedure will produce a multiplicative-error estimate $q_0(C)$ of the acceptance probability $p_0(C)$ that satisfies

$$p_0(C)c \leq q_0(C) \leq p_0(C)/c. \tag{2.67}$$

But since approximating $p_0(U)$ was a $\mathsf{GapP}$-hard task to begin with, this implies that the polynomial hierarchy collapses to $\Delta_3^p$. □

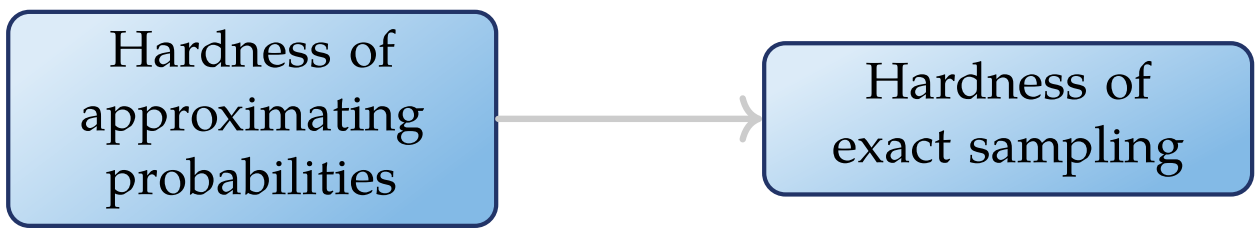

**Figure 2.4:** In the proof of Theorem 2.8 the idea is to relate the hardness of approximating the probabilities in a distribution to the hardness of sampling from that distribution.

We have now put together the pieces in a complexity theoretic argument that shows that it is computationally difficult to sample from distributions for which the output probabilities are $\mathsf{GapP}$-hard to approximate in the worst case. It turned out that if it was possible to efficiently sample from such a distribution, then this would imply that the polynomial hierarchy collapsed to its third level. The essence of the argument revolved around the question how computing probabilities and sampling from the output distribution are related. In particular, we saw that it is a necessary requirement for the argument to work that the probabilities are not only GapP-hard to compute exactly but also to approximate up to some constant relative error; see Fig. 2.4.

What happens, though, once the sampling algorithm is allowed to make some error as compared to the ideal target distribution? Indeed, while an ideal quantum device might be able to sample from the ideal distribution no such device can exist. Every physical realization of the ideal model will inevitably comprise noisy and imperfect components so that it may only be considered an *approximate sampler*. So does hardness of sampling still hold in the presence of errors? And if so, what types and magnitudes of errors are tolerated? Let us approach this question from two sides: first, we will ask the question how we can naturally relax the hardness proof to allow for errors. Coming from the opposite end we will approach the questions from the perspective of realistic quantum devices.

## Multiplicative-error sampling hardness

Quite naturally, the proof of sampling hardness can be extended from exact sampling, to sampling from a probability distribution $p$ that is *multiplicatively close* to the target distribution $p(C)$ in the sense that for some constant $d \in (0,1]$ each probability $p_x$ satisfies

$$d p_x(C) \leq p_x \leq p_x(C)/d. \tag{2.68}$$

We can then easily amend the proof of Theorem 2.8 for this case to prove multiplicative-error robustness.

*Multiplicative-error robustness of Thm 2.8.* Assume there was an efficient classical sampling algorithm $\mathcal{A}$ that, given as an input a description of a circuit $C \in \mathscr{C}$ could efficiently sample from a probability distribution $p$ that approximates the distribution $p(C)$ defined in Eq. (2.51) up to a multiplicative error $d$ as in Eq. (2.68). Then we could use Stockmeyer's algorithm to generate an approximation $q_0$ of the acceptance probability $p_0$ that is correct up to any constant multiplicative error $c$

$$c p_0 \leq q_0 \leq p_0/c. \tag{2.69}$$

But the probability $p$ was multiplicatively close to the ideal acceptance probability $p_0(C)$ to begin with so that we obtain

$$c d p_0(C) \leq c p_0 \leq q_0 \leq p_0/c \leq p_0(C)/(cd), \tag{2.70}$$

that is, an overall multiplicative-error approximation to the probability $p_0(C)$ with constant multiplicative error $cd$. If $c$ and $d$ are chosen such that the probability $p_0(C)$ is GapP-hard to approximate up to multiplicative error $cd$ the existence of an efficient sampling algorithm with multiplicative error guarantee $cd$ implies the collapse of the polynomial hierarchy. □

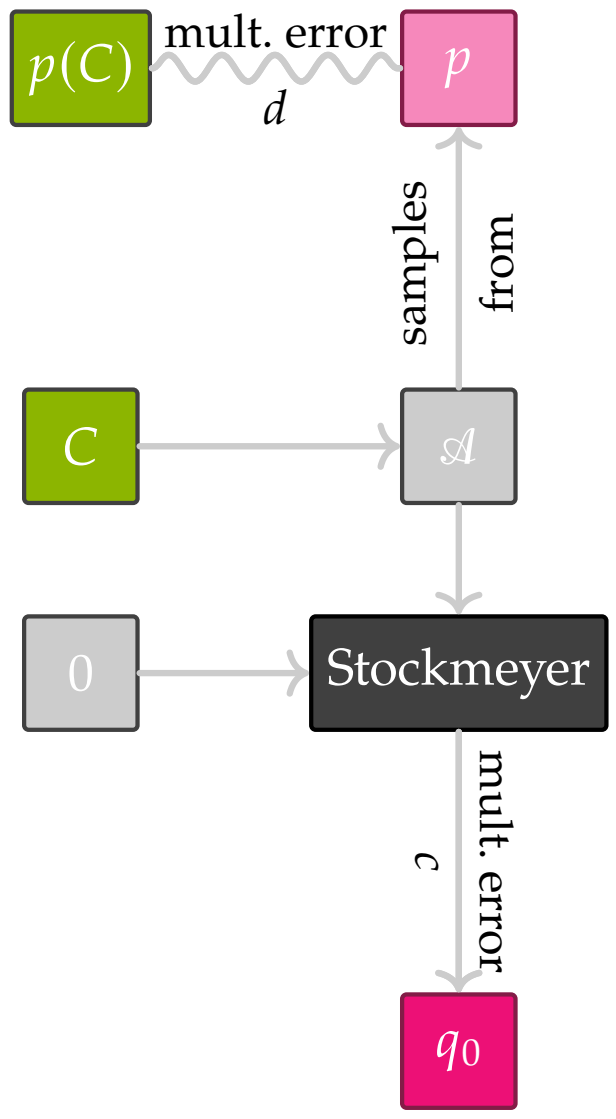

**Figure 2.5:** Outline of the proof strategy for additive-error sampling hardness: a derandomizable sampling algorithm $\mathcal{A}$, given $C$ as an input, samples from a distribution $p$ so that for all $x \in \Omega$ the probabilities $p_x$ are $d$-multipcliative approximations of the ideal probabillities $p_0(C)$. Given $\mathcal{A}$ and $x \in \Omega$, Stockmeyer's algorithm can infer a $c$-multiplicative approximation $q_x$ of the probabilities $p_x$.

We saw in our discussion about the approximability of GapP how extraordinarily demanding multiplicative errors are in the guise of Lemma 2.2.

There, we used that such approximations always preserve the sign of a quantity and, moreover, attain 100% accuracy if the quantity is 0. And this state of affairs reappears here: there is essentially no difference in the hardness of weak simulation as compared to the exact case when allowing for constant multiplicative errors. And indeed, to satisfy such a notion of approximation, a quantum device would need to account for the size of each of the exponentially many probabilities to begin with, a state of affairs that seems highly implausible.

What is a more plausible notion of approximation then? It turns out that additive-error approximation pop up naturally when allowing for realistic device errors. And indeed, the lever we used to prove GapP-completeness of multiplicatively approximating gaps of Boolean functions fails: in contrast to multiplicative errors, additive errors do not preserve the sign of a quantity as they are independent of the size of this quantity.

### Errors in quantum devices

Errors in quantum devices can occur at all stages of an experiment: at the level of state preparation, at the level of individual quantum gates and or the global circuit as so-called 'cross-talk' [Sar+19], and at the eventual measurement.

Both state preparation and measurement errors can be cast as errors in the quantum circuit, assuming an ideal preparation and measurements. Those errors can take on the shape of either coherent or incoherent errors. For *coherent errors*, the noisy variant of a target unitary transformation $U$ is given by a coherent over- or underrotation $V$ along some axis as

$$U \mapsto VU. \tag{2.71}$$

*Incoherent errors* occur when rather than implementing an ideal unitary quantum channel $\mathcal{U} = U \cdot U^\dagger$ erroneous unitary channels $\mathcal{V}_i,\ i = 1, \ldots, k$ are applied with probabilities $p_i$ so that

$$\mathcal{U} \mapsto p_0 \mathcal{U} + \sum_i p_i \mathcal{V}_i. \tag{2.72}$$

As a particularly simple example, the $\mathcal{V}_i$ might be equally probable random Pauli flips that transform a single-qubit state $\rho$ as $\rho \mapsto X\rho X + Y\rho Y + Z\rho Z$. This error model is called *depolarizing noise* and is one of the simplest models of errors in noisy devices. Depolarizing noise on the level of the $n$-qubit quantum state replaces a pure quantum state prepared by a unitary $U$ starting from the references state with the maximally mixed state with some probability $\epsilon$:

$$U|0\rangle\langle 0|U^\dagger \mapsto (1-\epsilon)U|0\rangle\langle 0|0U^\dagger + \epsilon\frac{1}{2^n}. \tag{2.73}$$

More generally, we can conceive of both coherent and incoherent errors as affecting the ideal state preparation in such a way that a mixture

$$\rho = (1-\epsilon)U|0\rangle\langle 0|U^\dagger + \epsilon\sigma \tag{2.74}$$

between the ideal target state $U|0\rangle\langle 0|U^\dagger$ and a noisy version thereof, $\sigma$, is prepared. Realistic errors therefore affect the quantum state as an *additive error* in that the noisy version $\rho_\epsilon$ of the pure target state satisfies

$$\|\rho_\epsilon - U|0\rangle\langle 0|U^\dagger\|_1 \leq 2\epsilon, \tag{2.75}$$

in trace norm. We will frequently use the Schatten-$p$ norms of an operator $X$, which are defined as

$$\|X\|_p := (\operatorname{Tr}|X|^p)^{1/p}, \tag{2.76}$$

where $|\cdot| : X \mapsto |X| = \sqrt{X^\dagger X}$ denotes the matrix absolute value.

What does the trace-norm difference (2.75) imply for the output distribution $p(\rho)$? We can easily see this from the operational characterization of the trace distance $d_1(\rho, \sigma) := \|\rho - \sigma\|_1/2$ as the probability that an optimal measurement $\{M_x\}_x$ distinguishes $\rho$ from $\sigma$.

$$\|\rho - \sigma\|_1 = \max_{\{M_x\}_x} |\operatorname{Tr}[(\rho - \sigma)M_x] \geq \sum_x |p_x(\rho) - p_x(\sigma)| = \|p(\rho) - p(\sigma)\|_{\ell_1}. \tag{2.77}$$

Here, we have defined the vector equivalent of the Schatten-$p$ norms, the $\ell_p$ norms, which for $0 < p < \infty$ and a vector $x = (x_1, \ldots, x_d) \in \mathbb{C}^d$ are defined as

$$\|x\|_{\ell_p} := \left( \sum_{i=1}^d |x_i|^p \right)^{1/p}. \tag{2.78}$$

We also define $\|x\|_{\ell_\infty} := \max_{i \in [n]} |x_i|$ and take $\|x\|_{\ell_0} := |\{i \in [n] : x_i \neq 0\}|$ to denote the number of non-zero elements of $x$. Thus, $\|\cdot\|_{\ell_p}$ is a standard norm whenever $p \geq 1$. For $p \in (0, 1)$, $\|\cdot\|_{\ell_p}$ no longer satisfies the triangle inequality, but it is obviously absolutely homogeneous and thus still defines a quasinorm.

The – for our purposes – most important of those norms is the $\ell_1$ norm, which measures statistical or *total-variation distance* between two probability distributions $p, q : \Omega \longrightarrow [0, 1]$ on a sample space $\Omega$

$$\|p - q\|_{TV} = \frac{1}{2}\|p - q\|_{\ell_1}. \tag{2.79}$$

Since the trace-distance maximizes over all possible measurements it upper-bounds the total variation distance between the outcome distributions of a fixed, say, computational-basis measurement on two quantum states.

To show the robustness of the hardness-of-sampling result to realistic noise we must therefore show that for any state $\rho$ satisfying

$$\|\rho - U|0\rangle\langle 0|U^\dagger\|_1 \leq \epsilon, \tag{2.80}$$

sampling from its output distribution is hard for classical computers. This endeavour is faced with the challenge that as $\epsilon$ increases, so does the legroom for classical simulation: to show hardness we have to prove that for any state within an $\epsilon$-trace norm neighbourhood, sampling from its output distribution is classically intractable. We are faced with a dramatically increased burden in the proof as hardness needs to be shown for an entire *volume* of probability distributions rather than a single point. As $\epsilon \longrightarrow 1$ the output state of the computation becomes classically simulable as, in particular, the uniform distribution is always within this error bound of the target distribution. But the uniform distribution is easy to sample from even on an exponentially large sample space.

### From multiplicative to additive robustness

Given what we have seen so far, there is a fundamental discrepancy between how the proof of exact sampling hardness can naturally be made robust to noise and the errors that naturally occur in realistic settings. The discrepancy is one between the utterly unrealistic notion of multiplicative errors on all probabilities and the more realistic notion of additive errors on the global outcome distribution. The question we will focus on now is whether we can overcome this hurdle?

In technical terms, what we would like to prove is that no efficient classical algorithm $\mathcal{A}$ taking as an input an efficient description of $C$ exists that samples from any distribution $p$ such that $\|p - p(C)\|_{\ell_1} \leq \epsilon$ for a constant $\epsilon > 0$. Again, we will try and use Stockmeyer's approximate counting algorithm with a derandomizable sampling algorithm as an input in order to take the step from hardness of approximating probabilities. So how can we take the leap from proving robust hardness-of-sampling results for multiplicative to ones for additive errors?

**Some preliminary observations** To approach an answer to this question let us conceive of the sampling algorithm $\mathcal{A}$ as an adversarial party that, given $U$ as an input, tries to adversarially obstruct the approximate counting algorithm in its goal to approximate specific probabilities. The adversarially acting sampling algorithm is, however, constrained to sample from a distribution satisfying the respective error bounds.

A few observations regarding the nature of additive errors in contrast to multiplicative ones turn out to be instructive.

i. When the sampling algorithm is constrained to multiplicative errors on individual probabilities, the total *additive error* it can make depends strongly on the shape of the distribution. In particular, every individual probability will be correct up to an error that depends on its size. In contrast, the additive-error constraint allows the adversarial party much more flexibility. An additive error can be viewed as a total *error budget* that may be distributed across the individual probabilities at will. In particular, a few probabilities can come with large (relative) errors suposing that the other ones are correct up to a very small additive error.
ii. When proving multiplicative-error robustness, the shape and volume of the region in the space of probability distributions on a sample space $\Omega$ of which hardness is proven depend heavily on the specific shape of the distribution. In contrast, for additive-error robustness volume and shape of this region are only sensitive to boundaries of this space.

iii. Approximating success probabilities of quantum computations up to an inverse polynomial additive error does not remain complete for GapP but only for BQP [De +11, Thm. 3]. Only for inverse exponentially small additive errors $\pm 1/2^n$ those approximations become again GapP-hard. This is easily seen using the fact that gaps of Boolean functions acting on $\{0,1\}^n$ only take on values that are integer multiples of $2/2^n$. Approximating those gaps up to an additive error $< 1/2^n$ is therefore just as hard as exactly computing them[30].

30: See also the Supplementary Material of Ref. [BMS16].

What can we take away from those observations?

Point iii. implies that to prove a polynomial-hierarchy collapse, we must still rely on the hardness to approximate output probabiliites of circuit families up to relative errors *or* exponentially small additive errors.

Points i. and ii. shine light on two sides of the same coin: in contrast to the case of multiplicative robustness, we cannot "hide hardness" in individual probabilities that might be very small. Instead, we must rely on circuit families for which not only single outcome probabilities of some members of the family are hard to compute, but rather a large – constant – fraction of *all output probabilities* of the circuit family must be hard to compute. In contrast to exact and multiplicative-error sampling hardness, it cannot be the case that only one of the circuits within $\mathcal{C}$ has a single output probability on which all classical algorithms fail. This idea is formalized within the notion of *average-case complexity*: approximating the outcome probabilities of quantum circuits must be hard for a large fraction of the instances, where an instance is defined by a specific quantum circuit.

From the same points, we can also learn that not all but very few of those hard probabilities can be tiny, i.e., smaller than, say, doubly exponentially small while very few large ones are easy to approximate. Indeed, if this were the case, since tiny quantities have tiny relative errors, the adversarially acting sampling algorithm could easily distribute the better part of its constant error budget on the few large probabilities but at the same time still surpass the relative-error threshold on the tiny probabilities. In this way they would meet the constraint imposed by the global additive error, but not achieve a provably hard task as the error on the computationally intractable probabilities would be too large. Rather, there must be a large fraction of hard instances that are reasonably large, say, at least as large as uniform probabilities $\sim 1/|\Omega|$ on the sample space $\Omega$. This idea is formalized within the notion of *anticoncentration*, which is a condition on the probability that a randomly drawn problem instance – again, specified by a circuit and an outcome string – is reasonably large. Anticoncentration constrains how the adversarial player can distribute their error budget: they can choose between getting many probabilities right with tiny errors, but making larger errors on a few outcomes, say, inverse polynomial errors on polynomially many probabilities, or getting all probabilities right with reasonably small inverse exponential errors.

These observations have been made by Aaronson and Arkhipov [AA13] who observed that the natural quantity in boson sampling, namely, computing a permanent is an average-case hard problem[31]: computing

31: We will get to the details of this in the next chapter (Chapter 3).

the permanent exactly (or with exponential precision) remains GapP-hard *on average* when the input to the problem is chosen at random from a Gaussian matrix ensemble [Lip91; AA13].

**Two notions of probability** In the previous discussion, we have been touching upon on a point that we had glossed over in our discussion of exact sampling hardness: it is key to random circuit sampling schemes that there are two notions of probability at play. First, there is the random choice of a circuit from the family $\mathscr{C}$, and second, there is the random choice of an outcome string $S$ that is distributed according to $p(U)$. Equally, there are two probability distributions – the distribution according to which random circuits are drawn, and the outcomes distribution of each such random circuit. These notions are crucially distinct.

We need to draw random circuit instances in order to make a complexity theoretic hardness argument. This is because in complexity theory we can never make a statement about a fixed instance, but only about a problem class certain instances of which can be reduced to problems for which no algorithm is known that solves all instances. Drawing a random circuit allows us to make use of this fuzziness in complexity theory. We need not specify a particular instance but rather, we are able to say that as we draw random instances, there is no efficient classical algorithm that solves our problem with 100% success probability. It might just hit a hard instance on which any such algorithm is doomed to fail.

The second notion of probability is intrinsic to our choice of problem. In the end, we aim to prove hardness of a sampling task. This is a task requiring randomness: we want to obtain a random sample from a distribution that we, in turn, chose at random from another ensemble.

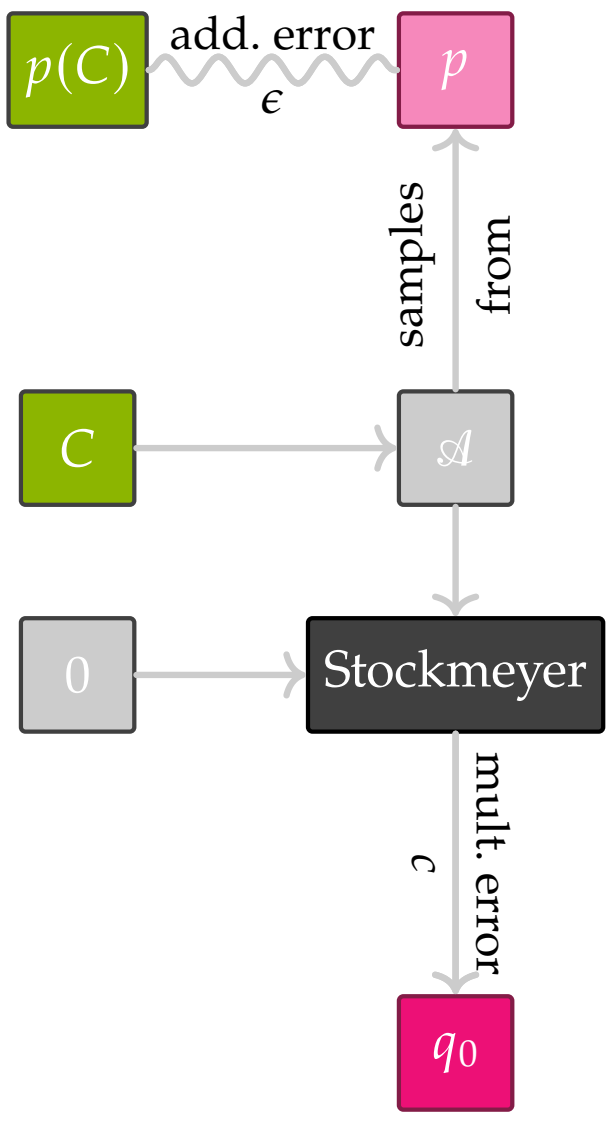

**Figure 2.6:** Outline of the proof strategy for additive-error sampling hardness: a derandomizable sampling algorithm $\mathscr{A}$, given C as an input, samples from a distribution $p$ that is $\epsilon$-close in total-variation distance to the target distribution $p(C)$. Given $\mathscr{A}$ and $x \in \Omega$, Stockmeyer's algorithm can infer a $c$-multiplicative approximation $q_0$ of the approximate acceptance probability $p_0$.

## Additive-error sampling hardness

As it turns out, if one is willing to make the assumptions of average-case hardness of approximating the output probabilities and their anticoncentration, we can prove a hardness-of-sampling result that is robust to constant additive errors. So without further ado, let us prove the hardness of additive error sampling. *En route*, we will encounter the precise definitions of those notions required for the proof to work. We will proceed analogously to the proof of multiplicative robustness, following the sketch in Fig. 2.6.

*Additive-error robustness of Thm 2.8.* Assume there is an efficient, derandomizable classical algorithm that takes as an input a description of a circuit instance $C$ from a family $\mathscr{C}$ and outputs samples distributed according to a probability distribution $p$ that satisfies

$$\|p - p(C)\|_{\ell_1} \leq \epsilon. \tag{2.81}$$

Here, $p(C)$ is the ideal target distribution defined in Eq. (2.51). We want to use this sampling algorithm in order to approximate a random problem instance as given by the success probability $p_0(C) = |\langle 0|C|0\rangle|^2$ of $C$. To generate such an instance, we draw $C \in \mathscr{C}$ at random.

To estimate the value of this instance, we can feed the algorithm $\mathscr{A}$ with input $C$ as well as the all-zero outcome string into Stockmeyer's approximate counting algorithm. Using access to its NP oracle, Stockmeyer's algorithm will output a multiplicative-error approximation $q_0$ of the noisy success probability $p_0$ satisfying

$$|q_0 - p_0| \leq c p_0, \tag{2.82}$$

in time $\mathsf{poly}(n, 1/c)$ within the third level $\Delta_3^p$ of the polynomial hierarchy.

Our goal is to bound the error

$$|q_0 - p_0(C)| \leq |q_0 - p_0| + |p_0 - p_0(C)|. \tag{2.83}$$

Eq. 2.82 already provides the first half of this bound. For the second bound we need to leverage the total-variation-distance bound (2.81) on the global distributions $p$ and $p(C)$ to an error bound on the individual probabilities $p_0$ and $p_0(C)$.

To obtain such a bound, consider again the sampling algorithm $\mathscr{A}$. Remember that *qua* being a derandomizable algorithm, on input $U, r$ with uniformly random $r \in \{0,1\}^{\mathsf{poly}(n)}$ it will output a random sample from $p$ so that

$$p_x(C) = \Pr[C \text{ outputs } x] \tag{2.84}$$

$$p_x = \Pr_r[\mathscr{A} \text{ outputs } x \text{ on input } C]. \tag{2.85}$$

Acting adversarially, the algorithm wants to maximize the error $|p_0 - p_0(C)|$. To do so, it needs to have some prior information about which of the outcome strings are more likely to be queried in Stockmeyer's algorithm given a certain input $C$ so that it can distribute more of its constant error budget on those outcomes. Such information would manifest itself in a distribution of outcomes $x$ that is non-uniform – and in fact concentrated on the single all-zero outcome – from the perspective of $\mathscr{A}$ given $C$ [AA13, p. 51]. This is because the all-zero outcome is the one we are *always* interested in since it defines the success probability.

**Hiding problem instances** To see how we can achieve that this distribution is uniform, consider the distribution over circuits $C_y$ obtained by drawing $C \in \mathscr{C}$ at random and then appending $X$-gates $X_1^{y_1} \cdot X_2^{y_2} \cdots X_n^{y_n}$ for uniformly random $y \in \{0,1\}^n$ to the end of the circuit [BMS16]. We can then reexpress the outcome probabilities of $C_y$ as

$$p_x(C_y) = |\langle x|C_y|0\rangle|^2 = |\langle x \oplus y|C|0\rangle|^2 = |\langle 0|C_{x\oplus y}|0\rangle|^2 = p_0(C_{x\oplus y}) \tag{2.86}$$

The upshot of this reformulation is that the very same problem instance $C$ can be equivalently obtained when providing the adversary $\mathscr{A}$ with an instance $C_y$ for uniformly random $y$ and then querying Stockmeyer's algorithm on the outcome $y$. Given $C_y$ as an input, the outcome $y$ therefore looks uniformly random from the perspective of $\mathscr{A}$. When aiming to estimate the problem instance $p_0(C)$ we can therefore *hide the instance* $C$ in the input unitary $C_y$ by randomly appending uniformly random $X$ gates according to $y$. The mechanism by which this hiding procedure works is the *one-time pad* : if $y \in \{0,1\}^n$ is uniformly random, so is $x \oplus y$ for any fixed string $x$.

This hiding procedure arises naturally if the distribution of random unitaries $C \in \mathscr{C}$ is invariant under a string of random $X$-gates at the end of the circuit. In fact, the *hiding property* holds very naturally for most random circuit families, and in particular so for universal random circuits where each gate is drawn from the Haar measure. This is because the Haar measure is left- and right-invariant under arbitrary unitaries and the Pauli-$X$ gate is one particular such unitary.

Another way of viewing the hiding property is that the adversary $\mathscr{A}$ cannot distinguish between whether we have directly generated a random problem instance $C$ querying on the all-zero outcome, or whether we have first drawn a random $C \in \mathscr{C}$ and then *hidden this instance* by constructing the unitary $C_y$ with uniformly random $y$ that occurs with equal probability as $C$, querying on the outcome $y$ [AA13, p. 51]). Without loss of generality we can therefore always query Stockmeyer's algorithm on the all-zero outcome of $C$, making use of the fact that this outcome is indistinguishable from a uniformly random one from the perspective of $\mathscr{A}$.

**The hiding property**

A qubit-circuit family $\mathscr{C}$ satisfies the *hiding property* if this distribution is invariant under appending a random string of Pauli-$X$ gates at the end of the circuit so that

$$\Pr_{C\sim\mathscr{C}}[C] = \Pr_{C\sim\mathscr{C},y\sim\{0,1\}^n}[C_y]. \tag{2.87}$$

If the hiding property holds, we can conceive of the outcomes of the circuits as being uniformly distributed from the perspective of $\mathscr{A}$. But in this case, we can apply Markov's inequality to obtain a bound on the error for individual probabilities. For uniformly random $x$ we obtain that

$$\begin{aligned}\Pr_{x\in\{0,1\}^n}&\left[|p_x - p_x(C)| \geq \frac{1}{\delta}\,\mathbb{E}_{x\in\{0,1\}^n}[|p_x - p_x(C)|]\right] \\ &= \Pr_{x\in\{0,1\}^n}\left[|p_x - p_x(C)| \leq \frac{\epsilon}{2^n}\right] \leq \delta,\end{aligned} \tag{2.88}$$

as

$$\begin{aligned}\mathbb{E}_{x\in\{0,1\}^n}\left[|p_x - p_x(C)|\right] &= \frac{1}{2^n}\sum_{x\in\{0,1\}^n}|p_x - p_x(C)| \\ &= \frac{1}{2^n}\|p - p(C)\|_{\ell_1} = \frac{\epsilon}{2^n}.\end{aligned} \tag{2.89}$$

We have now found that with probability at least $1-\delta$ over the inputs, the error of the estimate $q_0$ output by Stockmeyer's approximate counting

algorithm satisfies

$$|q_0 - p_0(C)| \le \frac{1}{\mathsf{poly}(n)} p_0 + \frac{\epsilon}{2^n \delta} \tag{2.90}$$

$$\le \frac{1}{\mathsf{poly}(n)} p_0(C) + \frac{\epsilon}{2^n \delta}\left(1 + \frac{1}{\mathsf{poly}(n)}\right). \tag{2.91}$$

This bound is a mixture of an exponentially small additive and inverse polynomially small relative error. However, the error bound does not hold for all possible inputs to Stockmeyer's algorithm, but only a $(1-\delta)$-fraction of the inputs.

In order to show a collapse of the polynomial hierarchy we need to show that achieving this error on a $1-\delta$ fraction of the instances is sufficient for a collapse of the polynomial hierarchy. Now, there are two issues to be resolved: first, we need to show that achieving an approximation within an error given by the mixture of multiplicative and additive error (2.91) is intractable for classical computers. Second, we need to show that within any fraction of instances on which the error bound holds, we are also guaranteed to find a hard instance.

This is where two properties of the distribution come in – anticoncentration and average-case hardness.

**Anticoncentration** To resolve the first issue, we will invoke the so-called *anticoncentration property*. This property is based on the following observation: if we could reduce the mixture of additive and relative errors (2.91) to a purely relative error (or an exponentially small additive error) with small enough constant, we would achieve the type of error bound for which worst-case hardness can be proven. Anticoncentration is a sufficient condition on the circuit family $\mathcal{C}$ that ensures that on a large fraction of the instances, Stockmeyer's algorithm is guaranteed to make a small relative error. To understand the idea behind this property observe that if in the error bound (2.91) the probability $p_0(C)$ is larger than $\alpha/2^n$ for some constant $\alpha > 0$ then (2.91) can be upper-bounded in terms of a relative error $\epsilon/(\alpha\delta)$. We formalize this idea in the following definition.

**Definition 2.12** (Anticoncentration) *We say that a circuit family $\mathcal{C}$ anticoncentrates if for constant $\alpha > 0$ there exists $\gamma(\alpha) > 0$ independent of n such that*

$$\Pr_{C\sim\mathcal{C},x\sim\{0,1\}^n}\left[p_0(C_x) \ge \frac{\alpha}{2^n}\right] \ge \gamma(\alpha). \tag{2.92}$$

Notice that the probability in Def. 2.12 runs over the choice of random circuit as well as uniformly appended $X$-gates. If $\mathcal{C}$ satisfies the hiding property the latter uniformly random choice of hiding $X$-gates becomes trivial, however, since $\Pr[U] = \Pr[U_y]$. Conversely, the probability in Eq. (2.88) runs only over the uniformly random choice of outcome. Hence, the probabilities are independent from one another and both bounds (2.88) and (3.16) are satisfied with probability at least $\gamma(\alpha)(1-\delta)$.

Applying anticoncentration to the error bound (2.91) we then that with probability at least $\gamma(\alpha)(1-\delta)$

$$|q_0 - p_0(C)| \le \left(\frac{\epsilon}{\delta\alpha} + \frac{1}{\mathsf{poly}(n)}\right) p_0(C). \tag{2.93}$$

If the probabilities $p_0(C)$ are GapP-hard to approximate up to a relative error $c$, we can set $\alpha = 1/c$, $\epsilon = \gamma(\alpha)/4$ and $\delta = \gamma(\alpha)/2$ to obtain a $(c/2 + o(1))$-relative approximation of $p_0(U)$ with probability at least $\gamma(1-\gamma/2)$ over the choice of instances[32].

In other words, whenever $\|p - p(C)\|_{\ell_1} \le \epsilon$, the error bound (2.93) holds for a $\gamma(1-\gamma/2)$ fraction of the instances[33].

32: Conversely, we can use concentration of the probability via Markov's inequality to reduce the mixture (2.91) to an additive error

$$\Pr\left[p_0(C) \ge \frac{\alpha}{2^n}\right] \le \frac{1}{\alpha} \tag{2.94}$$

$$\Rightarrow |q_0 - p_0(C)| \le \left(\frac{\epsilon}{\delta} + \frac{1}{\alpha} + o(1)\right)\frac{1}{2^n} \tag{2.95}$$

with probability $(1-\delta)(1-\alpha)$.

33: Notice a trade-off when using anti-concentration to obtain the multiplicative-error bound (2.93): we can increase our choice of $\epsilon$ as a function of $\gamma$.

**Approximate average-case hardness** What is left to show for a collapse of the polynomial hierarchy is that within this $\gamma(1-\gamma/2)$ fraction of the instances we are guaranteed to find a hard instance, too. While we know that *some* fraction of instances exists that contains a hard instance as the approximation problem is worst-case hard, we have no idea on which precise fraction of the instances the bound holds and thus guarantees that the outcome of Stockmeyer's algorithm is a suitable approximation to the target probability. So the machine $\mathscr{A}$ has only achieved a task it should not have been able to achieve if estimating the success probability $p_0(C)$ up to multiplicative error $c$ is GapP-hard on *any* $\gamma(1-\gamma/2)$ *fraction* of the instance; see Fig. 2.7. That is to say, a machine solving the estimation problem for $p_0(C)$ for any such fraction is as powerful as an arbitrary GapP machine.

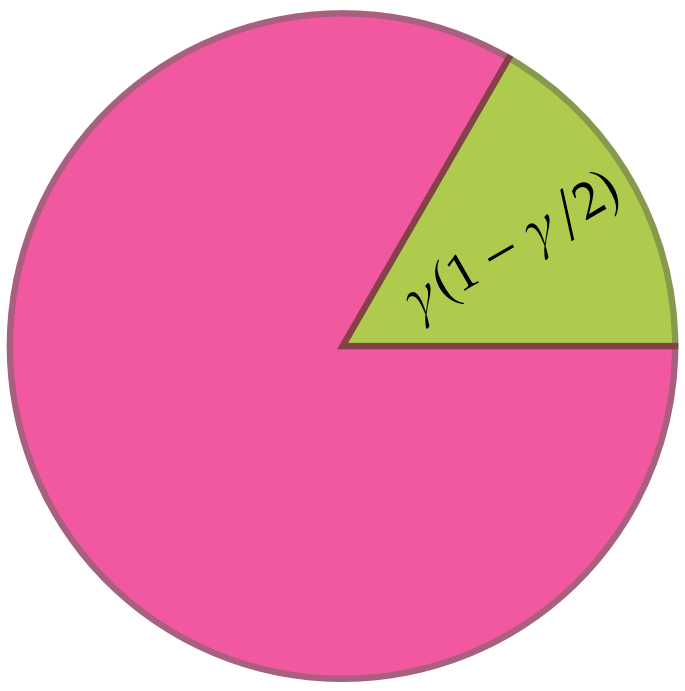

**Figure 2.7:** The error-bound (2.93) only holds for a $\gamma(1-\gamma/2)$ fraction of the instances. Approximate average-case hardness needs to be assumed for any such fraction. In the worst case, all instances in the complement of some specific such fraction are therefore hard instances.

Making this intuition rigorous is the idea of average-case hardness.

**Definition 2.13** (Approximate average-case hardness) *Let $c, \Gamma \in (0,1)$. A function class $\mathsf{F}$ is average-case hard with constant $\Gamma$, if approximating any $\Gamma$ fraction of the instances in $\mathsf{F}$ up to a constant relative error $c$ is $\mathsf{F}$-hard.*

If approximate average-case hardness holds, the existence of an efficient sampling algorithm $\mathscr{A}$ for the output distribution of a random instance $C \in \mathscr{C}$ implies that we can approximate #P-hard probabilities using Stockmeyer's algorithm. The polynomial hierarchy collapses. □

We have proven approximate sampling hardness; see Fig. 2.8.

**Theorem 2.9** (Additively robust sampling hardness) *Consider a circuit family $\mathscr{C}$ that satisfies*

*i. the hiding property,*
*ii. anticoncentration for $\alpha = 1$ with constant $\gamma$, and*
*iii. approximate average-case hardness up to a relative error $1/4$ on any $\gamma(1-\gamma/2)$ fraction.*

*Suppose there is an efficient classical sampling algorithm $\mathscr{A}$ exists that given $C \in \mathscr{C}$ drawn at random as an input, outputs samples from an additive approximation $p$ to the outcome distribution $p(C)$ satisfying $\|p - p(C)\|_{\ell_1} \le \gamma/8 = \epsilon$. Then the polynomial hierarchy collapses.*

Notice that in the exposition of the proof we could have also gone the other way. We could have simply conjectured average-case hardness up to the baroque error mixture (2.91) on any $1-\delta$ fraction and then conceive of anticoncentration as *evidence* for this conjecture to be true. We will turn to a more detailed discussion of hiding, anticoncentration, and average case hardness (i.-iii.) in Chapter 3. It is instructive to return to the generalized Hong-Ou-Mandel experiment and discuss those properties in the case of boson sampling. Not only did Aaronson and Arkhipov [AA13] introduce those ideas to the literature, but also significantly more work has to be done here in order to prove the hiding property. While for qubit-based schemes it arises naturally due to invariance under $X$-flips of the outcomes, this is not the case in boson sampling. Historically, the fact that permanents are known to be average-case hard to compute [Lip91] actually motivated Aaronson and Arkhipov [AA13] to prove approximate-sampling hardness for the permanent of random Gaussian matrices, and hence boson sampling.

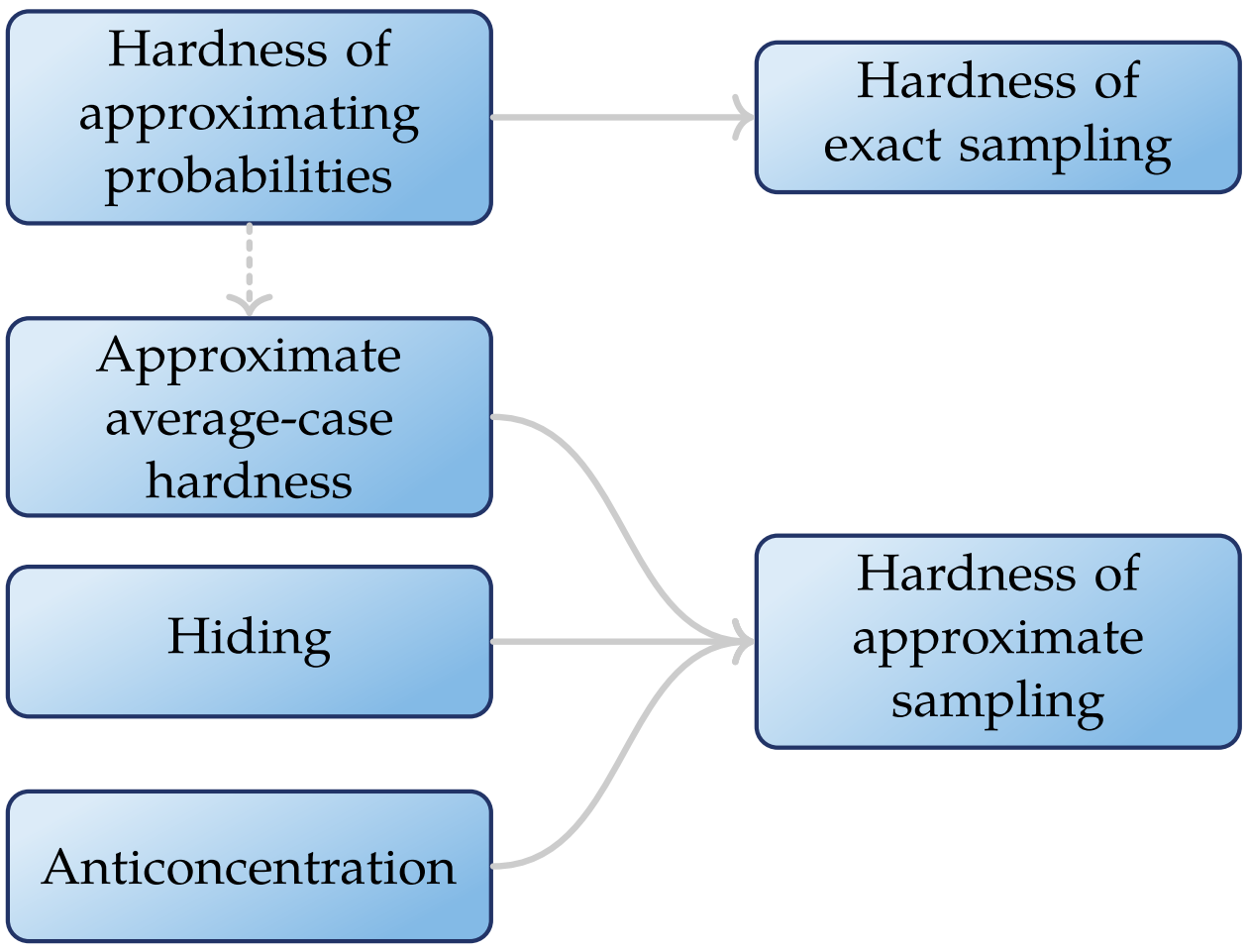

**Figure 2.8:** Similarly to the proof of exact sampling hardness, we can make use of the hardness of approximating the success probabilities of quantum circuits to prove approximate-sampling hardness. In addition, however, further properties of the circuit family $\mathscr{C}$ are required: approximate average-case hardness of computing the success probabilities, the hiding property and anticoncentration of the outcomes.

**Hiding and average-case hardness in boson sampling** We have already seen above in Sec. 2.4 that the output probabilities of boson sampling are given by permanents (6.36) of submatrices of Haar-random unitaries. Conceivably, though, there is some structure in such submatrices. To see this consider the case in which we obtain all bosons in a single mode as the outcome, i.e., $S = (n, 0, 0, \dots)$. In this case, all columns of the submatrix $U_S$ are equal and, plausibly, this can be exploited to approximate $|\operatorname{Perm}(U_S)|^2$.

On the other hand, Lipton [Lip91] famously proved average-case #P-hardness of exactly computing the permanent of random matrices over finite fields. And, in fact, the same result can be extended to complex-valued Gauss-random matrices [AA13]. Highly unstructured random matrices therefore seem ideally suited for proving additive-error sampling hardness in boson sampling. The question arises how one can make sure that in the setting that naturally arises in linear optics, where the outcome probabilities are given by permanents of submatrices $U_S$, fully unstructured (Gauss) random instances can be encoded.

*In the following discussion of boson sampling I follow and borrow from Ref. [Han+19].*

↓ ↓ ↓

The key to answering this question is to consider the *postselected boson sampling distribution* $P^*_{\mathrm{bs},U}$. The postselected distribution $P^*_{\mathrm{bs},U}$ is obtained from $P_{\mathrm{bs},U}$ by discarding all output sequences $S$ with more than one boson per mode, i.e., all $S$ which are not in the set of *collision-free* sequences

$$\Phi^*_{m,n} := \Big\{ S \in \Phi_{m,n} : \forall s \in S : s \in \{0,1\} \Big\}. \tag{2.96}$$

The hardness of sampling from the full boson sampling distribution follows from the fact that if $m$ is chosen to scale with $n$ in the right way, the postselection can be done efficiently in the sense that on average at least a constant fraction of the outcome sequences is collision-free (Theorem 13.4 in Ref. [AA13]).

Why are collision-free outcomes advantageous when proving hardness? Intuitively, this is because for collision-free outcomes, the submatrix $U_S$ has much less structure than for outcomes with collisions. If on top of that, the size of $U_S$ becomes sufficiently small compared to the full size of $U$, neither does there remain any of the structure in $U$ stemming from the orthogonality of its columns.

These intuitions become manifest in the following technical ingredient of their result: if $m$ grows sufficiently fast with $n$, namely if $m \in \Omega(n^5 \log(n)^2)$, the measure induced on $U \sim \mu_H$ by taking $n \times n$-submatrices of unitaries $U \in U(m)$ chosen with respect to the Haar measure $\mu_H$ is close to the complex Gaussian measure $\mu_G(\sigma)$ with mean zero and standard deviation $\sigma = 1/\sqrt{m}$ on $n \times n$-matrices.

Conversely, Aaronson and Arkhipov [AA13, Lemma 5.8] prove that given a Gauss-random instance $X \sim \mu_G(\sigma)$, there is a $\mathsf{BPP}^{\mathsf{NP}}$ algorithm[34] which, given $X$ hides this matrix in a large unitary matrix in the sense that it generates a Haar-random $U \in U(m)$ such that there is a uniformly random $S \in \Phi^*_{m,n}$ such that $X = U_S$. Hiding a Gauss-random instance $X$ is therefore possible by constructing a larger unitary matrix of which $X$ is a uniformly random submatrix, similarly to how we hid a qubit-circuit $C$ by appending uniformly random $X$-gates to it.

34: As BPP is contained in $\Sigma_2^p$, $\mathsf{BPP}^{\mathsf{NP}} \subset \Sigma_3^p$ and – like Stockmeyer's algorithm – is therefore in the third level of PH.

Given this result, one can apply Stockmeyer's algorithm to the samples obtained from $P^*_{\mathrm{bs},U}$ in order to infer the probabilities $P^*_{\mathrm{bs},U}(S)$ and thus solve a problem which is known to be exactly #P-hard, as these probabilities can be expressed as the permanent of a Gaussian matrix.

**Approximate average-case hardness** The key missing step in their proof as well as all other known proofs of approximate sampling hardness is average case hardness for *multiplicatively approximating* the outcome probabilities. As of today, this property is a conjecture in all such schemes. To see why this conjecture is plausibly true for GapP-functions, consider again the argument above. For typical #P functions the number of accepting paths is exponentially large and hence a multiplicative error is also of the same order of magnitude. In contrast, for typical GapP functions, being differences of #P functions, their number of accepting paths is a difference between two exponentially large numbers, which is often orders of magnitude smaller than each such number. This is why for #P functions we often do not expect approximate average-case hardness, while for GapP functions this conjecture seems reaonable.

Another argument in favour of approximate average-case hardness makes use of universal quantities such as the Ising partition function (2.57) [BJS10; GG14; BMS16; Boi+18], Tutte partition functions [GG14] or the Jones polynomial [Kup15b; MB17]. This argument observes that as we draw random instances of, say, an Ising partition function $Z_W$ no additional structure is present as compared to a worst-case instance which a hypothetical approximation algorithm might be able to exploit. While one might argue that these arguments are relatively weak, there have not been counterexamples to approximate average-case hardness either. As it stands, we must currently accept approximate average-case hardness as a conjecture. This does not mean, of course, that we cannot hope to make some progress towards proving this conjecture. This, as well as the same question for the anticoncentration property will be the topic of the next chapter. Before we get there, let us summarize.

## 2.6 Summary

In this chapter, we have walked a long route from the complexity-theoretic foundations of quantum speedups all the way to rigorous and approximate hardness-of sampling arguments relevant to near-term quantum technology.

The complexity-theoretic foundations of quantum speedups manifested themselves in the GapP vs. #P dichotomy a.k.a. the *quantum sign problem*: while multiplicatively approximating the success probabilities of classical circuits can be done in the third level of the polynomial hierarchy, this task remains GapP-complete for certain quantum circuit families. We then saw how the at-first-sight different tasks of sampling from a probability distribution (weakly simulating it) and approximating its outcome probabilities (strongly simulating it) are related on a rigorous level: Stockmeyer's approximate counting algorithm and the concept of the polynomial hierarchy proved key to this question. Building on those methods, we could show that the task of sampling from the output distribution of certain random quantum computations cannot be achieved by an efficient classical algorithm. In a last step, we aimed at making this result robust to realistic errors, that is, additive errors in total total-variation distance on the level of the output distributions. Several further properties were required for this leap: approximate average-case hardness, hiding, and anticoncentration.

---

We have already discussed worst-case hardness for quantum computations, making use of the complexity classes GapP and postBQP. But what about approximate average-case hardness and anticoncentration? When we touched on those issues in a discussion of boson sampling, it already became clear that approximate average-case hardness is out of reach for the available proof methods. Towards proving this conjecture, it is still desirable to close as many gaps or *loopholes* in the hardness argument as possible. This will be the theme of the next chapter.

# Closing gaps: anticoncentration and average-case hardness | 3

In the last chapter, we have seen the general strategy for proving the classical hardness of random circuit sampling schemes. In this chapter, we will turn to the question in which cases the properties required for an additively robust hardness result – anticoncentration and average-case hardness – can be proven, or at least given evidence for.

Despite significant efforts, closing those gaps or loopholes in the complexity-theoretic argument for hardness of classical simulation has remained a difficult task. As already mentioned, the elephant in the room is approximate average-case hardness which, while arguably quite plausible, bears no connections to any well-established results in the theory of computational complexity. Nevertheless, quite recently it was shown that *exact average-case hardness* can be proven for most of the proposed random circuit sampling schemes [Bou+19; Mov18; Mov19]. This constitutes a necessary first step towards proving approximate average-case hardness and at least serves as some additional evidence. However, as we will see below, the techniques used in the exact proofs are very unlikely to work for the approximate case as well.

We already mentioned above that one might consider anticoncentration to be another such piece of evidence as it reduces the baroque mixture of additive and relative errors (2.91) which the Stockmeyer estimate is guaranteed to satisfy to a purely relative error as is familiar from worst-case hardness results. Anticoncentration is, in fact, a property that can be handled much more easily and is by now known to hold for many circuit families proposed in this context [BMS16; Han+18; Haf+19]. To prove anticoncentration we merely need to derive statistical properties of the respective random ensembles, namely, the first and second moments of the entries of the global unitary when averaged over the circuit family. Anticoncentration is thus closely connected to the random matrix theory of the respective ensembles and, as it turns, their capacity to simulate quantum randomness. Quantum pseudorandomness is captured by the notion of *unitary or state designs*. As it turns out, if a circuit family generates a design it also anticoncentrates. This turns out to be the case for many circuit families. However, in some cases such as boson sampling, anticoncentration remains elusive.

In this chapter, we will discuss both properties and examples thereof in more detail. We will start with the somewhat easier-to-grasp notion of anticoncentration in Section 3.1. The result we will prove – that unitary 2-designs anticoncentrate – naturally gives rise to a simple 'recipe' for quantum supremacy schemes as we will discuss in Section 3.2. We proceed to a detailed discussion of average-case hardness in Section 3.3.

## 3.1 Anticoncentration

*The following two sections are based on joint work with Juani Bermejo-Vega, Martin Schwarz and Jens Eisert and closely follow the first part of our joint publication [Han+18].*

In this section we will prove that random quantum circuits circuits drawn from a unitary 2-design anticoncentrate in the sense of Def. 2.12. We then apply this result to both show that random circuits comprised of nearest-neighbour gates that are drawn from a universal gate set containing inverses anticoncentrate in linear depth, and propose two new schemes based on this insight. The application of our result to universal random circuits complies with the intuition that due to the ballistic spread of correlations anticoncentration will generically arise in a depth that scales linearly with the diameter of the system under consideration, and hence, linearly in the number of qubits in a one-dimensional architecture [Boi+18]. The schemes we consider are tailored to platforms in which achieving large numbers of qubits is expensive and local control feasible such that circuit-based schemes can be implemented at relative ease. Paradigmatic examples of such a platform are given by superconducting qubits [Nei+17] or ion traps [Deb+16].

Let us remind ourselves: to prove anticoncentration for a circuit family $\mathcal{C} = \{\mathcal{C}_n\}_{n\geq 1}$ we need to show that the problem instances as given by the acceptance probabilities $|\langle 0|C|0\rangle|^2$ for circuits $C \in \mathcal{C}_n$ are at least as large as the uniform distribution $1/2^n$ on a fraction of the instances that is independent of the problem size $n$. That is, for $\alpha > 0$ there exists $\gamma(\alpha) > 0$ independent of $n$ such that

$$\Pr_{C\sim\mathcal{C}_n}\left[|\langle 0|C|0\rangle|^2 \geq \frac{\alpha}{2^n}\right] \geq \gamma(\alpha). \tag{3.1}$$

The central tool that will help us to prove this inequality is the Paley-Zygmund inequality [BMS16], a lower-bound analogue of Markov-type tail bounds. The inequality can be stated as follows. For a random variable $Z \geq 0$ with finite variance, and $0 \leq \alpha \leq 1$

$$\Pr\left[Z > \alpha\,\mathbb{E}[Z]\right] \geq (1-\alpha)^2 \frac{\mathbb{E}[Z]^2}{\mathbb{E}[Z^2]}. \tag{3.2}$$

The Paley-Zygmund inequality lower bounds the probability that a positive random variable is small in terms of its mean and variance. In our case, the random variable $Z$ will of course be $|\langle 0|C|0\rangle|^2$ for random choices of $C \sim \mathcal{C}$. To prove anticoncentration for random circuit sampling schemes, we will therefore focus on circuit families for which we can upper-bound the second moment. A large class of circuit families that satisfy such an upper bound are – as we will see shortly – unitary designs.

### The Haar measure and designs on the unitary group

Unitary $k$-designs approximate the uniform (Haar) measure on the unitary group in the sense that the first $k$ moments of a unitary $k$-design and the Haar measure match (exactly or approximately).

**Definition 3.1** (Haar measure on groups) *The Haar measure $\mu_H$ on a locally compact Haussdorf group $\mathcal{U}$ is the unique (up to a strictly positive scalar factor) measure which is non-zero on non-empty open sets and is left- and* *right-translation invariant, i.e.*

$$\mu_H(U) > 0 \text{ for any non-empty open set } U \subset \mathcal{U} \tag{3.3}$$

*and*

$$\mu_H(B) = \mu_H(uB) = \mu_H(Bu) \tag{3.4}$$

*for any $u \in \mathcal{U}$ and Borel set $B$ of $\mathcal{U}$, where the left- and right-translate of $B$ with respect to $u$ is given by*

$$uB = \{u\, b : b \in B\} \qquad \text{and} \qquad Bu = \{b\, u : b \in B\}. \tag{3.5}$$

But to approximate the Haar measure on the $n$-qubit unitary group $U(2^n)$ requires a number of elementary quantum gates that scales exponentially in the number of qubits. The definition of a $k$-design captures the fact that, in experiments, only polynomially many gates can be realized so that one can only hope for *pseudorandomenss*. Such pseudorandomness is captured by designs in terms of the *moments* of the Haar measure. To define

We therefores need the notion of the $k^{\text{th}}$-moment operator that acts as a unitary twirl with respect to some measure $\mu$ on the unitary group maps on an operator.

**Definition 3.2** ($k^{\text{th}}$-moment operator) *Let $M_\mu^k$ be the k-th moment operator on $\mathcal{L}(\mathcal{H}^{\otimes k})$ with respect to a distribution $\mu$ on $U(D)$, $D = 2^n = \dim \mathcal{H}$ defined as*

$$\begin{aligned} X \mapsto M_\mu^k(X) &:= \mathbb{E}_\mu \left[ U^{\otimes k} X (U^\dagger)^{\otimes k} \right] \\ &= \int_{U(D)} U^{\otimes k} X (U^\dagger)^{\otimes k} \mu(U). \end{aligned} \tag{3.6}$$

We can now define unitary $k$-designs [GAE07; Dan+09].

**Definition 3.3** (Unitary $k$-design) *Let $\mu$ be a probability measure*[1] *on the unitary group $U(D)$. Then $\mu$ is an exact unitary k-design if*

$$M_\mu^k = M_{\mu_H}^k . \tag{3.7}$$

For the following, we need to relax this notion to that of an *approximate* unitary $k$-design. In such a definition we can allow for both relative and additive errors on the equality (3.7) [BHH16]:

**Definition 3.4** (Approximate unitary $k$-designs) *Let $\mu$ be a distribution on the unitary group $U(D)$. Then $\mu$ is*

1. *an additive $\epsilon$-approximate unitary k-design if*[2]

$$\| M_\mu^k - M_{\mu_H}^k \|_\diamond \leq \epsilon , \tag{3.11}$$

2. *a relative $\epsilon$-approximate unitary k-design if*

$$(1-\epsilon) M_{\mu_H}^k \leq M_\mu^k \leq (1+\epsilon) M_{\mu_H}^k . \tag{3.12}$$

Since the former definition is much more common in the literature, let us remark that the two definitions are closely related via the following Lemma of Brandão, Harrow, and Horodecki [BHH16] in which, however,

1: In the case we are concerned with here, this measure is induced by the uniform distribution over a discrete circuit family $\mathcal{C}_n \subset \mathcal{U}(2^n)$.

2: The distance between two quantum channels $\mathcal{U}, \mathcal{V}$ as measured by the *diamond norm* of their difference $\|\mathcal{U} - \mathcal{V}\|_\diamond$ quantifies the worst-case distinguishability between those channels with respect to state preparations and measurements. The diamond norm of a quantum channel $\mathcal{E} : L(\mathcal{A}) \to L(\mathcal{B})$ mapping from the linear operators on a complex vector space $\mathcal{A}$ to those on a space $\mathcal{B}$ is defined as the stabilized $(1 \to 1)$-norm

$$\|\mathcal{E}\|_\diamond = \|\mathcal{E} \otimes 1_{L(\mathcal{A})}\|_{1\to1}. \tag{3.8}$$

The $1 \to 1$ norm is the induced trace norm on the linear maps $L(\mathcal{A})$ and is given by

$$\|\mathcal{E}\|_{1\to1} = \sup_{\rho \in L(\mathcal{A})} \frac{\|\mathcal{E}(\rho)\|_1}{\|\rho\|_1} \tag{3.9}$$

$$= \sup_{\substack{\rho \in L(\mathcal{A}), \\ O \in L(\mathcal{B})}} \frac{|\operatorname{Tr}[O\mathcal{E}(\rho)]|}{\|\rho\|_1 \|O\|_\infty}. \tag{3.10}$$

a factor of the system dimension $D$ enters that renders it useless for our purposes.

**Lemma 3.1** (Additive and relative approximate designs) *If $\mu$ is a relative $\epsilon$-approximate unitary k-design then $\|M^k_\mu - M^k_{\mu_H}\|_\diamond \leq 2\epsilon$. Conversely, if $\|M^k_\mu - M^k_{\mu_H}\|_\diamond \leq \epsilon$, then $\mu$ is a relative $\epsilon D^{2k}$-approximate unitary k-design.*

It is a simple exercise to show that if $\mu$ is a unitary $k$-design, all up to the $k^{th}$ moments of $\mu$ equal the moments of the Haar measure.

**Lemma 3.2** ($k-1$ designs from $k$ designs) *Let $\mu$ be a distribution on the unitary group $U(D)$ that is an exact unitary k-design. Then $\mu$ is also a $(k-1)$-design.*

*Proof.* Let $\mu$ be a unitary $k$ design. That means that

$$\mathbb{E}_\mu \left[U^{\otimes k} X (U^\dagger)^{\otimes k}\right] = \mathbb{E}_{\mu_H} \left[U^{\otimes k} X (U^\dagger)^{\otimes k}\right] \tag{3.13}$$

holds for all operators $X$ acting on $\mathcal{H}^{\otimes k}$. Choose $X = Y \otimes 1$ with $Y$ being an arbitrary operator on $\mathcal{H}^{\otimes k-1}$. Then

$$\mathbb{E}_\mu \left[U^{\otimes k-1} Y (U^\dagger)^{\otimes k-1}\right] = \mathbb{E}_{\mu} \left[U^{\otimes k} X (U^\dagger)^{\otimes k}\right] \tag{3.14}$$

i.e., $\mu$ is a unitary $(k-1)$-design. □

**Corollary 3.3** (Approximate $k-1$ designs from approximate $k$ designs) *Let $\mu$ be an (additive or relative) approximate unitary k-design. Then $\mu$ is also an approximate unitary $(k-1)$-design, i.e.,*

$$\|M^k_\mu - M^k_{\mu_H}\|_\diamond \leq \epsilon \Rightarrow \|M^{k-1}_\mu - M^{k-1}_{\mu_H}\|_\diamond \leq \epsilon \tag{3.15}$$

*and likewise for relative errors.*

## Anticoncentration of 2-designs

Observing that Haar-random unitaries would trivially anticoncentrate (see below), we will now relax this unreasonable requirement on the circuit family to efficiently attainable designs: unitary 2-designs anticoncentrate.

**Theorem 3.4** (Anticoncentration of unitary 2-designs) *Let $\mu$ be a relative $\epsilon$-approximate unitary 2-design on the group $U(D)$. Then the success probabilities $|\langle 0|U|0\rangle|^2$ of a $\mu$-random unitary $U \in U(D)$ anticoncentrate in the sense that for $0 \leq \alpha \leq 1$*

$$\Pr_{U\sim\mu}\left(|\langle 0|U|0\rangle|^2 > \frac{\alpha(1-\epsilon)}{N}\right) \geq \frac{(1-\alpha)^2(1-\epsilon)^2}{2(1+\epsilon)}\,. \tag{3.16}$$

We point out that Theorem 3.4 also holds in exactly the same way for relative $\epsilon$-approximate state 2-designs. This is a weaker condition than the unitary design condition since any (approximate) unitary 2-design generates an (approximate) state 2-design via application to an arbitrary reference state. Also note that the fact that $\mu$ is a relative $\epsilon$-approximate 1-design is crucial for the bound (3.16) to become non-trivial. If instead $\mu$ was an additive design the lower bound would asymptotically tend to zero with $1/D$ and hence not stay larger than a constant as the number of qubits is increased. The 1-design condition still holds even exactly for many distributions $\mu$, even in cases in which the higher moments are only approximately given by the Haar moments.

*Proof of Theorem 3.4.* Our proof of the anticoncentration bound (3.16) uses the fact that 2-designs have bounded second moments in conjunction with the Paley-Zygmund inequality (6.20).

Let $\mu$ be a relative $\epsilon$-approximate unitary 2-design. Then for $l = 2, 4$, it holds that

$$\begin{aligned}(1-\epsilon)\,\mathbb{E}_{U\sim\mu_H}\left[|\langle a|U|b\rangle|^l\right] &\leq \mathbb{E}_{U\sim\mu}\left[|\langle a|U|b\rangle|^l\right]\\ &\leq (1+\epsilon)\,\mathbb{E}_{U\sim\mu_H}\left[|\langle a|U|b\rangle|^l\right]\ .\end{aligned} \tag{3.17}$$

This is due to the fact that for a unitary $k$-design $\mu$ the expectation value of an arbitrary polynomial of degree 2 in the matrix elements of both $U$ and $U^\dagger$ over $\mu$ equals the same expectation value but taken over the Haar measure up to a relative error $\epsilon > 0$ [HL09]. To see this, observe that averaging a monomial in the matrix elements of $U$ over the $k$-design $\mu$ can be expressed as $\langle i_1, \ldots, i_k|M^k_\mu(|j_1, \ldots, j_k\rangle\langle j'_1, \ldots, j'_k|)|i'_1, \ldots, i'_k\rangle$. Hence, if $M^k_\mu = M^k_{\mu_H}$, "then any polynomial of degree $k$ in the matrix elements of $U$ will have the same expectation over both distributions" [HL09]. This gives rise to

$$\begin{aligned}&\Pr_{U\sim\mu}\left(|\langle 0|U|0\rangle|^2 > \alpha(1-\epsilon)\,\mathbb{E}_{U\sim\mu_H}[|\langle 0|U|0\rangle|^2]\right)\\ &\quad\geq \Pr_{U\sim\mu}\left(|\langle 0|U|0\rangle|^2 > \alpha\,\mathbb{E}_{U\sim\mu}[|\langle 0|U|0\rangle|^2]\right)\\ &\quad\geq (1-\alpha)^2\frac{\mathbb{E}_{U\sim\mu}[|\langle 0|U|0\rangle|^2]^2}{\mathbb{E}_{U\sim\mu}[|\langle 0|U|0\rangle|^4]}\\ &\quad\geq (1-\alpha)^2\frac{(1-\epsilon)^2\,\mathbb{E}_{U\sim\mu_H}[|\langle 0|U|0\rangle|^2]^2}{(1+\epsilon)\,\mathbb{E}_{U\sim\mu_H}[|\langle 0|U|0\rangle|^4]}.\end{aligned} \tag{3.18}$$

**Lemma 3.5** (Output distribution of Haar-random unitaries) *The distribution $P_{\mu_H}$ of success probabilities $p = |\langle 0|U|0\rangle|^2$ of Haar-random unitaries $U$ is given by*

$$P_{\mu_H}(p) = (D-1)(1-p)^{D-2} \xrightarrow{D\gg 1} D\exp(-Dp). \tag{3.19}$$

*In particular, $P_{\mu_H}$'s first and second moments are given by*

$$\mathbb{E}_{\mu_H}[p] = \frac{1}{D}, \quad \mathbb{E}_{\mu_H}[p^2] = \frac{2}{D(D+1)}. \tag{3.20}$$

We prove this lemma below. Inserting the expressions (3.20) for the first and second moment $\mathbb{E}_{U\sim\mu_H}[|\langle 0|U|0\rangle|^2]$ and $\mathbb{E}_{U\sim\mu_H}[|\langle 0|U|0\rangle|^4]$, we find

$$\begin{aligned}&\Pr_{U\sim\mu}\left(|\langle 0|U|0\rangle|^2 > \frac{\alpha(1-\epsilon)}{D}\right)\\ &\quad\geq (1-\alpha)^2\frac{D(D+1)}{2D^2}\frac{(1-\epsilon)^2}{(1+\epsilon)} \geq (1-\alpha)^2\frac{(1-\epsilon)^2}{2(1+\epsilon)},\end{aligned}$$

which completes the proof. □

## Moments of the Haar measure

We now review two arguments showing Lemma 3.5. The first one is a direct calculation using random matrix theory, while the second one exploits Schur-Weyl duality.

Thanks go to Andreas Elben for fun discussions on random matrix theory in Bad Honnef.

**Argument from random matrix theory** Let us begin by introducing a few important ensembles of random matrices. We do so from a rather hands-on perspective.

- G($D$) (Ginibre Ensemble): The set of matrices $Z$ with complex Gaussian entries. G($D$) is characterized by the measure

$$\mathrm{d}\mu_G(Z) := \pi^{-N^2} \mathrm{e}^{-\operatorname{Tr}[Z^\dagger Z]} \mathrm{d}Z, \tag{3.21}$$

  i.e., each individual entry $z_{i,j}$ is distributed as $\exp(-|z_{i,j}|^2)/\pi$.
- GUE($D$) (Gaussian Unitary Ensemble): The set of $D \times D$ Hermitian matrices with complex Gaussian entries, i.e., $H \in \mathrm{GUE} \Leftrightarrow H = D + R + R^\dagger$, where $D$ is a diagonal matrix with real Gaussian entries and $R$ is an upper triangular matrix with complex Gaussian entries. GUE($D$) is characterized by the measure

$$\mathrm{d}\mu_{GUE} = Z_{\mathrm{GUE}(D)}{}^{-1} \mathrm{e}^{-D \operatorname{Tr}(H^2)/2}\, \mathrm{d}H, \tag{3.22}$$

  with $Z_{GUE} = \int \mathrm{e}^{-D \operatorname{Tr}(H^2)/2}\, \mathrm{d}H$ on the space of Hermitian matrices.
- CUE($D$) (Circular Unitary Ensemble): The set of Haar-random $D \times D$ unitary matrices. CUE($D$) is characterized by the Haar measure $\mathrm{d}\mu_H$.

All of $\mathrm{d}\mu_{GUE}$, $\mathrm{d}\mu_G$, and $\mathrm{d}\mu_H$ are left- and right-invariant under the action of $U(D)$. There are two ways of constructing Haar-random matrices.

1. Draw a Gaussian matrix $Z \in \mathrm{G}(D)$, and perform the unique QR decomposition such that $Z = QR$, with an orthogonal matrix $Q$ and $R$ is required to have positive diagonal entries. Setting $U = Q$, yields a Haar-random unitary [Ozo09; Mez06].
2. Draw a GUE matrix $Z \in \mathrm{GUE}$. Since $Z$ is Hermitian, the eigenvectors $v_i, i = 1, \ldots, D$ of $Z$ are orthonormal. Multiplying each eigenvector $v_i$ by a random phase $\mathrm{e}^{\phi_i}$ we can construct a Haar-random unitary matrix $U = (\mathrm{e}^{\phi_1} v_1\, \mathrm{e}^{\phi_2} v_2 \, \cdots \, \mathrm{e}^{\phi_n} v_D)$ writing those eigenvectors into the columns of $U$ [WH05].

Let us now derive the distribution of the amplitudes $|\langle a|U|b\rangle|^2$ of the matrix elements a Haar-random unitary $U$ [ZK94; PZK98; WH05; Haa10]. To this end we apply knowledge about the distribution of entries of eigenvectors of GUE matrices and their relation to Haar-random unitaries.

I will follow Ref. [Haa10, Chapter 4.9].

The eigenvectors $v_i$ of a given operator $H \in \mathrm{GUE}(D)$ have $D$ complex components $c_k$ and unit norm $\|v_i\|_2 = 1$. Since every eigenvector can be unitarily transformed into an arbitrary vector of unit norm, the only invariant characteristic of those eigenvectors is the norm itself. Thus, the joint probability for its components $\{c_k\}$ must read

$$P_{\mathrm{GUE}}(\{c_k\}) = \mathrm{const} \cdot \delta\left(1 - \sum_{k=1}^{N} |c_k|^2\right), \tag{3.23}$$

where $\delta(\cdot)$ is the $\delta$-distribution and the constant is fixed by normalization.

Assuming real entries for now (we can always go to complex ones by doubling $D$) we can calculate that normalization by evaluating the integral on the $D$-dimensional unit sphere

$$\begin{aligned}
\text{const} &= \int_{-\infty}^{\infty} \left(\prod_{i=1}^{D} \mathrm{d}c_i\right) \delta\left(1 - \sum_{k=1}^{D} |c_k|^2\right) && (3.24)\\
&= \int \mathrm{d}\omega^{D-1} \int_0^{\infty} \mathrm{d}R\, R^{D-1} \delta(1-R^2) && (3.25)\\
&= \int \mathrm{d}\omega^{D-1} \int_0^{\infty} \mathrm{d}R\, R^{D-1} \frac{1}{2R}[\delta(1-R) + \delta(1+R)] && (3.26)\\
&= \pi^{D/2}/\Gamma(D/2)\,. && (3.27)
\end{aligned}$$

Similarly, we can calculate the marginal distribution

$$P^{(D,l)}(c_1, \ldots, c_l) = \pi^{-D/2}\Gamma(D/2) \int_{-\infty}^{\infty} \left(\prod_{i=l+1}^{N} \mathrm{d}c_i\right) \delta\left(1 - \sum_{k=1}^{D} |c_k|^2\right) \tag{3.28}$$

$$= \int \mathrm{d}\omega^{D-l-1} \int_0^{\infty} \mathrm{d}R\, R^{D-l-1} \delta(1 - R^2 - \sum_{k=1}^{l} |c_k|^2) \tag{3.29}$$

$$= \pi^{-l/2} \frac{\Gamma(D/2)}{\Gamma((D-l)/2)} \left(1 - \sum_{k=1}^{l} |c_k|^2\right)^{(D-l-2)/2}. \tag{3.30}$$

For the GUE we then obtain the probability density for the amplitude $y = x_1^2 + x_2^2$ of a single complex entry $x_1 + \mathrm{i}x_2$ of an eigenvector to be the twofold integral over real and imaginary part

$$\begin{aligned}
P_{\mathrm{GUE}}(y) &= \int dx_1 dx_2 P^{(2D,2)}(x_1, x_2) \delta(y - x_1^2 - x_2^2)\\
&= (D-1)(1-y)^{D-2}. && (3.31)
\end{aligned}$$

Since the eigenvectors of a GUE matrix are identically distributed (up to a global phase) as the columns of a CUE matrix, we obtain the same distribution as (3.31) for the amplitudes of the matrix elements of a CUE matrix [WH05]. Notably, as $D$ becomes much larger than 1, we obtain the exponential or *Porter-Thomas distribution*

$$P_{\mu_H}(p) = (D-1)(1-p)^{D-2} \xrightarrow{D \gg 1} D \exp(-Dp)\,. \tag{3.32}$$

The first and second moments of $P_{\mathrm{CUE}}$ are then given by

$$\begin{aligned}
\mathbb{E}_{\mu_H}[p] &= \frac{1}{D}, && (3.33)\\
\mathbb{E}_{\mu_H}[p^2] &= \frac{2}{D(D+1)}. && (3.34)
\end{aligned}$$

**Argument from Schur-Weyl duality** The moments (3.20) can alternatively be obtained for both state and unitary 2-designs exploiting Schur-Weyl duality. Schur-Weyl duality invokes both the diagonal action

I am grateful to Richard Kueng and Emilio Onorati for introducing me into the wonders of Schur-Weyl duality. I will follow Zhu et al. [Zhu+16] in my sketch thereof.

of the unitary group $U(D)$ acting on $\mathcal{H} = \mathcal{C}^D$ on $k$ tensor copies thereof

$$U \mapsto \tau^k(U) : |\psi_1\rangle \otimes |\psi_2\rangle \otimes \cdots \otimes |\psi_k\rangle \mapsto U|\psi_1\rangle \otimes U|\psi_2\rangle \otimes \cdots \otimes U|\psi_k\rangle \quad \forall |\psi_j\rangle \in \mathcal{C}^D, \forall U \in U(D), \tag{3.35}$$

and the action on the symmetric group $S_k$ on $(\mathcal{C}^D)^{\otimes k}$ which permutes the tensor factors:

$$\pi(|\psi_1\rangle \otimes |\psi_2\rangle \otimes \cdots \otimes |\psi_k\rangle) = |\psi_{\pi(1)}\rangle \otimes |\psi_{\pi(2)}\rangle \otimes \cdots \otimes |\psi_{\pi(k)}\rangle \quad \forall |\psi_j\rangle \in \mathcal{C}^D, \forall \pi \in S_k. \tag{3.36}$$

Noting that the diagonal action of $U(D)$ and the permutation action of $S_k$ commute, Schur-Weyl duality states that the space $(\mathcal{C}^d)^{\otimes k}$ decomposes into multiplicity-free irreducible representations of $U(D) \times S_k$ [Zhu+16; GW09].

This yields an explicit expression of the $k^{\text{th}}$-moment operators $M^k_\mu(X)$ acting on a projector $|\psi\rangle\langle\psi|^{\otimes k}$ for some fixed reference state $|\psi\rangle \in \mathcal{H}$ as the projector onto the span of the symmetric group on $k$ tensor copies of the Hilbert space $\mathcal{H}$.

$$M^k_{\mu_H}(|\psi\rangle\langle\psi|^{\otimes 2}) = \mathbb{E}_{U\sim\mu_H}\left[U^{\otimes 2}|\psi\rangle\langle\psi|^{\otimes 2}(U^\dagger)^{\otimes 2}\right] = \frac{P_{[k]}}{D_{[k]}}, \tag{3.37}$$

where $P_{[k]}$ is the projector onto the symmetric subspace of $k$ tensor copies of $\mathcal{H}$ and

$$D_{[k]} = \binom{D+k-1}{k} \tag{3.38}$$

is the dimension of that space. We can apply the expression (3.37) to compute the moments of the output probabilities of unitary $k$-designs when applied to a reference state as

$$\mathbb{E}_{U\sim\mu_H}\left[|\langle 0|U|0\rangle|^{2k}\right] = \mathbb{E}_{U\sim\mu_H}\left[\langle 0|^{\otimes k}U^{\otimes k}|0\rangle\langle 0|^{\otimes k}(U^\dagger)^{\otimes k}|0\rangle^{\otimes k}\right] \tag{3.39}$$

$$= \frac{1}{D_{[k]}}\langle 0|^{\otimes k}P_{[k]}|0\rangle^{\otimes k} = \frac{1}{D_{[k]}}, \tag{3.40}$$

since the symmetric projector acts trivially on tensor copies $|\psi\rangle^{\otimes k}$ of the same state. Noting that $D_{[2]} = D(D+1)/2$ and $D_{[1]} = D$ this calculation reproduces the results of the one above.

As we noted, the output distribution (3.19) of a Haar-random unitary asymptotically approaches the exponential (Porter-Thomas) distribution (3.32). The very same behaviour has already been observed numerically in many different contexts involving pseudo-random operators [GAE07; Eme+03], non-adaptive measurement-based quantum computation [BWV08], and universal random circuits [Boi+18] and might even been viewed as a signature of non-simulability [Boi+18].

## 3.2 A quantum speedup recipe

Our anticoncentration results (Theorem 3.4) for approximate unitary and state 2-designs gives rise to a *generic recipe* for the identification of quantum circuit families and input states that are hard to simulate classically under plausible complexity-theoretic conjectures, building upon the approach of Refs. [AA13; BMS16]. The strategy goes in three steps parallel to ingredients (i-iii) of the proof of approximate sampling hardness (Theorem 2.5) based on Stockmeyer's algorithm (Thm. 2.4). In the following, we apply this general strategy to a few examples of random circuits, most prominently, universal random circuits.

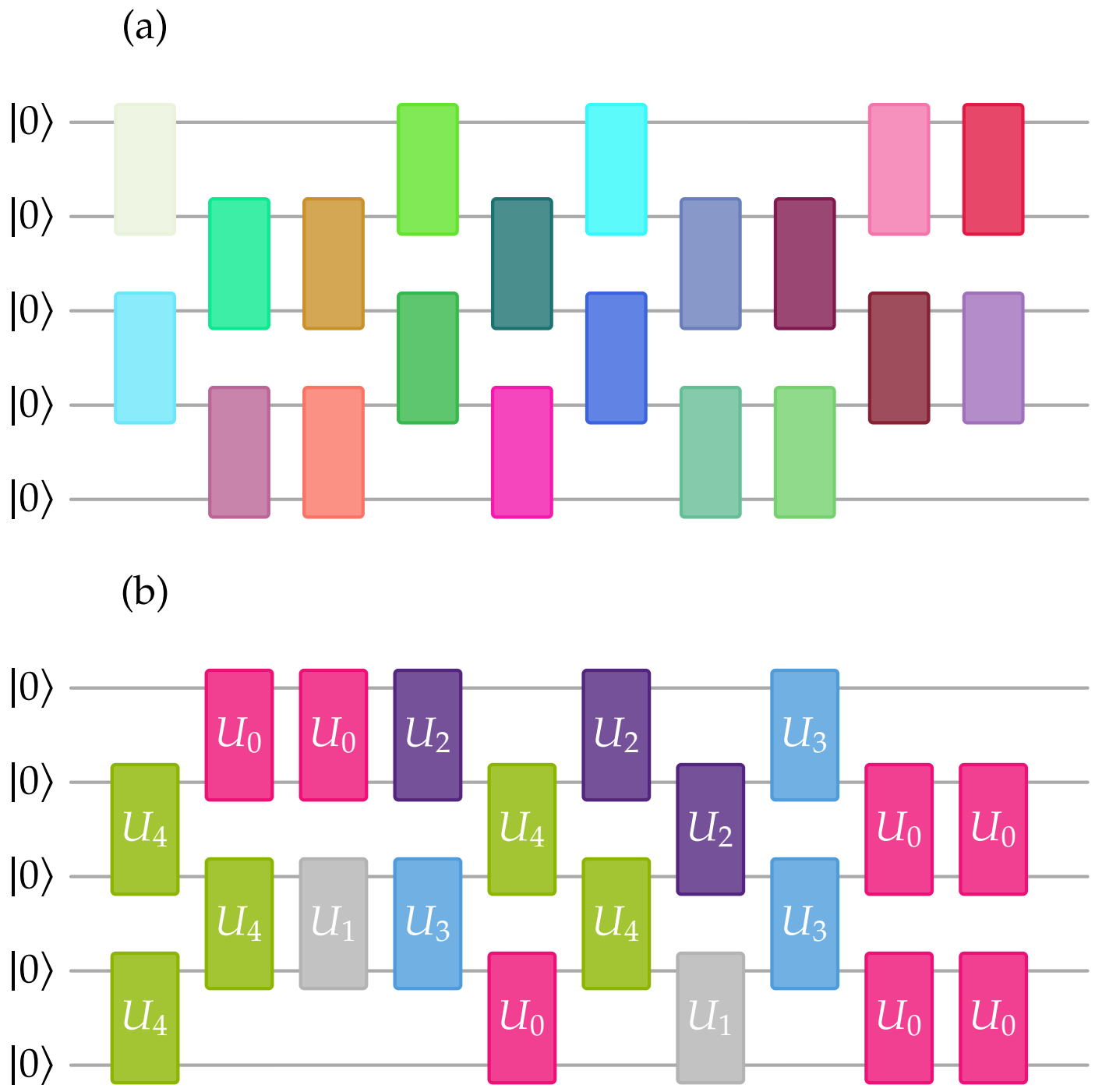

**Figure 3.1:** Layout of the parallel random circuit families. In each step either the even or odd configuration of parallel two-qubit unitaries is applied with probability 1/2. Every two-qubit gate is chosen from the respective measure on $U(4)$ – (a) the Haar measure, (b) the uniform distribution on the gate set $G$. Here we depict a five-qubit random instance of depth 10 where in (a) the colour choice represents different gates, and in (b) the gate set consists of 5 two-qubit unitaries $G = \{U_0, U_2, \ldots, U_4\}$.

**Universal random circuits.** The first example that we also focus on are random quantum circuits constructed from single- and two-qubit gates, most prominently, the gate set $\mathcal{G}_{\mathrm{BIS}} = \{CZ, H, \sqrt{X}, \sqrt{Y}, T\}$ studied by Boixo *et al.* [Boi+18; Nei+17]. When presenting this example we put particular emphasis on the *circuit depth* required to reach a scheme that shows a provable quantum speedup. As a first step (i) of the general strategy, the following corollary establishes that the output distribution of a random circuit formed in a particular fashion from $\mathcal{G}_{\mathrm{BIS}}$ anticoncentrates. This holds already in *linear depth*.

**Corollary 3.6** (Universal random circuits anticoncentrate) *The output probabilities of universal random circuits in one dimension from the following two circuit families (illustrated in Fig. 3.1) anticoncentrate in a depth that scales as $O(n\log(1/\epsilon))$ in the sense of Eq.* (3.16).

- Parallel local random circuits*: In each step either the unitary $U_{1,2} \otimes U_{3,4} \otimes \cdots \otimes U_{n-1,n}$ or the unitary $U_{2,3} \otimes U_{4,5} \otimes \cdots \otimes U_{n-2,n-1}$ is applied (each with probability* 1/2*), with $U_{j,j+1}$ independent unitaries drawn from the Haar measure on $U(4)$. (This assumes $n$ is even.)*
- Universal gate sets*: Let $G := \{g_i\}_{i=1}^m$ with each $g_i \in U(4)$ be a universal gate set containing inverses with elements composed of algebraic identities, i.e., a gate set G such that the group generated by G is dense in $U(4)$ and satisfying $g_i \in G \Rightarrow g_i^{-1} \in G$. In each step either the unitary $U_{1,2} \otimes U_{3,4} \otimes \cdots \otimes U_{n-1,n}$ or the unitary $U_{2,3} \otimes U_{4,5} \otimes \cdots \otimes U_{n-2,n-1}$* *is applied (each with probability* $1/2$*), with* $U_{j,j+1}$ *independent unitaries drawn uniformly from* $G$*.*

*Proof of Corollary 3.6.* The central ingredient of our proof of Corollary 3.6 is the result of Brandão, Harrow, and Horodecki [BHH16]. There, the authors show that the two random circuit families are relative $\epsilon$-approximate unitary $k$-designs on $U(2^n)$ in depth $\mathsf{poly}(k) \cdot O(n \log(1/\epsilon))$ [BHH16, Corollary 6 and 7]).

In particular, these random circuits are relative $\epsilon$-approximate unitary 2-designs in depth $O(n \log(1/\epsilon))$, i.e., linear in the number of qubits and logarithmic in $1/\epsilon$. Applying Theorem 3.4 to the output probabilities $|\langle x|C|0\rangle|^2$ of a random circuit $C$ applied to an initial all-zero state yields the claimed anticoncentration bound for the output probabilities of such circuits. □

To prove a quantum speedup using the Stockmeyer technique, the second required ingredient is #P-hardness of strong classical simulation of the output probabilities (ii). Indeed, since the gate set $\mathcal{G}_{\mathrm{BIS}}$ is universal, the $\mathsf{post}\mathcal{C} = \mathsf{postBQP}$ connection is immediate. Boixo *et al.* [Boi+18] moreover showed that the output probabilities can be expressed in terms of the imaginary-time partition function of a random Ising model, suggesting that the average-case conjecture for random circuits is a natural one (iii). It remains to be shown that random universal circuits are both not classically strongly simulable and anticoncentrate in *linear depth* in a one-dimensional setting. Lemma 3.7 (below) establishes this is indeed the case for a large class of finite gate sets with efficiently-computable matrix entries (so that they cannot artificially encode solutions to hard problems). It is an open question whether this can be improved to square-root depth in a two-dimensional setting such as that of Refs. [Boi+18; Nei+17][3].

3: Notice that since the publication of this work, Harrow and Mehraban [HM18] proved that random circuits in two dimensions give rise to an approximate unitary 2-design.

Given two $O(1)$-local gate sets $A$ and $B$, we say that $A$ *exactly synthesises* $B$ if every gate $V \in B$ can be exactly implemented via a polynomial-time computable, constant-size circuit comprising gates in $A$.

**Lemma 3.7** (Hardness of strong classical simulation) *Let* $G$ *be any finite universal gate set with algebraic efficiently computable matrix entries that can exactly synthesise either the* $\{e^{i\frac{\pi}{8}X}, e^{i\frac{\pi}{4}X\otimes X}, \mathrm{SWAP}\}$ *or* $\{e^{i\frac{\pi}{8}X\otimes X}, \mathrm{SWAP}\}$*. Then, approximating the output probabilities of* $O(n)$*-depth circuits of* $G$ *nearest-neighbour gates in one dimension up to relative error* $1/4$ *is* #P*-hard.*

Let us highlight that Lemma 3.7 applies to many well-studied universal gate sets, including $\mathcal{G}_{\mathrm{BIS}}$, the ubiquitous Clifford + $T$ [Boy+00], Hadamard + controlled-$\sqrt{Z}$ [KSV02], Hadamard + Toffoli [Shi02; PR13] and others [Sho96; ERW96; KLZ98]. Interestingly, Lemma 3.7 holds also for non-universal gate sets, though the latter may not always anticoncentrate. We now prove Lemma 3.7.

*Proof of Lemma 3.7.* We begin by showing that both given target gate sets can exactly implement subgroups of the 2-qubit dense IQP circuits of Ref. [BMS16]. Specifically, the first gate set gives us the group $\mathcal{G}_1$ generated by $\exp\left(i\frac{\pi}{8}X_i\right)$, $\exp\left(i\frac{\pi}{4}X_iX_j\right)$ gates acting on a complete graph, while the second gives the group generated by arbitrary long range $\exp\left(i\frac{\pi}{8}X_iX_j\right)$ gates. In both cases, long range interactions are obtained via the available SWAPs.

**Table 3.1:** Examples of random circuit families that exhibit a provable quantum speedup up to total-variation distance errors under an approximate average-case hardness assumption.

| Circuit family $\mathscr{C}$ | Input state $\|\psi_0\rangle$ | (State) 2-design property | Worst-case hardness (post$\mathscr{C}$ = postBQP) | Average-case conjecture in terms of universal quantity |
|---|---|---|---|---|
| $\mathscr{G}_{\mathrm{BIS}}$ | $\|0\rangle^{\otimes n}$ | [BHH16] | [Boi+18; MB17] | Ising part. func./Jones polynomial |
| Diagonal unitaries | $\|+\rangle^{\otimes n}$ | [NKM14] | [FM17; Kup15a] | Ising partition function |
| Clifford circuits | $(T\|0\rangle)^{\otimes n}$ | [Dan+09] | [JN13] | Ising partition function |

Next, we show that, like the circuits in Ref. [BMS16], both $\mathscr{G}_1$ and $\mathscr{G}_2$ are universal under postselection. Indeed, both can adaptively implement a single-qubit Hadamard via gate teleportation [GC99] (see also [RBB03a; BJS10]), and non-adaptively, if we can postselect. The claim follows from the universality of known gate sets [Boy+00; KSV02].

Last, due to Refs. [GG14; FM17], the output probabilities of postselected universal quantum circuits are #P-hard to approximate up to multiplicative error $\sqrt{2}$ (relative error $1/4$). The previous fact implies that this holds for the dense IQP circuits in $\mathscr{G}_1$ and $\mathscr{G}_2$. Furthermore, $n$-qubit dense IQP circuit can be exactly implemented in $O(n)$ depth on a 1D nearest-neighbour architecture using SWAP gates [Ber+18, Lemma 6]. It follows that the output probabilities of linear-depth circuits in $\mathscr{G}_1$ or $\mathscr{G}_2$ are #P-hard to approximate. This readily extends to any circuit family that can exactly synthesise either of the former, since this process only introduces a constant depth overhead. □

We do not know whether Lemma 3.7 extends to arbitrary gate sets since applying some Solovay-Kitaev type gate synthesis algorithms [Kit97; NC00; KSV02] might introduce a polynomial overhead factor in depth. This is because due to Chernoff-Hoeffding's bound #P-hard-to-approximate quantum probabilities need to be (at least) super-polynomially small, for otherwise they could be estimated up to an inverse polynomially small additive and hence constant relative error in quantum polynomial time by mere sampling, which is not believed possible [Ben+97; Aar05]. To approximate such small probabilities via the Solovay-Kitaev algorithm requires $\Omega(n^\alpha)$ overhead for some $\alpha > 0$ assuming the counting exponential time hypothesis (ETH)[4] [Del+14]. These issues are closely related to the open question of whether or not the power of postselected quantum circuits is gate set independent given some $\tilde{O}(n^\alpha)$ depth bound [Kup15a].

4: The exponential time hypothesis states that 3SAT on $n$ variables cannot be solved in time $\exp(o(n))$ [Del+14], i.e., the runtime of an algorithm for NP complete problems scales exponentially in the input size $n$. The counting ETH states that *counting* the satisfying assignments cannot be done in time $\exp(o(n))$.

**Commuting circuits.** As a second example, we consider circuits of diagonal unitaries composed of controlled-phase type one- and two-qubit gates of the form $\mathrm{diag}(1, 1, 1, \mathrm{e}^{\mathrm{i}\phi})$, and an input state $|+\rangle^{\otimes n}$. By the result of Ref. [NKM14] this gate set yields a state 2-design if the phases are picked from discrete sets ($\{0, \pi\}$ for the two-qubit gates, and $\{0, 2\pi/3, 4\pi/3\}$ for the single qubit gates), and thus satisfies anticoncentration in the sense of Eq. 3.16 (i). Adding the $S$-gate to the gate set and measuring all qubits in the $X$-basis we obtain post$\mathscr{C}$ = postBQP by Refs. [FM17; BMS16] as IQP circuits are an instance of diagonal unitaries (ii). Here, we have used the fact that adding the $S$-gate and postselection gives us access to the universal gate set Clifford + $\pi/12$ [CW15; BRS15]. Again, the average-case conjecture can be phrased in terms of an Ising partition function (iii).

I am grateful to Richard Kueng for pointing me to this example.

Last, the circuits can be implemented in linear depth if either long-range interactions or nearest-neighbour SWAPs are allowed [Ber+18].

**Clifford circuits with product-state inputs.** A similar argument can be applied to Clifford circuits which are known to be an exact 2-design [Dan+09; Cle+16] applied to magic input states. By the result of Ref. [KS14] an arbitrary element of the Clifford group in $2^n$ dimensions can be decomposed into $O(n^3)$ elementary Clifford gates. The result of Ref. [Cle+16] even achieves an exact 2-design using only quasi-linearly many one- and two-qubit Clifford gates. The post$\mathscr{C}$ =postBQP for this case is due to Ref. [JN13]. We summarize these examples in Table 3.1.

## 3.3 Average-case hardness

Our recipe for Stockmeyer-based quantum speedup schemes is guaranteed to satisfy anticoncentration by the 2-design property. Anticoncentration is an important part of the Stockmeyer argument for quantum supremacy. However, the recipe guarantees approximate worst-case hardness only. Approximate average-case hardness needs to be conjectured still, in all of the cases listed above (Table 3.1). If not prove[5], can we hope to provide further evidence for this property to hold?

5: Recall the discussion on approximate average-case hardness at the end of Sec. 2.5.

Average-case complexity is a suprisingly little studied question in complexity theory and comes with a number of intriguing peculiarities. The question of average-case hardness was first posed by Levin [Lev86] as a rigorous means to narrow down problem classes in which one can hope for simulation algorithms that work on average. What is the complexity of an instance drawn at random from some distribution $\mu$ over all possible problems? The most important question in the context of average-case complexity is one posed already by Levin [Lev86]: how does the average-case complexity of a problem class depend on the distribution? Clearly, if one defines a probability measure to be supported on hard problem instances only, average-case complexity equals worst-case complexity. Intriguingly, there even exists a single so-called 'universal distribution' for which the average-case complexity of any algorithm equals its worst-case complexity [LV92]. This is why average-case complexity under natural measures such as the uniform measure has remained largely elusive.

There are some negative results for the case of random satisfiability problems such as random 3SAT: if the ratio $m/n$ of the number of clauses $m$ to the number of variables $n$ is either quite large or quite small, naïvely guessing the problem as unsatisfiable or satisfiable, respectively, succeeds with high probability. Those regimes in which a good guess can be made efficiently can be significantly extended using the Lovász Local Lemma [Geb+09]. Indeed, it is believed that there exists a rather sharp threshold of $\leq m/n \sim 4.26$ [AB09] between those regimes. Except for instances very close to the threshold average-case hardness of 3SAT is thereby ruled out.

Positive average-case hardness results are only known for counting problems. The key conceptual idea underlying proofs of average-case hardness for such problems is the notion of *random self-reducibility*. We say that a computational problem is randomly self-reducible if we can polynomially reduce the problem of evaluating any fixed instance $x$ to evaluating random instances $y_1, \ldots, y_k$ with a bounded probability that is independent of the input. Random self-reducibility is therefore a particular type of *worst-to-average-case reduction*. We assume there was a machine that solved random instances with bounded probability and then use this machine to efficiently solve an arbitrary fixed instance. This shows that such a machine would allow one to solve any and in particular a worst-case instance in a time that is polynomially equivalent to the time it takes to solve a random such instance.

A first step towards proving approximate average-case hardness of quantum success probabilities is to prove average-case hardness of exactly computing those success probabilities for the respective circuit family. This is indeed possible and has been pioneered by Lipton [Lip91] for the permanent as it prominently features in boson sampling [AA13]. Quite recently, the very same method has also been adapted to a broad class of random-circuit sampling schemes [Bou+19; Mov18; Mov19]. However, the step from exact to multiplicatively robust average-case complexity remains wide open and indeed remains the central open question in the field of quantum supremacy from a complexity-theoretic viewpoint. While we have no new methods to offer on the question of average-case hardness, let us still review the state of the art for proving exact and near-exact average-case hardness[6].

6: My review is based on joint work on average-case hardness in different contexts with Jonas Haferkamp, Jens Eisert, and Marek Gluza [Haf+20] as well as Jonas Haferkamp, Adam Bouland, Bill Fefferman, Jens Eisert and Juani Bermejo-Vega [Haf+19]. I will also use material from Arora and Barak [AB09, Chpt. 19].

## Random self-reducibility of the permanent

Let us start from the simplest and historically original proof of average-case hardness for #P – random self-reducibility of the permanent over a finite field $\mathbb{F}$ for the uniform distribution over that field. Recall the definition of the permanent of an $n \times n$ matrix $X$ over $\mathbb{F}$ (2.64)

$$\operatorname{Perm}(X) = \sum_{\sigma \in S_n} \prod_{j=1}^{n} x_{j,\sigma(j)}. \tag{3.41}$$

The underlying structure in which the proof of random self-reducibility for the permanent is rooted is the algebraic fact that it is a degree-$n$ polynomial in the matrix entries of $X$. Specifically, the idea is the following: given a hard instance $A \in \mathbb{F}^{n\times n}$, draw a uniformly random matrix $B$ and for any $t \in \mathbb{F}$ define the matrix

$$E(t) := A + tB, \tag{3.42}$$

for $t \in \mathbb{F}$. Notice that for any fixed value of $t \neq 0$, $E(t)$ is distributed uniformly over $\mathbb{F}$. This is in spite of the fact that, of course, $E(t)$ and $E(t')$ are correlated for values $t, t' \in \mathbb{F}$. As the permanent is a degree-$n$ polynomial in the matrix entries of an $n \times n$ matrix, the permanent of the matrix $E(t)$ is a degree-$n$ polynomial $q(t) := \operatorname{Perm}(E(t))$ in $t$.

Let us now assume that their exists an efficient machine $\mathbb{O}$ that computes $\operatorname{Perm}(X)$ for uniformly random instances $X$ with failure proability $\delta$. Such an algorithm – while it may fail to evaluate $q(0) \equiv \operatorname{Perm}(A)$ – will, by assumption, likely correctly evaluate $q(t_i)$ for some choice of evaluation points $t_i$. The idea is to infer $q(0)$ from the values of $q$ at the points $\{t_i\}_i$ using polynomial interpolation; see Fig. 3.2.

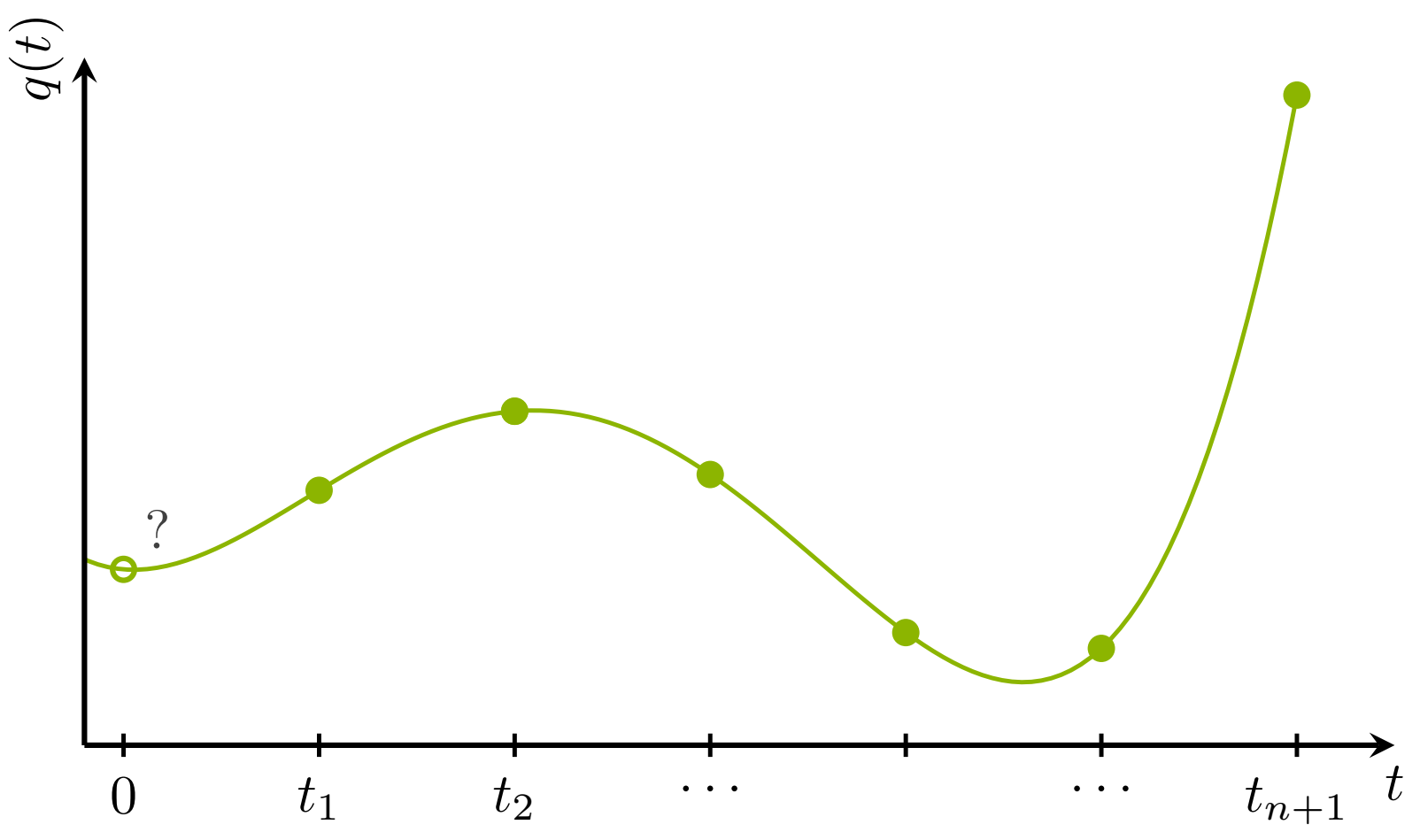

**Figure 3.2:** From at least $n + 1$ interpolation points $(t_i, q(t_i))$ one can efficiently interpolate the polynomial $q(t)$.

We can now query $\mathbb{O}$ on $n + 1$ distinct points[7] $t_1, \ldots, t_{n+1} \neq 0$ obtaining the values $q(t_i)$. Applying a union bound, the probability that all of those values are correct is lower-bounded by $1 - (n + 1)\delta$. We can now set $\delta = 1/3n$ in which case we obtain $n+1$ correct pairs $\{(t_i, q(t_i)), i \in [n+1]\}$ with probability at least $2/3 - 1/3n$. But $q$ is a degree-$n$ polynomial and hence those points uniquely determine $q$. We can now solve a linear system of equations to interpolate the polynomial $q$ and compute $q(0) = \mathrm{Perm}(A)$.

7: Notice that this requires the size of $\mathbb{F}$ to be at least $n + 2$ and hence Lipton's proof does not work for the field $\mathbb{F}_2$, for instance. Indeed, there are also known approximation schemes for the permanent [JSV04].

**Improving the success probability** So far, we have been able to prove #P hardness of evaluating any $1 - 1/3n$ fraction of the problem instances of the permanent. This is a rather strong requirement on the evaluation algorithm and, conversely, requires that only a $> 1/3n$-fraction to be indeed hard to compute. Naturally, it is desirable to lower this requirement as far as possible to assess average-case hardness as well as possible.

And indeed, we can bring down the requirement on $\mathbb{O}$ to work correctly only for a constant fraction $1/2 + \epsilon$ for any constant $\epsilon > 0$ of the instances [AB09, Sec. 8.7]. The idea is to use error-correction techniques for polynomial codes such as the famous Reed-Solomon code [RS60], where a string of $n$ symbols is identified with the coefficients of a degree-$(n-1)$ polynomial. Decoding algorithms for such codes output the correct polynomial even in the presence of some amount of errors.

An error-correction algorithm for Reed-Solomon codes that will be extremely useful for our purposes is the algorithm by Welch and Berlekamp [WB86] as it works over *arbitrary fields* and can even be extended to rational-function interpolation [Mov18; Mov19].

**Theorem 3.8** (Unique decoding for Reed-Solomon [WB86]) *Let $q$ be a degree-$r$ polynomial over any field $\mathbb{F}$. Suppose we are given $k$ pairs of elements $\{(t_i, y_i\})_{i\in[k]}$ with all $t_i$ distinct with the promise that $y_i = q(t_i)$ for at least* $\max(r + 1, (k + r)/2$ *points. Then one can uniquely recover $q$ exactly in* $\mathsf{poly}(k, r)$ *deterministic time.*

We illustrate the Berlekamp-Welch algorithm in Fig. 3.3. Notice that for polynomially large $k$ the Berlekamp-Welch decoding algorithm tolerates an error rate that is arbitrarily close to a half. The Berlekamp-Welch algorithm is thus optimal in that, clearly, as soon as less than half of the points are correct, no unique solution is guaranteed to exist[8].

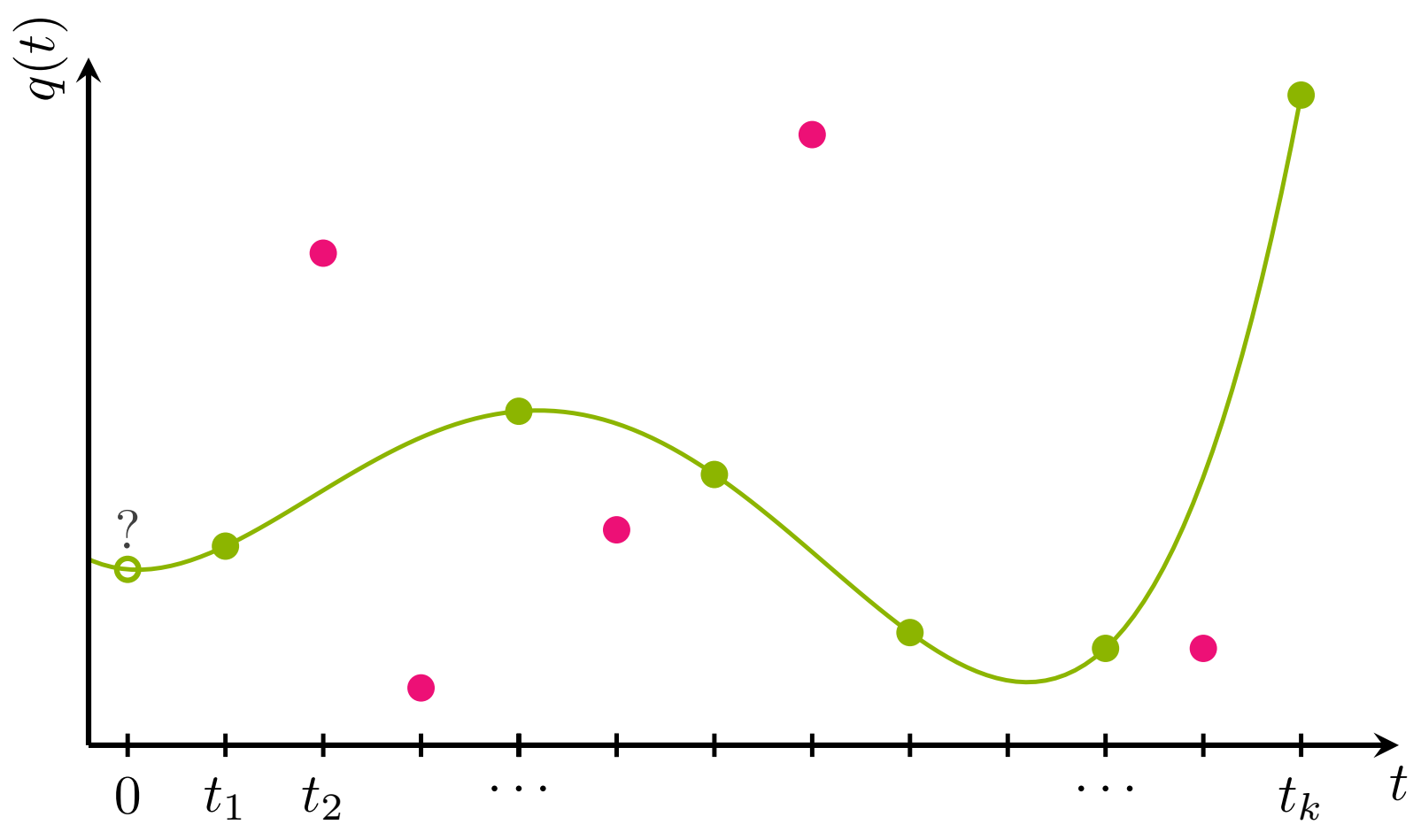

**Figure 3.3:** Using the Berlekamp-Welch decoding algorithm [WB86] for the Solomon-Reed code one can reconstruct a degree-$r$ polynomial from $k$ points $(t_i, y_i)$ if at least $(k+r)/2$ of those points are correct. This allows a worst-to-average case reduction for any $3/4 + o(1)$ fraction of the problem instances.

Using the Berlekamp-Welch algorithm, we can query $\mathcal{O}$ $k > 2(n+1)$ many times at distinct points $t_i$, obtaining pairs $(t_i, \mathcal{O}(t_i))$. We can then upper-bound the probability that less than $(k+n)/2$ of the obtained data points are correct

$$\Pr\left[|\{i, \mathcal{O}(t_i) \neq q(t_i)\}| > k - \frac{k+n}{2}\right] < \frac{2\delta k}{k-n}, \tag{3.43}$$

using Markov's inequality. This probability is at most $1/2$ if the failure probability of $\mathcal{O}$ satisfies

$$\delta < \frac{1}{4}\left(1 - \frac{k}{n}\right). \tag{3.44}$$

Hence the decoding procedure succeeds using $k$ samples as long as $\mathcal{O}$ works on a $3/4 + k/4n = 3/4 + 1/\mathsf{poly}(n)$ fraction of the instances.

In the case of boson sampling – and looking ahead also in the case of unitary quantum gates – we are concerned with matrices not over finite fields, but infinite ones, specifically the complex numbers $\mathbb{F} = \mathbb{C}$. In this case, we are faced with two additional technical difficulties[9]: first, there is no uniform or translation-invariant measure over the complex numbers. This means that when we construct the random matrix $E(t)$ as in Eq. (3.42) by drawing a random matrix $B$ from some distribution $\mu$, then $E(t)$ will be distributed according to another distribution $\mu'$ depending on the value of $t$ and the hard instance $A$. Assuming that we have found a solution to this problem, second, the polynomial interpolation and error-correction techniques that we have used above for the case of finite fields fail in case we only have a finite approximation of the values of $q(t_i)$. Numerically dealing with real numbers will, however, inevitably lead to finite-precision errors on the order $2^{-\mathsf{poly}(n)}$.

8: This issue is addressed by so-called *list-decoding algorithms* which output a list of compatible solutions, observing that there cannot be too many such solutions [AB09, Sec. 19.5]. Such algorithms have been developed [BF90; Lip91] for so called Reed-Muller codes [Ree54; Mul54] over finite fields of which Reed-Solomon is a special case [Sud97].

9: The technical strategies to deal with those arguments which I will sketch below are mostly due to Aaronson and Arkhipov [AA13].

**Distributions over infinite fields: the case of $\mathbb{F} = \mathbb{C}$** We can circumvent the first problem by choosing values of $t$ that are small such that the difference between $\mu'$ and $\mu$ in total-variation distance is small. As the total-variation distance upper-bounds the difference in probability that the two distributions assign to a specific event this difference translates to an additional contribution to the failure probability of $\mathfrak{O}$.

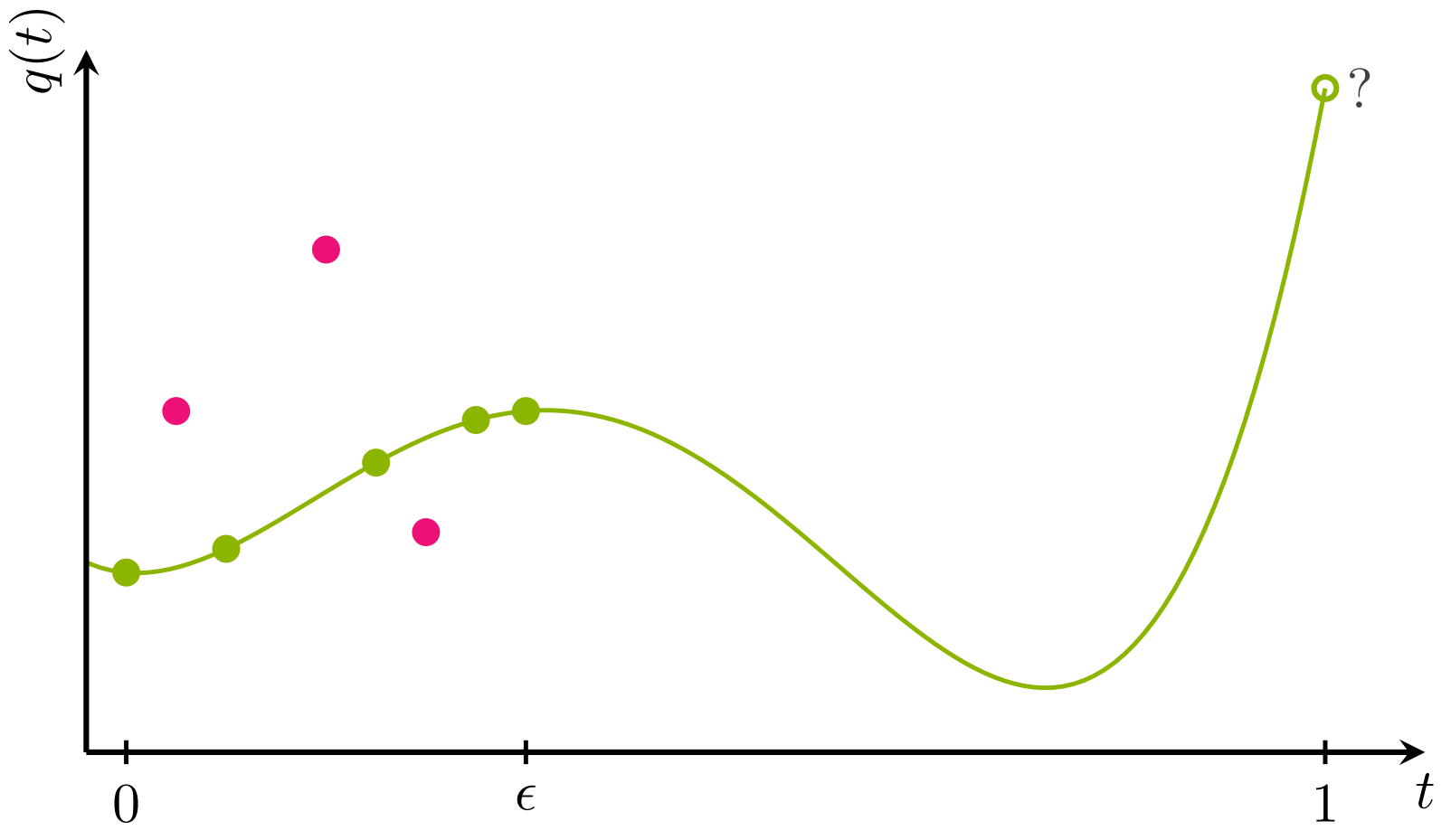

**Figure 3.4:** When drawing instances from a distribution on the infinite field $\mathbb{C}$ as opposed to the uniform measure over a finite field, the interpolation points are chosen from the interval $[0, \epsilon]$ for $\epsilon = 1/\mathsf{poly}(n)$ so that the distribution of $G(t)$ in Eq. (3.49) does not deviate too far from the original distribution.

The natural distribution over $\mathcal{C}$ that is also relevant to the case of boson sampling is the complex normal distribution $\mathcal{N}_{\mathbb{C}}(\mu, \sigma)$ with mean $\mu$ and variance $\sigma^2$. The following lemma, a variation of Ref. [AA13, Lemma 7.4], bounds the total-variation distance between slightly shifted and squashed products Gaussian distributions with products of the standard distribution.

**Lemma 3.9** (Autocorrelation of Gaussian distributions) *For the distributions*

$$\mathcal{D}_1 = \mathcal{N}_{\mathbb{C}}(0, (1-\epsilon^2)\sigma)^M, \tag{3.45}$$

$$\mathcal{D}_2 = \prod_{i=1}^{M} \mathcal{N}_{\mathbb{C}}(v_i, \sigma), \tag{3.46}$$

*with $v \in \mathbb{C}^M$ it holds that*

$$\|\mathcal{D}_1 - \mathcal{N}_{\mathbb{C}}(0,\sigma)^M\|_{TV} \leq 2M\epsilon \tag{3.47}$$

$$\|\mathcal{D}_2 - \mathcal{N}_{\mathbb{C}}(0,\sigma)^M\|_{TV} \leq \frac{1}{\sigma}\|v\|_{\ell_1}. \tag{3.48}$$

*The same result holds for the uniform distribution $\mathcal{U}_{\mathbb{C}}(\mu, \sigma)$ centered around $\mu$ with cutoff $\sigma$.*

For an arbitrary 0/1-matrix $A$ we now define the matrix

$$G(t) = tA + (1-t)B \tag{3.49}$$

similarly as above by drawing standard normal distributed instances $B \in \mathbb{C}^{n\times n}$. The matrix $E(t)$ is then distributed according to the new distribution

$$\mathcal{D} = \prod_{i,j=1}^{n} \mathcal{N}_{\mathbb{C}}(ta_{ij}, (1-t)^2), \tag{3.50}$$

where we have expressed in terms of its matrix entries as $A = (a_{ij})_{i,j}$. Choosing equidistant values of $t_i$ in the interval $(0, \epsilon]$ for some cutoff

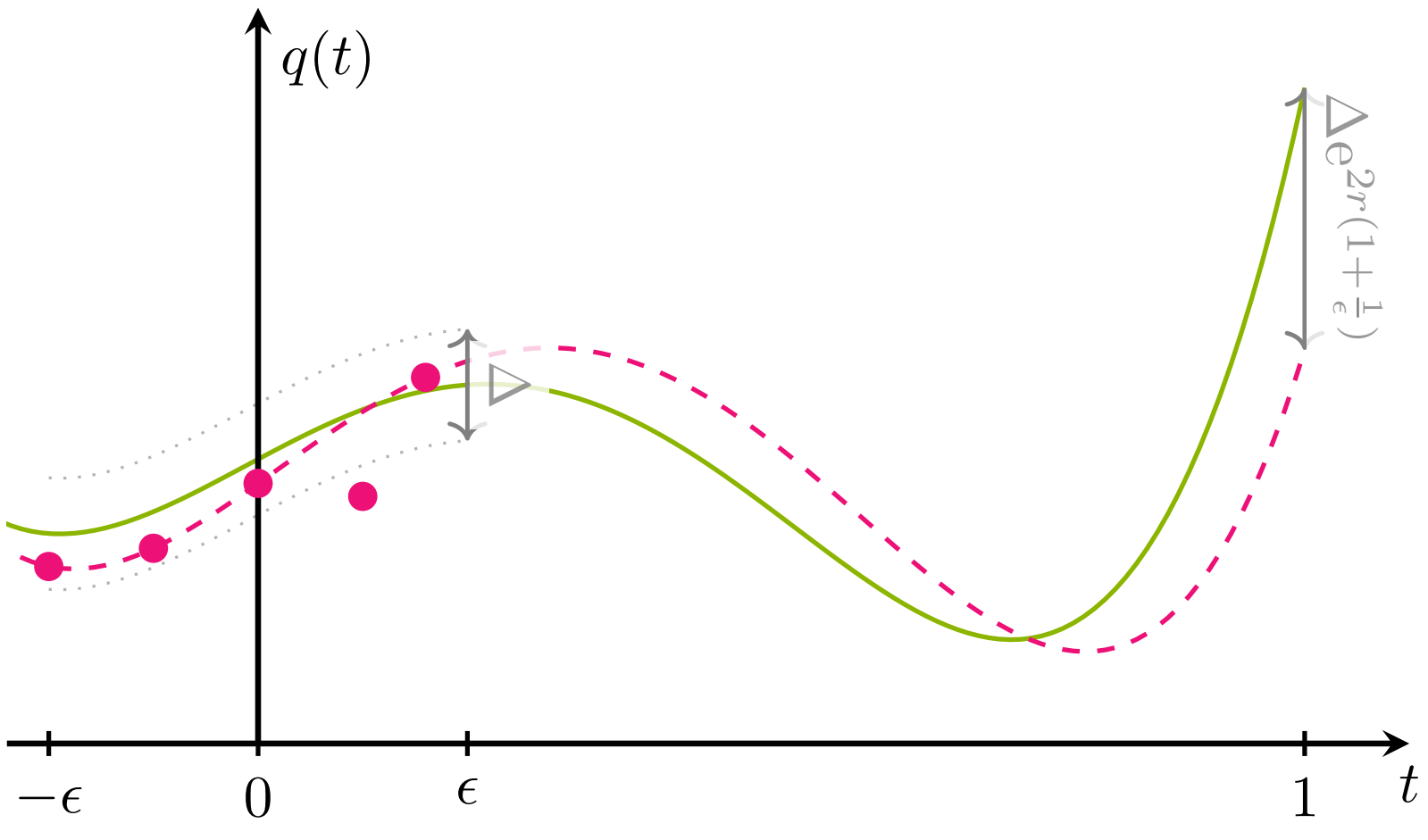

**Figure 3.5:** Using the results by Rakhmanov [Rak07] (Theorem 3.11) and Paturi [Pat92] (Lemma 3.10) one can bound the interpolation error of a degree-$r$ polynomial between approximate interpolation points that are correct up to an error $\Delta$ in the interval $(-\epsilon, \epsilon)$ as well as the extrapolation error when extrapolating to the hard problem instance at $t = 1$.

$\epsilon > 0$ will then result in a success probability of

$$\Pr\left[O(t_i) = q(t_i)\right] \geq 1 - \delta - \|\mathcal{D} - \mathcal{G}_{\mathbb{C}}(0,1)^{n^2}\|_{TV} \tag{3.51}$$
$$\geq 1 - \delta - 6n^2\epsilon. \tag{3.52}$$

The remainder of the argument follows analogously as above by choosing $\epsilon = \delta/(6n^2)$. We illustrate the procedure in Fig. 3.4

**Robustness to finite-precision errors** The finite-precision problem requires somewhat more powerful machinery: using bounds on the stable extrapolation and interpolation of polynomials, we can recover the original proof using polynomial interpolation. This comes at the cost, however, that we cannot make use of the powerful error-correction techniques of Berlekamp and Welch anymore.

The two results that have been identified [AA13, Sec. 9.1] as crucial to this effort are a Lemma by Paturi [Pat92] and a theorem by Rakhmanov [Rak07].

**Lemma 3.10** (Stable extrapolation [Pat92]) *Let $p : \mathbb{R} \to \mathbb{R}$ a polynomial of degree $r$ and suppose that $|p(x)| \leq \Delta$ for all $x$ such that $|x| \leq \epsilon$. Then $|p(1)| \leq \Delta e^{2r(1+1/\epsilon)}$.*

**Theorem 3.11** (Stable interpolation [Rak07]) *Let $E_k$ denote the set of $k$ equidistant points in $(-1, 1)$. Then for a polynomial $p : \mathbb{R} \to \mathbb{R}$ of degree $r$ such that $|p(x)| \leq 1$ for all $x \in E_k$, it holds that*

$$|p(x)| \leq C \log\left(\frac{\pi}{\arctan\left(\frac{k}{r}\sqrt{\mathcal{R}^2 - x^2}\right)}\right), \tag{3.53}$$

*for $|x| \leq \mathcal{R} := \sqrt{1 - r^2/k^2}$.*

We can now apply those results to the polynomial $p(t) = q(t) - q'(t)$, where $q'(t)$ is the polynomial defined by the slightly erroneous values $q'(t_i)$ of $q(t_i)$ satisfying $|q'(t_i) - q(t_i)| \leq 2^{-\mathsf{poly}(n)}$ for a sufficiently large polynomial. Using Rakhmanov's result [Rak07] we can bound the error between $q$ and $q'$ between the evaluation points; using Paturi's lemma [Pat92], we can then bound the error tolerance when extrapolating to $q(1)$; see Fig. 3.5.

### Average-case hardness of quantum success probabilities

Let us now turn to applying average-case hardness arguments to success probabilities of quantum circuits. Firstly, let us observe, that there is a natural polynomial structure on the success probabilities of quantum circuits. For a circuit $C = C_m \cdots C_2 C_1$ comprising $m$ gates $C_i$ acting on $n$ qubits the output amplitudes can be expressed in terms of a path integral

$$\langle 0|C|0\rangle = \sum_{\lambda_1,\ldots\lambda_{m-1}\in\{0,1\}^n} \langle 0|C_m|\lambda_{m-1}\rangle \cdots \langle \lambda_2|C_2|\lambda_1\rangle\langle\lambda_1|C_1|0\rangle \quad (3.54)$$

Consider that $C$ is drawn from some measure $\mu_{\mathscr{C}}$ that defines a circuit family $\mathscr{C}$ [Bou+19]. Some of the gates in $C$ might be randomly drawn from a gate set $\mathscr{G}$, others might be fixed across all $C \in \mathscr{C}$.

Now we are faced with a severe issue, however: when trying to construct an equivalent of $E(t)$ by choosing random instances $B$ for a fixed worst-case circuit $A$, the matrix given by $A + tB$ will not be unitary for $t \neq 0$ and therefore does not define a valid problem instance. Of course, this is because the unitary matrices do not form a group with respect to addition, but multiplication. How then can we perform a worst-to-average case reduction? We will have to make use of this structure by multiplying $A$ and $B$ in a gate-wise fashion. There are two distinct ways of doing this.

10: This approach is due to Bouland et al. [Bou+19]. We refer the interested reader to Refs. [Haf+19, App. D] and [Bou+19] for a more detailed exposition of the arguments sketched below.

**Taylor-series truncation approach** [10] The first approach is rooted in the idea to interpolate between a hard and a random instance as follows. For a hard instance of a circuit $C$ with random gates $C_1, \ldots, C_m$ drawn uniformly from a continuous subgroup $\mathscr{G}$ of the corresponding unitary group $U(d)$ we define a new circuit by setting each gate

$$C_i(t) := C_i H_i \mathrm{e}^{-\mathrm{i} t h_i}, \quad (3.55)$$

where $H_i$ is Haar random in $\mathscr{G}$ and $h_i = -\mathrm{i}\log H_i$ is its generator. Denote the resulting circuit as $C \times H(t)$ $C_i(0)$ is Haar-random in $\mathscr{G}$, while for $t = 1$ we recover the original gate $C_i$. Similarly to average-case hardness of Gauss-random permanents, for tiny $t$ the gate $H_i \mathrm{e}^{-\mathrm{i} t h_i}$ looks almost Haar-random. One can therefore hope to follow the same procedure as above to extrapolate to $t = 1$, given values of $|\langle 0|C \times H_t|0\rangle|^2$.

Unfortunately, the gates $C_i(t)$ and hence the output probability $|\langle 0|C \times H_t|0\rangle|^2$ is no longer a (low-degree) polynomial in $t$ so that the polynomial interpolation step does not work. An easy way to circumvent this problem is to consider Taylor-approximations of the deformed gates $C_i(t)$. While these will be slightly non-unitary, the resulting output probabilities are guaranteed to remain close to the ideal ones. Let us thus define the $(t, K)$-truncated and perturbed Haar measure on the circuit family $\mathscr{C}$ by replacing each Haar-random gate $H_i$ in a circuit $C$ by

$$G_i = H_i \left( \sum_{k=0}^{K} \frac{(-\mathrm{i} h_i t)^k}{k!} \right). \quad (3.56)$$

We can now use the standard (Suzuki) bound on Taylor truncations

$$|\langle\psi|C_i G_i - C_i H_i e^{-ith_i}|\psi\rangle| \leq \frac{\kappa}{K!}, \tag{3.57}$$

for a constant $\kappa > 0$, set $K = \mathsf{poly}(n)$, use an analogue of Lemma 3.9 and apply the stability results by Rakhmanov [Rak07] and Paturi [Pat92] to complete a worst-to-average case reduction that is robust to additive errors $2^{-\mathsf{poly}(n)}$.

A caveat of this proof is that in the reduction we have left the unitary group. This means that average-case hardness is not achieved for exactly evaluating the circuit success probabilities , but strictly speaking only for exactly evaluating numbers $p_0(C)'$ which are $2^{-\mathsf{poly}(n)}$ additive approximations thereof and which do not correspond to success probabilities of valid quantum circuits. Nevertheless, average-case hardness of those numbers is a necessary requirement for the approximate average-case hardness property (iii.) of the robust hardness proof in Theorem 2.9.

**Rational function interpolation approach** [11] An arguably more natural way to extrapolate in the unitary group is to directly extrapolate within the unitary group. A neat way to do this makes use of the *Cayley function*

11: This approach is due to and explained in detail by Movassagh [Mov18; Mov19].

$$f(x) = \frac{1 + ix}{1 - ix}, \tag{3.58}$$

for $x \in \mathbb{R}$, defining $f(-\infty) = -1$. This means that a Haar-random unitary matrix $H \in U(d)$ can be uniquely represented as

$$H = f(h), \quad h = h^\dagger, \tag{3.59}$$

and $H^\dagger = f(-h)$. For an arbitrary fixed quantum gate $C \in U(d)$ we can then construct the path

$$C(t) = CHf(-th), \tag{3.60}$$

which can be expressed as a fraction of degree-$d$ polynomials using the spectral decomposition of $h = \sum_{\alpha=1}^{d} h_\alpha |\psi_\alpha\rangle\langle\psi_\alpha|$ as

$$C(t) = \frac{1}{q(t)} \sum_{\alpha=1}^{d} p_\alpha(t) C|\psi_\alpha\rangle\langle\psi_\alpha|, \tag{3.61}$$

with

$$q(t) = \prod_{\alpha=1}^{d} (1 + ith_\alpha), \tag{3.62}$$

$$p_\alpha(t) = f(h_\alpha)(1 - th_\alpha) \prod_{\beta\in[d]\setminus\alpha} (1 + ith_\beta). \tag{3.63}$$

The problem this approach faces is that the techniques we have used so far worked only for polynomial interpolation. For the worst-to-average case we will now need to extrapolate a rational function. Fortunately, one can generalize the Berlekamp-Welch algorithm to rational functions with degrees $k_1, k_2$ in the numerator and denominator, respectively [Mov18]. This algorithm requires that the number of evaluation points $t_i$ is at least $k_1 + k_2 + 2e$, where $e$ is the number of errors made by the evaluation algorithm $\mathcal{O}$.

While the results on stable interpolation [Rak07] and extrapolation [Pat92] of low-degree polynomials do not apply here, one can directly compute the robustness of this approach as $\exp(-\Omega(m\epsilon^{-1}))$ [Mov19, Theorem 3], where $m$ is the number of gates in the circuit and $(0, \epsilon]$ defines the interval on which the success probabilities of $C(t)$ are evaluated. Putting everything together, for a two-dimensional circuit with depth $\sqrt{n}$ in the number of qubits as has been proposed [Boi+18; Aru+19] one obtains robustness of $\exp(-\Omega(n^3 \log n))$.

## 3.4 Discussion and conclusion

In this chapter, we have seen how to (partially) close loopholes in the Stockmeyer hardness proof. First, we have provided a simple proof of the anticoncentration property for circuit families that generate 2-designs[12]. Second, we have reviewed techniques for worst-to-average case reductions in the framework of random self-reducibility and their applications to average-case complexity of computing success probabilities of random quantum circuits.

12: Let us note here that subsequent to the preprint publication of our result, it was independently observed elsewhere [BFK18; MB17; HM18].

The anticoncentration result implies that local random quantum circuits in a one-dimensional parallel nearest-neighbour architecture anticoncentrate in linear depth. This is due to the fact that random quantum circuits form unitary $k$-designs in a depth that scales polynomially in $k$ and linearly in the number of qubits [BHH16].

*The following discussion of the depth scaling has previously been published in Ref. [Han+18].*

We conjecture that this linear scaling is optimal in a one-dimensional architecture. Indeed, on the one hand, the result agrees with the intuition that anticoncentration arises as soon as correlations have spread across the entire system, a process that occurs ballistically and thus scales with the diameter of the system. On the other hand, for one-dimensional random universal circuits to be intractable classically, the depth needs to be polynomial in the number of qubits. Hence, our result only leaves room for a sublinear improvement, since for circuits of polylogarithmic depth there is a quasi-polynomial time classical simulation based on matrix-product states. However, as is argued in Refs. [BMS17; LBR17], it would seem counter-intuitive that one can achieve sublinear depth. Indeed, standard tensor network contraction techniques would allow any output probabilities of a circuit of depth $t$ in one dimension to be computed in a time scaling as $O(2^t)$ [Joz06]. Hence, if the depth $t$ as a function of $n$ required for the classical hardness of generic circuits could be brought down to sublinear, this would violate the counting exponential time hypothesis (ETH) [IP99] and is therefore considered highly unlikely.

We can apply the same reasoning to argue that for a nearest-neighbour circuit in two dimensions, the circuit depth needs to scale in the diameter of the system, that is, as $\sqrt{n}$ for anticoncentration to arise. This intuition has also been put on a rigorous footing by Harrow and Mehraban [HM18], who proved that nearest-neighbour universal circuits in two dimensions form a unitary 2-design in depth $O(\sqrt{n})$. By Theorem 3.4, this implies anticoncentration of the output probabilities.

Let us also note that the ETH has been applied to derive more fine-grained results for various quantum-supremacy schemes including boson sampling, universal circuit sampling and IQP circuit sampling [Dal+18] as well as the DQC1 model and Clifford+$T$ circuits [MT19]. Those results are based on an ETH lower-bound of the type $e^{an}$ with $a > 0$ for computing a size-$n$ instance of the natural quantity in which the success probabilities can be expressed (e.g., the permanent for boson sampling; see Sec. 2.5). This yields a lower bound on the time required to weakly simulate the respective circuits. The argument is based on the idea that an algorithm that ran more efficiently than the ETH lower bound would constitute a nondeterministic algorithm that could decide whether the acceptance probability is zero or not. But this is already a hard problem as we have seen above in the guise of $\mathrm{ngap}(f)$ for the case of IQP circuits.

A key issue to note in the worst-to-average case reductions on the unitary group is that the random gates in the circuit families need to be drawn from *continuous subgroups* of the unitary group. Only if this is the case can one choose values of the interpolation parameter $t$ that are small enough such that the measure on the gate set is not perturbed too much in the interpolation step. In particular, this implies that the reduction does not apply to discrete gate sets and for some architectures the choice of random gates must be modified for the reduction to apply. For instance, in the case of the IQP family defined in Eq. (2.53), we need to choose the edge weights $w_{i,j}$ uniformly from the unit circle $S^1$ rather than from a discrete set of angles.

From the three examples discussed in Sec. 3.2, an average-case hardness reduction is therefore only possibly for the prime example of universal random circuits where the gates are drawn from the Haar measure on $U(4)$ and diagonal unitaries, where the angle $\phi$ is chosen from the unit circle. It does not trivially extend to the case of Clifford circuits, which is an intrinsically discrete subgroup of the unitary group.

Let us now address the elephant in the room: can we hope to prove approximate average-case hardness for quantum success probabilities? The key technical obstacle on the way to addressing this question is the instability of polynomials with respect to variations in the interpolation points. Indeed, we saw in Paturi's lemma (Lemma 3.10) that the extrapolation error of a bounded error polynomial scales exponentially in the degree $r$ and size of the interval $\epsilon$ on which the bound holds. As we have to make this interval inverse polynomially small to maintain closeness of the probability distributions, this results in a $2^{\mathsf{poly}(n)}$ increase of the Paturi bound, which can only be counter-weighted by an error bound of $2^{-\mathsf{poly}(n)}$ on the interval $(-\epsilon, \epsilon)$. Small variations of a polynomial at a few points can thus lead to very large variations far away from those points.

Random self-reducibility thus seems doomed when it comes to additive robustness of success probabilities on the order $2^{-n}$ as would be necessary for the quantum supremacy conjecture; see note 32 in Chapter 2. Indeed, for a somewhat baroque universal circuit architecture in two dimensions that is $\mathsf{GapP}$-hard to simulate strongly, it has recently been proven by Napp et al. [Nap+19] that no approximate worst-to-average case reduction is possible. Rather this architecture admits algorithms for both strong and weak simulation that are efficient on large fractions of the instances.

But for the same architecture, strong simulation is classically intractable unless GapP admits a polynomial time algorithm and the polynomial hierarchy collapses to its third level, respectively. More generally, Napp et al. [Nap+19] provide both numerical and analytical evidence that shallow universal circuits in a two-dimensional brickwork architecture can be efficiently simulated weakly. They do so in a twofold approach: first, they numerically demonstrate approximate simulation of random universal circuits in a $400 \times 400$ brickwork architecture using a tensor-network algorithm (which is worst-case hard to simulate strongly). They then provide analytical evidence for easiness using a mapping to a recently developed model consisting of alternating rounds of random unitaries and weak measurements [BCA19; Jia+19].

For an approximate worst-to-average case reduction hardness we would require, it seems, quantum success probabilities that are extremely robust to noise in generic instances. While techniques such as quantum error correction [RHG06] might at first sight seem ideally suited for this task, in such approaches errors need to be actively corrected. While in the framework of quantum sampling active correction can be bypassed using postselection [Fuj16; KD19], this means that only those probabilities corresponding to specific measurement outcomes on subsystems will be protected against errors. But by the hiding property every specific outcome probability is in one-to-one correspondence with the success probabilities of a circuit from the family. So postselected fault-tolerance is again in conflict with average-case hardness.

# The computational power of quantum simulations 4

In the previous chapters we have focused on the question how we can provide a rigorous footing to the promise that quantum computations exponentially outperform classical computations at all. To do so, we have invoked ideas from computational complexity theory and linked those to random circuit sampling architectures which can be efficiently implemented on quantum devices.

In a quest to experimentally disprove the complexity-theoretic Church-Turing thesis, not only is it necessary that quantum computations *can* outperform classical ones in principle, we also need to *be able to implement* such computations in a physical device or a laboratory. Even a universal quantum computer might come with specific desiderata such as the connectivity of the graph it can execute gates on, or the gates that are natural to it. Given that in the near term no large-scale quantum computer exists, for a prospective demonstration of quantum supremacy it is of paramount importance to tailor a sampling scheme of the type we are discussing to the specific architecture at hand. This has been done in the recent first demonstration of quantum supremacy by the Google AI Quantum team [Aru+19]. On their superconducting quantum processor, the natural two-qubit gate is the product of fractions of the `iSWAP` and the $CZ$ gate[1] .

1: The `iSWAP` gate swaps to qubits, applying a i phase to the first qubit when it is in the $|1\rangle$ state and a i phase to the second qubit when it is in the $|0\rangle$state. This gate is obtained from an evolution under an $XX$ interaction $\exp(\mathrm{i}\frac{\pi}{4}(X \otimes X + Y \otimes Y))$ and its matrix representation is given by

$$\mathtt{iSWAP} = \begin{pmatrix} 1 & 0 & 0 & 0 \\ 0 & 0 & \mathrm{i} & 0 \\ 0 & \mathrm{i} & 0 & 0 \\ 0 & 0 & 0 & 1 \end{pmatrix}. \quad (4.1)$$

Likewise, this is a prime motivation of boson sampling, which is naturally realized in linear-optical networks as they are routinely used in many laboratories around the world. This fact allowed the immediate demonstration of small-scale boson sampling experiments subsequent to the proposal [Spr+13; Til+13; Bro+13; Cre+13] which were significantly improved to achieve larger photon numbers of 6–7 photons [Car+14; Spa+14] and recently culminated in an experiment using 20 photons in 60 linear-optical modes [Wan+19]. But even such a large-scale experiment is prone to classical attacks: not only did the proposal of boson sampling promote experimental activity in linear optics, but it also prompted the development of better classical sampling schemes for exact [CC18; Nev+17] using which one can exactly simulate experiments with up to 50 photons [Nev+17] and algorithms that exploit imperfections in an experimental realization such as photon loss [OB18; GRS19] or partially distinguishable photons [RSG18; Ren+18; Moy+19].

To convincingly demonstrate a violation of the complexity-theoretic Church-Turing thesis, a reasonably large – once again architecture-specific – number of qubits or local degrees of freedom is required. In the case of universal circuit sampling this number is rather low at around 50 qubits [Boi+18; Aru+19; Ped+19], while in other cases such as IQP circuits it might be much larger [Dal+18]. Both the question of scalability of and the parsimonious use of the quantum resources offered by an experimental platform are therefore crucial in this quest. These questions can be put as follows: What is the best achievable compromise in a random circuit sampling scheme between the cost of the required quantum operations in a platform, the number of available qubits, and its computational complexity in terms of the minimal runtime required by a classical simulation algorithm?

Building on the existing strategies to prove quantum supremacy in restricted settings, our aim in this chapter will be to answer this question for quantum simulators. We do so in a twofold approach: On the one hand, we will assess the possibility of exploiting large-scale quantum simulation architectures in order to demonstrate quantum supremacy. On the other hand, we will ask the question whether the practical quantum advantage achieved in such quantum simulation architectures can be given a rigorous footing in terms of a complexity-theoretic argument. Both of these goals follow the overarching theme of this thesis, namely, to develop a rigorous computational perspective on processes occurring in nature and our quantum-mechanical description thereof. Such naturally occurring processes can be experimentally probed and manipulated in dynamical quantum simulations.

To achieve these goals, we will use the framework of quantum random sampling developed in Chapters 2 and 3. Invoking the complexity-theoretic argument for hardness of sampling, we will on the one hand substantiate the numerical evidence for the computational speedup offered in quantum simulators and on the other hand propose sampling schemes that can readily be realized in the available large-scale quantum simulation hardware. Specifically, we will design a quantum random sampling scheme that is tailored to specific experimental desiderata of a dynamical quantum simulation in an optical-lattice architecture (Section 4.2. Coming from a different vantage point, we will consider digital ion-trap quantum simulators. We observe that IQP circuits can be naturally realized in this architecture and develop an experimental protocol that uses the naturally available quantum gates in a highly efficient way (Section 4.3). We will show how to exploit a tradeoff offered by the hardness-of-sampling argument in order to further lower the experimental requirements (Section 4.7).

Before we get there, let us remind ourselves what the basic idea of quantum simulation is and how this idea relates to the rigorous hardness-of-sampling results of the previous chapters.

## 4.1 Simulating quantum systems using other quantum systems

Understood rather broadly, while not universal quantum computers quantum simulators are restricted quantum devices that exhibit some quantum behaviour. More concretely, the idea of quantum simulation, going back to Feynman [Fey82], is to use a (source) quantum system that can be manipulated and probed with a high degree of precision in order to *simulate* another (target) physical system[2]. Simulation can be understood in the following two senses, and it is the latter that will be of interest for our purposes: first, the simulator can be used to plainly *mimick* or reproduce the behaviour of the target system with the purpose to better understand, observe in more detail or manipulate in different regimes the physics of the target system. We might call this type of simulation, an

2: The following take on quantum simulation is joint work with Jacques Carolan and Karim Thébault [HCT17]. I am grateful for the fun and inspiring collaboration.

*emulation* of the target system. Second, a simulator can be used to realize a specific (idealized) model of the target system with the purpose to infer properties of that model that might be inaccessible by other means such as classical computation. We might call this type of simulation, a *computation* of properties of the target system. This distinction can also be understood in terms of the nature of the target system: while the target system of an emulation is a concrete and real physical system, a computation targets an abstract model.

Feynman observed that quantum simulators might be useful, efficient tools to compute the behaviour of large-scale quantum systems based on the observation that the target and source systems are in some sense of the same type. In contrast, when trying to compute properties of a quantum system using classical devices naïvely, the effort will scale in the exponentially large Hilbert space dimension. Given that out goal is to experimentally violate the complexity-theoretic Church-Turing thesis, it is precisely this aspect of quantum simulators that will be our concern here.

The most important type of presently available quantum simulators [CZ12] are so-called *dynamical analog quantum simulators*. Such simulators are able to simulate the time evolution of a certain class of Hamiltonians. This stands in contrast to digital quantum simulators, where the target time evolution is digitized into a discrete quantum circuit using the Trotter formula [Llo96]. When compared to digital quantum computation the computational foundations of dynamical quantum simulation stand on less firm grounds. While it has been proven that such dynamics can implement universal quantum computation [CGW13], the computational power of realistic quantum simulators is understood mostly in terms of intuition gained from numerically obtained practical advantages. The claim that those simulators are computationally superior to classical devices is grounded mainly in the absence of classical simulation algorithms, rather than complexity-theoretic considerations of the type we are concerned with here.

There are a large variety of analog quantum simulation platforms and models that can be realized using them. The arguably most important such platforms include superconducting qubits [Bar+15; HTK12; Aru+19; Rou+17], trapped ions [BR12; Fri+18; Zha+17], Rydberg atoms in optical tweezers [Ber+17b], cold atoms in optical-lattice potentials [BDZ08; Cho+16], and quantum-optical systems [Per+10]. Both superconducting qubits and ion traps can be used as digital quantum simulators with specific gate sets. While superconducting qubits are assembled in arrays [Bar+15; HTK12; Aru+19], ion traps admit long-range interactions [Fri+18] and even all-to-all connectivity [Ber+17a] between the ions in an individual trap[3]. At the same those platforms also admit the simulation of various Hamiltonian models including the Heisenberg model [Rou+17] and long-range transverse field XY and Ising models [Fri+18]. Large arrays of cold atoms in optical lattices realize the famous Hubbard model [Jak+98], which in its fermionic variant is the simplest model describing interacting fermions that features Coulomb repulsion, a nontrivial band structure, and incorporates the Pauli Principle [Hub63].

3: In order to scale up ion traps to a large number of qubits, presumably, several traps hosting tens of ions will have to be coupled, for example, optically [MK13; DM10].

Let us now put a focus on the optical-lattice architecture, which realizes the largest available analog quantum simulator and thus provides the ideal grounds to realize a large-scale quantum supremacy experiment. As an example of a digital quantum simulator suited to realize quantum sampling schemes, we will consider ion traps in Section 4.3

## Cold atoms in optical lattices

In a cold-atom quantum simulators [Jak+98; Gre+02; CZ12] an artificial lattice potential is created using counter-propagating laser beams: a crystal of light. The resulting intensity pattern acts as a space-dependent lattice potential for certain atoms via the dipole-dipole coupling between the light field and the dipole moment of these atoms. Such optical-lattice potentials can be combined with so-called magneto-optical traps (MOT) with which one can create a low-temperature state of a confined atomic cloud consisting of neutral atoms such as the bosonic $^{87}$Rb or the fermionic $^{40}$K. By adding the optical-lattice potential to the MOT potential one can realize a system in which hundreds to thousands of atoms evolve coherently while interacting among themselves and propagating through the lattice; see Fig. 4.1.

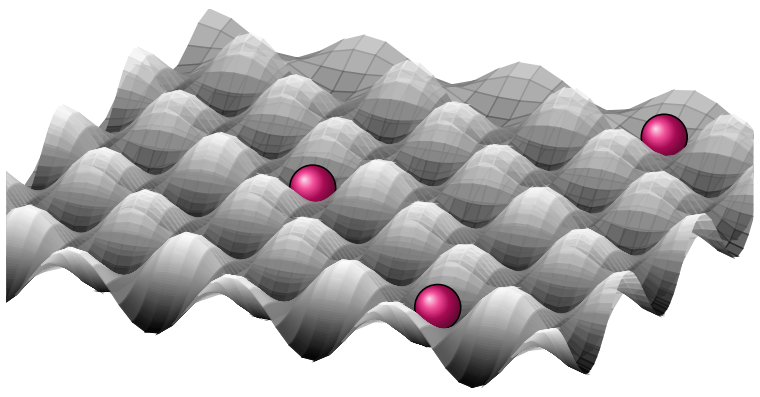

**Figure 4.1:** In a cold-atom quantum simulator ultracold neutral atoms such as $^{87}$Rb or $^{40}$K are trapped in an optical-lattice potential.

Cold-atom systems bear strong similarities to real solid-state systems, where one encounters the same lattice structure for the potential of electrons hopping between atoms. The Hamiltonian describing cold-atoms in optical lattices is the Hubbard Hamiltonian [Jak+98]

$$H_{BH} = -J \sum_{\langle j,k \rangle} \left( b_j^\dagger b_k + b_j^\dagger b_k \right) + U \sum_j b_j^\dagger b_j^\dagger b_j b_j \,. \tag{4.2}$$

Here, $b_j^\dagger$ ($b_j$) denotes a bosonic annihilation (creation) operator at site $j$, $U$ denotes the energy cost from having two atoms on the same site, and $J$ is the energy gain when hopping from one site to the next.

**Manipulating and probing** Cold atoms in optical lattices are particularly well suited for quantum simulation as the system can be manipulated with a high degree of precision, allowing the realization of a a broad parameter regime, while at the same time a large number of atoms is evolving coherently on time scales of several tunneling times. Using a Feshbach resonance [Fes58; Fes62] one can tune the interaction strength $U$ between the atoms, using amplitude and phase of the generating laser beams the hopping strength $J$. By superimposing several lattice potentials with different wavelengths one can even realise next-nearest-neighbour interactions and lattices with higher periodicity [Föl+07; Nas+12]. One can also use such double-well structures [Lee+07; Han+15] or the internal spin of the atoms in spin-dependent lattices [Man+03a; Lee+07; And+07] to process quantum information. This is achieved using controlled collisions between atoms [Jak+99], which realize Ising-type Hamiltonians and generate entanglement in a highly controlled fashion [Man+03b]. Using highly-focussed laser beams to manipulate the system at the single-atom level [Bak+10; She+10]. It is even possible to realise low (one and two) dimensional systems by increasing the potential barriers in the orthogonal directions to suppress tunnelling [Tro+10; Bra+15], and to prepare disordered and lattices by either superimposing incommensurate wavelengths [Sch+15] or using digital mirror devices to map an arbitrary intensity pattern onto the lattice plane [Cho+16].

Since the lattice spacing is several orders of magnitudes larger than in real solids, one can also probe the on the order of the wavelength of the generating laser beams as opposed to the order of picometers as would be the case for in real solid state systems. Using so-called time-of-flight imaging, and variants thereof, one can easily measure the free-space and quasi-momentum distribution of the atoms [Föl+07; BDZ08]. Using highly focused microscopes one can also probe individual atoms in the lattice with single-site and spin resolution [Bak+10; She+10].

**Outperforming classical computers using cold atoms** Using this toolbox [JZ05] one can simulate a large variety of phenomena including quantum phase transitions [Gre+02; Lan+16], magnetism [Str+11; Mur+15], metal-insulator transitions, and high-temperature superconductivity [Köh+05]. Most interesting from a computational perspective, however, is the simulation of the non-equilibrium quantum dynamics of the atoms such as the relaxation of an initial state to equilibrium [Tro+12], the dynamics of phase transitions [Bra+15], or the simulation of many-body localization phenomena [Sch+15; Lüs+16; Cho+16; Bor+17]. In all of those cases, we are faced with the situation that while the simulated phenomena can be computed using classical simulations in some regime, the quantum simulator can access regimes in which all classical methods fail. Most strikingly, this can be seen in situations in which both one- and two-dimensional systems are considered for system sizes of hundreds of atoms [Bra+15; Sch+15; Cho+16]. While in one dimensions classical simulations are often possible using matrix-product state methods [Sch11], those methods utterly fail in two dimensions and what is more, no other simulation method is available in those cases. For all practical purposes, cold atoms in optical lattices outperform classical simulations already today.

## 4.2 Short-time Hamiltonian dynamics in optical lattices

It will be the goal of this section, to find a setting in which the practical computational advantage that can already be achieved today can be grounded in a rigorous complexity-theoretic argument. To do so, we develop a quantum simulation protocol tailored to dynamical quantum simulators in a cold-atom architecture. We first formulate experimental desiderata and then state and analyze a quantum random sampling protocol that meets those desiderata in terms of its classical complexity. We find that standard-basis measurements after the short-time evolution of a translation-invariant Hamiltonian can not only be realized in a cold-atom architecture, but also that such a scheme provides a bone fide quantum speedup that meets the same standards as universal circuit sampling or IQP circuits.

The following section is based on joint work with Juani Bermejo-Vega, Martin Schwarz, Robert Raussendorf and Jens Eisert [Ber+18].

### Experimental desiderata

From the discussion above we can extract the following 'desiderata' of feasibility in a cold-atoms setting. What is particularly well-controlled and feasible in an optical-lattice setup is the coherent time-evolution under the Bose-Hubbard or Ising-type Hamiltonians that are translation-invariant with a low period, even for rather long times. Natural lattice architectures are simple cubic or hexagonal lattices; higher periods can be realized using double-well structures, but this becomes infeasibly challenging for periods beyond 2. While one can incorporate local disorder using the methods described above, the interactions will always be translation-invariant in this setting. This stands in contrast to, say, superconducting-qubit architectures, where one can locally address the interactions between neighbouring sites. But addressability comes with a tradeoff – being able to individually address sites also requires a highly accurate cross-calibration of all experimental knobs [Aru+19] Such issues of calibration currently limit superconducting-qubit architectures in their scalability, an issue that is less severe in cold atoms as it is not designed to be fully addressible in any case.

It is also feasible to prepare a translation-invariant initial product state that might, by the same token, be locally disordered by phases or the like. The most challenging part, and the one that allows least flexibility in optical-lattice experiments is arguably the measurement stage. Time-of-flight and real-space imaging is the standard measurement technique. While time-of-flight allows for a mapping of the quasimomentum *distribution* of the atoms, via real-space imaging one can measure the location of individual atoms. Measuring each atom individually in different bases can be done, for instance, via in-situ addressing using highly focused laser beams. This is experimentally possible but challenging.

When devising an analog simulation architecture that realizes a quantum sampling task, we thus start from the following physical desiderata:

- D1 The scheme should require translation-invariant interaction terms only and only locally distinct components.
- D2 The interaction graph of the scheme should be as simple as possible in terms of its periodicity, ideally a cubic or hexagonal lattice.
- D3 The measurement should be translation-invariant, requiring as little control as possible.

## A translation-invariant quantum sampling protocol

The key idea behind the protocol we propose is to exploit space-time trade-off: using the idea of measurement-based quantum computation [RB01; RBB03b] we can trade the depth of a universal random circuit for the number of qubits. Doing so, we arrive at a protocol that satisfies the desiderata D1–D3 as desired.

The dynamical quantum simulation, which we propose consists of the following three steps[4] [Haf+19], which we illustrate in Fig. 4.2:

4: We define a slightly modified protocol in which we draw angles Haar-randomly from $S^1$ and not discretely from $\{0, \pi/4\}$ as in the original protocol [Ber+18]. This choice allows us to apply the ideas from the previous chapter in Sec. 3.3 to prove exact average-case hardness.

**Protocol 4.1** (Sampling quantum simulators)

E1 **Preparation**: Arrange $N := nm$ qubits on an $n$-row $m$-column lattice $\mathscr{L}$, with vertices $V$ and edges $E$. Prepare the product state vector

$$|\psi_\beta\rangle = \bigotimes_{i=1}^{N} \frac{1}{\sqrt{2}} \left(|0\rangle + e^{\mathrm{i}\beta_i}|1\rangle\right), \ \beta \in [0, 2\pi)^N, \qquad (4.3)$$

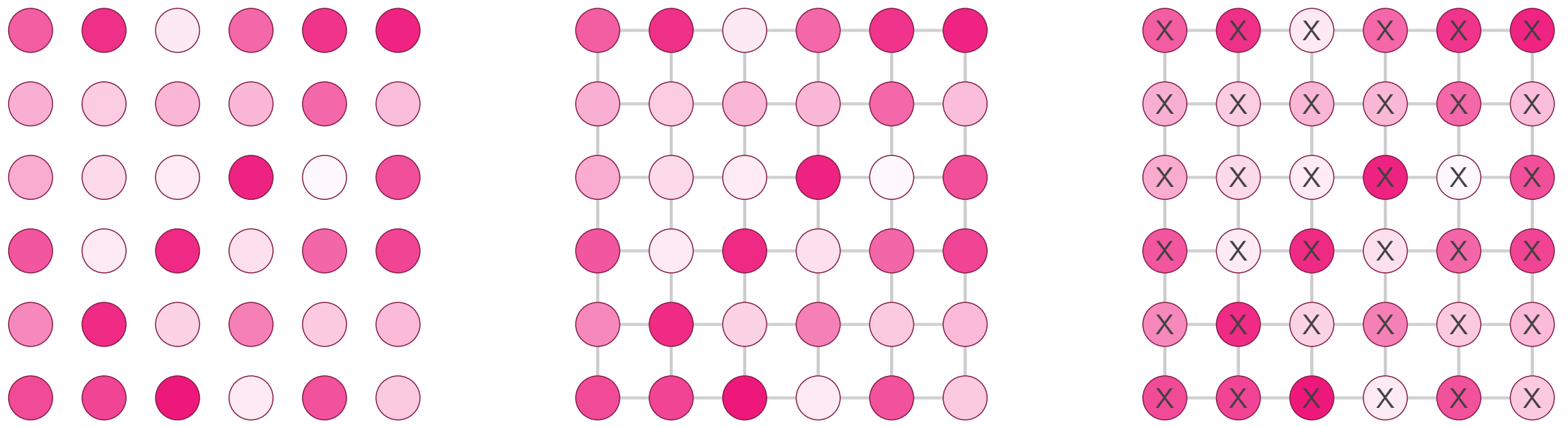

**Figure 4.2:** We illustrate the steps E1–E3 on a $6 \times 6$ square lattice. Circles represent qubits and different shades of pink indicate the continuous choices of phases $\beta_{ij} \in [0, 2\pi)$ in the initial state preparation. Connecting lines denote the application of a $CZ$-gate.

for $\beta$ chosen randomly from the Haar measure[5] on $(S^1)^{\times N}$.

E2 **Time evolution**: Quench the system to the nearest-neighbour and translation-invariant Ising Hamiltonian[6]

$$H_0 := \frac{\pi}{4} \sum_{(i,j)\in E} Z_i Z_j - \frac{\pi}{4} \sum_{i\in V} \deg(i) Z_i, \tag{4.4}$$

and evolve for constant time $\tau = 1$ to implement a unitary[7] $e^{iH_0}$.

E3 **Measurement**: Measure all qubits in the $X$ basis and report the results.

5: Recall that $S^1$ denotes the circle $[0, 2\pi]/\sim$, where $\sim$ identifies 0 and $2\pi$.

6: Here, $\deg(i) = |\{j : (i, j) \in E\}|$ denotes the degree of site $i$, i.e., the number of incoming edges in the graph.

7: This amounts to a constant depth circuit.

The quantum simulation protocol defines a family of quantum states $\{e^{iH_0}|\psi_\beta\rangle\}$ on $n \times m$-square lattices from which a random instance is drawn according to the Haar measure on $(S^1)^{\times N}$. While in the protocol the unitary is always fixed and all randomness remains in the initial state, one can of course recast the protocol as a random circuit sampling protocol, starting from the initial state $|0^N\rangle$, evolving under the unitary

$$U_\beta = H^{\otimes N} \left( \prod_{(i,j)\in E} CZ_{i,j} \prod_{i\in V} e^{-i\beta_i Z_i/2} \right) H^{\otimes N}, \tag{4.5}$$

and measured in the computational basis. To see this, notice that the $CZ$ gate acting on qubits $i$ and $j$ can be expressed as $\exp(i\frac{\pi}{4}(Z_i Z_j - Z_i - Z_j))$ up to a global phase of $e^{i\pi/4}$, which does not change the outcomes of the circuit . The quantum simulations architecture is thus a variant of an IQP circuit, and indeed, IQP circuits are largely motivated by measurement-based quantum computing. Equivalently, we can view the protocol as preparing a state $V_\beta|0^N\rangle$, where $V_\beta$ is implicitly defined by $U_\beta = H^{\otimes N} V_\beta$.

For this family of quantum states we find the following hardness-of-sampling result.

**Theorem 4.1** (Hardness of sampling in Protocol 4.1) *Suppose that the probability distributions $p_\beta$ defined by Protocol 4.1 satisfy the approximate average-case hardness conjecture 4.1. If there was an efficient classical sampling algorithm for approximately sampling from $p_\beta$ up to an additive error $1/22$ in $\ell_1$ norm, then the polynomial hierarchy would collapse to its third level $\Delta_3^p$.*

To state the approximate average-case hardness conjecture, we rephrase the success probabilities in terms of a universal quantity that is not specific to the sampling scheme at hand. Analogously to the case of the IQP circuits considered in Sec. 2.4 we find that such a universal quantity is given by an Ising partition function $Z_\beta$.

More precisely, using the expression (4.5) for the time evolution unitary, we can express the success probability $p_\beta(0)$ of Protocol 4.1 as

$$|\langle 0|U_\beta|0\rangle|^2 = |\operatorname{Tr}[e^{iH_\beta}]|^2 \equiv |Z_\beta|^2. \tag{4.6}$$

To see this notice that up to a global phase

$$\langle 0|H^{\otimes N} e^{iH_0}|\psi_\beta\rangle = \langle 0|U_\beta|0\rangle \tag{4.7}$$

$$= \frac{1}{2^N} \sum_{y,z\in\{0,1\}^N} \langle y| \exp\left( i\left( \sum_{(i,j)\in E} \frac{\pi}{4} Z_i Z_j + \sum_{i\in V} \left( \frac{\pi}{4} \deg(i) - \frac{\beta_i}{2} \right) Z_i \right) \right) |z\rangle, \tag{4.8}$$

$$= \frac{1}{2^N} \sum_{z\in\{-1,1\}^N} \exp\left( i\left( \sum_{(i,j)\in E} \frac{\pi}{4} z_i z_j + \sum_{i\in V} \left( \frac{\pi}{4} \deg(i) - \frac{\beta_i}{2} \right) z_i \right) \right) \tag{4.9}$$

$$\equiv \frac{1}{2^N} \operatorname{Tr}[\exp(iH_\beta)] \equiv \frac{1}{2^N} Z_\beta, \tag{4.10}$$

where we have implicitly defined the Ising Hamiltonian $H_\beta$ and its complex-temperature partition function $Z_\beta$. We can now state the approximate average-case hardness conjecture.

**Conjecture 4.1** (Average-case complexity) Approximating $|Z_\beta|^2$ up to relative error $1/4 + o(1)$ for any 0.3 fraction of the instances is GapP-hard.

Let us first observe that the hiding property (i.) in Theorem 2.9 trivially holds in the quantum simulation scheme. We have already translated the protocol to the time evolution of the state $|0^n\rangle$ under the unitary $U_\beta$ in Eq. (4.5). Adding a random string of $X$ gates with $x \in \{0,1\}^N$ at the end of the circuit results in

$$\prod_i X_i^{x_i} U_\beta = H^{\otimes N} \left( \prod_{(i,j)\in E} CZ_{i,j} \prod_{i\in V} Z_i^{x_i} e^{-i\beta_i Z_i/2} \right) H^{\otimes N} \tag{4.11}$$

$$= iU_{\beta+\pi x}. \tag{4.12}$$

But as $\beta$ is drawn from the Haar measure on $(S^1)^{\times N}$, $\beta + \pi x$ is distributed according to the same measure[8].

8: The global phase i stemming from the identity $Z = ie^{-i\pi Z/2}$ vanishes when moving to the probabilities and is therefore irrelevant for the proof.

We first prove GapP-hardness of approximating the success probabilities. Using the ideas from the previous chapters, we then translate this result on strong simulation to one on weak simulation. To do so, we provide evidence for approximate average-case hardness by proving exact average-case hardness. We then provide numerical evidence the fact the distribution $p_\beta$ anticoncentrates and is nearly Porter-Thomas distributed [Ber+18] and are also able to prove that the effective circuits generated in the scheme give rise approximate unitary 2-designs [Haf+19]. By the anticoncentration theorem for 2-designs (Theorem 3.4) this constitutes a proof of the anticoncentration property.

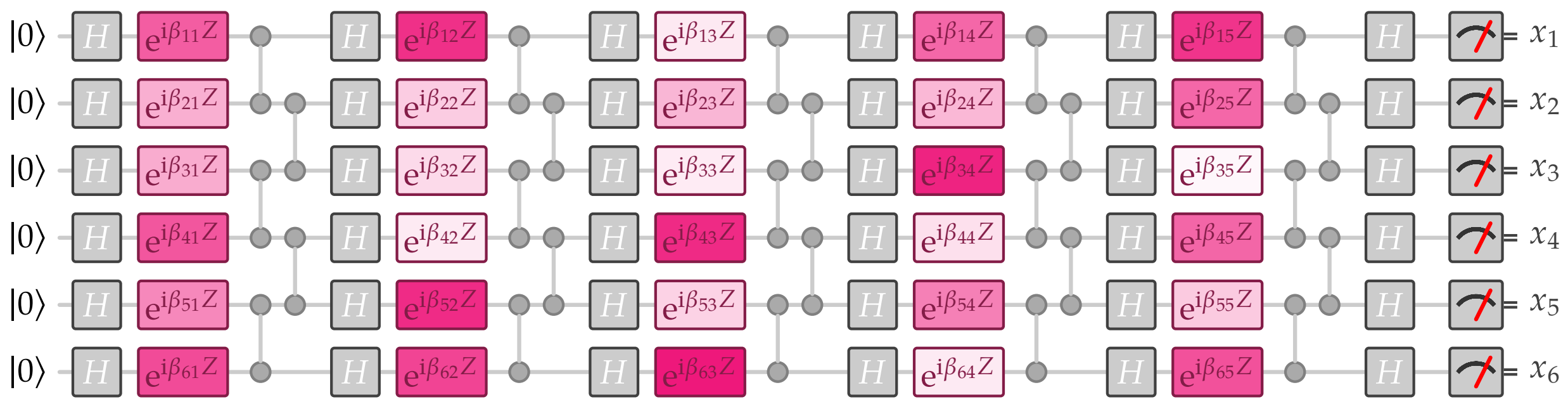

**Figure 4.3:** We show the logical circuit implemented by a measurement of the $6 \times 5$ sublattice of Fig. 4.2 for the case in which the outcome on all bulk qubits is 0. The connecting lines represent $CZ$-gates.

## Hardness of strong simulation

**Lemma 4.2** (GapP-hardness of strong simulation) *Approximating the success probabilities* $p_\beta(0) = |\langle 0|U_\beta|0\rangle|^2$ *up to relative error* $1/4 + o(1)$ *on an* $n \times m$ *square lattice is* GapP*-hard for* $m \in O(n^2)$.

*Proof sketch.* To prove the lemma, we make use of Lemma 2.3, stating that if a sampling scheme is complete for postBQP, then approximating the output probabilities is GapP-hard up to a multiplicative error $\sqrt{2}$. To show this, we prove that the ensemble $\{U|\psi_\beta\rangle\}$ is universal for postselected quantum computation using the idea of measurement-based quantum computation, which makes use of gate teleportation.

To see this, we conceptually divide the $n \times m$ lattice into a 'bulk' comprising the first $m-1$ columns and the last column, which will play the role of the output of the computation. The measurement outcome $S = (a, b)$ is then divided into bulk outcomes $a \in \{0,1\}^{n(m-1)}$ and final outcomes $b \in \{0,1\}^n$. In this mindset Protocol 4.1 is computationally equivalent to an encoded $1D$ nearest-neighbour circuit $C_\beta$ comprising random gates of the form

$$\left(\prod_{i=1}^{n-1} CZ_{i,i+1}\right)\left(\prod_{i=1}^{n} Z_i^{a_i} e^{-i\beta_i Z_i/2} H_i\right), \quad a_i \in \{0,1\}, \tag{4.13}$$

acting on the initial state $|0\rangle^n$. This is easily seen, observing that the protocol is equivalent to preparing a cluster state on the square lattice and measuring each qubit in the $X - Y$ plane with angle $\beta_i$ and applying gate teleportation from left to right.

But one can implement $n$-qubit, depth-$D$ Clifford+$T$ circuits using $X - Y$ plane measurements on cluster states of size $O(n) \times O(Dn)$ [MDF17, Lemma 3]. Given the power to postselect, one can therefore implement arbitrary $n \times D$ Clifford+$T$ circuits by postselecting the unitary family $\{U_\beta\}$ on values of $\beta$ and the measurement outcomes $a$ in the bulk. Hence, the scheme is universal for postBQP on an $O(n) \times O(n^2)$ lattice[9]. By the hiding property $|\langle x|U_\beta|0\rangle|^2 = |\langle 0|U_{\beta-\pi x}|0\rangle|^2$ postselection on specific measurement outcomes is equivalent to postselection on the all-zero string and values of $\beta$. Therefore computing $p_\beta(0)$ is GapP-hard to approximate up to relative error $1/4 + o(1)$ by Lemma 2.3.

□

9: In Ref. [Ber+18, Sec. VI C] we remove the linear overhead in the width of the lattice to an $n \times O(n)$ using a linear-depth implementation of IQP circuits.

Given GapP-hardness of approximating the success probabilities $p_\beta(0)$, we can apply the methods from Sec. 3.3 to prove exact average-case hardness of computing those success probabilities[10].

10: Recall that exact average-case hardness provides evidence for its multiplicatively robust version as it is necessary for it.

**Theorem 4.3** (Exact average-case hardness) *It is* GapP*-hard to exactly compute any* $3/4 + o(1)$ *fraction of the success probabilities* $p_\beta(0)$*, and to approximate any* $1 - 1/\mathsf{poly}(N)$ *fraction up to additive error* $1/2^{\mathsf{poly}(N)}$.

## Anticoncentration and closeness to the Porter-Thomas distribution

*The following section reproduces material from Ref. [Ber+18].*

Let us now approach the question whether the quantum simulation scheme in Protocol 4.1 anticoncentrates. We do so by first examining the output distribution $p_\beta$ in detail numerically. In a second step, we will then rigorously prove that the effective circuits generated by the scheme form a multiplicatively approximate unitary 2-design.

To start with, we notice that anticoncentration of the full output distribution $p_\beta(a,b) = |\langle a,b|U_\beta|0\rangle|^2$ can be reduced to anticoncentration of the conditional distribution $p_\beta(b|a)$ by the following property of $X$-teleporation circuits. The marginal distribution of each measurement outcome is uniform [ZLC00; CLN05] so that we obtain

$$p_\beta(a,b) = p_\beta(b|a)p_\beta(a) = p_\beta(b|a)\frac{1}{2^{N-n}}. \tag{4.14}$$

The anticoncentration property (3.16) for $p_\beta(0^N)$ thus reduces to anticoncentration of $p_\beta(0^n|0^{N-n})$, which is the output probability of an effective logical circuit with gates (4.13) on $n$ qubits.

We will show that the distribution $q_\beta(x) = p_\beta(x|0)$ is Porter-Thomas distributed, giving rise to the anticoncentration property

$$\Pr_{\beta\sim(S_1)^{\times N}}\left[q_\beta(0) \geq \frac{1}{2^n}\right] \geq \frac{1}{\mathrm{e}}. \tag{4.15}$$

The concrete circuit families associated to each architecture are derived below and depicted in Fig. 4.3. In our numerical simulations we simplify the scheme in that we choose the rotation angle discretely from the set $\{0, \pi/4, \pi, 5\pi/4\}$. This choice retains the hiding property while at the same time keeping the scheme postBQP-complete [Ber+18]. Without loss of generality, we can therefore consider the all-zero outcome on the bulk. To spell out the circuit protocol, let us label lattice sites by row-column coordinates $[i,j]$. The circuits are generated inductively, starting from the left column $j = 1$. Measurements are ordered from left to right. The computation begins on the $|+\rangle^{\otimes n}$ state and proceeds as follows:

i. Apply the gate $\exp(-\mathrm{i}\beta_{[i,j]}Z_{[i,j]}/2)$ to qubit $[i,j]$, with $\beta_{[i,j]}$ chosen uniformly from $\{0, \pi/4, \pi, 5\pi/4\}$.
ii. Apply $CZ$ on all neighboring qubits.
iii. Apply a Hadamard gate to each qubit.
iv. If $j = m$, measure in the standard basis and terminate; otherwise increase $j := j + 1$.

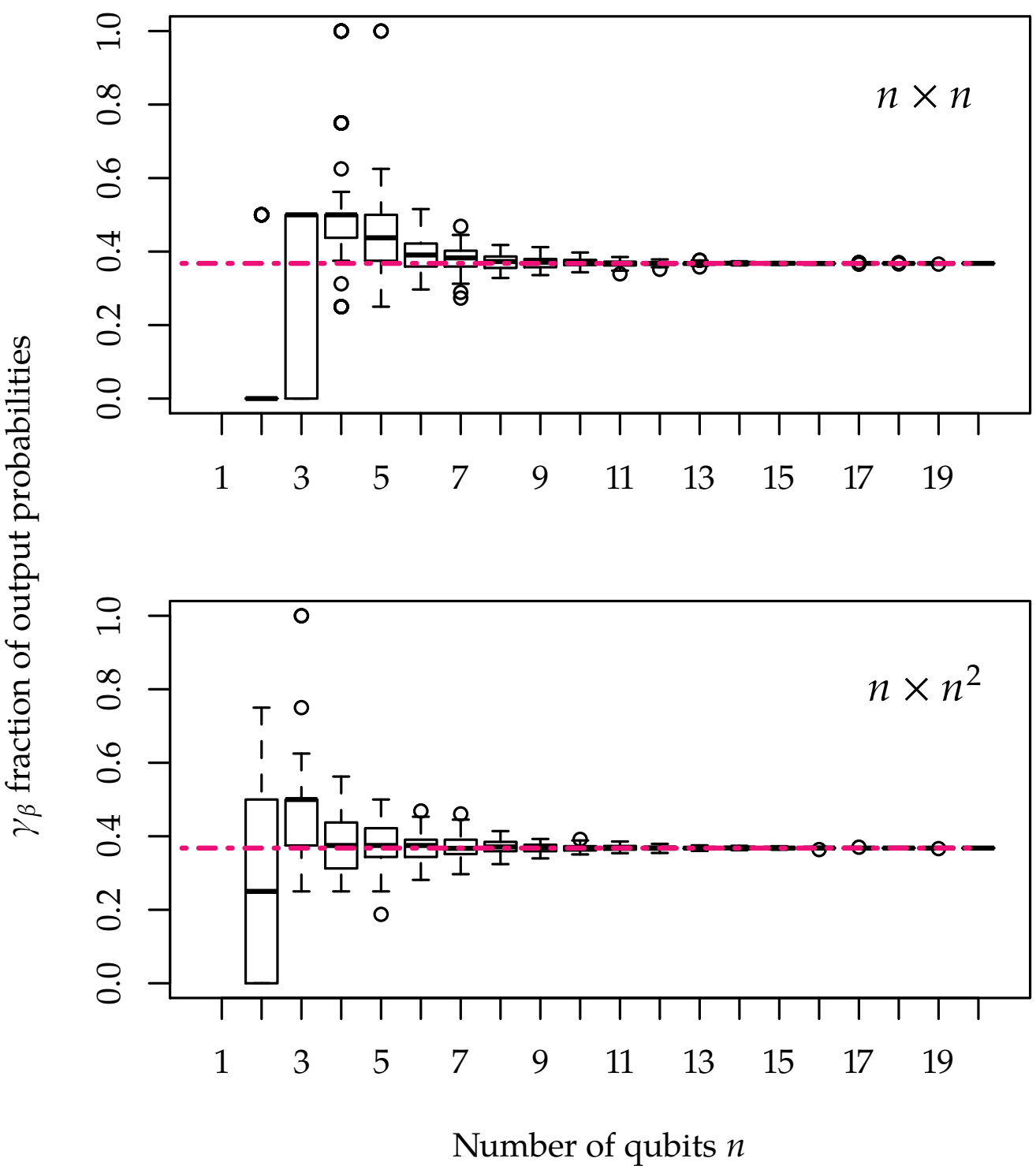

**Figure 4.4:** Fraction of output probabilities $\gamma_\beta$ that are larger than $1/2^n$ for random circuits drawn by choosing $\beta \in \{0, \pi/4, \pi, 5\pi/4\}$ uniformly at random for both linear (top) and quadratic (bottom) circuit depth in the number of qubits $n$ the circuit acts on. This corresponds to lattices of size $n \times n$ (top) and $n \times n^2$ (bottom). For each $n$ we draw 100 i.i.d. realizations $\beta$ and plot the resulting distribution in the form of a box plot according to [IS0 16269]. The pink dashed line shows the value of $1/e$, which is precisely the value to be expected if the output probabilities are Porter-Thomas distributed.

To numerically test Eq. (4.15) we have performed simulations of such randomly generated circuits of gates of form (4.13) (for each circuit family) in LIQUiD [WS14] with up to 20 logical qubits. For each system size, we generated 100 random instances for circuits associated to $n \times n$ and $n \times n^2$ lattices. For each instance, we exactly evaluated the fraction

$$\gamma_\beta = \frac{|\{x \in \{0,1\}^n : q_\beta(x) \geq 2^{-n}\}|}{2^n}. \tag{4.16}$$

of output-probabilities fulfilling (4.15). Our results are summarized in Fig. 4.4: therein, one can see that for circuits associated to both $n \times n$ and $n \times n^2$ lattices, this fraction quickly approaches a constant $\gamma = 1/e$ with rapidly decreasing variance with respect to the choice of circuits. We can conclude that, with very high probability, in a realization of the proposed experiment the amplitude of the final state of the computation anticoncentrates.

As discussed in Refs. [BMS16; BMS17; LBR17], it might seem a priori counter-intuitive that constant-depth nearest-neighbor architectures anticoncentrate. Even more so, the recent results of Napp et al. [Nap+19] strongly suggest that generic constant-depth circuits cannot be expected to yield an exponential speedup over classical computation or even to anticoncentrate. The above connections between our architectures and random circuits shed light into why this behavior is actually natural. As shown in Lemma 4.2 the random logical circuits of gates encoded in our architectures, taking the form (4.13), are universal for quantum computation. Universal random quantum circuits of increasing depth are known to approximate the Haar measure under various settings [Eme+03; ELL05; Bro+08; HL09; BHH16].

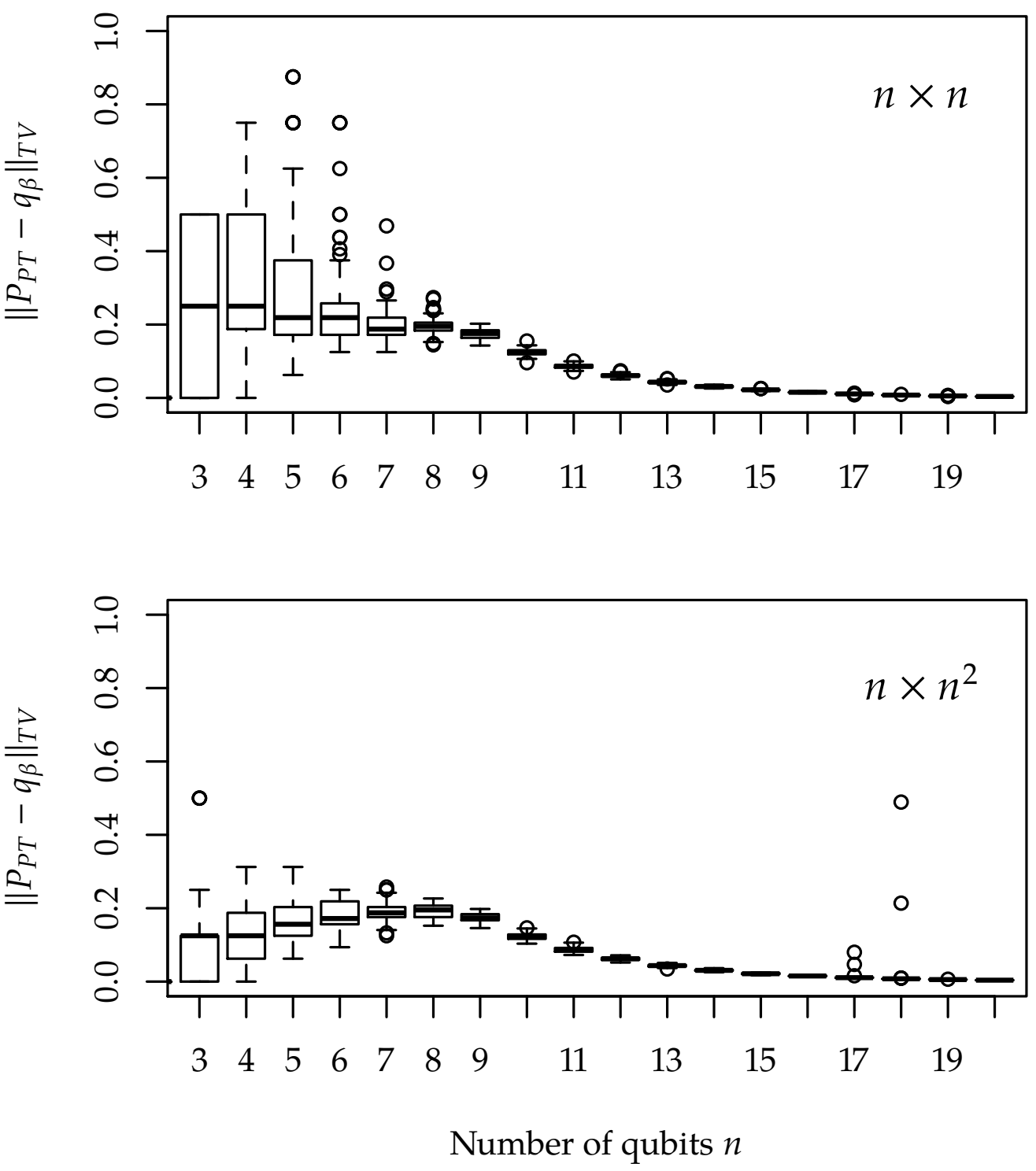

**Figure 4.5:** Total variation distance to the Porter-Thomas distribution of the empirical distribution of output probabilities of random circuits for both linear (top) and quadratic (bottom) circuit depth in the number of qubits the circuit acts on $n$, i.e., lattice of size $n \times n$ (top) and $n \times n^2$ (bottom). For each $n$ we draw 100 i.i.d. realizations $(\beta, y)$ and plot the resulting distribution in the form of a box plot accordin to [IS0 16269].

**Convergence to the chaotic regime.** For 1D nearest-neighbor layouts, they are expected to reach this regime in depth $D \in O(n)$ [KH13; Hos+16; BHH16] (cf. [Boi+18] for further discussion). This regime can equivalently characterized as the 'chaotic regime' in which the output distribution is exponentially (Porter-Thomas) distributed [PT56; Haa10; Eme+03]. Recall from Chapter 3, Eq. (3.32) that the Porter-Thomas distribution is given by

$$P_{PT}(p) = 2^n \exp(-2^n p)\,, \tag{4.17}$$

and thus anticoncentrates in precisely the fashion observed here. The Porter-Thomas regime is known to emerge in chaotic quantum systems for large system sizes [PT56; Haa10; Eme+03; Bro+08; Boi+18; AC17].

In Fig. 4.4, we observe that the anticoncentration fraction $\gamma_\beta$ quickly converges to the value $\gamma_\beta = 1/e$. This is already a signature of the exponential (Porter-Thomas) distribution. Going beyond providing evidence for anticoncentration, we numerically confirmed that the output probabilities of individual instances are actually close to being Porter-Thomas distributed in total-variation distance.

In Fig. 4.5 we show the total-variation distance between the empirical distributions of output probabilities of the random circuits generated in our numerical experiments and the discretized Porter-Thomas distribution. We can see that as the number of qubits increases, the output distributions of random circuits approach Porter-Thomas distribution.

To calculate the total-variation distance to the exponential distribution (4.17), we discretized the interval $[0, 1]$ into $m$ bins each of which contains probability weight $1/m$. In other words, the discretization $(p_0, p_1, \dots, p_m)$ is defined by given $p_0 = 0$, $p_m = 1$ and

$$\int_{p_i}^{p_{i+1}} P_{PT}(p)\mathrm{d}p = \frac{1}{m}\,. \tag{4.18}$$

Denote by $Q(p)$ the numerically observed distribution of output probabilities $p = |\langle x|C_\beta|0\rangle|^2$ over the set $\Omega = \{[p_{i-1}, p_i)\}_{i=1,\dots,m}$. The total variation distance between $P$ and the exponential distribution is then given by

$$\|P - Q\|_{\mathrm{TV}} = \frac{1}{2}\sum_{X\in\Omega} |P(X) - 1/m|\,. \tag{4.19}$$

Since the number of samples we obtain in each run is given by $2^n$we choose the number of bins $m$ depending on $n$. Specifically, we choose $m = \min\{\lceil 2^n/5\rceil, 100\}$ to allow fair comparison for small $n$. Nonetheless, we observe an inintially larger value of of the total-variation distance, which we attribute to the overall smaller sample size.

The same behavior was observed in previous work investigating random MBQC settings [Bro+08], as well as in recent work [Boi+18], which investigated random universal circuits on a 2D architecture. Notably, the finite and universal gate sets considered in these works are very similar to the ones considered here. Likewise, convergence to the exponential distribution was observed in Ref. [AC17] for approximately Haar-random two-qubit unitaries in a 2D setup.

**Convergence to unitary 2-designs** By significantly extending the proof strategy used by Brandão, Harrow, and Horodecki [BHH16] to show that universal random circuits are approximate unitary designs, we have also been able to prove that the effective circuits generated by the Hamiltonian dynamics are approximate unitary 2-designs [Haf+19]. Remarkably, we can prove this already for an $n \times O(n)$ square lattice[11].

11: For a proof of Theorem 4.4, we refer the interested reader to Ref. [Haf+19].

**Theorem 4.4** (Approximate 2-design) *Consider the quantum simulation scheme defined in Protocol 4.1 on an $n \times m$ lattice with $m \in \mathcal{O}\left(4n + \log(1/\varepsilon)\right)$. When measuring the first $m-1$ columns in the X-basis, the effective unitary acting on the last column forms a relative $\varepsilon$-approximate unitary 2-design.*

By the anticoncentration result from the previous chapter (Theorem 3.4), and the identity (4.14), Theorem 4.4 implies that the output distribution $p_\beta$ of Protocol 4.1 anticoncentrates.

## 4.3 Ion trap quantum simulation of IQP circuits

The following section is based on joint (unpublished) work with Paul K. Fährmann, Juani Bermejo-Vega, Jens Eisert, Thomas Monz and Martin Ringbauer [Han+]. I am particularly grateful to Paul Fährmann for working out the details of Lemma 4.5.

In the last section we saw an example of a sampling scheme that was particularly suited for an implementation in a two-dimensional, large-scale analogue quantum simulator. Indeed, in proving the hardness of the scheme using measurement-based quantum computing we exploited a space-time tradeoff between the number of qubits and the depth of the circuit. Thus, we were able to minimize the amount of local control required for the quantum speedup to a phase-disordered initial state.

In this section, we will consider the in some sense opposite end of the spectrum of quantum simulation architectures, namely, digital ion trap quantum simulators. In this platform both local control and long-range interaction gates can be performed with an extremely high fidelity and long coherence times. However, the number of qubits it can host is very limited; currently up to around 50 ions [Zha+17]. Demonstrating quantum supremacy in such an architecture thus requires very different features from a sampling scheme: rather long circuits and a high degree of local control are perfectly feasible, but we want to make as good use as possible of the available number of qubits. Ideally, the required interactions will be tailored to the available gate set.

A sampling scheme that appears to be very natural for ion-trap architectures are IQP circuits (cf. Sec. 2.4). In this section, we will derive a particularly efficient implementation of IQP circuits in an ion-trap architecture that exploits the available operations optimally. Let us begin by giving an overview of the available operations.

### Experimental methods in ion-trap architectures

In this platform, individual ions such as $^{40}Ca^{+}$ are trapped using, for instance, a quadrupole so-called Paul trap. Using internal electronic states one can encode individual qubits. The states of individual ions/qubits can be measured (in the computational basis) using state-dependent fluorescence measurements.

Both single and many-qubit quantum gates in ion trap architectures are implemented using atom-light interactions that may be driven by a global laser pulse close to resonance with the internal energy splitting of the two computational energy levels of the ions [Ber+17a]. In the regime in which the coupling of the internal degrees of freedom of the atoms to their motional state can be neglected, the phase, intensity and duration of the driving laser pulse can be used to drive Bloch sphere manipulations of the state of the two-level atoms around an axis in the $x - y$ plane via a Jaynes-Cummings interaction. In particular, this allows for the parallel execution of $x - y$-rotations on all ions:

$$S_{1,\ldots,n}(\phi) = \exp\left[-\mathrm{i}\phi \sum_{i=1}^{n} X_i\right] . \tag{4.20}$$

Single-qubit $Z$-rotations can be directly induced on individual ions by a strongly focused laser beam that induces a variable-duration AC-Stark shift on the ion.

The coupling of internal degrees of freedom to a global vibrational mode allows to drive entangling operations via the global 'Mølmer-Sørensen' (henceforth MS) gate

$$M_{1,\ldots,n}(\phi) = \exp\left[-\mathrm{i}\phi \sum_{i<j=1}^{n} X_i X_j\right] , \tag{4.21}$$

which couples all pairs of ions via the first order global vibrational mode. Here, the phase $\phi$ is controlled via the intensity and duration of a laser pulse. As a last ingredient, spectroscopic decoupling methods allow arbitrary individual addressing of the qubits via the above interactions. Similarly, one can implement a distance-dependent coupling, effectively realizing a long-range Ising interaction.

## An IQP protocol for trapped ions

Sampling from the output distribution of commuting (IQP) circuits might be considered the simplest circuit-based scheme that shows a quantum speedup. Recall from Sec. 2.4 that in this scheme, a commuting quantum circuit

$$C_W = \exp\left[\mathrm{i}\left(\sum_{i<j} w_{i,j} X_i X_j + \sum_i w_{i,i} X_i\right)\right] . \tag{4.22}$$

is drawn randomly from the circuit family $\mathscr{C}_{n,\mathrm{IQP}}$ on $n$ qubits defined by a uniform choice of angles $w_{i,j} \in A \equiv \{0, \pm\pi/8, \ldots, \pm 7\pi/8, \pi\}$[12]. In other words, every edge $(i,j)$ of the complete graph on $n$ qubits is weighted by $w_{i,j}$ and we apply a gate $\exp(\mathrm{i} w_{i,j} X_i X_j)$ on it; likewise for vertices with weights $w_{i,i}$ and gates $\exp(\mathrm{i} w_{i,i} X_i)$. The resulting weighted graph is then captured by the adjacency matrix $W = (w_{i,j})_{i,j=1,\ldots,n}$.

12: Notice that in contrast to Ref. [BMS16], we are distributing the angles uniformly-spaced across the unit sphere $S^1$

The laser-induced many-qubit (4.21) and single-qubit (4.20) interactions seem perfectly suited to create a fully connected interaction graph between all qubit ions with $XX$-couplings on the edges and on-site $X$-rotations in a single round of interaction. Thus, an ion-trap architecture seems to be the perfect candidate for an implementation of IQP circuits, which have precisely those properties.

What remains to be achieved if one wants to implement an IQP circuit (4.22), is to *decouple* the resulting interaction graph such that all vertex and edge weights are mutually uncorrelated and sampled from the correct distribution on $A$. This can be achieved in linear depth using combinations of single-qubit $Z$-gates and global Mølmer-Sørensen gates as we show in the following. The key idea of how we will achieve this mapping is the commutator primitive

$$\exp(\mathrm{i}\phi X_i X_j) Z_i = Z_i \exp(-\mathrm{i}\phi X_i X_j) , \tag{4.23}$$

for $i \neq j$. We now apply Eq. (4.23) iteratively, applying single-qubit $Z$-gates

$$Z^{z(k)} := \bigotimes_{i=1}^{n} Z_i^{z_i(k)}, \; z(k) \in \{0,1\}^n , \tag{4.24}$$

and global rotation angles $\phi_k$ in the $k^{\text{th}}$ step.

Our decoupling scheme now proceeds in two steps: first, we give a prescription that involves a constant number of iterations of single-qubit $Z$-gates and global Mølmer-Sørensen gates as described above such that afterwards, all marginal distributions of vertex and edge weights $P(w_{i,j})$ are uniform distributions over the set $A$. Second, we repeat this procedure a linear number of times (in $n$) to achieve a fully decoupled global distribution $P(\{w_{i,j}\}_{i,j}) = \prod_{i \le j} P(w_{i,j})$.

Formally, our object of investigation is the adjacency matrix $W$ on which

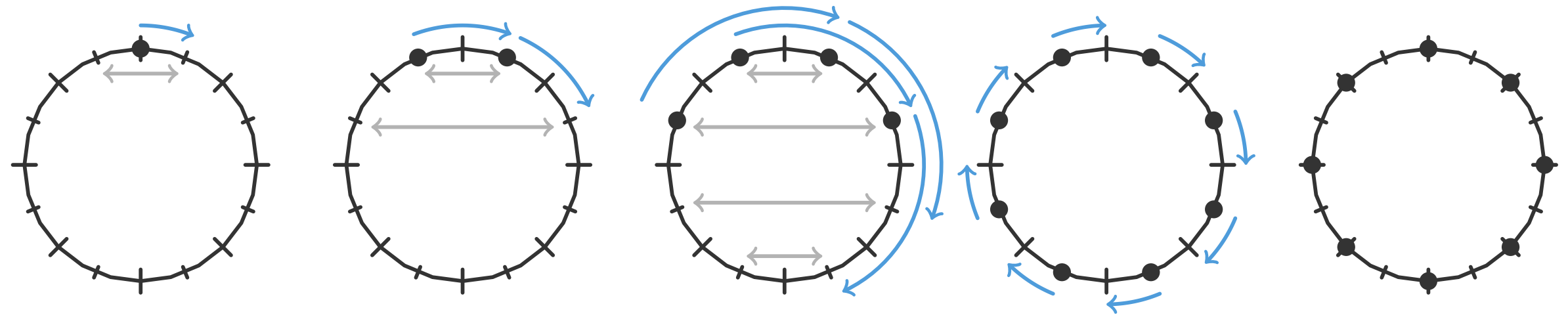

**Figure 4.6:** Illustration of the prescription described in Lemma 4.5 for the marginal distribution of weights $w_{i1}$ and the case of $k = 2$. The couplings $m\pi/2^k, m \in \mathbb{N}$ are represented by their position on the unit circle. Figs. (a-e) show all steps of the prescription in terms of the achievable value of the couplings (bullets). Blue arrows represent deterministic Mølmer-Sørensen rotations, while gray arrows represent random flips induced by the presence of a Pauli-$Z$ term. Starting out with the qubits prepared in $|0\rangle^{\otimes n}$ (a), in the first step, $z(1) \in \{0,1\}^n$ is chosen uniformly at random and the unitary $Z^{z(1)}SM(\pi/8)$ is applied. Subsequently, the distribution shown in (b) is attained in which there are two equally probable configurations $\pm\pi/8$. In the next step (c), $Z^{z(2)}SM(\pi/4)$ is applied resulting in a further doubling of the distribution. In the final step (e), a deterministic application of $SM(\pi/8)$ rotates the uniform $\pi/4$-spaced distribution in such a way that it takes on values $0, \pi/4, \ldots, 7\pi/4$.

we define the probability distribution

$$\begin{aligned} P : A^{n^2} &\to [0,1], \\ W &\mapsto P(\{w_{i,j}\}_{i,j}). \end{aligned} \tag{4.25}$$

The following lemma gives the experimental prescription for the first part of our protocol – marginal decoupling – for the simplest case that the angle set $A_k = \{0, \pm\pi/2^k, \ldots, \pm(2^k-1)\pi/2^k, \pi\}$ is the $\pi/2^k$-spaced uniform distribution over the unit circle $S^1$. In the lemma we denote the product of $X - Y$ rotations and MS gate around the angle $\phi$ by $SM(\phi) = S_{1,\ldots,n}(\phi)M_{1,\ldots,n}(\phi)$,

**Lemma 4.5** (Independent and identical distribution) *Fix an arbitrary $j \in [n]$ and let $z(i) \in \{0,1\}^n$ be iid. uniformly random for all $i \in [k+1]$. Then the couplings $w_{i,j}$ to qubit $j$ generated by the application of the unitary map*

$$U(z(1), \ldots, z(k+1)) = SM\left(\frac{\pi}{2^{k+1}}\right) \prod_{i=1}^{k+1} Z^{z(i)} SM\left(\frac{\pi}{2^i}\right), \tag{4.26}$$

*to a product input vector $|0^n\rangle$ are iid. distributed over the set $A_k$, i.e.,*

$$P\left(w_{1j}, \ldots, w_{nj}\right) = P(w_{1j}) \ldots P(w_{nj}) = \frac{1}{|A_k|^n}. \tag{4.27}$$

To prove the lemma, we will make use of the following intuition, which is illustrated in Fig. 4.6: Uniformly random Pauli-$Z$ gates give rise to a random flip of the couplings with a probability of $1/2$. Combining these random Pauli flips with deterministic Mølmer-Sørensen gates and $X - Y$ rotations with powers of 2 will lead to a uniform distribution of the probabilities on the circle. This explains why the constant rotation at the end is needed, for there is simply no way to flip probabilities between $\phi = 0$ and $\phi = \pi$. As a result, half of the desired couplings would always be left with a probability of zero. This problem is resolved by adding half the desired step size. Let us make this intuitive picture rigorous and prove Lemma 4.5.

*Proof.* First, consider the marginal distribution $P(w_{i,j})$ for some $i \in \{1,\ldots,n\}$ and recall Eq. (4.23). We start by noting that the action of the unitary can be split into two parts. The first part, consisting of repeatedly applying combinations of local Pauli-$Z$ and the global gate $SM(\phi)$, will lead to a uniform distribution on odd multiples of $\pi/2^{k+1}$. The second part, given by $SM(\pi/2^k)$ shifts all weights by $\pi/2^{k+1}$ so that the weights in the angle set $A_k$ can be generated using the unitary $U$. Since the deterministic application of $SM(\pi/2^k)$ does not change the probabilities of obtaining each coupling it is thus sufficient to show that all operations of the first part result in a uniform distribution on the set $A_k$.

To prove this, on the one hand, we make use of the identity (4.24) and obtain that the edge weights after a single application of $Z$-flips and the gate $SM(\phi)$ for some $\phi$ are given as follows. Fix $i, j \in \{1, 2, \ldots, n\}$ and choose $z \in \{0,1\}^n$ uniformly at random. This yields

$$Z^z SM(\phi) = \exp\left(\mathrm{i}\phi\left(\sum_{i<j}(-1)^{z_i+z_j}X_iX_j + \sum_i (-1)^{z_i}X_i\right)\right)Z^z \quad (4.28)$$

and hence the edge weights are distributed uniformly in $\{\pm\phi\}$. To see this observe that each $z_i$ is uniformly distributed over $\{0,1\}$ and hence adding an arbitrary bit (be it random or not) does not change this distribution. This results in an equal marginal probability of one half to set each $(i,j)^{\text{th}}$ entry of the adjacency matrix $W$ to $\pm\phi$, respectively. For ease of notation we abbreviate the resulting random gate as $SM(\pm\phi)$. On the other hand, the edge weights of

$$SM(\vartheta)Z^z SM(\phi), \quad (4.29)$$

are uniformly distributed over $\{\vartheta+\phi, \vartheta-\phi\}$. We apply this fact to sequences $\phi = \pi/2^{l+1}$, $\vartheta = \pi/2^l$ to obtain uniformly random elements of $\{\pm\pi/2^{l+1}, \pm 3\pi/2^{l+1}\}$. Since we choose powers of 2, in every step the size of this set will double; see Fig. 4.6.

We now apply those properties sequentially observing that $Z^{z(l)}SM(\pm\phi) = SM(\pm\phi)Z^{z(l)}$, i.e., that the random gate $SM(\pm\phi)$ commutes with random $Z$ flips[13]. This results in a sequence of rotations around an angle $\pi/2^l$ randomly in the positive and negative direction

13: Again, this follows from the property of the one-time-pad.

$$\prod_{l=1}^{k+1} Z^{z(l)}SM\left(\frac{\pi}{2^l}\right) = \prod_{l=1}^{k+1} SM\left(\pm\frac{\pi}{2^l}\right)Z^{\sum_{l=1}^{k+1} z(l)}. \quad (4.30)$$

The application of this sequence generates random weights

$$\begin{aligned} w_{i,j} &= (-1)^{x_{ij}(k+1)}\left(\frac{\pi}{2^1} + (-1)^{x_{ij}(k)}\left(\frac{\pi}{2^2} + \cdots\right.\right. && (4.31)\\ &\quad \left.\left.\left. + (-1)^{x_{ij}(2)}\left(\frac{\pi}{2^k} + (-1)^{x_{ij}(1)}\frac{\pi}{2^{k+1}}\right)\right)\right)\right) && (4.32)\\ &= \sum_{l=1}^{k+1}(-1)^{\sum_{m=1}^{l} x_{ij}(m)}\frac{\pi}{2^l} && (4.33)\end{aligned}$$

where each bit $x_{ij}(k) = \sum_{l=1}^{k} z_i(l) + z_j(l)$ is uniformly random in $\{0,1\}$ and hence each bit of $w_{i,j}/\pi$ is uniformly random. The first part of $U$ thus generates odd edge weights

$$w_{i,j} = \frac{\pi}{2^{k+1}}(\pm 2^0 \pm 2^1 \pm \cdots \pm 2^{k-1}), \tag{4.34}$$

that are uniformly distributed across $\{\pm 1/2^{k+1}, \pm 3/2^{k+1}, \ldots, \pm(2^k - 1)/2^{k+1}$. Consequently the marginal distribution of the edge weights $w_{i,j}$ after the deterministic application of $SM(\pi/2^{k+1})$ at the end of the sequence are uniformly random in $A_k$.

Applying $U$ will therefore set

$$w_{i,i} = \frac{\pi}{2^k} + \sum_{u=1}^{k}(-1)^{\sum_{l=1}^{u} z_i(l)}\frac{\pi}{2^u}, \text{ and} \tag{4.35}$$

$$w_{i,j} = \frac{\pi}{2^k} + \sum_{u=1}^{k}(-1)^{\sum_{l=1}^{u} z_i(l)+z_j(l)}\frac{\pi}{2^u}, \quad i < j. \tag{4.36}$$

What remains to be shown is that the full distribution decouples, i.e., $P(w_{1,j}, \ldots, w_{n,j}) = \prod_{i=1}^{n} P(w_{i,j})$. But this is easily seen from Eq. (4.36) since the values $z_i(l)$ are independent from $z_j(l)$: The $l$th bit of $w_{i,j}$ is determined by $z_i(l) + z_j(l)$ which is independent for different values of $i$. Therefore for fixed $j$ all weights $w_{i,j}$ for different $i \in [n]$ are independent from each other. □

As mentioned, the above prescription results in the correct distribution on the marginal distribution on a single column of $W$, i.e., the couplings $w_{i,j}, v_j$ to a fixed qubit $j$, and can be implemented in depth $2k + 1$. We conjecture the amount of randomness required by the protocol, namely $n(n + 1)/2 \cdot k$ random bits, and consequently the achieved circuit depth of Lemma 4.5 to be optimal. This is because a machine that ouputs i.i.d. samples from $A$ requires at least $\log_2 |A|$ many random bits per run. However, the columns will be mutually correlated, which is not surprising since the prescription requires only $n(k+1)$ random bits, while at the same time there are $n^2$ entries in $W$ that need to be uncorrelated. By the above argument, in order to fully decouple the distribution on $W$, $n^2 \log_2 |A|$ many random bits are required. Hence, repeating the prescription in Lemma 4.5 $n$ times supplies precisely the optimal number of random bits and we conjecture that the columns decouple in linear depth.

The following lemma gives an alternative procedure that is designed to decouple in depth $O(n)$ but is experimentally more expensive. This is because it requires the application of spectroscopic decoupling in order to exclude qubits from the laser-induced many-body interactions (4.20) and (4.21).

**Lemma 4.6** *Using the restricted MS gate*

$$M_S(\phi) = \exp(\mathrm{i}\phi \sum_{i<j\in S} X_i X_j) \tag{4.37}$$

*for arbitrary subsets* $S \subset [n]$ *and single qubit Z-gates an arbitrary IQP circuit* $C_W$ (4.22) *with* $w_{i,j} = 2r\pi/2^k$ *for* $r \in \{0, 1, \ldots, 2^k - 1\}$ *can be implemented in depth* $k \cdot n$ *using* $k \cdot n(n+1)/2$ *random bits.*

*Proof.* The idea is to apply Lemma 4.5 to $M_{1,2}$ first. The coupling $w_{i,j}$ is then distributed uniformly at random on the set $A$. In the $l$-th step we apply Lemma an to qubits $1, \ldots, l$ using the global gate $M_{1,\ldots,l}$. By assumption, all couplings $w_{i,j}, i, j < l$ are distributed i.i.d. uniformly over $A$. Applying Lemma 4.5 to the $l$ couplings $w_{i,l}, i \leq l$, these are now distributed i.i.d. uniformly at random. But the effect of the prescription on $W_{i,j}, i, j < l$ is merely a translation of the uniform distribution. In the final round the on-site couplings are chosen, requiring depth $2k + 1$ and $kn$ random bits. □

Let us note that Bremner, Montanaro, and Shepherd [BMS17] prove hardness of approximating the output distribution of sparse IQP circuits, i.e., IQP circuits for which $W$ is sparsely populated with $O(n \log n)$ many entries. Hence, these circuits require merely $n \log n \log_2 |A|$ many random bits. It is an interesting open question whether IQP circuits for which $W$ is densely populated using the prescription above $\log n$ times still satisfy the hardness and anticoncentration conditions required for a quantum speedup.

## 4.4 Reducing the experimental requirements

In the last two sections we have seen random sampling schemes that are suitably implemented in analog and digital quantum simulators, respectively. In any experimental realization of a quantum sampling scheme, we would of course want to lower the burden on the experimental precision as much as possible while still being able to make a riogorous statement about the hardness of sampling.

In all of the quantum random sampling schemes that we have discussed so far, the task is to sample from the output distribution of a random problem instance up to a constant total-variation distance error, say $1/22$ in the case of the quantum simulation protocol (Sec. 4.2) and $1/192$ in the case of IQP circuits [BMS16] (Sec. 4). A constant total-variation distance puts a tremendous burden on the experiment as it means that the gate and measurement errors need to scale as $1/(N + M)$, where $N$ is the total number of gates and $M$ the number of measurements. This is why it is desirable to maximize the total-variation distance necessary for a rigorous hardness result. To do so, we now leverage a trade-off in the hardness proof between the conjectured fraction of hard instances and the tolerated total-variation distance error.

Recall that in the proof for hardness of sampling we use Stockmeyer's algorithm, which lies in the third level of the polynomial hierarchy, in order to estimate #P-hard probabilities. To this end, we assume that there exists a classical algorithm $\mathcal{A}$ that samples from the output distribution $p_C \equiv p_{C|0\rangle}$ up to an additive total-variation distance error $\varepsilon$. We then feed $\mathcal{A}$ into Stockmeyer's algorithm and ask it to compute an approximation $q_C(0)$ of the success probability probability $p_C(0)$.

A crucial step in the hardness proof consists in balancing the error stemming from Stockmeyer's algorithm itself and the error incurred from the assumption to obtain a multiplicative approximation up to a factor $1/4$ with constant probability over the choice of $C$. The relevant expression (2.91) is then given by applying Markov's inequality yielding that with probability $1-\delta$

$$|p_C(0) - q_C(0)| \leq \frac{p_C(0)}{\mathrm{poly}(n)} + \frac{\varepsilon}{2^n\delta}\left(1 + \frac{1}{\mathrm{poly}(n)}\right). \tag{4.38}$$

We use that the distribution $p_C$ to anticoncentrates in the sense that

$$\Pr_C\left[p_C(0) \geq \frac{1}{2^n}\right] \geq \gamma, \tag{4.39}$$

for some constant $\gamma > 0$. As a result we obtain that with probability $\gamma(1-\delta)$ Stockmeyer's algorithm yields a relative-error $\varepsilon/\delta + o(1)$ approximation of $p_C(0)$. So in order for the hardness-proof to work, we need to conjecture that any $\gamma(1-\delta)$-fraction of the instances is #P-hard to approximate up to relative error $\varepsilon/\delta + o(1)$. So here, we can trade the fraction of instances we conjecture to be hard with the tolerated error $\epsilon$ of the classical algorithm by varying $\delta$

We can now trade the conjectured fraction of hard instances for a larger total-variation distance tolerated by the hardness proof: As in Sec. 4.2, we numerically find $\gamma = 1/\mathrm{e}$ and choose $\varepsilon = \gamma/8$, $\delta = \gamma/2$ so that with probability $\geq 0.3$ we obtain a relative-error $1/4 + o(1)$ approximation of the output probabilities. The average-case conjecture is a natural one as we prove that $1/4$-approximations of the output probabilities are #P-hard. Using $\gamma = 1/\mathrm{e}$ we can obtain a total-variation distance bound of $1/5$ at the cost of a strong average-case conjecture requiring #P hardness for any 0.07 fraction of the instances; see Fig. 4.7.

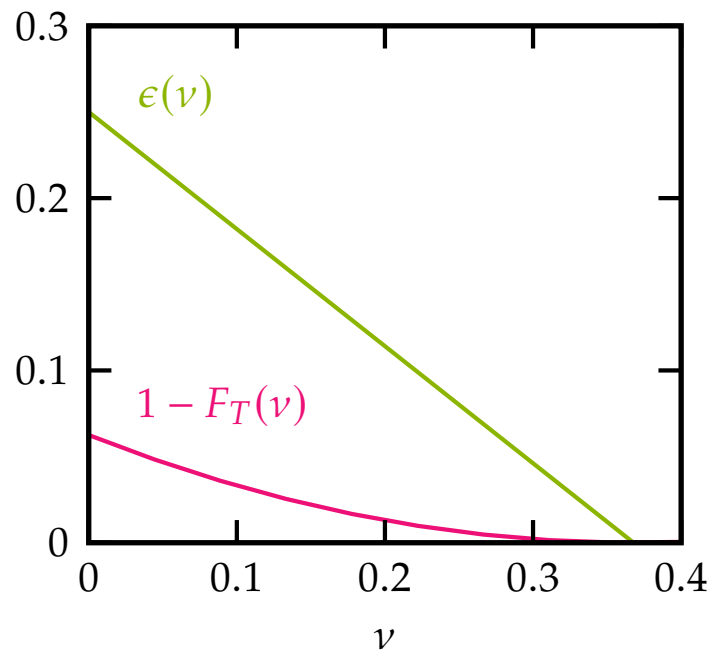

**Figure 4.7:** We can trade the fraction $\nu = \gamma(1-\delta)$ of instances conjectured to be approximate average-case hard for a larger total-variation distance bound $\epsilon$ while keeping the relative error tolerance at $\epsilon/\delta = 1/4$. We assume an anticoncentration fraction $\gamma = 1/e$. Assuming a rejection threshold $\epsilon$, we also show the corresponding infidelity bound $1 - F_T(\nu)$ obtained from the upper bound $d_1(\rho,\sigma) \leq \sqrt{1 - F(\rho,\sigma)} \leq \epsilon_T$ on the trace distance $d_1(\rho,\sigma)$ between ideal and imperfect state preparation.

## 4.5 Conclusion and outlook

**Sampling schemes for quantum simulators**

In this chapter, we have built bridges between rigorous hardness results for quantum random sampling schemes and quantum simulators that already now outperform classical simulation algorithms.

Based on experimental desiderata that are motivated by cold atoms in optical lattices, in Section 4.2, we developed a quantum sampling scheme for dynamical quantum simulators that provably outperforms classical simulations under an approximate average-case hardness conjecture. Cold atoms appear to be perfectly suited to aan experiment designed to disprove the complexity-theoretic Church-Turing thesis, mainly for two reasons: first, already now, the best classical-simulation algorithms are being outperformed using such architectures [Tro+12; Bra+15; Cho+16]. The scheme we develop is minimal in that it does not require any local control but only translation-invariant operations on a square lattice except for phase randomness in the intial state, a property that is well realizable in an optical-lattice setup. It merely comprises the constant-depth time evolution of an initially disordered product state, followed by an $X$-basis measurement of all qubits.

The key conceptual advantage of this scheme is that – as in measurement-based quantum computation – by a space-time trade-off all of the 'quan-tum power' can be transferred to single-qubit quantum measurements and two-local entangling gates. It is those measurements that drive a logical computation from left to right of a 2D square lattice. The measurement-based nature of the quantum simulation architecture will turn out to be crucial when we come to the question of thow to certify quantum sampling schemes in Part II.

At the same time, the presented scheme does not go the full way to a 'real' quantum simulation in that the output of the simulation is a standard-basis measurement rather than a physical property of the system. Physical properties that are often of interest in quantum simulations are of two-point correlators and the magnetization (or imbalance) of the system [Tro+12]. A practical but at the same time rigorous schemes demonstrating a computational supremacy over classical computations might involve quantum simulation schemes for such physical properties. But of course, it is precisely the fine-grained nature of computational-basis measurements that allowed us to leverage results on the hardness of computing individual output probabilities to the robust hardness of sampling. For such robust hardness of sampling results using Stockmeyer's algorithm, the exponential size of the sample space seems indispensible – for the better or the worse: only if the sample space is super-polynomially large do constant total-variation distance bounds require that most probabilities are correct up to inverse exponential errors. Otherwise, constant total-variation distances would allow constant or inverse polynomial bounds on each probability, making their approximate computation feasible within BQP, so that the collapse of the polynomial hierarchy would not follow as straightforwardly[14].

14: Nevertheless, it is widely believed that BQP is outside of the polynomial hierarchy. And in fact, only recently has an oracle separation between BQP and PH been proven [RT19].

Coming from a different vantage point, in Section 4.3 we have used digital ion-trap quantum simulators in order to efficiently implement IQP circuits. Ion traps are particularly well suited to implementations of circuit-based quantum supremacy schemes as they allow for high-precision local control as well as entanglement between arbitrary pairs of ions. Technically, the key challenge we faced was to use the available gate set comprising $X - Y$ rotations and the Mølmer-Sørensen global entangling gate, both with a fixed angle $\phi$, as efficiently as possible. To do so, we derived a decoupling procedure in which the only nondeterminisic ingredient are local random $Z$-flips in every round of the protocol (Lemma 4.5). The key challenge in using ion traps to outperform classical computers is the issue of scalability: as of today, the largest ion-trap quantum simulators comprise up to 50 ions [Zha+17].

---

With this, we conclude the discussion of rigorous hardness results for quantum random sampling schemes. Let us now move on to the next theme of this thesis and ask the question: Given a quantum sampling device, how can we certify its correct functioning?

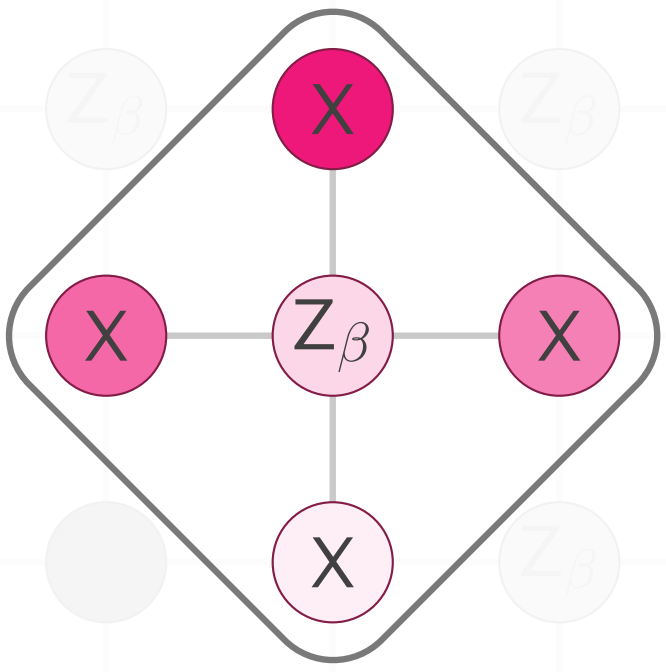

# Part II
Verifying quantum sampling schemes

# Verifying quantum sampling devices

# 5

In Part I of the thesis we have seen how and to some extent also why quantum samplers are able to outperform classical algorithms. We did so on a theoretical level within the model of quantum computation by invoking complexity-theoretic arguments. But to convincingly demonstrate that the advantage offered in theory by quantum devices it is not sufficient to design and build a quantum device that supposedly implements a certain quantum computation. Rather, to convincingly demonstrate quantum supremacy and violate the complexity-theoretic Church-Turing thesis we need to *verify* from a feasible number of uses of the device that it actually achieves the targeted task. In other words, we must *certify* that the device functions according to our specifications. In this part, we will ask the question:

> *Can we efficiently certify devices designed to perform quantum random sampling?*

What would it mean to certify a quantum sampling device? Clearly, a quantum device achieves a targeted sampling task if it returns samples which are – within the prescribed level of accuracy – distributed according to the target distribution defined by the probabilities $|\langle S|C|0\rangle|^2$ for a quantum circuit or unitary operation $C$ and output string $S$ . In the context of random quantum sampling schemes the question of verification can therfore be phrased as follows.

**The question of verification [Han+19]**

How can one convince a sceptical certifier that a quantum device, which supposedly achieves a task that no classical machine can do, actually samples from a distribution that is close enough to the ideal target distribution?

In this chapter, we will first discuss why traditional means of certifying quantum experiments on the one hand and certain computational tasks on the other hand fail (Section 5.1. Observing that this is the case, we will then set up a framework in which different settings and means of certification can be cast very naturally (Section 5.2. We will then identify the device-independent and what we call the 'measurement-device-dependent' scenario as meaningful settings when it comes to certifying quantum sampling devices designed to demonstrate quantum supremacy. In the next chapters (Chapter 6 and 7) we will assess the possibility of certifying quantum sampling devices in those settings.

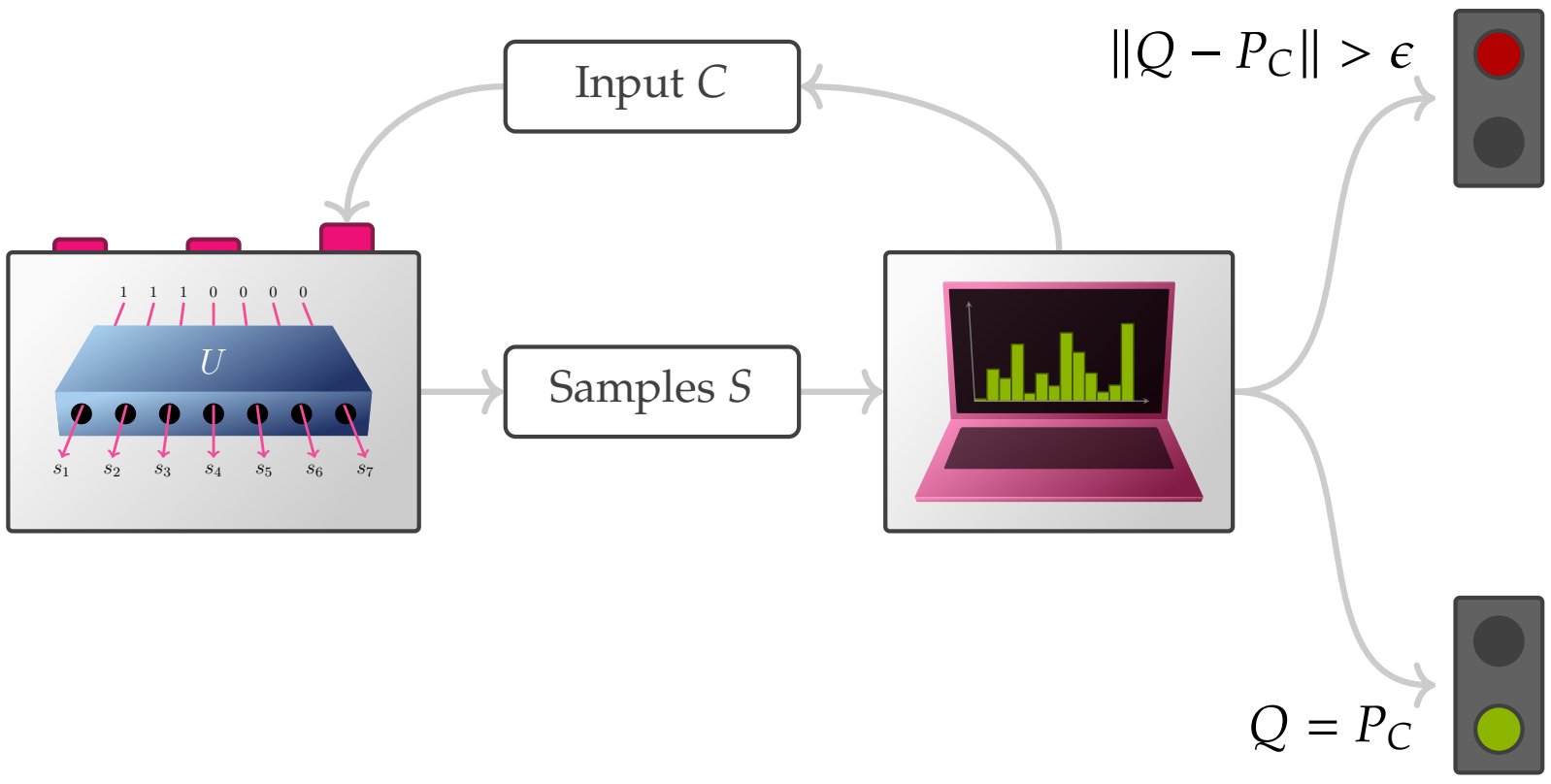

**Figure 5.1:** The question of verification asks: given an experimental prescription of a sampling task as defined by a quantum circuit $C$, can one verify that the outputs $S$ of the device are distributed according to the ideal target distribution $P_C$ with probabilities given by $P_C(S) = |\langle S|C|0\rangle|^2$? The inset image of the boson sampler is an adapted version of [Gog+13, Fig. 1].

Let us stress that the question of verification is a make-or-break issue. A negative answer to it would fatally thwart the quest to experimentally demonstrate quantum computational supremacy as the absence of a convincing certificate would always leave room for a sceptic to doubt that quantum computers are physically possible beyond small scales. Given its outstanding importance to the quest for quantum supremacy the question of verification has received relatively little attention in the literature – with some exceptions [SB09; Gog+13; AA14; Aol+15]. In part, this is due to the fact that it is an extremely difficicult question to answer in light of the fact that the very task of quantum random sampling is designed in such a way that we cannot reproduce the results of the computation.

## 5.1 Traditional approaches to certification

This is why the traditional means of certifying devices or experimental setups in physics fail. In an experiment, the goal may be either to test a theory or model in an experimental target system, or – as in analogue quantum simulators – to simulate such a model using the available experimental means of manipulating and probing a system. In either case, one wants to make sure that most if not all known sources of error with respect to the model are eliminated. In such a scenario, the standard means of verifying that this is indeed the case is to predict the outcome of the experiment and compare the results to the measured outcomes. Let us call this the 'predict-and-measure' method of verification.

But, of course, whenever the target model of the system dynamics is neither analytically solvable nor efficiently solvable using classical computing resources, this method will at some point inevitably reach its limits as the system size is increased. And indeed, in a quest to demonstrate quantum computational supremacy the very goal is to perform a task that *cannot* be reproduced on a classical computer. Hence, the predict-and-measure method of physics will fail.

Another approach that is common in physics experiments is to characterize and successively eliminate the sources of noise occurring in individual components of a larger quantum device, say, the individual quantum gates, single-qubit state preparations and the measurements. This approach would circumvent the exponential scaling of reproducing the outcomes of the entire experiment as a reasonable experiment comprises not too many individual components. But this route to making sure that a large-scale device operates correctly by certifying its individual components has the drawback that it cannot rigorously verify that the device actually performs the correct task: there is always the loophole that some unwanted effect might have happened between the application of the individual components. Any approach to verification that is based on certifying individual components of a large device is therefore doomed to run into this *composition loophole*.

But we can also come from the opposite end and ask how approaches to verification in the theory of computational complexity fare when put to the testbed of applicability in an experimental scenario. NP problems are already defined with the question of verification in mind: Remember that they are those problems which have efficient proofs or witnesses. If a computation of an NP problem accepts, then there will be some efficient way to prove that this is indeed the case.

The prime example of this idea in the context of quantum computing is Shor's factoring algorithm [Sho94; Sho97]. *Qua* solving the NP problem to decompose a large number into its prime factors[1], the outcome of applying Shor's algorithm is verifiable by the following simple algorithm: multiply the outputs and check whether the product equals the input. At the same time, it is strongly believed that factoring is an intractable problem for classical computers. While not a complete problem for the whole of NP its hardness is believed to be NP-intermediate, that is, strictly harder than any problem in P but also strictly easier than the hardest problems in NP such as 3SAT [Lad75][2]. Currently, numbers with description complexity 1024–4069 bits are used for public-key encryption via RSA [IEEE 1363]. Given the abundance of RSA, there is now overwhelming empirical evidence that no classical factoring algorithm exists, which is able to factor such numbers in a feasible runtime. A large-scale implementation of Shor's algorithm using which it is possible to factor large numbers of, say, 2048 bits would therefore constitute a demonstration of quantum supremacy by the book as factoring is both hard and classically verifiable.

1: Strictly speaking, prime factorization is not an NP-problem of course, since it is not a decision problem. The same ideas apply there nevertheless as the prime factors are unique and hence the outcome deterministic.

2: In a beautiful theorem, Ladner [Lad75] proved that if $\mathsf{P} \neq \mathsf{NP}$ then there exist problems that are NP-intermediate, that is, both strictly easier than NP-hard problems and strictly harder than problems in P.

However, it is not feasible on the quantum hardware that will be available any time soon. Performing Shor's algorithm beyond a proof-of-principle demonstration [Van+01] will require an error-corrected universal quantum computer. The resource estimates for factoring a 2048 bit integer on an error-correct computer require between 5000 and 10000 logical qubits and at least 20 million physical qubits [HRS17; OC17; GKK18; GE19]. Demonstrating a verified quantum supremacy via factoring is therefore out of reach in the near future.

It is also noteworthy that the complexity-theoretic grounding of factoring stands on much less firm grounds than the $\mathsf{P} \neq \mathsf{NP}$ conjecture. In particular, an efficient solution to factoring would not imply $\mathsf{P} = \mathsf{NP}$. Its hardness is mainly grounded in the evidence provided by the absence of efficient factoring algorithms in spite of (surely) extensive search for it (given the incentive of breaking large parts of present-day crypto systems). Theoretically, the hardness of factoring stands on a similar footing as the approximate average-case hardness conjectures discussed in the previous chapters, which, however, have been much less tested. Factoring is also a rare example of an NP problem which can also be solved efficiently on a fault-tolerant quantum computer. However, not all and in particular not the hardest problems in NP such as 3SAT can be efficiently solved using a quantum computer, so the common belief, posing further obstacles to demonstrating quantum supremacy via NP problems.

But maybe we can import ideas and results known from the study of NP to verifying quantum random sampling schemes. Again, there are a couple of obstacles to such an approach. First, while for NP tasks efficient proofs are guaranteed to *exist*, it may be computationally hard to *find them*. Second, NP is a class of decision problems rather than sampling problems of the type we are concerned with here. The output of NP algorithms is a single deterministic bit as opposed to samples from a potentially exponentially large sample space. It is an intriguing observation that, internally, Shor's algorithm comprises sampling from a probability distribution which cannot be efficiently certified by looking at the samples alone [SV13][3] [Han+19]. Nevertheless, in Chapter 7 we will get to know an idea for how to verify quantum sampling schemes that is very much in the spirit of factoring integers [SB09]: in this scheme, the idea is to *hide a secret* when setting the task for the quantum sampler. This secret can be recovered from the samples alone, if the quantum device has performed the correct task, and thus serves the purpose of a proof in NP.

3: We will elaborate this intriguing point in Chapter 6.

Already in our short discussion of traditional approaches to certification, we have seen the very different shapes that a certificate can come in. When verifying NP problems, a certificate is given by a proof, that is, a string of bits. When characterizing individual components of a device, a certificate comes as classical descriptions of those components. When verifying by classically predicting the outcome of an experiment, the certificate is given by that prediction.

A first step towards devising a certification protocol will therefore be to conceptualize the verification task for quantum experiments and devices more generally.

## 5.2 Classifying quantum certification

I developed the key ideas of the framework for verification that is presented in this section in joint work with Jens Eisert, Ingo Roth, Nathan Walk, Damian Markham, Rita Parekh, Ulysse Chabaud, and Elham Kashefi [EH+20]. I am also grateful for an insightful discussion with David Gross on this subject.

When verifying quantum samplers, depending on the experimental scenario at hand, we can conceive of verification in very different settings. It is instructive to conceptualize the quantum verification problem as a protocol between the *quantum device* (seen as being powerful) and its *user* who is severely restricted in their measurement devices and computational power. Depending on the setup, different parts of an experiment can be considered *trusted* and thus useful for verification, for instance, as we have previously verified the correct functioning of those parts, and which parts need to be verified and are therefore *untrusted*. When we make this choice, we can see this as *drawing a line* between the quantum device and its user, between what is part of the quantum device and what is its output[4].

4: In the theory of computational complexity those parties are typically referred to as a mighty *prover* and a restricted *verifier*.

As an example, consider the measurement of single photons in a quantum optics experiment. As in any quantum measurement, we need to draw a line between the qusantum part of the experiment and the 'classical' output of the measurement apparatus that results in a collapse of the wave function. In a single-photon measurement, a photon that arrives at the detector triggers an avalanche of further photons being emitted within the detector. The photocurrent is then then measured in a photo-diode by converting it to an electrical current, which is sent to the final readout computer for postprocessing. In this setting, we can place differing levels of trust in the individual parts of the measurement device. We are probably rather confident that the electrical current which is sent through a copper wire is barely suceptible to noise. But the trust situation may be very different in the initial stage of the measurement wherein an avalanche of photon emissions might easily be triggered by some noise process. Surely, the degree to which we trust the correct functioning of those parts – as we would in the case of a classical signal line – determines inhowfar we are willing to consider a device as pertaining to the user of the device who is post-processing the results of a quantum process.

In the same vein, we can regard certain of the quantum operations as being sufficiently trusted to become part of the resources available to the user in order to verify the correct functioning of the untrusted part of the device. To what extent we are willing to do this may to a large degree depend on the applications we have in mind and the physical setup at hand. While in a laboratory setting we would surely be rather confident that none other than certain measured effects influence the outcome, this might be quite different in a cloud computing setup in which we access a remotely operating quantum device via the internet.

It is therefore useful to conceptually separate different parts of a quantum verification scheme. The core part of such a scheme is, on the one hand, the *quantum (sampling) device* to be verified in its correct functioning. Internally, such a device will prepare a quantum state, execute some unitary operation on it, and finally measure the resulting state in some natural basis. Then, there is the user of the device who performs certain *control operations* on the device and processes the classical data produced by the device. This *classical data processing* consumes storage capacity and processing time on a classical computing device. In principle, they may perform several rounds of control and processing with adaptive operations in subsequent rounds. For the purpose of this thesis, however, we focus on a single-round setting: the user initiates the quantum device via a control sequence and processes the resulting data on a classical machine.

Depending on the trust situation at hand it may be convenient to conceive of the quantum device as either receiving or outputting quantum states, however. To this end, we distinguish the *state preparation* step which is used as a starting point of any quantum experiment. In many instances, state preparations can be performed with a high degree of control. Instances of this state of affairs include the preparation of coherent states in quantum optics [Mot+14] or certain ground states of simple Hamiltonians in a cold-atom setup [Tro+12]. Those might be used as a lever to perform uncalibrated tomography [RMH10]. Likewise, we can distinguish the *quantum measurement apparata* used to characterize the state after time evolution. Examples include single-photon detectors in an optical setup, or fluorescence detection as used in setups involving atoms such as ion traps or optical lattices [BDZ08; Ber+17a]. Conceptually speaking, a measurement apparatus might also include a small quantum computer, with the possibility to prepare quantum states on few qubits and perform short circuits.

In idealized settings, state preparations or measurement might be considered to be perfect. But of course, more relevant in practice are situations in which their functioning is well understood and characterized in terms of the potential sources of errors as well as their efficiency . In many physical architectures, certain quantum measurements or state preparations can be performed very accurately, while at the same time only certain *types* of states can be prepared or measurements be performed.

To summarize this discussion: depending on the *assumptions* one is willing to make regarding the state preparation, measurement apparata, or certain quantum operations, different parts of the quantum device may be conceived as pertaining to the user or verifier of the device. If a part is trusted to function correctly with a certain accuracy, it may be utilized in a verification protocol.

*The following passage has already been published in Ref. [EH+20].*

The effort or *complexity* of such a certification protocol can be divided into several distinct parts: This is the number of different settings or rounds in which data is obtained from the measurement device *(measurement complexity)*. Implementing those different settings might require different *quantum computational effort* as for example quantified by the length of the circuit that implements a certain measurement. Then, there is a minimal number of experiments and resulting samples that need to be obtained for a protocol to meaningfully succeed *(sample complexity)*[5]. Finally, one needs to process those samples involving classical computational effort in time and space *(post-processing complexity)*.

5: Speaking in the language of an interactive game between the verifier (the user) and the device (the prover), the measurement complexity of a certification protocol can be viewed as the total length of the messages sent from user to device, while the sample complexity is the total length of the messages sent from device to user, required in order for certification to be successful.

The goal of quantum verification is to establish the correct functioning of a quantum device. But different means of verification may well yield differing *amounts of information* about the actual state of the device. Such information proves to be crucial for designing and improving a concrete experimental setup. It may be less important when the user's goal is merely to check the correct functioning of, say, a newly bought device, or a remote server.

For instance, one can perform full tomographic recovery of a quantum state or probability distribution and thereby determine the deviation of the actually prepared state from the ideal target. Even tomographic schemes can be conceived of as a protocol that outputs '`Accept`' if the device functions correctly, and '`REJECT`' if it does not. Whether the protocol accepts or rejects is determined according to reasonable *measures of quality* that are appropriate for the respective property of the device being certified. We have already seen the total-variation distance (2.79) as a measure of quality on the level of probability distributions, the trace distance (2.77) on the level of quantum states, and the diamond norm distance (3.8) on the level of quantum channels. All of these measures have an operational interpretation as the optimal probability to distinguish the respective objects in an experiment.

Typically, which levels of trust on the quantum device are assumed also has an effect on the protocol's complexity, or even renders certification feasible in the first place. Likewise, at given trust levels the complexity of a protocol can be traded for the amount of information about the device that the user can extract when running the protocol.

**Framework for verification methods**

Different schemes for verifying quantum devices can be assessed in terms of the *assumed trust levels*, their overall *complexity*, and the *information gained* when using them.

At one end of this spectrum lie *tomographic methods* in which entire quantum states or processes are characterized in terms of their full density matrix [Hra97; Jam+01] or their Choi matrix [Kli19]. Such methods require full trust in one's measurement apparatus and the capability to perform a number of distinct measurements given by the number of degrees of freedom – the Hilbert space dimension squared for quantum states, and the Hilbert space dimension to the power of four for quantum processes. Moreover, those methods require the reproducibility of results in that the quantum states on which measurements are performed are independently and identically distributed (iid.). Tomographic schemes yield a large amount of information – namely a full characterization of a particular state – but at the same time require a lot of trust or assumptions on the correct functioning of the device already. In this sense they are *fully device-dependent*. Device-dependent methods are most relevant to applications in well-controlled laboratory environments. In those situations, one can often also develop tailor-made schemes that exploit the specific characteristics of an experimental setup. Examples of this include the idea of *self-verification* [Kok+19] and *cross-platform verification* [Elb+20] as well as the idea of compressed-sensing quantum state tomography [Gro+10], which exploits a high anticipated coherence of the state to be characterized in order to reduce the resource cost of tomography.

At the other end of the spectrum lie *device-independent* methods that do not make any assumptions at all about the quantum device and measurement apparata. Rather those methods conceive of the device as a black box and take into account only classical numbers as input and output of the device. Such methods typically consume a large amount of resources at little information gain. Famous examples include self-testing schemes[6] [MY04] which allow for device-independent certificates of quantum states [McK11], gates and instruments [Sek+18]. Loophole-free Bell inequality violations are famous applications of such tests to Bell states [Hen+15; Sha+15]. Device-independent methods put the lowest trust level on the device and thus allow a user who might not have any physical access to a device, for instance, because it can only be accessed remotely, to verify its correct functioning. This is why, when performing a test of the complexity-theoretic Church-Turing thesis by verifying a random quantum sampling scheme device-independent tests are the method of choice. Much like in Bell tests of the locality of nature, here we aim to close all possible loopholes in order to be as confident as possible in the outcome, providing an answer to the question: *Does nature permit computations that are systematically faster than classical computations?*

6: A comprehensive recent review of self-testing schemes can be found in Ref. [SB19].

Let us end this chapter by noting that one can relax both full device-dependence and stringent device-independence. On the one hand, one can often introduce mild assumptions such as bounds on the system dimension [Gal+10] in order to weaken the impractical stringency of full device independence to semi-device independence [PB11; LVB11;

Li+11; Li+12]. Such ideas are often interesting in realistic communication scenarios between physically distant servers.

Coming from the other end of the spectrum, one can soften the requirements of fully device-dependent schemes that crucially rely on perfect measurement apparata, giving rise to *semi-device-dependent schemes*. Famously, randomized benchmarking protocols are robust against state preparation and measurement (SPAM) errors [EAZ05; Dan+09; MGE12]. But one can also go further and allow for uncertainties in the calibration of the measurement apparatus in self-calibrating schemes [Bra+12; MRH12]. Randomized benchmarking protocols can even be extended to full quantum process tomography [Kim+14; Rot+18]. It is even possible to self-consistently characterize entire gate sets or their respective unitary errors from the observed statistics when applying different gate sequences [Mer+13; Blu+13; Gre15; Blu+17; COB20]. While such methods require less assumptions on the experiment, they typically require enormous resources in terms of the number of quantum samples and rather complex measurement prescriptions.

## 5.3 Verifying quantum samplers

Let us now return to the specific question how to verify quantum sampling devices. Put in complexity-theoretic terms, the question of verification for quantum sampling schemes amounts to the question whether or not the sampling equivalent of NP equals the sampling equivalent of P. Put in physics terms, it amounts to the question whether there are experimental scenarios in which one can certify the correct performance of an experiment, regardless of the fact that one cannot predict its outcome. In this section, we identify two meaningful settings in which we can rigorously analyze the certification task for quantum sampling schemes. Those settings are motivated from a cryptoraphic and a physics perspective, respectively

### Device-independently verifying quantum samplers

We begin with the complexity-theoretic or cryptographic perspective on verifying quantum sampling devices. From this perspective, a quantum sampler should be verified device-independently, that is, without making any assumptions whatsoever on the device or its component. In the device-independent scenario, a quantum sampling experiment is viewed as a large black box that takes as an input an experimental prescription and outputs classical numbers only. Such a perspective is adequate, for example, if a quantum sampling device is used to produce certified random numbers [Aar19], or if a paranoid sceptic wants to check whether quantum supremacy was actually achieved in the experiment by Arute et al. [Aru+19] just from by looking at the classical samples.

Verifying that a quantum device designed to sample from the output distribution of a randomly chosen $C \in \mathscr{C}$ functions correctly amounts to verifying the following: the distribution $Q$ of $s$ samples $S_1, \dots, S_s$ is $\epsilon$-close in total-variation distance to the distribution $P_C(S) = |\langle S|C|0\rangle|^2$. For a given problem size $n$, a verification test is a classical algorithm $\mathcal{T}$ that takes as an input a sequence $S \sim Q^s$ of $s$ samples drawn i.i.d. according to a distribution $Q$ over the respective sample space $\mathcal{E}$ and outputs 1 or 0 for '`Accept`' or '`REJECT`', respectively. The most general notion of classical device-independent verification can be phrased in terms of a weak-membership test [Aol+15; Han+17].

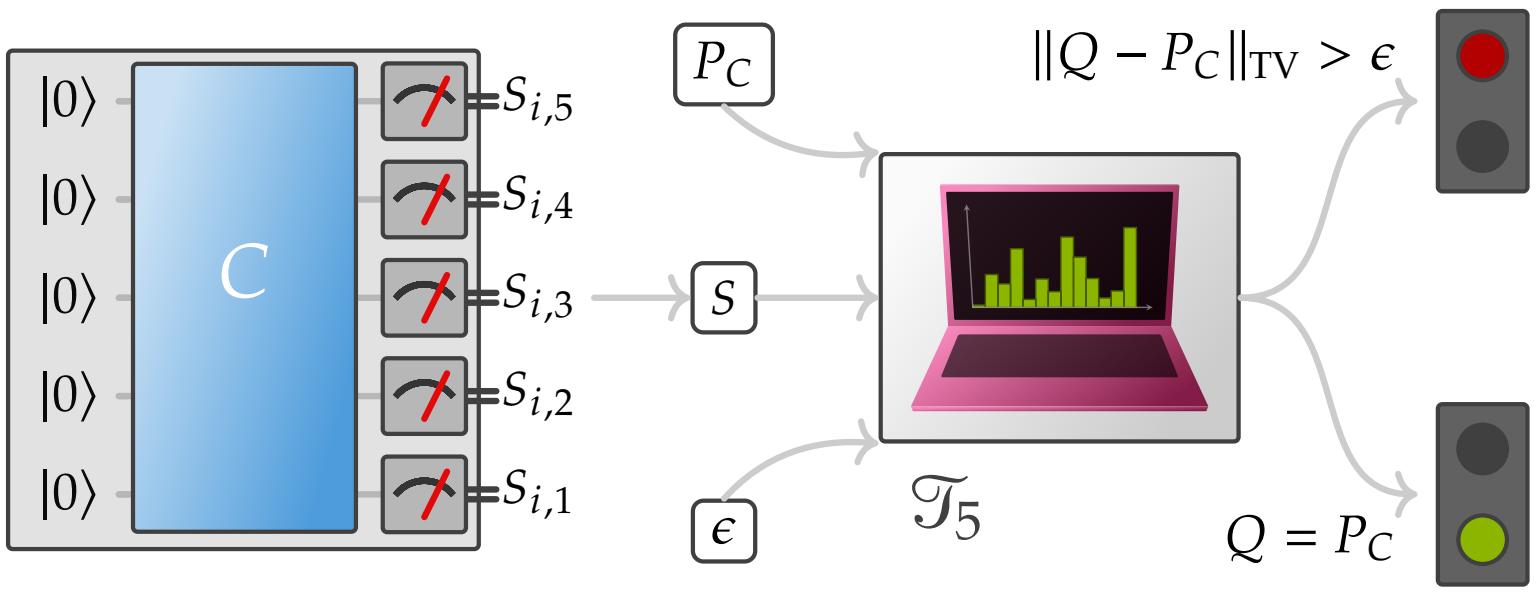

**Figure 5.2:** We consider the problem of certifying probability distributions of the form $P_C(S) = |\langle S|C|0\rangle|^2$ with an input state $|0\rangle = |0\rangle^{\otimes n}$ and a circuit $C \in \mathcal{C}$ drawn from some circuit family $\mathcal{C}$. Given $\epsilon > 0$ and access to an arbitrary-precision description of the target distribution $P_U$, the test $\mathcal{T}_n$ treats the sampler as a black box and receives a sequence $S = (S_i)_{i=1}^s \sim Q$ of $s$ samples from an unknown distribution $Q$. Given $S$ the test is asked to output '`Accept`' if $Q = P_C$ and '`REJECT`' if $\|Q - P_C\|_1 > \epsilon$ with high probability.

**Definition 5.1** (Weak-membership sampling test) *For any $n$ let $P$ be a target probability distribution over $\mathcal{E}$. An algorithm $\mathcal{T} : \mathcal{E}^s \longrightarrow \{0, 1\}$ is an $\epsilon$-*weak-membership test *for of a sampling device for $P$ if the following completeness and soundness conditions are satisfied on input of $s$ samples distributed according to any distribution $Q$ over $\mathcal{E}$.*

$$\|Q - P\|_{TV} \leq \epsilon - \alpha \ \Rightarrow \ \Pr_{\mathcal{S} \sim Q^s} [\mathcal{T}(\mathcal{S}) = 1] \geq \frac{2}{3}, \tag{5.1}$$

$$\|Q - P\|_{TV} > \epsilon \ \Rightarrow \ \Pr_{\mathcal{S} \sim Q^s} [\mathcal{T}(\mathcal{S}) = 1] \leq \frac{1}{3}. \tag{5.2}$$

*for some gap $\alpha > 0$. For a family $\{P_n\}_{n \in \mathbb{N}}$ of probability distributions we call a family of tests $\{\mathcal{T}_n\}$* sample-efficient *if, for every $n$, $\mathcal{T}_n$ is an $\epsilon$ weak-membership test from $s \in O(\mathrm{poly}(n, 1/\epsilon))$ samples.*

Notice that the gap $\alpha > 0$ between the regions of acceptance and rejection in state space ensures that such a test is computationally well defined in that it requires only finite precision. If there was no such gap, issues of infinite precision might arise on the boundary at which $\|Q - P\|_{TV} = \epsilon$. Weak-membership verification is thus the most demanding kind of verification: it requires that *all distributions* that satisfy the total-variation distance bound up to precision $\alpha$ are accepted, while all distributions that do not satisfy the bound are rejected. Except for a small region of width $\alpha$ the test thus explores the entire state space.

What are the minimal requirements that we need to impose on a meaningful verification test for sampling devices that uses only the classical output data? For once, it must not be the case that wrong distributions are accepted. So whenever $\|Q - P\|_{TV} > \epsilon$ the test should reject with high probability. Second, we must ensure that the test does not simply reject all distributions. The minimal condition on this is clearly that if $Q = P$ the test should accept with high probability. On all other distributions, that is, distributions for which $0 < \|Q - P\|_{TV} \leq \epsilon$, the test is allowed to perform arbitrarily. Of course, one might still hope that it actually accepts a small region around the ideal target distribution. We formalize these considerations in the notion of a *minimal verification test*; see Fig. 5.2.

**Definition 5.2** (Minimial verification test) *Fix a problem size $n$. A minimal verification test $\mathcal{T} : \mathcal{E}^s \longrightarrow \{0, 1\}$ for a sampling device with target distribution*

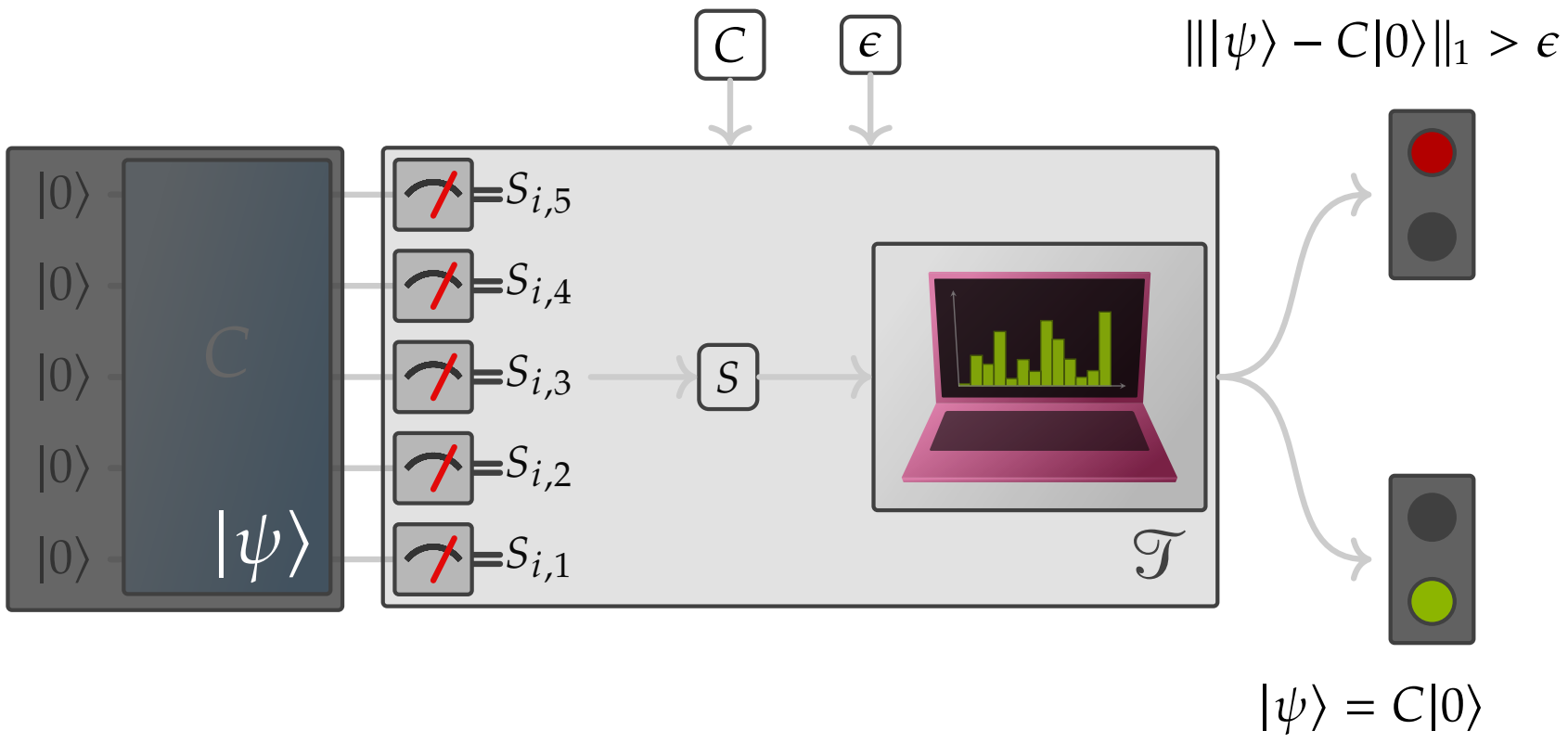

**Figure 5.3:** In some settings, for instance, in a well-controlled laboratory environment one might assign a high level of trust to parts of the quantum device such as the quantum measurement apparatus. In this case, we consider the input to a verification test to be the quantum state $|\psi\rangle$ produced by the quantum device. As measurements are trusted, the verifier $\mathcal{T}$ can then perform different measurements on $|\psi\rangle$ to verify that $|\psi\rangle$ is indeed $\epsilon$-close to the ideal target state $C|0\rangle$ in trace norm $\|\cdot\|_1$.

*$P$ using $s$ samples distributed according to $Q$ as an input must satisfy the completeness and soundness conditions*

$$Q = P \;\Rightarrow\; \Pr_{\mathcal{S}\sim Q^s}[\mathcal{T}(\mathcal{S}) = 1] \geq \frac{2}{3}, \tag{5.3}$$

$$\|Q - P\|_{\mathrm{TV}} > \epsilon \;\Rightarrow\; \Pr_{\mathcal{S}\sim Q^s}[\mathcal{T}(\mathcal{S}) = 1] \leq \frac{1}{3}. \tag{5.4}$$

### Measurement-device-dependent quantum certification

The device-independent setting may be much too paranoid from the perspective of an experimenter, who has spent much time calibrating and characterizing the correct functioning of the individual components of their experimental setup. And, indeed, device-indepence is a standard that is seldomly reached and rarely required for the confirmation of physical theories – the famous exception being loophole-free Bell tests [Hen+15; Sha+15]. From the perspective of a physicist who aims at confirming or falsifying a physical theory, we might therefore also be less demanding than in the device-independent scenario.

An experimentally well motivated setting is one in which certain measurements are characterized and calibrated to a high precision. In such a 'measurement-device-dependent' scenario, certain measurements are assumed to function with high fidelity. A particularly natural choice of trusted measuremnts in many settings such as quantum optics, superconducting qubits or trapped ions are single-qubit measurements in an arbitrary basis; see Fig. 5.2. Performing those measurements may require the application of single-qubit gates before the platform-specific natural measurrement can be carried out. Via single-qubit measurements it is possible to measure, for example, arbitrary Pauli correlators – arbitrary products of Pauli operators $1, X, Y, Z$ – and thus perform full quantum state tomography.

But if the quantum measurement apparata required for a quantum random sampling experiment are trusted to work with high fidelity, in a verification protocol we need only verify the quantum state on which those measurements are performed. And, in fact, the measurement-device-dependent scenario is a natural setting in many situations [see e.g. Lan+17]. In this setting, we can phrase a weak-membership certification test as follows.

**Definition 5.3** (Weak-membership quantum state certification) *For any $n$ let $|\psi\rangle$ be a target quantum state in a Hilbert space $\mathcal{H}$. An algorithm $\mathcal{T} : \mathcal{H}^{\otimes s} \longrightarrow \{0, 1\}$ is an $\epsilon$-*weak-membership quantum state certification* protocol for $|\psi\rangle$ if the following completeness and soundness conditions are satisfied, given $s$ iid. copies of an imperfect state preparation $\rho$*

$$\| |\psi\rangle\langle\psi| - \rho \|_1 \leq \epsilon - \alpha \;\Rightarrow\; \Pr[\mathcal{T}(\rho^{\otimes s}) = 1] \geq \frac{2}{3}, \tag{5.5}$$

$$\| |\psi\rangle\langle\psi| - \rho \|_1 > \epsilon \;\Rightarrow\; \Pr[\mathcal{T}(\rho^{\otimes s}) = 1] \leq \frac{1}{3}. \tag{5.6}$$

*for some gap $\alpha > 0$, where the probability runs over the internal randomness of the algorithm $\mathcal{T}$.*

In the following chapters, we will investigate the possibility of certification in the device-independent and the measurement-device-dependent scenario. We will see that fully device-independent verification of quantum random sampling in the supremacy regime is infeasible as it requires exponentially many samples from the quantum device (Chapter 6). We will discuss ways to circumvent this no-go result in Chapter 7. We will review settings in which certification from classical samples alone is possible from polynomially many samples if one is willing to make certain assumptions on the quantum state. Finally, we will show that rigorous verification is even *efficiently possible* in the measurement-device-dependent scenario. Surprisingly, it turns out that for the architecture developed in Section 4.2, the resource cost of measurement-device-dependent verification are asymptotically constant.

# Sample complexity of device-independently verifying quantum samplers

# 6

## 6.1 Introduction

In the previous chapter, we have established a framework for assessing different approaches to verification on the grounds of figures of merit that are relevant to applications of those approaches. We also identified settings within which the question of verification is meaningfully phrased – the device-independent and the measurement-device-dependent setting. In this chapter, we will return to the specific task of verifying quantum random sampling schemes designed to demonstrate quantum supremacy as they were introduced in Part I. We will begin by assessing the possibility of device-independent verification.

Above, we already argued that when aiming to demonstrate quantum supremacy we must rule out any alternative explanations of a possible violation of the complexity-theoretic Church-Turing thesis. The consequences of such a violation would be far reaching not only for theoretical computer science. It would certainly also boost the resources spent on developing practically useful algorithms. As such, a sceptic would not be willing to discard the complexity-theoretic Church-Turing thesis, which has been faring well for nearly a century, if there was room for doubt in a supremacy demonstration after all. Like the EPR paradoxon and subsequent experimental verification of nonlocality [FC72; ADR82; AGR82] fundamentally changed the way we think about the interactions between the (local) constituents of our world, the concept of a quantum computer and the eventual demonstration of quantum supremacy [Aru+19] overthrows the way we think about efficient computability using physical resources. But to overcome the inertia of established, well-confirmed and successful hypotheses about the nature of nature requires closing even the slightest possibility of doubt. Only recently – more than 30 years after the initial violation – has this been achieved for Bell inequality violations [Hen+15; Sha+15]. For demonstrations of quantum supremacy this challenge is outstanding both in terms of the complexity-theoretic loopholes, which we discussed in detail in Chapter 3, and the issue of fully convincing verification [SB09; Gog+13; AA14; Han+17; Boi+18; Bou+19; Han+19; Aru+19].

It may even be argued on economic grounds: if one is to invest large amounts of capital in the development of quantum computing devices, it would better be possible to actually build such devices. While *in theory* we have strong reasons to believe that quantum computing devices may outperform classical devices the risk may well be high enough to thwart spending the required amounts of money for their development. A convincing demonstration of quantum supremacy constitutes the first step to establishing their *physical possibility* and thus provides much stronger empirical support, reducing the risk of investment in the development of quantum computing devices.

*The remainder of this chapter is based on joint work with Martin Kliesch, Jens Eisert and Christian Gogolin and has already been published in Ref. [Han+19].*

The arguably most elegant and most convincing certification would be one based on purely classical data, ideally only the samples produced by the device and a description of the target distribution. Such certification would be free of additional complexity-theoretic assumptions and device-independent, in that it would be agnostic to all implementation details of the device and would directly certify that the classically defined sampling problem was solved.

In this chapter, we assess the possibility of such device-independent verification. We rigorously prove for a broad range of sampling problems, specifically for boson sampling [AA13], universal random circuit sampling [Boi+18; Bou+19], IQP circuit sampling [SB09; BMS16], and sampling from postselected-universal 2-designs [Han+18; NKM14; Nak+17; YJS19; BFK18] that they cannot be efficiently certified from classical samples and a description of the target probability distribution. Ironically, it turns out that the same property of a distribution that allows to prove the known approximate-hardness results also forbids their non-interactive sample-efficient device independent certification, to the effect that with the known proof methods both properties cannot be achieved simultaneously in such schemes.

More specifically, we directly bound the sample complexity of *minimal verification* (Def. 5.2 and Figure 5.2). This amounts to providing an answer to the question: How many samples from an unknown distribution are required to guarantee that this distribution is either identical to the target distribution or at least some preset distance away from it?

This notion of verification is *device-independent* in the sense that it does not assume anything about the internal working of the sampler (not even whether it is quantum or classical), but uses only the classical samples it outputs and a classical description of the target distribution. Among such device-independent certification scenarios, our scenario is the most general one in the sense that the certifier is given *all* the information contained in the target distribution. In particular, it is crucial that we explicitly allow the certification test $\mathcal{T}$ to depend on all details of the target distribution $P$.

As we are not concerned with the computational complexity of the test, but only its sample complexity, we allow the certification algorithm *unlimited* computational power. In particular, it does not matter how exactly $\mathcal{T}$ is given access to a description of $P$, but for the sake of concreteness $\mathcal{T}$ can be thought of as having access to an oracle that provides the probabilities $P(S)$ of all $S \in \mathcal{E}$ up to arbitrary precision. Sample-efficiency is clearly a necessary requirement for computational efficiency of a test, as any test takes at least the time it needs to read in the required number of samples, so that lower bounds on the sample complexity are stronger than such on the computational complexity. This means that our results cannot be circumvented by increasing the classical computational power of the certifier[1].

1: This makes our results conceptually different from the observation of Brandão. This observation is based on a result by Trevisan, Tulsiani, and Vadhan [TTV09], was reported by Aaronson and Arkhipov [AA14] and shows the following: For most unitaries $U$ drawn from the Haar measure, and any fixed circuit size $T$, there exists a classical "cheating" circuit of size polynomially larger than $T$, whose output distribution can not be distinguished from the corresponding boson sampling distribution by any "distinguisher" circuit of size $T$.

The notion of minimal certification corresponds to what in the literature on property testing is called *identity testing* with a fixed target distribution [Gol17]. Identity testing or minimal verification is an easier task than its robust (weak-membership) version (Def. 5.1) in which the certifier is moreover required to accept a constant-size region around the target distribution [VV10; Aol+15]. At the same time, it is much harder than mere *state-discrimination*, where the task is to differentiate between two fixed distributions [Gog+13; AA14]. Minimal certification in the sense of Definition 5.2 is more demanding in the sense that $P$ has to be distinguished from *all* distributions $Q$ such that $\|P - Q\|_{\ell_1} \geq \epsilon$. It is precisely this type of certification that is necessary to convince a sceptic of quantum supremacy via, say boson sampling, as the hardness results on approximate boson sampling only cover distributions within a small ball in $\ell_1$-norm around the ideal target distribution. A device sampling from a distribution further away from the ideal distribution, might still be doing something classically intractable, but this cannot be concluded from the hardness of approximate boson sampling.

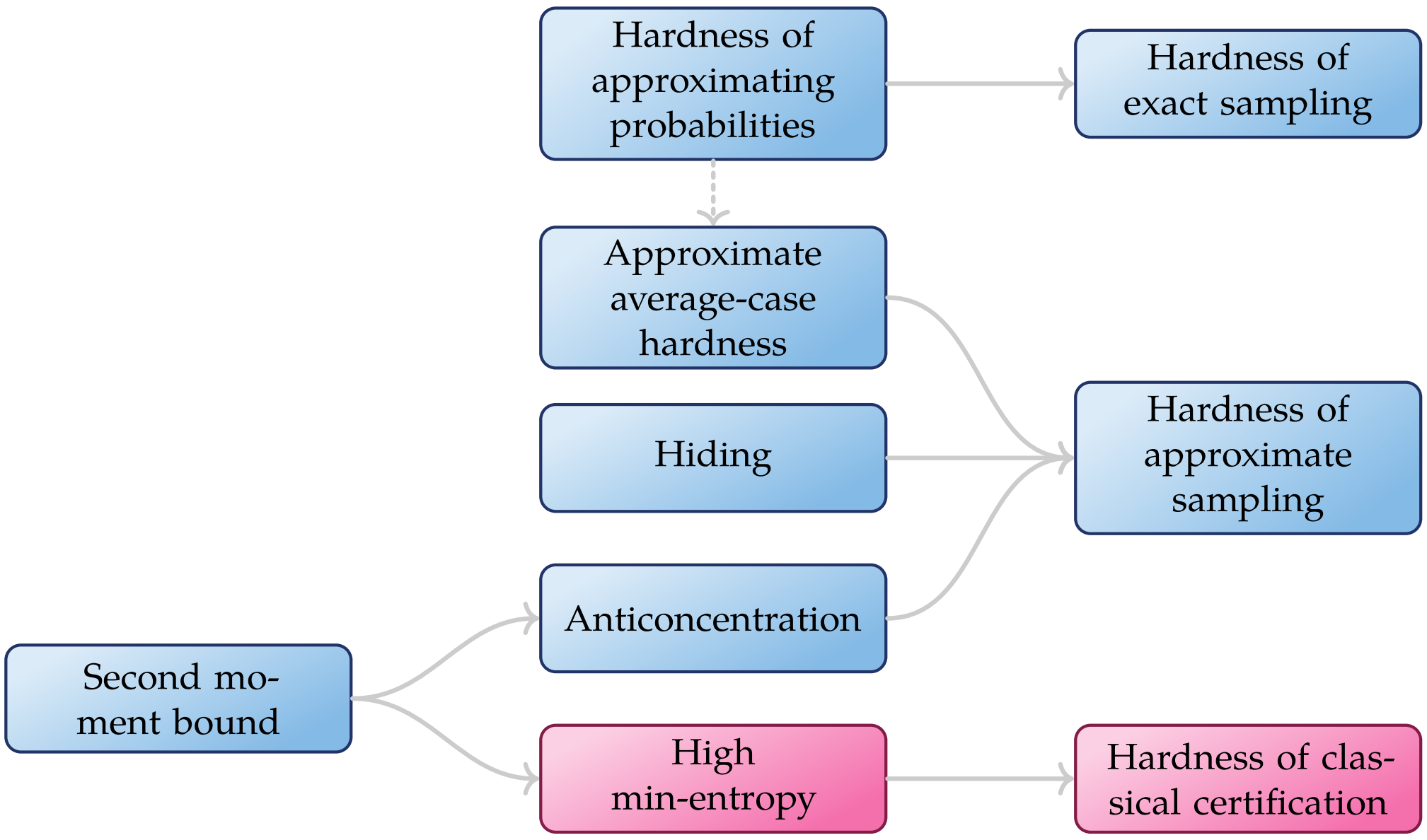

**Figure 6.1:** In this chapter, we supplement the high level overview (Fig. 2.5) of the approximate sampling quantum supremacy proofs of Refs. [AA13; BMS16; Boi+18; MSM17; GWD17; Ber+18; Han+18] using Stockmeyer's algorithm [Sto83; Sto85] with results on minimal verifiability. We show that the same property that is essential to arrive at a hardness result for approximate sampling via anticoncentration also makes it hard to certify from classical samples and a complete description of the target distribution, even with unbounded *computational power*.

Our result can be seen as an extension of lower bounds on the sample complexity of restricted state-discrimination scenarios [Gog+13], which prompted the development of schemes [AA14] that allow to corroborate and build trust in experiments [Spa+14; Car+14; Wal+16]. This helped spark interest in the problem of device-independent certification – on which there had not been much progress since [SB09]. In contrast to the restricted state-discrimination scenarios considered by Gogolin et al. [Gog+13], here, the certifier is given a full description of the target distribution[2] and unlimited computational power.

2: In particular, the certifier is given the value of all target probabilities to arbitrary precision.

We prove the impossibility of efficient classical verification of quantum random sampling schemes by making use of a key property for the proof of hardness of approximate sampling, namely an upper bound on the second moments of the output probabilities with respect to the choice of a random unitary specifying the instance of the sampling problem. Recall from Chapter 3 that the bound on the second moments implies that the probabilities are concentrated around the uniform distribution and hence an anticoncentration property. Recall from Chapter 2 that anticoncentration allows lifting results on the hardness of approximate sampling up to relative errors to ones for additive errors – provided relative-error approximation of the output probabilities is hard *on average*. It is thus a key property to prove hardness in the physically relevant case of approximate sampling that prevents a purely classical non-interactive certification of the output distribution; see Figure 6.1.

## Proof sketch

Before we delve into the technical details of our result, let us here sketch the main idea of its proof. A central ingredient to our proof is a recent result by Valiant and Valiant [VV17] specifying the optimal sample complexity of certifying a known target distribution $P$. It can be stated as follows. Fix a preset distance $\epsilon > 0$ up to which we want to certify. Now, suppose we receive samples from a device that samples from an unknown probability distribution $Q$. Then – for some constants $c_1, c_2$ – it requires at least

$$c_1 \cdot \max\left\{\frac{1}{\epsilon}, \frac{1}{\epsilon^2}\|P^{-\max}_{-2\epsilon}\|_{\ell_{2/3}}\right\} \tag{6.1}$$

and at most

$$c_2 \cdot \max\left\{\frac{1}{\epsilon}, \frac{1}{\epsilon^2}\|P^{-\max}_{-\epsilon/16}\|_{\ell_{2/3}}\right\} \tag{6.2}$$

many samples to distinguish the case $P = Q$ from the case $\|P - Q\|_{\ell_1} \geq \epsilon$ with high probability. Here $\|\cdot\|_{\ell_1}$ denotes the $\ell_1$-norm reflecting the total-variation distance (up to a factor of 2). The central quantity determining the sample complexity of certification is thus the quasi-norm $\|P^{-\max}_{-\epsilon}\|_{\ell_{2/3}}$ which is defined as follows. First, find the truncated distribution $P^{-\max}_{-\epsilon}$ by removing the tail of the target distribution $P$ with weight at most $\epsilon$ as well as its largest entry, see Figure 6.2. Then, take the $\ell_{2/3}$-norm as given by $\|x\|_{\ell_{2/3}} = (\sum_i |x_i|^{2/3})^{3/2}$ for a vector $x$ with entries $x_i$.

We now proceed in two steps. First, we show lower and upper bounds on the quantity $\|P^{-\max}_{-\epsilon}\|_{\ell_{2/3}}$ in terms of the largest probability $p_0$ occurring in $P$ and its support $\|P^{-\max}_{-\epsilon}\|_{\ell_0}$ as given by

$$\begin{aligned} p_0^{-\frac{1}{2}}\left(1 - \epsilon - p_0\right)^{3/2} &\leq \|P^{-\max}_{-\epsilon}\|_{\ell_{2/3}} \\ &\leq \left(1 - p_0\right)\|P^{-\max}_{-\epsilon}\|^{\frac{1}{2}}_{\ell_0}. \end{aligned} \tag{6.3}$$

Then it follows from Eqs. (6.1) and (6.3) that the sample complexity of minimally verifying a distribution $P$ up to a constant total-variation distance $\epsilon$ is essentially lower bounded by $1/\sqrt{p_0}$. Hence, if $P$ is exponentially flat in the sense that the largest probability is exponentially small in the problem size (here, the number of particles), $\epsilon$-certification requires exponentially many samples. Conversely, if $P^{-\max}_{-\epsilon/16}$ is supported on polynomially many outcomes only, sample-efficient certification is possible by the converse bound (7.3).

Second, we connect this result to the output distributions of quantum supremacy schemes. Specifically, we prove that with high probability over the choice of the random unitary in a scheme, the distribution over outputs associated with this unitary is exponentially flat.

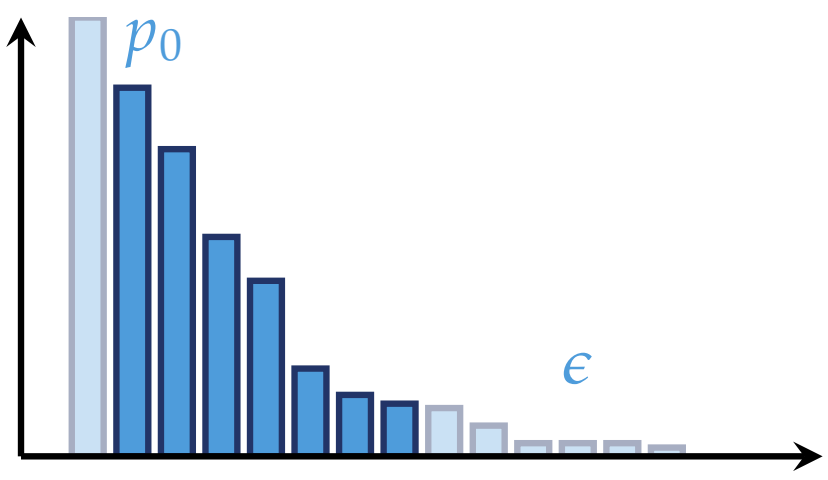

**Figure 6.2:** The vector $P_{-\epsilon}^{-\max}$ is obtained from $P$ by removing the largest element $p_0$ of $P$ as well as the smallest probabilities that accumulate to a total weight bounded by $\epsilon$.

Putting everything together we obtain lower bounds on the sample complexity of certification for boson sampling, IQP circuit sampling and random universal circuit sampling with (sufficiently many) $n$ particles. In all of these cases, the sample complexity scales at least as fast as

$$\frac{1}{\epsilon^2}(2^n\delta)^{1/4}, \tag{6.4}$$

with probability at least $1-\delta$ over the random choice of the unitary.

### Setup and definitions

Before we proceed with the technical part of this chapter let us set some more notation. We use the Landau symbols $O$ and $\Omega$ for asymptotic upper and lower bounds and $\Theta$ for their conjunction. We will make frequent use of the $\ell_p$ norms (defined in Eq. 2.78 in Chapter 2) and the $\alpha$-Rényi entropies, which for any probability vector $P=(p_1,\dots,p_n)$, $p_i \geq 0$, $\sum_i p_i = 1$ and $0 \leq \alpha \leq \infty$, $\alpha \neq 1$ are defined to be

$$H_\alpha(P) := \frac{\alpha}{1-\alpha}\log\|P\|_{\ell_\alpha}. \tag{6.5}$$

We refer to $H_\infty(P) = -\log\max_{i\in[n]} p_i$ as the *min-entropy* of $P$.

## 6.2 No certification of flat distributions

This section is concerned with the question of whether distributions with a high *min-entropy* can be certified in a sample-efficient way. The main insights into this question come from a work by Valiant and Valiant [VV17] on property testing, which gives a *sample-optimal* certification test (up to constant factors) for any fixed distribution $P$, as well as a lower bound on the sample complexity of certification. The result is stated in terms of an $\ell_{2/3}$-norm of a vector obtained from the distribution. Our main technical contribution is to find bounds on these quasi-norms that are relevant in the context of certifying quantum supremacy distributions.

To state the main result of Ref. [VV17], we adapt their following notation and illustrate it in Figure 6.2. For any vector of non-negative numbers $P$,

i. let $P^{-\max}$ be the vector obtained from $P$ by setting the largest entry to zero, and
ii. let $P_{-\epsilon}$ be the vector obtained from $P$ by iteratively setting the smallest entries to zero, while the sum of the removed entries remains upper bounded by $\epsilon > 0$.

It turns out that the optimal sample complexity for $\epsilon$-certifying any distribution $P$ is essentially given by $\frac{1}{\epsilon^2}\|P_{-\epsilon}^{-\max}\|_{\ell_{2/3}}$. The intuition is that any $\epsilon$ deviation from $P$ that is contained in either the largest probability or the tail of the distribution is easily detected. Intuitively, this is because a constant deviation in these parts of the distribution will be visible in the samples obtained with high probability [VV17]. More precisely, the following upper and lower bounds on the sample complexity of certification hold:

**Theorem 6.1** (Optimal certification tests [VV17]) *There exist constants* $c_1, c_2 > 0$ *such that for any* $\epsilon > 0$ *and any target distribution* $P$*, there exists an* $\epsilon$*-certification test from* $c_1 \max\{\frac{1}{\epsilon}, \frac{1}{\epsilon^2}\|P_{-\epsilon/16}^{-\max}\|_{\ell_{2/3}}\}$ *many samples, but there exists no* $\epsilon$*-certification test from fewer than* $c_2 \max\{\frac{1}{\epsilon}, \frac{1}{\epsilon^2}\|P_{-2\epsilon}^{-\max}\|_{\ell_{2/3}}\}$ *samples.*

We note that $\|P_{-\epsilon}^{-\max}\|_{\ell_{2/3}} \leq \|P\|_{\ell_{2/3}}$ for any $P$, and in many cases the former is only a constant factor away from the latter. We obtain the following general bounds on $\|P_{-\epsilon}^{-\max}\|_{\ell_{2/3}}$ in terms of the min-entropy and support of $P$.

**Lemma 6.2** (Bounds on $\|P_{-\epsilon}^{-\max}\|_{\ell_{2/3}}$)

$$\begin{aligned} 2^{\frac{1}{2}H_\infty(P)}\left(1-\epsilon-2^{-H_\infty(P)}\right)^{3/2} &\leq \|P_{-\epsilon}^{-\max}\|_{\ell_{2/3}} \\ &\leq \left(1-2^{-H_\infty(P)}\right)\|P_{-\epsilon}^{-\max}\|_{\ell_0}^{\frac{1}{2}}. \end{aligned} \tag{6.6}$$

To get a feeling for what these bounds imply, let us consider two special cases and sufficiently small $\epsilon$. If for some constant $\kappa$ it holds that $H_\infty(P) = \log(\kappa\,|\mathcal{E}|)$, they imply the following lower bound on the required minimal number of samples, $s_{\min}$:

$$s_{\min}^2 \geq c_2^2 \kappa \frac{|\mathcal{E}|}{\epsilon^4}\left(1-2\epsilon-\frac{1}{\kappa\,|\mathcal{E}|}\right)^3. \tag{6.7}$$

For all distributions whose min-entropy is essentially given by the logarithm of the size $|\mathcal{E}|$ of the sample space, the sample complexity for certification thus scales at least as the square root of that size. If, on the contrary, $P_{-\epsilon/16}$ has support on at most $s \geq \|P_{-\epsilon/16}\|_{\ell_0}$ many probabilities we have the following upper bound

$$s_{\text{suf}} \leq c_1 \frac{1-\frac{\epsilon}{16}}{\epsilon^2}\sqrt{s} \tag{6.8}$$

on the number of samples $s_{\text{suf}}$ that is sufficient for $\epsilon$-certification. This bound implies that distributions supported only on polynomially many outcomes can be certified from polynomially many samples.

*Proof of Lemma 6.2.* For the lower bound, we use that concavity of the function $x \mapsto x^{2/3}$ implies that for any fixed $x^* > 0$ and any $0 \leq x \leq x^*$ we have

$$x^{2/3} \geq \frac{x^{*2/3}}{x^*}x = x^{*-1/3}x \tag{6.9}$$

and thus for any (not necessarily normalized) $\tilde{P} := (\tilde{p}_1, \ldots, \tilde{p}_{\tilde{n}})$ with

$\tilde{p}_i \geq 0$

$$\|\tilde{P}\|_{\ell_{2/3}}^{2/3} = \sum_{i=1}^{\tilde{n}} \tilde{p}_i^{2/3} \geq \sum_{i=1}^{\tilde{n}} (\|\tilde{P}\|_{\ell_\infty}^{-1/3} \tilde{p}_i) \tag{6.10}$$

$$= \|\tilde{P}\|_{\ell_\infty}^{-1/3} \|\tilde{P}\|_{\ell_1}. \tag{6.11}$$

Using this for $\tilde{P} = P_{-\epsilon}^{-\max}$ and that both $\|P_{-\epsilon}^{-\max}\|_{\ell_\infty} \leq \|P\|_{\ell_\infty}$ and $\|P_{-\epsilon}^{-\max}\|_{\ell_1} \geq 1 - \epsilon - \|P\|_{\ell_\infty}$ finally implies the lower bound.

For the upper bound, we use that for any vector $v$ and $0 < p < q \leq \infty$ (see, e.g., Ref. [FR13, Eq. (A.3)])

$$\|v\|_{\ell_p} \leq s^{\frac{1}{p} - \frac{1}{q}} \|v\|_{\ell_q}, \tag{6.12}$$

where $s \geq \|v\|_{\ell_0}$. Inserting $p = 2/3$ and $q = 1$, one obtains for $v = P_{-\epsilon}^{-\max}$

$$\|P_{-\epsilon}^{-\max}\|_{\ell_{2/3}} \leq \|P_{-\epsilon}^{-\max}\|_{\ell_0}^{\frac{1}{2}} \|P_{-\epsilon}^{-\max}\|_{\ell_1} . \tag{6.13}$$

□

Valiant and Valiant's result [VV17] also has immediate consequences on the certifiability of postselected probability distributions, such as those arising in boson sampling [AA13]. A certification algorithm has to distinguish the target distribution $P$ from all probability distributions that are at least $\epsilon$-far away from $P$. That is true, in particular, for distributions that differ from $P$ by at least $\epsilon$ in $\ell_1$-norm only on some part $\mathcal{F}$ of the sample space, but are identical with $P$ on its complement $\mathcal{F}^c$. Intuitively one can expect that to distinguish such distributions, samples from $\mathcal{F}^c$ do not help. One might hence expect that it should be possible to lower bound the sample complexity of certifying the full distribution by the sample complexity of the postselected distribution on some subspace $\mathcal{F}$ of the sample space, at least as long as the postselection probability is not too low.

To make this intuition precise, define for any probability distribution $P$ and any subset $\mathcal{F} \subset \mathcal{E}$ the restriction $P_{\upharpoonright \mathcal{F}} := (p_i)_{i \in \mathcal{F}}$ of $P$ to $\mathcal{F}$ (no longer normalized), as well as the postselected probability distribution $P_{\mathcal{F}} := P_{\upharpoonright \mathcal{F}} / P(\mathcal{F})$, with postselection probability $P(\mathcal{F}) := \|P_{\upharpoonright \mathcal{F}}\|_{\ell_1}$.

**Lemma 6.3** (Lower bounds with postselected distributions) *Let $P$ be a probability distribution on $\mathcal{E}$. Then with $c_2$ the constant from Theorem 6.1 and for any $\epsilon > 0$ and $\mathcal{F} \subset \mathcal{E}$, there exists no $\epsilon$-certification test of $P$ from fewer than*

$$c_2 \cdot \max\left\{ \frac{1}{\epsilon}, \frac{1}{\epsilon^2} P(\mathcal{F}) \, \|(P_{\mathcal{F}})_{-2\epsilon/P(\mathcal{F})}^{-\max}\|_{\ell_{2/3}} \right\} \tag{6.14}$$

*many samples.*

*Proof of Lemma 6.3.* For any $\mathcal{F} \subset \mathcal{E}$ we have

$$\begin{aligned} \|P_{-\epsilon}^{-\max}\|_{\ell_{2/3}} &\geq \|(P_{-\epsilon}^{-\max})_{\restriction\mathcal{F}}\|_{\ell_{2/3}} && (6.15)\\ &\geq \|(P_{\restriction\mathcal{F}})_{-\epsilon}^{-\max}\|_{\ell_{2/3}} && (6.16)\\ &= P(\mathcal{F})\|(P_{\restriction\mathcal{F}})_{-\epsilon}^{-\max}/P(\mathcal{F})\|_{\ell_{2/3}} && (6.17)\\ &= P(\mathcal{F})\|(P_{\mathcal{F}})_{-\epsilon/P(\mathcal{F})}^{-\max}\|_{\ell_{2/3}}. && (6.18) \end{aligned}$$

Here, the first inequality becomes an equality in case $\mathcal{F}$ contains the support of $P_{-\epsilon}^{-\max}$. The second inequality becomes an equality whenever the smallest probabilities with weight not exceeding $\epsilon$ as well as the largest probability lie inside of $\mathcal{F}$. Finally, the last equality follows from the fact that when renormalizing $P_{\restriction\mathcal{F}}$ we also need to renormalize the subtracted total weight $\epsilon$ by the same factor. The claim then straightforwardly follows from Theorem 6.1. □

A non-trivial bound for the sample complexity is therefore achieved only in case the postselected subspace has at least weight $P(\mathcal{F}) > 2\epsilon$. This is due to the strength of Valiant and Valiant's result [VV17] in that a part of the distribution with total weight $2\epsilon$ does not influence the minimally required sample complexity of $\epsilon$-certification and this part might just be supported on $\mathcal{F}$.

## 6.3 Quantum supremacy distributions are flat

We will now apply the result of the previous section to the case of certifying quantum supremacy distributions. As a result, we find that prominent schemes aimed at demonstrating quantum supremacy, most importantly boson sampling and random universal circuit sampling, cannot be certified from polynomially many classical samples and a description of the target distribution alone. To this end we will make use of the anticoncentration property or, to be more precise, a bound on the second moments, which implies anticoncentration.

Recall from Chapter 2 that random circuit sampling schemes fit the following schema: Given the problem size $n$, start from a reference state vector $|0\rangle$ from a Hilbert space $\mathcal{H}_n$ and apply a unitary $U$ drawn with respect to some measure $\mu_n$ on the corresponding unitary group. The resulting state is then measured in the computational basis, thereby resulting in outcome $S \in \mathcal{E}_n = [\dim(\mathcal{H}_n)]$ with probability $P_U(S) := |\langle S|U|0\rangle|^2$. One then says that the distribution over $P_U$ induced by this procedure *anticoncentrates* if[3]

3: Notice that by the hiding property the following statement is equivalent to Def. 2.12.

$$\exists\, \alpha,\gamma > 0 : \ \forall n \in \mathbb{Z}_+ \text{ and } \forall S \in \mathcal{E}_n : \Pr_{U\sim\mu_n}\left(P_U(S) \geq \frac{\alpha}{|\mathcal{E}_n|}\right) \geq \gamma\,. \quad (6.19)$$

In words this roughly means: For any $n$ the induced distribution over $P_U$ has the property that for any fixed outcome $S$ it does not become too unlikely that the probability of getting that outcome is much smaller than it would be for the uniform distribution.

It is intuitive that due to normalization, anticoncentration also implies that not too much of the probability weight can be concentrated in few outcomes, hence the name. This is however slightly misleading, as anticoncentration does not in itself imply a that $P_U$ also needs to have a high min-entropy with high probability. To see this, take any $\alpha, \gamma(\alpha)$-anticoncentrating scheme of drawing probability distributions $P_U$. Construct $\tilde{P}_U$ from $P_U$ by dividing all probabilities in half and adding their joint weight to $P_U(0)$. The resulting scheme to construct $\tilde{P}_U$ is now still $\alpha/2, \gamma(\alpha)$-anticoncentrating, but has min-entropy $H_\infty(\tilde{P}_U) \leq \log(2)$ with probability one.

Recall from Chapter 3 that most known proofs of anticoncentration [BMS16; BMS17; AA14; Han+18] (with the notable exception of Morimae's hardness result[4] [Mor17]) rely on the Paley-Zygmund inequality

$$\Pr(Z > a\mathbb{E}[Z]) \geq (1-a)^2 \frac{\mathbb{E}[Z]^2}{\mathbb{E}[Z^2]}, \tag{6.20}$$

4: There, Morimae proves anticoncentration for the output distribution so-called one-clean qubit model (DQC1) in a direct manner.

for a random variable $Z \geq 0$ with finite variance and $0 \leq a \leq 1$. Anticoncentration is then proved by deriving a bound on the second moment of the distribution $\{P_U(S)\}_{U\sim\mu_n}$, and, as we will see in the next section, the same second moment bound that is used to derive anticoncentration implies a high min-entropy. We lay out the proof structure and the role that second moments play in Figure 6.1.

## Second moments bound the min-entropy

We now turn to showing that with high probability over the choice of $U$, the second moment of the distribution $\{P_U(S)\}_{U\sim\mu_n}$ (for fixed $S$) – implying the anticoncentration property (6.19) – yields a lower bound on the min-entropy of the output distribution $P_U$ of any fixed unitary $U$ with high probability. Lemma 6.2 then implies that distributions with exponentially (in $n$) small second moments cannot be certified from polynomially many samples with high probability. Thus, the *very property* that implies sampling hardness of a distribution $P_U$ up to an additive total-variation distance error also implies that the same distribution cannot be efficiently certified from classical samples only.

**Lemma 6.4** (Tail bound for the min-entropy) *For any $n \in \mathbb{Z}^+$, let $P_U$ be a distribution on $\mathscr{C}_n$ induced via $P_U(S) = |\langle S|U|0\rangle|^2$, $U \sim \mu_n$ by a corresponding measure $\mu_n$ on the unitary group. Then, with probability at least $1-\delta$ over the choice of $U \sim \mu_n$,*

$$H_\infty(P_U) \geq \frac{1}{2}\left(\log\delta - \log\sum_{S\in\mathscr{C}_n}\mathbb{E}_{U\sim\mu_n}[P_U(S)^2]\right). \tag{6.21}$$

The following arguments used to derive the lemma, in fact, hold for more general families of probability distributions $P_U$ where $U$ need not be unitary without any scaling in $n$.

*Proof of Lemma 6.4.* We proceed as follows: First, we prove a lower bound on the typical Rényi-2-entropy of $P_U$ using the second moment of $\{P_U\}_{U\sim\mu_n}$, and then use equivalence of the $\alpha$-Rényi entropies for $\alpha > 1$.

Analogously to Ref. [AA14, App. 11], we use Markov's inequality to obtain that with probability at least $1-\delta$ over the choice of $U$, we have

$$H_2(P_U) := -\log \sum_{S\in\mathscr{E}_n} P_U(S)^2 \tag{6.22}$$

$$\geq -\log\left(\frac{1}{\delta}\cdot \mathbb{E}_{U\sim\mu_n}\left[\sum_{S\in\mathscr{E}_n} P_U(S)^2\right]\right). \tag{6.23}$$

What is more, one can show that all $\alpha$-Rényi entropies for $\alpha > 1$ are essentially equivalent and in particular [Wil+19]

$$H_\alpha(P) \geq H_\infty(P) \geq \frac{\alpha-1}{\alpha} H_\alpha(P), \tag{6.24}$$

for any distribution $P$ on $\mathscr{E}_n$ from which the claim follows[5]. □

5: We show Eq. (6.24) as well as an alternative proof for Lemma 6.4 in Section D.1.

We note that, indeed, the notion of anticoncentration as formalized in Eq. (3.16) in itself does not necessarily imply that the output distribution of every (or most) fixed unitaries have high min-entropy. This is because anticoncentration merely requires that the tails of the distribution have sufficient (constant) weight, while allowing for few large probabilities. Nevertheless, in prominent cases, an anticoncentration result derives from bounds on the 2-Rényi entropy.

Let us now briefly recap the details on distributions arising from IQP circuits and universal random circuits (cf. Sec. 2.4) and in particular the second moment bounds which can be proven in both cases. We also prove lower bounds on the min-entropy of the boson sampling distribution. As the proof is more involved, however, we defer this deriviation Section 6.5. In Section 6.4, we then put all ingredients together to show that for all of these schemes with output distribution $P_U$ we have that $H_\infty(P_U) \in \Omega(n)$ and hence that the minimal sample complexity for certification scales exponentially in $n$.

## IQP circuits

An IQP circuit [SB09] is a quantum circuit of commuting gates that is drawn uniformly at random from the family $\mathcal{U}_{n,\mathrm{IQP}}$ on $n$ qubits. The sample space is therefore given by $\mathscr{E}_n = \{0,1\}^n$. This family as formulated by Bremner, Montanaro, and Shepherd [BMS16] is defined by a set of angles $A$, e.g., $A = \{0, \pi/8, \dots, 7\pi/8\}$. An instance $U_W \in \mathcal{U}_{n,\mathrm{IQP}}$ with $W := (w_{i,j})_{i,j=1,\dots,n}$ and $w_{i,j} \in A$ drawn uniformly at random, is then given by the following prescription

$$U_W = \exp\left[\mathrm{i}\left(\sum_{i<j} w_{i,j} X_i X_j + \sum_i w_{i,i} X_i\right)\right], \tag{6.25}$$

where $X_i$ is the Pauli-$X$ matrix acting on site $i$. In other words, on every edge $(i,j)$ of the complete graph on $n$ qubits a gate $\exp(\mathrm{i}w_{i,j}X_iX_j)$ with edge weight $w_{i,j}$ and on every vertex $i$ a gate $\exp(\mathrm{i}w_{i,i}X_i)$ with vertex weight $w_{i,i}$ is performed.

For the output distribution of IQP circuits, Bremner *et al.* [BMS16, Appendix F] prove the second-moment bound

$$\mathbb{E}_W[|\langle S|U_W|0\rangle|^4] \leq 3 \cdot 2^{-2n} . \tag{6.26}$$

By Lemma 6.4, this implies the following min-entropy bound

$$H_\infty(P_{U_W}) \geq \frac{1}{2}\left(n + \log\frac{\delta}{3}\right), \tag{6.27}$$

which holds with probability at least $1 - \delta$ over the choice of $U_W$.

### Universal random circuits and spherical 2-designs

A universal random circuit on $n$ qubits is defined by a universal gate set $\mathcal{G}$ comprising one- and two-qubit gates which give rise to the depth-$N$ family $\mathcal{U}_{\mathcal{G},N}$. A circuit $U \in \mathcal{U}_{\mathcal{G},N}$ is then constructed according to the standard prescription of choosing one- or two-qubit gates $G \in \mathcal{G}$ and the qubits they are applied to at random [BHH16], or according to some more specific prescription such as the one of Boixo et al. [Boi+18].

For the case of the random universal circuits of Ref. [Boi+18] there is evidence that the output distribution of fixed instances is essentially given by an exponential (Porter-Thomas) distribution $P_{\mathrm{PT}}$ whose second moment is given by [Han+18, Eq. (8)]

$$\mathbb{E}_{p\sim P_{\mathrm{PT}}}[p^2] = \frac{2}{|\mathcal{E}_n|(|\mathcal{E}_n|+1)} . \tag{6.28}$$

This is provably true for the local random universal circuits investigated by Brandão, Harrow, and Horodecki [BHH16] by the fact that the resulting circuit family forms a relative $\tilde{\varepsilon}$-approximate unitary 2-design $\mu$ in depth $O(n^2) + O(n \log 1/\tilde{\varepsilon})$ [BHH16] so that

$$\mathbb{E}_{U\sim\mu}[|\langle S|U|0\rangle|^4] \leq \frac{2(1+\tilde{\varepsilon})}{|\mathcal{E}_n|(|\mathcal{E}_n|+1)}. \tag{6.29}$$

Likewise, for any circuit family $\mathcal{U}_n$ on $n$ qubits such that $\{U|0\rangle\}_{U\sim\mathcal{U}_n}$ forms a relative $\tilde{\varepsilon}$-approximate spherical 2-design, the second moments are bounded as in Eq. (6.29). For all such circuit families, using Lemma 6.4, we thus obtain the min-entropy bound

$$H_\infty(P_U) \geq \frac{1}{2}\left(n + \log\frac{\delta}{2(1+\tilde{\varepsilon})}\right) , \tag{6.30}$$

which holds with probability at least $1 - \delta$ over the choice of $U$.

## 6.4 Quantum supremacy distributions cannot be efficiently certified

Let us now apply the above min-entropy bounds[6] to the most prominent examples of quantum supremacy schemes – boson sampling [AA13], IQP circuits [BMS16], and universal random circuits [Boi+18]. We will conclude from Lemmas 6.2–6.4 that these schemes cannot be efficiently

6: See Sec. 6.5 for the derivation of the min-entropy of the boson sampling distribution.

certified from polynomially many samples only. More precisely, we obtain the following lower bounds.

**Theorem 6.5** (Lower bounds on certifying random qubit schemes) *For* $0 < \epsilon < 1/2$ *and sufficiently large* $n$*, with probability at least* $1 - \delta$*, there exists no* $\epsilon$*-certification test from* $s < s_{\min}$ *many samples for*

*a. IQP circuit sampling on* $n$ *qubits, where*

$$s_{\min} \in \Omega\left(2^{n/4}\delta^{1/4}/\epsilon^2\right) . \tag{6.31}$$

*b.* $\tilde{\varepsilon}$*-approximate spherical 2-design sampling on* $n$ *qubits, and in particular, depth-*$(O(n^2) + O(n \log 1/\tilde{\varepsilon}))$ *local random universal circuits, where*

$$s_{\min} \in \Omega\left(\frac{2^{n/4}\delta^{1/4}}{\epsilon^2(1+\tilde{\varepsilon})^{1/4}}\right) . \tag{6.32}$$

The result of Theorem 6.5 applies to any circuit family $\mathcal{U}$ such that $\{U|0\rangle\}_{U\sim\mathcal{U}}$ forms a relative $\tilde{\varepsilon}$-approximate spherical 2-design, for which the second moments are upper bounded as in Eq. (6.29). This applies, in particular, to the random universal circuits of Refs. [BHH16; HL09; Boi+18; Bou+19] as well as other families of random circuits that have been proposed for the demonstration of quantum supremacy such as Clifford circuits with magic-state inputs [Han+18; YJS19], diagonal unitaries [Han+18; NKM14] and conjugated Clifford circuits [BFK18].

**Theorem 6.6** (Lower bounds on certifying boson sampling) *Let* $0 < \epsilon < 1/2$, $n \in \mathbb{Z}_+$ *sufficiently large and* $m \in \Theta(n^\nu)$*. Under the conditions on* $\nu$ *used in Ref. [AA13] to prove the hardness of approximate boson sampling, and with high probability over the random choice of the unitary, there exists no* $\epsilon$*-certification test of boson sampling with* $n$ *photons in* $m$ *modes from* $s < s_{\min}$ *many samples, where*

$$s_{\min} \in \Omega\left(2^n/\epsilon^2\right) . \tag{6.33}$$

In Section 6.5, we discuss in detail the conditions under which Aaronson and Arkhipov's hardness argument [AA13] holds and provide a full version of the theorem as Theorem 6.8. The key ingredient for this to be the case is the closeness of the measure obtained by taking $n \times n$-submatrices of Haar random unitaries $U \in U(m)$ and the Gaussian measure on $n \times n$-matrices. This is provably the case for $\nu > 5$, but is conjectured to hold even for $\nu > 2$ [AA13]. Our bound on $s_{\min}$ (see Theorem 6.8) holds with exponentially high probability (in $n$) for $\nu > 3$. In the case $\nu > 2$ our result holds only with polynomially high probability and fails to cover a small set of the instances. The argument proving Theorem 6.6 easily extends to certain variants of *quantum Fourier sampling* [FU15], the output probabilities of which are also given by permanents of nearly Gaussian matrices.

*Proofs of Theorems 6.6 and 6.5.* We use Theorem 6.1 and Lemmas 6.2–6.4 as well as the lower bounds (6.49), (6.27), and (6.30) on the min-entropy of the respective output distributions as given in the following sections. What is more, we use that for $0 < \epsilon < 1/2$ and sufficiently large $n$ the term $(1 - 2\epsilon - 2^{-H_\infty(P_U)})^{3/2}$ can be lower-bounded by a constant and, hence, be dropped inside the $\Omega$. □

## 6.5 Lower bounds for boson sampling

Let us now provide the missing details for the boson sampling problem.

In the boson sampling problem $n \geq 1$ photons are injected into the first $n$ of $m \in \mathsf{poly}(n)$ modes which are transformed in a linear-optical network via a mode transformation given by a Haar-random unitary $U \in U(m)$ and then measured in the Fock basis. The sample space of boson sampling is given by

$$\mathscr{E}_n := \Phi_{m,n} := \left\{(s_1, \dots, s_m) : \sum_{j=1}^{m} s_j = n\right\}, \tag{6.34}$$

i.e., the set of all sequences of non-negative integers of length $m$ which sum to $n$. Its output distribution $P_{\mathrm{bs},U}$ is

$$P_{\mathrm{bs},U}(S) := |\langle S|\varphi(U)|1_n\rangle|^2. \tag{6.35}$$

Here, the state vector $|S\rangle$ is the Fock space vector corresponding to a measurement outcome $S \in \Phi_{m,n}$, $|1_n\rangle$ is the initial state vector with $1_n := (1, \dots, 1, 0, \dots, 0)$, and $\varphi(U)$ the Fock space (metaplectic) representation of the implemented mode transformation $U$.

The distribution $P_{\mathrm{bs},U}$ can be expressed [Sch08; AA13] as

$$P_{\mathrm{bs},U}(S) = \frac{|\operatorname{Perm}(U_S)|^2}{\prod_{j=1}^{m}(s_j!)}, \tag{6.36}$$

in terms of the permanent of the matrix $U_S \in \mathbb{C}^{n\times n}$ constructed from $U$ by discarding all but the first $n$ columns of $U$ and then, for all $j \in [m]$, taking $s_j$ copies of the $j^{\text{th}}$ row of that matrix (deviating from Aaronson and Arkhipov's notation [AA13]). Here, the permanent for a matrix $X = (x_{j,k}) \in \mathbb{C}^{n\times n}$ is defined similarly to the determinant but without the negative signs as

$$\operatorname{Perm}(X) := \sum_{\tau\in\mathrm{Sym}([n])} \prod_{j=1}^{n} x_{j,\tau(j)}, \tag{6.37}$$

where Sym($[n]$) is the symmetric group acting on $[n]$. It is a known fact that calculating the permanent of a matrix to high precision is a problem that is #P-hard [Val79], while its close cousin, the determinant, is computable in polynomial time. In fact, computing the permanent exactly (or with exponential precision) is also #P-hard *on average* for randomly chosen Gaussian matrices [Lip91; AA13]. In Ref. [AA13] this connection is exploited to show that, up to plausible complexity-theoretic conjectures, approximately sampling from the boson sampling distribution is classically intractable with high probability over the choice of $U$ if $m$ is scaled appropriately with $n$.

The main part of the hardness proof of Ref. [AA13] is to prove the classical hardness of sampling from the *postselected boson sampling distribution* $P^*_{\mathrm{bs},U}$. The postselected distribution $P^*_{\mathrm{bs},U}$ is obtained from $P_{\mathrm{bs},U}$ by discarding all output sequences $S$ with more than one boson per mode, i.e., all $S$ which are not in the set of *collision-free* sequences

$$\Phi^*_{m,n} := \Big\{ S \in \Phi_{m,n} : \forall s \in S : s \in \{0,1\} \Big\}. \tag{6.38}$$

The hardness of sampling from the full boson sampling distribution follows from the fact that for the relevant scalings of $m$ with $n$ the postselection can be done efficiently in the sense that on average at least a constant fraction of the outcome sequences is collision-free (Theorem 13.4 in Ref. [AA13]).

More precisely, the actual result proved in Ref. [AA13, Theorem 1.3] states that unless certain complexity-theoretic conjectures fail, there exists no classical algorithm that can sample from a distribution $Q$ satisfying $\|Q - P_{\mathrm{bs},U}\|_{\ell_1} \leq \epsilon$ in time $\mathsf{poly}(n, 1/\epsilon)$. This result requires that $m \in \Omega(n^5 \log(n)^2)$, but it is conjectured that $m$ growing slightly faster than $\Omega(n^2)$ is sufficient for hardness. In fact, at the same time, a faster than quadratic scaling is necessary for the proof strategy to work.

The key technical ingredient in the proof strategy underlying these requirements is the following result: if $m$ grows sufficiently fast with $n$, the measure induced on $U \sim \mu_H$ by the map $g_S = (U \mapsto U_S)$ for collision-free $S \in \Phi^*_{m,n}$, i.e., the measure induced by taking $n \times n$-submatrices of unitaries $U \in U(m)$ chosen with respect to the Haar measure $\mu_H$ is close to the complex Gaussian measure $\mu_G(\sigma)$ with mean zero and standard deviation $\sigma = 1/\sqrt{m}$ on $n \times n$-matrices. Given this result, Stockmeyer's algorithm could be applied to the samples obtained from $P^*_{\mathrm{bs},U}$ in order to infer the probabilities $P^*_{\mathrm{bs},U}(S)$ and thus solve a #P-hard problem, as these probabilities can be expressed as the permanent of a Gaussian matrix. Since the closeness of those measures is the essential ingredient, also suitably large scaling of $m$ with $n$ is crucial for the hardness argument.

The formal statement of closeness of measures proved in Ref. [AA13] implies the following:

**Lemma 6.7** (implied by Ref. [AA13, Theorem 5.2]) *There exists a constant $C > 0$ such that for every $\nu > 5$ and every measurable $f : \mathbb{C}^{n \times n} \longrightarrow [0,1]$ and every $m \in \Omega(m^\nu)$ it holds that for all $S \in \Phi^*_{m,n}$*

$$\mathbb{E}_{U \sim \mu_H} f(U_S) \leq (1 + C)\, \mathbb{E}_{X \sim \mu_{G(1/\sqrt{m})}} f(X). \tag{6.39}$$

At the same time, it is known from Ref. [Jia06] (see also Ref. [AA13, Section 5.1 and 6.2]) that if $m \leq c\, n^\nu$ with $\nu \leq 2$ and $c \in O(1)$ the two measures $\mu_H \circ g_S^{-1}$ and $\mu_{G(1/\sqrt{m})}$ are no longer close for large $n$. One may hope [AA13] that there exists a constant $c > 0$ such that Theorem 5.2 in Ref. [AA13] and hence their hardness result as well as our Lemma 6.7 hold for any $m \geq c\, n^\nu$ with $\nu > 2$. What we show is that *even under this optimistic assumption* efficient certification from classical samples is impossible, if the postselection probability is large enough. This rules out many further cases for which one can hope to prove a hardness result by the same method.

**Theorem 6.8** (Lower bounds on certifying boson sampling (full version)) *Let $\nu > 2$, $0 < \epsilon < 1/2$, $n \in \mathbb{Z}_+$ sufficiently large and $m \in \Theta(n^\nu)$. Assume there exists a constant $C > 0$ such that the assertion* (6.39) *of Lemma 6.7 holds. Then:*

a. *With probability at least* $1-\delta-2n^2/(m\zeta)$ *over the choice of Haar-random unitaries* $U \sim \mu_H$ *there exists no* $\epsilon$*-certification test for boson sampling with* $n$ *photons in* $m$ *modes, from* $s < s_{\min}$ *many samples, where*

$$s_{\min} \in \Omega\left(n^{cn(\nu-1)/4}\delta^{1/4}\left(1-\zeta-2\epsilon\right)^{3/2}/\epsilon^2\right), \tag{6.40}$$

*and* $c > 0$ *is the implicit constant in* (6.47).

b. *For* $\nu > 3$*, with probability at least* $1-\exp(-\Omega(n^{\nu-2-1/n}))$ *over the Haar-random choice of* $U \sim \mu_H$*, there exists no* $\epsilon$*-certification test for boson sampling with* $n$ *photons in* $m$ *modes, where*

$$s_{\min} \in \Omega\left(2^n/\epsilon^2\right). \tag{6.41}$$

We remark that our results for the boson sampling distribution leave open the possiblity of sample-efficient $\epsilon$-certification for those instances of boson sampling with $2 < \nu \leq 3$ in the regime in which the probability weight of the collision-free subspace is very small. For instance, this is the case whenever $1/\mathsf{poly}(n) \leq P_{\mathrm{bs},U}(\Phi^*_{m,n}) \leq 2\epsilon$. This is because the bound (6.40) becomes trivial for $1-\zeta \leq 2\epsilon$.

However, our result fully covers the regime in which boson sampling is provably hard as shown in Ref. [AA13].

*Proof of Theorem 6.8a..* The proof proceeds along the same lines as the proofs of Theorem 6.5 and is based on direct applications of Lemma 6.3 to the collision-free subspace and the min-entropy bound (6.21) from Lemma 6.4 to the postselected boson sampling distribution $P^*_{\mathrm{bs},U}$ with postselection onto the collision-free subspace $\Phi^*_{m,n} \subset \Phi_{m,n}$.

To apply Lemmas 6.3 and 6.4 simultaneously we need to account both for the probability weight of the collision-free subspace and large probabilities, however. To account for the probability weight of the collision-free subspace we use a simple application of Markov's inequality to [AA13, Theorem 13.4] (restated as Lemma D.1 in Section D.1),

$$\Pr_{U\sim\mu_H}\left[P_{\mathrm{bs},U}[\Phi_{m,n}\setminus\Phi^*_{m,n}] > \zeta\right] < \frac{2n^2}{\zeta m}. \tag{6.42}$$

This shows that the total probability weight of the collision-free subspace is at least $1-\zeta$ with probability at least $1-2n^2/(m\zeta)$. We then apply a union bound argument to obtain

$$\Pr_{U\sim\mu_H}\Big[\left\{P_{\mathrm{bs},U} \text{ does not satisfy (6.21)}\right\} \cup \left\{P_{\mathrm{bs},U}(\Phi_{m,n}\setminus\Phi^*_{m,n}) > \zeta\right\}\Big] \leq \delta + \frac{2n^2}{\zeta m}. \tag{6.43}$$

In the next step, we use that the distribution of postselected boson sampling is given by $P^*_{\mathrm{bs},U} = (P_{\mathrm{bs},U})_{\restriction\Phi^*_{m,n}}/P_{\mathrm{bs},U}(\Phi^*_{m,n})$. Consequently, with probability at least $1-\delta-2n^2/(\zeta m)$ the boson sampling distribution $P_{\mathrm{bs},U}$ restricted to the collision-free subspace has both of the desired properties – a large min-entropy and a probability weight of at least $1-\zeta$ of the collision-free subspace.

Let us now compute the min-entropy for the collision-free subspace. For all samples $S \in \Phi^*_{m,n}$, Ref. [AA13, Lemma 8.8] implies that there exists $C > 0$ such that for $m \in \Theta(n^\nu)$ with any $\nu > 2$ for which the assertion of Lemma 6.7 holds, the following second moment bound also holds[7]:

$$\mathbb{E}_{U_S \sim \mu_H}[|\operatorname{Perm}(U_S)|^4] \leq (1 + C)(n!)^2 \,(n + 1)\, m^{-2n} \,. \tag{6.44}$$

7: The version of Ref. [AA13, Lemma 8.8] can be obtained from Eq. (6.44) from Lemma 6.7, normalizing the Gaussian measure $\mu_G$, and noting that $\mathbb{E}_{X \sim \mu_G(1)}[|\operatorname{Perm}(X)|^2] = n!$.

To obtain a lower bound on the min-entropy of the distribution $P^*_{\mathrm{bs},U}$ on the collision-free subspace we use that

$$H_\infty(P^*_{\mathrm{bs},U}) = \log P_{\mathrm{bs},U}(\Phi^*_{m,n}) + H_\infty((P_{\mathrm{bs},U})_{\restriction \Phi^*_{m,n}}). \tag{6.45}$$

Applying Lemma 6.4 together with the second moment bound (6.44), the union bound (6.43), the bound

$$|\Phi^*_{m,n}| = \binom{m}{n} = \frac{m(m-1)\cdots(m-n+1)}{n!} \leq \frac{m^n}{n!}, \tag{6.46}$$

on the size of the collision-free subspace and Stirling's formula yields

$$\begin{aligned} 2H_\infty(P^*_{\mathrm{bs},U}) &\geq 2\log(1-\zeta) + \log\delta \\ &\quad - \log\left(\frac{m^n}{n!}(1+C)(n!)^2(n+1)m^{-2n}\right) \\ &\in \Omega\left((\nu-1)n\log n\right) - \log\frac{1}{\delta} - 2\log\frac{1}{1-\zeta}, \end{aligned} \tag{6.47}$$

which holds with probability $1 - \delta - 2n^2/(m\zeta)$ over the choice of $U \sim \mu_H$.

We note that $2^{-H_\infty(P^*_{\mathrm{bs},U})} \in o(1)$; hence this term can be neglected when applying Eq. (6.6) in Lemma 6.2. Applying Lemmas 6.2 and 6.3, and the min-entropy bound (6.47) we obtain that the sample complexity for $\epsilon$-certifying boson-sampling scales as

$$s_{\min} \in \Omega\left(n^{cn(\nu-1)/4}\delta^{1/4}\,(1-\zeta-2\epsilon)^{3/2}\,/\epsilon^2\right) \tag{6.48}$$

with probability at least $1 - \delta - 2n^2/(\zeta m)$, where $c$ is the implicit constant in (6.47). This completes the proof of Theorem 6.8a.. □

Note that the bound (6.44) is essential for the hardness argument of Aaronson and Arkhipov [AA13]. Therefore a central ingredient to the hardness argument of Aaronson and Arkhipov [AA13] also prohibits sample-efficient certification of boson sampling.

It is important to stress that the boson sampling hardness proof [AA13] covers only those instances $U_S$ of boson sampling for which one can efficiently postselect on the collision-free outcomes. This is the case for those $U \sim \mu_H$ for which the probability weight of $\Phi^*_{m,n}$ is not smaller than polynomially small in $n$, i.e., $P_{\mathrm{bs},U}(\Phi^*_{m,n}) \in \Omega(1/\mathsf{poly}(n))$. Our proof method for Theorem 6.8a. thus permits sample-efficient certification for a small fraction of the instances, in particular, those instances of $U \sim \mu_H$ for which $2\epsilon \geq P_{\mathrm{bs},U}(\Phi^*_{m,n}) > 1/\mathsf{poly}(n)$.

In part b. of the theorem we can close this gap by extending the bound (6.47) on the min-entropy of the postselected distribution $P^*_{\mathrm{bs},U}$ to the full output distribution $P_{\mathrm{bs},U}$, however, at the cost of restricting to $\nu > 3$. This removes the need to use Lemma 6.3 and hence the dependence on the probability weight of the collision-free subspace. In the remaining case with $2 < \nu \leq 3$ hardness results have not been obtained, but it is conceivable that a hardness argument can be made.

*Proof of Theorem 6.8b..* Gogolin et al. [Gog+13] have proven the following strong lower bound on the min-entropy of the boson sampling distribution (see Theorem D.2 in Section D.1 for a restatement)

$$\Pr_{U \sim \mu_H} [H_\infty(P_{\mathrm{bs},U}) < 2\,n] \in \exp\left(-\Omega(n^{\nu-2-1/n})\right), \tag{6.49}$$

which holds whenever the condition of the theorem are fulfilled and in addition $\nu > 3$. In the proof, the probability measure induced on the matrices $U_S$ is related to a certain Gaussian measure $\mu_{G_S(\sigma)}$. Then, the min-entropy bound is proven using a trivial upper bound to the permanent as well as measure concentration for $\mu_{G_S(\sigma)}$. A simple application of Theorem 6.1 and Lemma 6.2 concludes the proof. □

## 6.6 Discussion

We have shown that probability distributions with a high min-entropy cannot be minimally certified from polynomially many samples, even when granting the certifier unlimited computational power and a full description of the target distribution. Our result applies to the problem of certifying quantum sampling problems as proposed to demonstrate quantum supremacy in a non-interactive device-independent fashion. We discuss the ironic situation that the very property that crucially contributes to the proof of approximate sampling hardness via Stockmeyer's algorithm and the Paley-Zygmund inequality – the second moments of the sampled distribution – forbids sample-efficient classical verification. This highlights the importance of devising alternative schemes of certification, or improved hardness results for more peaked distributions. Such schemes might allow for interaction between certifier and prover, invoke further complexity-theoretic assumptions or such on the sampling device, and/or grant the certifier some small amount of quantum capacities. We hope to stimulate research in such directions.

### Hardness, certification, and flatness

It is interesting to note the connection of our result with results on classical simulation. Similarly to our findings for the case of certification, Schwarz and Van den Nest [SV13] find that for certain natural families of quantum circuits (including IQP circuits) classical simulation is possible for highly concentrated distributions, but impossible for flat ones, see Figure 6.3. This again gives substance to the interesting connection between superior computational power, the flatness of the distribution and the impossibility of an efficient certification.

Curiously, at the same time, the property that prohibits sample-efficient certification is by no means due to the hardness of the distribution. It is merely the flatness of the distribution on an exponential-size sample space as effected by the random choice of the unitary that is required for the approximate hardness argument via Stockmeyer's algorithm and standard conjectures. The uniform distribution on an exponentially large sample space, which is classically efficiently samplable, can also not be sample-efficiently certified.

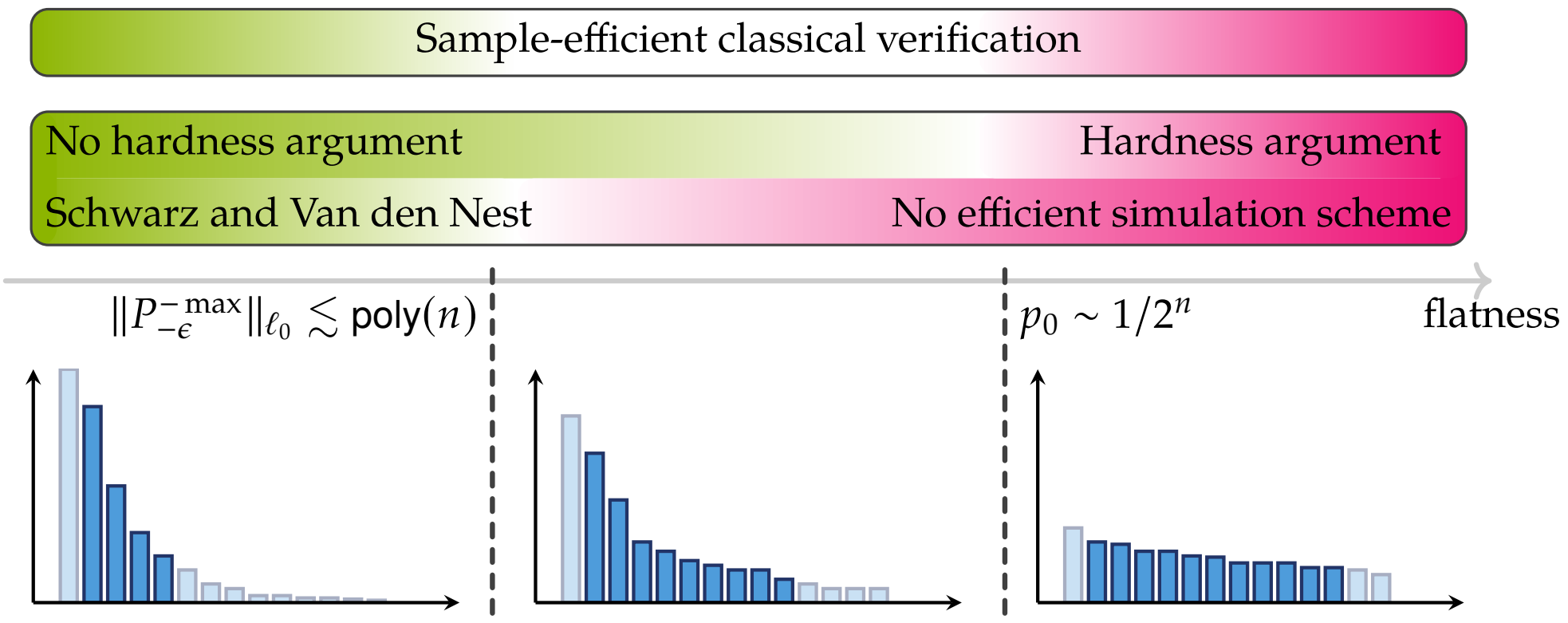

**Figure 6.3:** Hardness and certification in terms of the flatness of $P_{-\epsilon}^{-\max}$ for the example of IQP circuits [SB09; BMS16] on $n$ qubits as obtained from the present result and the classical simulation algorithm of Schwarz and Van den Nest [SV13]. There, it is shown that a certain natural family of quantum circuits (including IQP circuits) can be efficiently simulated on a classical computer if the output distribution is essentially concentrated on a polynomial number of outcomes only. In this case, i.e., for $\|P_{-\epsilon}^{-\max}\|_{\ell_0} \lesssim \mathsf{poly}(n)$, the output distribution is also sample-efficiently certifiable as the bounds (7.3) and (6.3) show. Their classical simulation algorithm breaks down if the distribution is essentially spread out over more than polynomially many outcomes, and we even have a rigorous hardness argument by Bremner, Montanaro, and Shepherd [BMS16] for exponentially flat distributions. Conversely, the number of samples required for certification becomes prohibitively large if the distribution is exponeentially spread out, as measured by the $\ell_{2/3}$-norm (6.1). Nevertheless, as we illustrate here, there could be "room in the middle" where, for reasonably but not exponentially flat distributions, one may hope to find tasks that are both classically intractable and sample-efficiently certifiable in a device-independent fashion.

A further noteworthy connection is that to Shor's algorithm. The output distribution of the quantum part of Shor's algorithm is typically spread out over super-polynomially many outcomes and can hence neither be efficiently simulated via the algorithm of Schwarz and Van den Nest [SV13], nor certified as we show here. However, after the classical post-processing, the output distribution is strongly concentrated on few outcomes – the factors – from which one can verify the correct working of the algorithm. A certification of the intermediate distribution is simply not necessary to demonstrate a quantum speedup in Shor's algorithm, as its speedup is derived from it solving a problem in $\mathsf{NP}$ and not from it sampling close to a hard distribution. This shows that while intermediate steps of a computation might not be certifiable, the final outcome may well be. Whether this is enough to demonstrate a speedup depends on the nature of the hardness argument. In fact, the abovementioned task HOG [AC17] bears many similarities to factoring and its certifiability from the outcomes of the algorithm.

We hope that our result will stimulate research into new ways of proving hardness of approximate sampling tasks that are more robust than those based on anticoncentration, as well as into devising alternative verification schemes possibly based on mild and physically reasonable assumptions on the sampling device or the verifier.

In this chapter, we have established the hardness of fully classical, device-independent verification of quantum random sampling schemes. In the next chapter, we will move beyond this no-go result and discuss possible routes to circumvent it – in the spirit of the framework presented in Chapter 5. What prevents fully classical verification is the overall *complexity* of a device-independent verification protocol, which we bounded in terms of the sample complexity. At the same time, the device-independent framework does not make any assumptions whatsoever on the device. It does not allow us to make use of any prior knowledge about or characterization of a given experimental setup, but conceptually remains in the framework of a fully untrusted quantum device. To reduce the overall complexity of the protocol, we will therefore have to introduce some assumptions or make the protocol qualitatively more difficult by allowing multiple rounds of interactions. It turns out that already slightly lowering the bar in this respect gives rise to practically feasible methods for verification.

# 7 Practical verification of quantum sampling

In the last chapter, we have seen how small second moments of a quantum random sampling scheme prohibit its efficient device-independent verification from samples alone: this task requires exponentially many runs of the experiment. Recall that, ironically, small second moments are an intrinsic property of such schemes which also provides evidence for approximate average-case hardness. In this chapter, we will proceed on a more positive note by showing up routes by which it is possible to circumvent this no-go result for device-independent verificiation. We will do so, following the general framework for verification methods of Chapter 5, by gradually reducing the complexity of verification protocols in exchange for additional assumptions.

First, in Section 7.1, I will review the optimal verification protocol due to Valiant and Valiant [VV17]. This protocol asymptotically achieves the necessary number of samples for identity testing.

We will then move on to sample-efficient but computationally inefficient protocols. In Section 7.2, I will review prominent ideas for how to verify or build confidence in a sampling device using an efficient amount of classical output data [Boi+18; AC17; Bou+19; Aru+19]. Those schemes are not device-independent in that they are either not guaranteed to verify the correct functioning of the device under arbitrary circumstances [Aru+19] or require some assumptions to be made on the device [Boi+18; Bou+19] or the protocol at hand [AC17].

It is even possible to come up with protocols that are fully efficient in their use of both samples and computational resources as we will see in Section 7.3. Such protocols lower the bar for what it means to certify a device: one can use additional cryptographic and average-case hardness assumptions to devise efficient tests of computational quantumness [SB09], or be content to distinguish the distribution of the samples against a particular adversary distribution in a *state discrimination* scheme [AA14].

While the above are valid approaches to certification, they are also not entirely satisfying for realistic experimental circumstances from the point of view of a physicist. This is because they do not exploit experimental knowledge about the setup at hand but rather introduce new assumptions on a theoretical level or the level of the device that are if at all hard to test [Kah19]. As argued in Chapter 5, in a laboratory environment, certain parts of the setup can often be characterized and calibrated to a very high degree of precision. This allows to build trust in the correct functioning of those parts of the experimental setup.

In a second step, we will therefore leverage natural trust assumptions on the device to develop fully efficient and yet rigorous certification protocols for quantum sampling schemes. The property of the experimental appartus that we exploit is that in many scenarios single-qubit measurements can be performed very accurately. Making use of this state of affairs gives rise to the measurement-device-dependent scenario (cf. Section 5.3). The first such protocol, introduced in Section 7.4, is a weak-membership certification protocol for the trace distance to the target state $C|0\rangle$ and is based on a natural *fidelity witness* for that state. The second protocol, introduced in Section 7.5, directly estimates the fidelity between the real state preparation $\rho$ and the target state $C|0\rangle$ and requires only a constant number of copies of the quantum state.

In Section 7.7, I will conclude the first two parts of this thesis that dealt with the classical hardness and the verification of quantum sampling schemes.

Before we move on to approach the quantum-classical divided from the classical side I will provide an outlook and discuss the major open questions in the field in the next chapter (Chapter 8).

## 7.1 Sample-optimal classical verification

Let us begin by recalling the no-go result derived in the previous chapter (Chapter 6). There, we made use of a result due to Valiant and Valiant [VV17], which showed that the necessary number of samples required for minimal certification or identity testing (Def. 5.2) up to (small enough) error $\epsilon$ scales as

$$\frac{1}{\epsilon^2}\|P^{-\max}_{-2\epsilon}\|_{\ell_{2/3}}. \tag{7.1}$$

The flipside of their result is that the necessary number of samples can (up to constant factors in $\epsilon$) be achieved and is therefore optimal. Here, I review this sample-optimal test. Recall that we are given $k$ samples $x_i, \ldots, x_k$ from the distribution $Q$ and wish to test whether $Q = P$ or whether $\|Q - P\|_{\mathrm{TV}} \geq \epsilon$ (Def. 5.2). Recall also the definition of $P_X$ as the distribution obtained by setting the probabilities $p_i$ for $i \in X \subset [2^n]$ to zero.

Let $X(x) = |\{x_i : x_i = x\}_i|$ be the number of occurences of outcome $x$, assume w.l.o.g. that the probabilities of $P$ are sorted in nonincreasing order, labeled by $p_i, i = 1, \ldots, 2^n$ and let $s = \min\{\sum_{j<i} p_j \leq \epsilon/8\}$, $M = \{2, \ldots, s\}$ and $S = \{s+1, s+2, \ldots\}$ . Valiant and Valiant [VV17] then prove that a simple variant of the $\chi^2$ statistic, which is given by $\sum_{x\in\{0,1\}^n}(X(x) - kP(x))^2/P_U(x)$, namely,

$$\chi_{2/3}(\{x_1, \ldots, x_k\}, P) = \sum_{x\in\{0,1\}^n} \frac{(X(x) - kP(x))^2 - X(x)}{P(x)^{2/3}} \tag{7.2}$$

gives rise to the following sample-optimal minimal certification test.

For Protocol 7.1 to be an $\epsilon$ minimal certification test (Def. 5.2)

$$O(\max\left\{\frac{1}{\epsilon}, \frac{1}{\epsilon^2}\|P^{-\max}_{-\epsilon/16}\|_{\ell_{2/3}}\right\}) \tag{7.3}$$

many samples from $Q$ are sufficient. Insofar the necessary number of samples scales in the same way up to constant factors in $\epsilon$ this test is sample optimal[1].

1: Recall the detailed discussion of their lower bound (6.1) in Chapter 6.

**Protocol 7.1** Instance-optimal identity test [VV17]

**Input:** $\epsilon > 0$, a set $\mathcal{X} = \{x_1, \ldots, x_k\}$ of $k$ samples drawn iid. from $Q$.
**Output:**
 **if** $\sum_{i \in M} \chi_{2/3}(\mathcal{X}, P) > 4k \|P_M\|_{2/3}^{1/3}$ , **then**
  'Reject',
 **else if** $\sum_{i \in S} X_i > 3\epsilon k/16$, **then**
  'Reject',
 **else**
  'Accept'.
 **end if**

Protocol 7.5 is neither efficient in terms of the number of samples nor the computation time or space required to evaluate the figure of merit. Indeed, evaluating the $\chi_{2/3}$ quantity requires computing *all* of the exponentially many probabilities $P(x)$. Nonetheless, it provides an assumption-free certificate for the correct functioning of a sampling device which may be a useful tool to test small or intermediate-sized instances as well as a standard to compare other certification tools to.

## 7.2 Sample-efficient classical verification

The most prominent ideas for verifying quantum sampling schemes relax the notion of verification from full-scale certification up to total-variation distance errors in the sense of Defs. 5.1 and 5.2 to cross-entropy like measures [Boi+18; AC17; Bou+19; Aru+19]. Such measures are weaker than the total-variation distance but at the same time are fairly good measures of similarity between distributions. A protocol of this type has been applied in the recent demonstration of quantum supremacy by the Google AI Quantum lab [Aru+19]. Cross-entropy type verification uses only polynomially many classical samples obtained from the device and thus circumvents the no-go result of the previous chapter. Nevertheless, it requires exponential post-processing time [Aru+19] or space [Ped+19]

The improvement over the sample-optimal verification of Valiant and Valiant [VV17] comes at the cost of not being able to distinguish the target distribution from arbitrary adversarial distributions, but only a certain subclass of distributions. This subclass can be characterized by certain conditions on the type of 'noise' which relates the actual 'noisy' distribution to the ideal target distribution [Boi+18; Bou+19; Aru+19].

Cross-entropy type measures between a 'noisy distribution' $Q$ and the ideal target distribution $P_U$ can be expressed in terms of a monotonously increasing function $f : [0, 1] \longrightarrow \mathbb{R}$

$$F_f(Q, P_U) := \sum_{x \in \{0,1\}^n} Q(x) f(P_U(x)), \tag{7.4}$$

and they provide measures of the correlations between the target distribution $Q$ and $P_U$.

The central intuition underlying cross-entropy type measures is the following: those distributions which get the *heavy outcomes* of a computation correct will score well on cross-entropy measures [AC17]. One can characterize 'heavy outcomes' as those bit strings $x \in \{0, 1\}^n$ for which the probability $P_U(x)$ of obtaining $x$ is large, for example, larger than the median of $P_U$; see Fig. 7.1.

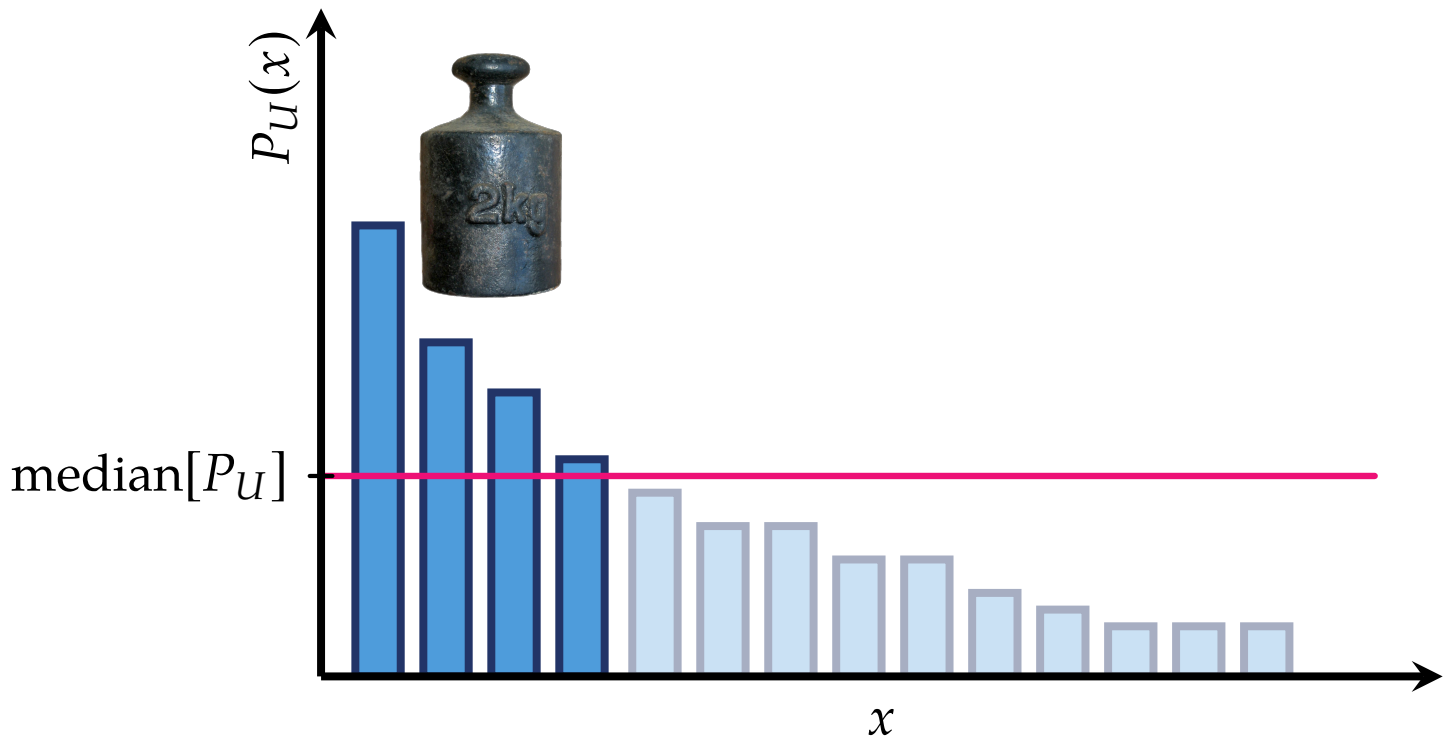

**Figure 7.1:** In the task *heavy outcome generation* (HOG) one is asked to output a list of strings $\{x_1, \ldots, x_k\}$ for which $P_U(x_i) \geq \text{median}(P_U)$. The image of the 2kg-weight by LoKiLeCh is licensed under [CC BY-SA 3.0].

## Heavy-outcome generation

The very task of outputting those 'heavy outcomes' has been dubbed 'heavy outcome generation' or HOG by Aaronson and Chen [AC17] and the same authors argued that HOG is computationally intractable for random quantum circuits. More precisely, HOG is formulated as follows.

**Problem 7.1** (HOG [AC17]) Given as input a random quantum circuit $C \in \mathscr{C}$ from a family $\mathscr{C}$, generate output strings $x_1, \ldots, x_k$, at least a $2/3$ fraction of which have a probability greater than median$[P_C]$.

The computational complexity of HOG can easily be reduced to the hardness of obtaining a guess whether $|\langle 0|C|0\rangle|^2$ is larger than median$[P_C]$ that is significantly better than a coin flip. The 'Quantum Threshold Assumption' (QUATH) [AC17] states that this task is computationally intractable for classical computers. On the other hand, the fact that HOG is defined in terms of the bias of the target distribution makes it sample-efficiently verifiable: given the samples $x_1, \ldots, x_k$, compute the probabilities $P_C(x_i)$ and compare them to the median. The median can be estimated very efficiently up to a small error from few samples. It is therefore no exaggeration to say that HOG is an intrinsically verifiable task.

But this comes at a cost: in verifying HOG one is not verifying that the correct distribution is sampled up to a total-variation distance bound $\varepsilon$. Rather, one is verifying that some task – HOG – is being achieved, which is conjectured to be hard. There is no complexity-theoretic evidence for the hardness of HOG beyond QUATH, however. Compared to, say, the conjectured non-collapse of the polynomial hierarchy which is grounded in significant evidence, both QUATH and the intractability of HOG are very strong conjectures with relatively little grounding.

Let us now go a step further and define HOG quantitatively. To this end, define $\theta : \mathbb{R} \longrightarrow \{0, 1\}$ as the step function which is 0 for $x < 0$ and 1 else. We can then formulate HOG as the task to sample from a distribution $Q$ for which the 'HOG fidelity'

$$F_{\mathsf{HOG}}(Q, P_U) = \frac{2}{\ln 2} \sum_{x \in \{0,1\}^n} Q(x) \left( \theta(P_U(x) - \mathrm{median}[P_U]) - \frac{1}{2} \right), \quad (7.5)$$

is bounded away from 0. Indeed, if $Q$ is maximally noisy – that is, the uniform distribution – then

$$F_{\mathsf{HOG}}(Q, P_U) = \frac{1}{\ln 2} \left( \frac{2}{2^n} |\{x : P_U(x) \geq \mathrm{median}[P_U]\}| - 1 \right) = 0, \quad (7.6)$$

as the median is defined as any number for which the fraction of outcome probabilities of $U$ exceeding that number is $1/2$. On the other hand, in an ideal implementation (in which $Q = P_U$) $F_{\mathsf{HOG}}(Q, P_U) > 0$ so long as $P_U$ is nonuniform. This is because, by definition, the probabilities above the median are larger than those below the median and hence the probability weight above the median is at least $1/2$. If, as is the case for Haar-random unitaries and as is heuristically shown for random universal circuits, the outcome probabilities $P_U(x)$ are exponentially (Porter-Thomas) distributed, then $F_{\mathsf{HOG}}(P_U, P_U) = 1$. This is because the median of the exponential distribution is given by $\ln 2/2^n$ and the total probability weight of $P_U$ above the median is then given by[2]

2: See also Ref. [AC17, footnote 3].

$$\sum_{x \in \{0,1\}^n} P_U(x)\theta(P_U(x) - \ln 2/2^n) \approx \int_{\ln 2/2^n}^{\infty} 2^n \mathrm{e}^{-2^n p} \, \mathrm{d}p = \frac{1 + \ln 2}{2}. \quad (7.7)$$

More generally, a distribution that scores well in terms of $F_{\mathsf{HOG}}$ will therefore tend to be closer to an ideal implementation of $P_U$ in terms of total-variation distance. This is rigorously true in case the noisy distribution

$$Q_\lambda(x) = (1 - \lambda) P_U(x) + \lambda \frac{1}{2^n}, \quad (7.8)$$

with $\lambda \in [0, 1]$ is a convex mixture of the ideal target distribution and the uniform distribution.

Clearly, there are also distributions, however, which score well on the HOG fidelity, but are far away from $P_U$. To see this, just take the distribution which is supported on $\{x : P_U(x) \geq \mathrm{median}[P_U]|\}$. This distribution will have a HOG fidelity of $1/\ln 2 > 1$ even though its total variation-distance to $P_U$ is at least $(1-\ln 2)/2$. But of course, it is extremely hard, in particular, at least as hard as QUATH, to find such a distribution. Therefore, scoring well on $F_{\mathsf{HOG}}$ may well be computationally hard even though it does not quantify the total-variation distance.

## Cross-entropy difference

To better capture correlations between the distribution $Q$ and $P_U$ a convenient measure of the same type as $F_{\mathsf{HOG}}$ is the *cross-entropy difference* [Boi+18][3]

3: Note that my definition of the cross-entropy differs from that of Boixo et al. [Boi+18], who define $d_{\mathrm{XE}}(Q, P_U) = \mathrm{CE}(1/2^n, P_U) - \mathrm{CE}(Q, P_U)$ as the deviation of cross-entropy between $Q$ and $P_U$ from the cross-entropy between the uniform distribution and $P_U$.

$$d_{\mathrm{CE}}(Q, P_U) = \mathrm{CE}(Q, P_U) - H(P_U) \tag{7.9}$$
$$= \sum_{x\in\{0,1\}^n} (Q(x) - P_U(x)) \log \frac{1}{P_U(x)} \tag{7.10}$$

where $\mathrm{CE}(Q,P) = -\sum_x Q(x)\log P(x)$ denotes the cross entropy between $Q$ and $P$ and $H$ denotes the Shannon entropy[4]. The cross-entropy is a well-known statistical measure of similarity between two distributions and measures correlations between the two distributions.

4: In Eq. (7.4) we thus write $f(x) = \log 1/x - \sum_x x \log 1/x$.

But how does the cross-entropy difference fare when applied to the task of verifying quantum supremacy distributions? Again, using the assumption that the ideal probabilities are exponentially distributed, we observe that it constitutes a good measure for distributions of the form $Q_\lambda$ (7.8)[5]:

5: This argument follows Boixo et al. [Boi+18, Sec. 2 of the SM].

$$d_{\mathrm{CE}}(Q_\lambda, P_U) = (1-\lambda) d_{\mathrm{CE}}(P_U, P_U) + \lambda d_{\mathrm{CE}}(1/2^n, P_U) \tag{7.11}$$
$$\approx (1-\lambda)\cdot 0 + \lambda \cdot 1 = \lambda. \tag{7.12}$$

To see why this is the case, we can compute the expectation value of $H(P_U)$ over the random choice of $U$ as [Boi+18]

$$\mathbb{E}_U\left[H(P_U)\right] = -\sum_x \mathbb{E}_U\left[P_U(x)\log P_U(x)\right] \tag{7.13}$$
$$= -2^n \int_0^\infty 2^n \mathrm{e}^{-2^n p} p \log p \, \mathrm{d}p \tag{7.14}$$
$$= n - 1 + \gamma, \tag{7.15}$$

where $\gamma \approx 0.5774$ is the Euler constant. Likewise, the cross-entropy between $P_U$ and the uniform distribution is in expectation given by

$$\mathbb{E}_U\left[\mathrm{CE}(1/2^n, P_U)\right] = -\frac{1}{2^n}\sum_{x\in\{0,1\}^n} \mathbb{E}_U\left[\log P_U(x)\right] \tag{7.16}$$
$$= -\int_0^\infty 2^n \mathrm{e}^{-2^n p} \log p \, \mathrm{d}p \tag{7.17}$$
$$= n + \gamma. \tag{7.18}$$

From this we obtain $\mathbb{E}_U[d_{\mathrm{CE}}(1/2^n, P_U)] = 1$. Since all probabilities $P_U(x)$ for a given $U$ are (pairwise) independently identically distributed according to the exponential distribution, with overwhelmingly high probability over the choice of $U$, $d_{\mathrm{CE}}(1/2^n, P_U) = 1$. Conversely, as the cross-entropy reduces to the Shannon entropy for $Q = P_U$ we also have $d_{\mathrm{CE}}(P_U, P_U) = 0$. To summarize, the cross-entropy difference attains the value 1 for the uniform distribution and vanishes for the ideal distribution, giving rise to linear interpolation (7.11) for states of the form $Q_\lambda$.

Notice that the same argument will work for any noisy distribution

$$Q'_\lambda = (1-\lambda)P_U + \lambda Q', \tag{7.19}$$

in which the uniform distribution is replaced by a distribution $Q'$ that is uncorrelated with $P_U$, i.e., $\mathbb{E}_U[\mathrm{CE}(Q', P_U)] = -\sum_x Q'(x)\,\mathbb{E}_U[\log P_U(x)]$.

Going yet a step further, one can easily derive a condition under which the cross-entropy difference provides a rigorous bound on the total variation distance [Bou+19]. To see this, notice that the definition of the cross-entropy difference is similar to that of the *Kullback-Leibler divergence*

$$D_{\mathrm{KL}}(Q\|P_U) = \mathrm{CE}(Q, P_U) - H(Q). \tag{7.20}$$

But the Kullback-Leibler divergence is know to bound the total-variation distance by Pinsker's inequality as

$$\|Q - P_U\|_{\mathrm{TV}} \leq \sqrt{D_{\mathrm{KL}}(Q\|P_U)/2}. \tag{7.21}$$

Hence, if the cross-entropy difference satisfies $d_{\mathrm{CE}}(Q, P_U) \leq \varepsilon$ and the noise is entropy-increasing such that $H(Q) \geq H(P_U)$ we have

$$\|Q - P_U\|_{\mathrm{TV}} \leq \sqrt{D_{\mathrm{KL}}(Q\|P_U)/2} \leq \sqrt{d_{\mathrm{CE}}(Q, P_U)/2} \leq \sqrt{\epsilon/2}. \tag{7.22}$$

The condition $H(Q) \geq H(P_U)$ is a fairly general condition on the type of noise under which the total-variation distance bound (7.22) holds. But it is also a condition that cannot be checked from fewer than exponentially many samples from $Q$. Moreover, one can easily construct examples of distributions that violate the inequality (7.22) [Bou+19]: those examples fare well on the cross-entropy difference, but are far from the ideal target distribution.

On the other hand, the cross-entropy difference can be efficiently estimated up to accuracy $\epsilon$ with failure probability $\alpha$ from

$$m \geq \frac{(n + O(\log n))^2}{2\epsilon^2} \log(2/\alpha) \tag{7.23}$$

many iid. samples from $Q$. To derive Eq. (7.23), we apply Hoeffding's inequality and assume that the probabilities $P_U(x)$ are Porter-Thomas distributed. We obtain that with probability at least $1 - 1/f(n)$ over the choice of $U$, the probabilities $P_U(x)$ satisfy

$$2^{-2n}/f(n) \leq P_U(x) \leq (n + \log f(n))2^{-n}, \tag{7.24}$$

so that their logarithms $\log P_U(x)$ differ only by a constant factor of $\sim (2 + O(\log(f(n)))$ from $n$.

## Cross-entropy fidelity

An alternative quantity with similar features as the cross-entropy difference is the *linear cross-entropy benchmarking (XEB) fidelity*[6] [Aru+19]

6: The XEB fidelity was utilized by Arute et al. [Aru+19] in the recent quantum supremacy demonstration.

$$F_{\mathrm{XEB}}(Q, P_U) = \sum_{x\in\{0,1\}^n} Q(x)(2^n P_U(x) - 1). \tag{7.25}$$

When averaged over the choice of random $U$ the cross-entropy fidelity evaluates to

$$\mathbb{E}_U[F_{\mathrm{XEB}}(Q, P_U)] = \begin{cases} \sum_x 2^n\, \mathbb{E}_U[P_U(x)^2] - 1 \approx 1 & Q = P_U, \\ \sum_x P_U(x) - 1 = 0 & Q = \frac{1}{2^n}, \end{cases} \tag{7.26}$$

in the extreme cases in which $Q$ is the ideal target distribution and the uniform distribution, respectively. Here, we have used that the second moments $\mathbb{E}[P_U(x)^2]$ are approximately given by the second moments of the Haar measure (3.20), and thus a much weaker property of the distribution of probabilities than assuming its full distribution to be exponential (as we did for the cross-entropy difference).

For Haar-random unitaries, using the bounds (7.24) on the size of the probabilities $P_U(x)$ we can estimate $\mathbb{E}_U[F_{\mathrm{XEB}}(Q, P_U)]$ up to error $2\epsilon$ with failure probability $\alpha$ from

$$m_U \geq \frac{1}{2\epsilon^2} \log \frac{2}{\alpha}, \tag{7.27}$$

many distinct random circuits and

$$m \geq \frac{(n + O(\log n))^2}{2\epsilon^2/m_U^2} \log(2/\alpha), \tag{7.28}$$

many samples per circuit. When the number of qubits is large, and the unitary $U$ is drawn Haar-randomly, Levy's lemma [Led01] implies that the fluctuations around the expectation value over $U$ (7.26) are expected to be on the order of $O(1/\sqrt{2^n})$. Consequently, for large numbers of qubits, the expectation value over the choice of random circuits might be neglected in practice [Aru+19].

In contrast to the cross-entropy difference, here, one cannot derive a general bound on the total-variation distance between $Q$ and $P_U$ in terms of $F_{\mathrm{XEB}}$, making a simple assumption on the noise. However, the same properties as for the cross-entropy difference hold in case the noisy distribution $Q$ can be written as a convex combintation of the ideal distribution and a distribution $Q'$ which is uncorrelated with $P_U$ (cf. (7.19)) so that [Aru+19, Eq. (25) in the SM]

$$\mathbb{E}_U\left[Q(x)P_U(x)\right] = Q(x)\,\mathbb{E}\left[P_U(x)\right]. \tag{7.29}$$

Of course, we may still build confidence in the validity of (7.29). Large parts of the analysis of the theoretical proposal of random circuit sampling [Boi+18] and the experimental realization thereof [Aru+19, Sec. IV B of the SM] was indeed dedicated to validating the assumption of uncorrelated noise. This can be done, for example, by numerically studying realistic error models such as random Pauli errors.

In a practical mindset, one can also use the estimate of $\mathbb{E}_U[F_{\mathrm{XEB}}(Q, P_U)]$ to estimate the depolarization error $p_c$ per cycle of the computation. This is based on the following model[7]. Consider the noisy quantum state

7: This brief explanation follows Arute et al. [Aru+19, Sec. IV A of the SM]; see there for more details.

$$\rho_U = \epsilon_m U|0\rangle\langle 0|U^\dagger + (1 - \epsilon_m)\chi_U, \tag{7.30}$$

after applying a random circuit $U$ with $m$ gate layers (cf. (2.91)). Here, $U|0\rangle$ is the ideal output state and the mixed state $\chi_U$ together with $\epsilon_m$ describes the effect of errors. As above, we will now assume that the erroneous state $\chi_U$ is uncorrelated with $U$ in the sense that the probabilities of a computational basis measurement are uncorrelated: $\mathbb{E}_U[\langle x|\chi_u|x\rangle\langle x|U|0\rangle\langle 0|U^\dagger|x\rangle] = \mathbb{E}_U[\langle x|\chi_u|x\rangle]\,\mathbb{E}_U[\langle x|U|0\rangle\langle 0|U^\dagger|x\rangle]$.

When averaging or 'twirling' over random unitaries that form a unitary design we would then expect to obtain a fully mixed state [Aru+19]

$$\mathbb{E}_U \left[U^\dagger \chi_U U\right] = \frac{1}{2^n}, \tag{7.31}$$

so that one might expect

$$\mathbb{E}_U \left[U^\dagger \rho_U U\right] = \overline{\epsilon_m}|0\rangle\langle 0| + (1 - \overline{\epsilon_m})\frac{1}{2^n}, \tag{7.32}$$

where $\overline{\epsilon_m}$ denotes the average of the $\epsilon_m$ over the random choice of unitaries. Eq. (7.32) precisely describes the effect of a depolarizing channel (2.73) acting in each cycle of the computation with depolarization fidelity $p_c$ such that $p_c^m = \overline{\epsilon_m}$.

We obtain an expression of the circuit-averaged XEB fidelity in terms of the depolarization fidelity

$$\mathbb{E}_U \left[F_{\mathrm{XEB}}(Q, P_U)\right] = p_c^m \left(2^n \sum_x \mathbb{E}_U[P_U(x)^2] - 1\right). \tag{7.33}$$

that we can use in order to estimate $p_c$ from $F_{\mathrm{XEB}}(Q, P_U)$. To do this, we classically estimate the quantity in brackets in Eq. (7.33) and obtain

$$p_c^m \approx \frac{\overline{\widehat{F_{\mathrm{XEB}}}(Q, P_U)}}{2^n \sum_x \mathbb{E}_U[P_U(x)^2] - 1}, \tag{7.34}$$

where $\widehat{F_{\mathrm{XEB}}}(Q, P_U)$ denotes the empirical estimate of $F_{\mathrm{XEB}}(Q, P_U)$ for a fixed circuit and $\overline{F_{\mathrm{XEB}}(Q, P_U)}$ denotes the empirical average over random circuits.

It can be shown that systematically cheating the linear cross-entropy fidelity is computationally difficult under complexity-theoretic assumptions that are directly analogous to those invoked in the argument for the hardness of HOG [AG19]. For a restricted class of circuits, namely, log-depth two-dimensional circuits, the XEB fidelity can be fooled in polynomial runtime, however [BCG20]. At the same time, recall from Section 4.2 that the evidence for computational intractability of such circuit families stands on the same complexity-theoretic footing as that for deep random circuits [Haf+19]. The result of Ref. [BCG20] thus constitutes evidence that passing a linear XEB test might be significantly easier than a full simulation of random quantum circuits.

### Binned outcome generation

A natural way to improve the properties of the HOG fidelity on the one hand, and the cross-entropy distance and fidelity on the other hand, is to bin probabilities in a more fine-grained fashion [Bou+19]. This retains the complexity-theoretic intuition behind HOG (that producing outcomes with the correct frequencies is hard) and the favourable feature of the cross-entropy distance (that it provides a conditional bound on the total-variation distance). A natural starting point for such a more fine-grained measure is to observe that HOG effectively divides the probabilities into two bins – those that are larger than the median and those that are smaller. The HOG benchmark is then obtained from testing whether the empirically obtained samples satisfy certain properties expected from ideally distributed samples on the respective bins.

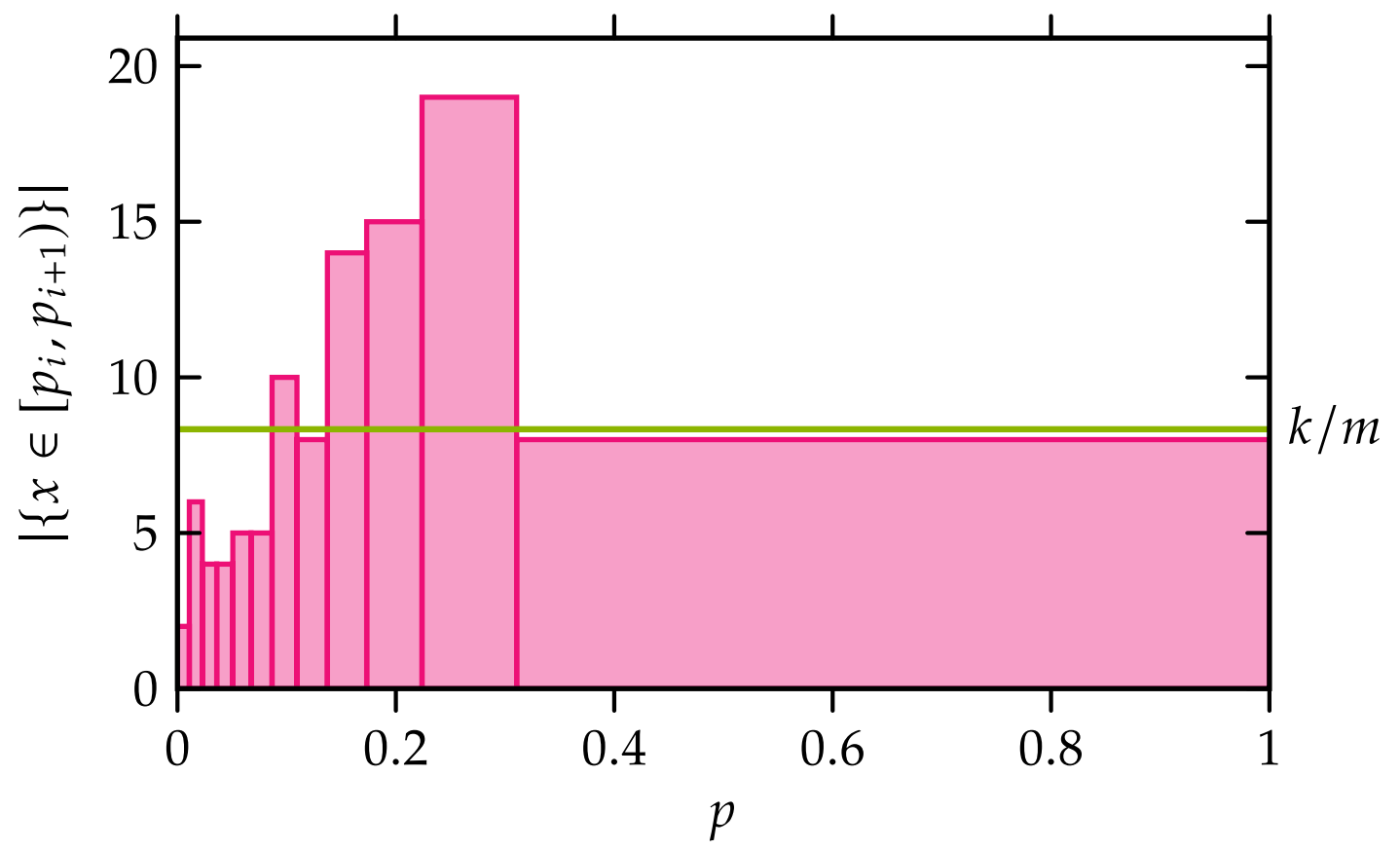

**Figure 7.2:** The fine-grained generalization of heavy outcome generation is to bin the samples $x_1, \ldots, x_k$ from the noisy distribution $Q$ according to the probabilities $P_U(x_i)$. This constitutes a coarse-grained estimator of the total-variation distance betwen $Q$ and $P_U$. Since $P_U$ is (nearly) exponentially distributed for random circuits, a suitable choice of $m$ bins $[p_i, p_{i+1})$ is such that they are equifilled with a $1/m$ fraction of the ideal samples. This is shown in the figure for a noisy exponential distribution on a $n = 3$-qubit sample space, with $m = 12$ bins and $k = 100$ samples.

The sample efficiency of computing this benchmark can be retained even when generalizing it to polynomially many bins and comparing the number of observed outcomes per bin with the number of expected outcomes[8].

8: This is why Bouland et al. [Bou+19] dubbed this idea 'binned outcome generation' or BOG.

Given that the distribution of outcome probabilities is expected to be an exponential distribution, the natural way to bin is to choose equifilled bins $[p_i, p_i + 1)$ satisfying

$$\int_{p_i}^{p_{i+1}} 2^n \mathrm{e}^{-2^n p}\, \mathrm{d}p = \frac{1}{m} . \tag{7.35}$$

for $i = 1, \ldots, m$ and setting $p_0 = 0,\ p_m = 1$. Define $\Omega = \{[p_i, p_i + 1)\}_{i \in [m]}$. The task is then to compute

$$d_{\mathsf{BOG}}(Q, P_U) = \sum_{X \in \Omega} \left| \frac{1}{2^n} \sum_{x \in \{0,1\}^n} (Q(x) - P_U(x)) \delta(P_U(x) \in X) \right| \tag{7.36}$$

$$= \sum_{X \in \Omega} \left| \frac{1}{2^n} \sum_{x \in \{0,1\}^n} Q(x) \delta(P_U(x) \in X) - \frac{1}{m} \right| , \tag{7.37}$$

where the last equality is true if $P_U$ is Porter-Thomas distributed. This is precisely a discretized estimator of the total-variation distance of the outcome distribution to the as we used it in Section 4 and can be estimated from polynomially many samples; see Fig. 7.2. Indeed, for $Q = P_U$ this measure is 0, while for any $Q \neq P_U$ it converges to $\|Q - P_U\|_{\mathrm{TV}}$ as $m \to \infty$. More precisely, Canonne and Wimmer [CW20] prove that *binned identity testing* with $k$ bins up to error $\epsilon$ is possible using $O(k/\epsilon^2)$ many samples, and moreover, that this is asymptotically optimal.

## 7.3 Efficient classical verification

In the last section, we have seen classical verification methods that are sample efficient in that they require only few (polynomially many) samples from the quantum device. But all of those methods were based on computing some type of overlap or coarse-grained distance of frequencies of the outcomes with the ideal probabilities and hence required actually computing those probabilities. In this section, we will try and loosen even this latter requirement to full efficiency. We will see one approach in this mindset that is based on cryptographic and additional complexity-theoretic assumptions. But we will also quickly run into limitations of this approach.

## Cryptographic tests of computational quantumness

A particularly promising avenue of this type of certification has been pioneered by Shepherd and Bremner [SB09]: By allowing the certifier to choose the classical input to the sampling device rather than drawing it fully at random, it is under cryptographic assumptions on the hardness of certain tasks possible to efficiently certify that a quantum device has performed a task that no classical device could have solved. This is facilitated by checking a previously hidden bias in the obtained samples and has been achieved for a certain family of IQP circuits [SB09].

I follow Ref. [SB09].

It is instructive to understand the idea behind such a test of computational quantumness. The protocol of Shepherd and Bremner [SB09] is formulated for a certain family of IQP circuits, called $X$-programs. An $X$-program acting on $n$ qubits is defined by a list of pairs $(\theta_p, p) \in [0, 2\pi] \times \{0,1\}^n$ and acts as

$$|0\rangle \mapsto \exp\left( \mathrm{i} \sum_p \theta_p \prod_{j=1}^n X_j^{p_j} \right) |0\rangle. \tag{7.38}$$

For the purposes of the quantumness test it is sufficient to choose a constant value of $\theta$ that is the same for every nonvanishing term in the Hamiltonian. In this case, an $X$-program with $k$ nonvanishing Hamiltonian terms acting on $n$ qubits can be represented by a 0/1 matrix $P \in \{0,1\}^{k \times n}$. Each row of this matrix specifies a Hamiltonian term and it is easy to see that the output distribution of such an $X$-program is given by

$$P_P(x) = \left| \sum_{a \in \{0,1\}^k :\, P^T \cdot a = x} \cos(\theta)^{k - wt(a)} \sin(\theta)^{wt(a)} \right|^2, \tag{7.39}$$

where $wt(a) = |\{l \in [k] : a_l = 1\}|$ is the *Hamming weight* of the binary string $a \in \{0,1\}^k$.

For a random variable $X$ taking values in $\{0,1\}^n$, and $s \in \{0,1\}^n$, the *bias* of $X$ in the direction of $s$ is just the probability that a sample $x \sim P_P$ is orthogonal to $s$, i.e., that $x^T s = 0$. The key idea of the test of computational quantumness is to hide a string $s$ the output probability distribution of an $X$-program in such a way that this string $s$ cannot be determined efficiently, but at the same time the bias of the output distribution of the $X$-program in direction $s$ is significantly larger than the bias of any cheating distribution that can efficiently be obtained using classical computing resources. In particular, the bias of the output distribution $P_P$ of the $X$-program defined by a matrix $P \in \{0,1\}^{k \times n}$ and angle $\theta$ is given

by

$$\Pr_{x\sim P_P}[x^T s = 0] = \sum_{x:\, x^T\cdot s=0} \left| \sum_{a\in\{0,1\}^k:\, P^T\cdot a = x} \cos(\theta)^{k-wt(a)} \sin(\theta)^{wt(a)} \right|^2 . \tag{7.40}$$

To achieve this, Shepherd and Bremner [SB09] notice that the matrix $P$ can be viewed as the *generator matrix* of a linear code. That is, the columns of $P$ span the code space $\mathscr{C} = \{P \cdot d : d \in \{0,1\}^n\}$. If we let $P_s$ be the $n_s \times n$- submatrix of $P$ obtained by deleting all rows $p$ for which $p^T s = 0$[9], and $\mathscr{C}_s$ be the code generated by $P_s$, then we can rewrite the bias (7.40) of $P_p$ as [SB09, Thm. 2.7]

9: Of course, this leaves only rows for which $p^T s = 1$.

$$\Pr_{x\sim P_P}[x^T s = 0] = \mathbb{E}_{c\sim\mathscr{C}_s}\left[\cos^2\left(\theta(n_s - 2\cdot wt(c)\right)\right] . \tag{7.41}$$

We can now set a quantum challenge that is intrinsically verifiable in the following way. To this end, we choose a code $\mathscr{C}_s$ and a value of $\theta$ in such a way that both the bias (7.41) is strictly larger than $1/2$, and that any classical strategy can only achieve a bias that is significantly lower, say, by a constant. We then choose a generating matrix $P_s$ for $\mathscr{C}_s$ such that $s$ is not orthogonal to any of the rows of $P_s$. Finally, we obfuscate this matrix by adding rows that are orthogonal to $s$, permuting all rows and potentially performing reversible column operations, giving rise to a matrix $P$. Given samples from the distribution $P_P$, we can now distinguish the hypothesis that the sampling device has quantum capacities from the hypothesis that it is cheating by comparing the frequencies of outcomes that are orthogonal to the hidden string $s$. Notice that this protocol *does not* certify that the samples are distributed according to the correct distribution. Therefore, it does not constitute a workaround to the no-go theorem of Chapter 6 based on cryptographic assumptions. Similarly to the HOG test (Problem 7.1), this cryptographic test of quantumness merely certifies that the device has the capacity to do something that presumably – under assumptions – no classical computing device could have achieved.

The particular suggestion of Shepherd and Bremner [SB09] is to use *quadratic residue codes* and a particular obfuscatio procedure that exploits specific properties of such codes (such as that the full-weight vector is always a codeword). They conjecture that recovering the matrix $P_s$ from the obfuscated matrix $P$ is NP-complete. Choosing $\theta = \pi/8$, this construction gives rise to a bias that serendipitously matches that of the Bell inequality: $\cos^2(\pi/8) \approx 0.854$ for the quantum value, and $3/4$ for the best known classical strategies[10].

10: There is no proof that this 3/4 is the optimal classical value.

Note also that besides the security assumption on the obfuscation procedure, additional conjectures need to be made [SB09, Conj.'s 4.2 & 4.3] for such a test to achieve its goal[11]: First, the distribution $P_P$ of a randomly selected $X$-program with constant $\theta = \pi/8$ should be hard to sample from so that only a quantum device can perform this task. Second, the output distribution should be sufficiently flat in the sense that its Renyi-2 entropy or *collision entropy* is close to maximal, i.e., $H_2(P_P) = \Omega(n)$ so that cheating becomes more difficult.

11: Indeed, for certain families of $X$-programs such as the family defined in Eq. (6.25), by now, there is significant evidence for those conjectures (cf. Section 2.4 and Chapter 6).

Iterating the importance of extensively testing cryptographic assumptions for their security, Kahanamoku-Meyer [Kah19] has recently developed a classical cheating strategy for the protocol proposed by Shepherd and Bremner [SB09]. Given a description of an $X$-program in the form of the matrix $P$, the cheating strategy extracts the secret vector $s$ with probability arbitrarily close to unity in an (empirically observed) average runtime of $O(n^3)$. Lately, this strategy has in turn been circumvented, making use of a different strategy than quadratic residue codes [YC20].

In the same mindset, albeit without restricting to sampling tasks for which there is strong complexity-theoretic evidence for hardness, cryptographic tests of quantumness were devised in Refs. [Bra+18; Bra+20]. There, the authors made use of so-called *trapdoor claw-free one-way functions*. The assumption is that those functions cannot be efficiently inverted even by a quantum computer. Using those functions, it is possible to classically delegate a quantum computation to a fully untrusted quantum server [Mah18].

### Efficient state discrimination

A yet weaker approach to the certification of quantum sampling devices than minimal certification (Def. 5.2) is the task of discriminating the imperfect distribution from fixed classically simulatable distributions. Indeed, one can see the minimal certification task as distinguishing the imperfect preparation against *all possible* distributions that are at least $\epsilon$-far away from the target distribution. This task was considered by Gogolin et al. [Gog+13] in a setting of a highly restricted client aiming to verify a boson sampler just from the histogram of outcomes without using the information about *which outcome was obtained*. They showed that in this setting, a boson sampling distribution cannot be distinguished from uniform and prompted the development of a fully efficient and simple state discrimination test that makes use of the actual outcomes [AA14].

The discriminating test is given by the so called 'row-norm estimator' for a matrix $X \in \mathscr{C}(n \times n)$

$$R^*(X) = \frac{1}{n^n} \prod_{i=1}^{n} R_i(X), \tag{7.42}$$

where $R_i(X) = \|x_i\|_2^2 := |x_{i1}|^2 + \cdots + |x_{in}|^2$ is the norm-squared of the $i$th row of $X$. Indeed, for a Gaussian normal matrix $X \sim \mathcal{N} \equiv \mathcal{N}_{\mathbb{C}}(0,1)^{n\times n}$ one expects $\mathbb{E}_{X\sim\mathcal{N}}[R^*(X)] = 1$. The differing fluctuations around this value depending on whether experimental samples are chosen from the boson sampling distribution or a uniform distributoin can be exploited to discriminate a device from uniform.

This is done by computing $R^*(U_S)$ for a few samples $S$ and comparing the outcome to one's expectation. To see this, let $\mathscr{H}$ be the distribution $\mathcal{N}$ with distribution function $p_{\mathcal{N}}(X)$ scaled by the probability of obtaining the corresponding outcome, i.e., $p_{\mathscr{H}}(X) = p_{\mathcal{N}}(X)P(X)$. When specializing to boson sampling, the matrix $X$ will be an approximately Gaussian-distributed submatrix $U_S$ of the linear-optical unitary $U$. Remember that the probability of obtaining this matrix, that is, the outcome $S$ is given by $P_U(S) = |\operatorname{Perm}(U_S)|^2/n!$ (cf. Eq. (6.36)).

One finds that [AA14, Corollary 18]

$$\Pr_{\mathcal{H}}[R^* \geq 1] - \Pr_{\mathcal{N}}[R^* \geq 1] = \frac{1}{2}\mathbb{E}_{\mathcal{N}}\left[|R^* - 1|\right] \geq 0.146 - O\left(\frac{1}{\sqrt{n}}\right), \quad (7.43)$$

In other words, the row norm estimator $R^*(X)$ is only ever so slightly correlated with $\mathrm{Perm}(X)$. An intuitive reason for this is that multiplying every row of $X$ by the same scalar $c$ also multiplies $\mathrm{Perm}(X)$ by $c$ [AA14]. At the same time, it can be computed in time $O(n^2)$.

To discriminate a boson sampler from the uniform distribution, one therefore needs to simply collect $k$ samples $S_1, \ldots, S_k$ from a device claimed to realize a boson sampler and compute

$$\frac{1}{k}\sum_{i=1}^{k}\left|R^*(U_{S_k}) - 1\right| \quad (7.44)$$

up to sufficiently high precision so as to confidently distinguish[12] the resulting value from 0.

12: This may be done in a Bayesian framework [Car+14].

In the same framework, one can distinguish a boson sampler against other – somewhat more informed – distributions such as a distribution of distinguishable particles that are sent through the linear-optical network [Car+14; Spa+14].

## 7.4 Efficient quantum fidelity witnessing

This section is based on joint work with Martin Kliesch, Martin Schwarz and Jens Eisert, published in Ref. [Han+17].

Until now we have conceived of the quantum sampler as a black box that takes classical numbers as input and output; see Figs. 5.1 and 5.2. But in an experimental scenario, parts of the experiment can often be characterized and calibrated to a very high precision so that those parts can be trusted with high confidence. Indeed, if this were not the case then it would be impossible to perform meaningful experiments in order to test specific functional dependencies in nature. For how could one know then whether a negative outcome of an experiment was due to a miscalibrated apparatus or whether the natural law to be tested did not hold in the first place?

Therefore, we will now leave the realm of fully classical certification in which the appartus is untrusted. We will instead make the assumption that the standard-basis measurement apparata can be trusted with high confidence, giving rise to the measurement device-dependent scenario. This assumption is entirely different in kind when compared to assumptions on the *global effect* of the noise on the outcome probability distribution $P_U$[13] : it is an assumption on single-qubit measurements and therefore *local*. This means that it can be verified to the same degree that one can characterize those measurement apparata. Still: there is a vicious circle lurking – how can we hope to verify the correct functioning of the quantum measurement apparata if this requires a trust assumption in the first place. It turns out, though, that using tools such as gate set tomography [Mer+13; Blu+13; Gre15; Blu+17; COB20] or the device-independent verification of quantum processes and instruments [Sek+18] this task can be achieved. Now, why do we not use those powerful tools to verify quantum samplers in the first place? The answer is simple: they

13: Recall, for instance the assumption $H(Q) \geq H(P_U)$ that was necessary for a cross-entropy based test to yield bounds on the total-variation distance.

are extremely inefficient and scale exponentially in the system size. But because we are only talking about single-qubit measurements there is no exponential scaling involved when characterizing individual single-qubit detectors and it can – at least in principle – be performed efficiently, namely, with an overall complexity that scales linearly in the number of measurement apparata to be certified.

It is not clear whether using such weak assumptions as trust in single-qubit measurements, it is possible to verify tasks that are not classically simulatable, not to speak of universal quantum computations. In the following, we will show that this is indeed possible using extremely simple protocols that – no surprises here – are entirely measurement-based. To this end, we will show that one can fully efficiently verify the ground state of an (at least inverse-polynomially) gapped Hamiltonian if the ground state energy of that Hamiltonian is known a priori. We do so by constructing a *fidelity witness* for this ground state. In a second step, we will then show how this protocol can be applied to universal quantum computations and, in particular, the quantum simulation architectures presented in Section 4.2.

Let us begin by discussing the fidelity as a distance measure on quantum states and defining the notion of a fidelity witness[14]. Recall from Section 2.4 that the natural and operationally meaningful distance measure on quantum state space is the trace distance defined in Eqs. (2.76) and (2.77). The *fidelity* between two quantum states $\rho$ and $\sigma$

14: To the best of my knowledge, the formalization of the idea of a fidelity witness is due to Ref. [Glu+18].

$$F(\rho,\sigma) = \mathrm{Tr}\left[\left(\sqrt{\rho}\sigma\sqrt{\rho}\right)^{1/2}\right]^2 \qquad (7.45)$$

is a weaker but somewhat more practical measure of distance. It is a measure of the overlap of the states. And, indeed, if $\rho$ or $\sigma$ is pure, then it reduces to $F(\rho,\sigma) = \mathrm{Tr}[\rho\sigma]$. While the fidelity itself does not have an operational interpretation in terms of distinguishability, it does provide – somewhat loose – upper and lower bounds on the trace distance by the Fuchs-van de Graaf inequality (see Fig. 7.3)

$$1 - \sqrt{F(\rho,\sigma)} \leq d_1(\rho,\sigma) \leq \sqrt{1 - F(\rho,\sigma)}. \qquad (7.46)$$

Analogously to the general definition of a weak membership quantum state certification (Def. 5.3), we can phrase quantum state certification as a weak-membership protocol in terms of the fidelity [Aol+15; Han+17].

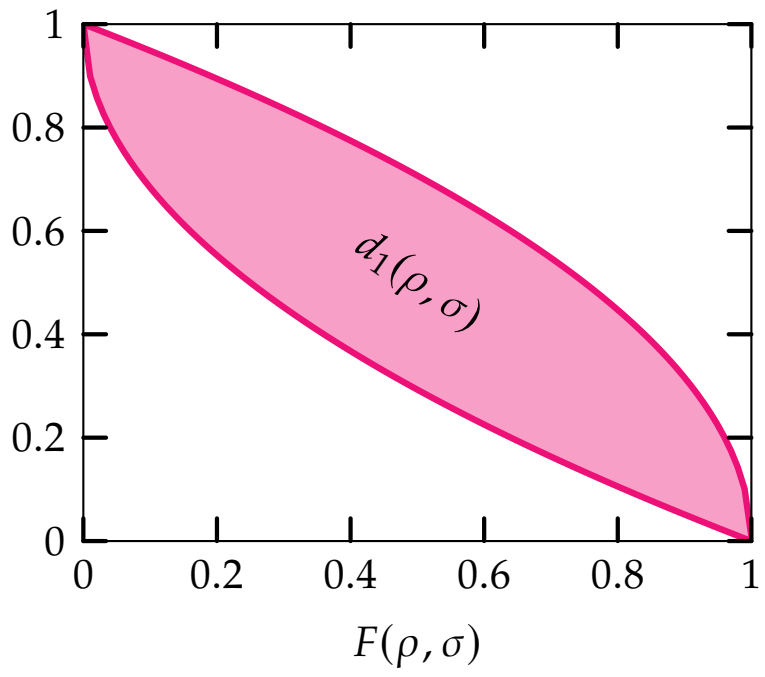

**Figure 7.3:** Lower and upper bounds to the trace distance $d_1(\rho,\sigma)$ in terms of the fidelity $F(\rho,\sigma)$ as given by the Fuchs-van de Graaf inequality (7.46).

**Definition 7.1** (Weak-membership quantum state certification) *Let* $F_T$ *be a threshold fidelity and* $0 < \alpha < 1$ *a maximal failure probability. A test which takes as an input a classical description of a target state* $\rho$ *and copies of a state preparation* $\sigma$ *and outputs 'reject' or 'accept' is a* weak-membership certification test *if with probability at least* $1-\alpha$ *it rejects every* $\sigma$ *for which* $F(\rho,\sigma) \leq F_T$ *and accepts every* $\sigma$ *for which* $F(\rho,\sigma) \geq F_T + \delta$ *for some fidelity gap* $\delta > 0$.

An easy way to construct such a weak-membership test is, of course, to estimate the fidelity $F(\rho,\sigma)$ up to a bounded error. But it turns out that even less suffices to construct a weak-membership quantum state certificate: one can construct a witness of the fidelity.

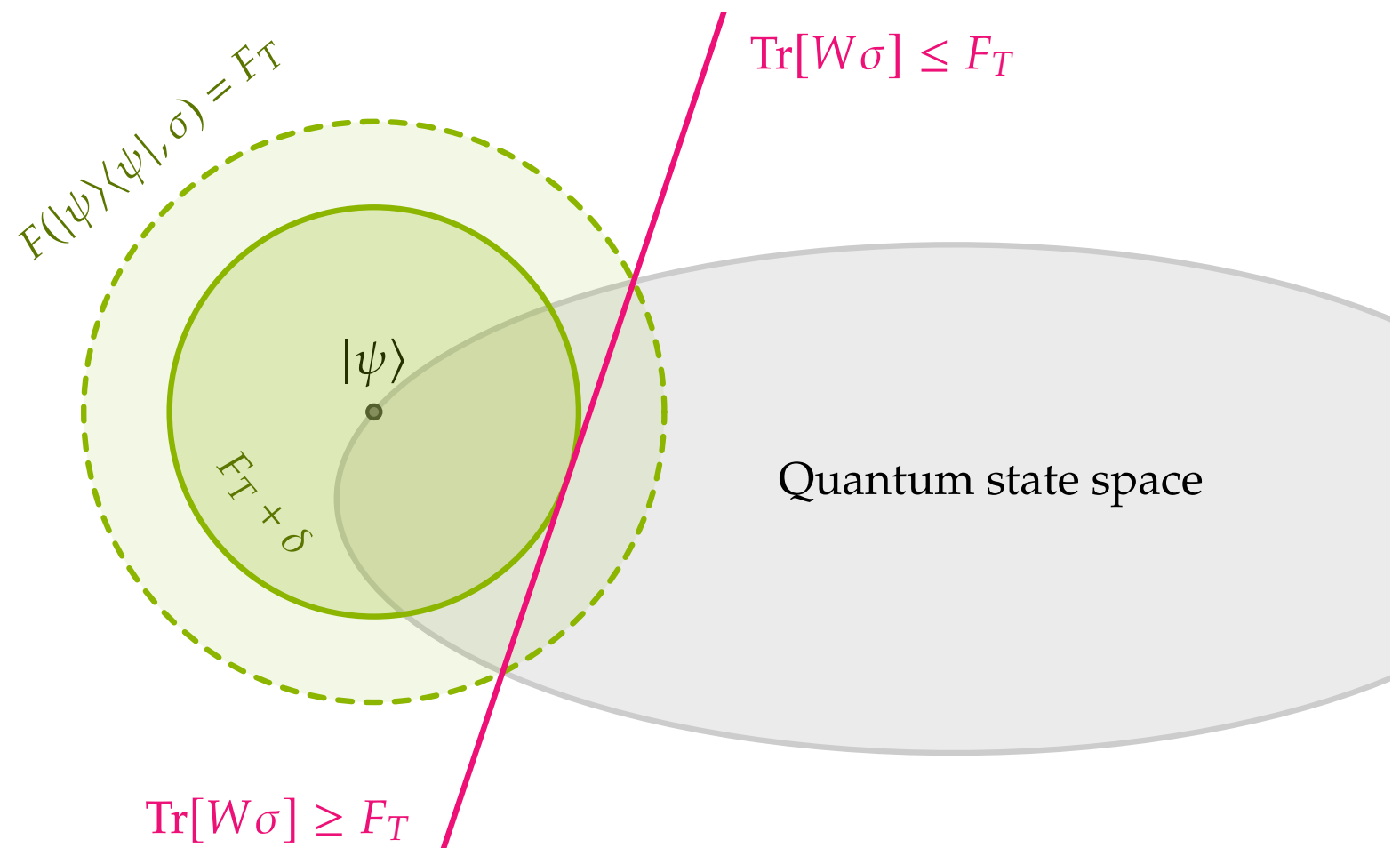

**Figure 7.4:** Given a target state $\rho = |\psi\rangle\langle\psi|$, a fidelity witness $W$ for $\rho$ provides a lower bound on the fidelity $F(\rho, \sigma) \geq \mathrm{Tr}[W\sigma]$ so that, in particular, all states $\sigma$ such that $F(\rho, \sigma) \leq F_T$ it also holds that $\mathrm{Tr}[W\sigma] \leq F_T$. Conversely, all states $\sigma$ satisfying $\mathrm{Tr}[W\sigma] \geq F_T$ will also satisfy $F(\rho, \sigma) \geq F_T$. There is a gap $\delta \geq 1 - F_T$ such that all states $\sigma$ with fidelity $F(\rho, \sigma) \geq F_T + \delta$ lie on the left side of the witness.

**Definition 7.2** (Fidelity witness [Glu+18]) *An observable $W$ is a fidelity witness for a target state $\rho$, if*

*i.* $\mathrm{Tr}[\sigma W] = 1$ iff $\rho = \sigma$
*ii.* $\mathrm{Tr}[\sigma W] \leq F(\rho, \sigma)$.

Conceptually speaking, fidelity witnesses are very much like entanglement witnesses [GT09] in that they cut a hyperplane through quantum state space, which detects a property of quantum states: Those states that lie on the left of the hyperplane defined by $\mathrm{Tr}[W\sigma] \geq F_T$ are guaranteed to have a high fidelity of at least $F_T$ since $\mathrm{Tr}[W\sigma]$ lower-bounds $F(\rho, \sigma)$. For those states on the right of the hyperplane – satisfying $\mathrm{Tr}[W\sigma] < F_T$ we cannot make a statement about their fidelity. Conversely though, all states $\sigma$ with low fidelity $F(\rho, \sigma) \leq F_T$ are guaranteed to lie to the right of the hyperplane as $\mathrm{Tr}[W\sigma] \leq F(\rho, \sigma) \leq F_T$. We illustrate the idea of a fidelity witness in Fig. 7.4.

## Efficiently verifying ground states of gapped Hamiltonians

We are now ready to state the certification protocol for ground states of gapped Hamiltonians. We consider a ground state $\rho$ of a Hamiltonian $H = \sum_\lambda h_\lambda$ with energy gap $\Delta$ above the ground state as the target state. This state is to be certified by a sequence of $m$ (local) measurements of each of the $n$ Hamiltonian terms $h_\lambda$ followed by efficient post-processing of the outcomes. For simplicity we normalize the ground state energy to 0.

*The following part has up to minor edits already been published in Ref. [Han+17].*

↓ ↓ ↓

Let us now prove that Protocol 7.2 is in fact a weak-membership quantum certification test in the Sense of Def. 5.3. The subsequent proposition makes the resources needed for such a certification more precise, and sets the fidelity gap $\delta$ for which Protocol 7.2 is a weak-membership test.

**Proposition 7.1** (Weak-membership certification) *Let $H = \sum_\lambda h_\lambda$ be a gapped Hamiltonian with unique ground state, ground state energy $E_0 := 0$, gap $\Delta$, and interaction strength*[15] $J = \max_\lambda \|h_\lambda\|$. *Then Protocol 7.2 is a*

15: Recall the definition of the operator norm $\|\cdot\|_\infty$ as the limit of (2.76) for $p \to \infty$. As is common, we will abbreviate the operator norm as $\|\cdot\|_\infty = \|\cdot\|$.

**Protocol 7.2** Certification protocol for gapped Hamiltonians

**Input:** A threshold fidelity $F_T < 1$, maximal failure probability $1 > \alpha > 0$, tolerated estimation error $\epsilon \leq (1 - F_T)/2$.

**Input:** $k = n \cdot m$ iid. copies of the prepared state $\sigma$, which is an imperfect preparation of the ground state $\rho$ of the (local) Hamiltonian $H = \sum_\lambda h_\lambda$.

1: Measure each of the $n$ Hamiltonian terms $h_\lambda$ $m$ times on the copies of $\sigma$ to determine an estimate $E^*$ of the expectation value $\mathrm{Tr}[H\sigma]$. Each measurement is performed on a new copy of $\sigma$.

2: From the estimate $E^*$ obtain an estimate $\langle W_\rho \rangle_\sigma^* = 1 - E^*/\Delta$ of the witness operator $W_\rho = 1 - H/\Delta$ such that $\langle W_\rho \rangle_\sigma^* \in [\langle W_\rho \rangle_\sigma - \epsilon, \langle W_\rho \rangle_\sigma + \epsilon]$ with probability at least $1 - \alpha$.

**Output:** If $\langle W_\rho \rangle_\sigma^* < F_T + \epsilon$ reject, otherwise accept.

*weak-membership certification test with fidelity gap*

$$\delta = (1 - F_T)\left(1 - \frac{\Delta}{\|H\|}\right) + \frac{2\epsilon\Delta}{\|H\|}, \tag{7.47}$$

*and requires*

$$m \geq \frac{J^2 n^2}{2\Delta^2 \epsilon^2} \ln\left[-\frac{n+1}{\ln(1-\alpha)}\right]. \tag{7.48}$$

*measurements of each of the $n$ local terms on $\sigma$ to determine the expectation value $\langle H \rangle_\sigma = \mathrm{Tr}[H\sigma]$.*

With Protocol 7.2 one is therefore able to efficiently certify ground state preparations of polynomially gapped Hamiltonians $H$ that are at least $1/\mathsf{poly}(n)$ close to the target state in fidelity. We show the acceptance and rejection region of the fidelity witness as a function of the total-variation distance bound in Fig. 7.5; compare also Fig. 4.7. We also show its dependence on the number of qubits for different values of the rejection threshold $1 - F_T$ in Fig. 7.6.

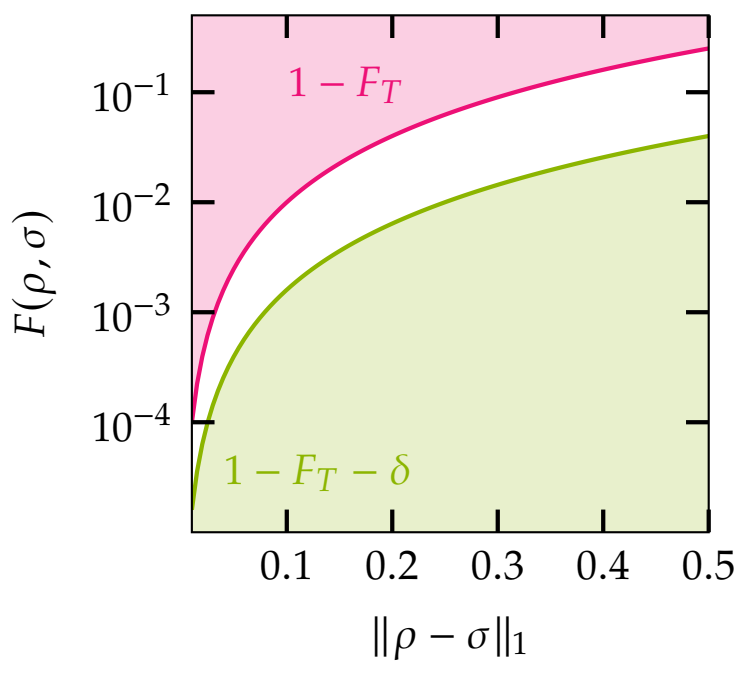

**Figure 7.5:** (a) We show the acceptance and rejection infidelity threshold of the fidelity witness (Protocol 7.2) as a function of the total-variation distance bound $\epsilon$, choosing $\epsilon = (1 - F_T)/10$, $\Delta = 2$, $\|H\| = 10$.

To keep the notation simple, we have assumed that the ground state energy $E_0 = 0$. Notice, though, that the general case with arbitrary $E_0$ can be reduced to it whenever the value of $E_0$ is known a priori. For instance, this is the case whenever $H$ is frustration-free, since in this case the global ground state energy can be obtained from the local ones. Moreover, in order to obtain the estimate $\langle W_\rho \rangle_\sigma^*$ of the expectation value $\langle W_\rho \rangle_\sigma$ of the witness, the value of the energy gap $\Delta$ needs to be known.

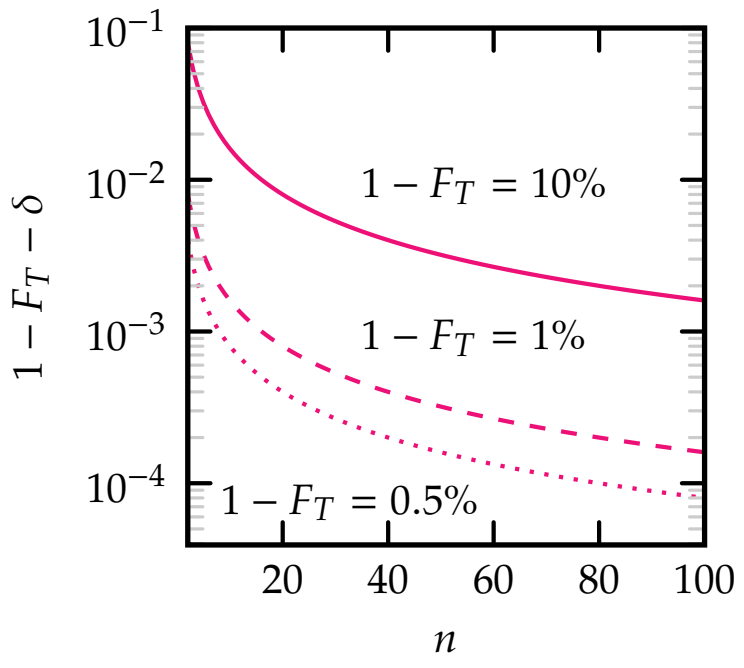

**Figure 7.6:** We show the infidelity acceptance threshold $1 - F_T - \delta$ of Protocol 7.2 as a function of the number of qubits $n$, assuming threshold infidelities of 0.5% (dotted), 1% (dashed), and 10% (solid), respectively. We have chosen an error tolerance of $(1 - F_T)/10$ and set the Hamiltonian paramters $\|H\| = n$ and $\Delta = 2$ to those for the cluster state (7.66); see below.

Note that *no* assumptions on the prepared state $\sigma$ are made, in particular, it need not be pure. Note also that it does not matter how the measurement of the total energy $\langle H \rangle$ is performed; depending on the setup at hand, it may be more suitable to measure the energy directly rather than measure all terms individually as insinuated in Step 3 of the protocol. One major advantage of this approach is that one can perform the output measurement on the same copies as the certification measurement. This means that our certification protocol can be simply integrated in the readout protocol of a random quantum sampling scheme: perform the certification measurements on the copies of $\sigma$ first and then use the same states for the readout measurements if the certification test accepts $\sigma$.

The key ingredient for the proof of Proposition 7.1 is the construction of a fidelity witness $W_\rho$, which provides a lower bound to the fidelity of $F(\rho, \sigma)$. We do so by upper- and lower-bounding the fidelity in terms of the expectation value[16] $\langle H \rangle_\sigma$.

16: Thereby, we are complementing the lower bound from Ref. [Cra+10].

**Lemma 7.2** (Bounds on the fidelity) *Given a Hamiltonian $H$ with gap $\Delta > 0$ above the unique ground state $\rho$ with energy $E_0 = 0$, and maximum energy $E_{\max} = \|H\|$. Suppose $\langle H \rangle_\sigma < E_1$, the energy of the first excited state, for some prepared state $\sigma$. Then the overlap $F(\rho, \rho)$ of $\sigma$ with the ground state is bounded as*

$$F_{\max}(\sigma) := 1 - \frac{\langle H \rangle_\sigma}{\|H\|} \geq F(\rho, \sigma) \geq 1 - \frac{\langle H \rangle_\sigma}{\Delta} =: F_{\min}(\sigma) . \tag{7.49}$$

*Proof.* Decompose $H = \sum_{i=0}^{n} E_i |E_i\rangle\langle E_i|$. Then,

$$\begin{aligned} \|H\|(1 - \mathrm{Tr}[\rho\sigma]) &= E_{\max} \sum_{i>0} \langle E_i, \sigma E_i \rangle \\ &\geq \mathrm{Tr}[H\sigma] \geq \Delta \sum_{i>0} \langle E_i, \sigma E_i \rangle \\ &= \Delta(1 - \mathrm{Tr}[\rho\sigma]) , \end{aligned} \tag{7.50}$$

(equivalently $\|H\|(1 - P_0) \geq H \geq \Delta(1 - P_0)$, where $P_0$ projects onto the ground space) yielding $1 - \mathrm{Tr}[H\sigma]/\|H\| \geq F(\rho, \sigma) \geq 1 - \mathrm{Tr}[H\sigma]/\Delta$. □

This means that the operator $W_\rho = 1 - H/\Delta$ is a fidelity witness of $\rho$: for $\sigma = \rho$ we have $\mathrm{Tr}[W_\rho \rho] = \mathrm{Tr}[(1 - H/\Delta)\rho] = 1 - 0/\Delta = 1$, and on the other hand, $\mathrm{Tr}[W_\rho \sigma] \leq F(\rho, \sigma)$ as shown in the lemma.

In order to show Proposition 7.1, we require some detail on how to estimate the global energy, or equivalently, the expectation value of the witness operator $W_\rho$ from measurements of local Hamiltonian terms, cast into the form of a large-deviation bound. Such a bound is given by the following Lemma, which is stated and proved along the same lines as Lemma S4 in Ref. [Aol+15].

**Lemma 7.3** (Estimation of the energy) *Decompose the local terms in their eigenbasis: $h_\lambda = \sum_\mu e_{\lambda,\mu} P_{\lambda,\mu}$, where the $P_{\lambda,\mu}$ are orthogonal projections onto the eigenspaces corresponding to eigenvalues $e_{\lambda,\mu}$. Let $X_\lambda^{(i)}$ be the random variable that takes the value $e_{\lambda,\mu}$ with probability $\mathrm{Tr}(\sigma P_{\lambda,\mu})$ by the measurement of $h_\lambda$ on the $i^{th}$ copy of $\sigma$. Moreover, let*

$$\langle h_\lambda \rangle_\sigma^* = \frac{1}{m} \sum_{i=1}^{m} X_\lambda^{(i)} \tag{7.51}$$

*be the estimate of $\langle h_\lambda \rangle$ on $\sigma$ by a finite-sample average of $m$ measurement outcomes, and $\langle H \rangle_\sigma^* = \sum_\lambda \langle h_\lambda \rangle_\sigma^*$ the resulting estimate of $\langle H \rangle_\sigma$. As above, define $J = \max_\lambda \|h_\lambda\|$. For $\bar{\alpha} \geq 1/2$ it holds that*

$$\Pr\left[ |\langle H \rangle_\sigma^* - \langle H \rangle_\sigma| \leq \epsilon \right] \geq \bar{\alpha} , \tag{7.52}$$

*whenever*

$$m \geq \frac{J^2 n^2}{2\epsilon^2} \ln\left[ \frac{n+1}{\ln(1/\bar{\alpha})} \right] . \tag{7.53}$$

*Proof.* By Hoeffding's inequality,

$$\forall \lambda \in [n]: \Pr\left[|\langle h_\lambda\rangle^*_\sigma - \langle h_\lambda\rangle_\sigma| \geq \frac{\epsilon}{m}\right] \leq 2\mathrm{e}^{-2\epsilon^2/m\|h_\lambda\|^2}, \tag{7.54}$$

since the $\{X^{(i)}_\lambda\}_\lambda$ are independent random variables and $0 \leq X^{(i)}_\lambda \leq \|h_\lambda\|$. Eq. (7.54) is equivalent to

$$\forall \lambda \in [n]: \Pr\left[|\langle h_\lambda\rangle^*_\sigma - \langle h_\lambda\rangle_\sigma| \leq \epsilon\right] \geq 1 - 2\mathrm{e}^{-2m\epsilon^2/\|h_\lambda\|^2}, \tag{7.55}$$

and since all measurements are independent

$$\Pr\left[\forall \lambda \in [n]: |\langle h_\lambda\rangle^*_\sigma - \langle h_\lambda\rangle_\sigma| \leq \epsilon\right] \tag{7.56}$$

$$\geq \prod_\lambda \left(1 - 2\mathrm{e}^{-\frac{2m\epsilon^2}{\|h_\lambda\|^2}}\right) \geq \left(1 - 2\mathrm{e}^{-2m\epsilon^2/J^2}\right)^n, \tag{7.57}$$

resulting in

$$\Pr\left[|\langle H\rangle^*_\sigma - \langle H\rangle_\sigma| \leq \epsilon\right] \geq \left(1 - 2\mathrm{e}^{-2m\epsilon^2/n^2J^2}\right)^n \geq \bar{\alpha}. \tag{7.58}$$

Eq. (7.58) holds whenever

$$m \geq \frac{J^2 n^2}{2\epsilon^2} \ln\left[\frac{2}{1-\bar{\alpha}^{1/n}}\right] =: m_{\mathrm{opt}}, \tag{7.59}$$

where we can upper-bound $m_{\mathrm{opt}}$ as $m_{\mathrm{opt}} \leq (Jn^2/2\epsilon^2)\ln[(n+1)/\ln(1/\bar{\alpha}])$ [Aol+15]. This shows the claim. □

From Lemma 7.3 we obtain that the estimator $\langle W_\rho\rangle^*_\sigma = 1 - \langle H\rangle^*_\sigma/\Delta$ of the expectation value $\langle W_\rho\rangle_\sigma = \mathrm{Tr}[W_\rho\sigma]$ of the witness operator $W_\rho = 1 - H/\Delta$ resulting from estimates of the energy estimates satisfies

$$\Pr\left[\left|\langle W_\rho\rangle^*_\sigma - \langle W_\rho\rangle_\sigma\right| \leq \epsilon\right] \geq 1 - \alpha, \tag{7.60}$$

whenever

$$m \geq \frac{J^2 n^2}{2\Delta^2\epsilon^2} \ln\left[-\frac{n+1}{\ln(1-\alpha)}\right]. \tag{7.61}$$

Putting everything together we can prove completeness and soundness for Proposition 7.1.

*Proof of Proposition 7.1.* (i) (Completeness) Let $\sigma$ be such that $F(\rho,\sigma) \geq F_T + \delta$ with $\delta$ given in Eq. (7.47). Then

$$F(\rho,\sigma) \geq F_T + (1-F_T)\left(1 - \frac{\Delta}{\|H\|}\right) + \frac{2\epsilon\Delta}{\|H\|}, \tag{7.62}$$

which is equivalent to

$$F_T + 2\epsilon \leq 1 - \|H\|(1 - F(\rho,\sigma))/\Delta. \tag{7.63}$$

On the other hand, the upper bound in Eq. (7.49) implies $\langle H\rangle_\sigma \leq \|H\|(1 - F(\rho,\sigma))$ and therefore $\mathrm{Tr}[W_\rho\sigma] = 1 - \langle H\rangle_\sigma/\Delta \geq 1 - \|H\|(1 - F(\rho,\sigma))/\Delta$.

From this, together with the bound (7.63), we obtain $F_T + 2\epsilon \leq \mathrm{Tr}[W_\rho \sigma] = F_{\min}$.

Finally, it follows from (7.60) that $\Pr[|\langle W_\rho\rangle_\sigma - \langle W_\rho\rangle^*_\sigma| \leq \epsilon] \geq 1-\alpha$, so that

$$\Pr[\langle W_\rho\rangle^*_\sigma \geq F_T + \epsilon] \geq 1 - \alpha\,. \tag{7.64}$$

Hence, with probability larger than $1-\alpha$ the test described in Protocol 7.2 accepts $\sigma$.

(ii) (Soundness) Let $\sigma$ be such that $F(\rho,\sigma) < F_T$. It then follows that $\langle W_\rho\rangle_\sigma \leq F(\rho,\sigma) < F_T$. Hence,

$$\Pr[\langle W_\rho\rangle^*_\sigma < F_T + \epsilon] > 1 - \alpha \tag{7.65}$$

is implied by (7.60) so that $\sigma$ is rejected with probability at least $1-\alpha$. □

## Application to universal computations

We have developed a weak-membership certification test for ground states of Hamiltonians with a known gap $\Delta \geq 1/\mathsf{poly}(n)$ and ground state energy $E_0$. But are such states able to encode classically intractable quantum computations at all? It turns out that, indeed, one can apply Protocol 7.2 to verify arbitrary universal quantum computations.

To see this, we once again make use of the framework of measurement-based quantum computation (MBQC) [RB01; RBB03b]. As we saw above (Section 4.2) a universal computation can be driven by certain single-qubit measurements in the $X$-$Y$ plane on a square lattice cluster state[17]

17: Recall that Mantri, Demarie, and Fitzsimons [MDF17] showed that those measurements are sufficient for universal cluster-state based quantum computations.

$$|\mathrm{CS}\rangle = \left(\prod_{\langle i,j\rangle} CZ_{i,j}\right) H^{\otimes N}|0^N\rangle, \tag{7.66}$$

where the symbol $\langle i,j\rangle$ denotes nearest neighbours in an $n\times m$ 2D square lattice and $N = n\cdot m$.

The remainder of the 'hard work' in the computation is done by single-qubit operations – adaptive measurements at the correct angles in the $X$-$Y$ plane (multiples of $\pi/8$ suffice). Assuming highly accurate single-qubit operations and measurements, we can now use Protocol 7.2 in order to verify the pre-measurement quantum state $|\mathrm{CS}\rangle$.

To do so, we need to derive a 'parent Hamiltonian' which has $|\mathrm{CS}\rangle$ as its ground state. This can be done easily by observing that the diagonal Hamiltonian

$$H_0 = -\sum_{i=1}^{N} Z_i\,, \tag{7.67}$$

has the all-zero state $|0^N\rangle$ as its ground state with ground state energy $E_0 = -N$ and gap $\Delta = 2$. Our strategy to derive a parent Hamiltonian $H$ of $|\mathrm{CS}\rangle$ is based on the observation that conjugation by unitary transformations $U$ preserves the eigenvalues so that $U|0^N\rangle$ is a ground state of $UH_0U^\dagger$ with ground state energy $E_0$ and gap $\Delta$.

Inserting $U = (\prod_{\langle i,j\rangle} CZ_{i,j})H^{\otimes N}$ and using the relation $CZ(X \otimes 1)CZ = X \otimes Z$ we obtain that the Hamiltonian

$$H = \sum_{i=1}^{N} \left( X_i \cdot \prod_{j \in \partial i} Z_j \right), \tag{7.68}$$

is a parent of $|\mathrm{CS}\rangle$ with ground state energy $E_0 = -N$ and gap $\Delta = 2$. Here, $\partial i = \{j \in V : (i,j) \in E\}$ denotes the neighbourhood of site $i$ on a graph $G = (V, E)$. The operators $S_i = X_i \otimes \sum_{j \in \partial\{i\}} Z_j$ are often called *star operators* as they act on the neighbours of the central site $i$. For square lattices they are 5 local; see Fig. 7.7 for an illustration.

Notice that we can measure the operators $S_i$ using single-qubit measurements of $X_i$ and the respective $Z_j$ in the neighbourhood of $i$ and multiplying the outcomes. Notice also that we can group the operators $S_i$ into two groups $A$ and $B$ within which the star operators are mutually commuting using a bicoloring of the square lattice. It follows that we can decompose the Hamiltonian $H$ into two terms with norm at most[18] $N/2$ each of which can be measured using single-qubit measurements only. An imperfect preparation $\sigma$ of the cluster state $|\mathrm{CS}\rangle$ can thus be witnessed using

$$m_{\mathrm{CS}} \gtrsim \frac{n^2}{2\epsilon^2} \ln \left[ -\frac{n+1}{\ln(1-\alpha)} \right], \tag{7.69}$$

many copies of $\sigma$.

18: Each $S_i$ clearly has operator norm $\|S_i\| = 1$.

Of course, the very same method gives rise to a parent Hamiltonian for the output state of arbitrary quantum computations. However, the resulting Hamiltonian will in general not only be highly non-local but will also involve local terms that are exponentially large sums of product operators. This means that the energy cannot be efficiently estimated using single-qubit measurements only.

## Application to quantum supremacy architectures

The fidelity witnessing protocol (Protocol 7.2) can be very naturally be applied to the quantum supremacy architecture presented in Section 4 as well as more general types of IQP circuits [BMS16; MSM17].

**Quantum simulation architectures** The quantum simulation architectures from Section 4 directly exploit the idea of measurement-based computation and are therefore verifiable via the protocol described above. Recall that the schemes were based on the short time evolution (4.5)

$$U_\beta = H^{\otimes N} \left( \prod_{(i,j)\in E} CZ_{i,j} \prod_{i \in V} \mathrm{e}^{-\mathrm{i}\beta_i Z_i/2} \right) H^{\otimes N}, \tag{7.70}$$

which incorporates the preparation of a cluster state and single-qubit unitaries followed by a final layer of Hadamard gates. The state $U_\beta|0^N\rangle$ can be considered a 'generalized cluster state' since it is equivalent to a cluster state up to the local unitary $R_\beta = H^{\otimes N}\prod_{i=1}^N e^{-i\beta_i Z_i/2}$.

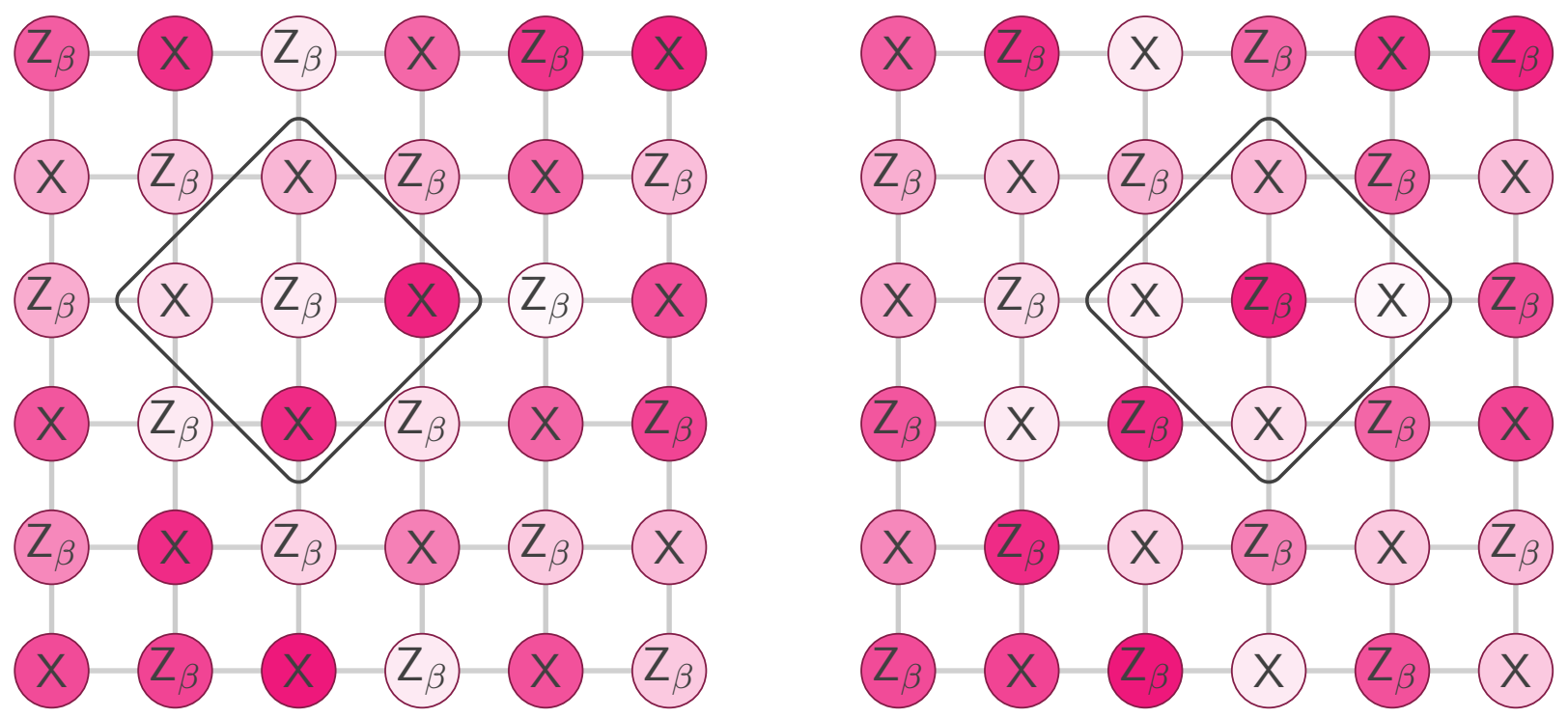

**Figure 7.7:** To certify the quantum simulation scheme of Fig. 4.2 as defined by the time evolution $U_\beta$in (7.70) two patterns of $X$ and locally rotated $Z$-measurements $Z_\beta = e^{-i\beta X/2} Z e^{i\beta X/2}$in the $Z$-$Y$ plane need to be performed.

As above, we can derive a parent Hamiltonian $H_\beta$ and corresponding witness $W_\beta = 1 - H_\beta/2$ for the state $U_\beta|0^N\rangle$ by conjugating $H_0$ with the time-evolution unitary $U_\beta$

$$H_\beta = U_\beta H_0 U_\beta^\dagger = -\sum_{i=1}^{N}\left(e^{-i\beta_i X_i/2} Z_i e^{i\beta_i X_i/2}\prod_{j\in\partial i} X_j\right) \equiv -\sum_{i=1}^{N}\left(Z_\beta \prod_{j\in\partial i} X_j\right), \tag{7.71}$$

which is a sum of star-shaped terms centred at site $i$; see Fig. 7.7. The state $U_\beta|0^N\rangle$ – like the cluster state itself – can be certified from $m_{CS}$ (cf. Eq. (7.69)) many copies of the imperfect state.

**IQP circuits** The protocol can also be directly applied to more general IQP circuits than the ones of the quantum simulation architecture. For instance, for the IQP circuits (6.25) on an arbitrary graph

$$C_W = \exp\left[i\left(\sum_{i<j} w_{i,j} X_i X_j + \sum_i w_{i,i} X_i\right)\right], \tag{7.72}$$

with interaction weights $w_{i,j} \in \{0, \pi/4\}$ for $i \neq j$ and local phases $w_{i,i} \in \pi/8 \cdot \mathbb{Z}$ we obtain the parent Hamiltonian

$$H_W = U_W H_0 U_W^\dagger = -\sum_{i=1}^{n}\left(e^{i\tilde{w}_{i,i}X_i} Z_i e^{-i\tilde{w}_{i,i}X_i}\prod_j X_j^{\frac{4}{\pi}w_{i,j}}\right), \tag{7.73}$$

where $\tilde{w}_{i,i} = w_{i,i} + \deg(i)\pi/4$. The energy of this Hamiltonian can be directly measured using single-qubit measurements of $Z$-$Y$ plane and $X$ measurements since each local term is a product of those operators.

Notice that in order to obtain a Hamiltonian $H_W$ it is required that the weights $w_{i,j}$ for $i \neq j$ are such that they give rise to two-qubit interactions of $CZ$ ($XX$) type. This is a straightforward generalization of the previously derived parent Hamiltonian, where we observe that up to a global phase $CZ_{i,j} = \exp(i\frac{\pi}{4}(Z_iZ_j - Z_i - Z_j))$ (cf. Sec. 4.2). Only if this is the case, does the conjugation of $H_0$ with the unitary $U_W$ give rise to products of single-qubit operators. For coupling strengths $w_{i,j}$ that are not integer multiples of $\pi/4$, the Hamiltonian terms will in general be exponentially large sums of product terms[19].

19: An efficient fidelity estimation protocol for arbitrary weighted graph states as they are generated by the IQP circuit $C_W$ with arbitrary weights $w_{i,j}$ has been developed in Refs. [MTH17; ZH19; HT19]. Those circuits can be seen to give rise to graph states in which not only vertices (as in the example above) but also edges can have arbitrary weights, so-called *weighted graph states*.

The presented protocol has also been applied to the verification of IQP circuits the diagonal part of which comprises $Z$, $CZ$ and the non-Clifford $CCZ$ gate [MSM17] as they are also considered by Bremner, Montanaro, and Shepherd [BMS16]. Such circuits are defined in terms of Boolean parameters $a_{ijk}, b_{ij}, c_i \in \{0, 1\}$ that are nonzero whenever a gate is applied. Their output probabilities are given by $|\,\mathrm{ngap}(f)|^2$ of the degree-3 Boolean polynomial[20]

$$f(x) = \sum_{ijk} a_{ijk} x_i x_j x_k + \sum_{ij} b_{ij} x_i x_j + \sum_i c_i x_i. \tag{7.74}$$

20: Recall that we introduced those circuit families in Eq. (2.52).

While the resulting non-local stabilizers $h_i$ are not directly products of Pauli operators in the same way as we obtained $CZ(X \otimes 1)CZ = X \otimes Z$, Miller, Sanders, and Miyake [MSM17] show that single-qubit Pauli-$X$ and $Z$ measurements suffice to measure those stabilizers. More precisely, a measurement of the stabilizer $h_i$ can be achieved by measuring $X_i \cdot \prod_{j \neq i} Z_j$ with outcome $v = (v_1, \ldots, v_n)$ and returning $(-1)^{\partial_i f(v) + v_i}$, where $\partial_i f(x) = f(x_1, \ldots, x_i + 1, \ldots, x_n) - f(x_1, \ldots, x_i, \ldots, x_n)$[21].

21: We refer the interested reader to Ref. [MSM17] for further details.

## 7.5 Rapid quantum fidelity estimation

The following section is based on joint work with Juani Bermejo-Vega [BH].

In the previous section, we have derived a certification protocol for ground states of gapped Hamiltonians that was based on an experimentally motivated trust assumption on the quantum measurement apparatus. Conceptually, the protocol was based on a fidelity witness that can be measured efficiently from measurements of local Hamiltonian terms. The protocol is fully efficient both in terms of the quantum sample complexity and the classical post-processing.

When it comes to its practical implementation, the fidelity witness has two major drawbacks, however. First, the fidelity gap $\delta$ (see Eq. (7.47)) is rather large so that the protocol is only guaranteed to accept imperfect state preparations $\sigma$, which have an extremely high fidelity; recall Fig. 7.5. Second, while the scaling is efficient in both the estimation error $\epsilon$ and the number of qubits $n$ in the complexity-theoretic sense of 'efficient', the sample complexity becomes unmanagably large already for reasonable-sized computations involving $n \sim 10 - 50$ qubits. For such computations and realistic parameters, on the order of $10^7 - 10^8$ many copies of the state preparation are required.

*En route* to the practical applicability of rigorous verification protocols, the aim of this section is to improve on the properties of the fidelity witness from the previous section (Protocol 7.2) in those two respects. We find a method to rapidly estimate the fidelity of a class of quantum states in a way that *does not scale* in the number of qubits in the ideal case. This method also improves on the fact that the fidelity witness has an extremely large acceptance threshold by performing *fidelity estimation* rather than witnessing. Fidelity estimation provides more information about the imperfect state preparation than a fidelity witness: while the information provided by a witness is merely whether a given state preparation is to the left or the right of some hyperplane in state space, fidelity estimation provides an estimate of *its distance* to the ideal target state as measured by the fidelity. Still, it can be achieved highly efficiently – with constant sample complexity – for certain restricted classes of quantum states.

Making use of the idea to directly esimate the fidelity via importance sampling due to Flammia and Liu [FL11], we derive a rapid fidelity estimation protocol for quantum states that can be expressed as convex mixtures of efficiently measurable operators. Among them are states which are local-unitary equivalent to graph states as they are required for the architectures for quantum supremacy presented in Section 4.2. We show that one can estimate the fidelity of imperfect state preparations to members of those restricted classes with high probability up to additive error $\epsilon$ with sample complexity $O(1/\epsilon^2)$. Rigorous quantum certification of the classical output distribution of the quantum simulation scheme for quantum supremacy can then be achieved with $\sim 10^5$ samples, a number that is now *independent* of the number of qubits.

## Fidelity estimation protocol

Our goal is to verify that an imperfect quantum state $\sigma$ generated in the laboratory is close to a pure target state $\rho = |\psi\rangle\langle\psi|$ of $n$ qubits from few iid. copies of $\sigma$. We restrict to states that can be decomposed as

$$\rho = \sum_{\lambda\in\Lambda} p_\lambda A_\lambda, \tag{7.75}$$

in terms of normal operators $\{A_\lambda\}_{\lambda\in\Lambda}$ weighted by probabilities $p_\lambda$. We can decompose $A_\lambda = \sum_{a\in\mathrm{spec}(A_\lambda)} aP^a_\lambda$ in terms of its eigenprojectors $P^a_\lambda$. We can show that for states of the form (7.75) certification is possible with *constant sample complexity* provided that the following conditions are satisfied:

i. For each $\lambda \in \Lambda$, $A_\lambda$ can be efficiently measured. In particular, this is the case if $A_\lambda = A_{\lambda_1} \otimes \cdots \otimes A_{\lambda_n}$ with $\lambda = (\lambda_1, \dots, \lambda_n)$ is a product of single-qubit operators $A_{\lambda_i}$.
ii. For each $\lambda \in \Lambda$, $\mathrm{spec}(A_\lambda) \subset [a_\lambda, b_\lambda]$ for constants $a_\lambda, b_\lambda \in \mathbb{R}$.
iii. The probability distribution $p = (p_\lambda)_{\lambda\in\Lambda}$ can be (classically) sampled efficiently.

For such states $\rho$ (7.75) we estimate the fidelity $F(\rho,\sigma) = \mathrm{Tr}[\rho\sigma]$, which gives upper and lower bounds on the trace distance $d(\rho,\sigma) = \frac{1}{2}\|\rho - \sigma\|_1$. This is achieved by the following protocol due to Flammia and Liu [FL11] and similar to Refs. [Bou+04; Kie+05; TG05; PLM18].

**Protocol 7.3** Rapid fidelity estimation

**Input:** $m$ iid. copies of $\sigma$, a classical description of $\{A_\lambda\}_\lambda$ and $(p_\lambda)_\lambda$.
1: **for** $i \in [m]$ **do**
2: Sample $\lambda_i$ with probability distribution $p : \lambda \mapsto p_\lambda$.
3: Measure $A_{\lambda^{(i)}}$ on the $i$-th copy of $\sigma$, obtaining outcome $a_{\lambda_i} \in \mathrm{spec}(A_{\lambda^{(i)}})$ with probability $\mathrm{Tr}[P^a_{\lambda^{(i)}}\sigma]$.
4: **end for**
5: Set $\hat{F}(\rho,\sigma) := \frac{1}{m}\sum_{i=1}^m a_i$.
**Output:** $\hat{F}(\rho,\sigma)$ as an estimate of $F(\rho,\sigma)$.

**Theorem 7.4** *Protocol 1 is an efficient certification protocol for $\rho$ in the sense*

*that* $|F(\rho,\sigma) - \hat{F}(\rho,\sigma)| \leq \epsilon$ *with probability at least* $1-\delta$ *provided that*

$$m \geq \log\left(\frac{2}{\delta}\right)\frac{c^2}{2\epsilon^2}, \tag{7.76}$$

*with* $c = \max_\lambda(b_\lambda - a_\lambda)$ *and conditions (i-iii) are satisfied.*

*Proof.* For the proof, notice that we can express the fidelity as

$$F(\rho,\sigma) = \sum_{\lambda\in\Lambda}\sum_{a\in\mathrm{spec}(A_\lambda)} p_\lambda \operatorname{Tr}[P^a_\lambda\sigma]\cdot a_\lambda. \tag{7.77}$$

We then apply the Hoeffding concentration bound

$$\Pr\left(\left|\hat{X} - \mathbb{E}[X]\right| \geq \epsilon\right) \leq 2\mathrm{e}^{-2m\epsilon^2/(b-a)^2}, \tag{7.78}$$

with $\hat{X} = \frac{1}{m}\sum_{i=1}^m X_i$ for bounded random variables $X_i \in [a,b]$ to the random variable that takes values $a_\lambda$ with probability $p_\lambda = \operatorname{Tr}[P^a_\lambda\sigma]$. □

## Application to generalized graph states

The rapid fidelity estimation protocol 7.3 becomes particularly simple for generalized graph states as they are applied in the quantum supremacy architectures from Section 4.2. Those states, prior to the measurement in the computational basis, are given by

$$|\varphi_\beta\rangle = U_\beta|0^N\rangle = H^{\otimes N}\prod_{\langle i,j\rangle} CZ_{i,j}\prod_{i=1}^N \mathrm{e}^{-\mathrm{i}\beta_i Z_i/2} H^{\otimes N}|0^N\rangle, \tag{7.79}$$

where $\langle\cdot,\cdot\rangle$ denotes nearest neighbours on an $n\times m = N$-qubit square lattice, and $\beta_i \in [0,2\pi)$ is picked randomly on each site. Those states are fully specified by the $N$ stabilizers[22] $S_{\beta,i} := \mathrm{e}^{-\mathrm{i}\beta_i X_i/2} Z_i \mathrm{e}^{\mathrm{i}\beta_i X_i/2}\prod_{j\in\partial i} X_j$ for each $i\in[N]$.

22: Recall from Section 7.4 that $\partial i = \{j \in V : (i,j)\in E\}$ denotes the neighbourhood of site $i$ on a graph $G=(V,E)$.

In fact, we can write the state $|\varphi_\beta\rangle\langle\varphi_\beta|$ in the form (7.75)

$$|\varphi_\beta\rangle\langle\varphi_\beta| = \frac{1}{2^N}\sum_{s\in\mathcal{S}_\beta} s, \tag{7.80}$$

where $\mathcal{S}_\beta$ is the group generated by the $N$ stabilizers $S_{\beta,i}$. Notice that all elements $s\in\mathcal{S}$ are tensor products of single-qubit Pauli-$Z$ and $X$ operators rotated by $\mathrm{e}^{-\mathrm{i}\beta_i X_i/2}$ as every stabilizer operator is, too. In particular, the following Lemma is true.

**Lemma 7.5** *Let* $\mathcal{S}$ *be the stabilizer group of the cluster state* (7.66) *on an* $n\times m = N$*-square lattice. Then*

$$\mathcal{S}_\beta = R_\beta\mathcal{S}R_\beta^\dagger, \tag{7.81}$$

*with* $R_\beta = H^{\otimes N}\prod_{i=1}^N \mathrm{e}^{-\mathrm{i}\beta_i Z_i/2}$, *i.e.,* $\mathcal{S}_\beta$ *is local-unitary equivalent to the cluster state.*

*Proof.* The generators of $\mathcal{S}$ are given by tensor products of Pauli-$X$ and $Z$ as well as single-qubit identity operators. The generators of $\mathcal{S}_\beta$ are given by tensor products of Pauli-$X$ and single-qubit identity operators which may appear on every lattice site, as well as operators $e^{-i\beta_i X_i} Z_i e^{i\beta_i X_i}$ on each site $i$. But $e^{-i\beta_i X/2}$ commutes with $X$ so that $X = e^{-i\beta_i X/2} X e^{i\beta_i X/2}$ from which the claim follows. □

This implies that one can generate all elements of $\mathcal{S}_\beta$ by generating an element of $\mathcal{S}$ and then conjugating it by $R_\beta$, which is efficient as $R_\beta$ is a tensor product of single-qubit unitaries. Moreover generating a uniformly random element of $\mathcal{S}$ can be done efficiently [Got97]: Since $\mathcal{S} = \langle S_1, \dots, S_N \rangle$ is generated by the $N$ linearly stabilizers, we can sample a uniformly random element $s \in \mathcal{S}$ by sampling $x \in \{0,1\}^N$ uniformly at random and setting $s = \prod_{i=1}^N S_i^{x_i}$. This procedure is efficiently doable within the stabilizer formalism. Moreover, due to the fact that $S_\beta$ is local-unitary equivalent to a stabilizer state, only single-qubit measurements of Pauli operators rotated by $He^{-i\beta_i X/2}$ need to be performed resulting in a uniformly random element $s_\beta = R_\beta s R_\beta^\dagger \in \mathcal{S}_\beta$.

We can now specify the rapid certification protocol (Protocol 7.3 for the state $|\varphi_\beta\rangle = U_\beta |0^N\rangle$.

**Protocol 7.4** Rapid certification of the supremacy architecture

**Input:** $m$ iid. copies of $\sigma$ and $\beta \in [0, 2\pi)^N$.

1: **for** $i \in [m]$ **do**
2: Sample $x \in \{0,1\}^N$ uniformly at random.
3: Set $s_i := \prod_{k=1}^N S_{\beta,k}^{x_k}$.
4: Measure $s_i \equiv \sum_{a \in \mathrm{spec}(s_i)} P_{s_i}^a a$ on the $i$-th copy of $\sigma$, obtaining outcome $a_{s_i} \in \mathrm{spec}(A_{\lambda_i})$ with probability $\mathrm{Tr}[P_{s_i}^a \sigma]$.
5: **end for**
6: Set $\hat{F}(|\varphi_\beta\rangle\langle\varphi_\beta|, \sigma) := \frac{1}{m} \sum_{i=1}^m a_{s_i}$.

**Output:** $\hat{F}(|\varphi_\beta\rangle\langle\varphi_\beta|, \sigma)$ as an estimate of $F(|\varphi_\beta\rangle\langle\varphi_\beta|, \sigma)$.

As each eigenvalue of a Pauli operator is either +1 or −1 the sample complexity required for an $\epsilon$-certification test is given by

$$m_{\mathrm{opt}} = \log\left(\frac{2}{\delta}\right)\frac{2}{\epsilon^2}. \tag{7.82}$$

Specifically, we can choose realistic parameters for the threshold fidelity, given the supremacy total-variation distance threshold $\epsilon$ that results from the hardness proof (cf. Section 4.4). Recall that we numerically find $\gamma = 1/e$ and choose $\varepsilon = \gamma/8$, $\delta = \gamma/2$ to obtain that with probability $\geq 0.3$ we obtain a relative-error $1/4 + o(1)$ approximation. Moreover, we choose the tolerated estimation error to be $\epsilon = (1 - F_T)/5$ and the success probability $\alpha = 0.01$. We then obtain the following numbers for the optimal sample complexity $m_{\mathrm{opt}}$ using Eq. (7.82).

| $\varepsilon$ | $\gamma(1-\delta)$ | $F_T$ | $m_{\mathrm{opt}}$ |
|---|---|---|---|
| 1/22 | 0.3 | 0.9979 | $7.8 \cdot 10^7$ |
| 1/5 | 0.07 | 0.96 | $7.5 \cdot 10^4$ |

Those numbers constitutes an improvement of around two orders of magnitude in terms of the sample complexity as compared to the sample complexity required for the fidelity witness (Protocol 7.2) on ∼ 50 qubits.

### Discussion and outlook

To even further reduce the experimental effort of verification one would need to improve the scaling in the tolerated estimation error $\epsilon$. For stabilizer states this has recently been achieved in Ref. [KKL19]. Since the states we are considering here are local-unitary equivalent to stabilizer states, this result directly applies, too. We leave the details as an exercise to the reader.

A potential drawback of the rapid fidelity estimation protocol (Protocol 7.3) as opposed to the fidelity witness (Protocol 7.2) is that it intrinsically requires a different *measurement setting* in each run of the experiment. In contrast, to evaluate the fidelity witness only two distinct measurement settings are repeated many times. So while the overall quantum sample complexity is dramatically reduced from $O(n^2)$ to $O(1)$ in the number of qubits, the measurement setting complexity is increased from $O(1)$ to $O(1/\epsilon^2)$ in the estimation error. Depending on the experimental setting at hand there may well be a trade-off between the time required to switch between settings and the time required for many repetitions of the same measurement setting.

It will be interesting to see which further classes of states can be rapidly certified via the importance sampling protocol 7.3. While this question is to some extent a theoretical question, it is also a practically minded one in that, ultimately, the operators in the expansion (7.75) need to be measurable in an experimental setup. Which types of operators beyond product operators are feasible in this respect strongly depends on the respective architecture at hand. Obvious candidates from a theoretical perspective are states with positive Wigner function representation [ME12] but also other classes such as the one considered in Refs. [NBV15; Van11]. But it is much less clear in which situations the respective point operators $A_\lambda$ can be efficiently measured.

## 7.6 Further approaches to the verification of quantum samplers

Let us conclude this chapter by mentioning a few alternative approaches to the certification of quantum states.

Optimal strategies for verifying arbitrary quantum states in the minimal sense (cf. Def. 5.2) were assessed by Pallister, Linden, and Montanaro [PLM18][23]. To this end[24], they consider 'measurement strategies' $(\mu_j, P_j)_j$ with probability vector $\mu = (\mu_1, \ldots, \mu_j)$ and $0 \leq P_j \leq 1$. In each run of the experiment one of the binary measurements $\{P_j, 1 - P_j\}$ with associated outcomes '`pass`' and '`fail`' is performed with probability $\mu_j$. The protocol then runs as follows [PLM18]

23: Similar protocols were considered in Ref. [DD18].

24: In my account of the protocol by Pallister, Linden, and Montanaro [PLM18] I follow the lecture notes of Kliesch [Kli19].

**Protocol 7.5** State certification protocol [PLM18]

**Input:** $m$ imperfect state preparations $\sigma_1, \ldots, \sigma_m$.
  **for** $i = 1, \ldots, m$ **do**
    Draw $j \sim \mu$ and measure the POVM $\{P_j, 1 - P_j\}$ on the copy $\sigma_i$.
    **if** the outcome is `'fail'`, **then**
      end the protocol and
**Output:** `'Reject'`.
    **end if**
  **end for**
**Output:** `'Accept'` if all outcomes are `'pass'`.

The probability that a state $\sigma$ passes the protocol is determined by the effective measurement operator $\Omega := \sum_j \mu_j P_j$

$$\Pr[\texttt{Accept}] = \sum_j \mu_j \operatorname{Tr}[P_j \sigma] = \operatorname{Tr}[\Omega\sigma]. \tag{7.83}$$

Any measurement strategy with nonzero gap $\Delta(\Omega) = 1 - \lambda_2(\Omega) > 0$ between the largest and second-largest eigenvalue that satisfies $\operatorname{Tr}[\Omega\rho] = 1$ gives rise to a minimal $\epsilon$-certification protocol. This is because, the probability that any state $\sigma$ satisfying $F(\rho, \sigma) < 1 - \epsilon$ is accepted is given by

$$\Pr[\texttt{Accept}] = \max_{\sigma:\operatorname{Tr}[\rho\sigma] \leq 1-\epsilon} \operatorname{Tr}[\Omega\sigma] = 1 - \Delta(\Omega)\epsilon. \tag{7.84}$$

The overall sample complexity of Protocol 7.5 is therefore determined by the gap $\Delta(\Omega)$ as well as the tolerated trace-distance error $\epsilon$ and maximal failure probability $\delta$ and given by [PLM18; Kli19]

$$m \geq \frac{\ln(1/\delta)}{\epsilon\Delta(\Omega)}. \tag{7.85}$$

The optimal strategy for verifying a state $|\psi\rangle\langle\psi|$ turns out the to be a projective measurement with measurement operators $\{|\psi\rangle\langle\psi|, 1 - |\psi\rangle\langle\psi|\}$ and has sample complexity $\log(1/\delta)/\epsilon$. However, in general, this measurement strategy will be extremely difficult to realize experimentally. Under restrictions on realizable projective measurement such as single-qubit measurements, the optimal sample complexity will be increased by $1/\Delta(\Omega)$. In many relevant settings the scaling can still be feasible. For instance, to certify $n$-qubit stabilizer states the optimal strategy is simply to measure all of the $n$ stabilizers with equal probability $1/n$ and the number of samples scales as $n \log(1/\delta)/(2\epsilon)$ since the gap of $\Omega$ in this case is simply $2/n$ [PLM18; TM18; MK20]. Likewise, to verify ground states of local Hamiltonians, one can choose energy measurements of the local Hamiltonian terms as in the fidelity witness approach.

A limitation of the optimal protocol is, however, that in practice no realistic state will achieve unit probability of being accepted by the protocol, since experimental imperfections are doomed to drive the state preparation even ever so slightly away from the ideal target state. The acceptance probability of the protocol then dramatically decreases with the infidelity $\epsilon$ leading to a state of affairs in which nearly all sequences of state preparations will be rejected. This is circumvented in weak-membership protocols as we presented them here. Bădescu, O'Donnell, and Wright [BOW19] provide such protocols for arbitrary quantum states and different distance measures with scaling roughly $O(d/\epsilon)$.

A complementary approach to verification of quantum states from measurements – blind verified quantum computation – has been developed by Broadbent, Fitzsimons, and Kashefi [BFK09] and Fitzsimons and Kashefi [FK17]. While the above protocols make use of the ability of the experimenter to measure single qubits with high fidelity, blind verified quantum computing presupposes the ability to accurately prepare single qubits. And indeed, blind verified quantum computing also applies measurement-based computation using cluster states, exploiting the property that single-qubit phase gates commute through the state preparation. While in our approaches, the imperfect state preparation is directly verified, however, in verified blind quantum computing so-called 'trap qubits' are made use of. The outcome of measurements on those qubits is deterministic and can thus be checked to build confidence in the correct functioning of an untrusted quantum server. By turning blind quantum computing upside down, a 'post hoc verification protocol' for quantum computations was developed in Ref. [FHM18].

In order to build trust in the correct functioning of a sampling device one may also resort to weaker types of verification than direct verification of the quantum state or output distribution. For instance, instead of directly running a randomly chosen unitary circuit, one can run specific computations on the device the output distribution of wich is highly structured such as the quantum Fourier transform [Tic+14]. One can also cross-verify the correct state preparation on subsystems of two distinct devices device by comparing the classical statistics of randomly chosen local unitaries [Elb+20]. While such approaches can indeed help to build trust in the device, their outcome also begging the question to a certain extent: after all one is trying to verify the quantum device *in the classically intractable regime* and the testing task was precisely chosen such that the exponential scaling of verification is circumvented. Finally, one can build trust in the device from certain efficiently computable benchmarks such as two-point correlation functions [Phi+19], or comparison to a coarse-grained distribution [WD16][25].

25: Recall also the idea of binning the outcomes of a random computation and comparing to the ideal distribution (Section 7.2).

## 7.7 Conclusion

The goal of this chapter was to show up routes by which the no-go result of the previous chapter (Chapter 6) on the impossibility of efficient device-independent verificiation could by circumvented. We first reviewed classical verification methods that overcome the hurdle of exponentially large sample complexity but incur restrictions in their power to verify closeness in total-variation distance. The cross-entropy which can be estimated from few experimental samples, for instance, provides rigorous bounds on the total-variation distance only if the noise in the device is such that it increases the entropy of the ideal distribution [Bou+19]. For other such measures no bound on the total-variation distance can be obtained. An approach that is often taken if this is the case is to develop hardness-of-classical-simulation arguments that are tailored towards the specific figure of merit at hand (or vice versa). An example of this is the sample-efficiently verifiable HOG-task for which it is argued that producing heavy outcomes of a quantum probability distribution is computationally intractable [AC17]. Evaluating cross-entropy based figures of merit, while sample-efficient, still incurs the exponential computational cost of estimating some of the target probabilities.

A way around this was presented in Section 7.3 and is – like HOG – also based on non-standard complexity-theoretic conjectures and cryptographic assumptions. One can tailor the random circuit ensemble such that it encodes a previously hidden secret in a quantum-secure way [SB09]. Checking the secret in the outcome of the computation gives confidence in the correctness of the device.

Finally, one can apply the recently developed interactive verification scheme for universal quantum computation to check the correctness of the sampled probability distribution [Mah18] that is based on the quantum security of certain one-way computational problems. Albeit being the strongest available method for verifying quantum devices device-independently based on (arguably minimal) cryptographic assumptions this method has the significant drawback of being experimentally unviable in near-term devices[26].

26: Big steps towards improving the practicality of this method have been taken in Ref. [ACH19].

In a second step, we took this as a motivation to develop verification protocols based on experimentally natural assumptions that are experimentally testable and often valid. More specifically, we assume that the single-qubit measurements can be performed with a high degree of precision (see Fig. 5.3) and that the quantum samples of the imperfect state preparation are independently and identically distributed according to some ensemble $\rho$.

This allowed us to derive an efficient fidelity witness for ground states of gapped Hamiltonians with known ground-state energy, requiring[27] $\tilde{O}(n^2/\epsilon^2)$ many samples from the quantum state. This protocol applies directly to the quantum simulation architectures presented in Section 4 and requires only single-qubit measurements of $X$ and $Z$-$Y$-plane observables. Drawbacks of the fidelity witness are its relatively small acceptance region (see Figs. 7.4 and 7.5) and the polynomial scaling in the number of qubits, making it practically infeasible already for moderate system sizes of $n \gtrsim 50$.

27: $\tilde{O}$, $\tilde{\Omega}$ and $\tilde{\Theta}$ denote the Landau-big-$O$ symbols up to log-factors.

Observing that the pre-measurement state of the quantum supremacy architecture is local-unitary equivalent to a graph state then allowed us to tailor a highly rapid verification protocol to this scheme. This 'direct fidelity estimation' scheme uses single-qubit measurements and iid. state preparations only and provides a *fidelity estimate* in constant sample complexity $O(1/\epsilon^2)$. The fidelity estimation protocol therefore not only yields more information about the imperfect state preparation, namely, an estimate of its fidelity with the target state, but also a better resource scaling. This comes at the cost of a much more limited applicability.

With the exception of the classical verification protocol due to Mahadev [Mah18] all of the verification protocols considered here require iid. state preparations which is an additional assumption – albeit a very realistic one. In order to relax this assumption to the non-iid. case, one can make use of de-Finetti arguments [Fin37; HM76; CFS02; KR05]. This was achieved by Takeuchi and Morimae [TM18] and optimized for an application to graph states in Refs. [Tak+19; MK20].

---

Let us now conclude the part of this thesis that dealt with the question how to experimentally test the computational nature of nature.

# 8 Outlook: Cloudy

The title of this chapter is inspired by an article in the Quanta Magazine from Jan. 24, 2018, available @ `https://www.quantamagazine.org/`: "The era of quantum computing is here. Outlook: Cloudy.".

*What is the computational nature of nature?*

More precisely: Can all naturally feasible computations be efficiently described within a Turing machine model? The complexity-theoretic Church-Turing thesis asserts that this is indeed the case, but it is challenged by the onset of realistic quantum computers. We have walked a long route, starting from theoretical arguments against the validity of the complexity-theoretic Church Turing thesis to the question how to verify those claims experimentally.

**Bringing quantum supremacy to the lab**

The goal of the first two parts of this thesis was to make tests of the computational complexity of nature

i. rigorously grounded in complexity theory,
ii. as practically feasible as possible using available quantum simulation hardware, and
iii. efficiently and practically verifiable while making only weak and experimentally motivated assumptions.

The technical contributions towards achieving those goals that I presented in Parts I and II of this thesis are the following.

First, we observed that wide range of quantum random sampling schemes, namely, families that form a 2-design satisfy an anticoncentration theorem (Chapter 3), which is a key ingredient in the hardness proof using Stockmeyer's argument (Chapter 2). Second, we developed a random quantum sampling scheme that is specifically tailored to experimental desiderata that are motivated by large-scale quantum simulation architectures. We provided complexity-theoretic evidence that this scheme outperforms classical computations (Section 4.2). In subsequent work, we rigorously proved that, in fact, it rises up to the highest available standard in terms of complexity-theoretic evidence for hardness of sampling [Haf+19]. We also showed how to realize general so-called IQP circuits in an ion-trap architecture using the naturally available resources in as economic a way as possible (Section 4.3).

In the second part of the thesis, we then turned to the question of verification and offered a general framework in which different certification tools can be assessed (Chapter 5). We assessed the possibility of efficiently verifying quantum random sampling schemes (Chapter 6). We found that exponentially many samples are required from the device, ironically, precisely due to the fact that those schemes showing a provable speedup over classical samplers have small second moments. Finally, we set out to circumvent this no-go result, again based on experimentally motivated assumptions. We showed that using only the ability to accurately perform single-qubit measurments in any basis suffices as a lever to perform fully efficient verification of the quantum output state (Sections 7.4 and 7.5).

## 8.1 Open questions for quantum random sampling

Over the course of this thesis, we have already mentioned several open problems and further directions of research in the context of quantum random sampling. Let us take the opportunity to take on a broader perspective on the field of quantum supremacy in general and quantum random sampling schemes, in particular.

### Complexity-theoretic foundations of sampling hardness

From a complexity-theoretic viewpoint, the prime and most pressing open problem for the existing schemes is to make the Stockmeyer hardness argument fully rigorous by proving approximate average-case hardness of computing output probabilities. As of now, approximate average-case hardness is a conjecture that is based solely on the lack of efficient classical simulation algorithms and the observation that random instances do not offer any additional structure that a classical simulation algorithm might exploit to perform better than in the worst case. As discussed extensively at the end of Chapter 3, however, the currently available methods for proving exact average-case hardness of #P problems with a polynomial structure fail as the error becomes too large. The development of new methods, while most pressing, is also elusive and constitutes a major challenge that reaches beyond the field of quantum random sampling schemes all the way into the midst of computational complexity theory.

Coming from the classical side, the question of approximate average-case hardness has recently been considered by Napp et al. [Nap+19]. There, the authors provided evidence that certain circuit families are in fact easy to sample from on average using a tensor-network based classical simulation algorithm. They also proved that certain baroque circuit families *cannot* satisfy approximate average-case hardness of computing the output probabilities.

### Physical supremacy

But even if this challenge could be solved, the state-of-the-art Stockmeyer argument as elaborated in Chapter 2 still allows only constant additive errors on the global output distribution of a quantum sampler. Going from relative to additive errors was a tremendous achievement, but still remains behind fully realistic errors in the following sense: in an experiment the total error of the global output distribution is accumulated from the local errors of individual gates and measurements. To achieve an overall constant additive error on the global distribution, the gate errors thus need to scale inversely as $1/N$ with the total number $N$ of gate applications, state preparations and measurements. As the number of qubits – and thus the total size of the circuit – is scaled up, one will inevitably hit this barrier sooner or later. The central challenge one has to overcome when realizing quantum supremacy is thus to bring this barrier down as far as possible such that the computing capabilities of classical computers can be surpassed before the barrier is hit. Arute et al. [Aru+19] have provided first evidence that this may indeed be possible.

Ultimately, however, one would like to make the hardness of quantum sampling robust to *constant local errors*. This can indeed be achieved for universal computations using *quantum error correction codes*: below a certain threshold all errors within a given error model can be corrected [AB08]. Now, one can run ahead and apply quantum (or in fact classical) error correction codes to quantum random sampling. However, quantum error correction is intrinsically based on the continuous measurement of *error syndromes* giving information about which errors have occurred during one cycle of the computation. Those errors then need to be actively corrected.

Coeherent errors do not constitute a deep problem for quantum random sampling schemes since, say, Pauli errors can often simply be absorbed in the random ensemble, giving rise to a different computation distributed according to the same ensemble [Fuj16; KD19]. However, to maintain hardness of sampling, we actually *need to know* how the circuit has changed due to the errors, that is, the errors need to be 'heralded'. But continuous measurements of syndromes complicates the computation significantly, and typically requires access to a universal fault-tolerant quantum computer. But the very goal of quantum random sampling schemes was to demonstrate quantum supremacy *before the advent* of such a device. Conversely, if the ongoing computation is not continuously measured in every gate cycle, it is not clear under which circumstances such a 'heralded noise model' is actually realistic. Finding ways around this obstacle, possibly using error detection and post-hoc corrections, is the major challenge towards making quantum random sampling schemes robust to physical noise and thus scalable.

Ultimately, the goal would be to prove a rigorous complexity-theoretic separation for a task that is natural in a physics mindset in general, and quantum simulations in particular. Quantities that first come to mind here are measurements of $k$-point correlation functions of the type $\langle b_i^\dagger b_j \rangle$. First steps towards this have been taken in Ref. [Bae+20], where a quantum advantage for the estimation of dynamical structure factors was shown, and Ref. [NBG19] where it was shown that one can run a Stockmeyer argument for the task of reproducing the statistics of an energy measurement of a local Hamiltonian. Both of these works show a much weaker complexity-theoretic consequence than a collapse of the polynomial hierarchy, namely that $\mathsf{BPP} = \mathsf{BQP}$. This reduces the complexity of the specific task (estimation of the structure factor, energy measurements) to the complexity of all of $\mathsf{BQP}$. Such results are important to bring complexity-theoretic ideas closer to experimental reality and to develop simple experimental prescriptions. From a complexity-theoretic perspective they are begging the question, though, as from this perspective, one would like to precisely collect evidence that $\mathsf{BPP} \neq \mathsf{BQP}$.

### Verification

We have dedicated significant efforts to answering the question how quantum sampling schemes may be verified. Nevertheless, this question has received little attention given its crucial role in a 'computational Bell test'. A major open problem is whether the no-go result of Chapter 6 can be circumvented by using anticoncentrating distributions with larger second moments. In other words: is there any 'room in the middle' between exponentially flat and polynomially concentrated distributions such that one can (sample-)efficiently verify the distribution from classical samples, while at the same time sampling is hard for classical computers but easy for quantum devices; cf. Fig. 6.3. Works such as the one by Morimae [Mor17] proving anticoncentration of the DQC1 model without resorting to a second-moment bound might yield some leeway in this direction.

Coming from the more practically minded end, there is the question under which circumstances cross-entropy type measures, and in particular the cross-entropy fidelity (7.25) used in the demonstration by the Google Quantum AI team [Aru+19] can yield rigorous certificates for the global distribution. Generally speaking, our results in Sections 7.4 and 7.5 indicate that highly efficient and practical verification tools may in many cases need to be custom-tailored to the requirements of a specific device at hand as well as the sampling scheme that is to be implemented on it.

## 8.2 Applications of quantum random sampling

While quantum random sampling schemes are not designed for specific applications, one might still wonder, whether one can exploit the provable speedup over classical sampling algorithms on the specific random sampling task for relevant applications. In the long term, quantum computers should be *useful*, after all, rather than mere demonstrations of the fact that classical computers do not exhaust the full computational capabilities offered by nature.

This question has mainly been asked for boson samplers and variants thereof. So-called Gaussian boson sampling [Lun+14; Ham+17] turns out to be particularly interesting in this respect. In contrast to the original boson sampling proposal by Aaronson and Arkhipov [AA13], in Gaussian boson sampling either the input state [Lun+14; Ham+17; Kru+19] or the measurement [LRR17; CC17] is done not in the Fock basis, but rather in a product of squeezed coherent states. If the input unitary is not chosen at random but bespokely, following a prescription that depends on the task at hand, one can use samples from a Gaussian boson sampler to estimate several interesting quantities.

For instance, one can estimate Franck-Condon factors which are properties of the spectra of molecules [Huh+15]. Since the output probabilities of Gaussian boson samplers are determined by a matrix Hafnian[1] [Kru+19], they can also naturally be applied to estimate certain graph properties. In particular, the Hafnian is related to the number of perfect matchings of a graph. The output probabilities of a Gaussian boson sampler count the number of perfect matchings in certain weighted subgraphs of the

1: Like the permanent, the Hafnian of a matrix is a certain polynomial of its matrix entries and defined for a matrix $A \in \mathbb{C}^{n\times n}$ as

$$\mathrm{Haf}(A) = \frac{1}{n!2^n} \sum_{\sigma\in S_{2n}} \prod_{j=1}^{n} A_{\sigma(2j-1),\sigma(2j)}, \tag{8.1}$$

where $S_{2n}$ is the symmetric group on $[2n]$. In particular, it holds that

$$\mathrm{Perm}(A) = \mathrm{Haf}\left[\begin{pmatrix} 0 & A \\ A^T & 0 \end{pmatrix}\right], \tag{8.2}$$

and hence approximating the Hafnian is just as hard as approximating the permanent.

complete graph. The bias of the full distribution towards subgraphs with more perfect matchings, that is, denser subgraphs [AB18; Brá+18] can be applied in quantum-enhanced machine learning [Sch+20] and to better solve optimization problems related to the probabilities [ABR18].

Taking another route, one can also exploit the fact that quantum supremacy distributions are very flat, that is, that they have a large min-entropy, to apply them in tasks that require samplers from such distributions. Examples of this are cryptography [Nik19; Hua+19; HKL20] and certified random number generation [Aar19].

A natural idea is also to use quantum random samplers as *benchmarking algorithms* for quantum devices, very much in the spirit of randomized benchmarking [EH+20]. In this mindset, performing quantum random sampling tasks might well play an important role in benchmarking later-generation fault-tolerant quantum devices on the hardware level.

## 8.3 Simulating quantum supremacy classically

The first two parts of this thesis focused on the question how to provide experimental evidence for the *classical hardness* of simulating paradigmatic tasks that can be perfomed efficiently on quantum devices. In the broader picture outlined in the introduction, this contributes to closing in on the boundary between efficient classical and efficient quantum computations from above.

The rigorous statements we could make always involved a separation in the *scaling* of classical versus quantum computations. Such statements show that, as systems are scaled up more and more, the speed of the respective quantum computations will at some point certainly surpass that of every classical algorithm. But how large does one actually have to make a quantum sampler such that it cannot be simulated classically, even when considering realistic amounts and sources of noise in the system? This question can only be answered for *specific classical algorithms* at a time. It is also a crucial part in the quest of violating the complexity-theoretic Church-Turing thesis: one has to demonstrate not only that the scaling is possible in principle, but also that the frontier determined by the best available classical algorithm running on the best available classical computing power can be surpassed using only relatively small quantum devices.

Violating the complexity-theoretic Church-Turing thesis, on a very practical level, thus also involves a competition between better and better classical algorithms on the one hand, and larger, less noisy quantum devices on the other hand. When the quantum supremacy barrier is reached is therefore to some extent also a sociological question that – before fundamental limitations are reached – heavily depends on the effort put into advancing either side. Only rather recently has the classical-simulation side of the medal received some more attention, again mostly so for boson samplers.

For boson sampling, Neville et al. [Nev+17] rather recently developed an approximate sampling algorithm that uses Markov chain Monte Carlo methods[2] that has a much better runtime than the naïve worst-case

2: We will discuss those methods in more detail in Part III.

complexity. Their results take into account the scaling not only in the number of photons, but also the probability of losing photons as they pass through the linear-optical network, which is the dominant error source in boson sampling. Their approximate algorithm was subsequently greatly improved by Clifford and Clifford [CC18], who provided an exact boson sampling algorithm with the same improved runtime of $O(n2^n + \mathsf{poly}(m, n))$ as compared to the worst-case runtime of $O(\binom{m+n-1}{n} n2^n)$. Altogether, these results indicate that boson samplers require at least $\sim 40$ photons before one can hope to surpass the capabilities of currently available classical computers.

The susceptibility of boson sampling to noise of different kinds has also been studied in more detail with respect to photon loss [OB18; GRS19; RSG18; RSG19; Moy+19], partial distinguishability of the photons [Ren+18; RSG18; Moy+19] and mode mismatching [RRC16]. The key observation in several of those proposals is that already low noise levels drive the output probability distribution closer to very simple, efficiently simulatable ones such as the output distribution of thermal states.

The same type of question has been pursued for qubit-based schemes such as IQP circuits with local depolarizing noise [BMS17], where the authors show a quantum-classical transition between an efficiently simulatable regime and one in which most likely there is no efficient classical simulation. Those methods can be extended to universal random circuits [YG17; GD18] but often the question remains, how realistic the considered type of noise actually is [BSN17].

---

As it turns out, some of the most efficient algorithms for simulating quantum-many body systems, quantum matter and materials are classical sampling algorithms, so called *Monte Carlo algorithms*. In the last part of this thesis, we will be dealing with classical sampling algorithms and approach the quantum-classical boundary from below by asking the question, under which conditions those algorithms are efficient. We will explore how far the applicability of Monte Carlo algorithm reaches into the ‘quantum realm’[3].

3: Monte Carlo algorithms were also used in Ref. [Nev+17] to simulate boson sampling. To some extent this effort can thus be seen as extending the realm of applicability of classical simulation algorithms for the simulation of quantum random sampling schemes.

In doing so, the question will arise, what it really is about quantum mechanical systems that makes them hard to simulate for classical computing devices. This general question can be made more specific in the case of sampling algorithms. And indeed, in Chapter 2, we have already caught a glimpse of the property that makes quantum sampling schemes hard to simulate classically from a complexity-theoretic viewpoint. Destructive interference or the appearance of alternating signs when expressing the output probabilities of quantum circuits made all the difference the complexity of sampling: the *quantum sign problem*. In Part III, we will see and study how the quantum sign problem manifests itself in classical Monte Carlo simulation algorithms for quantum sytems.

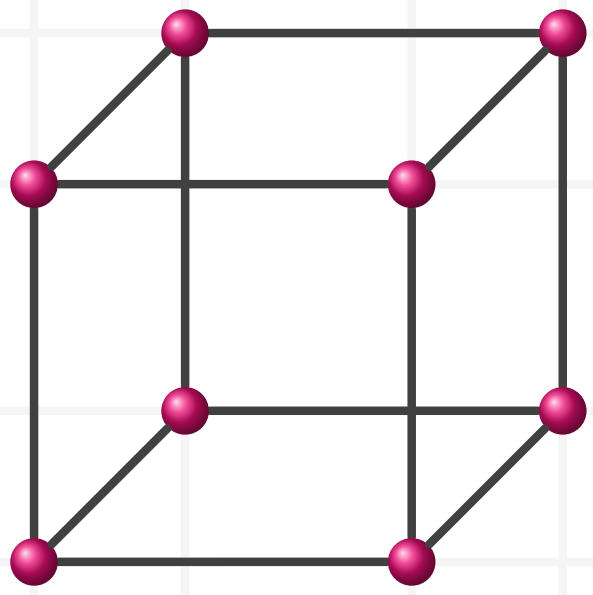

# Part III
## Simulating quantum systems using classical sampling

# Estimating physical properties via classical sampling | 9

In Parts I and II we studied the complexity and verifiability of quantum sampling devices. We asked and partially answered the question: How simple can one make a quantum sampling scheme in an experimentally relevant sense such that it is still provably hard to resimulate on a classical computer? We have therefore approached the boundary between the realms of quantum and classical computing – in terms of efficient sampling algorithms – from above.

In this last part of the thesis, we will move back to the classical realm. We will ask the converse question: How far can one push classical algorithms towards the quantum realm? To remain on the same field of play we will again consider sampling algorithms, only that now those algorithms have access to classical randomness and computing resources. More specifically, we will consider sampling-based algorithms for simulating large-scale quantum systems – so-called *Quantum Monte Carlo* algorithms.

This last part of the thesis is based on joint work with Ingo Roth, Daniel Nagaj and Jens Eisert [Han+20] that has kept me hooked since the very beginning of my PhD. I am grateful to Ingo, Daniel and Jens for the inspiring collaboration. I have also had the joy to discuss the questions we ask in this work at numerous occasions – with Martin Schwarz, Albert Werner, Juani Bermejo-Vega, and Christian Krumnow in early stages of the project; more recently with Barbara Terhal, Matthias Troyer, Joel Klassen, Marios Ioannou, Maria Laura Baez, Hakop Pashayan, Simon Trebst, Augustine Kshetrimayum, Alex Nietner and Paul Boës.

We will approach the quantum-classical boundary from below with the aim to further delineate this divide. The quantum sampling algorithms we considered focussed on sampling from a given probability distribution. One can devise classical Monte Carlo algorithms that achieve the same task of sampling from a probability distribution[1]. In contrast, the Monte Carlo algorithms considered from now on use sampling as a *tool* in order to obtain reliable and efficient approximations of properties of quantum systems, that is, to perform *strong simulations*.

1: Recall the notion of a derandomizable sampling algorithm introduced in Def. 2.9 or *weak simulation*. Such an algorithm takes as an input random bits and outputs a sample from a prescribed probability distribution.

One example of such a property that is motivated by the study of quantum sampling algorithms are probability amplitudes of random quantum circuits. Recall that in quantum random circuit sampling two notions of randomness played a role: first, the random choice of the circuit, and second the random sampling from the circuit's output distribution. In Quantum Monte Carlo yet another level of randomness comes into play. Using random samples from a specific probability distribution, the output probabilities of a quantum circuit can be approximated. It is this notion of randomness, used as a tool for computations, that will play the prominent role in this part of the thesis.

In Chapter 2 we leveraged the intractability of *computing outcome probabilities* of random quantum circuits to hardness of classical sampling. We observed that a key property of quantum systems that made the approximate computation of output probabilities intractable was interference, taking on its computational guise as the *quantum sign problem*[2]. In this chapter, we will deal with a manifestation of the quantum sign problem that appears as a practical obstruction when computing properties of quantum systems via sampling. This is famously known as the *Monte Carlo sign problem* [Hir+82; Sor+89; Loh+90; TW05]. It is an intriguing observation that the sign problem appears both in the computer-science

2: Recall that the quantum sign problem appeared when expressing output probabilities as gaps of Boolean polynomials.

perspective and the practical Quantum Monte Carlo perspective on large-scale quantum systems.

The Monte Carlo sign problem is particularly severe for fermionic Hamiltonians, as the particle-exchange anti-symmetry forces their matrix elements to have alternating signs in the standard basis [Hir+82]. In contrast, natural Hamiltonians are typically sign-problem free, and in fact, using the Perron-Frobenius theorem one can prove that the ground-state wave function of a bosonic Hamiltonian has nonnegative amplitudes. The sign problem also appears for spin Hamiltonians, however.

By exploring the Monte Carlo sign problem computationally, we will further delineate the boundary between quantum and classical computation with respect to what makes properties of quantum systems hard to compute. More specifically, in the work presented in the subsequent chapters we pose the question: When is the sign problem an intrinsic feature of quantum systems and in which cases is it merely an artifact of our description of those systems? Following the guiding principles of this thesis, we formulate this question in a computationally meaningful way and discuss different aspects of it. Central to this will be the notion of *easing* or alleviating the sign problem by changing to a physically equivalent but computationally inequivalent description of a system. Our work thus provides a fresh and interdisciplinary perspective on the sign problem and develops the first systematic, generally applicable and practically feasible framework for easing the sign problem computationally.

For practical purposes, we will focus on equilibrium properties of quantum many-body systems rather than output probabilities of quantum circuits as they were the focus of Parts I and II. Quantum Monte Carlo has long been the gold standard and one of the most powerful workhorses for computing such equilibrium properties, that is, expectation values of observables in ground and thermal states of various classes of many-body Hamiltonians [Hir+82; Tro+03; Pol12; Tro+10]. Quantum Monte Carlo techniques are therefore central to our understanding of the equilibrium physics of large-scale quantum systems.

In this chapter, we will set the stage. We will review some basic classical algorithms for sampling from a given probability distribution. We will get to know variants of Quantum Monte Carlo and see how the Monte Carlo sign problem obstructs efficient classical simulations of quantum systems. This allows us to pose the three central questions answering which will be the topic of the subsequent chapters.

## 9.1 Classical sampling algorithms

While this review is not essential to what follows, I include it with the intention to help the reader get an idea of how classical sampling algorithms work in practice. My selection hopes to convey some basic principles of classical sampling as well as practically relevant algorithms. I will follow the book by Devroye [Dev86] on non-uniform sampling.

Let us begin the discussion of Monte Carlo algorithms by reviewing three basic classical algorithms for sampling from a probability distribution: rejection sampling, marginal sampling, and Metropolis sampling. All of the methods assume access to uniformly random bits. The question they provide answers to is: How can uniform randomness be transformed into non-uniform randomness distributed according to a preset probaiblity distribution? As throughout this thesis, the focus lies on discrete probability distributions.

The methods outlined below therefore provide provably correct samples from the intended distribution only if input bits are perfectly uniformly random. What is more, the methods vary significantly in their *efficiency* and the conditions under which they are efficient. Depending on the given distribution, different methods may be most efficient in producing samples from that distribution[3][4].

3: How unbiased random bits are generated is another question that we will not discuss here. Suffice it to mention that, as an example, secure quantum random number generation [ROT94; Bra+18; Geh+20] is one way to produce random bits with certified quality. See Refs. [AM16; HG17] for recent reviews on the topic.

4: The selection I make here is by no means representative, nor does it aspire to paint a complete picture. Rather I will use those examples to highlight important features of classical sampling algorithms.

## Rejection sampling

We begin with one of the simplest methods to generate samples from a non-uniform probability distribution. The idea of the rejection method is best understood graphically: to generate samples from a non-uniform probability distribution $p$ on a discrete sample space $\Omega$ a uniformly random sample $(x, u)$ is drawn from an area that contains the (normalized) histogram of this distributionl. If the sample $(x, u)$ falls into the (normalized) histogram of $p$ it is accepted, otherwise it is rejected and the procedure is repeated until a sample is accepted; see Fig. 9.1.

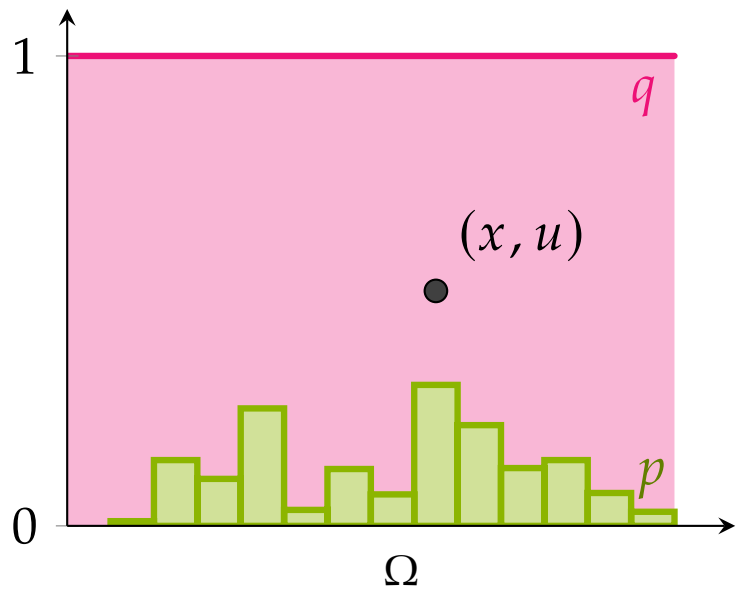

**Figure 9.1:** In the simplest instance of the rejection method a uniformly random sample $(x, u)$ is drawn from the rectangle $\Omega \times [0, 1]$. The sample is accepted if $u \leq p(x)$ and rejected otherwise.

In the simplest case, this area is a rectangle, corresponding to sampling uniformly on $\Omega \times [0, 1]$. Clearly, the rejection method is correct: For a uniformly random $(x, u) \in \Omega \times [0, 1]$, the overall acceptance probability of the method is given by the ratio of the areas. While the rectangle $\Omega \times [0, 1]$ has area $\Omega$, the area covered by the normalized histogram of $p$ is by definition 1. The acceptance probability in each iteration is therefore given by $1/|\Omega|$[5]. The probability of obtaining a sample conditioned on acceptance is then given by Bayes' rule[6] as

$$\Pr[x, u|\texttt{Accept}] = \frac{\Pr[\texttt{Accept}|x, u]\Pr[x]\Pr[u]}{\Pr[\texttt{Accept}]} \tag{9.1}$$

$$= \frac{\Pr[u \leq p(x)]\Pr[u]/|\Omega|}{1/|\Omega|} = p(x). \tag{9.2}$$

5: Hence, the overall runtime scales as $O(|\Omega|)$.

6: Recall that Bayes' principle states that the joint probability $\Pr[A, B]$ of two random variables $A$ and $B$ can be decomposed into conditional and marginal probabilities as $\Pr[A, B] = \Pr[A|B]\Pr[B] = \Pr[B|A]\Pr[A]$.

In order to improve the efficiency of the algorithm one therefore needs to maximize the overlap of the uniformly sampled area and the normalized histogram of $p$. This can be done both by scaling the rectangle so as to avoid redundant space, and by changing the shape of the area to match the target distribution as well as possible. The latter corresponds to replacing the uniform distribution over $\Omega$ with a prior distribution $q$ from which samples can be generated efficiently.

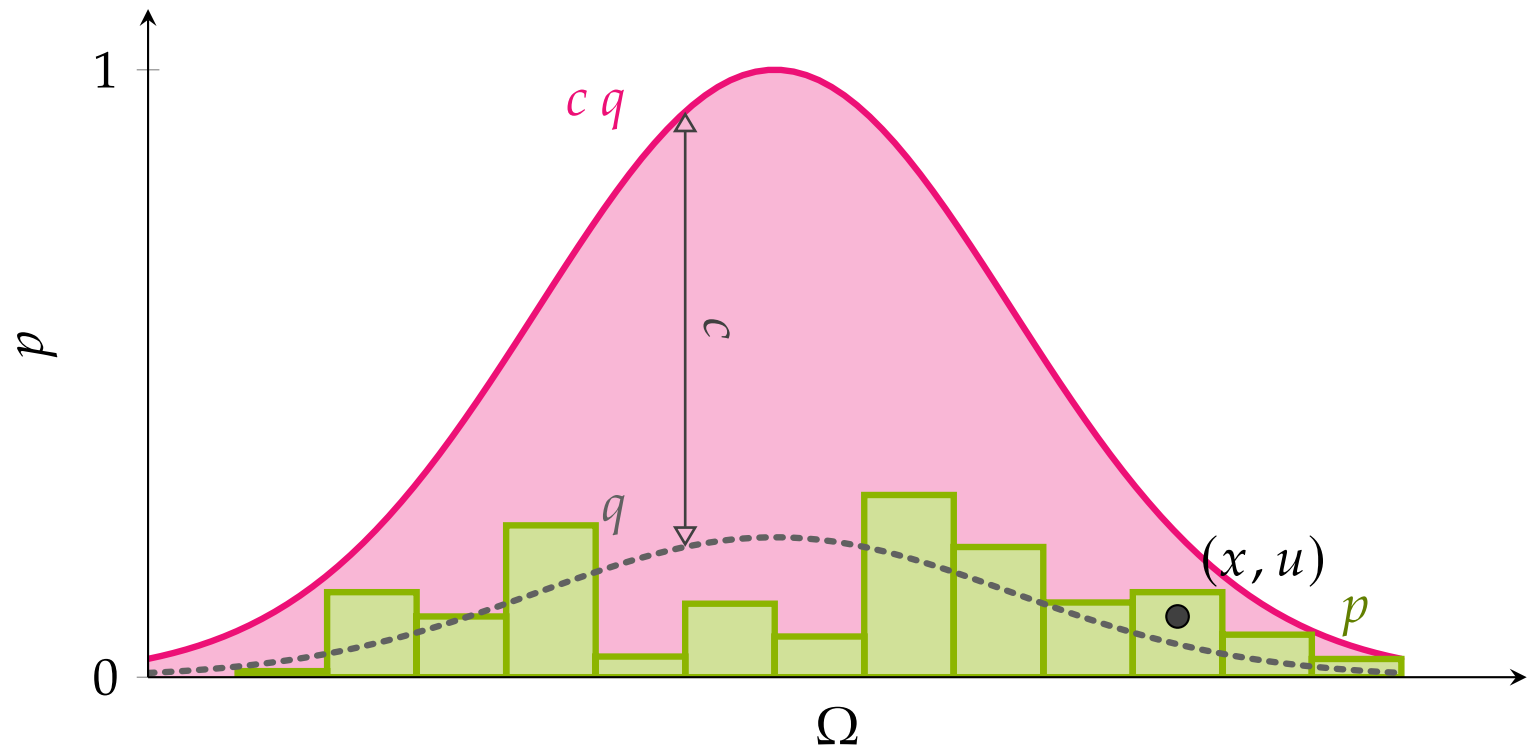

**Figure 9.2:** In rejection sampling the target is to sample from a distribution $p$ on a sample space $\Omega$. To this end, a random sample $x \sim q$ distributed according to a proposal distribution $q$ on $\Omega$ and a random sample $u \in [0, 1]$ is drawn. If $cq(x) \geq p(x)$ for all $x \in \Omega$ for some constant $c \geq 1$, and $cuq(x) \leq p(x)$ the sample is accepted and the $x \sim p$ returned as an output of the algorithm.

We can therefore generalize the basic idea of the plain rejection method in the following way, making use of an easy-to-sample distribution $q$ that for some $c \geq 1$ satisfies

$$p(x) \leq cq(x), \ \forall x \in \Omega. \tag{9.3}$$

**Protocol 9.1** Rejection sampling

**Input:** Target distribution $p$, $c \geq 1$ satisfying (9.3).
1: **repeat**
2: Generate a uniformly random number $u \in [0, 1]$.
3: Generate a sample $x \sim q$ distributed according to $q$.
4: **until** $ucq(x) \leq p(x)$
**Output:** $x$

The scaling constant $c$ is called the rejection constant, and determines the number of trials until a sample is accepted and thus the runtime of the algorithm[7]. For $c = 1$ the inequality (9.3) can only be met for $q = p$ by normalization of $q$ and $p$. In the worst case, the rejection constant is unbounded. This case is realized for uniform $p(x) = 1/|\Omega|$ and $q = \delta(x, x_0)$ supported on a single outcome only[8]. The rejection constant $c$ can thus be viewed as a measure of how much $q$ deviates from the target distribution $p$.

7: More precisely, the number of iterations required before the algorithm halts is given by a geometric distribution with parameter $p = 1/c$ [Dev86, p. 42].

8: Of course, this is a rather unwise choice of $q$ which might as well have been chosen uniform to begin with.

Indeed, in the example above, $q$ was chosen to be uniform with $q(x) = 1/|\Omega|$ for all $x \in \Omega$ and $c = |\Omega|$. This example also shows that one can always come up with a distribution $q$ that satisfies (9.3) for $c = |\Omega|$, namely the uniform distribution.

The rejection method is typically highly inefficient in the size of the sample space. As we discussed, this is because choosing an adequate $q$ with a small rejection constant $c$ that is at the same time efficiently sampleable turns out to be not straightforward at all. Instances in which this task is particularly demanding are given by probability distributions which are supported on an exponential number of outcomes that vary significantly in their probabilities. On the other hand, the method is suited well to distributions with largest probability $O(\mathsf{poly}(\log|\Omega|)/|\Omega|)$. In this latter case choosing $q$ uniform results in an inverse polynomially small rejection probability since a scaling constant $c \in \mathsf{poly}(\log(|\Omega|))$ suffices. However, in each iteration one also needs to compute $p(x)/cq(x)$ which may be demanding[9]. It goes without saying that more sophisticated variants of the rejection method have been developed over the years; see [Dev86, Ch. II.3] for details[10].

9: Recall that this is the situation for quantum supremacy schemes. Here, the distributions are flat in the sense that the largest probabilities scale as $1/\sqrt{|\Omega|}$ resulting in an acceptance probability of $1/\sqrt{\Omega}$ (which is still exponential however). Computing those probabilities is even harder still.

10: Let us only mention here *adaptive rejection sampling*, where the proposal distribution $q$ is updated dynamically within an ansatz class such as piecewise linear functions in order to obtain an improved approximation of the target distribution.

## Marginal sampling

In Chapter 2, we have already seen another method for sampling that works extremely well on (exponentially) large sample spaces[11] . This is the method of marginal sampling, where the idea is to sample an outcome (bit) string by sequentially sampling each individual bit. Given a probability distribution $p : \{0, 1\}^n \to [0, 1]$ on length-$n$ bit strings, we

11: The marginal sampling method appeared in our discussion of Terhal and DiVincenzo [TD04].

define the marginal probabilities on the first $k$ bits as

$$P_k(x_1, \dots, x_k) = \sum_{x_{k+1},\dots,x_n \in \{0,1\}} p(x_1, \dots, x_n), \tag{9.4}$$

and the conditional probabilites

$$P_k(x_k | x_{k-1}, \dots, x_1) = \frac{P_k(x_1, \dots, x_k)}{P_{k-1}(x_1, \dots, x_{k-1})}. \tag{9.5}$$

Marginal sampling then proceeds as follows:

---

**Protocol 9.2** Marginal sampling

**Input:** The target distribution $p : \{0,1\}^n \to [0,1]$

1: Set $k = 1$.
2: **while** $k \leq n$ **do**
3: Flip a coin with bias $P_k(x_k | x_{k-1}, \dots, x_1)$ and set $x_k$ to be the outcome.
4: Update $k \leftarrow k + 1$.
5: **end while**

**Output:** $x = (x_1, \dots, x_k)$

---

In marginal sampling, the challenging task of sampling from a distribution on an exponentially large sample space is broken up into biased coin flips. Those can be done with constant effort using rejection sampling, for instance. Marginal sampling is correct: The probability of obtaining a string $x$ as the output of the algorithm is simply given as

$$\begin{aligned}
\Pr[x] &= \prod_{k=1}^{n} P_k(x_k | x_{k-1}, \dots, x_1) && (9.6)\\
&= \frac{P_n(x_1, \dots, x_n)}{P_{n-1}(x_1, \dots, x_{n-1})} \cdot \frac{P_{n-1}(x_1, \dots, x_{n-1})}{P_{n-2}(x_1, \dots, x_{n-2})} \cdots \frac{P_2(x_1, x_2)}{P_1(x_1)} \cdot P_1(x_1) && (9.7)\\
&= P_n(x_1, \dots, x_n) = p(x). && (9.8)
\end{aligned}$$

Using marginal sampling one can thus sample out probability distributions on exponentially large sample spaces, provided that the respective marginal probabilities can be computed efficiently. This is often a severe obstacle to this method as not even the possibility to efficiently compute probabilities guarantees that it is possible to efficiently compute marginal probabilities. Those require the computation of exponentially large sums and therefore also exponentially precise estimations of the individual probabilities. Those conditions can still be satisfied if only polynomial accuracy can be achieved, however, provided that the output distribution is *sparse*, that is, close to a distribution that is supported on polynomially many outcomes only[12] [SV13]. In this case one can make use of an algorithm by Kushilevitz and Mansour [KM93] to efficiently find the (effective) support of the distribution.

12: Recall the discussion of flatness vs. simulatability in Chapter 6.

## Metropolis sampling

One of the, if not the, most important and efficient methods used to sample from a distribution $p$ on a exponentially large sample space is *Metropolis sampling* [Met+53] or, more generally, the *Metropolis-Hastings algorithm* [Has70].

My discussion of the Metropolis-Hastings algorithm is based on Landau and Binder [LB00, Sec. 4.2.1].

The key idea underyling this alogrithm is to construct a Markov chain of configurations $x_1 \to x_2 \to \cdots \to x_M$ with stationary distribution given by $p$. This Markov chain is specified by the probability $P_t(x)$ of being in state $x$ at time step $t$, and rates $W_{x\to x'}$ for the transition $x \to x'$ that determine the probability of moving from state $x$ to state $x'$. The time dependent behaviour of the Markov chain is then described by the Master equation

$$P_{t+1}(x) - P_t(x) = -\sum_{x\neq x'} \left[P_t(x)W_{x\to x'} - P_t(x')W_{x'\to x}\right], \tag{9.9}$$

as the population change of state $x$ when moving from time $t$ to $t+1$ is determined by the leaving and incoming population.

The Metropolis-Hastings algorithm is correct in that it produces samples from the target distribution $p$ after some threshold time $t_0$, provided that

i. the target distribution $p$ is a stationary distribution of the Markov chain,
ii. it is also the *unique* stationary distribution, and
iii. the Markov chain converges to that distribution.

The central question when devising a Metropolis-Hastings algorithm for a given distribution is therefore how to choose transition rates $W_{x\to x'}$ such that conditions i.-iii. are satisfied. A simple sufficient condition that ensures that the target distribution $p$ is in fact a stationary distribution is given by the so-called *detailed balance* on the transition rates:

$$p(x)W_{x\to x'} = p(x')W_{x'\to x}. \tag{9.10}$$

Indeed, suppose that $P_t = p$. Then stationarity requires that

$$\begin{aligned} 0 &\overset{!}{=} P_{t+1}(x) - P_t(x) \equiv P_{t+1}(x) - p(x) && (9.11)\\ &= -\sum_{x\neq x'} \left[p(x)W_{x\to x'} - p(x')W_{x'\to x}\right], && (9.12)\end{aligned}$$

which is fulfilled if the transition rates satisfy detailed balance (9.10).

The idea of Metropolis et al. [Met+53] for how to achieve this is to simulate a cooling process in order to converge to the stationary distribution. For such a process it is important not only to drive the Markov chain towards the target distribution, but also to allow for worse behaviour with some probability so that the dynamics can escape from local minima.

The overall idea, as in rejection sampling, is to construct the Markov chain based on a proposal distribution $q$ that determines the probability $q(x'|x)$ of moving to state $x'$ in step $i+1$ of the algorithm given that the

Markov chain is in state $x$ at step $i$. The transition probabilities are then given by

$$W_{x \to x'} = \Pr[\texttt{Accept}|(x'|x)]q(x'|x). \tag{9.13}$$

Given the condition of detailed balance (9.10), which may be rewritten as

$$\frac{W_{x \to x'}}{W_{x' \to x}} = \frac{p(x')}{p(x)}, \tag{9.14}$$

a simple choice of the acceptance probability is the Metropolis choice[13]

$$\Pr[\texttt{Accept}|(x'|x)] = \min\left\{ \frac{p(x')q(x|x')}{p(x)q(x'|x)}, 1 \right\}. \tag{9.15}$$

13: Notice that this choice satisfies detailed balance (9.14) as exactly one of $\Pr[\texttt{Accept}|(x'|x)] = 1$ or $\Pr[\texttt{Accept}|(x|x')] = 1$ will always hold.

This choice has the favourable property that it only depends on the *ratio* $p(x')/p(x)$. This means that one need not be able to compute those probabilities directly but only a function $f \propto p$. This is particularly relevant if the target distribution is given by a Boltzmann distribution $\mathrm{e}^{-\beta H(x)}/Z_\beta$ of a Hamiltonian $H$ at temperature $1/\beta$ with partition function $Z_\beta$. While computing $\mathrm{e}^{-\beta H(x)}$ is often feasible, computing the partition function is a hard task as it requires the summation of exponentially many terms.

Altogether, we can summarize the Metropolis sampling algorithm as follows.

**Protocol 9.3** Metropolis sampling

**Input:** Initial state $X$, target distribution $p$, proposal distribution $q$, chain length $M$.

Set $x_0 = X$.
**for** $i = 0, 1, \dots, M$ **do**
  Sample $x'$ from $q(x'|x_i)$.
  Sample $u$ uniformly from $[0, 1]$.
  **if** $u \leq \Pr[\texttt{Accept}|(x'|x_i)]$ **then**
    Let $x_{i+1} = x'$.
  **else**
    Let $x_{i+1} = x_i$.
  **end if**
**end for**

**Output:** $x_1, \dots, x_M$

So far, we have ensured that the target distribution $p$ is the stationary distribution of the Markov chain. A sufficient condition to ensure that the uniqueness condition (ii.) is satisfied is that the Markov chain is *irreducible*. The irreducibility condition is satisfied if $W_{x \to x'} > 0$ for all $x', x \in \Omega$ so that all configurations can be reached with non-vanishing probability. Simple choices of the proposal distribution that satisfy irreducibility for target distributions with full support include the uniform distribution distribution, or, when simulating spin-1/2 systems, flipping individual spins with uniform probability.

Finally, to satisfy the convergence condition (iii.), one has to ensure that in addition to being able to access all states, guaranteeing uniqueness, it is also impossible for cycles to occur. This is the case if the Markov chain is also aperiodic. Together, aperiodicity and irreducibility imply that the chain is *ergodic*, that is, that it explores the entire sample space. The convergence rate of the Markov chain to the stationary distribution is given by the gap $\Delta$ between largest (given by 1) and second largest eigenvalue $(1-\Delta)$ of the transition matrix $W$. If the gap is at most polynomially small in the number of degrees of freedom, the Markov chain converges efficiently[14] .

14: Proving that a Markov chain converges efficiently or *mixes rapidly* is a highly non-trivial task and only few methods exist. One is the *coupling lemma* [Ald83], another one the method of *canonical paths* that was pioneered by Jerrum and Sinclair [JS89].

## 9.2 Estimating physical properties via Monte Carlo methods

We are now ready to turn to applications of classical sampling algorithms to the task of estimating equilibrium properties of physical systems. Before we move to the realm of quantum systems, it is instructive to understand Monte Carlo algorithms in the classical setting.

### Classical Monte Carlo

The classic application of Metropolis sampling is the computation of partition functions and (local) observables in statistical physics. The expectation value of an observable $O : \Omega \longrightarrow \mathbb{R}$ on a configuration space $\Omega$ in the canonical ensemble of a classical Hamiltonian $H : \Omega \longrightarrow \mathbb{R}$ at inverse temperature $\beta$ is given by

$$\langle O \rangle_{\beta,H} = \sum_{x\in\Omega} p_{\beta,H}(x) O(x), \tag{9.16}$$

where $p_{\beta,H}(x) = \mathrm{e}^{-\beta H(x)}/Z_{\beta,H}$ with partition function $Z_{\beta,H} = \sum_x \mathrm{e}^{-\beta H(x)}$ is the canonical probability of the system being in state $x$. Via Metropolis importance sampling, this expectation value can be directly estimated without the need to compute the partition function, giving rise to *Markov-chain Monte Carlo (MCMC)*. Indeed, recall that in Metropolis sampling only the *ratios* $p_{\beta,H}(x')/p_{\beta,H}(x) = \mathrm{e}^{-\beta H(x')}/\mathrm{e}^{-\beta H(x)}$ were required. Given $s$ samples $x_1, \ldots, x_s$ from the distribution $p_{\beta,H}$ the expectation value $\langle O \rangle_{\beta,h}$ can be estimated as

$$\hat{O}_{\beta,H} = \frac{1}{s} \sum_{i=1}^{M} O(x_i), \tag{9.17}$$

up to error[15]

15: Apply Chebyshev's inequality.

$$\epsilon \leq \sqrt{\mathrm{Var}_{p_{\beta,H}}(O)/(s(1-\delta))}, \tag{9.18}$$

with probability at least $1-\delta$. To achieve any relative error $\tilde{\epsilon}$, the number of samples needs to grow with the variance $\mathrm{Var}(O)$ normalized by its expectation value $\langle O \rangle$.

It is less obvious how to use sampling to compute the partition function, however, as this requires precisely to learn the normalization. By an ingenious idea of Jerrum and Sinclair [JS89][16] this issue can be overcome,

16: See also [SJ89; JS90; JS93; JS97].

however. They rewrite the partition function $Z(\beta) := Z_{\beta,H}$ at temperature $\beta$ as a product

$$Z(\beta) = \frac{Z(\beta_r)}{Z(\beta_{r-1})} \cdot \frac{Z(\beta_{r-1})}{Z(\beta_{r-2})} \cdots \frac{Z(\beta_1)}{Z(\beta_0)} \cdot Z(\beta_0), \tag{9.19}$$

of $r$ ratios of partition functions at different temperatures $\beta_k$ with $0 = \beta_0 < \beta_1 < \cdots < \beta_{r-1} < \beta_r = \beta$. They observe that $Z(\beta_0) = Z_0 = |\Omega|$ and that the fractions $Z(\beta_k)/Z(\beta_{k-1})$ can be computed via Markov-chain Monte Carlo. To see this, observe that we can express the fraction

$$\frac{Z(\beta_{k-1})}{Z(\beta_k)} = \frac{1}{Z(\beta_k)} \sum_{x \in \Omega} \mathrm{e}^{-\beta_k H(x)} \left( \frac{\mathrm{e}^{-\beta_{k-1} H(x)}}{\mathrm{e}^{-\beta_k H(x)}} \right) \equiv \langle f_k \rangle_{p_{\beta_k}}, \tag{9.20}$$

as an expectation value of the estimator $f_k = \mathrm{e}^{-(\beta_{k-1}-\beta_k)H(x)}$ with respect to the canonical distribution $p_{\beta_k}(x) = \mathrm{e}^{-\beta_k H(x)}/Z(\beta_k)$. This fraction (9.20) can be estimated using Metropolis sampling[17].

17: In fact, Jerrum and Sinclair [JS90] show that the Markov chain *mixes rapidly* for specific Hamiltonian models (monomer-dimer systems), that is, converges in polynomial time to an inverse-polynomial relative approximation of $Z(\beta)$ for a specific choice of the sequence $(\beta_k)_{k=0,\ldots,r}$. This means that MCMC is a *fully polynomial randomized approximation scheme (FPRAS)* for $Z(\beta)$. Their algorithm uses a relation to perfect matchings. It has been applied to several other models such as ferromagnetic Ising models [BG17].

## Quantum Monte Carlo

When moving the realm of quantum systems, we are faced with a fundamental difficulty: in classical statistical physics, what defines an expectation value in equilibrium are precisely the probabilities of the canonical ensemble. Quantum mechanical expectation value in a Gibbs state $\rho_\beta = \exp(-\beta H)/Z_{\beta,H}$ of a Hamiltonian $H$ with partition function

$$Z_{\beta,H} = \mathrm{Tr}[\exp(-\beta H)] \tag{9.21}$$

are given by

$$\langle O \rangle_{\beta,H} = \mathrm{Tr}[\rho_\beta O]. \tag{9.22}$$

Those quantities are determined by a trace rather than a classical sum, however, which does not immediately give rise to an expression analogous to Eq. (9.16) that can be estimated using Monte Carlo sampling.

The goal of Quantum Monte Carlo is to recover the favourable properties of classical Monte Carlo by expressing quantum-mechanical expectation values of the form (9.22) as expectation values

$$\langle O \rangle_{\beta,H} = \mathbb{E}_p[f] = \sum_{\lambda \in \Lambda} p(\lambda) f(\lambda), \tag{9.23}$$

of a random variable $f : \Lambda \to \mathbb{R}$ distributed according to a *classical probability distribution* $p : \Lambda \to [0,1]$ on a configuration space $\Lambda$. This is nontrivial not only because quantum mechanics requires complex numbers but also because of the fact that states and observables are arbitrary complex-valued matrices. Diagonal matrices recover classical statistical mechanics.

If such a representation of the expectation values could be found, one could construct the linear estimator (9.17) via importance sampling from $p$ analogously to the classical case. The estimator $\hat{O}_{\beta,H}$ will efficiently approximate the true value $\langle O \rangle_{\beta,H}$ if two conditions are fulfilled

i. the variance $\mathrm{Var}_p(f)$ of the estimator scales at most polynomially with the number of degrees of freedoms, and
ii. samples from $p$ can be generated efficiently.

A sufficient condition for the latter condition when using Metropolis sampling – and the one that is typically referred to – is that a Markov chain can be constructed that mixes rapidly.

*This paragraph has previously been published as part of Ref. [Han+20].*

For the purpose of this work, we focus on the prominent *world-line Monte Carlo* method of calculating partition functions (9.21) and thermal expectation values (9.25) of a Hamiltonian $H$ at inverse temperature $\beta$ [LB00]. For a real Hamiltonian[18] $H$ acting on a $D$-dimensional Hilbert space $\mathcal{H}$, the idea at the heart of the most prominent variant of Quantum Monte Carlo is to sample out world-lines in a corresponding $(D+1)$-dimensional system, where the additional dimension is the (Monte Carlo) time dimension. These word lines correspond to paths through an $m$-fold expansion of $\mathrm{e}^{-\beta H} = (\mathrm{e}^{-\beta H/m})^m$ where an entry of the so-called *transfer matrix* $\mathrm{e}^{-\beta H/m}$ in a local basis is selected in each step. Each such path is associated with a probability which is proportional to the product of the selected entries.

18: We restrict – as is typically done in Quantum Monte Carlo – to the case of real Hamiltonians. The discussion for arbitrary Hermitian matrices is analogous.

More precisely, in *world-line Monte Carlo* the partition function (9.21) and expectation value (9.22) are expressed as[19]

$$Z_{\beta,H} \simeq \mathrm{Tr}[T_m^m] = \sum_{\vec{\lambda}\in\Lambda_{m+1},\, \lambda_{m+1}=\lambda_1} a(\vec{\lambda}) \tag{9.25}$$

$$\langle O\rangle_{\beta,H} \simeq \frac{1}{Z_{\beta,H}} \mathrm{Tr}[T_m^m O] = \frac{1}{Z_{\beta,H}} \sum_{\vec{\lambda}\in\Lambda_{m+1}} a(\vec{\lambda}) f_O(\vec{\lambda}), \tag{9.26}$$

19: An alternative is so-called *stochastic series expansion (SSE) Monte Carlo* that relies on the Taylor expansion of the matrix exponential:

$$Z_{\beta,H} = \sum_{i=1}^{\infty} \frac{(-\beta)^n}{n!} \mathrm{Tr}\,[H^n], \tag{9.24}$$

which can be expanded analogously to (9.25).

for large enough $m \in \mathbb{N}$ *Monte Carlo steps* in terms of the amplitudes

$$a(\vec{\lambda}) = T_m(\lambda_1|\lambda_2)T_m(\lambda_2|\lambda_3)\cdots T_m(\lambda_m|\lambda_{m+1}), \tag{9.27}$$

on the configuration space $\Lambda_{m+1} = [\dim \mathcal{H}]^{\times(m+1)}$. Here, we have defined the transfer matrix[20] $T_m(\lambda'|\lambda) = \langle \lambda'|1-\beta H/m|\lambda\rangle$ and in general denote the entries of a matrix $A$ as $A(\lambda_1|\lambda_2) = \langle\lambda_1|A|\lambda_2\rangle$. Notice that the computation of the partition function involves a summation over all closed paths of length $m$ (i.e., paths with periodic boundary conditions); the computation of observables involves a summation over all open paths. For diagonal observables $O$ we can choose $f_O(\vec{\lambda}) = O(\lambda_1|\lambda_1)$ and restrict the summation in (9.26) to all closed paths again, a property that is crucial to constructing an efficient estimator. For non-diagonal observables finding a suitable choice of $f_O$ is less trivial[21].

20: For a local Hamiltonian $H = \sum_x h_x$, the transfer matrix can alternatively be defined via the Trotter approximation of the operator $\mathrm{e}^{-\beta H/m}$ as

$$\mathrm{e}^{-\beta H/m} \approx \prod_x \mathrm{e}^{-\beta h_x/m} =: T'_m. \tag{9.28}$$

21: There are a few easy cases to note here: First, for local Hamiltonians and local observables $O$, one can always efficiently express the transfer matrix $T_m$ in a basis in which $O$ is diagonal. In general, we can try and set $f_O(\vec{\lambda}) = \langle\lambda_m|O|\lambda_1\rangle/\langle\lambda_m|T_m|\lambda_1\rangle$. But this only works if $\langle\lambda_m|T_m|\lambda_1\rangle = 0$ only if $\langle\lambda_m|O|\lambda_1\rangle = 0$, too. In other cases such as when measuring the kinetic energy, it is often possible to make clever approximations [LB00].

We can conceive of a sequence $\lambda_1, \lambda_2, \ldots, \lambda_m$ as a *path* following the imaginary-time dynamics of $\mathrm{e}^{-\beta H/m}$. We illustrate this idea in Fig. 9.3. For nonnegative path weights, the partition function (9.25) and thermal expectation value (9.26) may be rewritten as expectation values in a distribution $q$ on $\Lambda_{m+1}$ defined by

$$q(\vec{\lambda}) = \frac{1}{\sum_{\vec{\lambda}} a(\vec{\lambda})} a(\vec{\lambda}). \tag{9.29}$$

Notice that if the summation in Eq. (9.26) can be restricted to closed paths only, the amplitudes $a(\vec{\lambda})$ are naturally normalized by $Z_{\beta,H}$ so that

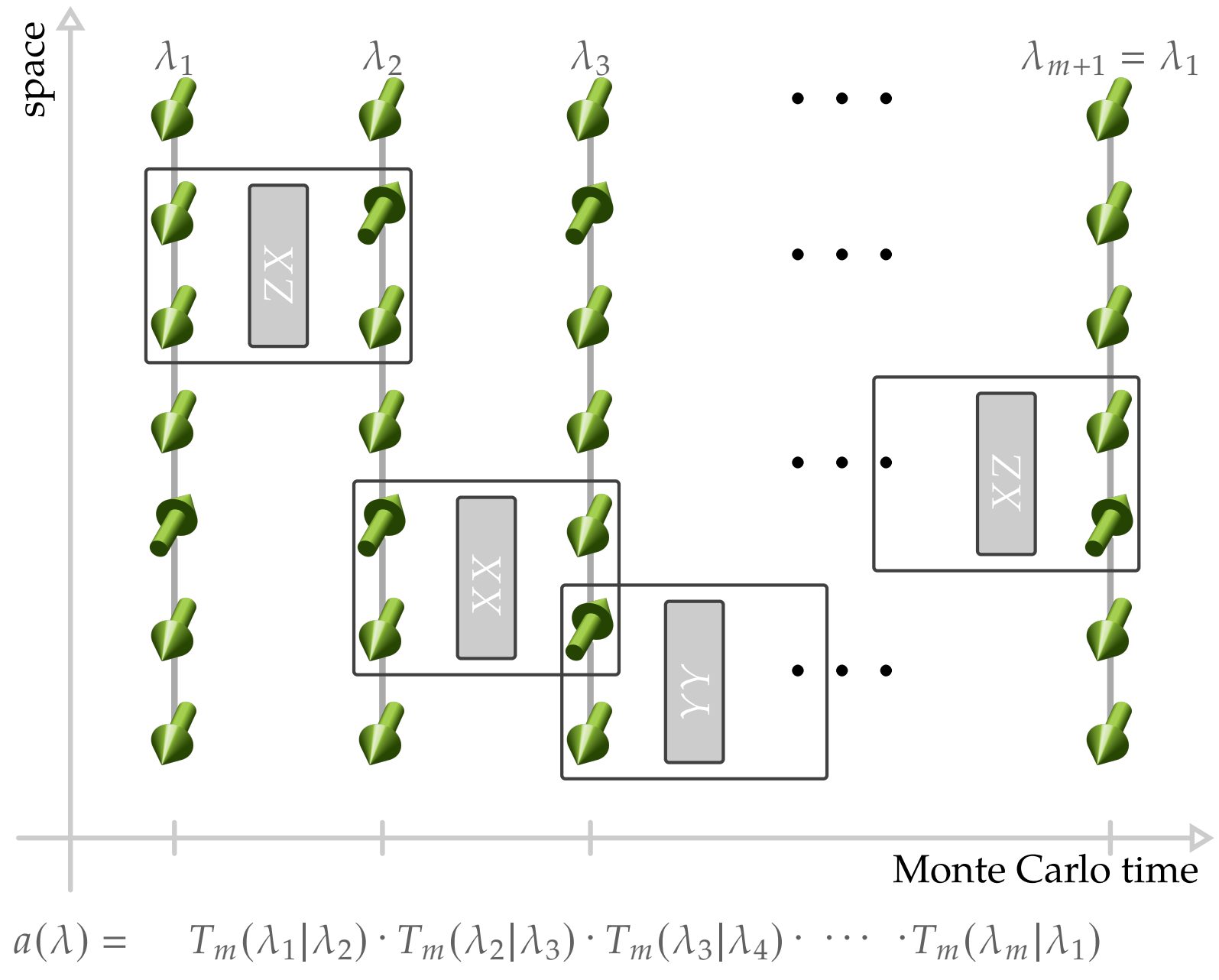

**Figure 9.3:** In world-line Monte Carlo a linear estimator is constructed via an $m$-fold expansion of the exponential $e^{-\beta H}$ that defines the estimator's probability distribution. This expansion is often viewed as a discretized imaginary-time dynamics with time step $\beta/m$ through the configuration space $\Lambda = [D]$ driven by the transfer matrix $T_m \approx e^{-\beta H/m}$.

in this case

$$\langle O \rangle_{\beta,H} = \sum_{\lambda \in \Lambda_{m+1}, \lambda_{m+1}=\lambda_1} q(\vec{\lambda}) f_O(\vec{\lambda}), \tag{9.30}$$

which can be estimated as in Eq. (9.17) with additive estimation error $1/\sqrt{s}$ (cf. (9.18)) via Metropolis sampling $s$ samples from $q$ . If the summation runs over open rather than closed paths alone as the observable $O$ is not diagonal, the estimator $f_O$ needs to be multiplied by $\sum_{\vec{\lambda}} a(\vec{\lambda})/Z_{\beta,H}$. This means that the normalization $\sum_{\vec{\lambda}} a(\vec{\lambda}) \neq Z_{\beta,H}$ needs to be computed efficiently, a task that seems challenging to say the least[22].

22: Of course, there are still ways to estimate non-diagonal observables; see Landau and Binder [LB00] for examples.

Before we now move on to the sign problem of Quantum Monte Carlo, let us remember that we have seen a similar expression in terms of world lines through a unitary time evolution given by a quantum circuit in Section 2.4 (Eq. (2.12)). There, we expressed the output amplitudes of a quantum circuit precisely in the form (9.23) via a GapP function. And indeed, the quantum sign problem of computing GapP sums, manifests itself in Monte Carlo simulations, too.

## 9.3 The sign problem of Quantum Monte Carlo

In the classical variant of Monte Carlo, the Hamiltonian is always diagonal, giving rise to positive weights[23] $a(\vec{\lambda})$ and therefore a probability distribution $q(\vec{\lambda})$ can be constructed. In Quantum Monte Carlo, in contrast, positive – in the most general case, even complex – off-diagonal matrix elements of $H$ potentially give rise to negative weights of the paths. This leads to what is famously known as the *sign problem* of Quantum Monte Carlo, namely, that now one is faced with the task of sampling the quasi-probability distribution $q$ (normalized but nonpositive) as opposed to a nonnegative probability distribution.

23: Indeed, notice that for a diagonal Hamiltonian matrix $H = \mathrm{diag}((H(x))_x)$ the transfer matrix can be computed exactly as $\exp(-\beta G) = (\exp(-\beta\lambda_x)_x$, recovering the Boltzmann distribution $p_{\beta,H}$.

*The following paragraphs have been published as part of Ref. [Han+20].*

↓ ↓ ↓

This task can be achieved by introducing a suitable probability distribution that reproduces the desired sampling averages but typically comes at the cost of an exponential increase in the sampling complexity and hence the runtime of the algorithm.

It can easily be shown that the variance-optimal linear estimator of the form (9.23) for the partition function $Z_{\beta,H}$ is given by the probability distribution[24]

24: Recall the definition of the $\ell_1$-norm (2.78): $\|a\|_{\ell_1} = \sum_{\vec{\lambda}} |a(\vec{\lambda})|$.

$$p(\vec{\lambda}) = \frac{1}{\|a\|_{\ell_1}} |a(\vec{\lambda})| \tag{9.31}$$

and the estimator

$$f(\vec{\lambda}) = \mathrm{sign}(a(\vec{\lambda})) \cdot \|a\|_{\ell_1}. \tag{9.32}$$

The variance of this estimator is given by

$$\mathrm{Var}_p(f) = \|a\|_{\ell_1}^2 (\|q\|_{\ell_1}^2 - 1) \tag{9.33}$$

and hence the relative error of the approximation by

$$\frac{\mathrm{Var}_p(f)}{\langle f \rangle_p^2} = \|q\|_{\ell_1}^2 - 1 \equiv \langle \mathrm{sign} \rangle_p^{-2} - 1. \tag{9.34}$$

Here, $\langle \mathrm{sign} \rangle_p = 1/\|q\|_{\ell_1}$ is called the *average sign* of the quasi-probability distribution $q$ for it may be reexpressed as

$$\langle \mathrm{sign} \rangle_p = \sum_{\vec{\lambda} \in \Lambda} p(\vec{\lambda}) \,\mathrm{sign}(q(\vec{\lambda})) = \frac{\sum_{\vec{\lambda} \in \Lambda} |q(\vec{\lambda})| \,\mathrm{sign}(q(\vec{\lambda}))}{\sum_{\vec{\lambda} \in \Lambda} |q(\vec{\lambda})|} = \frac{1}{\|q\|_{\ell_1}}. \tag{9.35}$$

By normalization of the quasiprobability distribution $q$, the average sign takes on its maximal value at unity and is always larger or equal than zero.

Using the Chebyshev bound (9.18) this implies that the optimal number of samples, $s$, required to achieve an error $\epsilon$ when estimating $\langle O \rangle_{\beta,H}$ or $Z_{\beta,H}$ via a linear estimator of the form (9.23) has to scale as

$$s \geq \frac{1}{\langle \mathrm{sign} \rangle_p^2 \epsilon^2}. \tag{9.36}$$

One may interpret the average sign as the ratio between the partition functions of the original system with Hamiltonian $H$ acting on $n$ qubits and a corresponding 'bosonic system' with stoquastic[25] Hamiltonian $H' = H - 2H_\neg$ as $\langle \mathrm{sign} \rangle_p = \mathrm{Tr}[\mathrm{e}^{-\beta H}]/\mathrm{Tr}[\mathrm{e}^{-\beta H'}]$. Here, as throughout the remainder of this thesis, $| \cdot |$ denotes the entrywise absolute value and $H_\neg$ the nonstoquastic part of the Hamiltonian which is defined by $(H_\neg)_{i,j} = h_{i,j}$ for $h_{i,j} > 0$ and $i \neq j$, and zero otherwise. . Generically, such a quantity is expected to scale as $\mathrm{e}^{-\beta n \Delta f}$, that is, inverse exponentially in the particle number $n$, the inverse temperature $\beta$, and the free energy density difference $\Delta f = f' - f \geq 0$ between 'bosonic' and original system [TW05]. This makes Quantum Monte Carlo simulations of systems with a sign problem infeasible.

25: We will define and discuss the notion of stoquasticity in detail below.

**The Monte Carlo sign problem**

The sign problem results in an exponential increase of the sampling complexity and hence the runtime of a Quantum Monte Carlo algorithm.

A basic but fundamental insight is that the Quantum Monte Carlo sign problem is a *basis-dependent* property [HS92; BF00; Has15; RK17]. For this reason, saying that 'a Hamiltonian does or does not exhibit a sign-problem' is meaningless without specifying a basis. Since physical quantities of interest are independent of the basis choice, the observation that the sign problem is basis-dependent gives immediate hope to actually mitigate the sign problem of Quantum Monte Carlo by expressing the Hamiltonian in a suitable basis.

## Curing the sign problem

In fact, it is known that one can completely *cure* the sign problem using basis rotations in certain situations. For specific models, sign-problem free bases can be found analytically, involving non-local bases, for example by using so-called auxiliary-field [WHZ03], Jordan-Wigner [OH14] or Majorana [LJY15; LJY16] transformations. One can also exploit specific known properties of the system such as that the system dimerizes [Nak98; ADP16; Hon+16; Wes+17]. Such findings motivate the quest for a more broadly applicable systematic search for basis changes that avoid the sign problem, in a way that *does not depend* on the specific physics of the problem at hand. After all, in a Quantum Monte Carlo simulation one wants to *learn about the physics* of a system in the first place and, indeed, the optimal basis choice may very well be closely related to that physics.

Clearly, a useful notion of curing has to restrict the set of allowed basis transformation such that expressing the Hamiltonian in the new basis is still computationally tractable. For example, in its eigenbasis every Hamiltonian is diagonal and thus sign-problem free, but even writing down this basis typically requires an exponential amount of resources. The *intrinsic sign problem* of a Hamiltonian is thus a property of its *equivalence classes* under conjugation with some suitable subgroup of the unitary group. The simplest examples of such choices include local Hadamard, Clifford or unitary transformations. Most generally, one can allow for quasi-local circuits which are efficiently computable [Has15], including short circuits and matrix product unitaries [Cir+17; Şah+18]. Going beyond orthogonal bases, one can in principle also allow for efficiently computable invertible transformations [DLA19], which are, while physically less motivated, mathematically perfectly alloed by the cylicity of the trace.

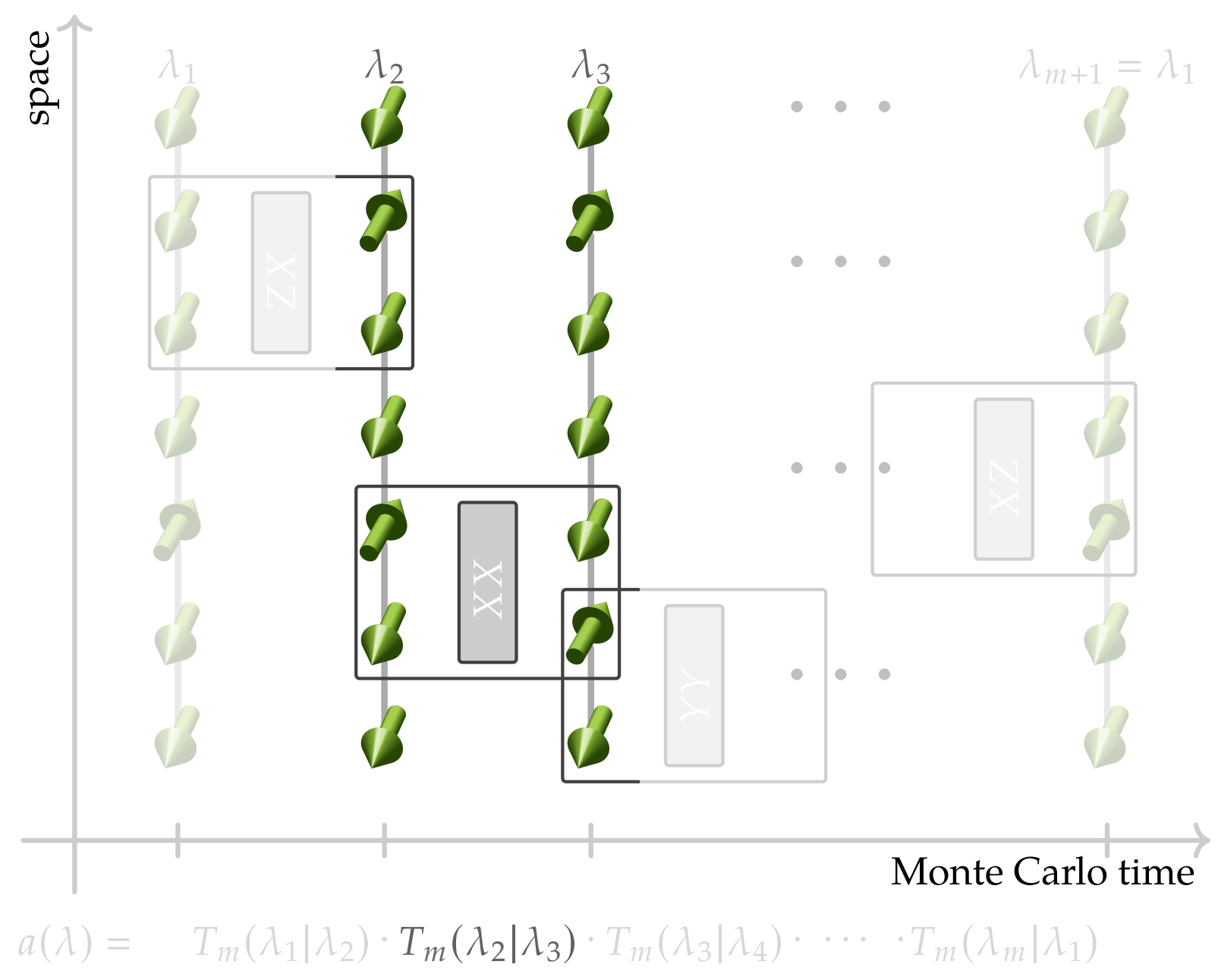

**Figure 9.4:** Rather than requiring all Monte Carlo world-lines to have nonnegative path weights, the stoquasticity condition ensures that every time slice of the world lines is nonnegative.

## Stoquastic Hamiltonians

Whether or not a specific Quantum Monte Carlo simulation suffers from a sign problem, though, not only depends on properties of the Hamiltonian, but also the observable for which the expectation value is evaluated and the remaining simulation parameters, including the inverse temperature $\beta$ and the number of Monte Carlo time steps $m$. When curing the sign problem by local basis choices, one might hope to find a Hamiltonian basis in which *every* Quantum Monte Carlo simulation is sign-problem free. A both useful and simple sufficient condition for the absence of a sign problem, independent of the specifics of a simulation, is that the Hamiltonian matrix is *stoquastic* [Bra+08].

A Hamiltonian matrix is stoquastic in a basis if it has only nonpositive off-diagonal entries in that basis. In world-line Monte Carlo, the transfer matrix $T_m$ of a stoquastic Hamiltonian only has nonnegative matrix entries for a suitably large-choice of the number of Monte Carlo time steps $m$,

$$T_m(\lambda_i|\lambda_j) = 1 - \frac{\beta}{m}\langle\lambda_i|H|\lambda_j\rangle \geq 0, \tag{9.37}$$

so that the amplitudes $a(\vec{\lambda}) \geq 0$ and corresponding quasiprobabilities $q(\vec{\lambda}) \geq 0$ are nonnegative; see Fig. 9.4.

Stoquasticity also provides a particularly simple and useful framework to assess the computational complexity of a systematic approach to curing the sign problem [TW05] that does not depend on the specifics of a simulation procedure. The curing problem is thereby reduced to the task of finding a stoquastic representative in the orbit of the Hamitonian under some efficient basis transformation. Only recently has the curing problem been shown to be an NP-complete task under unitary transformations for 2-local Hamiltonians with additional local fields [MLH19; Kla+19]. At the same time, it remains efficiently solvable for strictly 2-local Hamiltonians in the sense that in polynomial time it can be decided whether a stoquastic basis exists, and – should it exist – such a basis can be found [KT19; Kla+19]. It is also known, however, that not every local Hamiltonian has a stoquastic local basis that can be found efficiently unless AM = QMA [Bra+08], an unlikely complexity-theoretic equality[26].

26: If such a basis could be found then the complexity class StoqMA would equal the class QMA. Since AM is a class intermediate between those two, StoqMA = QMA implies that AM = QMA. Altogether we know the inclusions NP ⊂ MA ⊂ StoqMA ⊂ AM ⊂ QMA; see the appendix for definitions.

---

So is all hope lost for simulating a Hamiltonian problem via Quantum Monte Carlo more efficiently even when a stoquastic basis cannot be found in polynomial time?

## 9.4 A pragmatic approach: Easing the sign problem

In any Monte Carlo algorithm, computational hardness of the underlying problem is manifested in a super-polynomial increase in its sample complexity as the system size grows. Intuitively speaking, the sample complexity increases because the variance of the Monte Carlo estimator does. In this mindset, finding a Quantum Monte Carlo algorithm with feasible runtime for Hamiltonians with a sign problem does not require the much stronger task of finding a basis in which the Hamiltonian is fully stoquastic. Indeed, in many cases such a basis may not even exist within a given subgroup of the unitaries. Rather, often it is sufficient to merely find a basis in which the Hamiltonian is *approximately stoquastic* so that the scaling of the variance of the corresponding estimator with the system size is more favourable – ideally polynomial. Indeed, for the closely related case of calculating quantum circuit amplitudes via sampling, it has been shown that a small amount of negativity can be tolerated without losing efficiency of the algorithm [PWB15]. More pragmatically still, practitioners in Quantum Monte Carlo are increasingly less worried about small sign problems for which simulations are still feasible for reasonable system sizes using state-of-the-art computing power. This remains true even if the sampling effort may strictly speaking diverge exponentially with the system size. Consequently, we argue that practical computational approaches towards the sign problem, rather than focusing on *exactly curing* it, should target the less ambitious yet practically meaningful task of approximately solving or *easing* it in the best possible way.

In this work, we establish a comprehensive novel framework for assessing, understanding, and optimizing the sign problem computationally via basis rotations, asking the questions:

**Easing the Monte Carlo sign problem**

1. What is the optimal, computationally meaningful local basis choice for a Quantum Monte Carlo simulation of a Hamiltonian problem?
2. Can we find the sign-optimal basis?
3. How hard is the sign easing task in general?

An appealing feature of our framework is that it neither requires any *a priori* knowledge about the physics of a problem nor depends on specifics of a given simulation procedure, in contrast to other known refinements of Quantum Monte Carlo. At the heart of our approach lies a formulation of the easing problem in terms of a simple, efficiently computable measure of approximate stoquasticity that generically quantifies the sampling complexity.

## Overview

In the following chapters (Chapters 10, 11, 12) I will attempt to provide some answers to those questions. Those should not be conceived as full solutions but rather as potential starting points for future research. In Chapter 10 we develop efficiently computable measures for the sign problem. Those measures are based on the notion of stoquasticity and we will discuss the relation between the average sign and the nonstoquasticity. In Chapter 11 we perform a proof-of-principle numerical study showing that easing is both feasible and meaningful for translationally invariant models with a sign problem. In Chapter 12 we then study the fundamental limitations of a systematic approach to the sign problem in proving the computational hardness of easing the sign problem when allowing for both orthogonal Clifford (Theorem 12.3) and general orthogonal transformations (Theorem 12.4).

# Efficiently computable measures of the sign problem 10

*What is the optimal, computationally meaningful local basis choice for a QMC simulation of a Hamiltonian problem?*

In this chapter, we set out to provide an answer to this first question posed above. Key to answering this question is an understanding of the severeness of the sign problem in terms of the runtime of a Quantum Monte Carlo (QMC) algorithm. Recall from Sec. 9.3 that the optimal sample complexity of a QMC algorithm with a sign problem is determined by the inverse average sign, which directly bounds the variance of the QMC estimator (cf. Eq. (9.34)) [TW05].

*This paragraph has been published as part of Ref. [Han+20].*

In an attempt to ease the sign problem of a given Hamiltonian it is therefore natural to try and improve the average sign. For a few specific models such improvements have indeed been achieved by different means: for example, it is possible to exploit known physics to find bases with improved average sign [Shi+15; Wes+17]. Those bases are often induced by sparse representations [MA15; Tho+15; DLA19]. For particular observables, one can also exploit clever decompositions of the Monte Carlo estimator into clusters of paths such that the sum of path weights within a cluster has a nonnegative sign [BPW95; CW99; HS00; Nyf+08; HC16; HHC17; Hen19].

However, all such approaches suffer from the intrinsic flaw that they are inevitably tied to a particular instance of a QMC simulation of a Hamiltonian system – with fixed system size, inverse temperature, and number of Monte Carlo steps. For each instance the average sign has to be improved anew – a computationally challenging task. What is more, in many instances, known physics is exploited in order to find better bases. But in many situations it is precisely the goal to learn about the physics of a system so that such information is not available. Turning the above around, ideally, finding the sign-optimal basis would already tell us qualitative features of the physics of a system.

Let us now formulate the minimization of the inverse average sign in its most general form. We take this as a starting point into the search for efficiently computable measures of the sign problem.

## 10.1 Finding the optimal basis for Quantum Monte Carlo

More generally speaking, in order to minimize the relative approximation error of a QMC algorithm, we need to minimize the inverse average sign, or equivalently $\|q\|_{\ell_1}$, over the allowed set of basis choices which we denote by $\mathcal{U}$. A basis choice $U \in \mathcal{U}$ transforms the matrix representation of the Hamiltonian $H$ as $H \mapsto UHU^\dagger$ and likewise $T_m \mapsto UT_mU^\dagger$ as the transfer matrix $T_m$ is linear in $H$[1]. The inverse average sign or

1: Recall the definition of the transfer matrix of world-line Monte Carlo as $T_m = 1 - \beta H/m$.

$\ell_1$-norm of the quasiprobability distribution $q$ defined in Eq. (9.29) is then transformed as

$$\|q\|_{\ell_1} \mapsto \frac{\operatorname{Tr}\left[|UT_mU^\dagger|^m\right]}{\operatorname{Tr}[(UT_mU^\dagger)^m]}, \tag{10.1}$$

where as throughout the remainder of this thesis $|\cdot|$ denotes taking the entry-wise absolute value and not the matrix absolute value. Notice that the denominator is in fact unitarily invariant and hence always equal to $\operatorname{Tr}[(T^m)^m] \approx \operatorname{Tr}[\mathrm{e}^{-\beta H}]$. Consequently, in order to optimally ease the sign problem in terms of its sample complexity[2] the following minimization problem

2: The sample complexity gives a lower bound to the computational complexity; cf. Chapter 6.

$$\min_{U\in\mathcal{U}} \|q\|_{\ell_1}^2 - 1 = \min_{U\in\mathcal{U}} \frac{\operatorname{Tr}[|UT_mU^\dagger|^m]^2}{\operatorname{Tr}[T_m^m]^2} - 1, \tag{10.2}$$

has to be solved,

*The following sections have (up to small alterations) been published as Sec. II of Ref. [Han+20].*

↓ ↓ ↓

The difficulty in dealing with the minimization problem (10.2) is manifold. First, determining the quantity $\|q\|_{\ell_1} = \operatorname{Tr}[|T_m|^m]/\operatorname{Tr}[T_m^m]$ via QMC suffers from the very sign problem it quantifies: it can easily be checked that the relative variance of $\langle\text{sign}\rangle_p$ is precisely given by $\langle\text{sign}\rangle_p^{-2} - 1$. It thus inherits the complexity of computing the partition function $Z_{\beta,H}$ in the first place. Naïve optimization of the term $\operatorname{Tr}[|T_m|^m]/\operatorname{Tr}[T_m^m]$ even incurs the cost of diagonalizing the exponential-size matrices $T_m$ and $|T_m|$. Ironically, easing the sign problem by optimizing the average sign directly is therefore typically infeasible whenever there is a sign problem.

Second, the optimization problem is non-convex and highly non-linear in the unitary transformation $T \mapsto UTU^\dagger$ with $U \in \mathcal{U}$. While it might be possible to minimize the unitarily dependent term $\operatorname{Tr}[|T_m|^m]$ and its gradient stochastically via QMC in some cases [SBF12; LC19], such approaches cannot yield certificates for the quality of the obtained basis as the average sign itself is not computed. Moreover, they are dependent on the distribution defined by $|T_m|$ being well-behaved (i.e., ergodic and satisfying detailed balance) for QMC algorithms.

It therefore seems infeasible to find a converging and efficient algorithm for minimizing the average sign for general Hamiltonians directly. In analogy to the notion of stoquasticity, we would hence like to quantify the severeness of the sign problem in terms of a measure solely depending on the basis expression of the Hamiltonian. Such a measure could in particular be efficiently evaluated for physical Hamiltonians – a crucial property to be practically useful in a general approach to easing the sign problem. Ideally, we could thus find a simple measure of the nonstoquasticity of the Hamiltonian which can be connected to the inverse average sign in a meaningful way while at the same time admitting efficient evaluation. In the following, we explore the possibility of the existence of an (efficiently computable) nonstoquasticity measure that has a meaningful connection to the average sign.

To gain an intuition for the intricate interplay between the nonstoquasticity and the average sign, we begin by constructing examples of QMC procedures in Section 10.2[3]. We find that it is in fact impossible to directly connect such a continuous measure of nonstoquasticity to the average sign, which takes on its maximal value at unity and achieves

3: This question was also asked in the recent work by Gupta and Hen [GH20] in the context of the *off-diagonal series expansion* method [AWH17].

this value for stoquastic Hamiltonians: We can construct exotic examples of highly nonstoquastic Hamiltonians with large positive off-diagonal entries which also have unit average sign. Conversely, we provide an example of a Hamiltonian with arbitrarily small nonstoquasticity for which the average sign nearly vanishes.

On the one hand, our examples demonstrate a high sensitivity of the average sign to the Monte Carlo parameters. On the other hand, they also require a malicious interplay between the Hamiltonian matrix entries and highly fine-tuned Monte Carlo parameters. We therefore expect that in practically relevant situations a notion of approximate stoquasticity can be meaningfully connected to the sample complexity of QMC. In Section 10.3 we provide both analytical arguments and numerical evidence that this is indeed possible: for a real $D \times D$ Hamiltonian matrix $H$ we propose the sum of all nonstoquastic matrix entries[4]

$$\nu_1(H) := D^{-1} \|H_\neg\|_{\ell_1}, \tag{10.3}$$

as a natural measure of nonstoquasticity[5]. In principle, one can also conceive of other measures of nonstoquasticity such as the $\ell_{1\to 1}$-norm or the $\ell_2$-norm of the nonstoquastic part of $H$. We argue that the $\ell_1$-norm is the most meaningful measure that is agnostic to any particular structure of the Hamiltonian matrix and therefore the most versatile measure for a general approach to easing the sign problem. What is more, it acts as a natural regularizer promoting a sparse representation [FR13] in the spirit of Refs. [MA15; Tho+15; DLA19].

For local Hamiltonians on bounded-degree graphs such as regular lattices this measure can be efficiently computed from the nonstoquastic entries of the local terms themselves – for translation-invariant Hamiltonians even with constant effort. But we can also go beyond that and prove that for 2-local Hamiltonians acting on any graph the measure $\nu_1$ can be efficiently approximated up to any inverse polynomial error; see Theorem 10.2. The latter result renders our measure applicable to problems with long-range interactions as they arise for example in quantum chemistry.

4: Recall that we denote the nonstoquastic part of the Hamiltonian by $H_\neg$ which is defined by $(H_\neg)_{i,j} = h_{i,j}$ for $h_{i,j} > 0$ and $i \neq j$, and zero otherwise, and that $\|H\|_{\ell_1} = \sum_{i,j} |h_{i,j}|$ is the vector-$\ell_1$-norm.

5: In contrast to Ref. [MLH19] where *term-wise stoquasticity* is considered, our definition remains on the level of the global Hamiltonian. This is because positive matrix elements of the local Hamiltonian terms may cancel in the global matrix representation. Term-wise stoquasticity is thus meaningful (only) in the following sense: if a Hamiltonian admits a term-wise stoquastic local basis, it has no sign problem on an *arbitrary* lattice. However, for any given graph, it might well be possible to fully cure the sign problem using an allowed set of transformations even if the Hamiltonian *cannot* be made term-wise stoquastic.

## 10.2 Gaining intuition: Case studies

We now show that this hope is in vain in its most general formulation. Specifically, we provide an example of a Hamiltonian which has large positive entries but is nevertheless sign-problem free (has unit average sign) for specific choices of $\beta$ and $m$, as well as an example of an Hamiltonian that is close to stoquastic but incurs an arbitrarily small average sign for certain choices of $\beta$ and $m$ in a specific QMC procedure.

Here, as throughout this work, whenever we consider systems of multiple qubits, for $A \in \mathbb{C}^{2\times 2}$ we define

$$A_i = A \otimes 1_{\{i\}^c}, \tag{10.4}$$

to be the operator that acts as $A$ on qubit $i$ and trivially on its complement $\{i\}^c$.

**Example 10.1** (Highly nonstoquastic but sign-problem free Hamiltonians) Let us define a Hamiltonian term acting on two qubits with label $i, j$ as

$$h_{i,j} = -\frac{1}{2}(X_i X_j - Y_i Y_j) + X_i. \tag{10.5}$$

Then this Hamiltonian term is nonstoquastic with total weight $\nu_1(h_{i,j}) = 1$. What is more, the $n$-qubit Hamiltonian

$$H = 1 + \sum_{i<j}^{n} h_{i,j} \tag{10.6}$$

is highly nonstoquastic with total weight $\nu_1(H) = n$. At the same time, the QMC algorithm for computing the partition function of $H$ with parameters $\beta, m$, has average sign $\langle \text{sign} \rangle_{\beta,m} = 1$.

*Proof.* We first determine the nonstoquasticity of $H$ as

$$\nu_1(H) = \sum_i \nu_1(X_i) = n. \tag{10.7}$$

To see why the QMC algorithm has unit average sign, note that the transfer matrix $T_m = 1 - \beta H/m$ has negative entries $T_m(\lambda|\lambda') < 0$ only if the parity of $\lambda \oplus \lambda'$ is odd since for these terms only a single $X$ term contributes. Whenever $\lambda \oplus \lambda' = 0$, i.e., has even parity, we have $T_m(\lambda|\lambda') \geq 0$ since only $XX - YY$ terms or the diagonal contribute – both of which have non-negative matrix elements.

In the calculation of the partition function, the summation runs over closed paths only. But for any closed path $\lambda_1 \to \lambda_2 \to \cdots \to \lambda_m \to \lambda_1$, it is necessary that the total parity $(\lambda_1 \oplus \lambda_2) \oplus \ldots \oplus (\lambda_m \oplus \lambda_1)$ vanishes. In particular, this implies that every allowed path incurs an even number of odd-parity steps and therefore an even number of negative signs. Therefore, only non-negative paths contribute to the path integral and the average sign is attained at unity. □

**Example 10.2** (Barely nonstoquastic Hamiltonians with arbitrarily small average sign) Let us define the 2-qubit Hamiltonian

$$H_{a,b} = \frac{m}{\beta}\left(1 \otimes 1 - 1 \otimes X - \frac{1}{2}(X \otimes X + Y \otimes Y)\right. \tag{10.8}$$

$$\left. + \frac{1}{2}[(a+b)X \otimes Z + (b-a)X \otimes 1]\right), \tag{10.9}$$

with $b \geq a > 0$ positive numbers and $m \in 2\mathbb{N} + 1$ . The nonstoquasticity of $H_{a,b}$ is given by $\nu_1(H_{a,b}) = bm/(2\beta)$, the average sign of QMC with parameters $\beta$ and $m$ is dominated by $|\langle \text{sign} \rangle_{\beta,m}| \leq C(b-a)/a$, where $C$ is a constant. Thus, even for arbitrarily small nonstoquasticity we can make the sign problem unboundedly severe as we tune $a$ to be close to $b$.

*Proof.* We derive the bound on the average sign. For the given Hamiltonian, the corresponding transfer matrix for a QMC algorithm for inverse temperature $\beta$ with $m$ steps is given by

$$T_m \equiv T_{a,b} = \begin{pmatrix} 0 & 1 & -b & 0 \\ 1 & 0 & 1 & a \\ -b & 1 & 0 & 1 \\ 0 & a & 1 & 0 \end{pmatrix}. \tag{10.10}$$

Recall that the average sign can be written as

$$\langle \text{sign} \rangle_{\beta,m} = \frac{\mathrm{Tr}[T_m^m]}{\mathrm{Tr}[|T_m|^m]}. \tag{10.11}$$

We denote by $\overline{T}_m$ a matrix similar to $T_m$ but where the positions of $a$ and $-b$ are exchanged. Due to the symmetry of the problem we have that $\mathrm{Tr}[T_m^m] = \mathrm{Tr}[\overline{T}_m^m]$ and $\mathrm{Tr}[|T_m|^m] = \mathrm{Tr}[|\overline{T}_m|^m]$. Hence,

$$\mathrm{Tr}[T_m^m] = \frac{1}{2}\left[\mathrm{Tr}[T_m^m] + \mathrm{Tr}[\overline{T}_m^m]\right] \tag{10.12}$$

$$= \frac{1}{2}\sum_{\vec{\lambda}\in\Lambda_m}\Bigg[T_m(\lambda_1 \mid \lambda_2)\cdots T_m(\lambda_m \mid \lambda_1) \tag{10.13}$$

$$+ \overline{T}_m(\lambda_1 \mid \lambda_2)\cdots \overline{T}_m(\lambda_m \mid \lambda_1)\Bigg] \tag{10.14}$$

$$= \frac{1}{2}\sum_{\vec{\lambda}}\left[a^{f(\vec{\lambda})}(-b)^{g(\vec{\lambda})} + a^{g(\lambda)}(-b)^{f(\vec{\lambda})}\right], \tag{10.15}$$

where in the last line we have used the fact that every summand is a polynomial in the entries of $T_{a,b}$. The functions $g, f : \Lambda_m \to [m]$ describe the corresponding exponents. A little thought reveals that since all path are closed and $m$ is odd $g(\vec{\lambda}) + f(\vec{\lambda})$ is larger than 1 and also odd for all $\vec{\lambda}$. We thus find that one of the two terms for each $\vec{\lambda}$ must be negative and

$$|\,\mathrm{Tr}\, T_m^m| \leq \frac{1}{2}\sum_{\vec{\lambda}}|a^{f(\vec{\lambda})}b^{g(\vec{\lambda})} - b^{f(\lambda)}a^{g(\vec{\lambda})}| \tag{10.16}$$

$$\leq \frac{1}{2}\sum_{\vec{\lambda}}(2^{g(\vec{\lambda})} - 1)a^{f(\vec{\lambda})+g(\vec{\lambda})-1}|b - a|. \tag{10.17}$$

Furthermore, we have

$$|\mathrm{Tr}\,|T_m|^m| = \frac{1}{2}\sum_{\vec{\lambda}}(a^{f(\vec{\lambda})}b^{g(\vec{\lambda})} + b^{f(\lambda)}a^{g(\vec{\lambda})}) \tag{10.18}$$

$$\geq \sum_{\vec{\lambda}}\left(a^{f(\vec{\lambda})+g(\vec{\lambda})}\right). \tag{10.19}$$

Combining these two bounds and using $g(\vec{\lambda}) \leq m$, we conclude that

$$|\langle \text{sign} \rangle| \leq \left(2^{m-1} - \frac{1}{2}\right)\frac{|b - a|}{a}. \tag{10.20}$$

□

The second example shows that in principle also Hamiltonians with arbitrarily small positive entries can cause a severe increase of the sampling complexity of specific Monte Carlo algorithms. Interestingly, in this situation the sign problem cannot be eased by a basis change: the average sign vanishes since the *unitarily invariant* term $|\operatorname{Tr} T_m^m|$ is tuned to be small. On the contrary, the sign problem can be completely avoided in this example by choosing the Monte-Carlo path length to be even instead of odd.

These simple examples illustrate the following important observation: The sign problem as measured by the average sign can in certain situations be avoided or intensified by fine-tuning the problem and parameters of the QMC procedure independently of the actual magnitude of the positive entries of the Hamiltonian. But such examples seem to rely on an intricate conspiracy of the structure of the Hamiltonian and the chooen QMC procedure, e.g., the discretization. It is plausible to assume that the most pathological cases are unlikely to appear in practical applications, and can at least be rather easily overcome by slightly modifying the QMC algorithm.

## 10.3 Measures of nonstoquasticity

In this work, our goal is to develop a more general methodology for the task of easing the sign problem that is independent of the details of the QMC algorithm and the combinatorial properties of potential paths that can be constructed from the entries of the transfer matrix. Very much in the spirit of the notion of stoquasticity we aim at a property of the Hamiltonian in a given basis to measure its deviation from having a good sampling complexity. Natural candidates for such a nonstoquasticity measure of a Hamiltonian are entry-wise norms of its positive entries. For any $p \geq 1$ we define the nonstoquasticitiy of $H$ as

$$\nu_p(H) = D^{-1} \|H_\neg\|_{\ell_p}, \tag{10.21}$$

where $\|\cdot\|_{\ell_p}$ denotes the vector-$\ell_p$ norm. For every $p$, $\nu_p$ is efficiently computable for local Hamiltonians on bounded-degree graphs and therefore suitable for easing the sign problem of a large class of Hamiltonians by local basis choices. It is also obviously a measure of the nonstoquasticity in the sense that $\nu_p(H) = 0$ if and only if $H$ is stoquastic. We note that we have chosen our definition such that the nonstoquasticity measure $\nu_p$ scales extensively in the number of nonstoquastic terms of a local Hamiltonian. This is because every nonstoquastic local Hamiltonian term creates on the order of $2^n$ positive matrix entries in a global $n$-qubit Hamiltonian matrix due to tensoring with identities on the complement of its support.

Our examples in the previous section show that it is notoriously difficult if not impossible to connect any notion of nonstoquasticity to the actual sample complexity incurred by a QMC procedure as measured by the inverse average sign. This is due to the dependence of the average sign on the combinatorics of Monte Carlo paths. However, those examples involved a large degree of fine-tuning. Therefore, one might hope to find a connection between nonstoquasticity and average sign for *generic* cases.

So let us look at the connection between optimizing a nonstoquasticity measure $\nu_p$ and optimizing the QMC sampling complexity as in (10.2). Our measure can be expressed in terms of the transfer matrix $T_m$ as

$$\nu_p(H) = \frac{1}{D}\frac{m}{2\beta}\||T_m| - T_m\|_{\ell_p}, \tag{10.22}$$

where we assume that $\mathrm{diag}(\beta H/m) \leq 1$.

Due to the unitary invariance of the trace, the optimization of the sampling complexity via (10.2) is equivalent to minimising

$$S(U) = \mathrm{Tr}[|UT_mU^\dagger|^m] - \mathrm{Tr}[T_m^m]. \tag{10.23}$$

Let us for the sake of clarity, suppress the explicit dependence on the unitary $U$ and define $\hat{T}_m = UT_mU^\dagger$. If we define the positive and negative entries of the transfer matrix respectively as $\Delta_\pm = \frac{1}{2}\left(|\hat{T}_m| \pm \hat{T}_m\right)$, we can write

$$S(U) = \mathrm{Tr}[|\hat{T}_m|^m] - \mathrm{Tr}[\hat{T}_m^m] \tag{10.24}$$

$$= 2\sum_{\substack{\vec{s}\in\{\pm\}^m:\\ \vec{s}\text{ odd}}} \mathrm{Tr}[\Delta_{s_1}\cdots\Delta_{s_m}]. \tag{10.25}$$

The summation in the last line is restricted to all $\vec{s} \in \{\pm\}^m$ with an odd number of negative signs. The resulting expression thus involves a summation over closed paths that contain an odd number of negative contributions such that $\Delta_{s_1}(\lambda_1|\lambda_2)\cdots\Delta_{s_m}(\lambda_m|\lambda_1) < 0$. In particular, every such path contains at least one step with a negative contribution.

The size of $S(U)$ thus depends both on the number of 'negative paths' and their individual weight. Suppose the negative entries of $\hat{T}_m$ are small compared to the positive entries. In this case, the dominant contribution to $S(U)$ are paths with exactly one negative step so that we can approximate

$$S(U) = 2\sum_{\substack{\vec{s}\in\{\pm\}^m:\\ \vec{s}\text{ odd}}} \mathrm{Tr}[\Delta_{s_1}\cdots\Delta_{s_m}] \tag{10.26}$$

$$\approx 2m\sum_{\lambda_1,\lambda_2} \Delta_-(\lambda_1|\lambda_2)\Delta_+^{m-1}(\lambda_2|\lambda_1), \tag{10.27}$$

using the cyclicity of the trace. The expression in Eq. (10.27) is a weighted sum over the negative entries of $\hat{T}_m$, where the weights are given by the contribution $\Delta_+^{m-1}(\lambda_2|\lambda_1)$ of all positive paths of length $m-1$.

For a transfer matrix in which the positive entries do not significantly differ and their distribution relative to the negative entries is unstructured, we have constant $\Delta_+^{m-1}(\lambda_2|\lambda_1) \approx \|\Delta_+^{m-1}\|_{\ell_\infty}$. Therefore,

$$S(U) \approx 2m\|\Delta_-\|_{\ell_1}\|\Delta_+^{m-1}\|_{\ell_\infty} \propto D\,\nu_1(H). \tag{10.28}$$

We further observe that if the positive entries of $\hat{T}_m$ are more structured, the weights appearing in (10.27) might deviate from a uniform distribution. In such a case, other $\nu_p$-measures become meaningful as a measure of the inverse average sign since they saturate a corresponding Hölder bound. We argue that generically, the $\ell_1$ norm is the most meaningful measure of nonstoquasticity, however. This is because by minimizing the $\ell_1$ norm of $H_\neg$ one not only reduces the size of the positive matrix entries but also promotes a sparse matrix representation of the Hamiltonian and its transfer matrix [FR13]. Finding such a sparse matrix representation is vital to many implementations of Monte Carlo methods with a sign problem [MA15; Tho+15; DLA19].

Indeed, notice that when improving the overall average sign via basis transformations there is a tradeoff in terms ofbetween increased sparsity of the matrix representation of $H$ and decreased overall weight of its positive entries. On the one hand, for very sparse Hamiltonian matrices, there are only few paths with nonzero contribution to the overall path integrals (9.25) and (9.26). In this case the positive matrix entries giving rise to negative path weights cannot contribute to many paths. On the other hand, the size of the path weights depend strongly – namely as all odd powers between 1 and the number of Monte Carlo steps $m$ – on the size of the individual matrix entries. The tradeoff is therefore one between a combinatorial increase of possible paths incurring an odd number of positive matrix entries and an exponential decrease of the individual path weights.

We can also view the sparsity of the Hamiltonian matrix from the perspective of the probability distribution it gives rise to in a QMC algorithm[6]. For fixed nonstoquasticity $\|H_\neg\|_{\ell_1}$ a very sparse Hamiltonian gives rise to a sparse probability distribution which is supported on few paths $\lambda_1 \to \lambda_2 \to \cdots \to \lambda_m \to \lambda_1$ only. The individual probabilities are therefore reasonably large. In contrast, the probability distribution over the path weights of a dense Hamiltonian matrix is supported on a large fraction of the possible paths albeit with smaller probabilities. Dense Hamiltonians therefore give rise to distributions which much larger tails, making the sampling task increasingly difficult.

6: Recall that we encountered a similar contrast already when studying the simulatability and verifiability of quantum supremacy distributions in Part I, specifically, in Chapter 6. The evidence provided by the complexity-theoretic arguments points towards the following observation: distributions that are both far from uniform and extremely flat tend to be hard to sample from, while the complexity of this task is greatly reduce for sparse distributions (i.e., distributions with polynomially large support). In contrast, here, we are considering the question of computing the output probabilities of such distributions, and, interestingly, a similar phenomenon arises.

We leave a more thorough investigation of the intricate interplay between the sparsity and the nonstoquasticity that governs the average sign to future work.

*The following has been published as part of Sec. II in Ref. [Han+20].*

↓ ↓ ↓

For higher-order negative contributions, we expect that $S(U)$ or, correspondingly, the average sign scales as $\exp(c \cdot D\, \nu_1(H))$ for some $c > 0$. Our expectation is based on the following observation: in the calculation of the inverse average sign, all paths of length $m$ with an odd number of negative steps contribute. Potentially, in each step every negative entry of $T_m$ appears. Then the sum of all negative entries of $T_m$ contributes. But the number of paths with $k \in 2\mathbb{N}_0 + 1$ negative steps scales as $\binom{m}{k}$ which leads to an exponential growth in $\|H_\neg\|_{\ell_1}$ and hence $D\, \nu_1(H)$. In the following section, we provide a brief numerical analysis confirming this expectation.

## Numerical analysis

We now provide evidence that $\nu_1(H)$ is indeed a very much meaningful measure of the sample complexity and hence the inverse average sign by exactly calculating the inverse average sign as a function of $\nu_1(H)$. We do so by randomly drawing 2-local Hamiltonians on a line of $n$ qubits of the

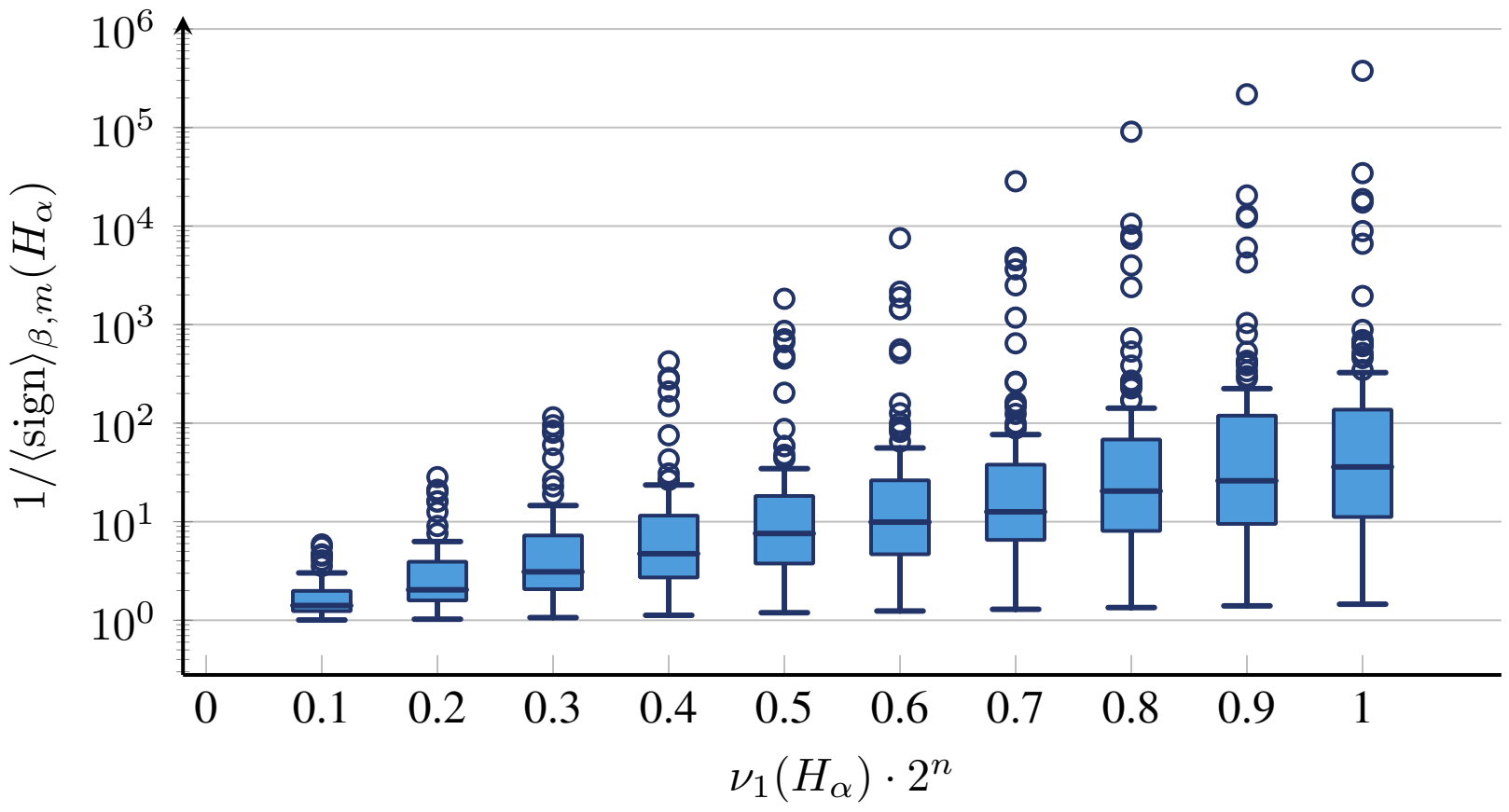

**Figure 10.1:** We show the inverse average sign for 100 randomly chosen instances of 5-qubit Hamiltonians $H_\alpha$ for $\beta = 1$ and $m = 100$ Monte Carlo steps as a function of $d\nu_1(H_\alpha)$ in terms of a box plot [IS0 16269, p. 2.16]. We find a roughly exponential dependence of the inverse average sign with $\nu_1(H_\alpha)$ as $1/\langle \text{sign} \rangle_{\beta,m}(H_\alpha) \propto \exp(a \cdot d\nu_1(H_\alpha))$ for $a > 0$.

form

$$H = \sum_{i=1}^{n} T_i(h), \tag{10.29}$$

where $h \in \mathbb{R}^{4\times 4}$ is a nearest-neighbour interaction term and the translation operator $T_i$ acts as $T_i(h) = 1_d^{\otimes(i-1)} \otimes h \otimes 1_d^{\otimes n-i-1}$. We choose each local term $h$ in an i.i.d. fashion from the zero-centered Gaussian measure and projecting to Hermitian matrices. For each random instance $H$, we then consider the one-parameter Hamiltonian family

$$H_\alpha = \frac{H - H_\neg + \alpha H_\neg}{2^n \nu_1(H_\neg)}. \tag{10.30}$$

Note that $\nu_1(H_\alpha) = \alpha/2^n$. Fig. 10.1 shows that, generically, the average sign monotonously depends on the nonstoquasticity. Indeed, as expected for large $m$, the dependence is an exponential one.

## 10.4 Efficiently computing the nonstoquasticity

I am grateful to Barbara Terhal for an insightful discussion on the question of how to decide whether a local Hamiltonian is stoquastic.

Above, we have argued that a key problem of the average sign lies in the fact that it is not efficiently computable whenever there is a sign problem. But how does the nonstoquasticity measure $\nu_1$ fare in this respect? We now show that for arbitrary 2-local Hamiltonians the nonstoquasticity measure $\nu_1$ can in fact be approximated up to an inverse polynomially small additive error in polynomial time. While this is not possible for arbitrary local Hamiltonians as a simple example shows, any $\nu_p$-measure can be efficiently computed exactly in polynomial time for local Hamiltonians acting on bounded-degree graphs.

We write a real 2-local Hamiltonian with 1-local terms as

$$\begin{aligned} H_{2+1} = \sum_{i<j} \big( & a_{i,j} X_i X_j + b_{i,j} Y_i Y_j + c_{i,j} Z_i Z_j \\ & + x_{i,j} X_i Z_j + x_{j,i} Z_i X_j \big) + \sum_i \big( \alpha_i X_i + \gamma_i Z_i \big) , \end{aligned} \tag{10.31}$$

parametrized by real coefficient vectors $a, b, c \in \mathbb{R}^{n(n-1)/2}$, $x \in \mathbb{R}^{n(n-1)}$ which are non-zero only on the edges $(i, j) \in E$ of the Hamiltonian graph $G$ as well as vectors $\alpha, \gamma \in \mathbb{R}^n$. Notice that obtaining an expression for the nonstoquasticity is non-trivial since several local Hamiltonian terms may contribute to a particular entry of the global Hamiltonian matrix.

**Lemma 10.1** (Non-stoquasticity of (2 + 1)-local Hamiltonians) *The non-stoquasticity measure $\nu_1$ of real 2-local Hamiltonians with 1-local terms of the form $H_{2+1}$ satisfies*

$$\begin{aligned} \nu_1(H_{2+1}) = &\sum_{i<j} \nu_1(a_{i,j} X_i X_j + b_{i,j} Y_i Y_j) \\ &+ \sum_i \nu_1\left(\alpha_i X_i + \sum_{j \in \mathcal{N}_{XZ}(i)} x_{i,j} X_i Z_j\right), \end{aligned} \tag{10.32}$$

*and it holds that*

$$\begin{aligned} &\nu_1(a_{i,j} X_i X_j + b_{i,j} Y_i Y_j) = \\ &\quad \frac{1}{2} \sum_{i<j} \left(\max\{a_{i,j} + b_{i,j}, 0\} + \max\{a_{i,j} - b_{i,j}, 0\}\right), \end{aligned} \tag{10.33}$$

$$\begin{aligned} &\nu_1\left(\alpha_i X_i + \sum_{j \in \mathcal{N}_{XZ}(i)} x_{i,j} X_i Z_j\right) = 2^{-\deg_{XZ}(i)} \\ &\quad \times \sum_{\lambda_{\mathcal{N}_{XZ}(i)}} \max\left\{\alpha_i + \sum_{j \in \mathcal{N}_{XZ}(i)} (-1)^{\lambda_j} x_{i,j}, 0\right\}. \end{aligned} \tag{10.34}$$

Here, we have defined the $XZ$ neighbourhood $\mathcal{N}_{XZ}(i) = \{j : x_{ij} \neq 0\}$ of site $i$ as all vertices $j$ connected to $i$ by an $X_i Z_j$-edge, as well as $\deg_{XZ}(i) = |\mathcal{N}_{XZ}(i)|$ and $\lambda_{\mathcal{N}_{XZ}(i)} = (\lambda_j)_{j \in \mathcal{N}_{XZ}(i)}$ as shorthands. We also conceive of summation over an empty set (the case that $\mathcal{N}_{XZ}(i) = \{\}$) as resulting in 0 so that the corresponding term in Eq. (12.7) vanishes.

Notice that the nonstoquasticity of an $XZ$ term does not depend on the sign of its weight, while for $XX$ and $X$ terms it crucially does.

*Proof.* We can determine the $\ell_1$-norm of the off-diagonal part of the Hamiltonian $H_{2+1}$ as

$$\begin{aligned} &\|H_{2+1} - \mathrm{diag}(H_{2+1})\|_{\ell_1} \\ &= \sum_\lambda \left\{ \sum_{i<j} |a_{i,j} - (-1)^{\lambda_i + \lambda_j} b_{i,j}| \right. \\ &\quad \left. + \sum_i \left| \alpha_i + \sum_{j \in \mathcal{N}_{XZ}(i)} (-1)^{\lambda_j} x_{i,j} \right| \right\} \\ &= 2^{n-1} \sum_{i<j} \left(|a_{i,j} + b_{i,j}| + |a_{i,j} - b_{i,j}|\right) \\ &\quad + \sum_i 2^{n - \deg_{XZ}(i)} \sum_{\lambda_{\mathcal{N}_{XZ}(i)}} \left| \alpha_i + \sum_{j \in \mathcal{N}_{XZ}(i)} (-1)^{\lambda_j} x_{i,j} \right|. \end{aligned} \tag{10.35}$$

From Eq. (10.35) we can then directly calculate the nonstoquasticity $\nu_1$ of $H_2$ by discarding all matrix entries with negative sign before taking the $\ell_1$-norm and dividing by $2^n$. □

Now, clearly we can efficiently compute the term (10.33) for arbitrary graphs as the sum runs over at most $n(n-1)/2$ many terms. In the term (10.34), in contrast, the sum over $\lambda_{\mathcal{N}_{XZ}(i)}$ runs over $2^{\deg_{XZ}(i)}$ many terms and hence this term is efficiently computable exactly in case the vertex degree $\deg_{XZ}(i)$ of any vertex $i$ grows at most logarithmically with $n$. This shows that for bounded-degree graphs such as regular lattices the nonstoquasticity can be computed efficiently.

But what if the degree of the graph grows faster than logarithmically with $n$ so that the sum runs over super-polynomially many non-trivial terms? The following Lemma shows that even in this case, that is, for 2-local Hamiltonians acting on arbitrary graphs, the nonstoquasticity can be efficiently approximated up to any inverse polynomially small additive error using Monte Carlo sampling.

**Theorem 10.2** *The sum* (10.34) *can be efficiently approximated up to additive error* $\epsilon$ *via Monte Carlo sampling with failure probability* $\delta$ *from*

$$16 \deg_{XZ}(i)(\max_j |x_{i,j}|)^2 \log(2/\delta)/\epsilon^2 \tag{10.36}$$

*many iid. samples.*

*Proof.* For the proof we will use concentration of measure for Lipschitz functions. To this end we begin by noticing that the sum (10.34) can be rewritten as a uniform expectation value over $k_i = \deg_{XZ}(i)$ many Rademacher (±1) random variables as

$$\begin{aligned}\sum_{\lambda_{\mathcal{N}_{XZ}(i)}} \max\left\{\alpha_i + \sum_{j\in\mathcal{N}_{XZ}(i)} (-1)^{\lambda_j} x_{i,j}, 0\right\} \\ = \mathbb{E}_{\sigma\in\{\pm1\}^{k_i}}[f^{(i)}_{\alpha,x}(\sigma)],\end{aligned} \tag{10.37}$$

where we have defined

$$\begin{aligned} f^{(i)}_{\alpha,x} &: \mathbb{R}^{k_i} \longrightarrow \mathbb{R} \\ s &\mapsto \max\left\{\alpha_i + \sum_{j\in\mathcal{N}_{XZ}(i)} s_j x_{i,j}, 0\right\}.\end{aligned} \tag{10.38}$$

It can easily be seen that the function $f^{(i)}_{\alpha,x}$ is Lipschitz with constant $\left(\max_j |x_{i,j}|\right)k_i^{1/2}$:

$$\left|f^{(i)}_{\alpha,x}(s) - f^{(i)}_{\alpha,x}(s')\right| = \left|\sum_{j=1}^{\deg_{XZ}(i)} x_{i,j}\left(s_i - s'_i\right)\right| \tag{10.39}$$

$$\leq \left(\max_j |x_{i,j}|\right)\|s - s'\|_{\ell_1} \tag{10.40}$$

$$\leq \left(\max_j |x_{i,j}|\right)k_i^{1/2}\|s - s'\|_{\ell_2}. \tag{10.41}$$

Here, we have used the fact that the $\ell_p$ norms on $\mathbb{R}^n$ satisfy Moreover, $f^{(i)}_{\alpha,x}$ is clearly *separately convex*, that is, for each $k = 1, 2, \dots, k_i$ the function $s_j \mapsto f^{(i)}_{\alpha,x}(s_1, s_2, \dots, s_{j-1}, s_j, s_{j+1}, s_{j+2}, \dots, s_n)$ is convex for each fixed $(s_1, s_2, \dots, s_{j-1}, s_{j+1}, s_{j+2}, \dots, s_n) \in \mathbb{R}^{k_i-1}$.

We can then apply [Wai19, Theorem 3.4] to obtain that the estimator

$$\hat{f}^{(i)}_{\alpha,x} = \frac{1}{m} \sum_{l=1}^{m} f^{(i)}_{\alpha,x}(\sigma^{(l)}), \tag{10.42}$$

for the $m$ Rademacher vectors $\sigma^{(l)} \in \{\pm 1\}^{k_i}$ drawn iid. uniformly at random satisfies

$$\mathbb{P}_\sigma \left[ \left| \hat{f}^{(i)}_{\alpha,x} - \mathbb{E}_\sigma [ f^{(i)}_{\alpha,x} \sigma ] \right| \geq \epsilon \right] \leq 2\mathrm{e}^{-\frac{m\epsilon^2}{16 k_i (\max_j |x_{i,j}|)^2}}. \tag{10.43}$$

This implies that with probability $1 - \delta$ the estimator satisfies

$$\left| \hat{f}^{(i)}_{\alpha,x} - \mathbb{E}_\sigma [ f^{(i)}_{\alpha,x} \sigma ] \right| \leq \epsilon \tag{10.44}$$

whenever the number $m$ of independently drawn Rademacher vectors satisfies

$$m \geq \frac{16\, k_i (\max_j |x_{i,j}|)^2}{\epsilon^2} \log(2/\delta). \tag{10.45}$$

□

In total we thus obtain a polynomial worst-case complexity of computing the nonstoquasticity of a (2 + 1)-local Hamiltonian of the form (10.31) given by

$$\frac{n(n-1)}{2} + \frac{16 \sum_i \deg_{XZ}(i) (\max_{i,j} x_{i,j})^2}{\epsilon^2} \log\left(\frac{2}{\delta}\right). \tag{10.46}$$

Let us recap. In this chapter, we investigated the question what might be a computationally meaningful measure of the sign problem. We observed the how the nonstoquastic entries of a Hamiltonian matrix are related to the average sign by means of counterexamples to such a relation. Observing that constructing such examples required a conspirative interplay between the matrix entries, we provided analytical and numerical evidence for the fact that nonstoquasticity and average sign go hand in hand in *generic instances*. Based on those observations, we propose the $\ell_1$ norm of the nonstoquastic part of a Hamiltonian as a measure of the sign problem. We show that for 2-local Hamiltonians this measure can be exactly evaluated efficiently on bounded-degree graphs such as regular lattices, and it can be efficiently approximated up to inverse polynomial additive error on arbitrary graphs.

In the next chapter, we will look at the practical question of how feasible and meaningful easing the sign problem in terms of the nonstoquasticity is in practice.

# Easing the sign problem in practice

# 11

*Can we find the sign-optimal basis?*

In this chapter, aim to find an answer to the second question posed above. Recall that in the last chapter, we motivated the $\ell_1$ norm of the Hamiltonian nonstoquasticity as a meaningful while at the same time efficiently computable measure of the sign problem. Now, we will carry out a stress test of this measure when applying it to practically relevant scenarios. More specifically, we ask the question: Can we practically ease the sign problem of physically relevant systems by minimizing their nonstoquasticity?

*The following sections have (up to small alterations) previously been published as part of Ref. [Han+20].*

To study this question, we consider translation-invariant nearest-neighbour Hamiltonians in a quasi one-dimensional geometry [MK04]. Quasi one-dimensional systems, such as anti-ferromagnetic Heisenberg Hamiltonians on ladder geometries [DR96; Tak96] (see Fig. 11.3) are the simplest non-trivial systems that exhibit a sign problem since they admit the phenomenon of geometric frustration [San10]. Frustration gives rise to a plethora of phenomena arising in quasi one-dimensional systems such as the emergence of quantum spin liquids [Men+15; Hua+17] and the interplay of spin-1/2 and spin-1 physics [Nie+17]. They are also somewhat more realistic descriptions of actual low-dimensional experimental situations than simple one-dimensional chains, serving as a model for small couplings in the transverse direction [Tak96; Yos+15; LV17]. Therefore quasi one-dimensional systems are often seen as a stepping stone towards studying higher dimensions [IKS15], where the sign problem inhibits QMC simulations [CHM15].

## 11.1 Sign easing for translation-invariant systems

Quasi one-dimensional systems are effectively described by a nearest-neighbour interaction term $h \in \mathbb{R}^{d^2 \times d^2}$ giving rise to the Hamiltonian (10.29)[1]

$$H = \sum_{i=1}^{n} T_i(h), \tag{11.1}$$

acting on $n$ $d$-dimensional quantum systems. For the sake of simplicity, we specialize here to closed boundary conditions, identifying $n + 1 = 1$. As a meaningful simple ansatz class, we consider on-site orthogonal transformations $O \in O(d)$ of the type

$$H = \sum_{i=1}^{n} T_i(h) \mapsto O^{\otimes n} H (O^T)^{\otimes n}. \tag{11.2}$$

1: Recall that we e have defined the translation of $h$ to site $i$ as $T_i(h) = 1^{\otimes(i-1)} \otimes h \otimes 1^{\otimes(n-i-1)}$.

On-site transformations can be handled particularly well as they preserve locality and translation-invariance of local Hamiltonians. In particular, for such transformations, the global nonstoquasticity measure can be expressed locally in terms of the transformed term $O^{\otimes 2} h (O^T)^{\otimes 2}$ so that the optimization problem has constant complexity in the system size. This constitutes an exponential improvement over approaches that directly optimize the average sign.

More precisely, for Hamiltonians of the form (11.1) we can express the nonstoquasticity measure $\nu_1(H) = n 2^{n-3} \tilde{\nu}_1(h)$ in terms of an effective local measure

$$\tilde{\nu}_1(h) = \sum_{\substack{ijk;lmn: \\ j \neq m, k=n}} \max \left\{ 0, (h \otimes 1 + 1 \otimes h)_{ijk;lmn} \right\}. \tag{11.3}$$

We derive the effective nonstoquasticity measure (11.3) in the Appendix D.2.

Optimizing $\nu_1(H)$ for the global Hamiltonian is thereby reduced to the much smaller problem of minimizing $\tilde{\nu}_1(h)$. While the nonstoquasticity measure $\nu_1$ can be efficiently evaluated, thus satisfying a necessary criterion for an efficient optimization algorithm, minimizing $\nu_1$ may and in fact *will* still be a non-trivial task in general – an intuition we make rigorous below (see Chapter 12). This is because in optimizing the basis-dependent measure $\nu_1$ over quasi-local basis choices one is faced with a highly non-convex optimization problem of a high-order polynomial over the sphere of orthogonal matrices.

## 11.2 Algorithmic procedure

To practically minimize the nonstoquasticity $\tilde{\nu}_1$ over the orthogonal group $O(d)$, we have implemented a geometric optimization method suitable for group manifolds, namely, a conjugate gradient descent algorithm over the orthogonal group $O(d)$ [AEK09; HR19]. Those algorithms are among the best developed multi-purpose methods for optimization over group manifolds such as the orthogonal group [AEK09]. Compared to simple gradient-descent algorithms, conjugate gradient algorithms are capable to better incorporate the underlying group structure to the effect that they achieve much faster runtimes and better convergence properties. Our implementation is publicly available [HR19].

The key ingredient for the conjugate gradient descent algorithm is the derivative of the objective function $\tilde{\nu}_1(h(O))$ with respect to the orthogonal matrix $O$. We derive this gradient in Appendix D.2

We start our conjugate gradient algorithm either at the identity matrix (with or without a small perturbation) or a Haar-randomly chosen orthogonal matrix as indicated at the respective places in the main text.

Since the minimization of the $\tilde{\nu}_1$-measure is numerically not well behaved, we improve the performance of the algorithm in several ways. First, we observe that the measure $\tilde{\nu}_2$ given by the Frobenius norm of the nonstoquastic part of the Hamiltonian is numerically much better behaved. This is due to the $\ell_2$-norm being continuously differentiable while the $\ell_1$-norm is only subdifferentible. In particular, at its minima the gradient of the $\ell_1$-norm is discontinuous and never vanishes. For this reason, rather than optimizing the $\ell_1$-norm of the nonstoquastic part, we optimize a smooth approximation thereof as given by [SFR07]

$$\nu_{1,\alpha}(H) := \sum_{i \neq j} f_\alpha\left(H_{i,j}\right), \tag{11.4}$$

with $f_\alpha(x) = x + \alpha^{-1}\log(1+\exp(-\alpha x))$.

To achieve the best possible performance, we then carry out a hybrid approach: First, we pre-optimize by minimizing $\tilde{\nu}_2$ using our conjugate gradient descent algorithm. Second, starting at the minimizer obtained in the Frobenius norm optimization, we minimize the smooth nonstoquasticity measure $\tilde{\nu}_{1,\alpha}{}^2$. We then compare the result to a direct minimization of the nonstoquasticity $\tilde{\nu}_{1,\alpha}$ starting from the original initial point and choose whichever of the minimizations performed best. The exact details of our optimization algorithms can be found at Ref. [HR19] together with code to reproduce the figures shown here. Our code framework can be easily adapted for optimization of other Hamiltonian models and more general circuit architectures.

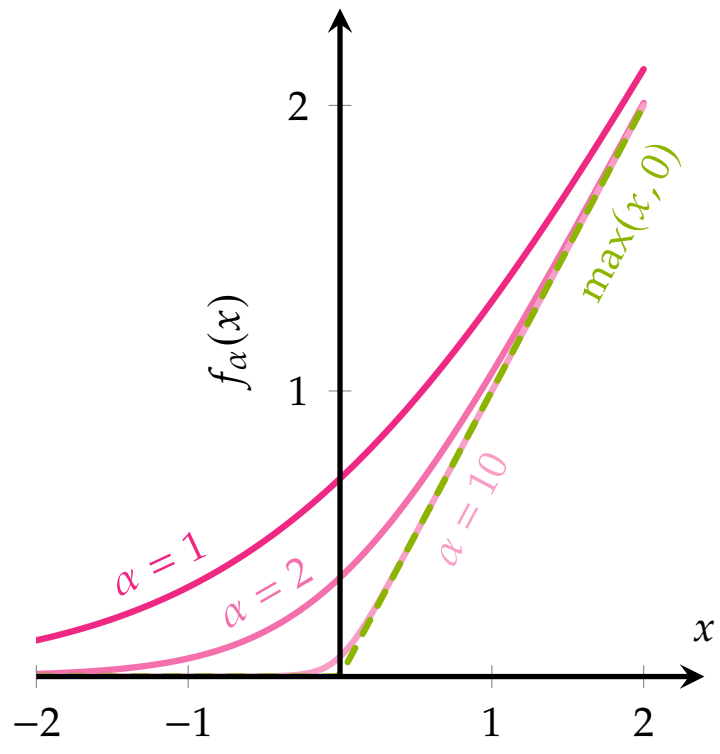

**Figure 11.1:** Using the smooth function $f_\alpha$ (defined below Eq. (11.4)) as opposed to the exact but discontinuous $\max(\cdot, 0)$ greatly improves the performance of the conjugate-gradient descent algorithm. With increasing $\alpha > 0$, shown here for $\alpha = 1, 2, 10$ (solid) $f_\alpha$ approaches $\max(\cdot, 0)$ (dashed).

2: We choose the value $\alpha = 50$, $\alpha = 100$ and $\alpha = 40$, for the random stoquastic Hamiltonians, the $J_0$-$J_1$-$J_2$-$J_3$-model, and the frustrated ladder model, respectively. See below for details.

## 11.3 Results

We first benchmark the algorithm on Hamiltonians which we know to admit an on-site stoquastic basis. Specifically, we apply the algorithm to recover an on-site stoquastic basis of the random translation-invariant Hamiltonian

$$H = \sum_{i=1}^{n} T_i \left(O^{\otimes 2}(h - h_\neg)(O^T)^{\otimes 2}\right) \tag{11.5}$$

on $n$ qudits where the two-local term $h \in \mathbb{R}^{d^2 \times d^2}$ is a Hamiltonian term with uniformly random spectrum expressed in a Haar-random basis and $O \in O(d)$ is a Haar-random on-site orthogonal matrix. In Fig. 11.2(a) we show the performance of the algorithm on randomly chosen instances of (11.5) for different values of the local dimension $d$. In all but very few instances our algorithm essentially recovers the stoquastic basis of the random Hamiltonian. This shows that, generically, the algorithm recovers an on-site stoquastic basis for random Hamiltonians which are known to admit such a basis a priori.

By construction, in the instances in which no stoquastic basis could be recovered the algorithm gets stuck in local minima, indicating a potential limitation of first-order optimization techniques as a tool for easing the sign problem of general Hamiltonians.

We now apply the algorithm to frustrated anti-ferromagnetic Heisenberg Hamiltonians, i.e., Hamiltonians with positively weighted interaction terms $\vec{S}_i \cdot \vec{S}_j$, on different ladder geometries. Here, $\vec{S}_i = (X_i, Y_i, Z_i)^T$ is the spin operator at site $i$. Ladder geometries are not only interesting for the reasons described above, but also because in spite of frustration effects they often admit sign-problem free QMC methods [Nak98; Hon+16; Wes+17]. This is achieved by going to a dimer basis [Nak98; Hon+16; Wes+17]. However – and this is important for our approach – in those cases the sign problem is not removed by finding a *stoquastic local basis*, but rather by exploiting specific properties of the Monte Carlo simulation at hand, for example, that no negative paths occur in the simulation [Nak98] or by exploiting specific properties of the Monte Carlo implementation at hand [Hon+16; Wes+17]. Therefore, frustrated Heisenberg ladders constitute the ideal playground to explore the methodology of easing the sign problem by (quasi-)local basis choices.

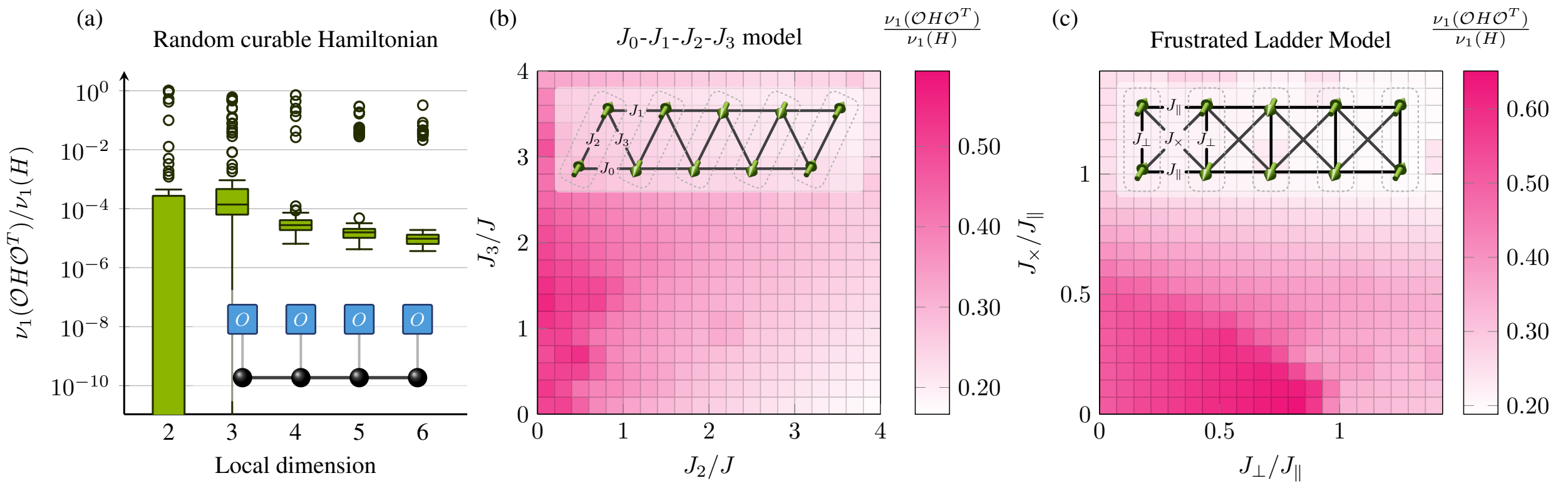

**Figure 11.2:** We optimize the non-stoquasticity $\nu_1$ of translation-invariant, two-local Hamiltonians over on-site orthogonal transformations $\mathcal{O} = O^{\otimes n}$ using a conjugate gradient method for manifold optimization [AEK09; HR19]. Figure *(a)* shows the relative non-stoquasticity improvement of random two-local Hamiltonians that are known to admit an on-site stoquastic basis. For each local dimension 100 instances are drawn and the results displayed as a box plot according to Ref. [IS0 16269, p. 2.16]. This serves as a benchmark of our algorithm, which for almost all instances accurately recovers a stoquastic on-site basis. Figure *(b)* displays the optimized non-stoquasticity of the anti-ferromagnetic $J_0$-$J_1$-$J_2$-$J_3$-Heisenberg model relative to the computational basis as a function of $J_2/J$, $J_3/J$, where $J_0 = J_1 = J$. The algorithm is initialized in a Haar random orthogonal on-site basis. This model is known to admit a *non-local* stoquastic basis for $J_3 \geq J_0 + J_1$ [Nak98]. Figure *(c)* shows the optimized non-stoquasticity of the anti-ferromagnetic Heisenberg ladder illustrated in the inset with couplings $J_\parallel$, $J_\perp$, $J_\times$ relative to the computational basis as a function of $J_\perp/J_\parallel$ and $J_\times/J_\parallel$. We initialized the algorithm at the identity matrix (that was randomly perturbed by a small amount). The phase diagram of the non-stoquasticity qualitatively agrees with the findings of Ref. [Wes+17], where the stochastic series expansion (SSE) QMC method was studied. There, it was found that the sign problem can be completely eliminated for a completely frustrated arrangement where $J_\times = J_\parallel$, while the sign problem remains present for partially frustrated couplings $J_\times \neq J_\parallel$. However, throughout the parameter regime the stoquasticity remains non-trivial, which may be due to the fact that the optimization algorithm converges to local minima.

The first model we study is the $J_0$-$J_1$-$J_2$-$J_3$-model [Nak98] whose Hamiltonian is given by (see Fig. 11.3(a))

$$H_{\vec{J}} = \sum_{i=1}^{n} \left( J_0 \vec{S}_i^1 \vec{S}_{i+1}^1 + J_1 \vec{S}_i^2 \vec{S}_{i+1}^2 + J_2 \vec{S}_i^1 \vec{S}_i^2 + J_3 \vec{S}_{i+1}^1 \vec{S}_i^2 \right), \tag{11.6}$$

where $\vec{S}_i^1$ denotes the spin operator at site $i$ on the lower rung and $\vec{S}_i^2$ on the upper rung of the ladder, respectively, and $J_i \geq 0$ for all $i$. Intriguingly, this Hamiltonian does not have a sign problem in the singlet-triplet dimer basis even though the Hamiltonian is not stoquastic in that basis. However, there exists a *non-local* stoquastic basis for values of $J_3 \geq J_0 + J_1$ [Nak98]. We show the results of optimizing the nonstoquasticity of $H_{\vec{J}}$ with $J_0 = J_1 = J$ over a translation-invariant dimer basis (see Fig. 11.3(a)) in Fig. 11.2(b). We initialize our simulations in a Haar random orthogonal on-site basis. We find an improvement of the nonstoquasticity under on-site orthogonal basis choices of factors 2 to 5 depending on the region in the phase diagram. Interestingly, that region does not seem to correlate with the region in which a non-local stoquastic basis was found in Ref. [Nak98]. It may well be the case that no stoquastic dimer basis exists *even though* other variants of QMC do not incur a sign problem for such basis choices: in Ref. [Nak98] a stoquastic but *non-local* basis of the $J_0$-$J_1$-$J_2$-$J_3$-model is identified for values of $J_3 \geq J_0 + J_1$. We view this as an indication that less local ansatz classes such as quasi-local circuits can help to further improve the nonstoquasticity for this model.

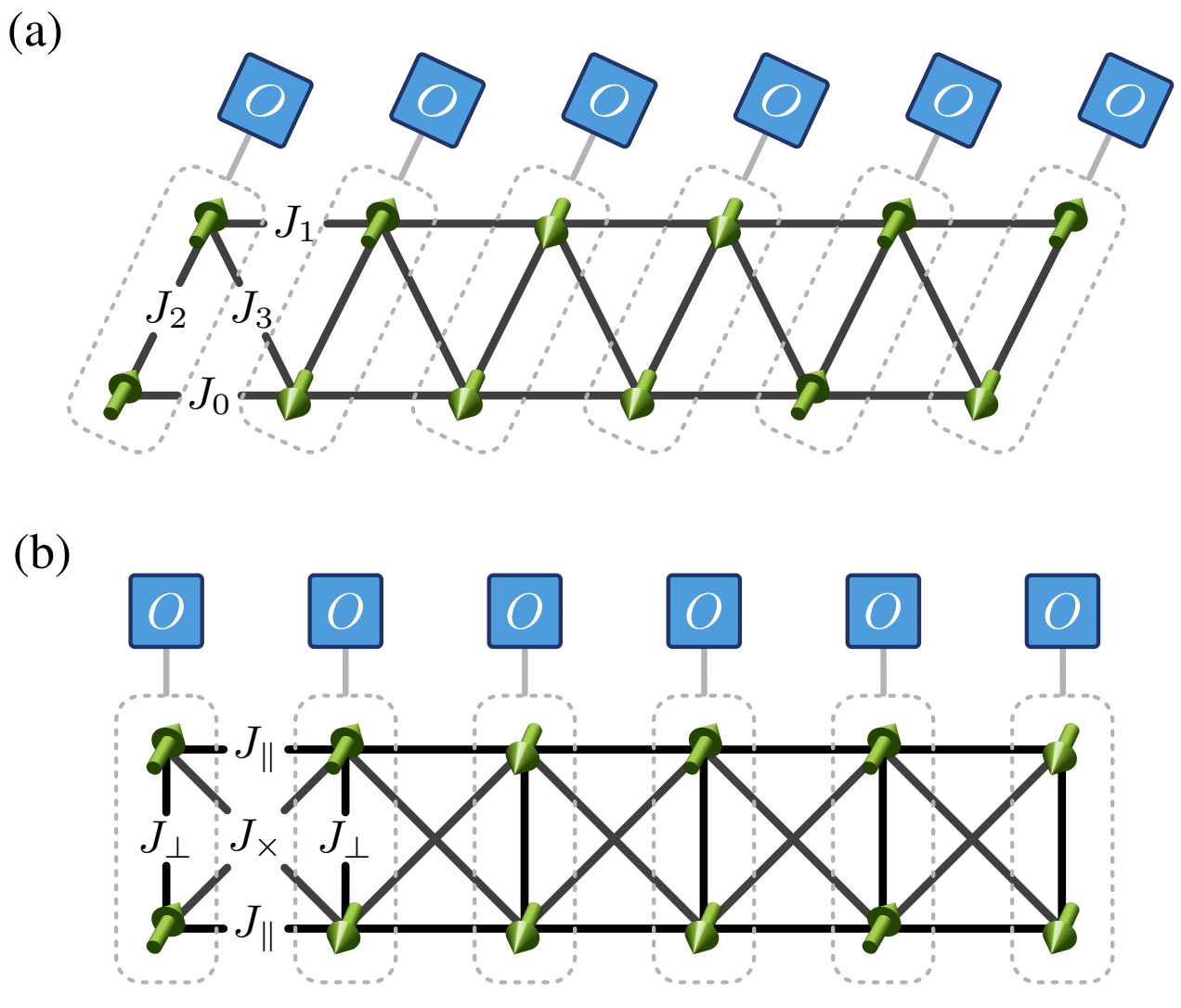

**Figure 11.3:** Figure *(a)* shows the lattice of the $J_0$-$J_1$-$J_2$-$J_3$-Heisenberg model on a triangular quasi one-dimensional lattice as given in Eq. (11.6). Figure *(b)* shows the lattice structure of the frustrated Heisenberg model (11.7) with couplings $J_\perp$, $J_\parallel$ and $J_\times$ on a square-lattice ladder with cross coupling. In our simulations, we group sites to dimers as indicated in the figures and then optimize the measure $\tilde{\nu}_1(h)$ of the effective 2-local terms $h$ over on-site orthogonal transformations $O^{\otimes 2}$.

We now apply the algorithm to the anti-ferromagnetic Heisenberg ladder studied in Refs. [Hon+16; Wes+17]. The Hamiltonian of this system is given by (see Fig. 11.3(b))

$$H_{J_\parallel, J_\perp, J_\times} = \sum_{i=1}^{n} \Big( J_\parallel \left( \vec{S}_i^1 \vec{S}_{i+1}^1 + \vec{S}_i^2 \vec{S}_{i+1}^2 \right) + J_\perp \vec{S}_i^1 \vec{S}_i^2 \\ + J_\times \left( \vec{S}_i^1 \vec{S}_{i+1}^2 + \vec{S}_{i+1}^1 \vec{S}_i^2 \right) \Big), \tag{11.7}$$

with interaction constants $J_\parallel, J_\perp, J_\times \geq 0$. For this geometry, the situation is somewhat more involved: Refs. [Hon+16; Wes+17] find that a sign-problem free QMC procedure exists, albeit for a slightly different QMC procedure than we consider here, namely stochastic series expansion (SSE) Monte Carlo [San10]. Similar to the world-line Monte Carlo method discussed here, SSE is based on an expansion of the exponential $\exp(-\beta H)$ albeit via a Taylor expansion as opposed to a product expansion. Specifically, for the partially frustrated case in which $J_\times \neq J_\parallel$ their solution of the sign problem exploits a specific sublattice structure of the Hamiltonian in combination with the SSE approach.

We optimize the nonstoquasticity of dimer basis choices as shown in Fig. 11.3(b) when starting from a random initial point that is close to the identity. Our results, shown in Fig. 11.2(c), qualitatively reflect the findings of Ref. [Wes+17] for SSE in terms of stoquasticity in that the nonstoquasticity can be significantly reduced for the fully frustrated case $J_\parallel = J_\times$, while it can be merely slightly improved for the partially frustrated case.

At the same time, the algorithm does not recover a fully stoquastic basis for the frustrated ladder model $H_{J_\parallel, J_\perp, J_\times}$ as might be expected. There may be several reasons for this: either the nearly sign-problem free QMC procedure found in Refs. [Hon+16; Wes+17] is in fact specific to SSE in that no stoquastic dimer basis and hence no sign-problem free world-line

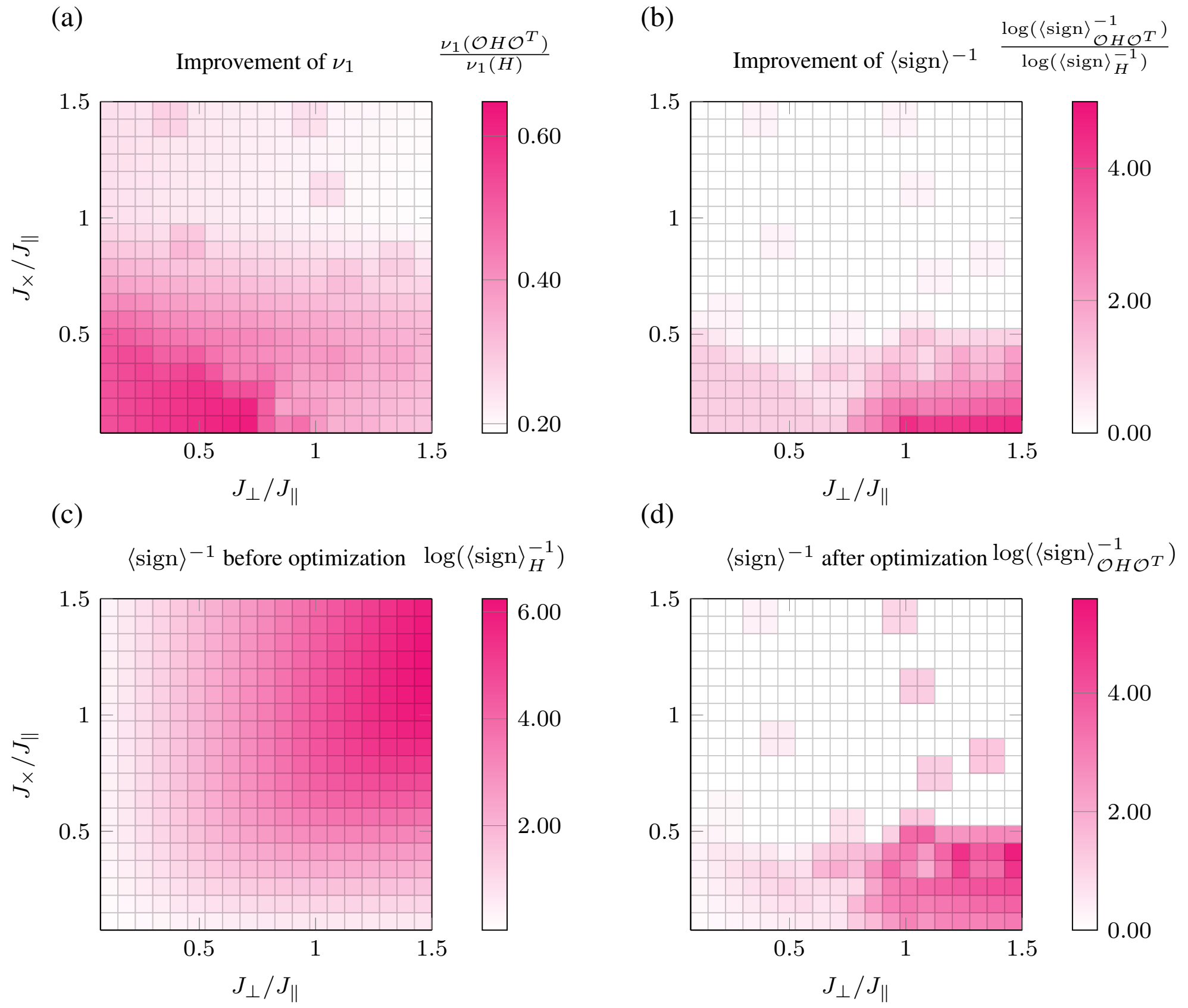

**Figure 11.4:** Figure *(a)* shows the optimized nonstoquasticity $\nu_1$ in terms of its relative improvement compared to the computational basis. *(b)* We expect the inverse average sign to scale exponentially in the nonstoquasticity. Therefore, we plot the ratio of the logarithm of the inverse average sign before optimization to that after optimization. We compute the average sign via exact diagonalization for a ladder of $2 \times 4$-sites, $m = 100$ Monte Carlo steps and inverse temperature $\beta = 1$. We also plot the logarithm of the inverse average sign *(c)* before and *(d)* after optimization of a local orthogonal basis.

Monte Carlo method exists in the orbit of orthogonal dimer bases, or the conjugate gradient algorithm gets stuck in local minima.

Generally speaking, we observe that first-order optimization algorithms such as the employed conjugate gradient method encounter obstacles. This demonstrates that the optimization landscape is generically an extremely rich and rugged one. Intuitively, this landscape is governed by the combinatorial increase of possible assignments of signs to the Hamiltonian matrix elements, resulting in the computational hardness of the optimization problem in general – an issue that we will deal with in detail in the next chapter.

The question we set out to answer in the previous chapter (Chapter 10) was: Is the nonstoquasticity related to the average sign of a QMC simulation and, if so, how? We found evidence that in generic instances the relation is monotonous so that the average sign may be improved by optimizing nonstoquasticity. But what is the state of affairs when we move away from generic instances to specific, physically relevant ones such as the ladder geometries considered here? Importantly and in spite of those seemingly moderate improvements of nonstoquasticity, we find that the sample complexity of QMC as governed by the inverse average sign is greatly diminished to approximate unity in large regions of the parameter space for the frustrated ladder model; see Fig. 11.4.

We first observe that the improvement of the average sign concomitant with the improvement in nonstoquasticity shown Figs. 11.4(a) and (b) is compatible with an exponential dependence of the inverse average sign on the nonstoquasticity $\langle \text{sign} \rangle^{-1} \propto \exp(c\nu_1(H))$ as conjectured above: in the regions in which a significant improvement of the nonstoquasticity could be achieved by local basis choices, the inverse average sign could also be strongly improved. Importantly, while the Hamiltonian could not be made entirely stoquastic, the improvement in the inverse average sign reaches an extent to which nearly no sign problem remains in those regions. This shows that also moderate improvements in nonstoquasticity can yield tremendous improvements of the average sign. At the same time, a severe sign problem remains – and actually becomes worse – in a small region of the parameter space (around $J_\perp/J_\parallel \gtrsim 3/4$ and $J_\times/J_\parallel \lesssim 1/2$) even though the nonstoquasticity could be reduced to some extent in that region. This may reflect upon open questions about the relation between average sign and nonstoquasticity that arose in our earlier discussion in Chapter 10: while in generic cases the two notions of severeness of the sign problem are expected to be closely related, there is no general simple correspondence between them.

## 11.4 Discussion

Our findings show that one can in fact efficiently optimize the nonstoquasticity for translation-invariant problems that admit a stoquastic basis lying within the ansatz orbit. They also further substantiate the claim that optimizing nonstoquasticity typically eases the sign problem and dampens the increase in sampling complexity. They demonstrate both the feasibility of minimizing the nonstoquasticity in order to ease the sign problem by optimizing over suitably chosen ansatz classes of unitary/orthogonal transformations and potential obstacles to a universal solution of the sign problem. In particular, for translation-invariant problems – while it may well be computationally infeasible – the complexity of the optimization problem only scales with the locality of the Hamiltonian, the local dimension and the depth of the circuit.

What is more, they indicate that more general ansatz classes such as quasi-local circuits yield the promise to further reduce the nonstoquasticity of ladder models. We therefore expect that optimizing nonstoquasticity is a feasible and promising means to reduce the sign problem for many different systems, including two-dimensional lattice systems, by exploiting the flexibility offered by larger ansatz classes within our framework.

Different classes of orthogonal transformations can be straightforwardly incorporated in our algorithmic approach. A detailed study of different ansatz classes and their potential for easing the sign problem is, however, beyond the scope of this work. Furthermore, it will be interesting to study the optimal bases of different ($\ell_p$) nonstoquasticity measures in more detail, in particular, with respect to their sparsity. It is the subject of ongoing and future efforts to study the optimal basis choice in terms of the nonstoquasticity for both deeper circuits and further models as well as the connection between the average sign and the nonstoquasticity.

But already in our small study we also encountered obstacles preventing efficient optimization of the nonstoquasticity in the guise of a complicated and rugged optimization landscape. In the next chapter we will further investigate this landscape in a setting that is amenable to complexity-theoretic analysis. Rather than considering translation-invariant Hamiltonians with varying local dimension, we will consider qubit Hamiltonians with differing nearest-neighbour interactions.

# The computational complexity of easing the sign problem | 12

*How hard is the sign easing task in general?*

This last question will occupy us in the last chapter of this part and therefore this thesis as a whole. This question is an instance of the more general questions whether it is possible to find the optimal representation of quantum-mechanical quantities mechanical quantities with respect to a certain algorithm. Finding an answer to this question is indispensable when aiming to better understand the origin of computational complexity in quantum-mechanical systems. Is the increased complexity of simulating such systems merely rooted in the fact that our *representation* of those systems – for example in terms of a Monte Carlo path integral – is badly chosen and ill-suited to classical simulation techniques? Or is it fundamental in that for some systems no representation that is amenable to classical computation exists at all?

The computational complexity of the sign easing task can provide insight into those questions: if this task was possible efficiently for systems encompassing the full computational power of quantum devices, it would at least be possible to *efficiently check* whether or not a given Hamiltonian system has an intrinsic and severe sign problem. Studying it is therefore an indispensable part of the endeavour to challenge the computational power of quantum computing devices but also of natural quantum-mechanical systems in general.

*The remainder of this chapter has (up to small alterations) previously been published as part of Ref. [Han+20].*

So how far can an approach to easing the sign problem using optimization over local bases carry in principle? In our main complexity-theoretic result, we systematically study the fundamental limits of minimizing the nonstoquasticity as a means to ease the sign problem. To do so, we complement the pragmatic mindset of this work with the rigorous machinery of computational complexity theory, asking the question: What is the computational complexity of optimally easing the sign problem?

In order to formalize this question, we introduce the corresponding decision problem:

**Definition 12.1** (SignEasing) *Given an $n$-qubit Hamiltonian $H$, constants $B > A \geq 0$ with $B - A \geq 1/\mathsf{poly}(n)$, and a set of allowed unitary transformations $\mathcal{U}$, decide which of the following is the case:*

$$YES: \quad \exists U \in \mathcal{U} : \nu_1(UHU^\dagger) \leq A, \text{ or} \tag{12.1}$$

$$NO: \quad \forall U \in \mathcal{U} : \nu_1(UHU^\dagger) \geq B. \tag{12.2}$$

We derive the computational complexity of the sign easing problem in simple settings, namely for 2-local Hamiltonians, allowing for on-site orthogonal Clifford operations as well as for on-site general orthogonal transformations. We prove that under both classes of transformations SignEasing is NP-complete. Intriguingly, this holds true even in cases in which the curing problem can be decided efficiently, namely, for strictly 2-local XYZ Hamiltonians of the type considered in Refs. [KT19; Kla+19].

**Table 12.1:** The satisfiability equivalent of curing the sign problem is to decide whether a given sentence is satisfiable, while the equivalent of easing is to find the minimal number of clauses that are violated by a sentence. Similarly, results on the computational complexity of curing and easing the nonstoquasticity of a local Hamiltonian $H$ are in one-to-one correspondence with the hardness of satisfiability problems.

| Satisfiability | Stoquasticity | Complexity | Refs. |
|---|---|---|---|
| 3SAT | Curing 2+1-local Hamiltonians | NP-complete | [MLH19; Kla+19] |
| 2SAT | Curing strictly 2-local Hamiltonians | in ¶ | [KT19; Kla+19] |
| MAX2SAT | Easing strictly 2-local Hamiltonians | NP-complete | *here* |

**Theorem 12.1** (Complexity of SignEasing) SignEasing *is* NP-*complete for 2-local (XYZ) Hamiltonians under*

*i. on-site orthogonal Clifford transformations, and*
*ii. on-site general orthogonal transformations.*

From a practical perspective, our results pose limitations on the worst-case runtime of algorithms designed to find optimal QMC bases for the physically relevant case of 2-local Hamiltonians. From a complexity-theoretic perspective, they manifest a sign problem variant of the dichotomy between the efficiently solvable 2SAT-problem to decide whether there exists a satisfying assignment for a 2-local sentence, and the NP-complete MAX2SAT-problem asking what is the least possible number of broken clauses. They thus complete the picture drawn by Refs. [KT19; MLH19; Kla+19] regarding the connection between satisfiability problems and the problems of curing and easing the sign problem on arbitrary graphs, a state of affairs which we illustrate in Table 12.1. It is natural to ask the question how far this connection extends and what we can learn from it about efficiently solvable instances. For example, one may ask, whether results about the hard regions of 3SAT and MAX2SAT carry over to the problems of curing and easing the sign problem.

We prove Theorem 12.1i. and ii. as Theorems 12.3 and 12.4 in Section 12.2 and 12.3, respectively. The essential idea of our proof, sketched below and illustrated in Fig. 12.1, is to design a corresponding Hamiltonian such that if the sign problem could be optimally eased for this Hamiltonian under the respective ansatz class, one could also find the ground state energy of the original anti-ferromagnetic Ising Hamiltonian, a task that is NP-hard to begin with. It is straightforward to prove versions of Theorem 12.1 for any $\ell_p$-norm of the nonstoquastic part of $H$ with finite $p$ as a measure of nonstoquasticity. Our result is therefore independent of the particular choice of ($\ell_p$) nonstoquasticity measure.

*Proof sketch.* SignEasing for arbitrary 2-local Hamiltonians is contained in NP – given a basis transformation, we can approximate the measure of nonstoquasticity from the transformed local terms up to any inverse polynomial error and hence verify the YES-case (12.1); see Theorem 10.2.

The key idea of the harder direction of the proof is to encode the promise version of the MAXCUT-problem into the SignEasing-problem. An instance of MAXCUT is given by a graph $G = (V, E)$, and the problem

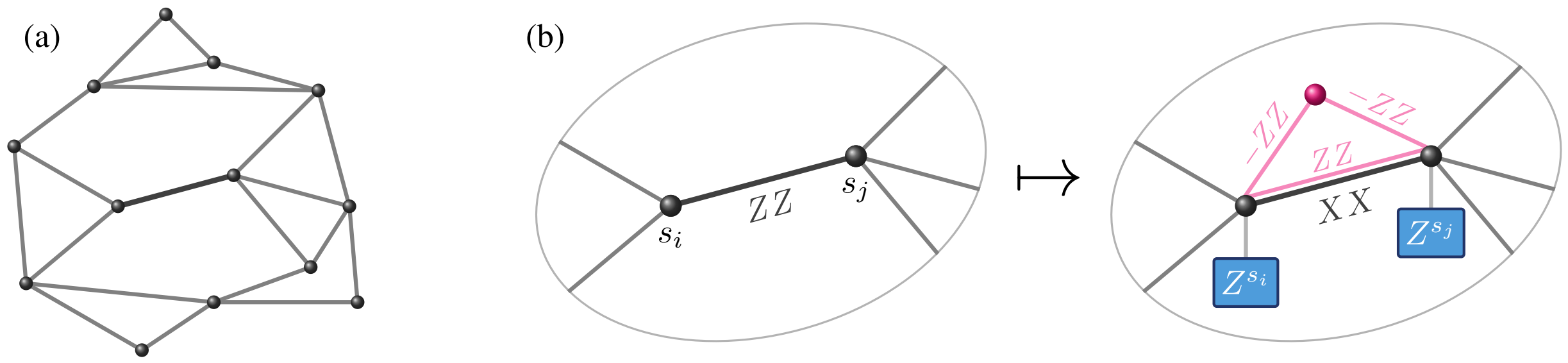

**Figure 12.1:** *(a)* To prove NP-completeness of SignEasing, we reduce it to the MAXCUT-problem which asks for the ground-state energy of an anti-ferromagnetic Ising Hamiltonian $H$ on a graph $G$. *(b)* In our encoding, we map $H$ to a Hamiltonian $H'$ in which all $ZZ$-interactions are replaced by $XX$-interactions and translate the spin configurations $(s_1, \dots, s_n) \in \{0,1\}^n$ of the anti-ferromagnetic Ising model to on-site transformations $Z_1^{s_1} \cdots Z_n^{s_n}$. To achieve this restriction, we penalize all other transformations by adding an ancilla qubit $a_{i,j}$ for every edge $(i,j)$ of $G$ and adding the interaction term $C(Z_iZ_j - Z_iZ_{a_{i,j}} - Z_jZ_{a_{i,j}})$ with a suitably chosen constant $C > 0$. We obtain that $\nu_1(H')$ can be eased below a certain value if and only if the ground state energy of $H$ is below that value to begin with, thus establishing the reduction.

is to decide whether the ground-state energy of the anti-ferromagnetic (AF) Ising Hamiltonian

$$H = \sum_{(i,j)\in E} Z_i Z_j \,, \tag{12.3}$$

is below a constant $A$ or above $B$. Here, $Z_i$ is the Pauli-$Z$-operator acting on site $i$. We now define a Hamiltonian $H'$ in which we replace every $Z_iZ_j$ interaction of $H$ by an $X_iX_j$ interaction as we illustrate in Fig. 12.1. To understand our embedding, suppose that we perform basis changes only by applying $Z$ or $1$ at every site. In this case a Hamiltonian term can be made stoquastic if and only if $X_iX_j \mapsto -X_iX_j$ which is achieved by a transformation $Z^{s_i}Z^{s_j}$ with $(s_i, s_j) = (0,1) \vee (1,0)$. A term remains stoquastic for $(s_i, s_j) = (1,1) \vee (0,0)$. This provides a direct mapping between spin configurations $(1,0)$ and $(0,1)$, which do not contribute to the ground state energy of the anti-ferromagnetic Ising model and transformations that make local terms in $H'$ stoquastic and thus decrease the nonstoquasticity.

To prove the theorem for arbitrary on-site Clifford and orthogonal transformations, we introduce an ancilla qubit $a_{i,j}$ for every edge $(i,j)$ and add interaction terms $C(Z_iZ_j - Z_iZ_{a_{i,j}} - Z_jZ_{a_{i,j}})$ to $H'$ with constant $C$, see Fig. 12.1(b). These terms penalize all other transformations such that the optimal nonstoquasticity of $H'$ is always achieved for transformations of the form $Z_1^{s_1} \cdots Z_n^{s_n}$ with $(s_1, \dots, s_n) \in \{0,1\}^n$. For example, suppose that we apply Hadamard transformations to all sites $i, j, a_{i,j}$, then the $ZZ$ interactions and $XX$ interactions change roles so that the nonstoquasticity cannot be decreased by such a transformation. Showing this for all possible transformations constitutes the main technical part of the proof. □

Since MAXCUT is a variant of the MAX2SAT-problem our results not only manifest but also crucially utilise the 2SAT-MAX2SAT dichotomy. Notice also that since the MAXCUT-problem is NP-hard already on subgraphs of the double-layered square lattice [Bar82] (see Fig. 12.2), one can find hard instances of the sign-easing problem already for quasi two-dimensional lattices and graphs with low connectivity.

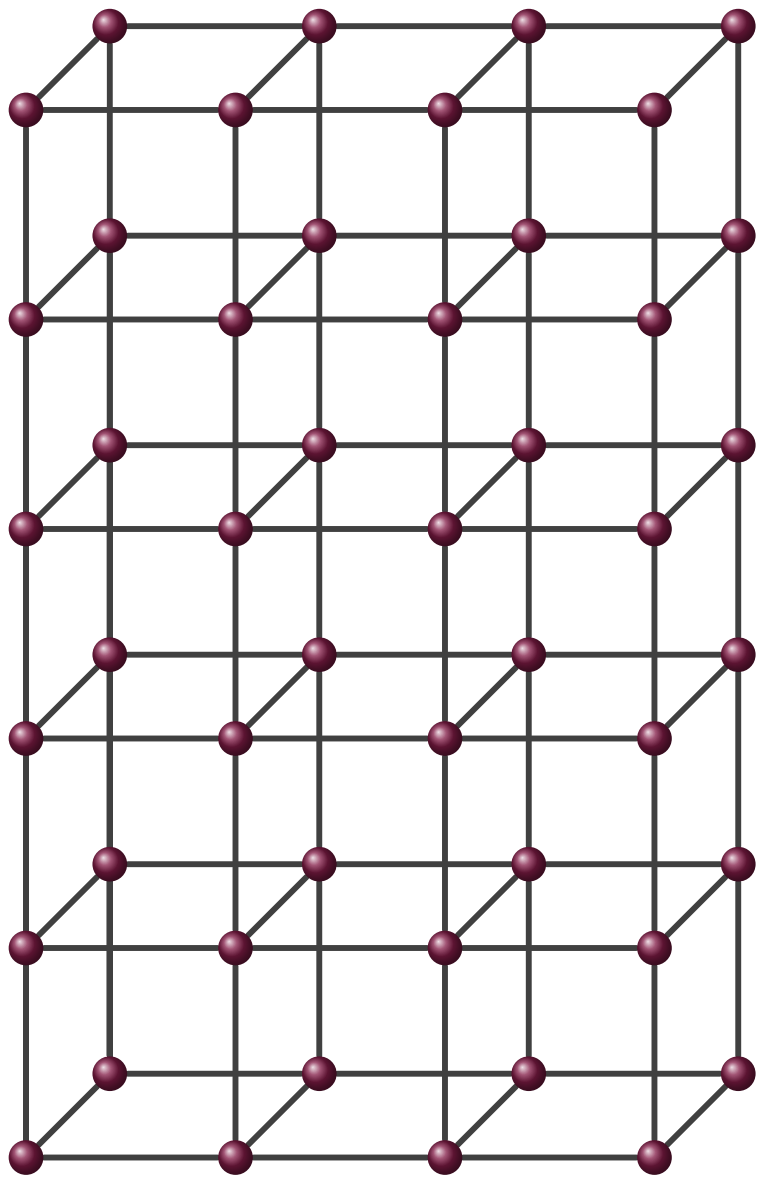

**Figure 12.2:** The SignEasing problem is NP-complete already on subgraphs of the double-layered square lattice due to the fact that MAXCUT remains hard when restricting to such lattices [Bar82].

In our complexity-theoretic analysis, we have focused on the computa-tional complexity of easing the sign problem as the size of an arbitrary input graph is scaled up, in the same mindset as Refs. [TW05; MLH19; KT19; MLH19]. We expect, however, that the complexity of SignEasing scales similarly in the size of the lattice unit cell and the local dimension of translation-invariant systems such as those discussed above.

Our results underpin the computational hardness of systematic approaches towards alleviating the sign problem in that we show its hardness in the fourfold simplest case in which the problem can be understood: (i) We consider the less demanding and practically relevant task of *easing* the sign problem as opposed to curing it. (ii) We consider a particularly simple *measure of nonstoquasticity*, namely, one that depends in the simplest possible fashion on the Hamiltonian in question and in particular can be efficiently evaluated for local Hamiltonians. (iii) We consider the smallest meaningful sets of *allowed basis transformations*, namely, on-site (orthogonal and Clifford) operations which preserve the locality of the Hamiltonian. (iv) We ask the question for the smallest non-trivial *Hamiltonian locality*, namely, 2-local (XYZ) Hamiltonians.

---

Let us now turn to proving Theorem 12.1 in detail. Building on the results in Section 10.4, we begin by deriving a general expression for the nonstoquasticity that will be essential to the proof of the Theorem. In Sections 12.2 and 12.3 we then prove the result for orthogonal Cliffords and general orthogonal on-site transformations, respectively. In the presentation of the proof we focus on the essential ideas when moving to progressively richer ansatz classes, while deferring the detailed calculation necessary for the most general (orthogonal) case to Appendix D.3. Subsequently, in Section 12.4 we conclude this last part of the thesis with an outlook.

## 12.1 The nonstoquasticity of 2-local Hamiltonians

A central ingredient in proving Theorem 12.1 is an expression for the nonstoquasticity measure $\nu_1$ of strictly 2-local Hamiltonians of the form

$$H_2 = \sum_{i<j} \left( a_{i,j} X_i X_j + c_{i,j} Z_i Z_j + x_{i,j} X_i Z_j + x_{j,i} Z_i X_j \right) . \tag{12.4}$$

It is sufficient to restrict to Hamiltonians of the form (10.31) because the orbit of XYZ Hamiltonians under on-site orthogonal (Clifford) transformations does not reach $YY$ terms.

It is a direct consequence of Lemma 10.1 that

$$\nu_1(H_2) = \sum_{i<j} \nu_1(a_{i,j} X_i X_j) + \sum_i \nu_1 \left( \sum_{j \in \mathcal{N}_{XZ}(i)} x_{i,j} X_i Z_j \right), \tag{12.5}$$

where we the $XZ$ neighbourhood $\mathcal{N}_{XZ}(i)$ of a vertex $i$ and related notions are defined in Sec. 10.4. More specifically, following Eqs. (10.33) and (10.34) we find that

$$\nu_1(a_{i,j}X_iX_j) = \sum_{i<j} \max\{a_{i,j}, 0\}, \tag{12.6}$$

$$\nu_1\left(\sum_{j\in\mathcal{N}_{XZ}(i)} x_{i,j}X_iZ_j\right) = 2^{-\deg_{XZ}(i)} \times \sum_{\lambda_{\mathcal{N}_{XZ}(i)}} \max\left\{\sum_{j\in\mathcal{N}_{XZ}(i)} (-1)^{\lambda_j}x_{i,j}, 0\right\}. \tag{12.7}$$

Since for the proof of hardness we need analytical expressions of the nonstoquasticity, we cannot resort to the sampling algorithm to evaluate the nonstoquasticity of $XZ$ terms as proposed in Sec. 10.4. We analytically bound the contribution of a vertex with non-trivial $XZ$ neighbourhood with the following lemma.

**Lemma 12.2** ($XZ$ nonstoquasticity) *The following bound is true for* $k \in \mathbb{N}$

$$\sum_{\lambda\in\{0,1\}^k} \max\left\{\sum_{j=1}^{k}(-1)^{\lambda_j}x_j, 0\right\} \geq \max_j |x_j| \cdot 2^{k-1}. \tag{12.8}$$

*Proof.* Let us assume wlog. that $x_1 \geq x_2 \geq \ldots \geq x_k \geq 0$, all terms being positive and non-increasingly ordered. This does not restrict generality as all possible combinations of signs appear in the sum (12.7). We prove the claim by induction. For $k = 1$, the statement is true by immediate inspection. For the induction step, we use the following inequality for $a, b \in \mathbb{R}$

$$\max\{a + b, 0\} + \max\{a - b, 0\} \geq 2\max\{a, 0\}, \tag{12.9}$$

which can be easily checked by checking the three cases $a \geq |b|$, $a \leq -|b|$ and $-|b| < a < |b|$. We then calculate

$$\sum_{\lambda\in\{0,1\}^k} \max\left\{\sum_{j=1}^{k}(-1)^{\lambda_j}x_j, 0\right\} \tag{12.10}$$

$$= \sum_{\lambda_1,\ldots,\lambda_{k-1}\in\{0,1\}} \max\left\{x_k + \sum_{j=1}^{k-1}(-1)^{\lambda_j}x_j, 0\right\} + \sum_{\lambda_1,\ldots,\lambda_{k-1}\in\{0,1\}} \max\left\{-x_k + \sum_{j=1}^{k-1}(-1)^{\lambda_j}x_j, 0\right\}. \tag{12.11}$$

$$\geq 2 \sum_{\lambda'\in\{0,1\}^{k-1}} \max\left\{\sum_{j=1}^{k-1}(-1)^{\lambda'_j}x_j, 0\right\} \tag{12.12}$$

$$\overset{\text{I.H.}}{\geq} 2 \cdot 2^{k-2}|x_1| = 2^{k-1}|x_1|, \tag{12.13}$$

where we have used (12.9) in the second to last and the induction hypothesis in the last step. This proves the claim. □

In the proof of Theorem 12.1 we will use that Lemma 10.1 implies that every term $a_{i,j}X_iX_j$ contributes an additional cost $\max\{a_{i,j}, 0\}$ to the nonstoquasticity of $H_2$. Moreover, since $\max_{j\in[k]} |x_j| \geq \sum_{j=1}^{k} |x_j|/k$,

Lemmas 10.1 and 12.2 imply that every term $x_{i,j}X_iZ_j$ of $H_2$ contributes at least a cost $|x_{i,j}|/(2\deg(G'))$ to the nonstoquasticity of $H_2$.

## 12.2 SignEasing under orthogonal Clifford transformations

We are now ready to show that with respect to the nonstoquasticity measure $\nu_1$ easing the sign problem for 2-local XYZ Hamiltonians with on-site Cliffords is NP-complete on arbitrary graphs. We restate Theorem 12.1i. here.

**Theorem 12.3** (SignEasing under orthogonal Clifford transformations) SignEasing *is* NP*-complete for* 2*-local Hamiltonians on an arbitrary graph, in particular for XYZ Hamiltonians, under on-site orthogonal Clifford transformations, that is, the real group generated by* $\{X, Z, W\}$ *with* $W$ *the Hadamard matrix.*

*Proof.* Clearly the problem is in NP, since one can simply receive a (polynomial-size) description of the transformation in the Yes-case, and then calculate the measure of nonstoquasticity efficiently for XYZ Hamiltonians, verifying the solution.

To prove NP-hardness, we encode the MAXCUT problem in the SignEasing problem. A MAXCUT instance can be phrased in terms of asking whether an anti-ferromagnetic (AF) Ising model on a graph $G = (V, E)$ with $e = |E|$ edges on $v = |V|$ spins

$$H = \sum_{(i,j)\in E} Z_i Z_j, \tag{12.14}$$

has ground-state energy $\lambda_{\min}(H)$ below $A$ or above $B$ with constants $B - A \geq 1/\mathsf{poly}(v)$. This is because in the Ising model one gets energy $-1$ for a $(0,1)$ or $(1,0)$ -edge and $+1$ for a $(0,0)$ or $(1,1)$ edge.

Let us now encode the MAXCUT problem phrased in terms of the AF Ising model problem into SignEasing for the XYZ Hamiltonian. We will design a Hamiltonian $H'$, and ask if on-site orthogonal Clifford transformations can decrease its measure of nonstoquasticity $\nu_1$ below $A$, or whether it remains above $B$ for any Clifford basis choice.

For each AF edge between spins $i, j$ in the AF Ising model, the new Hamiltonian $H'$ will have an edge

$$h_{i,j} = X_i X_j. \tag{12.15}$$

On top of that, for every edge $(i, j) \in E$ we add one ancilla qubit $a_{i,j}$ as shown in Figure 12.1, and interactions

$$h^{(a)}_{i,j} = C\left(Z_i Z_j - Z_i Z_{a_{i,j}} - Z_{a_{i,j}} Z_j\right), \tag{12.16}$$

where $C = 4\deg(G)$. Note that the additional terms are diagonal and hence stoquastic. The total new Hamiltonian then reads

$$H' = \sum_{(i,j)\in E} \left[ X_i X_j + C \left( Z_i Z_j - Z_i Z_{a_{i,j}} - Z_{a_{i,j}} Z_j \right) \right], \tag{12.17}$$

and acts on $n = v + e$ qubits. We construct $H'$ so that an attempt to decrease the nonstoquasticity $\nu_1$ by swapping $Z$ and $X$ operators via Hadamard transformations will fail, and so the best one can do is to choose a sign in front of each local $X$ operator. Of course, this then becomes the original, hard, MAXCUT problem in disguise. Let us prove this.

We start the proof with fixing some notation: We call $\mathcal{N}(i) = \{j : (i,j) \in E\}$ the neighbourhood of site $i$ on the original graph $G$ and $\mathcal{N}'(i) = \mathcal{N}(i) \cup \{a_{i,j} : (i,j) \in E\}$ the neighbourhood of site $i$ on the augmented graph on which $H'$ lives. Moreover $\deg(i) = |\mathcal{N}(i)|$ is the degree of site $i$ on the original graph $G$, whereas $\deg'(i) = |\mathcal{N}'(i)|$ the degree of site $i$ on the augmented graph $G' = (V', E')$. Note that $\deg'(i) = 2\deg(i)$.

## Orthogonal Clifford transformations

First, let us note that any element of the orthogonal Clifford group can be written as

$$C = \pm W^w X^x Z^z, \tag{12.18}$$

where we denote the Hadamard matrix with $W$ and $w, x, z \in \{0,1\}$. Since the global sign is irrelevant, a real $n$-qubit Clifford of the form $C = C_1 \otimes \cdots \otimes C_n$ is parametrized by binary vectors $\vec{w}, \vec{x}, \vec{z} \in \{0,1\}^n$.

How does $H'$ transform under real single-qubit Clifford transformations? By definition $CZC^\dagger \in \{\pm Z, \pm X\}$ and likewise for $X$. Therefore, the transformed Hamiltonian will be of the form (10.31). Throughout the proof, we will use that every term $a_{i,j} X_i X_j$ contributes at least $\max\{a_{i,j}, 0\}$ to the nonstoquasticity, while every term $x_{i,j} X_i Z_j$ contributes at least $|x_{i,j}|/(2\deg(G')) = |x_{i,j}|/(4\deg(G))$ as shown by Lemmas 10.1 and 12.2 above.

We now show that MAXCUT can be embedded into SignEasing under on-site orthogonal Clifford transformations. To do so, we need to show two things: first, that in the (yes)-case that $\lambda_{\min}(H) \le A$, the nonstoquasticity of $H'$ can be brought below $A$ using on-site orthogonal Clifford transformations. Second, we show that in the (no)-case that $\lambda_{\min}(H) \ge B$, the nonstoquasticity of $H'$ cannot be brought below $B$ using on-site orthogonal Clifford transformations.

## Yes-case: (Diagonal) transformations that map $X$ to $\pm X$ ($\vec{w} = 0$).

These transformations only change the sign in front of $X_i$, keeping its form. At the same time they only change the signs of the $Z_i Z_j$ terms, keeping them diagonal and hence stoquastic. The transformed $X_i X_j$ terms (12.15) will be stoquastic if and only if exactly one of the transformations of the $X$ at sites $i, j$ is a $Z$-flip.

We can view the coefficient $z_i$ as a spin $s_i$ in the original AF Ising model: for $z_i = 1$, $X_i \longrightarrow -X_i$, corresponding to a spin $s_i = 1$ in the original AF Ising model, while for $z_i = 0$, $X_i \longrightarrow X_i$, which we view it as the Ising spin $s_i = 0$.

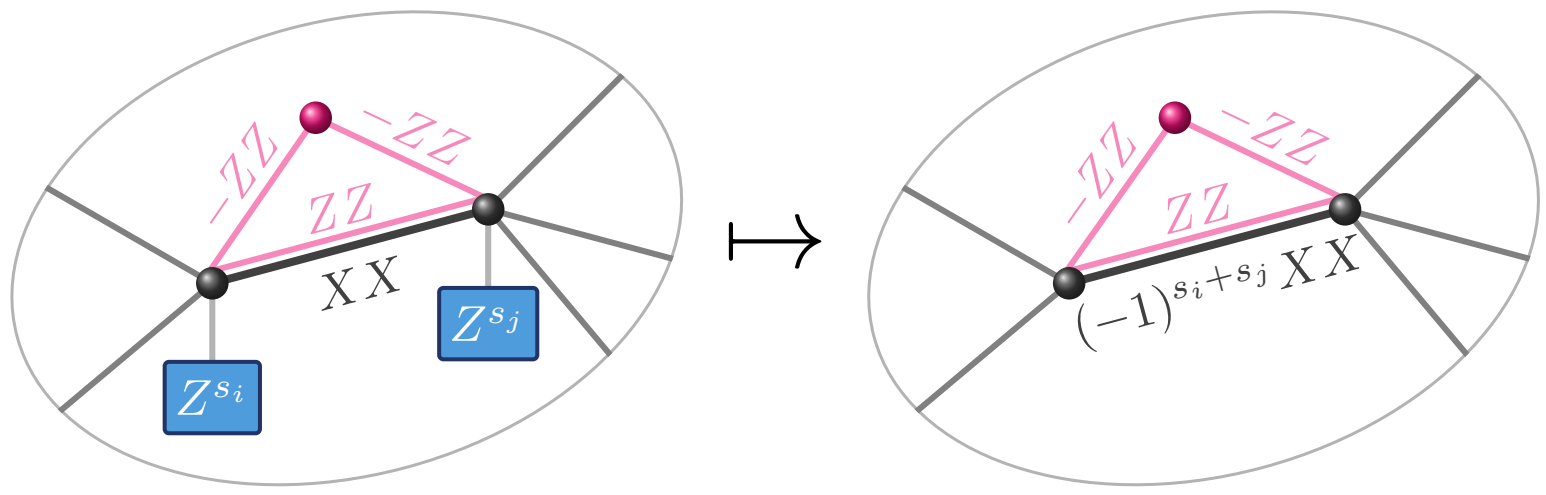

Each such Clifford transformation thus corresponds to a particular state of the original AF Ising model as given by a spin configuration $\vec{s} \in \{0,1\}^v$. Whenever the transformations on neighbouring sites result in a stoquastic interaction $-X_iX_j$ in the transformed XYZ Hamiltonian, we have a $(0,1)$ or $(1,0)$ anti-ferromagnetic Ising edge with cost 0. On the other hand, each nonstoquastic $X_iX_j$ term in the XYZ Hamiltonian has cost 1, while the corresponding edge in the Ising model is $(0,0)$ or $(1,1)$ also with cost 1.

What is the amount of sign easing we can hope to achieve? We have argued above that only diagonal transformations which map $X_i \mapsto \pm X_i$ potentially ease the sign problem since we designed the interactions so that a Hadamard transformation always incurs a larger cost than keeping an $X_iX_j$ term nonstoquastic. For those transformations, we have a one-to-one correspondence with the ground state of the original AF Ising model. Hence, the original AF Ising model ground state energy $\lambda_{\min}(H)$ is also the optimal number of nonstoquastic terms $X_iX_j$ which one can achieve via on-site orthogonal Clifford transformations, each adding an additional cost 1 to the nonstoquasticity measure $\nu_1$.

In the yes-case we can therefore achieve nonstoquasticity

$$\nu_1(yes) \leq A, \tag{12.19}$$

by choosing $\vec{x}, \vec{w} = 0$ and $(z_1, \ldots, z_v)^T = \vec{s}_0$, the ground state of $H$.

We now show that in the *no*-case, the nonstoquasticity measure will be at least

$$\nu_1(no) \geq B. \tag{12.20}$$

### No-case: (Hadamard) transformations that map $X$ to $\pm Z$ $(\vec{w} \neq 0)$.

We have designed (12.16) so that such transformations result in large nonstoquasticity. Specifically, we show that for any choice of $\vec{z}$, choosing $\vec{x} = \vec{w} = 0$ achieves the optimal nonstoquasticity in the orbit of orthogonal Clifford transformations.

It is sufficient to show that any Clifford transformation on an edge $(i, j)$ (and its ancilla qubit $a_{i,j}$) that is nonstoquastic for a given choice of $\vec{z}$ can only increase the nonstoquasticity.

To begin with, note that choosing $x_i = 1$ results in $Z_i \mapsto -Z_i$, $X_i \mapsto X_i$, and choosing $w_i = 1$ maps $X_i \mapsto Z_i$ and $Z_i \mapsto X_i$. We obtain the following transformation rules of Pauli $Z_i$ and $X_i$, given choices of $x_i$ and $w_i$:

| $x_i$ | $w_i$ | $Z_i$ | $X_i$ |
|---|---|---|---|
| 0 | 0 | $Z_i$ | $X_i$ |
| 0 | 1 | $X_i$ | $Z_i$ |
| 1 | 0 | $-Z_i$ | $X_i$ |
| 1 | 1 | $-X_i$ | $Z_i$ |

First, suppose that for an edge $(i, j)$, a Hadamard transformation is performed on qubit $i$, but not $j$ so that we have $w_i = 1$, $w_j = 0$. Then for some choice of $x_i, x_j$ the transformed edge is given by

$$W_i(X_iX_j \pm CZ_iZ_j)W_i = Z_iX_j \pm CX_iZ_j, \tag{12.21}$$

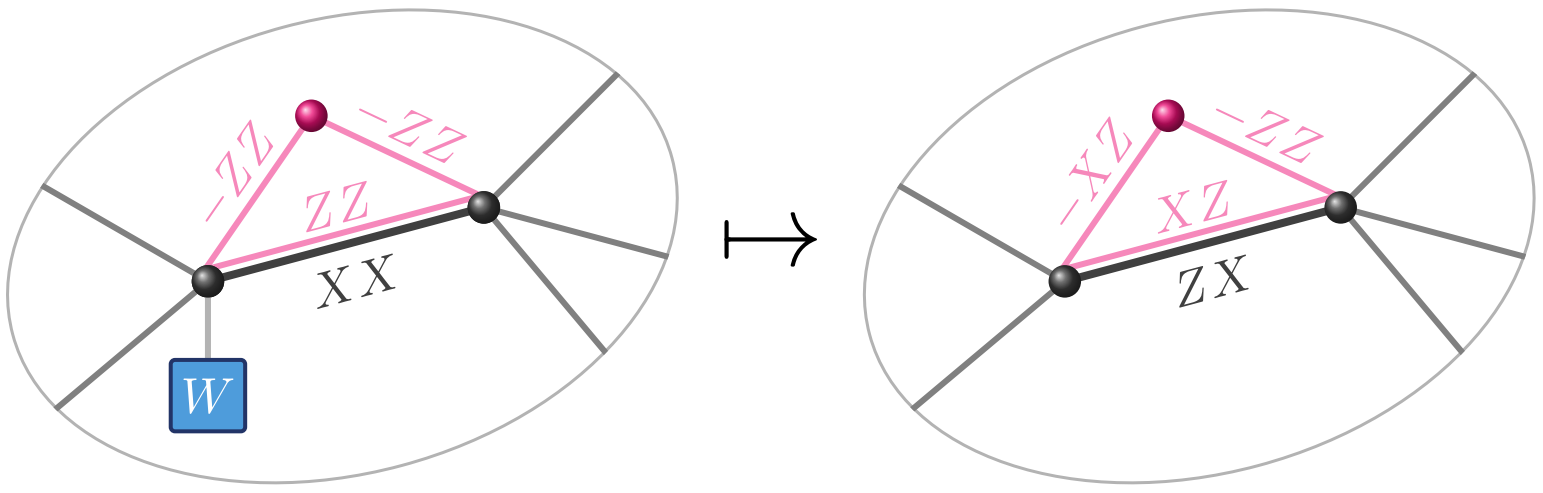

and has nonstoquasticity at least $(C + 1)/(2 \deg G')$.

Now, suppose that a Hadamard transformation is performed on both qubit $i$ and $j$ but not its ancilla qubit. Then the transformed term is given by

$$\begin{aligned} &W_iW_jX_i^{x_i}X_j^{x_j}X_{a_{i,j}}^{x_{a_{i,j}}}(h_{i,j} + h_{i,j}^{(a)})X_i^{x_i}X_j^{x_j}X_{a_{i,j}}^{x_{a_{i,j}}}W_iW_j \\ &= \pm Z_iZ_j + C\left(\pm X_iX_j \pm X_iZ_{a_{i,j}} \pm Z_{a_{i,j}}X_j\right), \end{aligned} \tag{12.22}$$

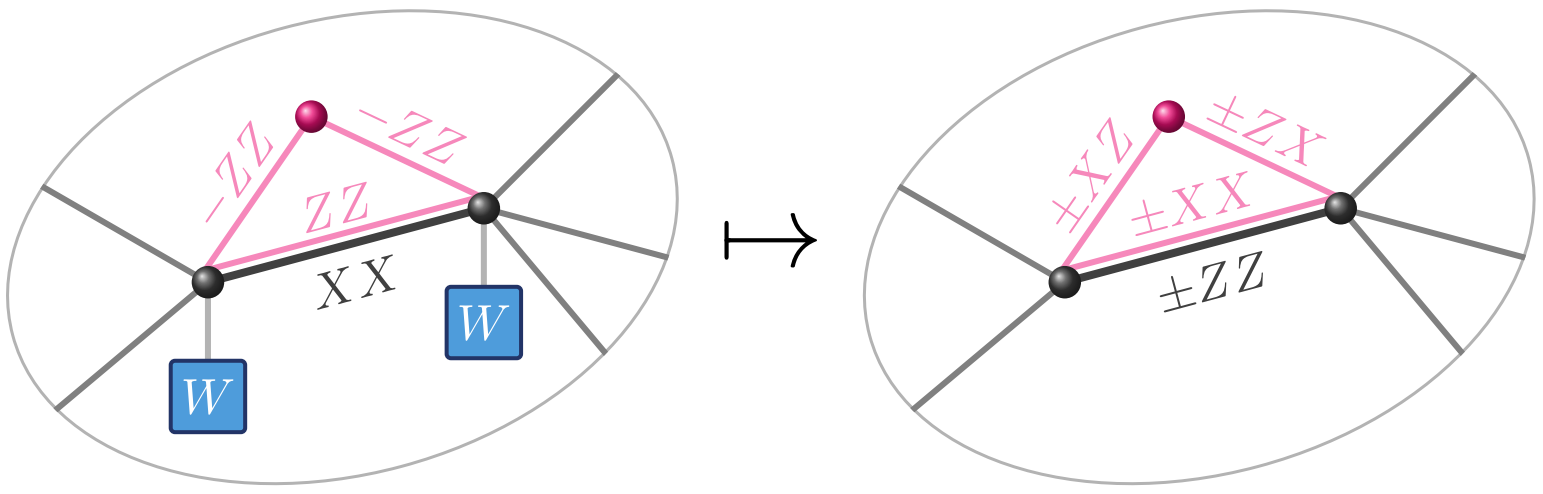

with nonstoquasticity cost at least

$$\nu_1\left(C(\pm X_i Z_{a_{i,j}} \pm Z_{a_{i,j}} X_j)\right) = 2C/(2\deg G'). \tag{12.23}$$

Could the edge be possibly cured by performing a Hadamard transformation on the ancilla qubit as well? In this case, we get

$$\begin{aligned} &W_i W_j W_{a_{i,j}} (h_{i,j} + h_{i,j}^{(a)}) W_i W_j W_{a_{i,j}} \\ &\quad = Z_i Z_j + C\left(X_i X_j - X_i X_{a_{i,j}} - X_{a_{i,j}} X_j\right), \end{aligned} \tag{12.24}$$

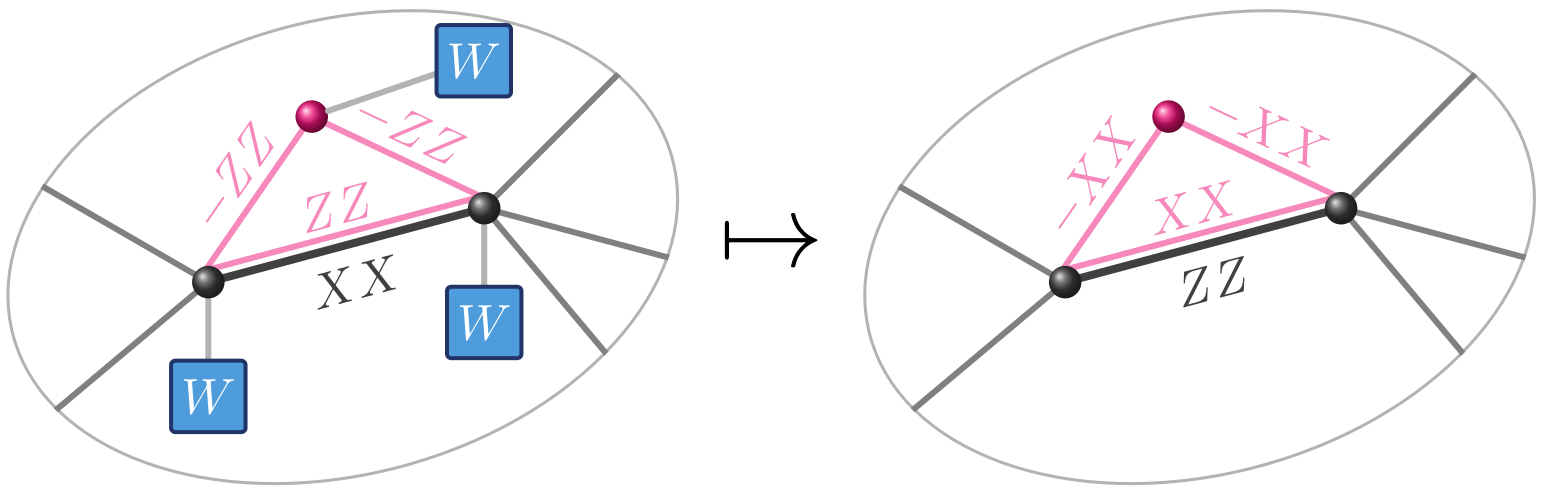

with nonstoquasticity

$$\nu_1(W_i W_j W_{a_{i,j}} (h_{i,j} + h_{i,j}^{(a)}) W_i W_j W_{a_{i,j}}) = C. \tag{12.25}$$

Because of the frustrated arrangement of the signs of the $ZZ$ terms, no local sign flip of those terms (achieved by choices of $x_i, x_j, x_{a_{i,j}} \neq 0$) can cure the sign problem of an ancillary triangle, leaving it lower bounded by (12.25).

On the other hand, the original cost incurred from local sign flips via $Z$-transformations, is given by

$$\nu_1(X_i X_j) = 1, \tag{12.26}$$

which is always smaller than the cost incurred if additional $X$ or $W$ transformations are applied since we chose $C$ such that $C/(2\deg G') = 1$. Therefore, in the no-case of the original AF Ising model the nonstoquasticity of $H'$ cannot be brought below

$$\nu_1(no) \geq B, \tag{12.27}$$

with the optimal choice achieved for $\vec{x}, \vec{w} = 0$ and $(z_1, \ldots, z_v)^T = \vec{s}_0$.

□

## 12.3 SignEasing under orthogonal transformations

**Theorem 12.4** (SignEasing under orthogonal transformations) SignEasing *is* NP*-complete for* 2*-local Hamiltonians on an arbitrary graph, in particular, for XYZ Hamiltonians under on-site orthogonal transformations.*

*Proof.* We proceed analogously to the proof for orthogonal Clifford transformations, showing that in the yes-case, there exists a product orthogonal transformation $O = O_1 \cdots O_n$ such that $\nu_1(OH'O^T) \leq A$, while in the no-case there exists no such transformation $P$ with $\nu_1(PH'P^T) \leq B$.

The yes-case is clear: In this case, the energy of $\vec{s}$ is below $A$. Then by our construction, the sign problem can be eased below $A$ with the orthogonal transformation

$$O = \prod_{i \in V} Z_i^{s_i}. \tag{12.28}$$

We now need to show that in the no-case, any orthogonal transformation incurs nonstoquasticity above $B$. We first remark that the orthogonal group $O(2)$ decomposes into two sectors with determinant $\pm 1$, respectively. Therefore, any $2 \times 2$ orthogonal matrix can be written as

$$O_a(\theta) = \begin{pmatrix} \cos\theta & -a\sin\theta \\ \sin\theta & a\cos\theta \end{pmatrix}, \tag{12.29}$$

which for $a = \det(O_a(\theta)) = -1$ is a reflection and for $a = \det(O_a(\theta)) = +1$ a rotation by an angle $\theta$. Note that the following composition laws hold

$$O_{-1}(\theta)O_1(\phi) = O_{-1}(\theta - \phi), \tag{12.30}$$
$$O_1(\theta)O_{-1}(\phi) = O_{-1}(\theta + \phi), \tag{12.31}$$
$$O_1(\theta)O_1(\phi) = O_1(\theta + \phi), \tag{12.32}$$
$$O_{-1}(\theta)O_{-1}(\phi) = O_1(\theta - \phi). \tag{12.33}$$

Now observe three facts: First, any reflection by an angle $\theta$ can be written as a product of a reflection across the $X$-axis and a rotation as

$$\begin{pmatrix} \cos\theta & \sin\theta \\ \sin\theta & -\cos\theta \end{pmatrix} = \begin{pmatrix} \cos\theta & -\sin\theta \\ \sin\theta & \cos\theta \end{pmatrix} Z = R(\theta)Z, \tag{12.34}$$

where $R(\theta)$ is the rotation by an angle $\theta$. Second, for any Hermitian matrix $H$ and any angle $\theta$, it holds that

$$O(\theta)HO(\theta)^T = O(\theta + \pi)HO(\theta + \pi)^T, \tag{12.35}$$

so that it suffices to restrict to angles $\theta \in [-\pi/4, 3\pi/4]$ in an interval of length $\pi$. Third, a rotation by an angle $\pi/2$ can be decomposed into two reflections as

$$R(\pi/2) = XZ. \tag{12.36}$$

Taken together, these facts imply that an arbitrary single-qubit orthogonal transformation is given by

$$O(\theta, z, p) = R\left(\theta + (\pi/2)^p\right) \cdot Z^z = R(\theta) \cdot X^p Z^{p+z}, \tag{12.37}$$

where the rotation angle is given by $\theta \in [-\pi/4, \pi/4]$, $z \in \{0, 1\}$ fixes whether or not a $Z$-flip is applied, and $p \in \{0, 1\}$ mods out a rotation by an angle $\pi/2$. Now define $O(\vec{\theta}, \vec{z}, \vec{p}) = \prod_i O_i(\theta_i, z_i, p_i)$.

We now need to show that, in the no-case, for any choice of $\vec{\theta} \in$

$[-\pi/4, \pi/4]^n$, $\vec{z}, \vec{p} \in \{0,1\}^n$ it holds that

$$\nu_1\left(O(\vec{\theta}, \vec{z}, \vec{p})H'O(\vec{\theta}, \vec{z}, \vec{p})^T\right) \geq b. \tag{12.38}$$

To complete the proof, we use that the action of $R(\theta)$ on Pauli-$X$ and $Z$ matrices is given by

$$R(\theta)ZR(\theta)^T = \cos(2\theta)Z + \sin(2\theta)X, \tag{12.39}$$
$$R(\theta)XR(\theta)^T = \cos(2\theta)X - \sin(2\theta)Z. \tag{12.40}$$

In Appendix D.3, we show that given any choice of $\vec{z}$ and $\vec{p}$, no choice of $\vec{\theta}$ can decrease the nonstoquasticity of an edge with non-zero nonstoquasticity below 1. That is, we analyze transformations of the following form:

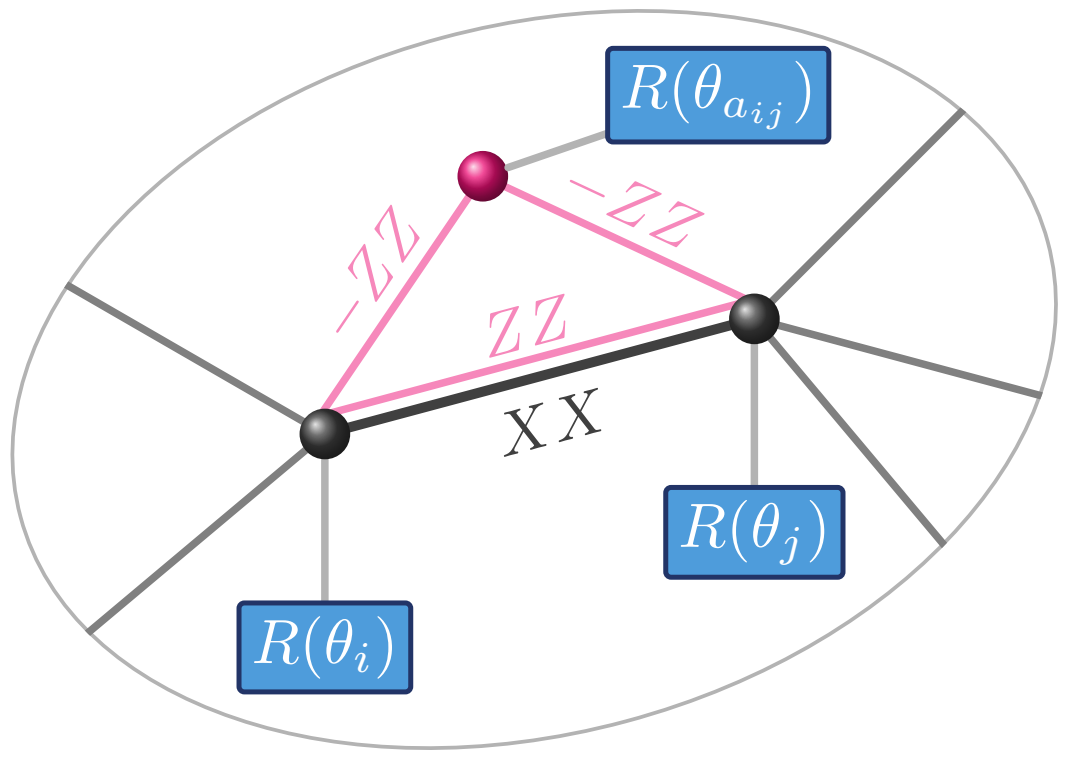

We do this by using the standard form (12.37) of the local transformations in terms of $X$, $Z$ and restricted rotation matrices with angles in $[-\pi/4, \pi/4]$. We split the proof into three parts, analyzing three different (continous) regions for rotation angle choices. The only difference in the construction when compared to the Clifford-case is that here we choose $C = (2\deg G')^2$. This completes the proof. □

We remark that one can easily extend the proofs of Theorems 12.3 and 12.4 for different nonstoquasticity measures $\nu_p$ with $1 < p < \infty$. To see this, note that the decision problem for $\nu_p(H)$ is equivalent to the problem for $\nu_p(H)^p$.

For terms $x_{i,j}X_iZ_j$ we can then use the trivial bound

$$\sum_{\lambda\in\{0,1\}^k} \max\left\{\sum_{j=1}^{k}(-1)^{\lambda_j}x_{i,j}, 0\right\}^p \geq 2^{p(k-\deg(i))}\sum_j |x_j|^p \tag{12.41}$$

instead of Lemma 12.2. Thus, every term $x_{i,j}X_iZ_j$ contributes at least $2^{-p\deg G'}|x_{i,j}|^p$ to $\nu_p^p$, while a term $a_{i,j}X_iX_j$ contributes

$$\nu_p(a_{i,j}X_iX_j)^p = (\max\{a_{i,j}, 0\})^p, \tag{12.42}$$

to the total nonstoquasticity of $H'$.

For general $\ell_p$-nonstoquasticity measures $\nu_p$ one therefore need merely choose $C = 2^{\deg G'}$ to prove Theorem 12.3 and $C = 2^{2\deg G'}$ for Theorem 12.4.

With this, we conclude the complexity-theoretic analysis of the SignEasing problem.

## 12.4 Easing the sign problem: Summary and Outlook

Let us now recap Part III as a whole. Our work introduces the sign easing methodology as a systematic novel paradigm useful for assessing and understanding the sign problem of QMC simulations. We ask and answer three central questions using complementary methods from theoretical and applied computer science as well as from physics. First, we define a measure of non-stoquasticity suitable for easing the sign problem and extensively discussed its relation to the average sign. Second, we demonstrate that one can feasibly optimize this measure over local bases in simple settings by applying geometric optimization methods. Finally, we establish the computational complexity of sign easing in a broader but still simple setting. In this way, our work not only identifies a means of easing the sign problem and demonstrates its feasibility and potential, but also shows up its fundamental limitations in terms of computational complexity. Even more so, we are confident that the framework of our work provides both valuable guidance and the practical means for future research on systematically easing the sign problem of Hamiltonians that are particularly interesting and relevant in condensed-matter applications.

### Outlook

As a first general and systematic attempt to easing the sign problem, we have restricted the focus of this work in several ways. As such, a number of questions, generalizing our results in different directions, are left open.

First, we have restricted our discussion to the prominent world-line Monte Carlo method to maintain clarity throughout the manuscript. We are confident, however, that our results find immediate application for other Monte Carlo methods such as stochastic series expansion Monte Carlo and determinantal Monte Carlo [LB00; San10] as well as diffusion Monte Carlo techniques such as full-configuration-interaction Monte Carlo [BTA09]. Similar sign problems involving the sampling from quasi-probability distributions also appear in different contexts, for example, in approaches to the classical simulation of quantum circuits [Daw+05; JGL10; PWB15] or high-energy physics [AN02]. In these contexts, too, the problem of finding better bases in which to perform the sampling appears. While the framework developed in this work uses the specific features of QMC, the general idea and mindset behind it applies to all basis-dependent sign problems. Our work thus paves the way towards easing sign problems in a plethora of contexts.

Second, we have only considered real-valued Hamiltonians and transformations which preserve this property. For general complex-valued Hamiltonians, the sign problem takes the form of a *complex phase problem*. A natural follow-up of our work is to explore how our results on easing the sign problem generalize to the complex phase problem.

Third, we have put an emphasis on the conjugation of Hamiltonians under on-site Clifford and orthogonal circuits. In principle, one may also allow for arbitrary quasi-local circuits, as long as the conjugation can be efficiently computed; albeit of exponentially increasing effort with the support of the involved unitaries. This leads to the interesting insight that within the trivial phase of matter, one can always remove the sign problem: One has to conjugate the Hamiltonian with the quasi-local unitary that brings a given Hamiltonian into an on-site form of a *fixed point Hamiltonian*. For given Hamiltonians, this may be impractical, of course. In this sense, one can identify trivial quantum phases of matter as *efficiently computable phases of matter*, an intriguing state of affairs from a conceptual perspective. Conversely, for topologically ordered systems, there may be topological obstructions to curing the sign problem by any quasi-local circuit [Has15], giving rise to an entire phase of matter that exhibits an intrinsic sign problem [RK17; SGR20; GSR20]. For example, the fixed point Hamiltonians of the most general class of non-chiral topologically ordered systems, the Levin-Wen models [LW05], are associated with 12-local Hamiltonians, many of which are expected to not be curable from their sign problem. This insight further motivates to study the sign easing problem for efficiently computable subgroups of local unitaries from a perspective of topological phases of matter.

Our work also opens up several paths for future research. The immediate and practically most relevant direction is of course to find the best possible way of minimizing the non-stoquasticity of translation-invariant systems and to explore how well the sign problem can be eased in systems that are not yet amenable to QMC. We have already introduce a flexible optimization approach which can be straightforwardly applied to a wide range of translation-invariant systems and ansatz classes in any dimensionality.

Furthermore, in our hardness proof we have shown that the easing problem is intricately related to satisfiability problems. Building on this connection, an exciting direction of research is to combine highly efficient SAT-solvers that are capable of exploring combinatorically large sets, with manifold optimization techniques that are able to handle rich geometrical structures, in the spirit of recent work [Sho+17]. While our hardness result shows up fundamental limitations of SignEasing in the general case, it thus also opens the door to potentially solve the sign easing problem in relevant instances by applying methods well known in computer science to relaxed versions of the easing problem. One may thus hope that for large classes of relevant instances for which minimizing non-stoquasticity is actually tractable.

A question closely related to the sign easing problem is the following: How hard is it to find the ground state energy of a stoquastic Hamiltonian

– a sub-problem of the so-called local Hamiltonian problem. The computational complexity of this *stoquastic local Hamiltonian problem* poses fundamental limitations on the classical simulatability of Hamiltonians which do not suffer from a sign problem and are therefore amenable to QMC simulations. It has been shown that the 2-local stoquastic Hamiltonian problem is complete for the class StoqMA [Bra+08; BT09], a class intermediate between AM and MA that also functions as a genuinely intermediate class in the complexity classification of local Hamiltonian problems [CM16], even when extending to the full low-energy spectrum [CMP18]. But the complexity of stoquastic local Hamiltonian problem not only limits QMC procedures: for example, it is also reflected in the improved runtime of quantum annealing algorithms that exploit non-stoquastic interaction [Hor+17], and finds applications in long-standing open classical complexity-theoretic questions [AB19].

Indeed, for efficiently curable Hamiltonians, the local Hamiltonian problem is reduced to a stoquastic local Hamiltonian problem. Conversely, both the easing problem and the stoquastic local Hamiltonian problem contribute to the hardness of a QMC procedure. For a given Hamiltonian QMC may thus be computationally intractable for two reasons: it is hard to find a basis in which the Hamiltonian is stoquastic, or cooling to its ground state is computationally hard in its own right. In a QMC algorithm, the latter hardness is manifested as a Markov chain Monte Carlo algorithm not converging in polynomial time. This may be the case even for classical models such as Ising spin glasses [Bar82]. An important open question is whether this connection runs deeper in that a Hamiltonian simulation can be intractable *only* due to the hardness of finding a suitable basis, or whether the hardness of classical simulation remains in the guise of a ground state problem.

# Conclusion 13

One of the most puzzling features of quantum theory is its intrinsic randomness: The outcome of any quantum experiment will be a random sample from a probability distribution that is determined by the quantum state and measurement. The generation of random samples is therefore a natural technological application of quantum devices.

In this thesis, we explored quantum sampling from a computational viewpoint and, vice versa, studied the computational manifestations of physical effects via the notion of sampling, asking the questions.

> *What is the computational complexity of nature?*
> *What is the physical nature of computational complexity in quantum theory?*

In order to study these questions, we took on different standpoints with the goal to illuminate the computational nature of nature from different angles and using a variety of methods.

In the first part of this thesis, we studied the computational complexity of randomly chosen quantum computations. We reviewed the complexity-theoretic foundations for the computational power of quantum sampling and the hardness of reproducing quantum sampling experiments on a classical computer. The key technical tool that allowed us to separate classical sampling algorithms from quantum ones in terms of their computational complexity was the approximate counting algorithm due to Stockmeyer [Sto83]. En route, we identified the *quantum sign problem* as a root of the computational complexity of quantum sampling: in order to express outcome probabilities of quantum circuits, signed functions are required while for outcome probabilities of classical randomized computation Boolean functions are sufficient. This could be traced back to the possibility of destructive interference in quantum theory. Vice versa, the argument shows that quantum sampling from randomly chosen computations provides a meaningful way to experimentally violate the complexity-theoretic Church-Turing thesis and demonstrate so-called quantum supremacy over classical computation.

We then worked to further close the gaps or loopholes in the complexity-theoretic argument for hardness, which required certain structural assumptions, called *anticoncentration*, on the sampled probability distribution and average-case hardness conjectures for the task of *approximating outcome probabilities* of quantum circuits (Chapter 3). We reviewed strategies for how to prove exact average-case hardness of computing the output probabilities of quantum computations by means of polynomial interpolation and extrapolation, and discussed why this argument fails when it comes to *approximate average-case hardness* as would be required for a loophole-free proof of quantum supremacy Closing a further loophole for a wide range of quantum supremacy schemes, including universal circuit sampling, we proved anticoncentration for circuit families that form a 2-design. We sketched a 'recipe' for how a variety of such circuit families give rise to sampling-based quantum supremacy schemes with a rigorous complexity-theoretic underpinning.

In Chapter 4, we made the link back to experiments in physics by developing quantum random sampling schemes that are tailored to specific quantum simulation architectures. In particular, we developed a scheme tailored to optical-lattice architectures in which local control is expensive but large numbers of particles evolve coherently. Making use of a space-time tradeoff, the proposed scheme involves the constant time evolution of a translation-invariant Hamiltonian as well as single-qubit rotations and measurements.

In the second part of the thesis, we turned to the issue of certification: How, if at all, can quantum random sampling be verified in the absence of an efficient classical resimulation algorithm? We proved that this task is impossible using few (polynomially many) samples from the target probability distribution (Chapter 6).

We then saw different means to practically circumvent this no-go result classically by making different kinds of assumptions (Chapter 7). In particular, we developed fully efficient verification schemes for universal quantum computations in what we called the measurement-device-dependent setting. In this setting, single-qubit operations are required to function reliably with high accuracy. We proved that a simple fidelity witness gives rise to a full quantum state certificate from few samples. Even more so, the fidelity of a restricted class of quantum states can be estimated rapidly – in constant sample complexity. My discussion of the different routes of circumventing the no-go result as an obstacle to verification highlights that in the absence of error correction, verifying quantum devices will require custom-tailored techniques. Only by optimally exploiting the specifics of an architecture or a scheme at hand can the resources required for verification be made manageable.

In the third and final part of this thesis we turned to investigating the quantum-classical divide, as demarcated by the quantum sign problem, from the classical side. We discussed various methods to classically generate samples from nonuniform probability distributions, as well as Quantum Monte Carlo methods that use random samples in order to estimate properties of quantum systems. The quantum sign problem finds it direct practical manifestation in Monte Carlo methods.

> *In which cases, though, is the sign problem intrinsic to a physical problems and when is it a mere artifact of our description?*

In Chapters 10, 11 and 12 we studied this question from a computational viewpoint, observing that the Monte Carlo sign problem is a basis-dependent property. We explored the relation between the sign problem of a Monte Carlo simulation and the nonstoquasticity of a Hamiltonian matrix and proposed a measure of nonstoquasticity as an efficiently computable, generic measure of the sign problem. We showed that this nonstoquasticity measure can be optimized in practice and demonstrated fundamental limitations of any such approach by proving NP hardness of the SignEasing task.

Not only do a number of open questions remain unanswered but the results in this thesis also raise new questions. We discussed some of those questions in Chapter 8 and Section 12.4. Let me therefore close this thesis by mentioning broad themes only.

Coming from a foundational perspective, a central theme raised in this thesis is the fundamental nature of the quantum sign problem. The notion of easing, which we introduced here, opens the door for the study of manifestations of the sign problem in different contexts, seeking answers to questions such as: Is the intuitive connection between destructive interference and the sign problem one that can be made rigorous? How does the quantum sign problem relate to negativity of quasiprobability distributions? What are meaningful choices of ansatz classes to ease phase-space negativity?

A large chunk of this work was devoted to closing loopholes for demonstrations of quantum supremacy using quantum random sampling. The results on closing the complexity-theoretic loopholes as well as on the verification loophole constitute first landmarks in a largely uncharted landscape. After the first demonstration of quantum supremacy by Arute et al. [Aru+19], it will be important to map out the complexity-theoretic foundations of quantum random sampling more clearly. As quantum computers grow, the question of verification will become even more pressing. I hope that the discussion of verification in this thesis provides a useful guideline for future research into efficient verification methods.

What is certain is that exciting times lie ahead of us. As classical simulation algorithms on the one hand and quantum devices on the other hand advance to new frontiers of computing, we will surely discover new facets of the question asked in this thesis.

*What is the complexity of nature?*

# Appendices

# Acknowledgements — A

I am grateful to Jens Eisert for creating a unique research environment in Dahlem in which I could flourish – an environment that is as inspiring as it is diverse, that brings together people, colleagues, friends like no other, that is nourished and sustained by Jens's enduring enthusiasm and optimism. Working together with Jens has been joy and challenge, a challenge that made me grow and learn the art of research.

Thank you, Jens.

Research is a joint endeavour and over the course of my PhD I have had the pleasure to work together with a number of collaborators both in Dahlem and around the world – Martin Kliesch, Martin Schwarz, Juani Bermejo-Vega, Robert Raussendorf, Christian Gogolin, Ingo Roth, Daniel Nagaj, Jonas Haferkamp, Marek Gluza, Nathan Walk, Damian Markham, Rhea Parekh, Ulysse Chabaud, Elham Kashefi, Adam Bouland, Bill Fefferman, Jacques Carolan, Karim Thébault, and of course Jens Eisert. I am grateful for the hours we spent together, the many discussions we had, and the inspiration you gave me.

I want to especially thank Ingo. For our coffee breaks, our after-work beers, our brainstorming, the many hours in discussion and conversation. Your inspiring way of doing research, the next idea just around the corner, has often kept me up and running. Equally so, I want to thank Juani for her unwavering support, for the ride we have taken together in the past years, and for the pink. From both of you I learned complementing ways of doing research. While working together with you was not always easy, it has enriched my time in Dahlem, made it colourful and worthwhile. Thank you!

To the two Martins – Martin Kliesch and Martin Schwarz – I am grateful for teaching me so much in the beginning of my PhD. Both of you have uniquely shaped my way of thinking.

My time in Dahlem would not have been the same without the wonderful people gathered there, friends with whom I could have snowball fights as well as long walks, colleagues at whose office I could drop in whenever I needed distraction and have Veggie Mensa lunch with. I am particularly grateful to Frederik with whom I could share my thoughts, fold paper stars and improve my paper basketball skills. Frederik made coming into the office something to look forward to every day. Thank you. My thanks also go to Alex and the ObstOffice, Marek, Jonas, Marcel, Ingo, Paul, Christian, Nathan and Jadwiga.

For comments on a draft of this thesis I am grateful to Nathan, Marcel, Jonas and Juani, especially so to Tobias, who is the only person who made it through the entire draft. Thank you for your endurance. I am also grateful for the Latex and Tikz Stackexchange communities for providing an answer to each and every one of my questions.

Writing this thesis in the times of isolation would not have been possible without the support of my flatmates – Aştî, Carlotta, Charlotte, Leon and Nicole. Thank you for the table tennis sessions and many hours of Qwixx and Set that helped me switch off my mind.

Nicht fehlen dürfen auf dieser Liste auch meine Eltern ohne die ich nicht wäre, wo ich bin und wer ich bin, deren Unterstützung ich mir zu jedem Zeitpunkt immer sicher sein konnte, die mir diesen Weg ermöglicht haben. Danke.

Mein großes Glück entdeckte ich mit Nicole. Zusammen mit Dir zu leben und lieben, Gedanken zu teilen, die Welt zu erkunden, macht mein Leben bunt und reich, voller Freude und Liebe, und gibt Raum für Wut, Ärger und Traurigkeit. Auch das war für diese Arbeit unverzichtlich. Danke.

*Dominik Hangleiter*
*Berlin, May 2020*

# Details and proofs

# D

## D.1 Sample complexity of device-independently certified quantum samplers

In the following, we (re)state some facts and slightly improved earlier results on verifying quantum random sampling schemes.

*This section has already been published as the appendix of Ref. [Han+19], which is joint work with Martin Kliesch, Jens Eisert and Christian Gogolin.*

### Proofs for bounding the min-entropy

Here, we provide some details and proofs to statements made in Section 4. First, we show the equivalence of the Rényi entropies (6.24) proceeding analogously to Ref. [Wil+19]: we simply use that for $\alpha > 1$ and $p_0 = \|P\|_\infty$ we have $p_0^\alpha \leq \sum_i p_i^\alpha$. Hence,

$$\frac{\alpha}{\alpha-1}\log(p_0) \leq \frac{1}{\alpha-1}\log\sum_{i=0}^{|\mathscr{E}_n|-1} p_i^\alpha \tag{D.1}$$

$$\Leftrightarrow -\frac{\alpha}{\alpha-1}H_\infty(P) \leq -H_\alpha(P) \tag{D.2}$$

$$\Leftrightarrow H_\infty(P) \geq \frac{\alpha-1}{\alpha}H_\alpha(P)\,. \tag{D.3}$$

We also provide an alternative proof of Lemma 6.4 based on the proof of Ref. [Gog+13, Theorem 13].

*Alternative proof of Lemma 6.4.* We begin the proof by noting that

$$\Pr_{U\sim\mu_n}\left[H_\infty(P_U) \leq \log\frac{1}{\delta}\right] = \Pr_{U\sim\mu_n}\left[\exists S \in \mathscr{E}_n : P_U(S) \geq \delta\right]. \tag{D.4}$$

Using the union bound (also known as Boole's inequality) we obtain that for every $\delta > 0$

$$\Pr_{U\sim\mu_n}\left[\exists S \in \mathscr{E}_n : P_U(S) \geq \delta\right] \leq \sum_{S\in\mathscr{E}_n}\Pr_{U\sim\mu_n}\left[P_U(S) \geq \delta\right]. \tag{D.5}$$

Next, using Markov's inequality we can bound

$$\Pr_{U\sim\mu_n}\left[P_U(S) \geq \delta\right] \leq \frac{1}{\delta^2}\mathbb{E}_{U\sim\mu_n}\left[P_U(S)^2\right], \tag{D.6}$$

which concludes the proof. □

### Probability weight of the collision-free subspace

We recapitulate a bound of Aaronson and Arkhipov [AA13] on the probability weight of the collision-free subspace.

**Lemma D.1** ([AA13, Theorem 13.4]) *Let $\mu_H$ be the Haar measure on $U(m)$ and $m \geq n$. Then*

$$\mathbb{E}_{U\sim\mu_H}\left[P_{\mathrm{bs},U}(\Phi_{m,n} \setminus \Phi^*_{m,n})\right] \leq \frac{2n^2}{m}\,. \tag{D.7}$$

### The min-entropy bound for boson sampling

Here, we provide a slightly improved proof of the following min-entropy bound for boson sampling from [Gog+13, Theorem 12].

**Theorem D.2** (Min-entropy bound for boson sampling [Gog+13, Theorem 12]) *Let $\nu > 3$ and assume that the assertion* (6.39) *of Lemma 6.7 holds. Then, the boson sampling output distribution $P_{\mathrm{bs},U}$ satisfies for $n$ bosons in $m \in \Theta(n^\nu)$ modes*

$$\Pr_{U\sim\mu_H}\left[H_\infty(P_{\mathrm{bs},U}) < 2\,n\right] \in \exp\left(-\Omega(n^{\nu-2-1/n})\right). \tag{D.8}$$

The proof crucially uses the closeness of the Gaussian measure to the post-selected Haar measure as expressed by Lemma 6.7. Lemma 6.7, however, is not quite strong enough for proving Theorem 6.8, as we must be able to control all of $\Phi_{m,n}$ and not only the collision-free subspace $\Phi^*_{m,n}$. Fortunately, the above lemma extends naturally to all $S \in \Phi_{m,n}$ for the same scaling of $m$ with $n$ for which a version of Lemma 6.7 holds.

To state this extension we need some notation first: For every sequence $S$, let $\tilde{S}$ be the sequence obtained from $S$ by removing all the zeros, i.e,

$$\tilde{S} = (\tilde{s}_1, \dots, \tilde{s}_{|\tilde{S}|}) := (s \in S : s > 0). \tag{D.9}$$

Further, let $\mu_{G_S(\sigma)}$ be the probability measure on $\mathbb{C}^{n\times n}$ obtained by drawing the real and imaginary part of every entry of a $|\tilde{S}| \times n$ matrix independently from a Gaussian distribution with mean zero and standard deviation $\sigma$ and then for all $j \in [|\tilde{S}|]$ taking $\tilde{s}_j$ copies of the $j^{\text{th}}$ row of this matrix. We can prove the following multiplicative error bound on the closeness of this measure and the Haar measure $\mu_H$ for all $S \in \Phi_{m,n}$:

**Lemma D.3** (Multiplicative error bound) *Let $f : \mathbb{C}^{n\times n} \to [0,1]$ be measurable, then for any $m, n$ such that*

$$\forall S \in \Phi^*_{m,n} : \quad \mathbb{E}_{U\sim\mu_H} f(U_S) \leq (1 + C)\, \mathbb{E}_{X\sim\mu_{G(1/\sqrt{m})}} f(X), \tag{D.10}$$

*is true for some constant $C > 0$, it holds that*

$$\forall S \in \Phi_{m,n} : \quad \mathbb{E}_{U\sim\mu_H} f(U_S) \leq (1 + C)\, \mathbb{E}_{X\sim\mu_{G_S(1/\sqrt{m})}} f(X). \tag{D.11}$$

*Proof.* Let $S \in \Phi_{m,n}$, define $\tilde{S}$ as in Eq. (D.9) and $m' := |\tilde{S}|$. Define $v$ to be the sequence containing $\tilde{s}_j$ times the integer $j$ for every $j \in [m']$ in increasing order and $w$ the sequence containing the positions of each of the first of the repeated rows in $U_S$, i.e.,

$$v := (\underbrace{1,\dots,1}_{\tilde{s}_1},\underbrace{2,\dots,2}_{\tilde{s}_2},\dots,\underbrace{m',\dots,m'}_{\tilde{s}_{m'}}) \in (\mathbb{Z}^+)^n, \tag{D.12}$$

$$w := (1, 1+\tilde{s}_1, 1+\tilde{s}_1+\tilde{s}_2, \dots, 1+\sum_{j=1}^{m'-1}\tilde{s}_j) \in (\mathbb{Z}^+)^{m'}. \tag{D.13}$$

The sequence $v$ defines a linear embedding $\eta : \mathbb{C}^{m'\times n} \to \mathbb{C}^{n\times n}$ component wise by

$$\eta(Y)_{i,j} := Y_{v_i,j} \quad \forall i,j \in [n], \tag{D.14}$$

i.e., $\eta(Y)$ has $s_j$ copies of the $j$-th row of $Y$. The sequence $w$, in turn, defines a linear projection $\pi : \mathbb{C}^{n\times n} \to \mathbb{C}^{m'\times n}$ by

$$\pi(X)_{i,j} := X_{w_i,j} \quad \forall i \in [m'],\ j \in [n], \tag{D.15}$$

in particular, $\pi(U_S)$ contains only the first out of each series of the repeated rows in $U_S$. Note that $\eta \circ \pi : \mathbb{C}^{n\times n} \to \mathbb{C}^{n\times n}$ is a projection onto the subspace of matrices that have the same repetition structure as $U_S$. Let

$$f_S := f \circ \eta \circ \pi, \tag{D.16}$$

then $f_S(U_S) = f(U_S)$ only depends on the first of the repeated rows in $U_S$ and is independent of all the other rows. Since the Haar measure is permutation-invariant we have

$$\mathbb{E}_{U\sim\mu_H} f_S(U_S) = \mathbb{E}_{U\sim\mu_H} f_S(U_{1_n}). \tag{D.17}$$

Hence, using Lemma 6.7 in the second step, we obtain

$$\begin{aligned}\mathbb{E}_{U\sim\mu_H} f(U_S) &= \mathbb{E}_{U\sim\mu_H} f_S(U_{1_n}) && \text{(D.18)}\\ &\le (1+C)\,\mathbb{E}_{X\sim\mu_{G(1/\sqrt{m})}} f_S(X) && \text{(D.19)}\\ &= (1+C)\,\mathbb{E}_{X\sim\mu_{G_S(1/\sqrt{m})}} f(X), && \text{(D.20)}\end{aligned}$$

which finishes the proof. □

In addition to the multiplicative error bound we need the following concentration result for the Gaussian measure $\mu_{G_S(\sigma)}$, which implies that even the largest entry of a matrix drawn from $\mu_{G_S(\sigma)}$ is unlikely to be much larger than $\sigma$.

**Lemma D.4** (Concentration of the Gaussian measure $\mu_{G_S(\sigma)}$) *For all $n, m \in \mathbb{Z}^+$, all $S \in \Phi_{m,n}$ and all $\xi > 0$ it holds that*

$$\Pr_{X\sim\mu_{G_S(\sigma)}}\left[\max_{j,k\in[n]} |x_{j,k}| \ge \xi\right] \le 1 - \left(1 - \operatorname{Erfc}\left(\frac{\xi}{\sqrt{2}\,\sigma}\right)\right)^{n^2}, \tag{D.21}$$

*where*

$$\operatorname{Erfc}\left(\frac{\xi}{\sqrt{2}\,\sigma}\right) := 2\int_{\xi}^{\infty} \frac{\mathrm{e}^{-\frac{x^2}{2\sigma^2}}}{\sqrt{2\pi\sigma^2}}\,\mathrm{d}x \tag{D.22}$$

*is the complementary error function.*

*Proof.* For Gaussian random variables we have

$$\forall \xi > 0,\ j,k \in [n]: \quad \Pr_{X \sim \mu_{G(\sigma)}} \left[ |x_{j,k}| \geq \xi \right] = \operatorname{Erfc}\left( \frac{\xi}{\sqrt{2}\,\sigma} \right). \tag{D.23}$$

This implies that

$$\forall \xi > 0: \quad \Pr_{X \sim \mu_{G(\sigma)}} \left[ \forall j,k \in [n] : |x_{j,k}| \leq \xi \right] = \left( 1 - \operatorname{Erfc}\left( \frac{\xi}{\sqrt{2}\,\sigma} \right) \right)^{n^2}. \tag{D.24}$$

At the same time, for all $S \in \Phi_{m,n}$ and $\xi > 0$ it holds that

$$\Pr_{X \sim \mu_{G_S(\sigma)}} \left[ \forall j,k \in [n] : |x_{j,k}| \leq \xi \right] \geq \Pr_{X \sim \mu_{G(\sigma)}} \left[ \forall j,k \in [n] : |x_{j,k}| \leq \xi \right], \tag{D.25}$$

because the repetition of entries in $X \sim \mu_{G_S(\sigma)}$ only increases the chance of not having an exceptionally large entry. □

As a last ingredient we need to bound the size

$$|\Phi_{m,n}| = \binom{m+n-1}{n} \tag{D.26}$$

of the sample space $\Phi_{m,n}$ of boson sampling (recall Eq. (6.34)). It grows faster than than exponentially with $n$, but if for some $\nu \geq 1$ and $c \geq 0$ it holds that $m \leq c\, n^{\nu}$, then

$$|\Phi_{m,n}| \leq \frac{(m+n-1)^n}{n!} \leq \left( \frac{(m+n-1)\,\mathrm{e}}{n} \right)^n \tag{D.27}$$

$$\leq \mathrm{e}^n\, (c\, n^{\nu-1} + 1 - 1/n)^n \leq (2\,(c+1)\,\mathrm{e})^n\, n^{(\nu-1)\,n}. \tag{D.28}$$

We now have all the ingredients rederive the desired min-entropy bound in Theorem D.2.

*Proof of Theorem D.2.* Using the union bound (also known as Boole's inequality) in the first step we obtain that for every $\epsilon > 0$

$$\Pr_{U \sim \mu_H} \left[ \exists S \in \Phi_{m,n} : P_{\mathrm{bs},U}(S) \geq \epsilon \right] \tag{D.29}$$

$$\leq \sum_{S \in \Phi_{m,n}} \Pr_{U \sim \mu_H} \left[ P_{\mathrm{bs},U}(S) \geq \epsilon \right] \tag{D.30}$$

$$\leq |\Phi_{m,n}| \max_{S \in \Phi_{m,n}} \Pr_{U \sim \mu_H} \left[ P_{\mathrm{bs},U}(S) \geq \epsilon \right] \tag{D.31}$$

$$= |\Phi_{m,n}| \max_{S \in \Phi_{m,n}} \Pr_{U \sim \mu_H} \left[ \frac{|\operatorname{Perm}(U_S)|^2}{\prod_{j=1}^{m} (s_j!)} \geq \epsilon \right]. \tag{D.32}$$

We now apply Lemma D.3 to the indicator function

$$f(U_S) = \begin{cases} 1 & \text{if } \frac{|\operatorname{Perm}(U_S)|^2}{\prod_{j=1}^{m}(s_j!)} \geq \epsilon \\ 0 & \text{otherwise} \end{cases}, \tag{D.33}$$

and the $S$ for which the maximum in Eq. (D.32) is attained, to obtain

$$\Pr_{U\sim\mu_H}\left[\exists S\in\Phi_{m,n} : P_{\mathrm{bs},U}(S)\geq\epsilon\right]$$
$$\leq (1+C)\,|\Phi_{m,n}|\max_{S\in\Phi_{m,n}}\Pr_{X\sim\mu_{G_S(1/\sqrt{m})}}\left[\frac{|\operatorname{Perm}(X)|^2}{\prod_{j=1}^m(s_j!)}\geq\epsilon\right]. \tag{D.34}$$

The definition of the permanent (recall Eq. (6.37)) implies that

$$\frac{|\operatorname{Perm}(X)|^2}{\prod_{j=1}^m(s_j!)}\leq|\operatorname{Perm}(X)|^2\leq(n!)^2\left(\max_{j,k\in[n]}|x_{j,k}|\right)^{2n}. \tag{D.35}$$

Hence, for every $S\in\Phi_{m,n}$ and every $\epsilon>0$

$$\Pr_{X\sim\mu_{G_S(1/\sqrt{m})}}\left[\frac{|\operatorname{Perm}(X)|^2}{\prod_{j=1}^m(s_j!)}\geq\epsilon\right]\leq\Pr_{X\sim\mu_{G_S(1/\sqrt{m})}}\left[\max_{j,k\in[n]}|x_{j,k}|\geq\left(\frac{\sqrt{\epsilon}}{n!}\right)^{1/n}\right]. \tag{D.36}$$

Plugging this into Eq. (D.34), using Lemma D.4 with $\xi=\left(\sqrt{\epsilon}/n!\right)^{1/n}$ and the bound on $|\Phi_{m,n}|$ from Eq. (D.28) we arrive at

$$\Pr_{U\sim\mu_H}\left[\exists S\in\Phi_{m,n} : P_{\mathrm{bs},U}(S)\geq\epsilon\right]$$
$$\leq(1+C)\,(2\,(c+1)\,\mathrm{e})^n\,n^{(\nu-1)n}\left(1-\left(1-\operatorname{Erfc}\sqrt{\frac{c}{2}\frac{\epsilon^{1/n}\,n^\nu}{(n!)^{2/n}}}\right)^{n^2}\right). \tag{D.37}$$

Bounding the complementary error function by [EH04]

$$\operatorname{Erfc}(x)\leq\mathrm{e}^{-x^2}, \tag{D.38}$$

we obtain

$$1-(1-\operatorname{Erfc}(x))^{n^2}\leq1-\left(1-\mathrm{e}^{-x^2}\right)^{n^2}=1-\sum_{k=0}^{n^2}\binom{n^2}{k}(-\mathrm{e}^{-x^2})^k \tag{D.39}$$
$$=\sum_{k=1}^{n^2}\binom{n^2}{k}\mathrm{e}^{-x^2k}\,(-1)^{k-1}\leq\sum_{k=1}^{n^2}(n^2\mathrm{e}/k)^k\,\mathrm{e}^{-x^2k} \tag{D.40}$$
$$\leq\sum_{k=1}^{n^2}(n^2\,\mathrm{e}^{-x^2+1})^k. \tag{D.41}$$

If $x$ is large enough such that

$$n^2\,\mathrm{e}^{-x^2+1}\leq\frac{1}{2}<1, \tag{D.42}$$

the geometric series in Eq. (D.41) converges and we get the simple bound

$$1-(1-\operatorname{Erfc}(x))^{n^2}\leq\sum_{k=1}^{n^2}(n^2\mathrm{e}^{-x^2+1})^k\leq\frac{n^2\mathrm{e}^{-x^2+1}}{1-n^2\mathrm{e}^{-x^2+1}}\leq2\,n^2\,\mathrm{e}^{-x^2+1}. \tag{D.43}$$

To satisfy Eq. (D.42) for large $n$, it is sufficient that $x$ grows slightly faster than $\sqrt{\log(n^2)}$ and we hence need to demand a growth slightly faster than $\log(n^2)$ from the argument of the square root in the error function in Eq. (D.37). Because of the bound $n!\leq\mathrm{e}^{1-n}\,n^{n+1/2}$ (a variant of Stirling's approximation) we have for the argument of that square root in Eq. (D.37)

$$\frac{c}{2}\frac{\epsilon^{1/n} n^{\nu}}{(n!)^{2/n}} \geq \frac{c}{2}\frac{\epsilon^{1/n} n^{\nu}}{\mathrm{e}^{2/n-2} n^{2+1/n}} = \frac{c}{2}\frac{\epsilon^{1/n}}{\mathrm{e}^{2/n-2}} n^{\nu-2-1/n}, \tag{D.44}$$

Demanding $\nu > 2$ is hence all we need to be able to use the bound (D.43) for large $n$. With the convenient choice $\epsilon = 2^{-2n}$ it hence follows that for all $\nu > 2$

$$\Pr_{U\sim\mu_H}\left[\exists S \in \Phi_{m,n} : P_{\mathrm{bs},U}(S) \geq 2^{-2n}\right]$$
$$\in O\left(n^2\,(2\,(c+1)\,\mathrm{e})^n\, n^{(\nu-1)n} \exp(-c\,\mathrm{e}^{-2/n+2}\, n^{\nu-2-1/n}/8)\right). \tag{D.45}$$

The argument of the $O(\cdot)$ is dominated by the product

$$n^{(\nu-1)n} \exp\left(-c\,\mathrm{e}^{-2/n+2}\, n^{\nu-2-1/n}/8\right), \tag{D.46}$$

which decays for large increasing $n$ only for $\nu > 3$. More precisely, there are constants $n_0 \in \mathbb{N}$ and $C_1, C_2, C_3 > 0$ such that for $n \geq n_0$

$$n^2\,(2\,(c+1)\,\mathrm{e})^n\, n^{(\nu-1)n} \exp(-c\,\mathrm{e}^{-2/n+2}\, n^{\nu-2-1/n}/8) \tag{D.47}$$
$$= \exp\big(2\ln(n) + n\ln(2\,(c+1)\,\mathrm{e}) + n(\nu-1)\ln n \tag{D.48}$$
$$- c\,\mathrm{e}^{-2/n+2}\, n^{\nu-2-1/n}/8\big) \tag{D.49}$$
$$\leq \exp\left(C_1 n(\nu-1)\ln n - c\,\mathrm{e}^{-2/n+2}\, n^{\nu-2-1/n}/8)\right) \tag{D.50}$$
$$\leq \exp\left(C_1 n(\nu-1)\ln n - C_2 n^{\nu-2-1/n}\right) \tag{D.51}$$
$$\stackrel{\nu>3}{\leq} \exp\left(-C_3 n^{\nu-2-1/n}\right) \in \exp\left(-\Omega(n^{\nu-2-1/n})\right). \tag{D.52}$$

where the last inequality holds only for $\nu > 3$ since the logarithm grows slower than any power law with positive exponent. This completes the proof. □

## D.2 Easing the sign problem of translation-invariant Hamiltonians

### Derivation of the translation-invariant formulation of $\nu_1$

*The following has been published as part of Appendix A in Ref. [Han+20].*

To derive Eq. (11.3) we re-express the global nonstoquasticity measure as follows:

$$\nu_1(H) = \sum_{\substack{(i_1,\dots,i_n)\neq\\(j_1,\dots,j_n)}} \max\left\{0, \langle i_1,\dots,i_n| \sum_{l=1}^{n} T_l(h)|j_1,\dots,j_n\rangle\right\} \tag{D.53}$$

$$= \sum_{p=1}^{n} \sum_{\substack{i_1,\dots,i_n,\\ j_p,j_{p+1}:\\ i_{p+1}\neq j_{p+1}}} \max\left\{0, \langle i_1,\dots,i_n|(T_p(h)+ + T_{p+1}(h))|i_1,\dots,i_{p-1},j_p,j_{p+1},i_{p+2},\dots,i_n\rangle\right\} \tag{D.54}$$

$$= 2^{n-3} \sum_{p=1}^{n} \sum_{\substack{i_p,\dots,i_{p+2},\\ j_p,j_{p+1}:\\ i_{p+1}\neq j_{p+1}}} \max\left\{0, \langle i_p,\dots,i_{p+2}|(h\otimes 1 + 1\otimes h)|j_p,j_{p+1},i_{p+2}\rangle\right\} \tag{D.55}$$

$$= n 2^{n-3} \sum_{\substack{i_1,\dots,i_3,\\ j_1,j_2:\\ i_2\neq j_2}} \max\left\{0, \langle i_1,i_2,i_3|(h\otimes 1 + 1\otimes h)|j_1,j_2,i_3\rangle\right\}. \tag{D.56}$$

In the first step, we have used that the condition $(i_1,\dots,i_n)\neq(j_1,\dots,j_n)$ implies that at least one of the indices differs. Summing over $p$, we can let this index be the $(p+1)^{\text{st}}$ one. To avoid double-counting, we then divide the sum over all strings which differ on at most two nearest neighbours into a sum over all strings which potentially differ on two nearest neighbours *left* of the $(p+1)^{\text{st}}$ index. Patches which differ on more than two nearest neighbours vanish for nearest-neighbour Hamiltonians. In the second step, we use that all terms with support left of the $p^{\text{th}}$ qubit vanish, since the basis strings differ on the $(p+1)^{\text{st}}$ qubit. In the last step, we use the translation-invariance again to account for the sum over $p$, incurring a factor $n$.

## Gradient of the objective function

As an ansatz class we choose on-site orthogonal transformations in $O(d)$, where $d$ is the dimension of a local constituent of the system. More precisely, we consider transformations of the type (see Eq. (11.2))

$$H = \sum_{i=1}^{n} T_i(h) \mapsto O^{\otimes n} H (O^T)^{\otimes n}, \tag{D.57}$$

which locally amounts to

$$h \mapsto h(O) := (O\otimes O)h(O^T\otimes O^T). \tag{D.58}$$

The key ingredient for the conjugate gradient descent algorithm is the derivative of the objective function $\tilde{\nu}_1(h(O))$ with respect to the orthogonal matrix $O$. The gradient is given by

$$\frac{\partial}{\partial O}\tilde{\nu}_1(h(O)) = \sum_{i,j}\frac{\partial \tilde{\nu}_1}{\partial h(i|j)}\cdot\frac{\partial h(i|j)}{\partial O}. \tag{D.59}$$

We can further expand the terms of the global gradient (D.59): in particular, we can express the gradient of the effective local terms as a conjugation $h(O) = \mathscr{C}(O)h\mathscr{C}^\dagger(O)$ of the local term $h$ by the orthorgonal circuit $\mathscr{C}(O) = O \otimes O$. We will now derive expressions for the measure and the gradients that the algorithm has to evaluate.

**The objective function gradient** We first determine the gradient of the objective function. Since we will also make use of different measures $\nu_p$ as defined in Eq. (10.22), we write the objective function for different $p \geq 1$ as

$$\tilde{\nu}_p^p(h) = \sum_{\substack{i,j,k;l,m,n:\\ k\neq l, m=n}} \max\left\{(h\otimes 1 + 1\otimes h)_{i,k,m;j,l,n}\,,0\right\}^p. \tag{D.60}$$

Then, the gradient of the objective function is given by

$$\left.\frac{\partial \tilde{\nu}_p^p}{\partial h(x|y)}\right|_{h(0)} = p\sum_{\substack{i,j,k;l,m,n:\\ k\neq l, m=n}} \left(|x\rangle\langle y|\otimes 1 + 1\otimes |x\rangle\langle y|\right)_{i,k,m;j,l,n}\cdot$$
$$\cdot\max\left\{(h\otimes 1 + 1\otimes h)_{i,k,m;j,l,n}\,,0\right\}^{p-1} \tag{D.61}$$

**Gradient of the transformed Hamiltonian** We can expand the gradient of the transformed Hamiltonian term by the orthogonal matrix as

$$\frac{\partial h(i|j)}{\partial O} = \sum_{m,n}\frac{\partial\,\mathrm{adj}_h(\mathscr{C})(i|j)}{\partial\mathscr{C}(m|n)}\frac{\partial\mathscr{C}(m|n)}{\partial O,} \tag{D.62}$$

where by $\mathrm{adj}_h(\mathscr{C})$ we denote the adjunction map $h \mapsto \mathscr{C}h\mathscr{C}^T$. The derivative of the adjoint action of $\mathscr{C}$ on $h$ is given by

$$\frac{\partial\,\mathrm{adj}_h(\mathscr{C})(i|j)}{\partial\mathscr{C}(m|n)} = \sum_{k,l}\frac{\partial}{\partial\mathscr{C}(m|n)}\mathscr{C}(i|k)h(k|l)\mathscr{C}(j|l) \tag{D.63}$$
$$= \sum_{k,l}\left[\delta_{m,i}\delta_{n,k}h(k|l)\mathscr{C}(j|l) + \mathscr{C}(i|k)h(k|l)\delta_{m,j}\delta_{n,l}\right] \tag{D.64}$$
$$= \langle m|i\rangle\langle j|\mathscr{C}h^T|n\rangle + \langle m|j\rangle\langle i|\mathscr{C}h|n\rangle \tag{D.65}$$

From this expression, we can directly read off its matrix form

$$\frac{\partial\,\mathrm{ad}_h(\mathscr{C})(i|j)}{\partial\mathscr{C}} = |i\rangle\langle j|\mathscr{C}h^T + |j\rangle\langle i|\mathscr{C}h. \tag{D.66}$$

It remains to compute the gradient of the circuit with respect to the orthogonal matrix. We obtain with $\langle m| = \langle m_1| \otimes \langle m_2|$

$$\frac{\partial \mathscr{C}(m|n)}{\partial O(k|l)} = \sum_{i=1}^{2} \delta_{m_i,k}\delta_{n_i,l} O(m_1|n_1)O(m_2|n_2) \tag{D.67}$$

$$= \sum_{i=1}^{2} \langle m_i|k\rangle\langle l|n_i\rangle O(m_1|n_1)O(m_2|n_2), \tag{D.68}$$

which expressed in matrix form is then given by

$$\frac{\partial \mathscr{C}_{l_e}(m|n)}{\partial O} = |n_1\rangle\langle m_1|\langle m_2|O|n_2\rangle + |n_2\rangle\langle m_2|\langle m_1|O|n_1\rangle. \tag{D.69}$$

## D.3 Orthogonal transformations of the penalty terms (proof of Theorem 12.4)

*Proof of Theorem 12.4 (continued).* As in the proof for orthogonal Clifford transformations, we will show that for any given choice of $Z$-transformations one cannot further decrease the non-stoquasticity by exploiting the additional freedom offered by the full orthogonal group. Above, we have argued that applying an arbitrary orthogonal transformation at a single site can be reduced to applying $R(\theta)X^p Z^{p+z}$ with $\theta \in [-\pi/4, \pi/4]$ and $z, p \in \{0, 1\}$. We will now show that for any choice of $\vec{z}$ and $\vec{p}$, a rotation by angles $\vec{\theta} \in [-\pi/4, \pi/4]^n$ cannot decrease the non-stoquasticity any further.

Analogously to the proof for Clifford-transformations, we discuss all possible transformations by dividing them into different cases. In each case the non-stoquasticity of an uncured edge $(i, j)$ and its ancilla qubit $a_{i,j}$ cannot be eased below its previous value of 1. The additional difficulty we encouter here is that the orthogonal group is continuous as opposed to the orthogonal Clifford group, which is a discrete and rather 'small' group.

Given a choice of $\vec{z}$, consider an edge $(i, j)$ with a non-trivial contribution to $\nu_1$ and its corresponding ancilla qubit $a_{i,j}$. We begin, supposing that $p_i = p_j = p_{a_{i,j}} = 0$ so that the $X$-flips act trivially on all three qubits. We now analyze the effect of the remaining rotations $R(\theta)$ on each of the qubits.

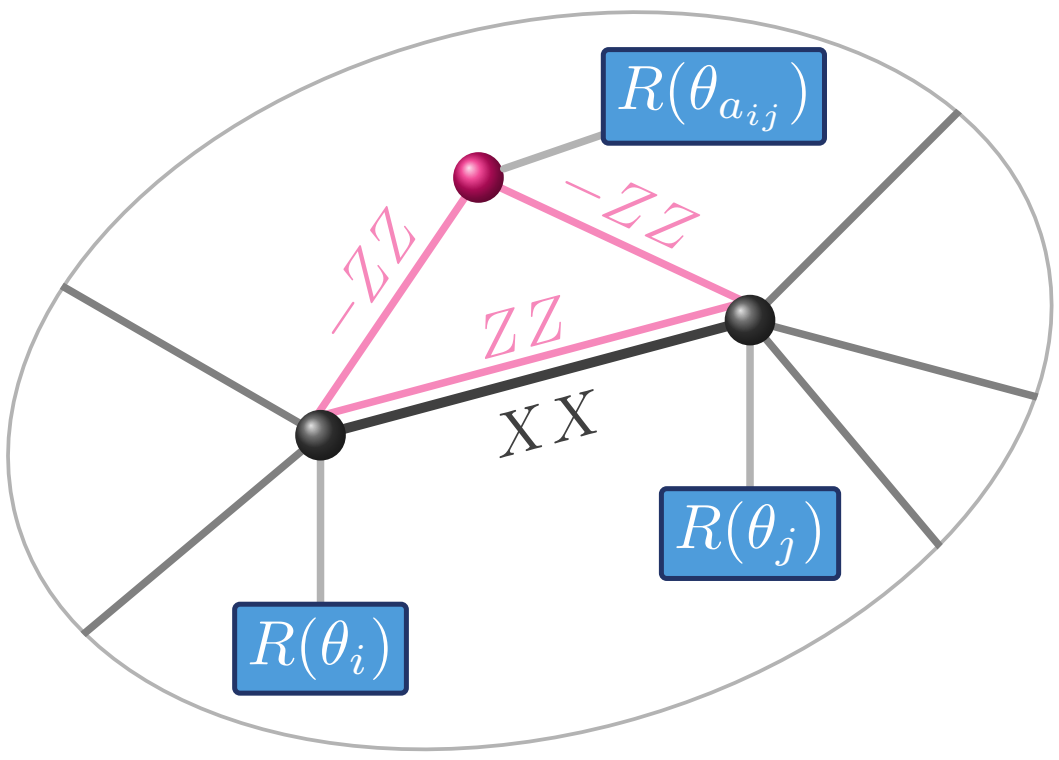

Specifically, we apply rotations with angles $\theta_i/2$, $\theta_j/2$, $\theta_{a_{i,j}}/2$ with $\theta_i, \theta_j, \theta_{a_{i,j}} \in [-\pi/2, \pi/2]$ to the three qubits. Note that we consider rotations by half-angles $\theta \longrightarrow \theta/2$ while at the same time doubling the interval $[-\pi/4, \pi/4] \rightarrow [-\pi/2, \pi/2]$ to ease notation later in the proof. The effect of rotations on each vertex of an edge $(i, j)$ is given by

$$
\begin{aligned}
X_i X_j + C Z_i Z_j \mapsto \\
&[C \cos(\theta_i)\cos(\theta_j) - \sin(\theta_i)\sin(\theta_j)] Z_i Z_j \\
&+ [C \cos(\theta_i)\sin(\theta_j) - \sin(\theta_i)\cos(\theta_j)] Z_i X_j \\
&+ [C \sin(\theta_i)\cos(\theta_j) - \cos(\theta_i)\sin(\theta_j)] X_i Z_j \\
&+ [C \sin(\theta_i)\sin(\theta_j) + \cos(\theta_i)\cos(\theta_j)] X_i X_j,
\end{aligned} \tag{D.70}
$$

and likewise for the edges $(i, a_{i,j})$ and $(j, a_{i,j})$. The non-stoquasticity of the transformed Hamiltonian terms corresponding to the three qubits is then given by

$$
\nu_1\Big( R_i(\theta_i/2) R_j(\theta_j/2) R_{a_{i,j}}(\theta_{a_{i,j}}/2) \big( X_i X_j \tag{D.71}
$$

$$
+ C(Z_i Z_j - Z_i Z_{a_{i,j}} - Z_j Z_{a_{i,j}}) \big) R_i(\theta_i/2)^T R_j(\theta_j/2)^T R_{a_{i,j}}(\theta_{a_{i,j}}/2)^T \Big) \tag{D.72}
$$

$$
\geq \max\left\{ C \sin(\theta_i)\sin(\theta_j) + \cos(\theta_i)\cos(\theta_j), 0 \right\} \tag{D.73}
$$

$$
\begin{aligned}
&+ (2 \deg G')^{-1} \cdot \Big( |C \sin(\theta_i)\cos(\theta_j) - \cos(\theta_i)\sin(\theta_j)| \\
&\quad + |C \sin(\theta_j)\cos(\theta_i) - \cos(\theta_j)\sin(\theta_i)| \Big)
\end{aligned} \tag{D.74}
$$

$$
+ \Big( \max\left\{ -C \sin(\theta_i)\sin(\theta_{a_{i,j}}), 0 \right\} + \max\left\{ -C \sin(\theta_j)\sin(\theta_{a_{i,j}}), 0 \right\} \Big) \tag{D.75}
$$

$$
\begin{aligned}
&+ (2 \deg G')^{-1} \cdot C\Big( |\cos(\theta_i)\sin(\theta_{a_{i,j}})| + |\sin(\theta_i)\cos(\theta_{a_{i,j}})| \\
&\quad + |\cos(\theta_j)\sin(\theta_{a_{i,j}})| + |\sin(\theta_j)\cos(\theta_{a_{i,j}})| \Big).
\end{aligned} \tag{D.76}
$$

Note that the terms (D.73) and (D.75) stem from the $XX$ interactions with a positive sign and therefore depend on the signs of sin terms. Conversely, the terms (D.74) and (D.76) stem from the $XZ$ interactions and therefore involve absolute values.

We divide the allowed rotations into different sectors corresponding to the different combinations of the signs of $\sin(\theta_i)$ and $\sin(\theta_j)$ as shown in Fig. D.1. Which of the terms in (D.73) and (D.75) are non-trivial precisely depends on these combinations. We divide the cases as follows: first, the sectors in which $\operatorname{sign}(\theta_i) = \operatorname{sign}(\theta_j)$. Second, the sectors in which $\operatorname{sign}(\theta_i) = -\operatorname{sign}(\theta_j)$ and $|\theta_i| + |\theta_j| \leq \pi/2$. Third, the sectors in which $\operatorname{sign}(\theta_i) = -\operatorname{sign}(\theta_j)$ and $|\theta_i| + |\theta_j| \geq \pi/2$. Taken together, the three cases cover the entire range of allowed angles. Moreover, in all cases we allow arbitrary choices of $\theta_{a_{i,j}} \in [-\pi/2, \pi/2]$. We now proceed to lower-bound the non-stoquasticity of the Hamiltonian terms acting on the three qubits, where in each case we will use different terms of Eqs. (D.73)-(D.76).

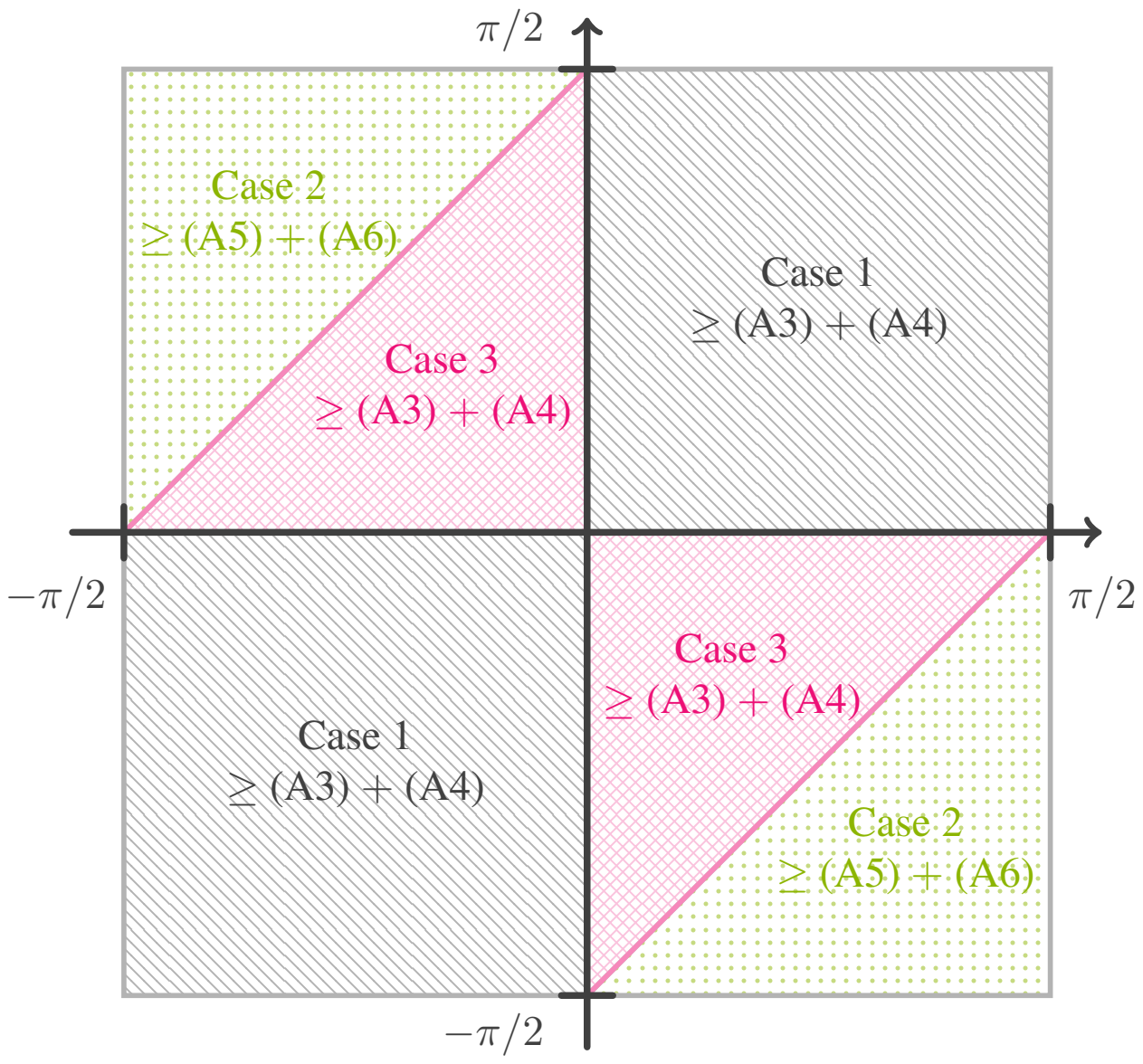

**Figure D.1:** We divide the derivation of a lower bound on the non-stoquasticity of an edge $(i, j)$ and its ancilla qubit $a_{i,j}$ on which we apply rotations $R_i(\theta_i/2)R_j(\theta_j/2)R_{a_{i,j}}(\theta_{a_{i,j}}/2)$ with $\theta_i, \theta_j, \theta_{a_{i,j}} \in [\pi/2, \pi/2]$ into three distinct cases. In each case we use different terms of the expression (D.73)-(D.76) to lower bound the non-stoquasticity.

We first discuss the case in which $\operatorname{sign}(\theta_i) = \operatorname{sign}(\theta_j)$. In this case, it suffices to consider the terms (D.73) and (D.74). Observe that in this case both $C\sin(\theta_i)\sin(\theta_j) \geq 0$ and $\cos(\theta_i)\cos(\theta_j) \geq 0$. Moreover, to our choice of $C$ and noting that $(|C\sin(\theta_i)\cos(\theta_j) - \cos(\theta_i)\sin(\theta_j)| + |C\sin(\theta_j)\cos(\theta_i) - \cos(\theta_j)\sin(\theta_i)|) \geq C|\sin(\theta_i - \theta_j)|$, we obtain

$$\text{(D.73)} + \text{(D.74)} \geq \cos(\theta_i - \theta_j) + C/(2\deg G') \cdot |\sin(\theta_i - \theta_j)| \geq 1. \tag{D.77}$$

Second, we discuss the case in which $\operatorname{sign}(\theta_i) = -\operatorname{sign}(\theta_j)$ and $\pi/2 < |\theta_i| + |\theta_j| \leq \pi$. In this case, we consider the terms (D.75) and (D.76):

$$\text{(D.75)} + \text{(D.76)} \geq \; C/(2\deg G') \cdot \Big( |\cos(\theta_{a_{i,j}})|(\sin(|\theta_i|) + \sin(|\theta_j|))$$
$$+|\sin(\theta_{a_{i,j}})| \left(2\deg(G')\min\{\sin(|\theta_i|), \sin(|\theta_j|)\} + |\cos(\theta_i)| + |\cos(\theta_j)|\right) \Big)$$
$$\geq 1. \tag{D.78}$$

Here, we have used that

$$\sin(|\theta_i|) + \sin(|\theta_j|) = 2\sin\left(\frac{|x| + |y|}{2}\right)\cos\left(\frac{|x| - |y|}{2}\right) \geq 1, \tag{D.79}$$

for $\pi/2 \leq |\theta_i| + |\theta_j| \leq \pi$ so that $|x| - |y| \leq \pi/2$ and the definition of $C = (2\deg G')^2$.

Finally, we have the remaining case $\operatorname{sign}(\theta_i) = -\operatorname{sign}(\theta_j)$ and $|\theta_i| + |\theta_j| \leq \pi/2$. In this case, it is again sufficient to consider terms (D.73) and (D.74). This is the hardest case since the sin terms in (D.73) increase much faster than the cos terms decrease due to the factor of $C > 1$. Therefore, we cannot find a bound in terms of a sum-of-angles rule as in the previous cases. Instead, to lower-bound the terms terms (D.73) and (D.74) in this case, we proceed as follows: We reduce the problem of minimizing the sum (D.73) + (D.74) $\geq 1$ to a one-dimensional problem by noting two facts: first, the term (D.74) depends only on the sum $|\theta_i| + |\theta_j|$. Moreover,

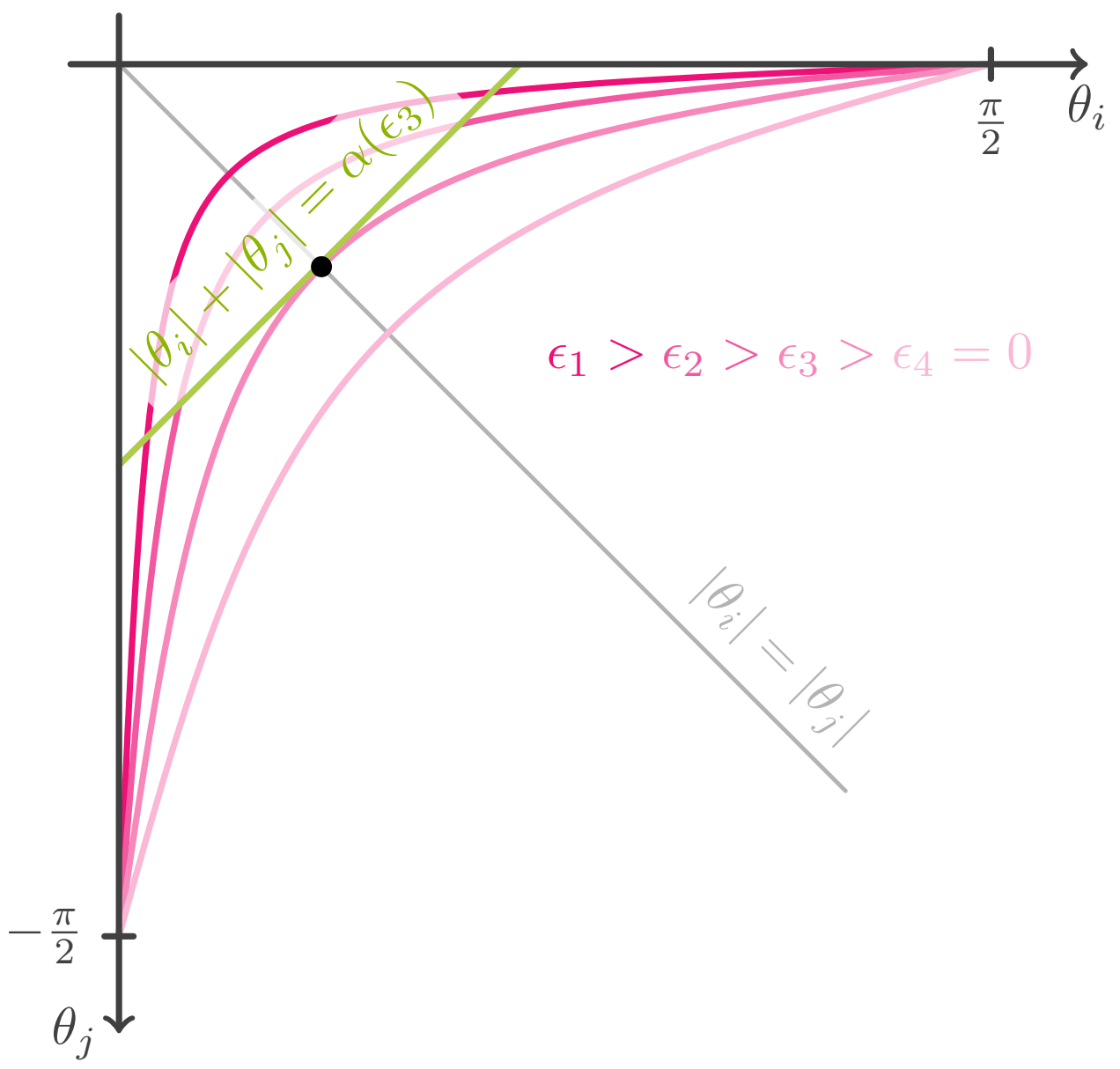

**Figure D.2:** Illustration of the lower bound in case 3: Every $\theta_i$ and $\theta_j$ such that $|\theta_i| + |\theta_j| \leq \pi/2$ defines a contour line $\epsilon(\theta_i, \theta_j) = \epsilon_1, \epsilon_2, \epsilon_3, \epsilon_4 = 0$ (shades of pink). Given $\theta_i, \theta_j$ and defining $\overline{\epsilon} = \epsilon(\theta_i, \theta_j)$, Lemma D.5 implies the lower bound $|\theta_i| + |\theta_j| \geq \alpha(\overline{\epsilon})$ as defined in Eq. (D.90). We then obtain (D.73) + (D.74) $\geq \overline{\epsilon} + (C+1)\sin(\alpha(\overline{\epsilon})) \geq 1$.

it increases monotonously in this sum. Second, for every choice of $\theta_i, \theta_j$, the value of (D.73) takes on its minimal value $\overline{\epsilon}$ at $|\theta_i| = |\theta_j| =: \alpha(\overline{\epsilon})/2$. Hence, the sum (D.73) + (D.74) is lower bounded by the sum of $\overline{\epsilon}$ and (D.74) evaluated at $\alpha(\overline{\epsilon})$. The proof is concluded by a lower bound on the latter term. We now elaborate those steps one-by-one.

Let us begin by expressing (D.74) as a function of $|\theta_i| + |\theta_j|$

$$|C\sin(\theta_i)\cos(\theta_j) - \cos(\theta_i)\sin(\theta_j)| + |C\sin(\theta_j)\cos(\theta_i) - \cos(\theta_j)\sin(\theta_i)| \tag{D.80}$$

$$= C\sin(|\theta_i|)\cos(\theta_j) + \cos(\theta_i)\sin(|\theta_j|) + C\sin(|\theta_j|)\cos(\theta_i) + \cos(\theta_j)\sin(|\theta_i|) \tag{D.81}$$

$$= (C+1)\sin(|\theta_i| + |\theta_j|), \tag{D.82}$$

where we have used the fact that $\text{sign}(\sin(\theta_i)) = -\text{sign}(\sin(\theta_j))$ and that the cosines are non-negative. For $|\theta_i| + |\theta_j| \leq \pi/2$ this is a monotonously increasing function in $|\theta_i| + |\theta_j|$. We now define the value of the term (D.73) to be

$$\epsilon(\theta_i, \theta_j) := \max\{-C\sin|\theta_i|\sin|\theta_j| + \cos|\theta_i|\cos|\theta_j|, 0\} \geq 0. \tag{D.83}$$

For every choice of $\alpha = \alpha(\theta_i, \theta_j) := |\theta_i| + |\theta_j|$, the minimal value of

$$\text{(D.73)} + \text{(D.74)} = \epsilon(\theta_i, \theta_j) + (C+1)/(2\deg G') \cdot \sin(\alpha(\theta_i, \theta_j)), \tag{D.84}$$

is therefore attained at the minimal value of $\epsilon(\theta_i, \theta_j)$ subject to the constraint $|\theta_i| + |\theta_j| = \alpha \leq \pi/2$. Covnersely, the value is attained at the minimal value of $\alpha(\theta_i, \theta_j)$ subject to the constraint (D.83). This reduces the problem to a one dimensional problem, which we exploit explicitly in the following lemma. The intuition behind this lemma is shown in Fig. D.2.

**Lemma D.5** *For any fixed value $\pi/2 \geq |\theta_i| + |\theta_j| = \alpha \geq 0$, the minimal value $\overline{\epsilon}(\alpha)$ of $\epsilon(\theta_i, \theta_j)$ is achieved at $|\theta_i| = |\theta_j| = \alpha/2$. Moreover, for every $\theta_i, \theta_j$ such that $\epsilon(\theta_i, \theta_j) \geq \overline{\epsilon}(\alpha)$ it holds that $|\theta_i| + |\theta_j| \geq \alpha$.*

*Proof.* Let $|\theta_i| = (\alpha - \delta)/2$, $|\theta_j| = (\alpha + \delta)/2$ for $0 \leq \delta \leq \pi/2$. Then

$$\epsilon(\theta_i, \theta_j) = -C \sin\left(\frac{\alpha-\delta}{2}\right) \sin\left(\frac{\alpha+\delta}{2}\right) + \cos\left(\frac{\alpha+\delta}{2}\right) \cos\left(\frac{\alpha-\delta}{2}\right) \tag{D.85}$$

$$= \frac{C}{2}(\cos\alpha - \cos\delta) + \frac{1}{2}(\cos\alpha + \cos\delta) \tag{D.86}$$

$$= \frac{1}{2}\left((C+1)\cos\alpha - (C-1)\cos\delta\right), \tag{D.87}$$

which is minimal at $\delta = 0$.

The second part of the lemma can be be seen by contraposition: Assume $|\theta_i| + |\theta_j| < \alpha(\overline{\epsilon})$. Then

$$\epsilon(\theta_i, \theta_j) = \frac{1}{2}\left((C+1)\cos(|\theta_i| + |\theta_j|) - (C-1)\cos(|\theta_i| - |\theta_j|)\right) \tag{D.88}$$

$$\leq \frac{1}{2}(C+1)\cos(|\theta_i| + |\theta_j|) < \frac{1}{2}(C+1)\cos\alpha \equiv \overline{\epsilon}(\alpha), \tag{D.89}$$

where we have used the assumption and the monotonicity of the cosine in the interval $[0, \pi/2]$ in the last inequality and its non-negativity in the interval $[-\pi/2, \pi/2]$ in the second to last inequality. □

Now, for every choice of $\theta_i, \theta_j$, the minimal value of $\overline{\epsilon}(\alpha)$ of $\epsilon(\theta_i, \theta_j)$ corresponding to $\alpha = |\theta_i| + |\theta_j|$ is attained at $|\theta_i| = |\theta_j| = \alpha/2$. Correspondingly, we can re-express $\alpha$ in terms of $\overline{\epsilon}$ as

$$\overline{\epsilon} \equiv \overline{\epsilon}(\alpha) = -C\sin^2(\alpha/2) + \cos^2(\alpha/2) \quad \Leftrightarrow \quad \alpha(\overline{\epsilon}) = 2\arctan\sqrt{\frac{1-\overline{\epsilon}}{C+\overline{\epsilon}}}. \tag{D.90}$$

The second part of Lemma D.5 states that for all $\theta_i, \theta_j$ such that $\epsilon(\theta_i, \theta_j) \geq \overline{\epsilon}(\alpha) \geq 0$ we have $|\theta_i| + |\theta_j| \geq \alpha(\overline{\epsilon})$ and consequently $\sin(|\theta_i| + |\theta_j|) \geq \sin(\alpha(\overline{\epsilon}))$. Given $\theta_i, \theta_j$ and defining $\overline{\epsilon} := \epsilon(\theta_i, \theta_j)$ this implies the lower bound

$$(\text{D.73}) + (\text{D.74}) \geq \overline{\epsilon} + (C+1)/(2\deg G') \cdot \sin(\alpha(\overline{\epsilon})), \tag{D.91}$$

where we have used the equivalence (D.90).

It remains to lower-bound $\sin(\alpha(\overline{\epsilon}))$. Define $x = \sqrt{(1-\overline{\epsilon})/(C+\overline{\epsilon})}$. We can then rewrite

$$\sin(\alpha(\overline{\epsilon})) = \sin(2\arctan x) = 2\sin(\arctan x)\cos(\arctan x) = 2\frac{x}{1+x^2} \geq x, \tag{D.92}$$

for $x \leq 1$, where we have used that $\sin(\arctan(x)) = x\cos(\arctan x) = x/\sqrt{1+x^2}$. We can also bound

$$x = \sqrt{\frac{1-\overline{\epsilon}}{C+\overline{\epsilon}}} \geq \sqrt{\frac{1}{C}}\sqrt{\frac{1-\overline{\epsilon}}{1+\overline{\epsilon}/C}} \geq \sqrt{\frac{1}{C}}\sqrt{\frac{1-\overline{\epsilon}}{1+\overline{\epsilon}}} \geq \frac{1-\overline{\epsilon}}{\sqrt{C}}, \tag{D.93}$$

where the last inequality can be seen by squaring both sides and using

$0 \le \overline{\epsilon} < 1$. Combining everything we obtain

$$(D.73) + (D.74) \ge \overline{\epsilon} + \frac{\sqrt{C}}{2\deg G'} \cdot (1 - \overline{\epsilon}) \tag{D.94}$$

$$= \overline{\epsilon}\left(1 - \frac{\sqrt{C}}{2\deg G'}\right) + \frac{\sqrt{C}}{2\deg G'} \tag{D.95}$$

$$= 1. \tag{D.96}$$

due to our choice of $C = (2\deg G')^2$.

To conclude the proof, we discuss the effect of applying $X$-flips to each of the sites. Applying $X_i X_j$ or $X_{a_{i,j}}$ merely alters the signs of the terms in (D.75). But since we did not constrain the sign of $\theta_{a_{i,j}}$ in the proof, everything remains unchanged. Suppose an $X$-flip is applied to either qubit $i$ or $j$, or either both qubit $i$ and $a_{i,j}$ or either both qubit $j$ and $a_{i,j}$ . Assuming wlog. that qubit $i$ is $X$-flipped, we achieve the same lower bounds as before by identifying $\theta_i \mapsto -\theta_i$. □

# Summary | S

Randomness is an intrinsic feature of quantum theory. The outcome of any measurement will be random, sampled from a probability distribution that is defined by the measured quantum state. The task of sampling from a prescribed probability distribution therefore seems to be a natural technological application of quantum devices. And indeed, certain random sampling tasks have been proposed to experimentally demonstrate the speedup of quantum over classical computation, so-called "quantum computational supremacy".

In the research presented in this thesis, I investigate the complexity-theoretic and physical foundations of quantum sampling algorithms. Using the theory of computational complexity, I assess the computational power of natural quantum simulators and close loopholes in the complexity-theoretic argument for the classical intractability of quantum samplers (Part I). In particular, I prove anticoncentration for quantum circuit families that give rise to a 2-design and review methods for proving average-case hardness. I present quantum random sampling schemes that are tailored to large-scale quantum simulation hardware but at the same time rise up to the highest standard in terms of their complexity-theoretic underpinning. Using methods from property testing and quantum system identification, I shed light on the question, how and under which conditions quantum sampling devices can be tested or verified in regimes that are not simulable on classical computers (Part II). I present a no-go result that prevents efficient verification of quantum random sampling schemes as well as approaches using which this no-go result can be circumvented. In particular, I develop fully efficient verification protocols in what I call the measurement-device-dependent scenario in which single-qubit measurements are assumed to function with high accuracy. Finally, I try to understand the physical mechanisms governing the computational boundary between classical and quantum computing devices by challenging their computational power using tools from computational physics and the theory of computational complexity (Part III). I develop efficiently computable measures of the infamous Monte Carlo sign problem and assess those measures both in terms of their practicability as a tool for alleviating or easing the sign problem and the computational complexity of this task.

An overarching theme of the thesis is the quantum sign problem which arises due to destructive interference between paths – an intrinsically quantum effect. The (non-)existence of a sign problem takes on the role as a criterion which delineates the boundary between classical and quantum computing devices. I begin the thesis by identifying the quantum sign problem as a root of the computational intractability of quantum output probabilities. It turns out that the intricate structure of the probability distributions the sign problem gives rise to, prohibits their verification from few samples. In an ironic twist, I show that assessing the intrinsic sign problem of a quantum system is again an intractable problem.

# Zusammenfassung | Z

Ihre Zufälligkeit ist eine faszinierende Eigenschaft von quantenmechanischen Messungen: Das Ergebnis einer jeden solchen Messung ist zufällig, gezogen aus einer Verteilung, die durch den gemessenen Quantenzustand bestimmt ist. Insofern ist es kein Wunder, dass eine der naheliegendsten Anwendungen von Quantentechnologien im Allgemeinen und Quantenrechnern im Speziellen die Generierung von Samples aus einer vorgegebenen Zufallsverteilung ist. In der Tat, Samples aus bestimmten zufälligen Quantenrechnungen zu ziehen ist für den experimentellen Nachweis einer sogenannten Quantenüberlegenheit vorgeschlagen.

In der in dieser Dissertation vorgestellten Forschung beschäftige ich mich mit den komplexitätstheoretischen und physikalischen Grundlagen von Quantenzufallsalgorithmen. Unter Verwendung von Komplexitätstheorie bewerte ich die Rechenleistung von natürlichen Quantensimulatoren und schließe wichtige Hintertüren im komplexitätstheoretischen Beweis für die Schwierigkeit, Quantenzufallsgeneratoren klassisch zu simulieren (Teil I). Insbesondere beweise ich sogenannte 'Anticoncentration' von Familien von Quantenrechnungen die ein 2-Design bilden. Ich stelle neue Familien zufälliger Quantenrechnungen vor, die einerseits hocheffizient auf großskaliger Quantensimulationshardware ausgeführt werden können, und andererseits den höchsten komplexitätstheoretischen Standard erreichen. Mit Methoden aus den Gebieten der Eigenschaftstests und Quantensystemidentifikation beleuchte ich die Frage, wie und unter welchen Umständen Quantenzufallsgeneratoren in Regimes getestet oder verifiziert werden können, die nicht klassisch simulierbar sind (Teil II). Der Dreh- und Angelpunkt dieser Diskussion ist ein Beweis dafür, dass die effiziente Verifizierung von bestimmten zufälligen Quantenrechnungen unmöglich ist. Dennoch ergeben sich Schlupflöcher und Schleichwege: ich entwickle hocheffiziente Protokolle für die Verifzierung von Quantenrechnungen, die geeignet sind für Situationen, in denen einzelne Qubits mit sehr hoher Verlässlichkeit und Genauigkeit gemessen werden können. Schließlich versuche ich die physikalischen Mechanismen zu verstehen, die die Grenze zwischen Quantenrechnern und klassischen Computern bestimmen. Dazu fordere ich die Quantenrechner gewissermaßen heraus: mit klassischen Rechenmethoden, die auf dem Ziehen von Samples aus bestimmen Zufallsverteilungen basieren – sogenannten Monte-Carlo Algorithmen (Teil III). Ich entwickle effizient berechenbare Maße des sogenannten Vorzeichenproblems von Quanten-Monte-Carlo Algorithmen und bewerte diese Maße bezüglich ihrer Anwendbarkeit dafür, das Vorzeichenproblem abzuschwächen oder zu lindern, sowie der Komplexität dieser Aufgabe.

Ein roter Faden, der sich durch diese Arbeit zieht, ist das Vorzeichenproblem, das sich mit der destruktiven Interferenz zwischen möglichen Rechenpfaden erklären lässt – ein spezifisch quantenmechanischer Effekt. Die (Nicht-)Existenz eines Vorzeichenproblems übernimmt die Rolle eines Maßstabs, an dem wir die Grenze zwischen klassischen und quantischen Rechnungen festmachen können. Ich beginne die Arbeit damit, das Quantenvorzeichenproblem als Ursprung der Schwierigkeit zu identifizieren, die Ausgangswahrscheinlichkeiten zufälliger Quantenrechnungen zu berechnen. Es stellt sich heraus, dass die komplizierte Struktur der vom Vorzeichenproblem befallenen Ausgangswahrscheinlichkeiten ihre effiziente Verifzierung verhindert. In einer ironischen Volte zeige ich dann zum Schluss, dass das intrinsische Vorzeichenproblem überhaupt nur abzuschätzen nicht effizient möglich ist.

# C Author contributions

Diese Arbeit basiert auf den Publikationen [Han+17; Ber+18; Han+18; Han+19; EH+20; Han+20], sowie noch unveröffentlichten Arbeiten [Han+; BH], die in enger Zusammenarbeit mit anderen Wissenschaftlern entstanden sind. Im Folgenden führe ich meinen Anteil an der Konzeption, Durchfürhung und dem Verfassen dieser Publikationen auf.

[Han+17] D.H. war federführend bei dieser Arbeit und war maßgeblich für ihre Gestaltung. Er führte die Beweise und verfasste das Manuskript.

[Ber+18] D.H. war maßgeblich an der Konzeption und Herleitung der Hauptergebnisse, sowie der Formulierung des Manuskripts beteiligt. Insbesondere hat er die numerischen Rechnungen zur 'Anticoncentration' durchgeführt.

[Han+18] D.H. war federführend bei dieser Arbeit. Insbesondere hat er die Hauptergebnisse im ersten Teil er Arbeit hergeleitet, sowie wesentliche Teile des Manuskripts formuliert.

[Han+19] D.H. war federführend bei diesem Projekt. Er hat maßgeblich bei der Konzeption dieses Projekts, sowie der Herleitung der zentralen Ergebnisse beigetragen. Insbesondere hat er die Beweise der Hauptergebnisse ausgearbeitet, die Grafiken erstellt, sowie weite Teile des Manuskripts formuliert.

[EH+20] D.H. war gemeinsam mit J.E. federführend bei diesem Projekt. Insbesondere hat er das Klassifikationsschema für Zertifizierungsmethoden ausgearbeitet, weite Teile des Manuskripts formuliert, sowie maßgeblich die tabellarische Zusammenfassung verschiedener Methoden ausgearbeitet.

[Han+20] D.H. war federführend bei diesem Projekt und hat das Projekt maßgeblich gestaltet. Insbesondere hat er die Hauptergebnisse hergeleitet, die numerischen Rechnungen durchgeführt, die Grafiken erstellt, sowie weite Teile des Manuskripts verfasst.

[Han+] D.H. ist federführend bei diesem Projekt. Insbesondere hatte er die Beweisidee von Lemma 4.5 sowie die Abschätzungen zur Reduktion der experimentellen Anfordernisse auf ihn zurück.

[BH] D.H. ist maßgeblich an der Herleitung des Hauptergebnisses beteiligt.

# E Eigenständigkeitserklärung

Ich erkläre gegenüber der Freien Universität Berlin, dass ich die vorliegende Dissertation selbstständig und ohne Benutzung anderer als der angegebenen Quellen und Hilfsmittel angefertigt habe. Die vorliegende Arbeit ist frei von Plagiaten. Alle Ausführungen, die wörtlich oder inhaltlich aus anderen Schriften entnommen sind, habe ich als solche kenntlich gemacht. Diese Dissertation wurde in gleicher oder ähnlicher Form noch in keinem früheren Promotionsverfahren eingereicht.

..........................................................................

Dominik Hangleiter *Berlin, den 29. Mai 2020*

# Important concepts

## Complexity classes

P Decision problems that can be solved on a classical computer in polynomial time.

BPP Decision problems that can be solved on a probabilistic classical computer in polynomial time with bounded bounded error (Def. 2.3).

BQP Decision problems that can be solved on a quantum computer in polynomial time with bounded error (Def. 2.3).

NP Decision problems such that if the answer is 'YES' there exists a proof of this fact that can be verified in P (Def. 2.2).

PP Decision problems that accept if the fraction of accepting paths to an NP problem is larger or equal ot 1/2 (Def. 2.10).

#P Function problems that count the number of accepting paths to an NP problem (Def. 2.4).

GapP Function problems that count the difference between the number of accepting and rejecting paths to an NP problem, i.e., compute the *gap* of #P functions (Def. 2.5).

MA Decision problems for which an answer 'YES' can be verified in a one-round interactive proof in which a powerful prover (Merlin) sends a BPP verifier (Arthur) a polynomial-size proof based on which Arthur decides to accept or reject. Merlin can generate a proof which Arthur accepts with bounded failure probability if and only if the answer to the problem is 'YES'.

AM Decision problems for which an answer 'YES' can be verified in a two-round interactive proof in which a BPP verifier (Arthur) generates a 'challenge' based on the input, which he sends to Merlin, who sends back a response based on which Arthur decides to accept or reject. Merlin can generate a response which Arthur accepts with bounded failure probability if and only if the answer to the problem is 'YES'.

QMA Decision problems which can be verified by a one-message quantum interactive proof: if the answer is 'YES', there exists a quantum state (the proof) which a BQP machine accepts with bounded failure probability, while it rejects all states with bounded failure probability if the answer is 'NO'.

StoqMA Decision problems which can be verified by a one-message quantum interactive proof: if the answer is 'YES', there exists a quantum state (the proof) which a *stoquastic* BQP *machine* with access to the gates $X$, CNOT, TOF accepts with bounded failure probability, while it rejects all states with bounded failure probability if the answer is 'NO'.

## Distance measures

$\| \cdot \|_{\ell_p}$ The vector-$\ell_p$ norms for $0 < p < \infty$ and a vector $x = (x_1, \dots, x_d) \in \mathbb{C}^d$ are defined as (2.78)

$$\|x\|_{\ell_p} := \left( \sum_{i=1}^{d} |x_i|^p \right)^{1/p} .$$

$\|\cdot\|_{\mathrm{TV}}$ The *total-variation distance* between two probability distributions $p, q : \Omega \longrightarrow [0,1]$ on a sample space $\Omega$ is defined as (2.79)

$$\|p - q\|_{TV} = \frac{1}{2}\|p - q\|_{\ell_1}.$$

$\|\cdot\|_p$ The Schatten-$p$ norms of an operator $X$ are defined as (2.76)

$$\|X\|_p := (\operatorname{Tr}|X|^p)^{1/p},$$

where $|\cdot| : X \mapsto |X| = \sqrt{X^\dagger X}$ denotes the matrix absolute value.

$F(\rho,\sigma)$ The *fidelity* between two quantum states $\rho, \sigma$ is given by

$$F(\rho,\sigma) = \operatorname{Tr}\left[\left(\sqrt{\sigma}\rho\sqrt{\sigma}\right)^{1/2}\right]^2. \tag{1}$$

In the special case that either $\rho$ or $\sigma$ is a pure state, the fidelity reduces to the overlap $F(\rho,\sigma) = \operatorname{Tr}[\rho\sigma]$ between the two states. The fidelity relates to the trace distance via (7.46)

$$1 - \sqrt{F(\rho,\sigma)} \le \frac{1}{2}\|\rho - \sigma\|_1 \le \sqrt{1 - F(\rho,\sigma)}. \tag{2}$$

$\|\cdot\|_\diamond$ The *diamond norm* of a quantum channel $\mathscr{E} : L(\mathscr{A}) \longrightarrow L(\mathscr{B})$ mapping from the linear operators on a complex vector space $\mathscr{A}$ to those on a space $\mathscr{B}$ is defined as the stabilized $(1 \longrightarrow 1)$-norm (defined in Eq. (3.9))

$$\|\mathscr{E}\|_\diamond = \|\mathscr{E} \otimes 1_{L(\mathscr{A})}\|_{1\to 1}.$$

## Asymptotics

$O$ Asymptotic upper bound (Big $O$). For two functions $f, g$ mapping to $\mathbb{R}$,

$$f \in O(g) \Leftrightarrow \lim \sup_{x\to\infty} |f(x)|/g(x) < \infty.$$

$\Omega$ Asymptotic lower bound (Big $\Omega$. For two functions $f, g$ mapping to $\mathbb{R}$,

$$f \in \Omega(g) \Leftrightarrow \lim \inf_{x\to\infty} f(x)/g(x) > 0.$$

$\Theta$ Asymptotic upper and lower bound (Big $\Theta$). For two functions $f, g$ mapping to $\mathbb{R}$,

$$f \in \Theta(g) \Leftrightarrow f \in O(g) \wedge f \in \Omega(g).$$

$o$ Asymptotically dominated (Small $o$). For two functions $f, g$ mapping to $\mathbb{R}$,

$$f \in O(g) \Leftrightarrow \lim_{x\to\infty} f(x)/g(x) = 0.$$

$\omega$ Asymptotically dominating (Small $\omega$). For two functions $f, g$ mapping to $\mathbb{R}$,

$$f \in \omega(g) \Leftrightarrow \lim_{x\to\infty} |f(x)/g(x)| = \infty.$$

$\sim$ Scales as: For two functions $f, g$ mapping to $\mathbb{R}$,

$$f \sim g \Leftrightarrow \lim_{x \to \infty} f(x)/g(x) = 1.$$

$\tilde{O}, \tilde{\Omega}, \tilde{\Theta}$ denote the Landau-big-$O$ symbols up to log-factors.